\documentclass[titlepage,numbers=noenddot,headinclude,twoside,openright,
    bibliography=totoc,paper=17cm:24cm,fontsize=10pt,cleardoublepage=empty,
    abstractoff,footinclude=true,dutch,english]{scrreprt}

\usepackage[utf8]{inputenc}
\usepackage[dutch,english]{babel}

\ifdefined\printversion
    \usepackage[colorlinks=false,hidelinks]{hyperref}
\else
    \usepackage[colorlinks,linkcolor=MidnightBlue,urlcolor=Fuchsia,citecolor=OliveGreen]{hyperref}
\fi

\usepackage[beramono,parts,manychapters,dottedtoc]{classicthesis}

\ifdefined\printversion
    \usepackage[verbose,tmargin=2.5cm,bmargin=2.5cm,lmargin=2.0cm,rmargin=2.0cm,footskip=1.2cm]{geometry}
\else
    \usepackage[verbose,tmargin=2.5cm,bmargin=2.5cm,lmargin=2.0cm,rmargin=2.0cm]{geometry}
\fi

\titleformat{\chapter}[display]
    {\relax}{\vspace*{-3\baselineskip}\makebox[\linewidth][r]{\color{halfgray}\chapterNumber\thechapter}}{10pt}{\raggedright\spacedallcaps}[\normalsize\vspace*{.8\baselineskip}\titlerule] 

\AtBeginDocument{} 
\usepackage{etoolbox}
\pretocmd{\oldpart}{\cleardoublepage}{}{} 

\clearscrheadfoot 
\rofoot[\pagemark]{\pagemark}
\lefoot[\pagemark]{\pagemark}

\usepackage{amsmath}
\usepackage{amssymb}
\usepackage{graphicx}

\usepackage{color}

\usepackage{slashed}
\usepackage{braket}

\usepackage{empheq}
\usepackage{multirow}
\usepackage{adjustbox}
\usepackage{subcaption}
\usepackage[font=small]{caption} 

\usepackage{epigraph} 
\usepackage{verbatim}
\usepackage{url}

\usepackage{stackengine} 

\usepackage{cite}
\usepackage{bibentry}
\nobibliography* 

\newcommand*{\GtrSim}{\smallrel\gtrsim}
\newcommand*{\LessSim}{\smallrel\lesssim}
\makeatletter
\newcommand*{\smallrel}[2][.8]{%
  \mathrel{\mathpalette{\smallrel@{#1}}{#2}}%
}
\newcommand*{\smallrel@}[3]{%
  \sbox0{$#2\vcenter{}$}%
  \dimen@=\ht0 %
  \raise\dimen@\hbox{%
    \scalebox{#1}{%
      \raise-\dimen@\hbox{$#2#3\m@th$}%
    }%
  }%
}
\makeatother

\usepackage[normalem]{ulem} 
\renewcommand\sout{\bgroup \color[rgb]{0.59,0.44,0.84} \ULdepth=-.5ex \ULset} 

\usepackage[topthumbmargin=5.1cm,eventxtindent=10.4pt,oddtxtexdent=7pt]{thumbs} 
\def\thumbbg{lightgray} 
\def\thumbfg{white}

\usepackage{feynmp-auto}
\usepackage{array}

\newcolumntype{C}{>{\centering\arraybackslash}p{.22\textwidth}}
\newcolumntype{L}{>{\raggedright\arraybackslash}p{.20\textwidth}}
\newcommand{\mybox}[1]{\makebox[\dimexpr.84\textwidth+\columnsep\relax]{\centering #1}}

\usepackage{longtable}
\usepackage{arydshln}

\newcommand{\marrow}[5]{%
    \fmfcmd{style_def marrow#1
    expr p = drawarrow subpath (1/4, 3/4) of p shifted 6 #2 withpen pencircle scaled 0.4;
    label.#3(btex #4 etex, point 0.5 of p shifted 6 #2);
    enddef;}
    \fmf{marrow#1,tension=0}{#5}}

\newcommand{\Marrow}[6]{%
    \fmfcmd{style_def marrow#1
    expr p = drawarrow subpath (1/4, 3/4) of p shifted #6 #2 withpen pencircle scaled 0.4;
    label.#3(btex #4 etex, point 0.5 of p shifted #6 #2);
    enddef;}
    \fmf{marrow#1,tension=0}{#5}}

\newcommand{\p}{\partial}
\newcommand{\bm}{\boldsymbol}

\newcommand{\sst}{{\scriptscriptstyle T}} 

\def\nn{\nonumber}
\def\cd{{\cdot}}

\DeclareMathOperator{\tr}{Tr}

\makeatletter 
\DeclareOldFontCommand{\rm}{\normalfont\rmfamily}{\mathrm}
\DeclareOldFontCommand{\sf}{\normalfont\sffamily}{\mathsf}
\DeclareOldFontCommand{\tt}{\normalfont\ttfamily}{\mathtt}
\DeclareOldFontCommand{\bf}{\normalfont\bfseries}{\mathbf}
\DeclareOldFontCommand{\it}{\normalfont\itshape}{\mathit}
\DeclareOldFontCommand{\sl}{\normalfont\slshape}{\@nomath\sl}
\DeclareOldFontCommand{\sc}{\normalfont\scshape}{\@nomath\sc}
\makeatother

\setcapindent{0pt} 
\allowdisplaybreaks
\numberwithin{equation}{chapter} 
\numberwithin{figure}{chapter}
\numberwithin{table}{chapter}
\begin{document}

\begin{fmffile}{z}
\fmfset{dot_size}{1thick}
\fmfset{curly_len}{1.5mm}
\fmfset{arrow_len}{3mm}
\fmfpen{thin}

\selectlanguage{english}
\pagenumbering{roman}

\thispagestyle{empty}

\vspace*{2.5cm}

\begin{center}

\textbf{\huge Mapping the internal structure of\\[5pt] hadrons through color and spin effects}

\vspace{2cm}

{\Large Tom van Daal}

\end{center}

\newpage
\vspace*{2.5cm} 

\begin{tabular}{l c l}
\textbf{ISBN:} & & 978-94-034-1172-9 (printed version), \\
& & 978-94-034-1171-2 (electronic version) \\
& & \\
\textbf{Printed by:} & & Gildeprint - Enschede \\
& & \\
\textbf{Front cover:} & & ArSciMed/Science Photo Library (image credit), \\
& & Roy van den Heuvel (design)
\end{tabular}

\vfill

\begin{figure}[!htb]
    \centering
    \includegraphics[width=0.35\textwidth]{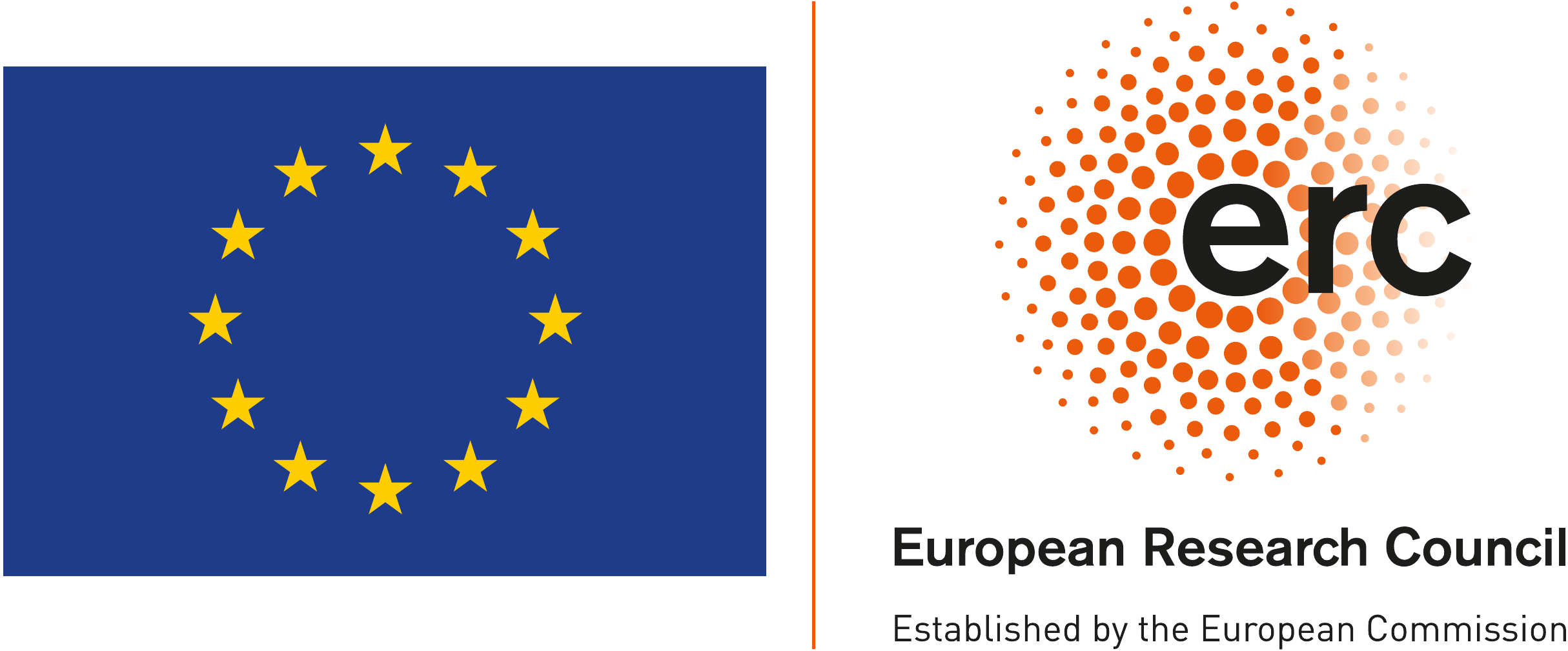}
\end{figure}

\begin{figure}[!htb]
\minipage{0.28\textwidth}
  \includegraphics[width=\linewidth]{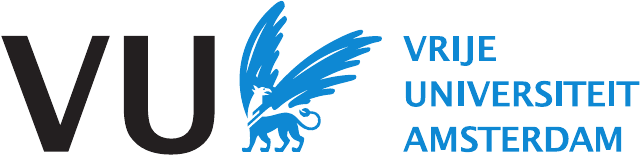}
\endminipage \hfill
\minipage{0.28\textwidth}
  \includegraphics[width=\linewidth]{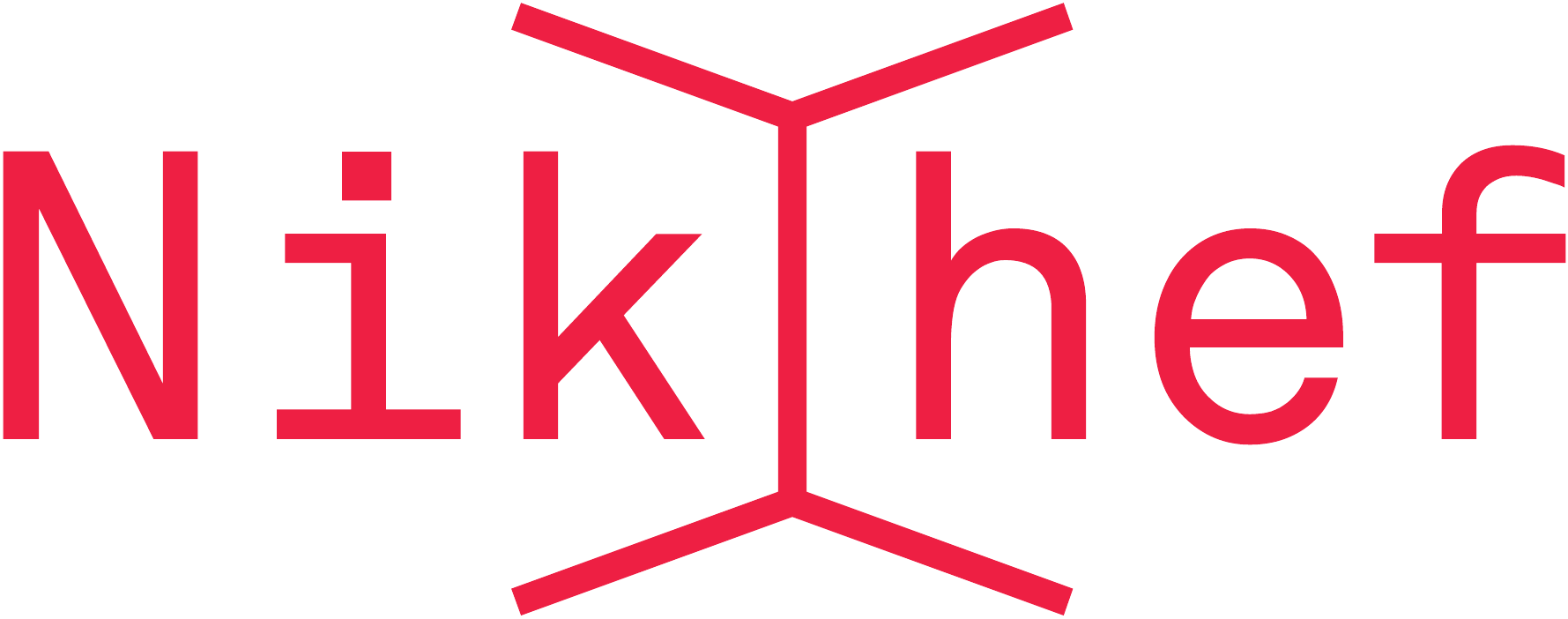}
\endminipage \hfill
\minipage{0.28\textwidth}
  \includegraphics[width=\linewidth]{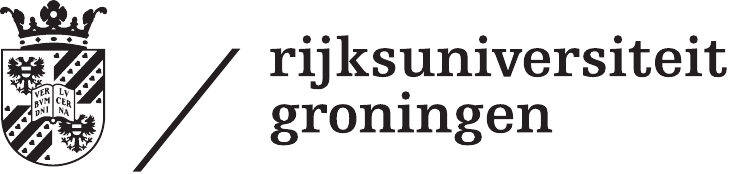}
\endminipage \hfill
\end{figure}

\noindent This work has received funding from the European Research Council (ERC) under the European Union's Seventh Framework Program (grant agreement no. 320389). 

\clearpage 

\selectlanguage{dutch}
\thispagestyle{empty}

\begin{figure}[htb]
\begin{flushleft}
    \includegraphics[height=1.75cm]{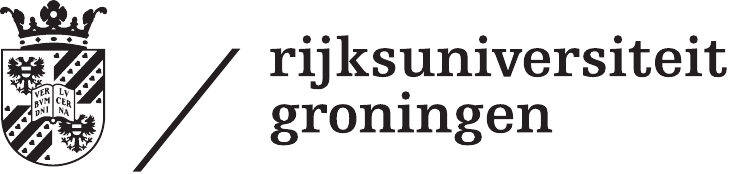}
\end{flushleft}
\end{figure}

\vspace{0.3cm}

\begin{center}

\textbf{\LARGE Mapping the internal structure of hadrons\\[5pt] through color and spin effects}

\vfill

\textbf{Proefschrift}

\vfill

ter verkrijging van de graad van doctor aan de\\
Rijksuniversiteit Groningen\\
op gezag van de\\
rector magnificus prof.\ dr.\ E.\ Sterken\\
en volgens besluit van het College voor Promoties.\\ $ $\\

De openbare verdediging zal plaatsvinden op\\ $ $\\

donderdag 13 december 2018 om 12.45 uur

\vfill

door

\vfill

\textbf{Tom Adrianus Antonius van Daal}

\vfill
\vspace{-0.5cm}

geboren op 25 oktober 1989\\
te Nijmegen

\end{center}

\newpage
\thispagestyle{empty}

\noindent
\textbf{Promotores:} \\
Prof.\ dr.\ D.\ Boer\\
Prof.\ dr.\ P.J.G.\ Mulders\\ \\

\noindent
\textbf{Beoordelingscommissie:} \\
Prof.\ dr.\ A.\ Sch\"afer\\
Prof.\ dr.\ I.\ Scimemi\\
Prof.\ dr.\ R.G.E.\ Timmermans

\newpage
\noindent
The material presented in this thesis is partly based on the following publications:

\begin{itemize}
    \item \bibentry{Boer:2016xqr}
    \item \bibentry{Boer:2017hqr}
    \item \bibentry{Cotogno:2017puy}
    \item \bibentry{Boer:2018vdi}
\end{itemize}

\noindent
A full up-to-date list of my publications, including contributions to conference proceedings, can be found at the high-energy physics literature database called \href{http://inspirehep.net/author/profile/Tom.Van.Daal.1}{INSPIRE-HEP}.

\cleardoublepage 

\selectlanguage{english}

\tableofcontents
\addtocontents{toc}{\vspace{0.4cm}} 

\cleardoublepage 

\pagenumbering{arabic}

\chapter[\stackunder{Introduction}{}]{Introduction} \label{c:Introduction}
\addthumb{\thechapter}{\Large{\thechapter}}{\thumbfg}{\thumbbg}

One of the main goals in physics is to unravel the fundamental structure of the universe in terms of its most elementary constituents. In the 19th century, the elementary building blocks of nature were thought to be those appearing in Mendeleev's periodic table of elements. Today, however, we know that the chemical elements are not elementary but composite; they consist of protons and neutrons that make up the atomic nuclei, surrounded by electrons. 

Protons and neutrons are examples of hadrons. In the 1960s, a plethora of different kinds of hadrons were identified in particle detectors, all of which were believed to be elementary particles, albeit with a finite size. This multitude of newly discovered particles with similar properties, dubbed the `particle zoo', led Gell-Mann~\cite{GellMann:1964nj} and Zweig~\cite{Zweig:1981pd,Zweig:1964jf} to propose the quark model in 1964. They postulated that hadrons are composite objects, composed of more elementary particles called quarks. To explain the physical properties of the hadrons known at that time, their model contained three flavors of quarks, called up, down, and strange. The existence of quarks was experimentally verified in 1968 at SLAC~\cite{Bloom:1969kc,Breidenbach:1969kd}; through deep-inelastic scattering of electrons off protons and neutrons the up and down quark flavors were discovered. For various reasons, theorists predicted the existence of more flavors. A total of six quark flavors was proposed in 1973 by Kobayashi and Maskawa~\cite{Kobayashi:1973fv}, which was not experimentally verified until the discovery of the much heavier top quark in 1995 at Fermilab~\cite{Abe:1995hr}. Soon after the conception of the quark model, it was proposed that quarks are bound together inside hadrons through the strong force mediated by so-called gluons. The existence of gluons was confirmed in 1979 at DESY~\cite{Brandelik:1979bd,Barber:1979yr,Berger:1979cj,Bartel:1979ut}. So far, there are no experimental indications that quarks and gluons have any substructure -- they are considered to be elementary particles. 

The dynamics of quarks, antiquarks, and gluons (collectively known as partons) is described by the theory of the strong interaction called quantum chromodynamics (QCD). QCD is a non-Abelian gauge theory (also called Yang-Mills theory) whose fields carry a quantum number called color. Unlike the situation in quantum electrodynamics where the gauge field is electrically neutral, the gluons carry color charge. This implies that they not only interact with quarks and antiquarks, but also with each other. A consequence of the non-Abelian nature of QCD is that the strong force between partons increases as they are being separated. When the distance between two partons grows, it becomes at some point energetically favorable to create a new $q\Bar{q}$ pair. As a result, we do not observe individual partons in experiments -- rather, we detect jets of colorless combinations of partons, the hadrons. This phenomenon is called (color) confinement. There is no mathematical proof yet of confinement. 

On the other hand, at smaller length scales, QCD becomes weakly coupled and the partons essentially behave as free particles. This is called asymptotic freedom and was discovered in 1973 by Gross, Wilczek, and Politzer~\cite{Gross:1973id,Politzer:1973fx} -- an important finding for which they were awarded the 2004 Nobel Prize in Physics. While the weakly coupled regime can be described using perturbation theory, we rely on fits to data for the nonperturbative part where the theory is strongly coupled. In this thesis we will focus on the latter regime.

Especially at hadron colliders, QCD plays a vital role in describing the high-energy collisions. When there is at least one hadron involved in either the initial or final state, the calculation of a cross section entails the perturbative calculation of the hard scattering event (`hard' means that $Q \gg M$, with $Q$ the hard scale of the process and $M$ the hadron mass), combined with unknown functions that parametrize the hadron structure that is governed by soft processes (`soft' means that $Q \sim M$ or smaller). Those functions are called parton distribution functions (PDFs) for partons inside initial-state hadrons and fragmentation functions (FFs) for partons decaying into final-state hadrons. Both types of functions can be constrained by fits to data. The simplest and most well-known type of PDF is the so-called collinear PDF, introduced by Feynman in 1969~\cite{Feynman:1969ej}. These PDFs depend only on the motion of partons \emph{collinear} to their parent hadron (the other momentum components are integrated over). A collinear PDF $f_i(x)$ represents the probability density to find a parton of type $i$ with longitudinal momentum fraction $x$ inside a given hadron.

The experimental extraction of PDFs relies on the possibility to \emph{factorize} cross sections into separate pieces that describe short- and long-distance physics. When factorization applies, a cross section can be expressed in terms of a hard scattering factor and one or more collinear PDFs. The latter are \emph{universal}, i.e.\ independent of the hard scattering event, which implies that they can be extracted from one process and used in others -- a key ingredient of QCD phenomenology. Besides being a useful tool to describe collisions at hadron colliders, PDFs can teach us more about the inner workings of hadrons. However, the one-dimensional collinear PDFs are not sensitive to various interesting physics phenomena that are at play inside hadrons. To that end, we must go beyond a collinear description of partons inside hadrons.

To obtain more insight into the internal dynamics of hadrons, one can study \emph{multidimensional} PDFs instead. The phenomenology of multidimensional quark and gluon PDFs forms the main subject of this thesis. The type of PDFs that we will discuss most are the so-called transverse momentum dependent (TMD) PDFs, often simply referred to as TMDs. Besides $x$, these functions also depend on the partonic transverse momentum $\bm{k}$ (the transverse directions are perpendicular to the collision axis), making them three-dimensional. For example, TMDs can encode interesting correlations between $\bm{k}$ and the spin polarization of the parton or hadron. 

Typically, TMD effects manifest themselves in cross sections through angular correlations between the produced particles. To understand how those correlations arise, let us consider the production of two jets in a hadron-hadron collision. If the partons inside the hadrons would have no transverse momentum,\footnote{For simplicity we only consider here `intrinsic' transverse momentum and not perturbatively generated transverse momentum.} i.e.\ they move perfectly collinear to their parent hadrons, then the produced jets would appear back-to-back in the transverse plane due to momentum conservation. However, if at least one parton participating in the hard scattering process has a nonzero transverse component, then the jets are no longer produced back-to-back in the transverse plane. The latter is referred to as an azimuthal asymmetry and can be measured in experiment. Observables can be sensitive to TMDs (and thus the three-dimensional inner structure of hadrons) if they are differential in sufficiently many variables, particularly (directions of) transverse momenta of final-state particles.

As it turns out, unlike the collinear PDFs, TMDs are not universal -- they are \emph{process dependent}. The process dependence of TMDs originates from the color structure of the hard scattering event. Fortunately, as the color structure of the hard scattering event can be calculated, the process dependence of TMDs is calculable event-by-event, implying that not all universality (and hence predictability) is in fact lost. Compared to collinear PDFs, our current knowledge of TMDs is rather limited; this is because the extraction of TMDs requires measurements of more complex processes. 

In general, greater knowledge on multidimensional PDFs teaches us more about the inner structure of hadrons. Through multidimensional PDFs we could deepen our understanding of the angular momentum structure of hadrons. For example, it is unclear how the proton spin can be decomposed into quark and gluon contributions. Furthermore, multidimensional PDFs could shed light on the origins of hadron mass. After all, the Higgs mechanism only explains quark masses that add up to at most tens of MeVs, which is a few percent of the proton mass. The actual proton mass is much larger due to QCD dynamics that leads to the hadronic scale of several hundreds of MeVs, which is also the order of magnitude of intrinsic partonic transverse momenta. On the other hand, many theoretical and experimental efforts on PDFs over the last decades have set the stage to test some important fundamental predictions of QCD, for example ones related to the process dependence of TMDs. Furthermore, an improved understanding of PDFs can lead to more accurate descriptions of certain processes that are sensitive to physics beyond the Standard Model. Thus, the study of PDFs not only enriches our knowledge on the dynamics of quarks and gluons inside hadrons, it can also serve to check the consistency of the QCD framework and possibly identify deviations from the Standard Model.

In this thesis we will discuss multidimensional PDFs from a theoretical perspective. To this end, we will first provide an introduction to PDFs in chapter~\ref{c:PDFs}. After discussing the basics of QCD, we will define various types of PDFs, both collinear and multidimensional ones. Since we will mainly focus on TMDs, we will set up the necessary formalism in detail for the TMD case.

The next two chapters comprise part~\ref{p:1} of this thesis and discuss factorization of Drell-Yan (DY) scattering in terms of a perturbatively calculable hard factor and two TMDs, called TMD factorization. The DY process describes lepton-antilepton production through $q\Bar{q}$ annihilation in hadron-hadron collisions. For certain TMDs that give rise to particular azimuthal asymmetries in the DY cross section, it is unclear whether the (standard) factorization theorem holds due to seemingly nontrivial color structures. In chapter~\ref{c:CSS} we review how TMD factorization has been established for DY, and in chapter~\ref{c:Disentanglement} we perform a factorization study largely along the lines of the original proof for the potentially problematic TMDs, investigating whether factorization could be violated. 

Part~\ref{p:2} of this thesis is about multidimensional gluon PDFs. In chapter~\ref{c:Polarized_gluon_TMDs} we introduce TMDs that can be used to describe gluons inside polarized hadrons up to spin $1$. We also derive bounds for those TMDs. Subsequently, in chapter~\ref{c:Small_x}, we study gluon TMDs and other multidimensional gluon PDFs at small $x$. In the small-$x$ kinematic region the hadron content is dominated by gluons. In that regime, we show that the number of independent gluon PDFs significantly drops, thereby establishing a simplified picture of the gluonic structure of hadrons. We will also demonstrate that correlations between gluons inside nuclei can give rise to (odd-harmonic) angular correlations between two final-state hadrons in $pA$ collisions. 

In chapter~\ref{c:Conclusions} we summarize the main conclusions of this thesis. In appendices~\ref{a:conventions}--\ref{a:feynman_rules} we list the most important conventions, and in the remaining appendices we provide additional technical details. Finally, we provide a lay summary of this thesis (including a Dutch translation).
\chapter{Mapping the internal structure of hadrons} \label{c:PDFs}
\addthumb{\thechapter}{\Large{\thechapter}}{\thumbfg}{\thumbbg}

Since all research presented in this thesis is related to the strong interaction, we first review the most important aspects of QCD. Subsequently, in section~\ref{s:quark_correlator} we discuss how the internal structure of hadrons can be described in terms of quark PDFs. After reviewing the collinear case, we discuss the more involved multidimensional PDFs that will play a central role in later chapters. Special attention is given to the three-dimensional TMDs in section~\ref{s:TMD_formalism}. We will set up the TMD formalism for quarks inside polarized hadrons, needed for part~\ref{p:1} of this thesis. Fortunately, most of the discussion on quark distributions in this chapter can be easily extended to the multidimensional gluon distributions that are discussed in part~\ref{p:2}.

\section{Quantum chromodynamics in a nutshell}

QCD is a sector of the Standard Model gauge theory. The underlying symmetry group is the non-Abelian Lie group $\text{SU}(N_c)$, with $N_c = 3$ (the number of colors). Invariance under local $\text{SU}(3)$ transformations gives rise to eight massless gauge boson fields called gluons. The gluons belong to the adjoint representation of the gauge group, while the fermions that couple to the gluons, the quarks, belong to the fundamental representation.

\subsection{The Lagrangian}
A quantum field theory is usually described by a Lagrangian density. The Lagrangian density of QCD is given by~\cite{Fritzsch:1973pi}
\begin{equation}
    \mathcal{L}_\text{QCD} = -\frac{1}{4} F_{\mu\nu}^a F^{a\mu\nu} + \sum_q \,\overline{\psi}_q \!\left( i\slashed{D} - m_q I \right) \!\psi_q \,,
    \label{e:QCD_Lagrangian}
\end{equation}
where the field $\psi_q$ with mass $m_q$ is a color triplet quark field pertaining to one of the six quark flavors labeled by $q$. In this thesis, $I$ denotes the identity matrix in \emph{any} space. The covariant derivative is given by
\begin{equation}
    D_\mu \equiv I \,\partial_\mu + igt^a \!A_\mu^a \,,
\end{equation}
where $A$ denotes the color octet gluon field and $g$ is the gauge coupling. The gluon field strength tensor is defined as
\begin{equation}
    F_{\mu\nu}^a \equiv \partial_\mu A_\nu^a - \partial_\nu A_\mu^a - g f^{abc} A_\mu^b A_\nu^c \,,
    \label{e:fieldstrength}
\end{equation}
where the last term gives rise to three- and four-gluon vertices in the first term of eq.~\eqref{e:QCD_Lagrangian}.

The color matrices $t^a$ are the generators of the fundamental representation; they are traceless and Hermitian, and satisfy the following (anti)commutation relations:
\begin{equation}
    [t^a,t^b] = if^{abc} t^c , \qquad \{t^a,t^b\} = \frac{1}{N_c} \,\delta^{ab} + d^{abc} t^c .
    \label{e:Lie_bracket}
\end{equation}
Furthermore, they are normalized as
\begin{equation}
    \tr(t^a t^b) = \tfrac{1}{2} \,\delta^{ab} .
\end{equation}
The structure constants of $\text{SU}(3)$ are given by
\begin{equation}
    f^{abc} = -2i \tr\left( [t^a,t^b] t^c \right) , \qquad d^{abc} = 2 \tr\left( \{t^a,t^b\} t^c \right) .
    \label{f_and_d}
\end{equation}
Both $f$ and $d$ are real and they are, respectively, totally antisymmetric and symmetric under the interchange of any of their indices. The numerical values of these constants, as well as explicit representations for the color matrices, can be found in e.g.~\cite{GellMann:1962xb}.

\subsection{The running coupling}
Despite its name, the QCD coupling constant $\alpha_s \equiv g^2/(4\pi)$ is \emph{not} a constant. The value that should be used in calculations depends on the energy scale involved. Since $\alpha_s$ satisfies a first-order differential equation (a renormalization group equation), its value at any scale $Q$ can be related to one particular reference scale. For this reference scale we could take the Landau pole, which is the energy scale at which the coupling constant becomes infinite, also referred to as the QCD scale ($\Lambda_\text{QCD}$). At leading order (i.e.\ at one-loop order), the running coupling is given by\footnote{Today, the running of $\alpha_s$ is known up to five loops~\cite{Baikov:2016tgj,Herzog:2017ohr}.}
\begin{equation}
    \alpha_s(Q) = \frac{2\pi}{b_0 \ln(Q/\Lambda_\text{QCD})} \,, \qquad b_0 \equiv 11 - \frac{2}{3} \,n_f \,,
    \label{e:running}
\end{equation}
with $n_f$ the number of active flavors at the scale $Q$. The value of the QCD scale is not uniquely determined; besides a dependence on $n_f$ and the perturbative accuracy, this scale also depends on the renormalization scheme (this dependence starts at three loops) -- typically $\Lambda_\text{QCD} \sim 0.2$ GeV~\cite{Patrignani:2016xqp}. Note that for $n_f \leq 16$ we have $b_0 > 0$, which implies that the coupling decreases with increasing energy. As a consequence, the strong interaction becomes asymptotically weaker at decreasing length scales, a property of QCD called asymptotic freedom. This phenomenon is a feature of Yang-Mills theories in general~\cite{Zee:1973gn,Coleman:1973jx}.

Whenever the energy scale is somewhat larger than $\Lambda_\text{QCD}$, perturbative calculations are allowed; typically one takes for this scale $Q = 1$ GeV (which is approximately the mass of the proton), where $\alpha_s(Q) \approx 0.4$. At lower energies, however, the theory becomes strongly coupled, leading to the \emph{confinement} of quarks and gluons inside hadrons -- this is the nonperturbative regime. When quarks are produced in particle accelerators, they cannot be isolated in the detectors; we instead observe jets that contain many color-neutral mesons and baryons. Hence, the low- and high-energy particle degrees of freedom of QCD are constituted by hadrons and partons, respectively.

\subsection{Deep-inelastic scattering} \label{s:QCD_phenomenology}
QCD is essential in providing descriptions of collisions that feature the strong interaction. For a given process that involves hadrons, the hard scattering event can be calculated using perturbation theory, while the hadronic structure can be parametrized in terms of PDFs or FFs that need to be fitted to data (in this thesis we will only consider PDFs). The simplest and most well-known type of PDF is the collinear one. As already mentioned in the introduction, a collinear PDF $f_i(x)$ represents the probability density for finding a parton of type $i$ with longitudinal momentum fraction $x$ inside a given hadron.

To illustrate how a collinear PDF enters a cross section calculation, we need to consider a process that involves at least one hadron in the initial state, e.g.\ deep-inelastic scattering (DIS) given by $e^-H \to e^-X$. DIS is the simplest and most-studied process in QCD. By $H$ we denote a hadron and $X$ represents all unobserved final-state particles; \emph{only} the electron is measured after the scattering. For that reason, this process is also referred to as `inclusive' DIS. In contrast, the process $e^-H \to e^-H'X$, where also one hadron $H'$ is measured in the final state, is called semi-inclusive DIS (SIDIS).

\begin{figure}[htb]
\centering
    \includegraphics[height=2.89cm]{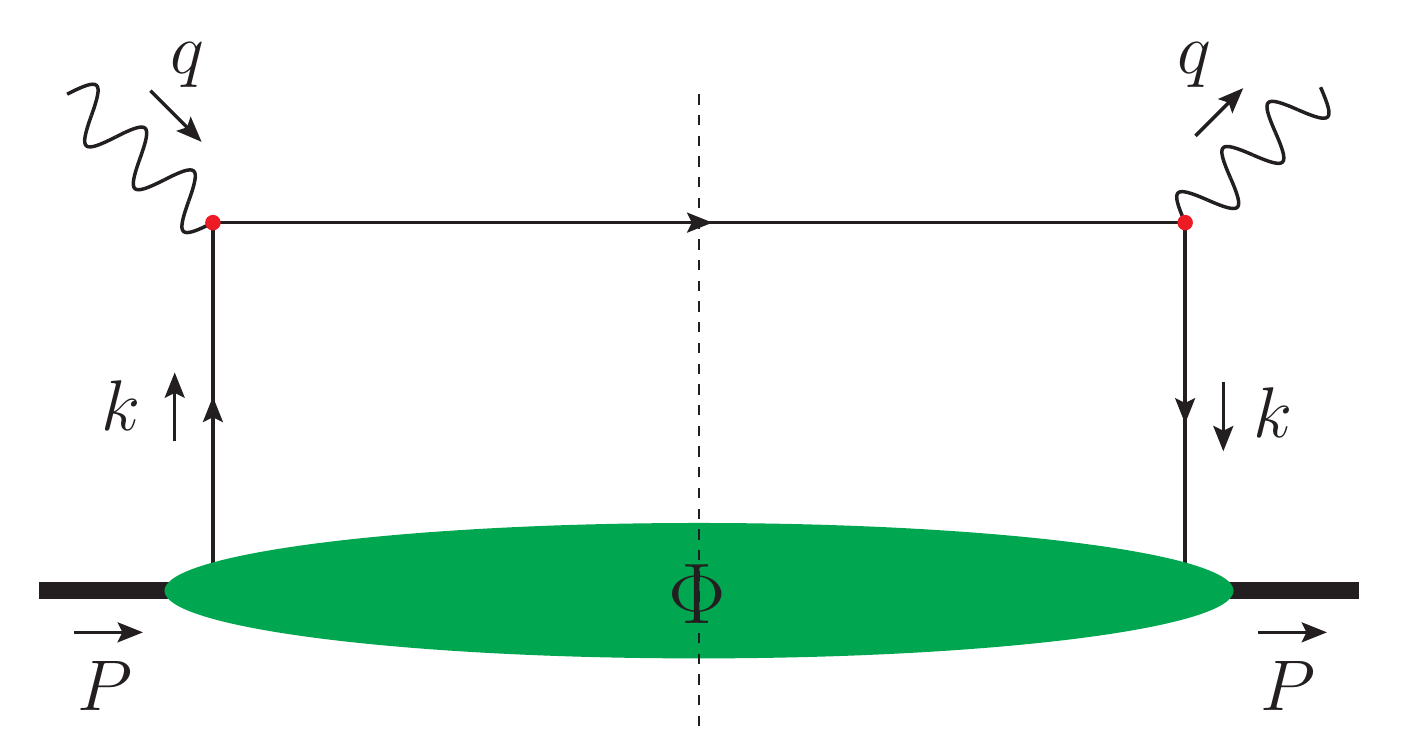}
    \caption{The tree-level (amplitude squared) graph for DIS: an electron emits a photon that subsequently scatters off a quark inside a hadron (the electrons are not shown). The green `blob' represents the quark-quark correlator, and the dotted line in the middle represents the final-state cut.}
\label{f:DIS}
\end{figure}

The leading-order (also referred to as tree-level) diagram for DIS is given in figure~\ref{f:DIS}. The hadron and quark momenta are denoted by $P$ and $k$ respectively, and the photon momentum is denoted by $q$. The hard scale of the process is set by the photon virtuality $Q^2 \equiv -q^2 > 0$. The graph in figure~\ref{f:DIS} is split by a dotted line, called the final-state cut, into an amplitude (all that is to the left of the cut) and a conjugate amplitude (all that is to the right of the cut). The green `blob', denoted by $\Phi$, represents the quark-quark correlation function, also referred to as the correlator (the operator definition of this quantity will be given in the next section). Since $\Phi$ encodes the internal dynamics of the hadron, it cannot be calculated using perturbation theory; rather, it is parametrized in terms of a collinear PDF called $f_1$ (see eq.~\eqref{e:PDF_parametrization} in the next section).

The cross section for DIS has the following form:
\begin{equation}
    \sigma = \sum_i \int_0^1 dx \,f_{1,i}(x) \,\hat{\sigma}_i(x) ,
    \label{e:DIS_factorization}
\end{equation}
where the sum only includes contributions from quarks and antiquarks. The hard-scattering coefficient $\hat{\sigma}_i$ describes the partonic process $e^-q_i \to e^-q_i$ and can be calculated using perturbation theory. Since $f_{1,i}$ provides the probability for finding a parton $i$ inside the hadron, it must be normalized in a way such that the valence quark content of the hadron is respected (this is called a number sum rule). The range of allowed values for $x$ ($0 \leq x \leq 1$) will be commented on in the next section.

The result in eq.~\eqref{e:DIS_factorization} is an approximation -- it is given at leading order in $\alpha_s$ and at leading power in $M/Q$. This is based on the fact that the accuracy of observables in QCD is determined by \emph{two different} expansions. On the one hand, there is an expansion in the coupling constant $\alpha_s$, and, on the other hand, there is an expansion in the ratio $M/Q$. The latter is referred to as the twist expansion and higher order terms in this expansion are called power corrections.\footnote{The precise definition of `twist' is not relevant for our purposes. In fact, there exist various definitions of twist in the literature, see e.g.\ the review in~\cite{Signori:2016lvd}.} Whereas the coupling expansion quantifies the accuracy of the perturbative structure of the theory, the twist expansion quantifies the accuracy of its nonperturbative structure. In contrast to the expansion in $\alpha_s$, the full twist expansion is always limited to a certain power. In this thesis we will only consider quantities at leading twist, which is justified as long as $Q \gg M$. Power corrections to eq.~\eqref{e:DIS_factorization} modify the hard-scattering coefficient and include PDFs of subleading twist.

Let us also briefly comment on higher-order $\alpha_s$ corrections to eq.~\eqref{e:DIS_factorization}. Perturbative QCD corrections modify both the hard-scattering coefficient and the PDF. Already at the one-loop level, several types of divergences appear. The ultraviolet divergences can be taken care of using renormalization, which implies the introduction of a renormalization scale $\mu$. Since a physical observable cannot depend on $\mu$, this leads to renormalization group equations. For the collinear PDFs they are called the Dokshitzer-Gribov-Lipatov-Altarelli-Parisi (DGLAP) equations~\cite{Gribov:1972ri,Dokshitzer:1977sg,Altarelli:1977zs}; these coupled differential equations can be used to evolve PDFs from one scale to another (as evolution will not play a role in this thesis, the scale dependence of PDFs will always be implicit). Other divergences that appear are infrared divergences, reflecting the sensitivity of the cross section to long-distance effects. They can arise when a gluon is emitted collinear to a quark or when a gluon has a vanishing momentum. For DIS, the latter type of divergence cancels upon summing over all real and virtual diagrams~\cite{Collins:2011zzd}. The former type, however, also referred to as the collinear divergence, does not cancel and can be absorbed by the nonperturbative PDF.

The DIS cross section in eq.~\eqref{e:DIS_factorization} shows a very important feature, namely the \emph{factorization} of the short-distance physics (captured by $\hat{\sigma}$) and the long-distance physics (captured by $f_1$). Factorization is a key ingredient of QCD phenomenology. As already mentioned in the introduction, a consequence of factorization is that the collinear PDFs are \emph{universal} -- the internal dynamics of hadrons is the same for all processes. As it turns out, proving factorization is highly nontrivial due to for instance gluon exchanges between the different parts of a process. Only for a limited number of processes proofs are available up to all orders in $\alpha_s$. Factorization will be discussed in detail in chapter~\ref{c:CSS}.

\section{The quark-quark correlation function} \label{s:quark_correlator}

In this section we provide the precise field-theoretical definition of the quark-quark correlator $\Phi$ which we encountered already in the DIS graph given in figure~\ref{f:DIS}. In subsection~\ref{s:collinear_PDFs} we describe the collinear correlator that can be parametrized in terms of the collinear PDF $f_1$, and in subsection~\ref{s:multidimensional_PDFs} we discuss more complex correlators that can be parametrized in terms of multidimensional PDFs. First, however, let us discuss some important conventions.

Throughout this thesis we make extensive use of light-cone coordinates, which are convenient for describing high-energy scattering processes. In terms of those coordinates, a four-vector $a$ can be represented as $a = (a^+,a^-,\bm{a})$, where $a^\pm \equiv (a^0 \pm a^3)/\sqrt{2}$ and $\bm{a} \equiv (a^1, a^2)$. Adopting the `mostly minus' convention for the Minkowski metric $g$, a dot product between two four-vectors $a$ and $b$ is given by $a \cd b \equiv g^{\mu\nu} a_\mu b_\nu = a^+b^- + a^-b^+ - \bm{a} \cd \bm{b}$. Furthermore, we employ the lightlike vectors $\bar{n} \equiv (1,0,\bm{0})$ and $n \equiv (0,1,\bm{0})$, such that $a$ can be decomposed as $a^\mu = a^+ \bar{n}^\mu + a^- n^\mu + a_\sst^\mu$. We have defined the quantity $a_\sst \equiv (0,0,\bm{a})$, so that $a_\sst^2 = - \bm{a}^2$. An overview of relevant conventions is provided in  appendix~\ref{a:conventions}. In terms of $\bar{n}$ and $n$, the hadron and quark momenta $P$ and $k$ are given by:
\begin{align}
    P^\mu &= P^+ \bar{n}^\mu + \frac{M^2}{2P^+} \,n^\mu , \label{e:P} \\[2pt]
    k^\mu &= xP^+ \bar{n}^\mu + k^- n^\mu + k_\sst^\mu \,. \label{e:k}
\end{align}
Note that we have chosen the hadron to move along the $z$-axis, satisfying the on-shell condition $P^2 = M^2$. Furthermore, we have defined the longitudinal momentum fraction as $x \equiv k^+/P^+$. Typically, $P^+ \sim Q$ in cross section calculations, where $Q$ is the relevant hard scale in the process. In the high-energy limit, the minus component of $P$ becomes negligible compared to its plus component. Hence, in this limit, $P$ becomes lightlike and the hadron momentum is directed along the plus direction.

\subsection{The unintegrated correlator} \label{s:unintegrated_correlator}
\begin{figure}[htb]
\centering
    \begin{subfigure}{.445\textwidth}
    \begin{flushleft}
        \includegraphics[height=2.0cm]{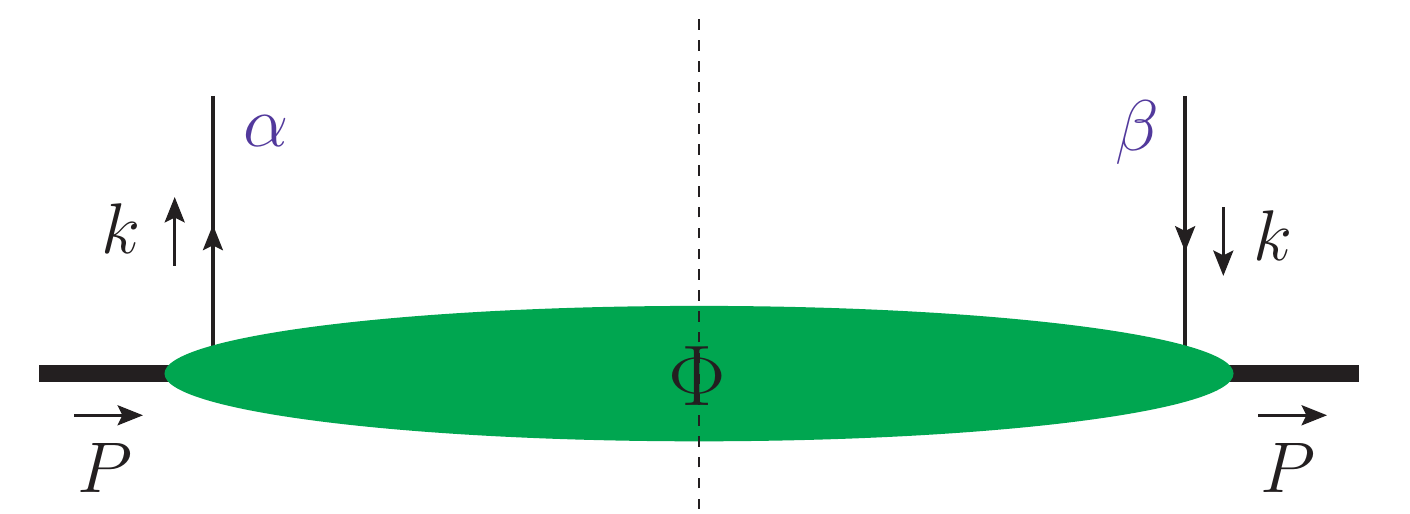}
    \end{flushleft}
    \end{subfigure}
    \begin{subfigure}{.445\textwidth}
    \begin{flushright}
        \includegraphics[height=2.0cm]{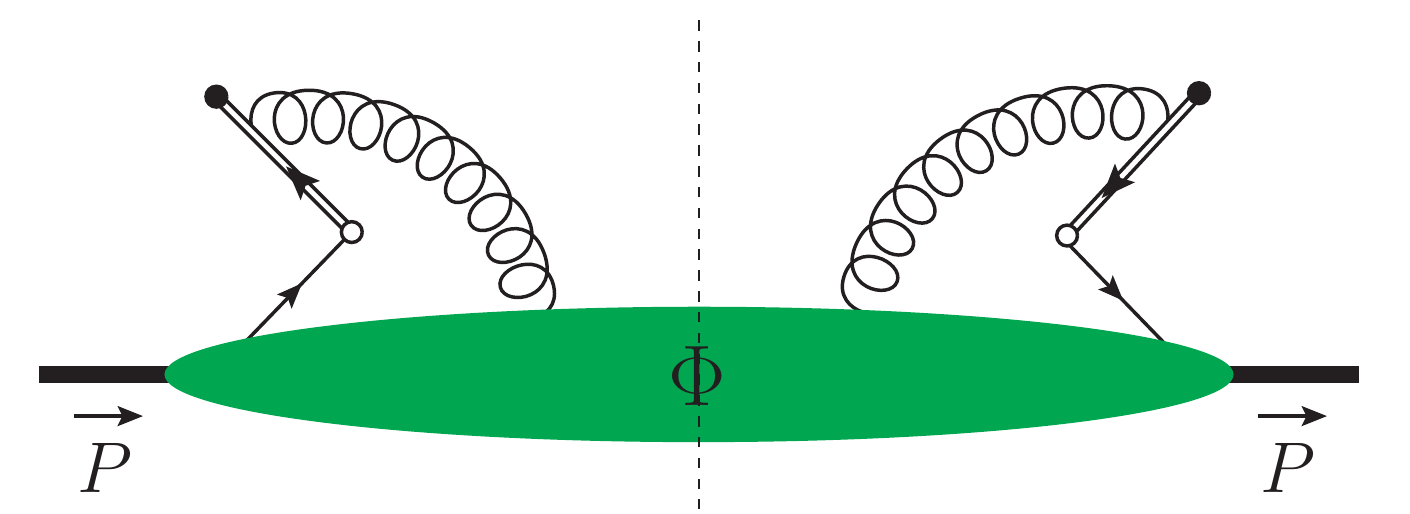}
    \end{flushright}
    \end{subfigure}
    \caption{The definition of the quark-quark correlator $\Phi$. The graph on the left is not gauge invariant; the necessary Wilson line is only included in the graph on the right (infinitely many gluon emissions are implied). See appendix~\ref{a:feynman_rules} for the notation regarding eikonal lines and vertices.}
\label{f:quark_correlators}
\end{figure}

The field-theoretical definition of the quark-quark correlation matrix $\Phi$ is given by
\begin{equation}
    \Phi_{\alpha\beta}(k) \equiv \sum_X \int \frac{d^3\vec{P}_X}{(2\pi)^3 \,2P_X^0} \,\bra{P} \overline{\psi}_\beta(0) \ket{X} \!\bra{X} \psi_\alpha(0) \ket{P} \delta^{(4)}(P-k-P_X) ,
    \label{e:quark-quark_correlator_definition}
\end{equation}
where $\alpha,\beta$ are Dirac indices. A summation over color triplet indices is implicitly understood. The dependence of $\Phi$ on $P$ is implicit (as this is not the dependence we are ultimately interested in), and, to lighten the notation, we suppress flavor labels for the most part in this thesis. Any dependence on hadron polarization will not be discussed until section~\ref{s:TMD_formalism}, so here we consider unpolarized hadrons only. The product of matrix elements in eq.~\eqref{e:quark-quark_correlator_definition} gives the probability density for the extraction of a quark with momentum $k$ from the hadron, yielding a remnant $X$ with momentum $P_X$. As $X$ is a final state we integrate over its phase space, and because it is not measured experimentally we also perform a sum over all possibilities. The delta function reflects momentum conservation; its Fourier transform to the position variable $z$ can be used to perform a translation from the point $0$ to $z$ of the field operator $\psi$. Then, upon using the completeness relation for the unobserved final states, we obtain
\begin{equation}
    \Phi_{\alpha\beta}(k) = \int \frac{d^4z}{(2\pi)^4} \;e^{ik\cdot z} \bra{P} \overline{\psi}_\beta(0) \psi_\alpha(z) \ket{P} .
    \label{e:unintegrated_quarkcorrelator}
\end{equation}
Since this correlator depends on the full quark four-momentum $k$, it is also referred to as the \emph{unintegrated} correlator. The pictorial definition of $\Phi$ is given by the graph on the left in figure~\ref{f:quark_correlators}.

The expression for $\Phi$ in eq.~\eqref{e:unintegrated_quarkcorrelator} is not completely correct -- one important ingredient is missing. Since the two quark fields reside at two different spacetime positions (the matrix element is said to be bilocal), the correlator is \emph{not} gauge invariant. After all, under local $\text{SU}(3)$ transformations $V$, the quark fields transform as
\begin{equation}
    \psi_\alpha(z) \;\to\; V(z) \,\psi_\alpha(z) , \qquad \overline{\psi}_\beta(0) \;\to\; \overline{\psi}_\beta(0) \,V^\dag(0) ,
\end{equation}
so that the operator combination $\overline{\psi}_\beta(0) \psi_\alpha(z)$ changes. The gauge dependence of the correlator is concerning. Since the PDFs appearing in the parametrization of $\Phi$ can be extracted from experiments, they should be independent of arbitrary gauge transformations. This problem can be fixed as follows: if we could insert in between the quark fields a (color) matrix-valued quantity $U_{[0,z]}$ that transforms as
\begin{equation}
    U_{[0,z]} \;\to\; V(0) \,U_{[0,z]} V^\dag(z) ,
\end{equation}
then the correlator would be gauge invariant. A quantity with these properties exists -- it is called a Wilson line (or gauge link), given by~\cite{Wilson:1974sk}
\begin{equation}
    U_{[0,z]} = \mathcal{P} \exp \left( ig \int_0^z d\eta^\mu A_\mu(\eta) \right) ,
    \label{e:WL_def}
\end{equation}
where $A_\mu \equiv A_\mu^a t^a$. The symbol $\mathcal{P}$ denotes path ordering; for two fields $A$ and $B$ at different positions, path ordering is defined as
\begin{equation}
    \mathcal{P} A(x)B(y) \equiv \theta(x-y) \,A(x) B(y) + \theta(y-x) \,B(y) A(x) ,
    \label{e:path_ordering}
\end{equation}
where $\theta$ is the Heaviside step function. Eq.~\eqref{e:path_ordering} can be easily generalized to include more fields.

For a correlator to be gauge invariant, one needs to include a Wilson line. The gauge-invariant correlator corresponds to the graph on the right in figure~\ref{f:quark_correlators}. Physically, a Wilson line reflects a sum over infinitely many gluon emissions which can be calculated explicitly~\cite{Efremov:1978xm}. The latter is discussed in detail for the TMD case in e.g.~\cite{Buffing:2015tja} (in section~\ref{s:BM_function} we calculate the $\mathcal{O}(g)$ contribution to a Wilson line).

One might wonder whether there are any restrictions on the integration path of the Wilson line in eq.~\eqref{e:WL_def}. In principle, any possible path running between $0$ and $z$ guarantees gauge invariance of $\Phi$. However, since in QCD a closed path does not vanish, different paths correspond to physically different situations. As it turns out, for certain types of correlators different paths correspond to different processes in which the correlator appears. The latter has far-reaching consequences and forms one of the main themes of this thesis. We will discuss this matter in detail in section~\ref{s:TMD_formalism}.

\subsection{The collinear correlator} \label{s:collinear_PDFs}
\begin{figure}[htb]
\centering
    \includegraphics[height=2.5cm]{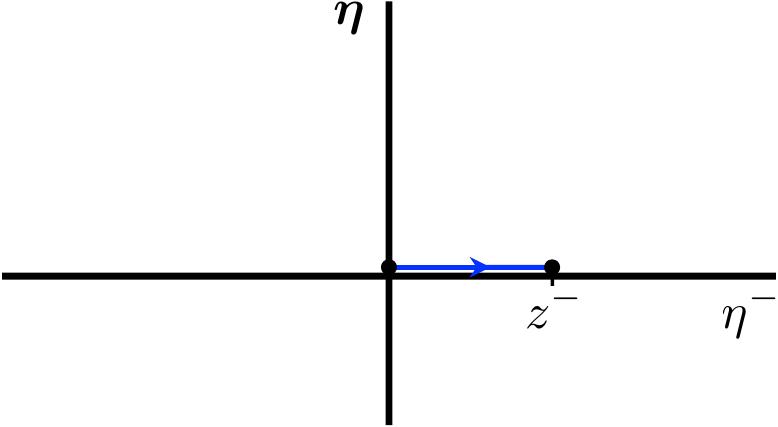}
    \caption{The integration path of the gauge link $U_{[0,z^-;\bm{0}]}^n$. The path runs along the light-front $\eta^+=0$ at transverse position $\bm{\eta} = \bm{0}$.}
\label{f:collinear_WL_path}
\end{figure}

Since the unintegrated correlator in eq.~\eqref{e:unintegrated_quarkcorrelator} does not appear in any known process, we are led to study certain projections. The most well-known projection is the collinear correlator, obtained after integration over $k^-$ and $\bm{k}$; it is given by~\cite{Collins:1981uw}
\begin{equation}
    \Phi_{\alpha\beta}(x) \equiv \int \left. \frac{dz^-}{2\pi} \;e^{ik\cdot z} \bra{P} \overline{\psi}_\beta(0) \,U_{[0,z^-;\bm{0}]}^n \,\psi_\alpha(z) \ket{P} \vphantom{\int} \right|_{z^2=0} .
    \label{e:PDF_quarkcorrelator}
\end{equation}
As this correlator is evaluated at $z^+=0$, $\bm{z}=\bm{0}$ (implying $z^2=0$), it is also sometimes called the light-cone correlator. The gauge link that bridges the nonlocality runs along the minus direction (indicated by the superscript $n$) and is given by
\begin{equation}
    U_{[0,z^-;\bm{0}]}^n \equiv \mathcal{P} \exp\left( ig \int_{0}^{z^-} d\eta^- A^+(\eta^+=0,\eta^-,\bm{\eta}=\bm{0}) \right) .
    \label{e:U^n}
\end{equation}
Note that the integration path is fixed, as illustrated in figure~\ref{f:collinear_WL_path}. This implies that, in the collinear case, the gauge link is uniquely determined. It might seem strange at this point to indicate the transverse position of the link ($\bm{\eta} = \bm{0}$) on the left-hand side of eq.~\eqref{e:U^n}, but this notation will be useful later when we will also consider links that are not unique, involving integration paths at $\bm{\eta} \neq \bm{0}$. In the $A^+ = 0$ light-cone gauge, the collinear Wilson line reduces to unity and can hence be omitted. By contrast, this is \emph{not} possible in the more complex situations that will be considered later.

The correlator in eq.~\eqref{e:PDF_quarkcorrelator} is precisely the one that we encountered in the DIS example in subsection~\ref{s:QCD_phenomenology}. As mentioned earlier, it can be parametrized in terms of one collinear PDF at leading twist (this is indicated by the subscript $1$):
\begin{equation}
    \Phi(x) = \frac{1}{2} \,\slashed{\bar{n}} f_1(x) .
    \label{e:PDF_parametrization}
\end{equation}
At subleading twist, i.e.\ at $\mathcal{O}(M/P^+)$ or beyond, there are more functions, each appearing with a different Dirac structure~\cite{Tangerman:1994eh,Mulders:1995dh}.

Let us comment on the range of allowed values for $x$. The intermediate state in the correlator in figure~\ref{f:quark_correlators} carries momentum $P-k$, and, since it is cut, its energy must be positive. Thus, it follows that $P^+-k^+ \geq 0$, which translates into the condition $x \leq 1$. Similarly, for negative values of $x$ it follows that $x \geq -1$. For negative $x$, though, a quark PDF can be related to an antiquark PDF (indicated by a bar) at positive $x$: $f_1(x) = -\bar{f}_1(-x)$. Conventionally, only positive values of $x$ are considered, both for quarks and antiquarks, and so we always take $0 \leq x \leq 1$.

\begin{figure}[htb]
\centering
    \includegraphics[height=7.0cm]{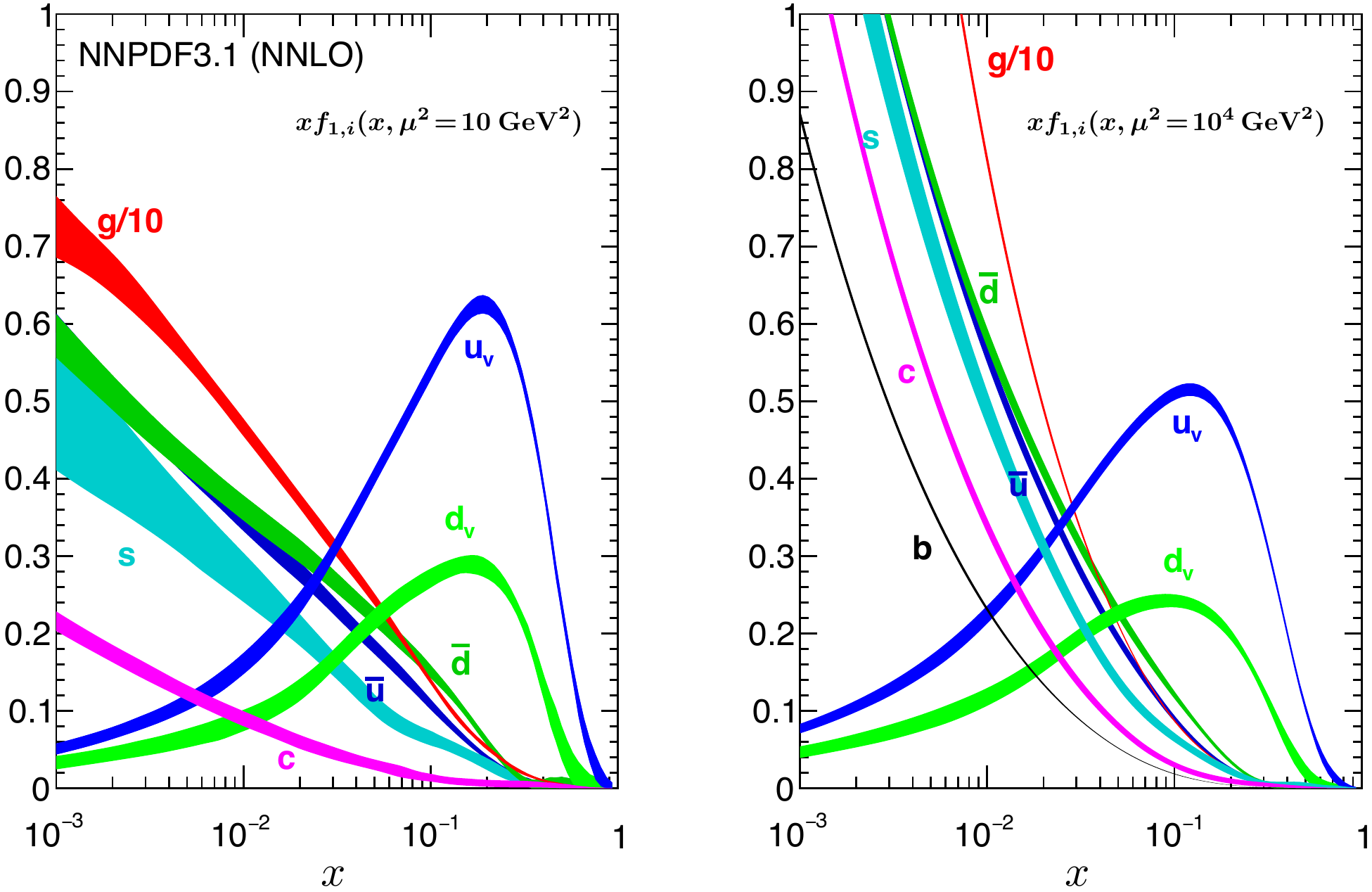}
    \caption{The collinear PDFs $xf_{1,i}(x,\mu^2)$ for the proton at two different energy scales $\mu$; in the left plot $\mu^2 = 10$ GeV$^2$ and in the right plot $\mu^2 = 10^4$ GeV$^2$~\cite{Ball:2017nwa}. The subscript $i$ labels all valence quarks, sea quarks, and gluons. Note that the gluon contributions have been divided by a factor of $10$.}
\label{f:proton_PDFs}
\end{figure}

Besides number sum rules, PDFs also satisfy a momentum sum rule. This rule says that the total fractional momentum carried by partons is unity:
\begin{equation}
    \sum_i \int_0^1 dx \,xf_{1,i}(x) = 1 ,
    \label{e:momentum_sumrule}
\end{equation}
the total momentum of the parent hadron in the plus direction. For the case of a proton, the momentum contributions from each parton are given in the plot in figure~\ref{f:proton_PDFs}. As it turns out, when $Q \gg M$, quarks and antiquarks only account for about half of the total momentum -- the other half is carried by gluons. Furthermore, note that at small $x$ the proton content is dominated by gluons. For that reason, we will study gluon distributions at small $x$ in chapter~\ref{c:Small_x}.

\subsection{Multidimensional correlators} \label{s:multidimensional_PDFs}
\begin{figure}[htb]
\centering
    \includegraphics[height=5.0cm]{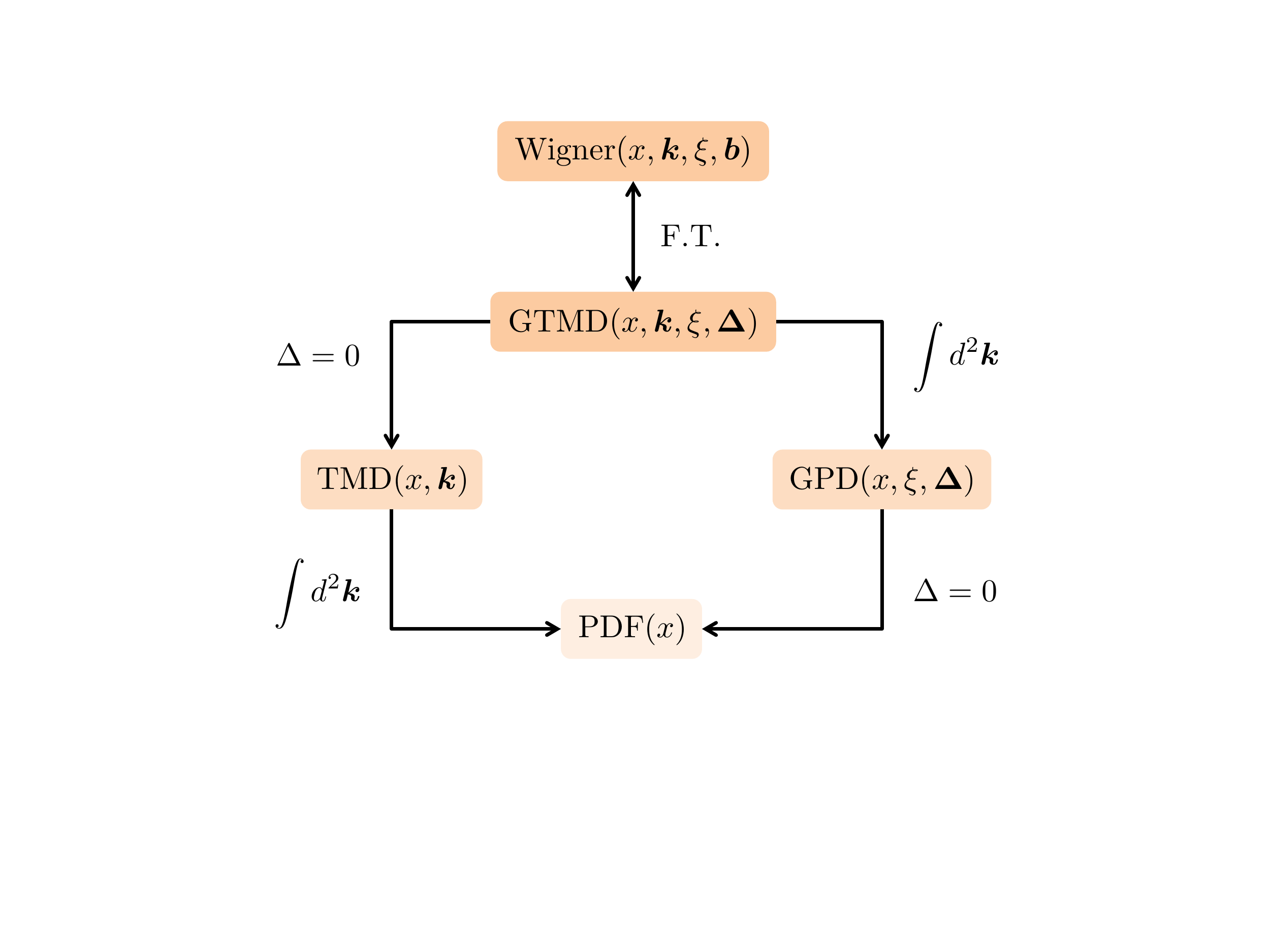}
    \caption{Relations between Wigner distributions, GTMDs, GPDs, TMDs, and (collinear) PDFs. In between brackets it is indicated which kinematic variables the functions depend on.}
\label{f:projections}
\end{figure}

So far we have described the hadron structure only in terms of the one-dimensional collinear PDFs that appear in the parametrization of the collinear correlator defined in eq.~\eqref{e:PDF_quarkcorrelator}. The collinear correlator can be obtained through certain projections or limits from more general correlators, see figure~\ref{f:projections}. Generally speaking, correlators that depend on more kinematical variables contain more information on the internal structure of hadrons. Similarly to the one-dimensional case in eq.~\eqref{e:PDF_parametrization}, those correlators can be parametrized in terms of \emph{multidimensional} PDFs. In this subsection we introduce the various types of PDFs that will be discussed in later chapters. Recent reviews on multidimensional distribution functions include~\cite{Burkardt:2015qoa,Diehl:2015uka,Bacchetta:2016ccz}.

\begin{figure}[htb]
\centering
    \includegraphics[height=1.78cm]{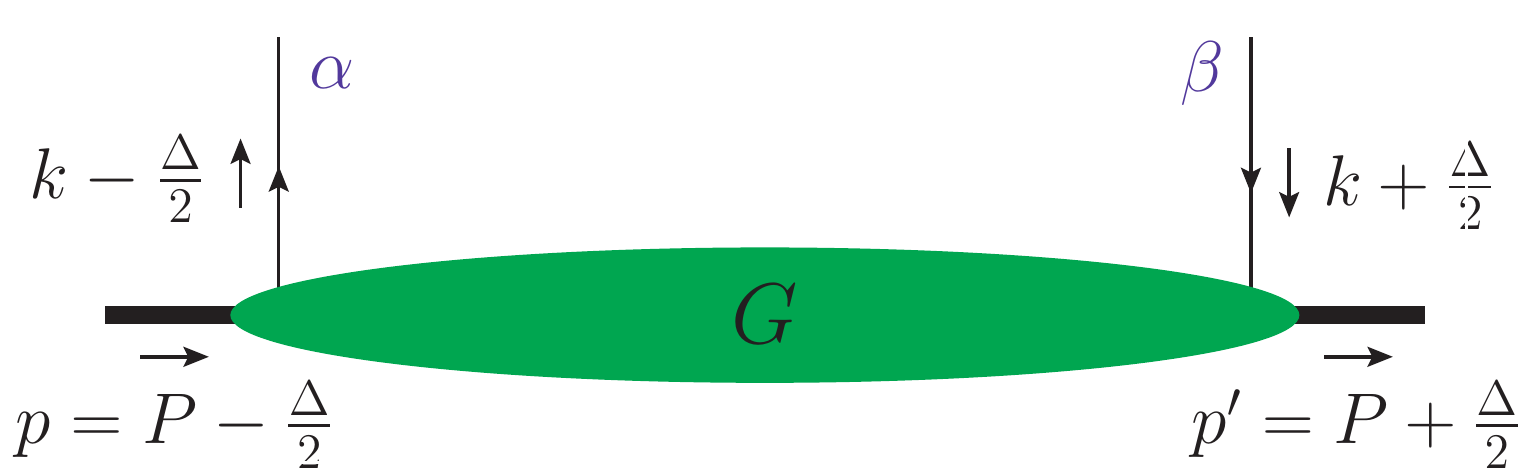}
    \caption{The definition of the off-forward quark-quark correlator $G$ (the contributions to the necessary gauge link are not shown).}
\label{f:GTMD_correlator}
\end{figure}

Let us start our discussion on the various distribution functions from the so-called generalized TMDs (GTMDs). GTMDs depend on six momentum variables, and of all functions we will consider here, they encode most information about the distributions of partons inside hadrons. These functions appear in the parametrization of the GTMD correlator. The quark-quark GTMD correlator is defined as\footnote{It was recently shown that a proper definition of the GTMD correlator includes additional dependence on soft radiation described by a so-called soft factor~\cite{Echevarria:2016mrc}. However, as this will not play any role in this thesis, it is not discussed here (a similar dependence on soft radiation arises in the TMD case; this will be briefly discussed in section~\ref{s:final_steps_factorization}).}
\begin{equation}
    G_{\alpha\beta}(x,\bm{k},\xi,\bm{\Delta}) \equiv \int \left. \frac{dz^- d^2\bm{z}}{(2\pi)^3} \;e^{ik\cdot z} \bra{p'} \overline{\psi}_\beta\left( -\tfrac{z}{2} \right) U_{\left[-\tfrac{z}{2},\tfrac{z}{2}\right]} \,\psi_\alpha\left( \tfrac{z}{2} \right) \ket{p} \vphantom{\int} \right|_{z^+=0} .
    \label{e:qq_GTMDcorrelator}
\end{equation}
The term `generalized' in the name refers to the fact that this is the Fourier transform of an \emph{off-forward} matrix element. The off-forwardness is quantified by the momentum transfer $\Delta \equiv p'-p$ (the momenta $p$ and $p'$ can be taken to have large plus components). The average hadron and parton momenta are respectively given by $P \equiv (p'+p)/2$ and $k$, see also eqs.~\eqref{e:P} and~\eqref{e:k}. The kinematics for off-forward distributions is clarified in figure~\ref{f:GTMD_correlator}. Note that there is no final-state cut -- an off-forward correlator represents an amplitude rather than a squared amplitude. The momentum transfer $\Delta$ can be decomposed as
\begin{equation}
    \Delta^\mu = -2\xi P^+ \bar{n}^\mu + \frac{\xi M^2}{P^+} \,n^\mu + \Delta_\sst^\mu \,,
    \label{e:Delta}
\end{equation}
where $\xi$ denotes the skewness parameter, given by $2\xi \equiv -\Delta^+/P^+$ (its physical region corresponds to $-1 \leq \xi \leq 1$, see e.g.~\cite{Diehl:2003ny}).

For a spin-$\tfrac{1}{2}$ hadron, there are sixteen quark GTMDs at leading twist~\cite{Meissner:2009ww}. In contrast to all other distributions appearing in figure~\ref{f:projections}, GTMDs are complex-valued functions. Very recently it was shown for the first time that observables can be sensitive to quark GTMDs. It was found that they could be accessed (in principle) through exclusive double DY scattering~\cite{Bhattacharya:2017bvs}.

The so-called Wigner distributions can be obtained from GTMDs by a Fourier transformation from $\bm{\Delta}$ to $\bm{b}$:
\begin{equation}
    W_{\alpha\beta}(x,\bm{k},\xi,\bm{b}) \equiv \int d^2\bm{\Delta} \;e^{-i\bm{\Delta}\cdot \bm{b}} \,G_{\alpha\beta}(x,\bm{k},\xi,\bm{\Delta}) .
\end{equation}
The variable $\bm{b}$ is called the impact parameter and represents the average partonic transverse position. The six-dimensional Wigner functions for the parton-hadron system were introduced in~\cite{Ji:2003ak,Belitsky:2003nz} and are analogous to classical phase-space distributions. Whereas the latter are probability densities, the former are \emph{quasiprobability} densities -- Wigner distributions are for example not positive definite, but \emph{can} be used to compute expectation values of physical observables.

Quark GTMDs and Wigner distributions have not been measured yet in experiment and for that reason one currently has to rely on model studies, see e.g.~\cite{Lorce:2011dv,Lorce:2011ni,Mukherjee:2014nya,Liu:2014vwa,Liu:2015eqa,Lorce:2015sqe,More:2017zqq,Chakrabarti:2017teq,Kaur:2018dns}. In section~\ref{s:pA_collisions} we will study an observable that is sensitive to gluon Wigner distributions.

Upon integrating the GTMD correlator over $\bm{k}$, we obtain the so-called generalized parton distribution (GPD) correlator that depends on $x$, $\xi$, and $\bm{\Delta}$. At leading twist, there are eight quark GPDs for a spin-$\tfrac{1}{2}$ hadron~\cite{Diehl:2001pm}. Interest in quark GPDs is for example related to the observation that they could be used to obtain more information about the total angular momentum carried by quarks~\cite{Ji:1996ek}. GPDs can be accessed through various exclusive processes, albeit very challenging. We will not discuss GPDs further in this thesis; for more information, see e.g.~\cite{Diehl:2003ny}.

When we set $\Delta=0$ in the GTMD correlator (this is called the `forward' limit), we obtain the TMD correlator. As mentioned already in the introduction, TMDs depend on $x$ as well as on the partonic transverse momentum $\bm{k}$. Since the phenomenology of quark and gluon TMDs forms one of the main topics of this thesis, we provide an extensive introduction to the TMD formalism in the next section. Among other things, we will see that there are eight leading-twist quark TMDs for a spin-$\tfrac{1}{2}$ hadron~\cite{Mulders:1995dh}.

Since GPDs and TMDs can be obtained from GTMDs (or, equivalently, from Wigner distributions) through certain projections or limits, the latter are sometimes referred to as the `mother' distributions. It is important to note that GTMDs contain \emph{more information} on hadron structure than can be acquired from the combined knowledge of GPDs and TMDs -- several GTMDs vanish for the GPD and TMD limits of the correlator~\cite{Meissner:2009ww}. The reason for this is that GTMDs can contain information on correlations between $\bm{k}$ and $\bm{\Delta}$, which cannot be described by GPDs or TMDs alone.

Both the GPD and TMD correlators reduce to the collinear correlator in eq.~\eqref{e:PDF_quarkcorrelator} upon setting $\Delta=0$ and integrating over $\bm{k}$, respectively. For a spin-$\tfrac{1}{2}$ hadron, there are three collinear PDFs at leading twist~\cite{Ralston:1979ys}, all of which have been extracted from experiment.

\section{The TMD formalism} \label{s:TMD_formalism}

In this section we set up the TMD formalism. We discuss in detail the definition of the quark-quark TMD correlator and its parametrization in terms of quark TMDs (the gluon case is very similar and is extensively discussed in chapter~\ref{c:Polarized_gluon_TMDs}). Since hadron polarization forms a key component of TMD phenomenology, we start this section with a description of hadron spin. Although we will only consider spin-$1$ hadrons explicitly, spin-$\tfrac{1}{2}$ and spin-$0$ systems are trivially covered as well. In subsection~\ref{s:process_dependence} we explain the origin of the process dependence of TMDs. Most of the discussions in this section straightforwardly apply to GTMDs as well, which we encounter in chapter~\ref{c:Small_x} for the case of gluons. Introductions to the TMD formalism are also found in e.g.~\cite{Barone:2010zz,Buffing:2015tja,Signori:2016lvd}.

\subsection{Inclusion of hadron spin} \label{s:inclusion_spin}
So far we have ignored any dependence of the correlator on the spin polarization of the hadron. Whereas spin-$0$ hadrons are unpolarized (by definition), an ensemble of spin-$\tfrac{1}{2}$ hadrons can be unpolarized or vector polarized. To describe the degree of vector polarization, we make use of a spacelike spin vector $S$. For spin-$1$ hadrons, we need besides $S$ also a symmetric traceless spin tensor $T$ to describe tensor polarized states. The simplest and hence most-studied spin-$1$ system is the deuteron (this is a deuterium nucleus which consists of one proton and one neutron). Satisfying $P \cd S = 0$ and $P_\mu T^{\mu\nu} = 0$, the spin vector and spin tensor can be parametrized as follows~\cite{Bacchetta:2000jk,Bacchetta:2002}:
\begin{align}
    S^\mu &= S_L \frac{P^+}{M} \,\bar{n}^\mu - S_L \frac{M}{2 P^+} \,n^\mu + S_T^\mu \,, \label{e:spin_vector} \\[5pt]
    T^{\mu\nu} &= \frac{1}{2} \left[ \frac{4}{3} S_{LL} \frac{(P^+)^2}{M^2} \,\bar{n}^\mu \bar{n}^\nu + \frac{P^+}{M} \,\bar{n}_{\phantom{LT}}^{\{\mu} S_{LT}^{\nu\}} - \frac{2}{3} S_{LL} \left( \bar{n}^{\{\mu} n^{\nu\}} - g_\sst^{\mu\nu} \right) \right. \nn \\
    &\quad\, \left. + \,S_{TT}^{\mu\nu} - \frac{M}{2P^+} \,n_{\phantom{LT}}^{\{\mu} S_{LT}^{\nu\}} + \frac{1}{3} S_{LL} \frac{M^2}{(P^+)^2} \,n^\mu n^\nu \right] , \label{e:spin_tensor}
\end{align}
where curly brackets denote symmetrization of the indices, and $g_\sst^{\mu\nu} \equiv g^{\mu\nu} - \bar{n}^{\{\mu} n^{\nu\}}$ denotes the metric tensor in transverse space with nonzero components $g_\sst^{11} = g_\sst^{22} = -1$. The spin vector has three independent parameters, namely $S_L$ and the two transverse components of $S_T$; they characterize longitudinal and transverse polarization, respectively. The spin tensor, on the other hand, contains five independent parameters, namely $S_{LL}$, the two transverse components of $S_{LT}$, as well as the two independent components of the symmetric traceless transverse tensor $S_{TT}$. The physical interpretations of the various components of the spin tensor are discussed in e.g.~\cite{Bacchetta:2000jk,Bacchetta:2002}.

From the unintegrated correlator in eq.~\eqref{e:unintegrated_quarkcorrelator} we can obtain the TMD correlator upon integration over $k^-$. The quark-quark TMD correlator for spin-$1$ hadrons is given by\footnote{To obtain the correlator for a spin-$0$ or spin-$\tfrac{1}{2}$ hadron, one simply sets $S = T = 0$ or $T = 0$, respectively.}
\begin{equation}
    \Phi_{\alpha\beta}^{[U]}(x,\bm{k},S,T) \equiv \int \left. \frac{dz^- d^2\bm{z}}{(2\pi)^3} \;e^{ik\cdot z} \bra{P;S,T} \overline{\psi}_\beta(0) \,U_{[0,z]} \,\psi_\alpha(z) \ket{P;S,T} \vphantom{\int} \right|_{z^+=0} .
    \label{e:TMD_quarkcorrelator}
\end{equation}
Since this correlator is evaluated at $z^+ = 0$, it is also referred to as the light-front correlator. The gauge link that is required for gauge invariance is discussed in detail in subsection~\ref{s:process_dependence} (the dependence on the gauge link will be implicit for the most part). As will be briefly discussed in section~\ref{s:final_steps_factorization}, eq.~\eqref{e:TMD_quarkcorrelator} is actually not the full definition of the TMD correlator -- there is also a dependence on soft radiation. More specifically, the bilocal matrix element needs to be accompanied by a soft factor that removes so-called rapidity divergences~\cite{Collins:2011zzd,GarciaEchevarria:2011rb}.\footnote{The inclusion of a soft factor brings in a dependence on a rapidity scale. The dependence of TMDs on both renormalization and rapidity scales will be implicit throughout this thesis.} With regards to hadron spin, the correlator in eq.~\eqref{e:TMD_quarkcorrelator} is defined as
\begin{equation}
    \Phi_{\alpha\beta}(x,\bm{k},S,T) \equiv \tr \left( \rho(S,T) \,M_{\alpha\beta}(x,\bm{k}) \right) ,
    \label{e:spin_averaged}
\end{equation}
where the information on the spin states of the hadron is encoded in the $3\times3$ density matrix $\rho$ and the combined information on the hadron and quark spins is contained in $M$. The latter quantity can be regarded as a $6\times6$ matrix in parton $\otimes$ hadron spin space, which is discussed in detail for the case of gluons in section~\ref{s:positivity_bounds}. The density matrix can be parametrized in terms of $S$ and $T$ as follows:
\begin{equation}
    \rho(S,T) = \frac{1}{3} \left( I + \frac{3}{2} \,S^i \Sigma^i + 3 \,T^{ij} \Sigma^{ij} \right) ,
    \label{e:rho}
\end{equation}
where we have chosen a Cartesian basis of $3\times3$ matrices consisting of the identity matrix $I$, the three matrices $\Sigma^i$ (these are a generalization of the Pauli matrices to the three-dimensional case), and the five matrices $\Sigma^{ij} \equiv \tfrac{1}{2} (\Sigma^i \Sigma^j + \Sigma^j \Sigma^i) - \tfrac{2}{3} I \,\delta^{ij}$. Making use of eqs.~\eqref{e:spin_vector} and~\eqref{e:spin_tensor}, the density matrix takes the explicit form~\cite{Bacchetta:2000jk,Bacchetta:2002}:
\begin{equation}
    \rho(S,T) = \left( \begin{array}{ccc}
    \frac{1}{3} + \frac{S_L}{2} + \frac{S_{LL}}{3} & \frac{S_T^1 - iS_T^2}{2\sqrt{2}} + \frac{S_{LT}^1 - iS_{LT}^2}{2\sqrt{2}} &
    \frac{S_{TT}^{11} - iS_{TT}^{12}}{2} \\[5pt]
    \frac{S_T^1 + iS_T^2}{2\sqrt{2}} + \frac{S_{LT}^1 + iS_{LT}^2}{2\sqrt{2}} & \frac{1}{3} - \frac{2 S_{LL}}{3} & \frac{S_T^1 - iS_T^2}{2\sqrt{2}} - \frac{S_{LT}^1 - iS_{LT}^2}{2\sqrt{2}} \\[5pt]
    \frac{S_{TT}^{11} + iS_{TT}^{12}}{2} & \frac{S_T^1 + iS_T^2}{2\sqrt{2}} - \frac{S_{LT}^1 + iS_{LT}^2}{2\sqrt{2}} & \frac{1}{3} - \frac{S_L}{2} + \frac{S_{LL}}{3} \\[5 pt]
    \end{array} \right) .
    \label{e:density_matrix}
\end{equation}

As we will see in the next subsection, the inclusion of hadron spin in the correlator gives rise to a wealth of different TMDs. This is because TMDs can encode correlations between spin components and partonic transverse momentum.

\subsection{Parametrization of the TMD correlator} \label{s:parametrization_TMD_correlator}
Since a hadronic correlator cannot be calculated using perturbative methods, it is parametrized in terms of PDFs. The correlator in eq.~\eqref{e:TMD_quarkcorrelator} can be parametrized in terms of quark TMDs. For unpolarized and vector polarized hadrons, the first parametrizations were given in the 1990s in~\cite{Tangerman:1994eh,Boer:1997nt}, and in 2000 also tensor polarized hadrons were considered~\cite{Bacchetta:2000jk} (the collinear case was studied already in 1988 in~\cite{Hoodbhoy:1988am}). For the quark TMDs we will adopt the notation of~\cite{Bacchetta:2000jk}, which coincides with the notation in~\cite{Tangerman:1994eh,Boer:1997nt} for the unpolarized and vector polarized cases. This is often referred to as the `Amsterdam' notation.

It is easiest to start the parametrization at the unintegrated level and then perform the integration over $k^-$. The correlator is a matrix in Dirac space and hence it can be parametrized in terms of Dirac matrices. A convenient basis of Dirac structures is given by the set
\begin{equation}
    \{ I, \gamma^\mu, \sigma^{\mu\nu}, \gamma^\mu \gamma^5, i\gamma^5 \} ,
\end{equation}
where $\sigma^{\mu\nu} \equiv \tfrac{i}{2} [\gamma^\mu,\gamma^\nu]$. The various Dirac structures can be contracted with the available vectors and tensors $k$, $P$, $S$, $T$, and $n$.\footnote{The four-vector $n$ was not included in the original parametrizations in~\cite{Tangerman:1994eh,Boer:1997nt,Bacchetta:2000jk} but, as was argued in~\cite{Goeke:2003az}, it should be included as it is ultimately connected to the direction of the Wilson line. However, as it turns out, it will not give rise to any new functions at leading twist~\cite{Goeke:2005hb}.} The correlator is constrained by requirements that arise from Hermiticity and parity. For the unintegrated correlator, they are respectively given by~\cite{Bacchetta:2000jk}:
\begin{align}
    \Phi(k,P,S,T,n) &= \gamma^0 \,\Phi^\dag(k,P,S,T,n) \,\gamma^0 , \label{e:hermiticity} \\
    \Phi(k,P,S,T,n) &= \gamma^0 \,\Phi(\bar{k},\bar{P},-\bar{S},\bar{T},\bar{n}) \,\gamma^0 , \label{e:parity}
\end{align}
where we have introduced the notation $\bar{a}^\mu \equiv \delta^{\mu\nu} a_\nu$ and $\bar{b}^{\mu\nu} \equiv \delta^{\mu\rho} \delta^{\nu\sigma} b_{\rho\sigma}$. By contrast, time reversal (T) does \emph{not} constrain the form of the parametrization; this will be explained in the next subsection.

Parametrizations of hadronic correlators in terms of TMDs are usually given in momentum space (or $k_\sst$-space). However, for certain applications such as the implementation of TMD evolution, correlators are studied in coordinate space instead (or $z_\sst$-space, where $z_\sst$ is Fourier conjugate to $k_\sst$). To ensure a one-to-one correspondence between TMDs in $k_\sst$-space and $z_\sst$-space, it is essential that a correlator is parametrized in terms of TMDs of \emph{definite rank}, which has not been done consistently in existing parametrizations. To this end, a correlator in $k_\sst$-space needs to be parametrized using symmetric traceless tensors in $k_\sst$ (see appendix~\ref{a:STTs} for the definitions of these tensors up to rank $4$). Here we will present the parametrization of the TMD correlator in terms of symmetric traceless tensors given in our work in~\cite{vanDaal:2016glj}, which is based on the original parametrization in~\cite{Bacchetta:2000jk}.
Separating the various possible hadronic polarization states, the correlator in eq.~\eqref{e:TMD_quarkcorrelator} can be parametrized in terms of leading-twist quark TMDs of definite rank as follows:\footnote{In most of this thesis, the dependence of correlators on $P$, $S$, $T$, or $n$ is implicit. }
\begin{align}
    \Phi(x,\bm{k}) &= \Phi_U(x,\bm{k}) + \Phi_L(x,\bm{k}) + \Phi_T(x,\bm{k}) \nn \\
    &\quad\, + \Phi_{LL}(x,\bm{k}) + \Phi_{LT}(x,\bm{k}) + \Phi_{TT}(x,\bm{k}) ,
    \label{e:quark_TMD_parametrization}
\end{align}
with
\begin{align}
    \Phi_U(x,\bm{k}) &= \frac{1}{2} \left[ \slashed{\bar{n}} \,f_1(x,\bm{k}^2) + \frac{\sigma_{\mu\nu} k_\sst^\mu \bar{n}^\nu}{M} \,h_1^\perp(x,\bm{k}^2) \right] , \label{e:quark_TMD_parametrization_unpolarized} \\[5pt]
    \Phi_L(x,\bm{k}) &= \frac{1}{2} \left[ \gamma^5 \slashed{\bar{n}} \,S_L \,g_1(x,\bm{k}^2) + \frac{i\sigma_{\mu\nu} \gamma^5 \bar{n}^\mu k_\sst^\nu S_L}{M} \,h_{1L}^\perp(x,\bm{k}^2) \right] , \\[5pt]
    \Phi_T(x,\bm{k}) &= \frac{1}{2} \left[ \frac{\slashed{\bar{n}} \,\epsilon_\sst^{S_T k_\sst}}{M} \,f_{1T}^\perp(x,\bm{k}^2) + \frac{\gamma^5 \slashed{\bar{n}} \,\bm{k} \cd \bm{S}_T}{M} \,g_{1T}(x,\bm{k}^2) \right. \nn \\
    &\quad\, \left. + \,i\sigma_{\mu\nu} \gamma^5 \bar{n}^\mu S_T^\nu \,h_1(x,\bm{k}^2) - \frac{i\sigma_{\mu\nu} \gamma^5 \bar{n}^\mu k_\sst^{\nu\rho} {S_T}_\rho}{M^2} \,h_{1T}^\perp(x,\bm{k}^2) \right] , \label{e:quark_TMD_parametrization_vpolarized} \\[5pt]
    \Phi_{LL}(x,\bm{k}) &= \frac{1}{2} \left[ \slashed{\bar{n}} \,S_{LL} \,f_{1LL}(x,\bm{k}^2) + \frac{\sigma_{\mu\nu} k_\sst^\mu \bar{n}^\nu S_{LL}}{M} \,h_{1LL}^\perp(x,\bm{k}^2) \right] , \\[5pt]
    \Phi_{LT}(x,\bm{k}) &= \frac{1}{2} \left[ \frac{\slashed{\bar{n}} \,\bm{k} \cd \bm{S}_{LT}}{M} \,f_{1LT}(x,\bm{k}^2) + \frac{\gamma^5 \slashed{\bar{n}} \,\epsilon_\sst^{S_{LT}k_\sst}}{M} \,g_{1LT}(x,\bm{k}^2) \right. \nn \\
    &\quad\, \left. + \,\sigma_{\mu\nu} \bar{n}^\nu S_{LT}^\mu \,h_{1LT}(x,\bm{k}^2) - \frac{\sigma_{\mu\nu} \bar{n}^\nu k_\sst^{\mu\rho} {S_{LT}}_\rho}{M^2} \,h_{1LT}^\perp(x,\bm{k}^2) \right] , \\[5pt]
    \Phi_{TT}(x,\bm{k}) &= \frac{1}{2} \left[ \frac{\slashed{\bar{n}} \,k_\sst^{\mu\nu} {S_{TT}}_{\mu\nu}}{M^2} \,f_{1TT}(x,\bm{k}^2) - \frac{\gamma^5 \slashed{\bar{n}} \,{\epsilon_\sst}_{\mu\nu} k_\sst^{\mu\rho} {S_{TT}^\nu}_\rho}{M^2} \,g_{1TT}(x,\bm{k}^2) \right. \nn \\
    &\quad\, \left. - \,\frac{\sigma_{\mu\nu} \bar{n}^\nu k_\sst^\rho {S_{TT}^\mu}_\rho}{M} \,h_{1TT}(x,\bm{k}^2) + \frac{\sigma_{\mu\nu} \bar{n}^\nu k_\sst^{\mu\rho\sigma} {S_{TT}}_{\rho\sigma}}{M^3} \,h_{1TT}^\perp(x,\bm{k}^2) \right] ,
\end{align}
where we have employed the notation $\epsilon_\sst^{ab} \equiv \epsilon_\sst^{\mu\nu} a_\mu b_\nu$, with $\epsilon_\sst^{\mu\nu} \equiv \epsilon^{\mu\nu-+}$ (its nonzero components are $\epsilon_\sst^{12} = -\epsilon_\sst^{21} = 1$). To avoid clutter, we suppress in the names of the functions a reference to quarks as it should be clear from the context whether we are discussing quark or gluon TMDs. As we will motivate in the next subsection, TMDs depend on the precise form of the gauge link that is used. In appendix~\ref{a:coordinate_space} we provide the equivalent parametrization of the TMD correlator in coordinate space.

In table~\ref{t:overview_quark_tmds} we have organized the quark TMDs appearing in eq.~\eqref{e:quark_TMD_parametrization} by hadron and quark polarizations. We have also indicated the behavior under time reversal (the difference between T-even and T-odd functions will be explained in subsection~\ref{s:process_dependence}) and which TMDs have a collinear counterpart. The polarization of the quarks is projected out by the Dirac structures that come with the functions; we use the letters $f$, $g$, and $h$ to indicate unpolarized, longitudinally polarized, and transversely polarized quarks, respectively. \\

\begin{table*}[!htb]
\begin{center}
{\renewcommand{\arraystretch}{1.4}
\begin{tabular}{|c|c|c|c|}
\hline
& $\gamma^+$ & $\gamma^+ \gamma^5$ & $i \sigma^{j+} \gamma^5$ \\ \hline \hline
U & $\bm{f_1}$ & & $\color{red}h_1^\perp$ \\ \hline
L & & $\bm{g_1}$ & $h_{1L}^\perp$ \\ \hline
T & $\color{red}f_{1T}^\perp$ & $g_{1T}$ & $\bm{h_1}$, $h_{1T}^\perp$ \\ \hline
LL & $\bm{f_{1LL}}$ & & $\color{red}h_{1LL}^\perp$ \\ \hline
LT & $f_{1LT}$ & $\color{red}g_{1LT}$ & $\color{red}h_{1LT}$, $\color{red}h_{1LT}^\perp$ \\ \hline
TT & $f_{1TT}$ & $\color{red}g_{1TT}$ & $\color{red}h_{1TT}$, $\color{red}h_{1TT}^\perp$ \\
\hline
\end{tabular}}
\caption{An overview of the leading-twist quark TMDs for unpolarized (U), vector polarized (L or T), and tensor polarized (LL, LT, or TT) hadrons. The functions indicated in boldface also occur as collinear PDFs, and the ones in red are T-odd. The Dirac structures $\gamma^+$, $\gamma^+ \gamma^5$, and $i\sigma^{j+} \gamma^5$ correspond to unpolarized, longitudinally polarized, and transversely polarized quarks, respectively.}
\label{t:overview_quark_tmds}
\end{center}
\end{table*}

As with respect to existing literature our parametrization has been written completely in terms of symmetric traceless tensor structures, the functions that were not of definite rank have been replaced by functions that are of definite rank. Employing the common shorthand notation
\begin{equation}
     f^{(n)}(x,\bm{k}^2) \equiv \left( \frac{\bm{k}^2}{2M^2} \right)^n f(x,\bm{k}^2) ,
     \label{e:moment_notation}
\end{equation}
the functions $h_{1T}$, $h_{1LT}'$, and $h_{1TT}'$ that appear in the original parametrizations in~\cite{Tangerman:1994eh,Bacchetta:2000jk} have been replaced by the following definite-rank TMDs:\footnote{These definitions were already proposed in~\cite{Bacchetta:2000jk}, so there is no conflict of notation here.}
\begin{equation}
    h_1 \equiv h_{1T} + h_{1T}^{\perp(1)} , \qquad h_{1LT} \equiv h_{1LT}' + h_{1LT}^{\perp(1)} , \qquad h_{1TT} \equiv h_{1TT}' + h_{1TT}^{\perp(1)} ,
\end{equation}
where, as we will do often, we suppressed the arguments of the functions to lighten the notation. An additional benefit of using definite-rank TMDs is that the functions also appearing in the collinear case, i.e.\ those that survive integration over $\bm{k}$, are simply the rank-$0$ functions (the function $h_{1LT}$ is an exception to this, though; this will be commented on in the next subsection). Furthermore, we have defined $g_1 \equiv g_{1L}$.

Some of the TMDs also go by a name, especially those appearing in the description of a spin-$\tfrac{1}{2}$ hadron. The most well-known T-even TMDs are $f_1$, $g_1$, and $h_1$, referred to as the unpolarized, helicity, and transversity functions, respectively. The unpolarized distribution represents the number density of unpolarized quarks with momentum fraction $x$ and transverse momentum $\bm{k}$ inside an unpolarized hadron. The helicity distribution, on the other hand, represents the number density of quarks with momentum fraction $x$ and transverse momentum $\bm{k}$ and polarization parallel to that of the hadron \emph{minus} the number density of quarks with the same momentum but antiparallel polarization. The interpretation of $h_1$ is similar to that of $g_1$, but involves transversely polarized quarks instead. These three functions also have collinear counterparts~\cite{Ralston:1979ys} that bear the same names. We will discuss T-odd functions in the next subsection.

As will be explained in detail in section~\ref{s:positivity_bounds} for the case of gluons, in general TMDs satisfy certain bounds (also called positivity bounds) that follow from the operator structure of the correlator. In the quark case they arise from the positive semidefiniteness of the matrix $M$ appearing in eq.~\eqref{e:spin_averaged}. Bounds on quark TMDs can be found in~\cite{Bacchetta:1999kz} for spin-$\tfrac{1}{2}$ hadrons and in~\cite{Bacchetta:2001rb} for spin-$1$ hadrons. For example, the two TMDs appearing in the parametrization for unpolarized hadrons in eq.~\eqref{e:quark_TMD_parametrization_unpolarized} satisfy the following inequality:
\begin{equation}
    \frac{|\bm{k}|}{M} \,|h_1^\perp| \leq f_1 \,,
\end{equation}
which provides an upper bound for the magnitude of the function $h_1^\perp$.

\subsection{Process dependence} \label{s:process_dependence}
As explained in subsection~\ref{s:unintegrated_correlator}, ensuring gauge invariance of the correlator requires the inclusion of a gauge link. In the collinear case, discussed in subsection~\ref{s:collinear_PDFs}, we saw that the path of the gauge link is fixed; it runs along the minus direction at transverse position $\bm{\eta}=\bm{0}$. However, now the situation is different. In the TMD case, the gauge link is no longer unique as the quark fields are not only separated in the minus direction but $\emph{also}$ in the transverse directions. This means that the gauge link necessarily includes a transverse piece~\cite{Belitsky:2002sm,Boer:2003cm}.

\begin{figure}[htb]
\centering
    \begin{subfigure}{.42\textwidth}
    \begin{flushleft}
        \includegraphics[height=2.5cm]{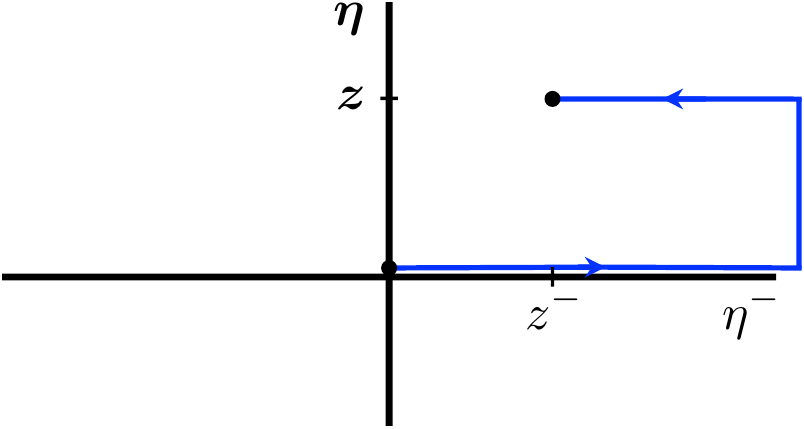}
    \end{flushleft}
    \end{subfigure}
    \begin{subfigure}{.42\textwidth}
    \begin{flushright}
        \includegraphics[height=2.5cm]{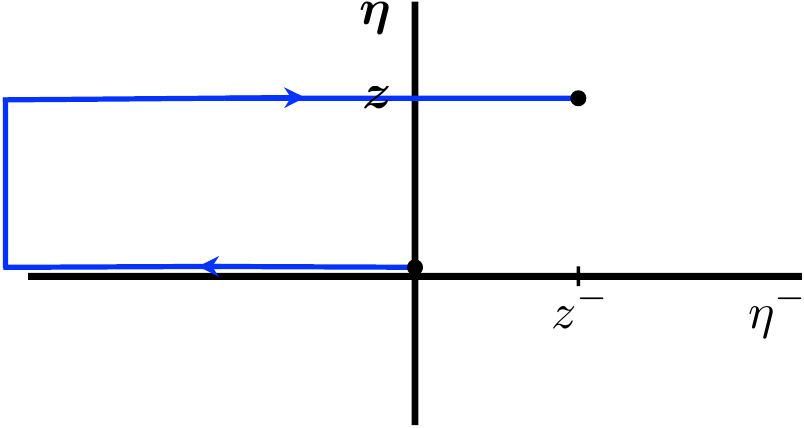}
    \end{flushright}
    \end{subfigure}
    \caption{The integration paths of the future-pointing gauge link $U_{[0,z]}^{[+]}$ (on the left) and of the past-pointing gauge link $U_{[0,z]}^{[-]}$ (on the right). The paths run along the light-front $\eta^+=0$ via plus or minus light-cone infinity, respectively.}
\label{f:plusminus_links}
\end{figure}

The two most important gauge links that will be discussed in this thesis are the so-called future- and past-pointing gauge links (also referred to as the `staple-like' gauge links) denoted by $U_{[0,z]}^{[+]}$ and $U_{[0,z]}^{[-]}$, respectively. They are defined as
\begin{equation}
    U_{[0,z]}^{[\pm]} \equiv U_{[0,\pm \infty;\bm{0}]}^n \,U_{[\pm \infty;\bm{0},\bm{z}]}^T \,U_{[\pm \infty,z^-;\bm{z}]}^n \,.
    \label{e:WLs}
\end{equation}
The links along the minus direction are defined in eq.~\eqref{e:U^n} and the link along the transverse directions (indicated by the superscript $T$) is given by
\begin{equation}
    U_{[\pm \infty;\bm{0},\bm{z}]}^T \equiv \mathcal{P} \exp\left( ig \int_{\bm{0}}^{\bm{z}} d\bm{\eta} \cdot \bm{A}(\eta^+=0,\eta^-=\pm \infty,\bm{\eta}) \right) .
    \label{e:U^T}
\end{equation}
The integration paths of the staple-like gauge links are illustrated in figure~\ref{f:plusminus_links}.

Any path connecting the fields ensures gauge invariance, however, as mentioned already in subsection~\ref{s:unintegrated_correlator}, different paths reflect different physics and turn out to correspond to different processes. The process dependence of the gauge link (and thus of the TMD correlator) originates from the color structure of the hard scattering process. In general, the more partons participate in the hard scattering, the more complicated the resulting gauge links~\cite{Bomhof:2006dp}. The future- and past-pointing gauge links of eq.~\eqref{e:WLs} are obtained in the SIDIS and DY processes, respectively. In DY the color flows into the initial state (the gluon interactions happen \emph{before} the quarks are annihilated), whereas in SIDIS the color flows into the final state (the gluon interactions happen \emph{after} the quark is struck by the photon). For this reason, the associated gauge links are also referred to as initial- and final-state Wilson lines, respectively.

In the previous subsection we discussed the constraints on the correlator that arise from Hermiticity and parity, see eqs.~\eqref{e:hermiticity} and~\eqref{e:parity}. Now, let us discuss the effects of time reversal transformations on the correlator (we will only consider staple-like gauge links here). Time reversal leads to the following constraint~\cite{Bacchetta:2000jk}:
\begin{equation}
    \Phi^{[\pm]*}(k,P,S,T,n) = i\gamma^1\gamma^3 \,\Phi^{[\mp]}(\bar{k},\bar{P},\bar{S},\bar{T},\bar{n}) \,i\gamma^1\gamma^3 .
    \label{e:time_reversal}
\end{equation}
Note that under time reversal a future-pointing gauge link becomes a past-pointing one and vice versa. This means that time reversal does not constrain a single correlator -- it relates \emph{two different} correlators with different gauge links. Hence, unlike Hermiticity and parity, time reversal does not actually constrain the parametrization of the TMD correlator in terms of TMDs.

From eq.~\eqref{e:time_reversal} it follows that we can define T-even and T-odd contributions to the correlator according to the behavior
\begin{align}
    \Phi^{(\text{T-even})\ast}(k,P,S,T,n) &= i\gamma^1\gamma^3 \,\Phi^{(\text{T-even})}(\bar{k},\bar{P},\bar{S},\bar{T},\bar{n}) \,i\gamma^1\gamma^3 , \\
    \Phi^{(\text{T-odd})\ast}(k,P,S,T,n) &= - i\gamma^1\gamma^3 \,\Phi^{(\text{T-odd})}(\bar{k},\bar{P},\bar{S},\bar{T},\bar{n}) \,i\gamma^1\gamma^3 ,
\end{align}
where the T-even and T-odd parts of $\Phi$ are given by
\begin{align}
    \Phi^{(\text{T-even})}(k,P,S,T,n) &= \frac{1}{2} \left[ \Phi^{[+]}(k,P,S,T,n) + \Phi^{[-]}(k,P,S,T,n) \right] , \label{e:T-even} \\[5pt]
    \Phi^{(\text{T-odd})}(k,P,S,T,n) &= \frac{1}{2} \left[ \Phi^{[+]}(k,P,S,T,n) - \Phi^{[-]}(k,P,S,T,n) \right] \label{e:T-odd}.
\end{align}
This means that we can classify TMDs as being either T-even or T-odd, see table~\ref{t:overview_quark_tmds}. The most well-known T-odd functions are $h_1^\perp$, called the Boer-Mulders (BM) function~\cite{Boer:1997nt}, and $f_{1T}^\perp$, called the Sivers function~\cite{Sivers:1989cc,Sivers:1990fh}. The former describes a transversely polarized quark inside an unpolarized hadron, whereas the latter describes an unpolarized quark inside a transversely polarized hadron.

The existence of T-odd functions completely relies on the gauge link structure -- after all, if we were to naively set the gauge links to unity,\footnote{In a light-cone gauge, the transverse links do \emph{not} reduce to unity~\cite{Belitsky:2002sm}.} these functions would be \emph{forbidden} by time reversal symmetry. For that reason, T-odd functions were initially thought to vanish (for more than a decade), until a model calculation by Brodsky, Hwang, and Schmidt (BHS) in 2002 showed that the Sivers function could be nonzero~\cite{Brodsky:2002cx}. A few years later, in 2004, a nonzero quark Sivers function was measured for the first time by the HERMES experiment at DESY~\cite{Airapetian:2004tw}. The quark BM and Sivers functions will play an important role in chapter~\ref{c:Disentanglement}.

Since T-odd TMDs change sign under time reversal (which amounts to swapping future- and past-pointing gauge links), their signs actually depend on whether they are probed in SIDIS or in DY. This was first pointed out by Collins in a seminal paper~\cite{Collins:2002kn} a few months after the BHS calculation came out. The sign flip of T-odd TMDs between SIDIS and DY is one of the most striking predictions of the TMD formalism. By contrast, T-even quark TMDs are the same in both processes. For example, the unpolarized and BM functions that appear in the parametrization for unpolarized hadrons in eq.~\eqref{e:quark_TMD_parametrization_unpolarized} are related in SIDIS and DY as
\begin{equation}
    f_1^{[+]} = f_1^{[-]} , \qquad h_1^{\perp[+]} = -h_1^{\perp[-]} .
    \label{e:SIDISvs.DY}
\end{equation}
These relations follow from the parity and time reversal constraints given in eqs.~\eqref{e:parity} and~\eqref{e:time_reversal}, respectively. The process dependence of TMDs (both T-even and T-odd ones) becomes more involved in processes with more complicated gauge links~\cite{Buffing:2012sz,Buffing:2013kca,Boer:2015kxa}. Although for a given process the gauge link can always be calculated, it is not always possible to relate TMDs between different processes through a constant factor as in eq.~\eqref{e:SIDISvs.DY}~\cite{Bomhof:2006dp}. We will consider an example of the latter for the case of gluons in section~\ref{s:process_dependence_gluons}.

One might wonder whether the described nonuniversality of TMDs can spoil factorization. As it turns out, the factorized structure of a cross section can remain valid as long as TMDs are defined with the appropriate process-dependent gauge links (this is sometimes referred to as \emph{generalized} factorization). This for example means that we take future-pointing links for SIDIS and past-pointing ones for DY. In section~\ref{s:possible problems} we mention a process for which this procedure potentially fails.

A different way in which PDFs in general and TMDs in particular depend on the hard process under consideration is through the energy scale at which the hard scattering occurs. The energy scale dependence (as well as the dependence on the rapidity scale) can be calculated perturbatively, though. This means that TMDs that are extracted in one process at a particular scale can be used in other processes by evolving them to the appropriate scale. Hence, the violations of universality, both due to the dependence on the color structure and the energy scale of the hard process, are calculable -- they are, in a sense, `under control', enabling global extractions of TMDs.

Upon integrating the TMD correlator over $\bm{k}$, we recover the collinear situation, in which case the staple-like gauge links reduce to the unique \emph{process-independent} link $U_{[0,z^-;\bm{0}]}^n$ given in eq.~\eqref{e:U^n}. It then follows from eq.~\eqref{e:T-odd} that all T-odd functions vanish. Consequently, all collinear PDFs must be T-even. Even though the T-odd, rank-$0$ Bacchetta function $h_{1LT}$ survives integration over $\bm{k}$, it in fact vanishes due to the combined effects of Hermiticity and time reversal constraints~\cite{Signori:2016lvd}, as it should.

Finally, let us briefly comment on the gauge link (in)dependence of the other multidimensional distributions that appear in the scheme in figure~\ref{f:projections}. Not only TMDs suffer from complications regarding universality -- in fact, \emph{all} distributions that have a dependence on $\bm{k}$ are in principle process dependent. Thus, besides TMDs, also GTMDs and Wigner distributions come with a process-dependent gauge link. Similarly to the collinear correlator, the GPD correlator features a Wilson line with a unique straight path directed along the minus direction at $\bm{\eta}=\bm{0}$. Hence, just as collinear PDFs, also GPDs are universal.

\subsection{Status of extractions} \label{s:phenomenology}
So far, our discussion on TMDs has been rather theoretical. In this subsection we briefly comment on the experimental side of TMD phenomenology. We will focus here on quark rather than gluon TMDs (the latter will be covered in section~\ref{s:experimental possibilities}).

As mentioned already in the introduction, TMD effects typically give rise to angular correlations between the produced particles. An example of an angular correlation is the well-known $\cos(2\phi)$ azimuthal asymmetry in both the SIDIS and DY cross sections as induced by the quark BM function $h_1^\perp$ (see section~\ref{s:DY_scattering} for more details). The BM function can be extracted through a measurement of the azimuthal angular distribution of the final-state particles. Since $h_1^\perp$ encodes a correlation between partonic transverse momentum and partonic transverse spin, its extraction provides access to the internal (transverse) structure of hadrons. The experimental status of the BM function will be discussed in section~\ref{s:BM_effect}.

One of the vital tests of the TMD formalism is to experimentally verify the predicted sign change of T-odd functions between SIDIS and DY. A nonzero quark Sivers function $f_{1T}^\perp$ has been extracted from SIDIS measurements by various experiments, see e.g.~\cite{Airapetian:2009ae,Ageev:2006da,Alekseev:2008aa,Adolph:2012sp,Adolph:2014zba,Adolph:2016dvl,Qian:2011py}. Recently, $f_{1T}^\perp$ was also extracted from $W$ boson production and DY measurements (both processes come with a future-pointing gauge link): both the STAR experiment at BNL~\cite{Adamczyk:2015gyk} and the COMPASS experiment at CERN~\cite{Aghasyan:2017jop} show results that favor a sign change, but more data is needed to provide conclusive evidence. Due to its profound origin (gauge invariance and factorization theorems), a falsification of the expected sign flip would imply that there is something very general about QCD that we do not understand, or (probably more likely) that something has been overlooked in the factorization proofs.

Of all TMDs, the most well-known one is the unpolarized quark TMD $f_1$. In fact, recently $f_1$ was extracted from a fit to available SIDIS and DY data~\cite{Bacchetta:2017gcc} -- the first \emph{global} extraction of a TMD. To this end, TMD evolution at next-to-leading logarithmic accuracy was implemented to connect the data at different scales.

Most TMDs are (virtually) unknown, though. The future Electron-Ion Collider (EIC), which is currently considered in the United States, might change this and revolutionize the field of TMD phenomenology~\cite{Accardi:2012qut,Ent:2016lod}. At high luminosity and high energy, it could scatter electrons off polarized protons and light nuclei, including deuterons. Since it will be able to probe small values of $x$, also sea quark and gluon TMDs could be extracted with much higher precision than what has been achieved so far. More information on the experimental status of TMDs can be found in e.g.~\cite{Boer:2011fh,Aidala:2012mv,Angeles-Martinez:2015sea,Boglione:2015zyc,Bacchetta:2016ccz,Signori:2016lvd}.

\stopthumb
\part{Factorization of Drell-Yan scattering} \label{p:1}
\chapter{Elements of CSS factorization} \label{c:CSS}
\addthumb{\thechapter}{\Large{\thechapter}}{\thumbfg}{\thumbbg}

Part~\ref{p:1} of this thesis is centered around the concept of factorization, which forms the cornerstone of QCD phenomenology. In particular, we discuss factorization of the DY cross section differential in the transverse momentum of the lepton pair, involving one TMD for each colliding hadron -- this will be referred to as TMD factorization. The DY cross section integrated over transverse momentum was shown to factorize in terms of a hard scattering factor and collinear PDFs (called collinear factorization) to all orders in perturbation theory in the 1980s by Collins, Soper, and Sterman (CSS)~\cite{Collins:1985ue,Collins:1988ig}. This proof of factorization is highly nontrivial and contains many ingredients -- in~\cite{Collins:1988ig} it was remarked: ``The complexity of such a proof will be comparable to that of proofs of renormalizability''. In the TMD case, the factorization proof proceeds largely along the same lines and is extensively covered in~\cite{Collins:2011zzd}. The latter is \emph{also} referred to as the CSS proof and forms the main topic of this chapter. Other approaches to TMD factorization, particularly in the context of soft-collinear effective theory (SCET)~\cite{Becher:2010tm,GarciaEchevarria:2011rb,Echevarria:2012js}, will not be discussed here.

It is unclear whether TMD factorization of DY also holds for T-odd TMDs that are associated to azimuthal asymmetries; no explicit proof for this has been provided yet. In~\cite{Buffing:2013dxa} it was suggested that in this case the factorization theorem, if it holds, would come with a modified color factor compared to the naive prediction based on the CSS works. In chapter~\ref{c:Disentanglement} we will investigate this issue in detail by performing a factorization calculation largely along the lines of the CSS proof. To facilitate that calculation, we provide in this chapter the necessary ingredients. In particular, we discuss how leading-power regions of loop momenta can be identified and how contributions from those regions need to be combined; the latter procedure is given by the Collins subtraction scheme. Furthermore, we review the complications that arise from so-called Glauber modes. A significant part of the CSS proof is devoted to the unitarity cancellation of Glauber exchanges, which turns out to be an important aspect in our factorization study in chapter~\ref{c:Disentanglement}.

Because of its importance, factorization is a vast topic for which a lot of literature is available. The material presented in this chapter is largely based on treatments in Collins's book ``Foundations of perturbative QCD''~\cite{Collins:2011zzd}, our review in~\cite{Boer:2017hqr}, as well as on reviews provided in~\cite{Diehl:2011yj,Diehl:2015bca,Gaunt:2014ska}; other useful resources include~\cite{Collins:1989gx,Sterman:1995fz,Dominguez:2011saa,Rogers:2015sqa,Bonocore:2016wur}.

\section{Concept of factorization}

Factorization is a consequence of the approximation that long-distance dynamics describing the hadron structure decouples from short-distance dynamics which describes a hard scattering event. The most well-known and well-established form of factorization is collinear factorization. This type of factorization applies to scattering processes in which all involved kinematic scales are large and of the same order; other energy scales, such as $\Lambda_\text{QCD}$ and $M$, characterize the long-distance physics. An example of such a process is inclusive DY scattering in which the virtuality of the intermediate gauge boson sets the hard scale. The transverse momentum of the boson should either be of the same order as the hard scale or integrated over. Collinear factorization of DY means that its cross section factorizes into two collinear PDFs (one for each hadron), and a perturbatively calculable hard factor that describes the partonic scattering.

The idea of factorization arises very naturally. In DY the colliding hadrons and associated soft gluon fields are Lorentz contracted in the center-of-mass (CM) frame of the collision to very thin objects such that they essentially do not `feel' each other prior to the collision. Furthermore, a large virtuality of the gauge boson implies that a hard scattering event happens on a very short time scale, while the soft physics that governs the internal structure of hadrons involves much longer time scales. Thus, intuitively, the hadrons appear `frozen' with respect to the hard scattering event and therefore should remain unaffected. Hence, the short-distance dynamics is decoupled from the long-distance dynamics, with any interference effects suppressed by inverse powers of the hard scale. This picture also illustrates why the PDFs are expected to be universal -- they pertain to hadrons and are independent of any hard scattering event. This makes factorization the cornerstone of QCD phenomenology; the PDFs that are extracted from one process can be used in others. Without factorization, QCD loses its predictive power in scattering processes.

Since factorization is such a vital component of QCD, a solid proof of this is essential. As it turns out, proving factorization is very challenging, for instance due to the fact that arbitrarily many soft gluons can be exchanged between the two hadrons. However, in the 1980s CSS managed to establish collinear factorization of DY scattering to all orders in perturbation theory~\cite{Collins:1985ue,Collins:1988ig}. The CSS proof holds up to leading power in the inverse hard scale. A few years later, the proof was extended to the first subleading power~\cite{Qiu:1990xxa,Qiu:1990xy}.

In most of this thesis and in particular in part~\ref{p:1}, we focus on another type of factorization called TMD factorization. This type of factorization, unlike collinear factorization, takes into account transverse momentum effects. TMD factorization is different due to the presence of two kinematic scales in the process that are of different order. One scale is provided by the virtuality of the gauge boson, $Q^2$, and the other by its noncollinearity, $Q_\sst^2$. If the latter is substantially smaller than the former, one could expand in their ratio and only consider the leading term (also logarithms of the form $\alpha_s \ln(Q/Q_\sst)$ need to be resummed). At leading power and to all orders in perturbation theory, TMD factorization for DY was established in~\cite{Collins:2011zzd}. This proof proceeds largely along the same lines as the one for the collinear case. Besides DY, TMD factorization has also been proven to hold for $e^+ e^-$ annihilation into two hadrons~\cite{Collins:1981uk} and SIDIS~\cite{Ji:2004wu}, and has largely been completed for the double DY process~\cite{Diehl:2011yj,Diehl:2015bca,Buffing:2017mqm}. Note that all these processes involve two hadrons only.

\section{The Drell-Yan process} \label{s:DY_scattering}

\begin{figure}[htb]
\centering
    \includegraphics[height=4.1cm]{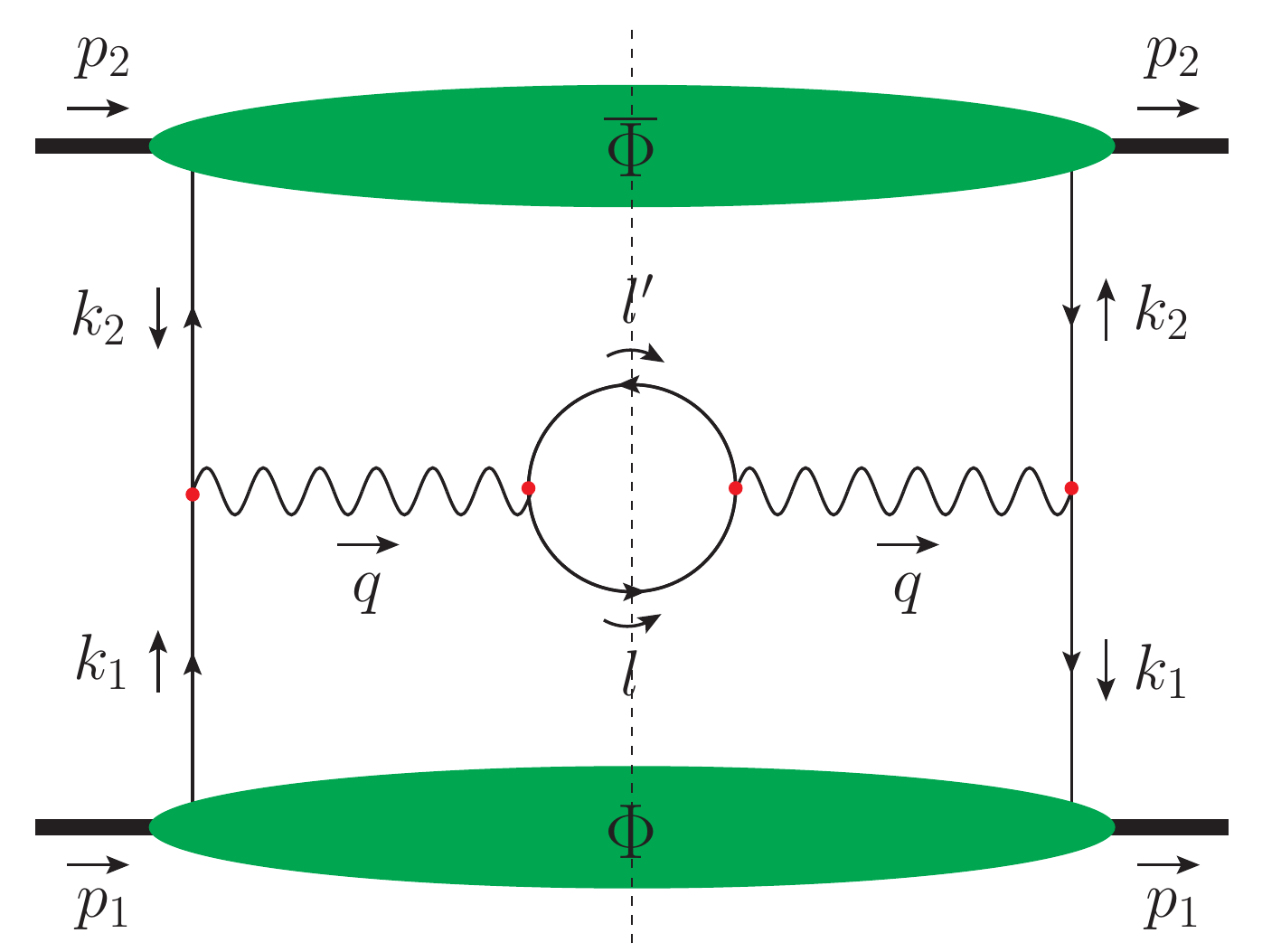}
    \caption{The tree-level graph for the DY process: a quark and antiquark from the colliding hadrons produce a virtual photon that subsequently decays into a charged lepton-antilepton pair. The green `blobs' represent the quark and antiquark correlators.}
\label{f:DY}
\end{figure}

In part~\ref{p:1} of this thesis we focus on inclusive DY scattering. The DY process was proposed in 1970 to describe the production of lepton-antilepton pairs in high-energy hadron-hadron collisions~\cite{Drell:1970wh}. In the collision of two hadrons with momenta $p_1$ and $p_2$, a quark and antiquark with momenta $k_1$ and $k_2$ interact to produce a virtual photon (or $Z$ boson) with momentum $q$. This gauge boson in turn decays into a charged lepton-antilepton pair with momenta $l$ and $l'$. The leading-order diagram for this process is schematically illustrated in figure~\ref{f:DY}. Although for definiteness we will consider $p\bar{p}$ collisions in this part of the thesis, most discussions trivially apply to other hadrons as well.

The distribution of quarks and antiquarks inside the colliding protons and antiprotons is encoded by the quark and antiquark correlators $\Phi$ and $\overline{\Phi}$, respectively. As discussed in detail in subsection~\ref{s:parametrization_TMD_correlator}, these correlators can be parametrized in terms of TMDs. For unpolarized hadrons, the quark-quark TMD correlator can be parametrized in terms of two leading-twist quark TMDs, namely the T-even unpolarized function $f_1$ and the T-odd BM function $h_1^\perp$, see eq.~\eqref{e:quark_TMD_parametrization_unpolarized} (we will denote the antiquark analogues with a bar). These TMDs depend on the longitudinal momentum fraction $x_1 \equiv k_1^+/p_1^+$, as well as on the transverse momentum $\bm{k}_1^2$.

\begin{figure}[htb]
\centering
    \includegraphics[height=3.0cm]{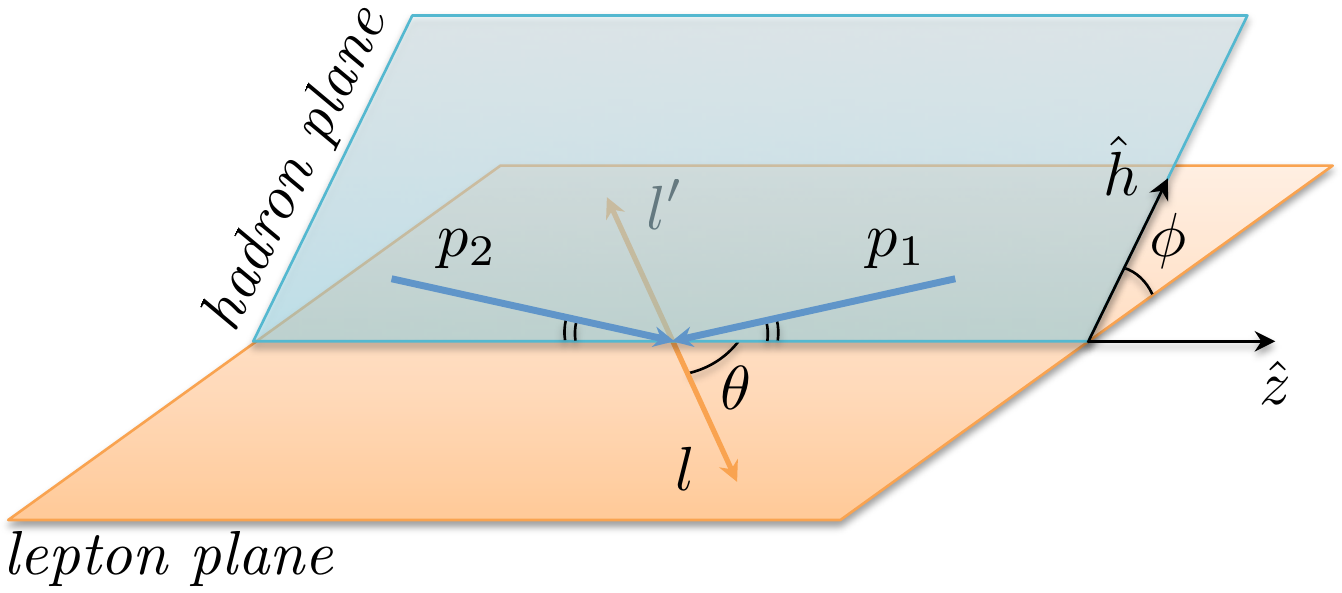}
    \caption{The definition of the angles $\theta$ and $\phi$ in the CS frame. This frame is a lepton CM frame in which the hadrons come in symmetrically with respect to the $z$-axis.}
\label{f:CS}
\end{figure}

The TMD factorization theorem by CSS holds up to leading power in $\Lambda/Q$, where $Q^2 \equiv q^2 > 0$ represents the hard scale of the process and $\Lambda$ represents the transverse momentum of the electroweak boson, $Q_\sst \equiv \sqrt{\bm{q}^2}$, or the scale of nonperturbative QCD interactions. For unpolarized hadrons, the factorization formula at leading order in the hard scattering takes the following form~\cite{Boer:1999mm,Boer:2002ju}:
\begin{align}
    \frac{d\sigma}{d\Omega \,dx_1 dx_2 \,d^2\bm{q}} &= \frac{\alpha^2}{N_c \,q^2} \,\sum_q e_q^2 \left\{ A(\theta) \,\mathcal{F}\left[ f_1 \bar{f}_1 \right] \vphantom{\left[ w(\bm{k}_1,\bm{k}_2) \,h_1^\perp \bar{h}_1^\perp \right]} \right. \nn \\
    &\quad\, \left. + \,B(\theta) \cos(2\phi) \,\mathcal{F}\left[ w(\bm{k}_1,\bm{k}_2) \,h_1^\perp \bar{h}_1^\perp \right] \right\} ,
    \label{e:fac}
\end{align}
with the convolution of TMDs defined as:
\begin{equation}
    \mathcal{F}\left[ f_1 \bar{f}_1 \right] \equiv \int d^2\bm{k}_1 \int d^2\bm{k}_2 \;\delta^{(2)}(\bm{k}_1+\bm{k}_2-\bm{q}) \,f_{1,q}(x_1,\bm{k}_1^2) \,\bar{f}_{1,q}(x_2,\bm{k}_2^2) .
\end{equation}
The functions $A$ and $B$ are given by
\begin{equation}
    A(\theta) = \frac{1}{4} (1 + \cos^2\theta) , \qquad B(\theta) = \frac{1}{4} \sin^2\theta ,
\end{equation}
and the BM weight function reads
\begin{equation}
    w(\bm{k}_1,\bm{k}_2) = \frac{2(\hat{h} \cd \bm{k}_1)(\hat{h} \cd \bm{k}_2) - \bm{k}_1 \cd \bm{k}_2}{M^2} \,.
\end{equation}
The cross section is given in terms of the Collins-Soper (CS) angles $\theta$ and $\phi$~\cite{Collins:1977iv}, see figure~\ref{f:CS}. The unit vector $\hat{h}$ is defined in the hadron CM frame as $\hat{h} \equiv \bm{q}/|\bm{q}|$, and $p_1^2 = p_2^2 = M^2$, where $M$ is the mass of the hadron, as before. The sum in eq.~\eqref{e:fac} runs over the different quark flavors labeled by $q$. Furthermore, the electrical charge $e_q$ is given in units of the elementary charge $e$, and $\alpha$ denotes the fine-structure constant.

We will refer to the first term in eq.~\eqref{e:fac} as the `double unpolarized' contribution, because it involves unpolarized quarks, and the second term will be referred to as the `double BM' (dBM) contribution and involves transversely polarized quarks. The dBM term comes with a $\cos(2\phi)$ dependence which can be used to isolate this contribution in the experimental data. The BM function $h_1^\perp$, as opposed to the unpolarized TMD $f_1$, is called chiral-odd as it flips the quark helicity between amplitude and conjugate amplitude. Since helicity is conserved (in the massless limit), the BM function needs to be coupled to another chiral-odd function in the cross section. Hence, for the unpolarized DY process, $h_1^\perp$ can only appear together with $\bar{h}_1^\perp$.

The quark TMDs $f_1$ and $h_1^\perp$ appearing in eq.~\eqref{e:fac} have the following operator definitions:
\begin{align}
    f_{1,q}(x_1,\bm{k}_1^2) &\equiv \int \left. \frac{dz^- d^2\bm{z}}{(2\pi)^3} \;e^{ik_1 \cdot z} \bra{p_1} \overline{\psi}_q(0) \,U_{[0,z]}^{[-]} \,\Gamma_U \,\psi_q(z) \ket{p_1} \vphantom{\int} \right|_{z^+=0} , \label{e:f_1} \\
    \frac{\widetilde{k}_{1\sst}^j}{M} \,h_{1,q}^\perp(x_1,\bm{k}_1^2) &\equiv \int \left. \frac{dz^- d^2\bm{z}}{(2\pi)^3} \;e^{ik_1 \cdot z} \bra{p_1} \overline{\psi}_q(0) \,U_{[0,z]}^{[-]} \,\Gamma_T^j \,\psi_q(z) \ket{p_1} \vphantom{\int} \right|_{z^+=0} , \label{e:h_1}
\end{align}
where we have employed the notation $\widetilde{a}_\sst^{\nu} \equiv \epsilon_\sst^{\mu\nu} {a_\sst}_\mu$. The necessary soft factors are not considered explicitly as they will not play a role in our factorization calculation in chapter~\ref{c:Disentanglement}. As appropriate for the DY process (see section~\ref{s:process_dependence}), both TMDs come with the past-pointing gauge link $U_{[0,z]}^{[-]}$, defined in eq.~\eqref{e:WLs}. Furthermore, $\Gamma_U$ and $\Gamma_T^j$ are Dirac projectors that, respectively, select unpolarized and transversely polarized quarks:
\begin{equation}
    \Gamma_U \equiv \tfrac{1}{2} \gamma^+ , \qquad \Gamma_T^j \equiv \tfrac{1}{2} i \,\sigma^{j+} \gamma^5 ,
    \label{e:projector}
\end{equation}
where $j$ is a transverse index. In the TMD definitions~\eqref{e:f_1} and \eqref{e:h_1} a summation over color is implicitly assumed. Since only quarks and antiquarks of the same color can annihilate, the factorization formula, eq.~\eqref{e:fac}, appears with a $1/N_c$ color factor. The definitions of the unpolarized and BM functions for the antiquark are analogous to the quark case.

In the next sections we discuss various ingredients of the CSS factorization proof that are essential for our factorization calculation in chapter~\ref{c:Disentanglement}. Although we do not discuss different quark or hadron polarizations explicitly, the CSS proof is intended to apply to all polarized cases. In section~\ref{s:possible problems} we will come back to this statement.

\section{Momentum regions analysis} \label{s:momentum_regions_analysis}

The TMD factorization approach is based on an expansion in powers of the small parameter $\Lambda/Q$ and, today, factorization of the DY cross section has only been established up to leading power in this parameter. In this section we discuss how leading graph contributions are found. For each graph that contributes to the cross section, the Libby-Sterman method can be used to identify the leading regions of loop momentum~\cite{Libby:1978bx,Sterman:1978bi}. Since the details of this method are not relevant for our factorization study in chapter~\ref{c:Disentanglement}, we will only summarize its main results here; a detailed review of this method can be found in~\cite{Collins:2011zzd}. Subsequently, in subsection~\ref{s:subtraction_method}, we will discuss the Collins subtraction scheme that describes how the various leading region contributions should be combined to reconstruct the full graph.

\subsection{Identifying leading regions} \label{s:leading_regions}
Naively, one expects dominant contributions to loop integrations for a given graph to originate from momentum regions around points at which certain lines go on shell, as in that case the propagator denominators vanish.\footnote{An exception to this is the hard region defined in~\eqref{e:Hscaling}, which can contribute at leading power due to the volume of phase space~\cite{Collins:2011zzd}.} However, if the pole structure of the propagators is such that the Feynman integrals (that run along real axes) can avoid these singularities, then, according to Cauchy's theorem, the integration contours can be deformed into the complex plane away from the poles, giving finite integrals. Hence the following condition arises: the leading momentum regions must correspond to situations where the integration contours cannot be deformed away from the on-shell poles. This is the case when two poles on opposite sides of the contour are merged into a single point upon setting to zero all quantities of order $\Lambda$ in the diagram; the contour is then `pinched', prohibiting a deformation away from the singularity. The conditions on the loop momenta that lead to a pinch define a surface in multidimensional momentum space, called the pinch surface. Leading momentum regions are each associated to such a surface. For a given diagram, the pinch surfaces can be found with the Landau equations~\cite{Landau:1959fi}, aided by the Coleman-Norton method~\cite{Coleman:1965xm}. Since we will not make use of these techniques in chapter~\ref{c:Disentanglement}, we will not discuss them further here.

Subsequently, a power counting analysis is used to determine which momentum scalings give leading contributions to the DY graphs. We choose a coordinate system in the hadron CM frame where both incoming hadrons have zero transverse momentum, with one hadron moving fast in the $+\hat{z}$ direction (the `right-moving' hadron) and the other moving fast in the $-\hat{z}$ direction (the `left-moving' hadron). For the DY process, it follows from the Libby-Sterman method that the potentially relevant regions of loop momentum $\ell$ are given by~\cite{Libby:1978bx,Sterman:1978bi,Collins:2011zzd}:\footnote{The term `central' refers to the central rapidity region. The rapidity $y$ of a four-vector $\ell$ is defined as $y \equiv \tfrac{1}{2} \ln(|\ell^+/\ell^-|)$.}
\begin{alignat}{2}
    &\text{hard ($H$):} \qquad &&\ell \sim (1,1,1)Q, \label{e:Hscaling} \\
    &\text{right-moving collinear ($C_1$):} \qquad &&\ell \sim (1,\lambda^2,\lambda)Q, \label{e:C1scaling} \\
    &\text{left-moving collinear ($C_2$):} \qquad &&\ell \sim (\lambda^2,1,\lambda)Q, \label{e:C2scaling} \\
    &\text{central soft ($S$):} \qquad &&\ell \sim (\lambda,\lambda,\lambda)Q, \label{e:Sscaling} \\
    &\text{central ultrasoft ($U$):} \qquad &&\ell \sim (\lambda^2,\lambda^2,\lambda^2)Q, \label{e:USscaling} \\
    &\text{Glauber:} \qquad &&|\ell^+\ell^-| \ll \bm{\ell}^2 \ll Q^2 , \label{e:Gscaling_generic}
\end{alignat}
where $\lambda$ is a small parameter of order $\Lambda/Q$. In the CSS approach, the soft and ultrasoft regions are treated together and in this chapter we often use `soft' to refer to both the central soft and central ultrasoft regions simultaneously; in chapter~\ref{c:Disentanglement}, however, we distinguish between these two modes. The Glauber condition~\eqref{e:Gscaling_generic} allows for various possible scalings that are in fact all treated together in the CSS methodology. The Glauber scalings that are important in the analysis of chapter~\ref{c:Disentanglement} are:
\begin{alignat}{2}
    &\text{right-moving Glauber ($G_1$):} \qquad &\ell &\sim (\lambda,\lambda^2,\lambda)Q, \label{e:G1scaling}\\
    &\text{left-moving Glauber ($G_2$):} \qquad &\ell &\sim (\lambda^2,\lambda,\lambda)Q, \label{e:G2scaling} \\
    &\text{central Glauber ($G$):} \qquad &\ell &\sim (\lambda^2,\lambda^2,\lambda)Q. \label{e:Gscaling}
\end{alignat}

\begin{figure}[htb]
\centering
    \includegraphics[height=4.03cm]{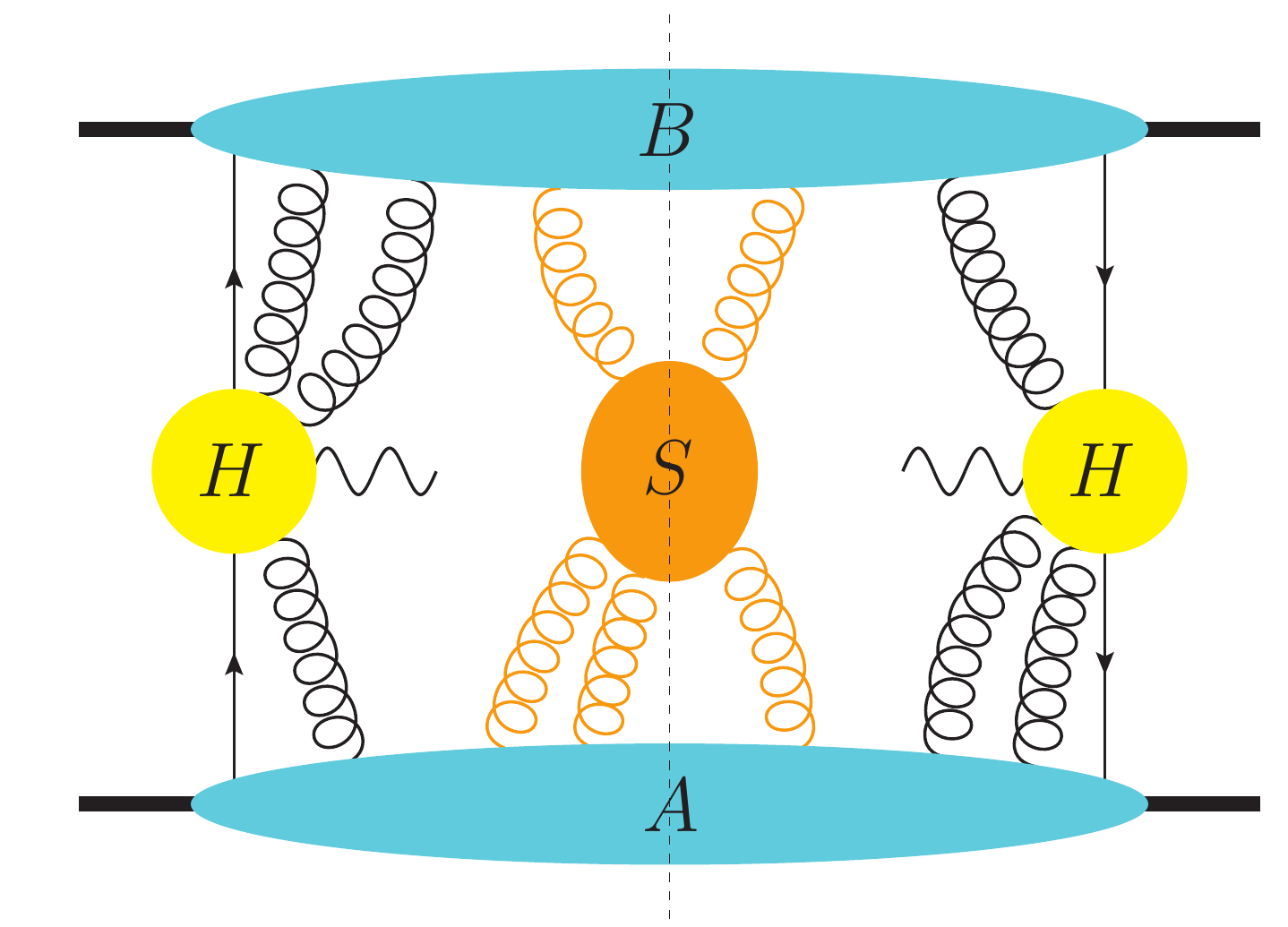}
    \caption{The partitioning of the leading-power DY graphs in the TMD case into various subgraphs (represented by `blobs') that are each characterized by a particular momentum scaling. The right- and left-moving collinear subgraphs are respectively denoted by $A$ and $B$, the soft (plus Glauber) subgraph by $S$, and the hard subgraph by $H$. For convenience we have suppressed the final-state leptons.}
\label{f:subgraphs}
\end{figure}

The DY graphs can be organized into subgraphs that each characterize a particular momentum scaling. From the Libby-Sterman method it follows which connections between subgraphs give a leading contribution. In the TMD case, the dominant graphs for DY have the structure shown in figure~\ref{f:subgraphs}. There are two collinear subgraphs: the one corresponding to the right-moving hadron is denoted by $A$ and the other one by $B$. Furthermore, there is a hard subgraph denoted by $H$ that is connected to both $A$ and $B$ by one fermion line and an arbitrary number of gluons. Lastly, there is a subgraph $S$ that initially contains both soft and Glauber modes, which connect via soft and Glauber gluon attachments to either of the collinear subgraphs.

\begin{figure}[htb]
\centering
    \begin{subfigure}{.46\textwidth}
    \begin{flushleft}
        \includegraphics[height=4.1cm]{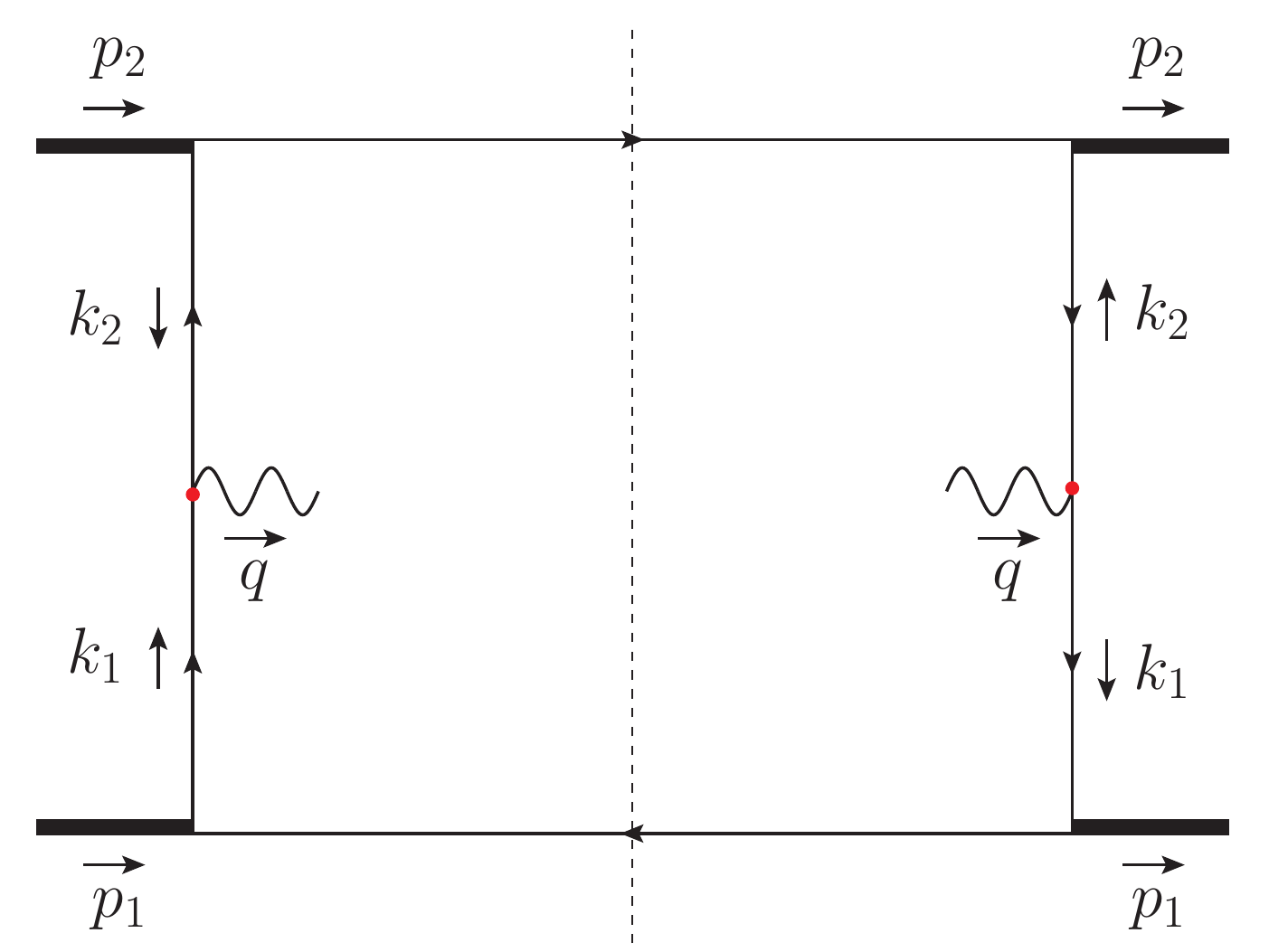}
    \end{flushleft}
    \end{subfigure}
    \begin{subfigure}{.46\textwidth}
    \begin{flushright}
        \includegraphics[height=4.1cm]{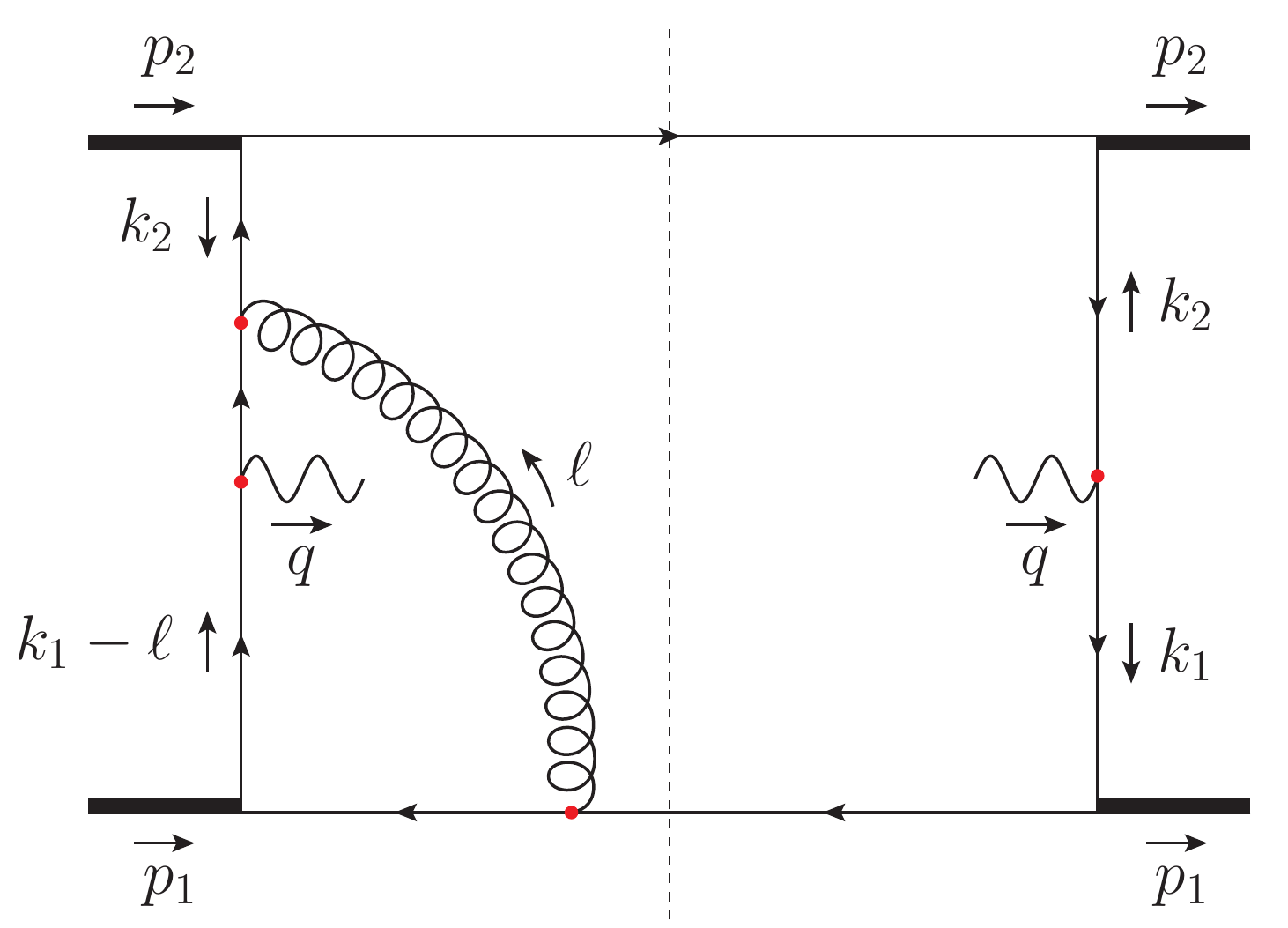}
    \end{flushright}
    \end{subfigure}
    \caption{The diagram on the right features an active-spectator gluon exchange. Various gluon momentum regions contribute at leading power, which is set by the tree-level graph given on the left.}
\label{f:power_counting}
\end{figure}

To get a feel for how the power counting works in practice for a simple DY graph, we now consider an example. To this end, we use the model that is discussed in section 14.3 of~\cite{Collins:2011zzd}: the hadrons are treated as being composed of a quark and an antiquark (the details are not relevant here). Let us consider a single-gluon exchange graph where the gluon is connected between an `active' and a `spectator' line (an active line is involved in the hard scattering while a spectator line is not).\footnote{A more general definition of active and spectator lines is given in~\cite{Rothstein:2016bsq}.} The proton and active quark carry $C_1$ momenta, whereas the antiproton and active antiquark carry $C_2$ momenta; to be precise,
\begin{alignat}{2}
    p_1 &\sim (1,\lambda^2,0)Q , &\qquad p_2 &\sim (\lambda^2,1,0)Q , \label{e:scaling_p12} \\
    k_1 &\sim (1,\lambda^2,\lambda)Q , &\qquad k_2 &\sim (\lambda^2,1,\lambda)Q . \label{e:scaling_k12}
\end{alignat}
To determine which gluon momentum scalings give a leading contribution to the graph, one first needs to define what is meant by `leading'. By definition, the leading power is set by the tree-level graph.\footnote{Sometimes a region gives a super-leading contribution to an individual graph. However, as it turns out, those contributions cancel generally upon summing over all graphs~\cite{Collins:2011zzd}.} Both the tree-level and single-gluon exchange graphs are given in figure~\ref{f:power_counting}. With respect to the tree-level graph there are three additional propagators in the single-gluon exchange graph, namely those pertaining to the momenta $\ell$, $k_2+\ell$, and $p_1-k_1+\ell$. Moreover, one propagator is modified in the amplitude: $k_1 \to k_1-\ell$, which means that its scaling might change with respect to the tree-level graph (depending on the scaling of $\ell$). In fact, it is easy to see that the scaling of this line is unchanged if $\ell^-$ is at most of order $\lambda^2 Q$ -- then there is no `penalty' factor. In that case, the exchanged gluon could give a leading contribution if
\begin{equation}
    \int \frac{d^4\ell}{(2\pi)^4} \,\frac{i}{(p_1-k_1+\ell)^2 + i\epsilon} \,\frac{i}{\ell^2 + i\epsilon} \,\frac{i}{(k_2+\ell)^2 + i\epsilon}
    \label{e:power_counting}
\end{equation}
is of order $1$ (for convenience we will set $Q=1$ in this example).

Let us now systematically consider all relevant regions defined in~\eqref{e:Hscaling}--\eqref{e:USscaling} for the gluon loop momentum $\ell$, including the various Glauber scalings given in~\eqref{e:G1scaling}--\eqref{e:Gscaling}, to investigate which ones give a leading-power contribution to the graph:
\begin{itemize}
    \item \textbf{The hard region ($\bm{H}$).} Since $\ell \sim (1,1,1)$, we have $d^4\ell \sim 1$ and $\ell^2 \sim 1$. Furthermore, $(p_1-k_1+\ell)^2 \sim 1$ and $(k_2+\ell)^2 \sim 1$  (note that in this example there are no extra factors of $\lambda$ coming from propagator numerators), and there is a penalty factor of order $\lambda^2$ coming from the $k_1-\ell$ line. This means that the $H$ region is suppressed with respect to the tree-level graph by a factor of order $\lambda^2$.
    \item \textbf{The collinear regions ($\bm{C_1}$ and $\bm{C_2}$).} For $C_1$, $\ell \sim (1,\lambda^2,\lambda)$, so that $d^4\ell \sim \lambda^4$ and $\ell^2 \sim \lambda^2$. Furthermore, $(p_1-k_1+\ell)^2 \sim \lambda^2$ and $(k_2+\ell)^2 \sim 1$, which means that the expression in~\eqref{e:power_counting} is of order $1$. Hence, the $C_1$ region gives a leading contribution. For the $C_2$ region, however, there is a penalty factor of order $\lambda^2$ coming from the $k_1-\ell$ line, leading to an overall $\lambda^2$ suppression.
    \item \textbf{The central soft region ($\bm{S}$).} Since $\ell \sim (\lambda,\lambda,\lambda)$, we have $d^4\ell \sim \lambda^4$ and $\ell^2 \sim \lambda^2$. Furthermore, $(p_1-k_1+\ell)^2 \sim \lambda$ and $(k_2+\ell)^2 \sim \lambda$, and there is a penalty factor of order $\lambda$ coming from the $k_1-\ell$ line. This means that the $S$ region is suppressed with respect to the tree-level graph by a factor of order $\lambda$.
    \item \textbf{The central ultrasoft region ($\bm{U}$).} Since $\ell \sim (\lambda^2,\lambda^2,\lambda^2)$, we have $d^4\ell \sim \lambda^8$ and $\ell^2 \sim \lambda^4$. Furthermore, $(p_1-k_1+\ell)^2 \sim \lambda^2$ and $(k_2+\ell)^2 \sim \lambda^2$, which means that the expression in~\eqref{e:power_counting} is of order $1$. Hence, the $U$ region gives a leading contribution.
    \item \textbf{The noncentral Glauber regions ($\bm{G_1}$ and $\bm{G_2}$).} For $G_1$, $\ell \sim (\lambda,\lambda^2,\lambda)$, so that $d^4\ell \sim \lambda^5$ and $\ell^2 \sim \lambda^2$. Furthermore, $(p_1-k_1+\ell)^2 \sim \lambda^2$ and $(k_2+\ell)^2 \sim \lambda$, which means that the expression in~\eqref{e:power_counting} is of order $1$. Hence, the $G_1$ region gives a leading contribution. For the $G_2$ region, however, there is a penalty factor of order $\lambda$ coming from the $k_1-\ell$ line, leading to an overall $\lambda$ suppression.
    \item \textbf{The central Glauber region ($\bm{G}$).} Since $\ell \sim (\lambda^2,\lambda^2,\lambda)$, we have $d^4\ell \sim \lambda^6$ and $\ell^2 \sim \lambda^2$. Furthermore, $(p_1-k_1+\ell)^2 \sim \lambda^2$ and $(k_2+\ell)^2 \sim \lambda^2$, which means that the expression in~\eqref{e:power_counting} is of order $1$. Hence, the $G$ region gives a leading contribution.
\end{itemize}
From this power counting analysis it follows that the $C_1$, $U$, $G_1$, and $G$ regions give rise to a leading-power contribution, whereas the other region contributions are power suppressed. However, unlike $C_1$ and $U$, the $G_1$ and $G$ regions do \emph{not} correspond to pinch surfaces. As will be discussed in detail in subsection~\ref{s:trapped_Glauber}, the $\ell^+$ contours are not pinched in the Glauber regions, which allows for contour deformations into the $C_1$ region. This implies that there is no need to consider the Glauber regions separately from $C_1$.

To sum up, the (nontrivial) leading contributions to the single-gluon exchange diagram come from the $U$ and $C_1$ regions. They are associated to the pinch surfaces defined by $\ell=0$ and $\ell \propto k_1$, respectively. Furthermore, we note that the single-gluon exchange graph in figure~\ref{f:power_counting} in fact contributes to \emph{all} subgraphs defined in figure~\ref{f:subgraphs}: the $C_1$ gluon (being a part of $A$) connects the $A$ and $H$ subgraphs (the $k_2+\ell$ line is hard), whereas the $U$ gluon (being a part of $S$) connects the $A$ and $B$ subgraphs. The left-moving lines constitute the $B$ subgraph.

Knowing the leading regions, we can approximate the complicated integral over the loop momentum $\ell$. Denoting the single-gluon exchange graph by $\Gamma$ (only the relevant part is considered here), we can write
\begin{equation}
    \Gamma \equiv \int \frac{d^4\ell}{(2\pi)^4} \,J(\ell) ,
\end{equation}
with
\begin{equation}
    J(\ell) \equiv \frac{i}{(k_1-\ell)^2 + i\epsilon} \,\frac{i}{(p_1-k_1+\ell)^2 + i\epsilon} \,\frac{i}{\ell^2 + i\epsilon} \,\frac{i}{(k_2+\ell)^2 + i\epsilon} \,.
\end{equation}
Up to power corrections, $\Gamma$ is given by the sum of the two leading region contributions:
\begin{equation}
    \Gamma = \int_{U} \frac{d^4\ell}{(2\pi)^4} \;J_U(\ell) + \int_{C_1} \frac{d^4\ell}{(2\pi)^4} \;J_{C_1}(\ell) .
\end{equation}
The subscript $U$ on the integral denotes the restriction of the integration to the central ultrasoft region, and the same subscript on the integrand denotes the application of the leading-power approximation as appropriate for the central ultrasoft region (and likewise for the $C_1$ region). The restriction of the integration domain to a certain region turns out to be inconvenient in practice; we will come back to this in the next subsection. If we were instead to integrate $J_U(\ell)$ over \emph{all} values of $\ell$, then also contributions are picked up from regions where $J_U(\ell)$ fails to be a good approximation (and similarly for the $C_1$ region). In the next subsection we discuss how this problem can be solved by introducing appropriate subtraction terms to avoid any double counting.

\subsection{Collins subtraction method} \label{s:subtraction_method}
To obtain the leading-power contribution to the DY cross section, one needs to sum over all leading momentum regions for a given graph and subsequently sum over all graphs. In this subsection we focus on the former; we discuss how a graph can be reconstructed from its leading region contributions. This procedure is described by the Collins subtraction method that was introduced in~\cite{Collins:2011zzd} and reviewed in~\cite{Diehl:2015bca}; we follow the discussion of the latter.

Let us denote by $\Gamma$ a particular Feynman graph (integrated over all its loop momenta) and by $R$ a leading momentum region of $\Gamma$, specified by the set of scalings for all loop momenta. A region $R$ can contain any of the hard, collinear, soft, and Glauber scalings as defined in~\eqref{e:Hscaling}--\eqref{e:Gscaling_generic}. With each region $R$ we associate an approximator $T_R$ that approximates the integrand of $\Gamma$ as appropriate for that region. We will employ an approximator that drops all terms in numerators and denominators that are power suppressed compared to the terms that are leading in the considered region. In general, approximators must satisfy $T_R \Gamma_R = \Gamma_R$ (up to power corrections), where the subscript $R$ on $\Gamma$ denotes the restriction of the loop integrals to the `design' region $R$. This restriction, however, complicates the analysis as the introduction of momentum cut-offs breaks Lorentz invariance. Hence, simply adding up the various contributions $\Gamma_R$ to approximate the full graph is not what we are after.

The problem of restricting the loop integrals to the appropriate design regions can be overcome by instead integrating the approximated integrands over the \emph{full} range of loop momenta. Consequently, the `unwanted' contributions that are included in this way need to be properly subtracted. This is achieved by the Collins subtraction method~\cite{Collins:2011zzd}.\footnote{A similar method exists in the context of SCET called zero-bin subtraction~\cite{Manohar:2006nz}.} For a given region $R$, contributions from so-called smaller regions need to be subtracted to avoid double counting. The hierarchy for the various possible regions is determined by the dimensions of the associated pinch surfaces: we say that $R'$ is smaller than $R$ (or $R'<R$) if $R'$ is associated to a lower-dimensional pinch surface. The contribution from a particular region $R$ to a graph $\Gamma$, including subtractions to avoid double counting with smaller regions, is denoted by $C_R \Gamma$ and given by
\begin{equation}
    C_R \Gamma \equiv T_R \left( \Gamma - \sum_{R'<R} C_{R'} \Gamma \right) .
    \label{e:subtr-def}
\end{equation}
This is a recursive definition of $C_R \Gamma$, starting from the smallest leading region (note that for the smallest region $R=R_0$ one simply has $C_{R_0} \Gamma = T_{R_0} \Gamma$). We refer to the first and second terms on the right-hand side of eq.~\eqref{e:subtr-def} by `naive graph' and `subtraction' terms, respectively. Finally, the correct leading-power approximation to the graph $\Gamma$ can be obtained by summing over all leading region contributions $C_R \Gamma$:
\begin{equation}
    \Gamma = \sum_R C_R \Gamma .
    \label{e:subtr-sum}
\end{equation}
We refer to~\cite{Collins:2011zzd} for a proof of this result.

Now let us apply the Collins subtraction method to the example discussed in the previous subsection. We found that the loop momentum $\ell$ in the single-gluon exchange graph in figure~\ref{f:power_counting} receives leading-power contributions from two (nontrivial) momentum regions, namely from the $U$ and $C_1$ regions. Since the associated pinch surfaces are, respectively, zero- and one-dimensional, we have $U < C_1$. Hence, it follows from eqs.~\eqref{e:subtr-def} and~\eqref{e:subtr-sum} that we can approximate $\Gamma$ at leading power as
\begin{equation}
    \Gamma = C_U \Gamma + C_{C_1} \Gamma = T_U \Gamma + T_{C_1} \left( 1 - T_U \right) \Gamma .
    \label{e:subtractions_example}
\end{equation}
The naive graph terms are given by
\begingroup 
\thinmuskip=1.5mu
\medmuskip=2.0mu
\begin{align}
    \hspace{-0.10cm} T_U \Gamma &= \int \frac{d^4\ell}{(2\pi)^4} \,\frac{i}{2k_2^-\ell^+ + i\epsilon} \,\frac{i}{2k_1^+\ell^- + i\epsilon} \,\frac{i}{2(p_1^+ - k_1^+)\ell^- + i\epsilon} \,\frac{i}{2\ell^+\ell^- - \bm{\ell}^2 + i\epsilon} \,, \\[5pt]
    \hspace{-0.15cm} T_{C_1} \Gamma &= \int \frac{d^4\ell}{(2\pi)^4} \,\frac{i}{2k_2^-\ell^+ + i\epsilon} \,\frac{i}{2(k_1^+ - \ell^+)(k_1^- - \ell^-) - (\bm{k}_1 - \bm{\ell})^2 + i\epsilon} \nn \\
    &\quad\, \times \frac{i}{2(p_1^+ - k_1^+ + \ell^+)(p_1^- - k_1^- + \ell^-) - (\bm{k}_1 - \bm{\ell})^2 + i\epsilon} \,\frac{i}{2\ell^+\ell^- - \bm{\ell}^2 + i\epsilon} \,.
\end{align}
\endgroup
It is easy to see that in this simple example the subtraction term $T_{C_1} T_U \Gamma$ in fact equals the naive graph term $T_U \Gamma$. Hence, the graph can be approximated as $\Gamma = T_{C_1} \Gamma$; we say that the $U$ region can be `absorbed' into the $C_1$ region.

Finally, let us make an important remark about the subtraction procedure. Due to the presence of the subtraction terms in eq.~\eqref{e:subtr-def} it is sufficient to consider only the design region of momentum for a particular region contribution to a given graph. We will make use of this fact in the factorization calculation in chapter~\ref{c:Disentanglement}. For example, the $C_1$ contribution to a graph $\Gamma$ that receives leading contributions from both the $S$ and $C_1$ regions is given by $C_{C_1} \Gamma = T_{C_1} (1 - T_S)\Gamma$. However, in the region $S$ this contribution vanishes, $C_{C_1} \Gamma_S = 0$, as $T_S$ then reduces to the unit operator. To establish factorization it is necessary to verify that the various steps described in this chapter hold for both the naive graph and subtraction terms.

\subsection{Rapidity divergences} \label{s:rapidity_divergences}
A complication of the momentum regions analysis is that it can come with so-called rapidity divergences, which need to be properly regulated~\cite{Manohar:2006nz}. Rapidity divergences can arise if the observable under consideration receives contributions from different momentum regions that are of the same virtuality but have distinct rapidities. For example, both the central soft and collinear regions have the same number of powers of $\lambda$ in their phase space $\int d^4\ell$ (and are thus of the same virtuality), but their rapidities are parametrically different -- soft modes have small rapidities as $\ell^+/\ell^- \sim 1$, while collinear modes have very large rapidities as $\ell^\pm/\ell^\mp \gg 1$. When both of these regions contribute to an observable, rapidity divergences can appear; this is illustrated below.

To illustrate how rapidity divergences arise, let us consider the following (finite) integral:
\begin{equation}
    I \equiv \int_{Q_\sst}^Q \frac{d\ell^+}{\ell^+} = \int \frac{d\ell^+}{\ell^+} \;\theta\left( \ell^+ - Q_\sst \right) \theta\left( Q - \ell^+ \right) .
    \label{e:rap_div_example_start}
\end{equation}
The momentum component $\ell^+$ scales differently in different momentum regions; for example, $\ell^+ \sim Q_\sst$ in $S$ and $\ell^+ \sim Q$ in $C_1$. Hence, the integral range of $I$ covers both the $S$ and $C_1$ momentum regions. According to eqs.~\eqref{e:subtr-def} and~\eqref{e:subtr-sum}, $I$ can be approximated at leading power as
\begin{equation}
    I = C_S I + C_{C_1} I = T_S I + T_{C_1} \left( 1 - T_S \right) I .
    \label{e:I_subtraction}
\end{equation}
With the naive graph and subtraction terms given by
\begin{align}
    T_S I &= \int \frac{d\ell^+}{\ell^+} \;\theta\left( \ell^+ - Q_\sst \right) , \\[5pt]
    T_{C_1} I &= \int \frac{d\ell^+}{\ell^+} \;\theta\left( \ell^+ \right) \theta\left( Q - \ell^+ \right) , \\[5pt]
    T_{C_1} T_S I &= \int \frac{d\ell^+}{\ell^+} \;\theta\left( \ell^+ \right) ,
\end{align}
it follows from eq.~\eqref{e:I_subtraction} that
\begin{equation}
    I = \int_{Q_\sst}^\infty \frac{d\ell^+}{\ell^+} + \int_0^Q \frac{d\ell^+}{\ell^+} - \int_0^\infty \frac{d\ell^+}{\ell^+} \,.
    \label{e:rap_div_example}
\end{equation}
It is clear that in this split-up into different rapidity regions each contribution diverges -- the integrals are \emph{ill-defined}. Hence, we require a regularization prescription.

The fact that our regions analysis can introduce rapidity divergences is okay as long as they are properly regulated to give well-defined integrals. Rapidity divergences are \emph{not} regulated by dimensional regularization and thus require their own regularization prescription. Multiple rapidity regulators have been described in the literature and in the factorization study of chapter~\ref{c:Disentanglement}, we choose a regulator of the following type~\cite{Chiu:2011qc,Chiu:2012ir}:
\begin{equation}
    \left| \frac{\ell^+}{\nu} \right|^{-\eta} .
    \label{e:regulatorbasic}
\end{equation}
This regularization prescription can be used for all the collinear, soft, and Glauber scalings discussed in subsection~\ref{s:leading_regions} and could be included in the definition of the region approximator $T_R$. The rapidity scale $\nu$ is a quantity of mass dimension $1$ analogous to the renormalization scale $\mu$ in dimensional regularization, and the rapidity regulator $\eta$ is analogous to the fractional dimension $\varepsilon$ in dimensional regularization. Rapidity divergences manifest themselves as poles in $\eta$. Note that a full graph $\Gamma$ does not have rapidity divergences -- after all, they are merely an artifact of the regions analysis. This means that any rapidity regulator dependence must drop out on a graph-by-graph basis upon summing over all regions according to eq.~\eqref{e:subtr-sum}. Hence, at the end one can safely take the limit $\eta \to 0$ for each graph individually.

Let us now come back again to the example. With the regularization prescription in~\eqref{e:regulatorbasic}, the three contributions to $I$ in eq.~\eqref{e:rap_div_example} are given by
\begin{alignat}{2}
    \int_{Q_\sst}^\infty \frac{d\ell^+}{\ell^+} &\;\to\;\;& &\nu^\eta \int_{Q_\sst}^\infty d\ell^+ \,\frac{|\ell^+|^{-\eta}}{\ell^+} = \frac{1}{\eta} + \ln\left( \frac{\nu}{Q_\sst} \right) + \mathcal{O}(\eta) , \\[5pt]
    \int_0^Q \frac{d\ell^+}{\ell^+} &\;\to\;\;& &\nu^\eta \int_0^Q d\ell^+ \,\frac{|\ell^+|^{-\eta}}{\ell^+} = - \frac{1}{\eta} - \ln\left( \frac{\nu}{Q} \right) + \mathcal{O}(\eta) , \\[5pt]
    \int_0^\infty \frac{d\ell^+}{\ell^+} &\;\to\;\;& &\nu^\eta \int_0^\infty d\ell^+ \,\frac{|\ell^+|^{-\eta}}{\ell^+} = 0 .
\end{alignat}
Since the subtraction term is given by a so-called scaleless integral, it vanishes, see e.g.~\cite{Becher:2014oda}. We have expanded the first two integrals in the regulator $\eta$. Indeed, the rapidity divergences are now properly regulated, giving rise to well-defined integrals. Note that upon summing together both contributions, the poles in $\eta$ precisely cancel. After performing the limit $\eta \to 0$, the sum gives the finite result $\ln(Q/Q_\sst)$, consistent with the original integral in eq.~\eqref{e:rap_div_example_start}.

\section{Complications from Glauber modes} \label{s:complications_from_Glaubers}

We have discussed how to partition the leading-power DY graphs into various subgraphs that each characterize a particular momentum scaling, see figure~\ref{f:subgraphs}. To establish factorization, one needs to sum over all possible gluon attachments between those subgraphs. Before doing so, so-called Grammer-Yennie approximations~\cite{Grammer:1973db} can be applied to soft gluons and unphysically polarized collinear gluons. These approximations are essential to give rise to the necessary Wilson line operators through non-Abelian Ward identities. Unfortunately, the Grammer-Yennie approximation is not valid for partons with Glauber scaling. In this section we review the complications that arise from Glauber modes and how they are dealt with in the CSS proof. In subsection~\ref{s:unitarity_cancellation} we explain that to establish factorization one is not only required to sum over all possible graphs and leading regions, but also over all possible final-state cuts.

\subsection{Grammer-Yennie approximations}
For gluons that connect the different subgraphs one can make certain approximations. The following discussion summarizes the reviews presented in~\cite{Diehl:2011yj,Diehl:2015bca}. Let us first consider a right-moving collinear gluon with momentum $\ell$ connecting the $A$ and $H$ subgraphs, which is of the form $A^{\mu}(\ell) \,H_{\mu}(\ell)$, where $\mu$ is the polarization index of the gluon. For this collinear line it is sufficient to only keep the large light-cone component in $H$, i.e.\ $\ell \to \hat{\ell} \equiv (\ell^+,0,\bm{0})$. Since $|A^+| \gg |A^-|,|\bm{A}|$ ($A^\mu$ scales as a right-moving collinear momentum~\cite{Luo:1994np,Collins:2011zzd}), and all components of $H$ are generically of the same order, we approximate $A^{\mu}(\ell) \,H_{\mu}(\ell)$ as follows:
\begin{equation}
    A^{\mu}(\ell) \,H_{\mu}(\hat{\ell}) \approx A^+(\ell) \,\frac{\ell^+ v^-}{\ell^+ v^- + i\epsilon} \,H^-(\hat{\ell}) \approx A_{\mu}(\ell) \,\frac{v^{\mu}}{\ell \cd v + i\epsilon} \;\hat{\ell}^{\nu} H_{\nu}(\hat{\ell}) ,
    \label{e:GY-collinear}
\end{equation}
where $v \equiv (v^+,v^-,\bm{0})$ is an auxiliary vector with either $|v^-| \gg |v^+|$ or $|v^-| \sim |v^+|$. Note that in the second step we added an $i\epsilon$ prescription to the denominator to regulate the pole at $\ell \cd v = 0$ (this prescription is commented on in section~\ref{s:final_steps_factorization}). Eq.~\eqref{e:GY-collinear} is the Grammer-Yennie approximation appropriate for right-moving collinear gluons with unphysical polarization,\footnote{In their factorization proof, CSS adopt the Feynman gauge (in part~\ref{p:1} of this thesis we do the same), and, as it turns out, in this gauge only collinear gluons with unphysical polarization are leading. In a light-cone gauge, however, also physically polarized gluons can contribute at leading power~\cite{Boer:1999si}, giving rise to transverse gauge links at minus light-cone infinity~\cite{Belitsky:2002sm,Boer:2003cm}.} and the right-hand side is suitable for the use of a Ward identity. Naturally, analogous approximations can be made for left-moving collinear gluons connecting the $B$ and $H$ subgraphs. Upon summing over all collinear gluons, the collinear subgraphs are stripped from the hard subgraph, and the former end up with past-pointing Wilson line operators as appropriate for DY scattering (see e.g.~\cite{Diehl:2011yj} and section~\ref{s:BM_function} for more details).

One can also apply approximations to gluons that connect the soft and collinear subgraphs. Let us consider a soft gluon with momentum $\ell$ that connects the $S$ and $A$ subgraphs, which is of the form $S^{\mu}(\ell) \,A_{\mu}(\ell)$. We can neglect the plus component of $\ell$ compared to the large plus components in $A$, i.e.\ $\ell \to \tilde{\ell} \equiv (0,\ell^-,\bm{\ell})$. In this case the Grammer-Yennie approximation is given by
\begin{equation}
    S^{\mu}(\ell) \,A_{\mu}(\tilde{\ell}) \approx S^-(\ell) \,\frac{\ell^- w^+}{\ell^- w^+ + i\epsilon} \,A^+(\tilde{\ell}) \approx S_{\mu}(\ell) \,\frac{w^{\mu}}{\ell \cd w + i\epsilon} \;\tilde{\ell}^{\nu} A_{\nu}(\tilde{\ell}) ,
    \label{e:GY-soft}
\end{equation}
where $w \equiv (w^+,w^-,\bm{0})$ is an auxiliary vector with $|w^+| \gg |w^-|$. Again, we have provided an $i\epsilon$ prescription to the denominator in the second step (see section~\ref{s:final_steps_factorization}). The right-hand side of eq.~\eqref{e:GY-soft} is appropriate for the use of a Ward identity. Analogous approximations can be applied to soft gluons that connect the $S$ and $B$ subgraphs. Upon summing over all soft gluon attachments, the soft subgraph is stripped from the collinear subgraphs and is given by the vacuum expectation value of a product of Wilson lines (see e.g.~\cite{Diehl:2011yj,Collins:2011zzd} for more details).

If this were the full story then one would have now established factorization of the DY cross section into soft, collinear, and hard subgraphs. However, there is a complication that arises from Glauber modes that are initially contained in the soft subgraph. The Grammer-Yennie approximation for soft gluons, eq.~\eqref{e:GY-soft}, does \emph{not} hold for Glauber modes. If $\ell^- \sim \lambda^2 Q$ and $|\bm{\ell}| \sim \lambda Q$, then the approximation $\ell^- A^+ \approx \tilde{\ell}^{\nu} A_{\nu}$ fails as the transverse contribution to $\tilde{\ell}^{\nu} A_{\nu}$ cannot be neglected. A similar complication arises when considering Glauber attachments to $B$.

These problems can be overcome if the integrations over Glauber momenta could be deformed into a region of the complex plane (soft or collinear) where the Grammer-Yennie approximation is valid. Contour deformations are only possible if (some of) the integrations over the Glauber momentum components are \emph{not} pinched. If no deformations into a different momentum region are possible, we say that the integration is `trapped' in the Glauber region. In the next subsection we consider the pole structure associated to Glauber exchanges and provide examples of both a trapped and an untrapped Glauber gluon.

\subsection{Trapped in the Glauber region?} \label{s:trapped_Glauber}
Let us consider again the single-gluon exchange diagram given on the right in figure~\ref{f:power_counting}. In subsection~\ref{s:leading_regions} we showed with a power counting analysis that the $C_1$, $U$, $G_1$, and $G$ regions contribute at leading power. Here we show explicitly that the loop integration contours are not trapped in the two Glauber regions $G_1$ and $G$, implying that they could be deformed into different momentum regions. To this end, we investigate the poles of this diagram in the light-cone components of $\ell$.

\begin{figure}[htb]
\centering
    \begin{subfigure}{.42\textwidth}
    \begin{flushleft}
        \includegraphics[height=2.5cm]{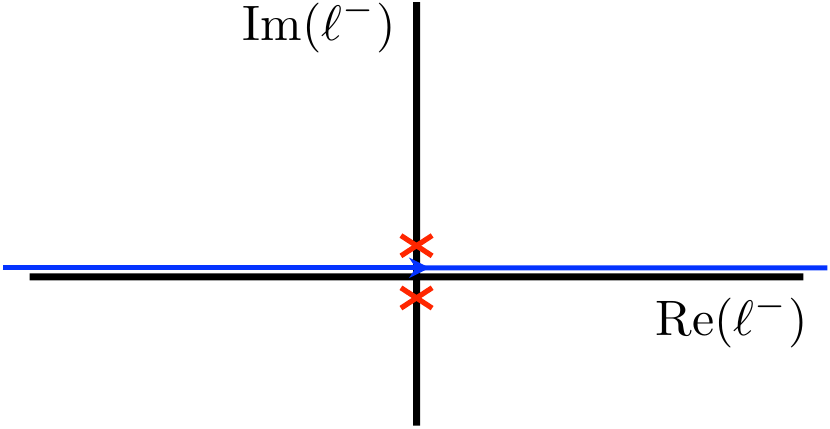}
    \end{flushleft}
    \end{subfigure}
    \begin{subfigure}{.42\textwidth}
    \begin{flushright}
        \includegraphics[height=2.5cm]{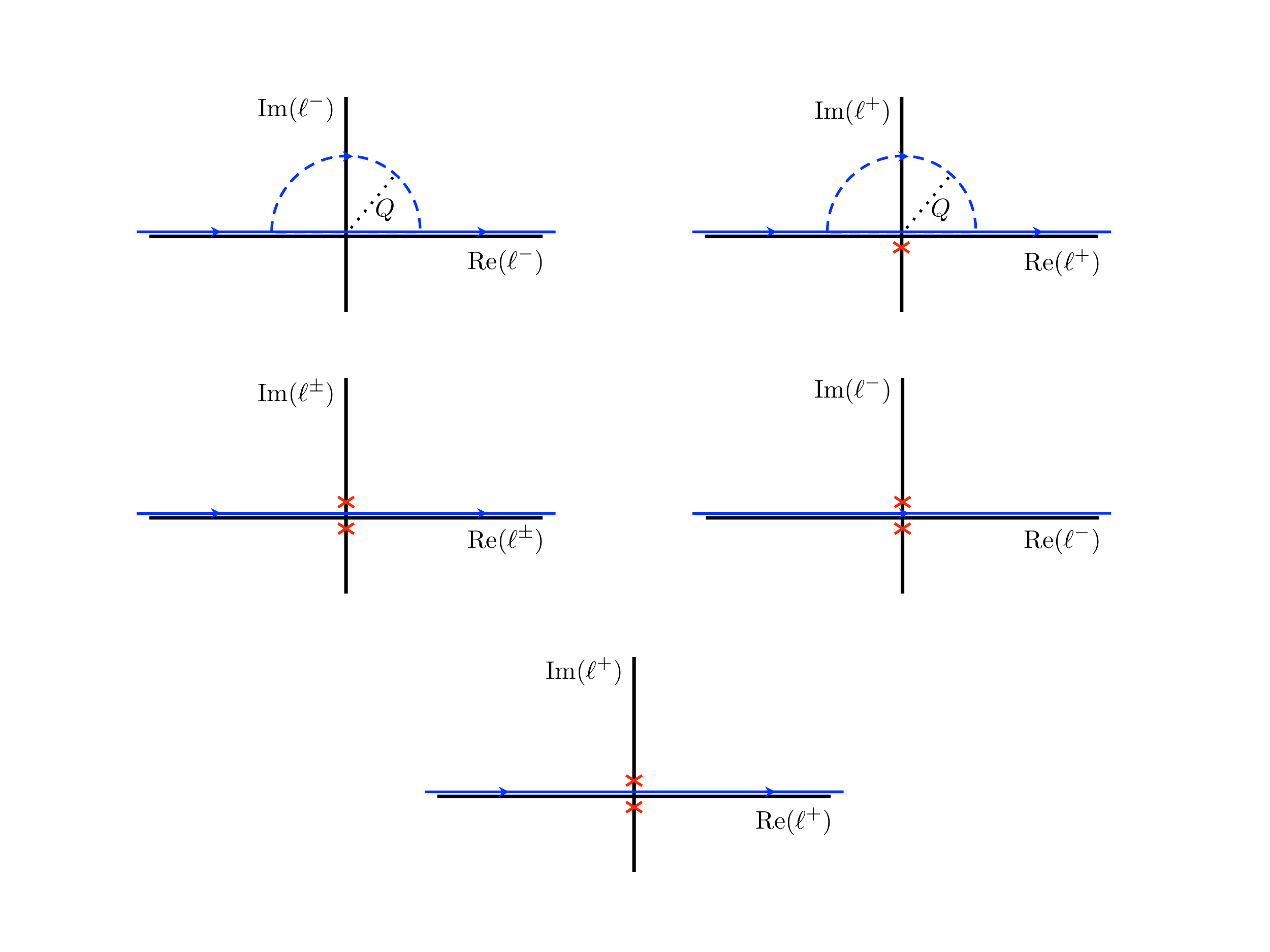}
    \end{flushright}
    \end{subfigure}
    \caption{The pole structure for the light-cone components of the gluon momentum $\ell$ appearing in the active-spectator gluon exchange graph on the right in figure~\ref{f:power_counting}. The integration contour for the minus component is pinched between two poles (indicated by red crosses), whereas the one for the plus component is not. The latter contour can be deformed in the upper half plane into a semicircle around the origin with radius of order $Q$ (indicated by the dashed line).}
\label{f:untrapped_glauber_contours}
\end{figure}

At leading power in the $G$ region, the active-spectator gluon exchange graph has the following structure (we only consider the relevant pieces here, i.e.\ the factors that depend on $\ell^+$ and $\ell^-$):
\begin{align}
    &\int \frac{d\ell^+ d\ell^-}{(2\pi)^2} \,\frac{i}{(k_1-\ell)^2 + i\epsilon} \,\frac{i}{(p_1-k_1+\ell)^2 + i\epsilon} \,\frac{i}{(k_2+\ell)^2 + i\epsilon} \nn \\[3pt]
    = &\left[ \int \frac{d\ell^-}{2\pi} \,\frac{i}{2k_1^+(k_1^--\ell^-) - (\bm{k}_1-\bm{\ell})^2 + i\epsilon} \right. \nn \\
    &\times \left. \!\frac{i}{2(p_1^+-k_1^+)(p_1^--k_1^-+\ell^-) - (\bm{k}_1-\bm{\ell})^2 + i\epsilon} \right] \nn \\
    &\times \left[ \int \frac{d\ell^+}{2\pi} \,\frac{i}{2k_2^-(k_2^++\ell^+) - (\bm{k}_2+\bm{\ell})^2 + i\epsilon} \right]
    \label{e:untrapped_example} .
\end{align}
For the $G_1$ region the result is the same, except that on the last line the $k_2^+$ component can in that case be neglected compared to $\ell^+$. From the structure of the propagators it follows that for both Glauber regions there are two poles in $\ell^-$ (labeled by $a,b$) and there is one pole in $\ell^+$:
\begin{equation}
    \ell_a^\pm = \mathcal{O}(\lambda^2) - i\epsilon , \qquad \ell_b^- = \mathcal{O}(\lambda^2) + i\epsilon .
\end{equation}
Whereas the $\ell^-$ contour is pinched at the origin between two poles, the $\ell^+$ contour is not pinched, see also figure~\ref{f:untrapped_glauber_contours}. The pole $\ell_a^+$ is also referred to as an `initial-state' pole as it is consistent with the ultimate formation of an initial-state Wilson line (a pole on the opposite side of the real axis is also referred to as a `final-state' pole). Since the $\ell^+$ component only has a pole in the lower half plane, we can freely deform the $\ell^+$ contour into the upper half plane according to Cauchy's theorem. In fact, if we deform it into a semicircle around the origin with radius of order $Q$, we enter the $C_1$ momentum region for which $\ell \sim (1,\lambda^2,\lambda)Q$. Hence, the two Glauber modes can be absorbed into a collinear region contribution where the appropriate Grammer-Yennie approximation can be applied. Note that since the $\ell^-$ contour is pinched, a deformation into the central soft region is not possible in this case.

\begin{figure}[htb]
\centering
    \includegraphics[height=4.1cm]{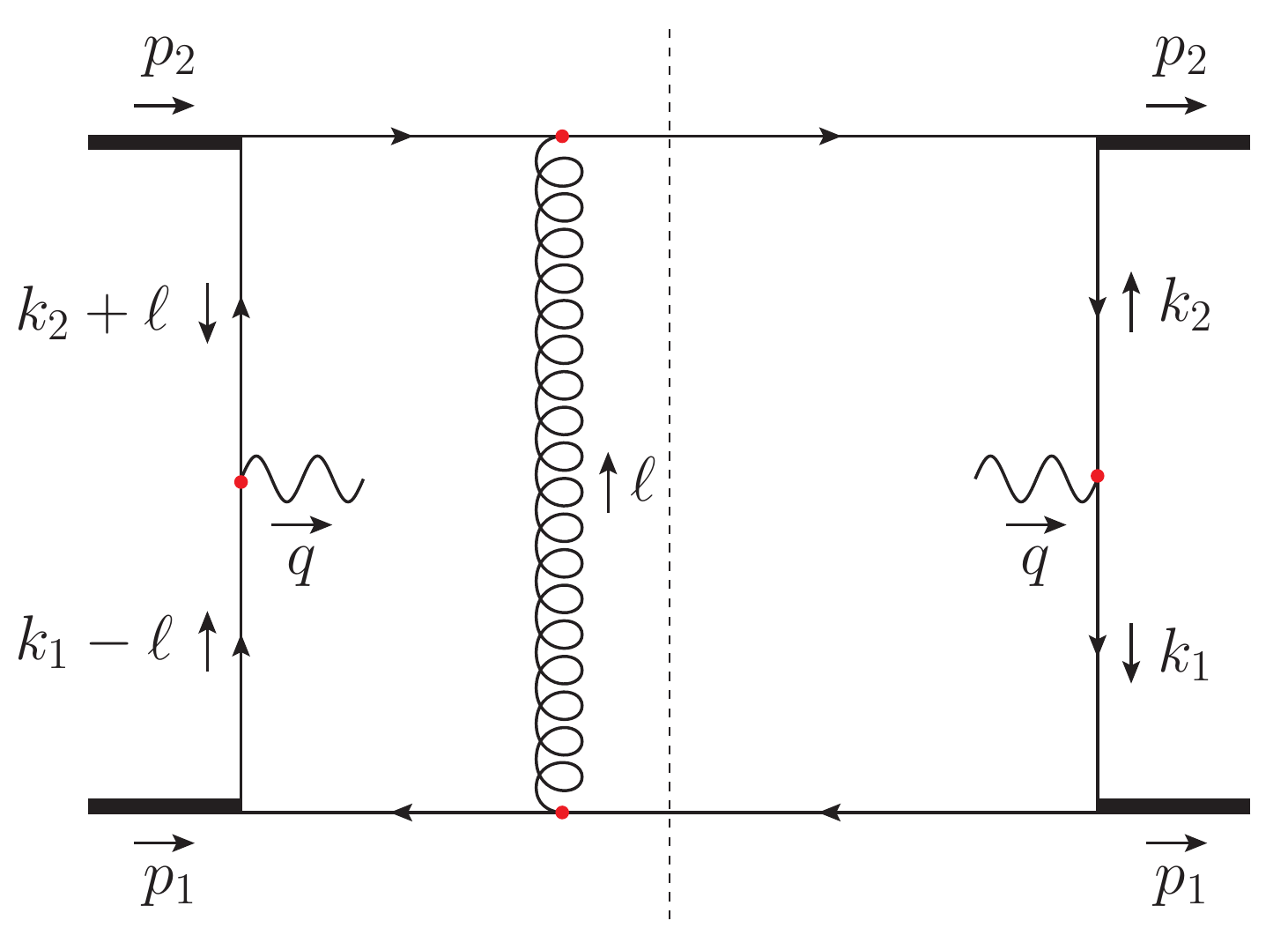}
    \caption{A spectator-spectator gluon exchange diagram that receives a leading contribution when the gluon momentum $\ell$ has (central) Glauber scaling. The integration over this loop momentum is trapped in the Glauber region.}
\label{f:trapped}
\end{figure}

For the next example, let us consider the spectator-spectator gluon exchange diagram in figure~\ref{f:trapped} (we employ the same model). Using the power counting techniques from subsection~\ref{s:leading_regions}, it is straightforward to show that this graph receives a leading-power contribution from the central Glauber region $G$; there are three additional propagators with respect to the tree-level graph and combined they give rise to a $1/\lambda^6$ enhancement, and, at the same time, the Glauber phase space $\int d^4\ell$ gives a $\lambda^6$ suppression. Hence, the overall scaling of the graph is $\lambda^0$, implying that the region $G$ contributes at leading power.

\begin{figure}[htb]
\centering
    \includegraphics[height=2.5cm]{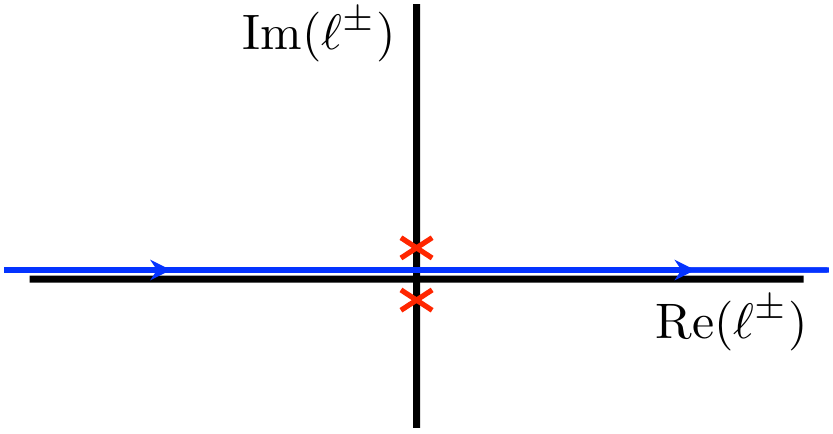}
    \caption{The pole structure for the light-cone components of the gluon momentum $\ell$ appearing in the spectator-spectator gluon exchange graph in figure~\ref{f:trapped}. For each component, the integration contour is pinched between two poles at the origin.}
\label{f:trapped_glauber_contours}
\end{figure}

At leading power in the $G$ region, the spectator-spectator exchange graph has the following structure (we again consider the relevant pieces only):
\begingroup 
\thinmuskip=2.4mu
\medmuskip=3.2mu
\begin{align}
    &\int \frac{d\ell^+ d\ell^-}{(2\pi)^2} \,\frac{i}{(k_1-\ell)^2 + i\epsilon} \,\frac{i}{(p_1-k_1+\ell)^2 + i\epsilon} \,\frac{i}{(k_2+\ell)^2 + i\epsilon} \,\frac{i}{(p_2-k_2-\ell)^2 + i\epsilon} \nn \\[3pt]
    = &\left[ \int \frac{d\ell^-}{2\pi} \,\frac{i}{2k_1^+(k_1^--\ell^-) - (\bm{k}_1-\bm{\ell})^2 + i\epsilon} \right. \nn \\
    &\times \left. \!\frac{i}{2(p_1^+-k_1^+)(p_1^--k_1^-+\ell^-) - (\bm{k}_1-\bm{\ell})^2 + i\epsilon} \right] \nn \\
    &\times \left[ \int \frac{d\ell^+}{2\pi} \,\frac{i}{2k_2^-(k_2^++\ell^+) - (\bm{k}_2+\bm{\ell})^2 + i\epsilon} \right. \nn \\
    &\times \left. \!\frac{i}{2(p_2^--k_2^-)(p_2^+-k_2^+-\ell^+) - (\bm{k}_2+\bm{\ell})^2 + i\epsilon} \right]
    \label{e:trapped_example} .
\end{align}
\endgroup
From the structure of the propagators we infer that there are two poles for each light-cone component of $\ell$, namely
\begin{equation}
    \ell_a^\pm = \mathcal{O}(\lambda^2) - i\epsilon , \qquad \ell_b^\pm = \mathcal{O}(\lambda^2) + i\epsilon .
\end{equation}
This means that both the $\ell^+$ and $\ell^-$ contours are pinched at the origin between two poles situated on either side of the real axis, see also figure~\ref{f:trapped_glauber_contours}. Hence, neither contour can be deformed freely into the complex plane -- the loop integration is \emph{trapped} in the Glauber region.

With these two examples we have demonstrated that in DY both trapped and untrapped Glauber exchanges occur. The former situation generally results from spectator-spectator interactions~\cite{Collins:2011zzd} and seems problematic. Since the Grammer-Yennie approximations are key to establish factorization through Ward identities, we need to be able to deform \emph{all} Glauber modes into soft or collinear regions where those approximations are valid. In the next subsection we will explain how the CSS approach takes care of this issue.

\subsection{Unitarity cancellation} \label{s:unitarity_cancellation}
The Glauber modes prevent one from factorizing the soft, collinear, and hard subgraphs upon summing over all soft and collinear attachments. The problem lies in the fact that in DY graphs both initial- and final-state poles appear, which is \emph{not} the case in for example $e^+e^-$ annihilation into two hadrons and in SIDIS~\cite{Ji:2004wu,Collins:2011zzd}. This extra difficulty in DY was overcome by CSS by summing over all kinematically allowed final-state cuts -- a sum over different final states is indeed appropriate for \emph{inclusive} DY scattering. Using light-cone perturbation theory, CSS proved to all orders that in fact all final-state poles cancel in the sum over cuts~\cite{Collins:1985ue,Collins:1988ig,Collins:2011zzd}. The sum over cuts is performed on a graph-by-graph basis and the pole cancellation is connected to unitarity of the theory through the Cutkosky cutting rules. Hence, the physical reason underlying this cancellation is conservation of probability. The unitarity cancellation implies that all pinches are lifted and that the Glauber modes become untrapped, allowing for contour deformations into soft or collinear regions where the Grammer-Yennie approximations are valid. As the details of the proof are not relevant for our factorization analysis in chapter~\ref{c:Disentanglement}, we will simply clarify the sum-over-cuts argument by providing an explicit example.

Let us consider again the single-gluon exchange graph in figure~\ref{f:trapped}. In the previous subsection we showed that for this spectator-spectator exchange the loop integration is trapped in the Glauber region. However, according to CSS, we need to sum over all unobserved final states for any given graph and leading region. For this particular graph, in the Glauber region, there are two possible final-state cuts, namely one directly to the left and one directly to the right of the exchanged gluon. Note that we should disregard a cut through the Glauber gluon itself, as a Glauber mode is an \emph{off-shell} momentum mode.

\begin{figure}[htb]
\centering
    \begin{subfigure}{.46\textwidth}
    \begin{flushleft}
        \includegraphics[height=4.1cm]{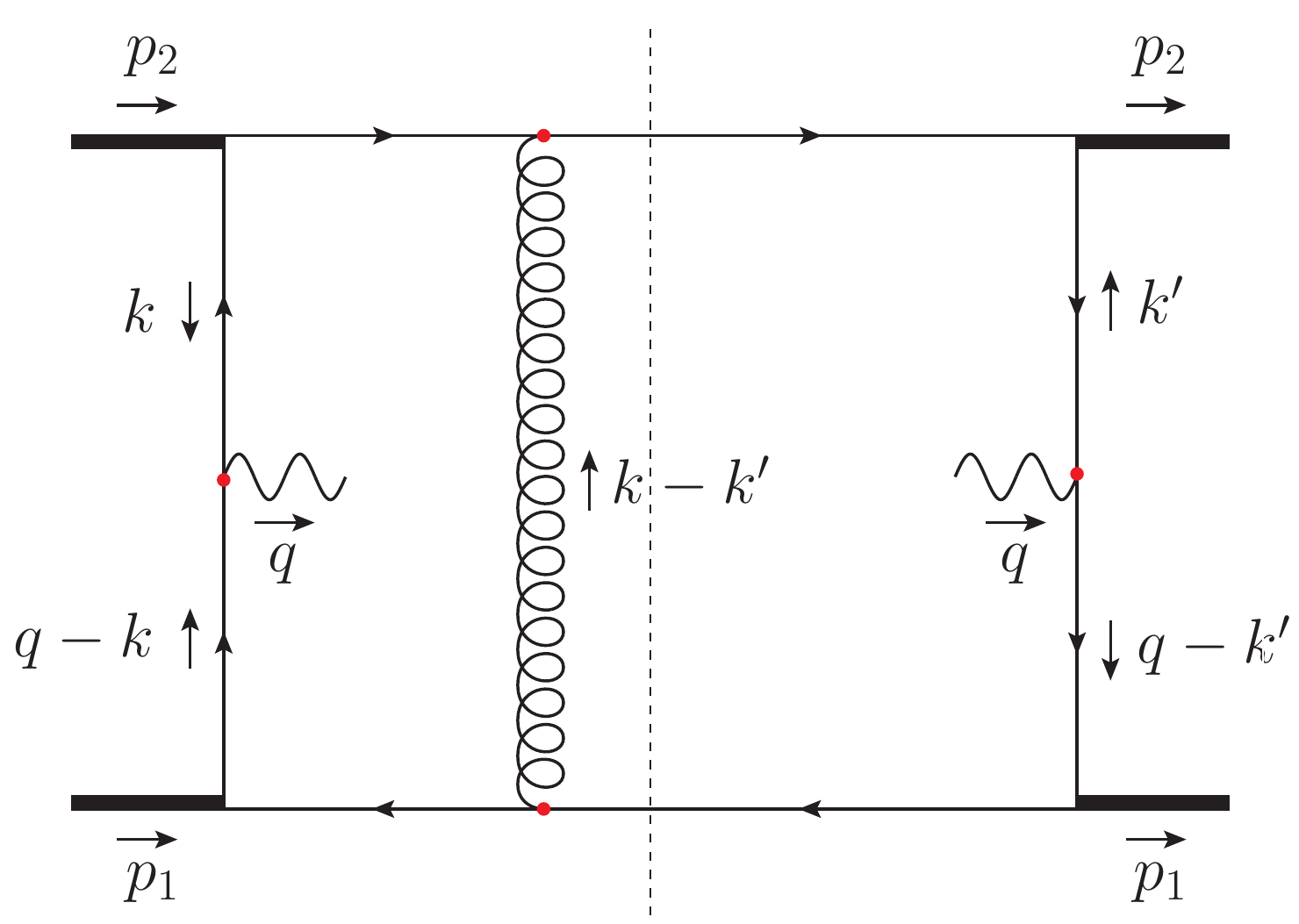}
    \end{flushleft}
    \end{subfigure}
    \begin{subfigure}{.46\textwidth}
    \begin{flushright}
        \includegraphics[height=4.1cm]{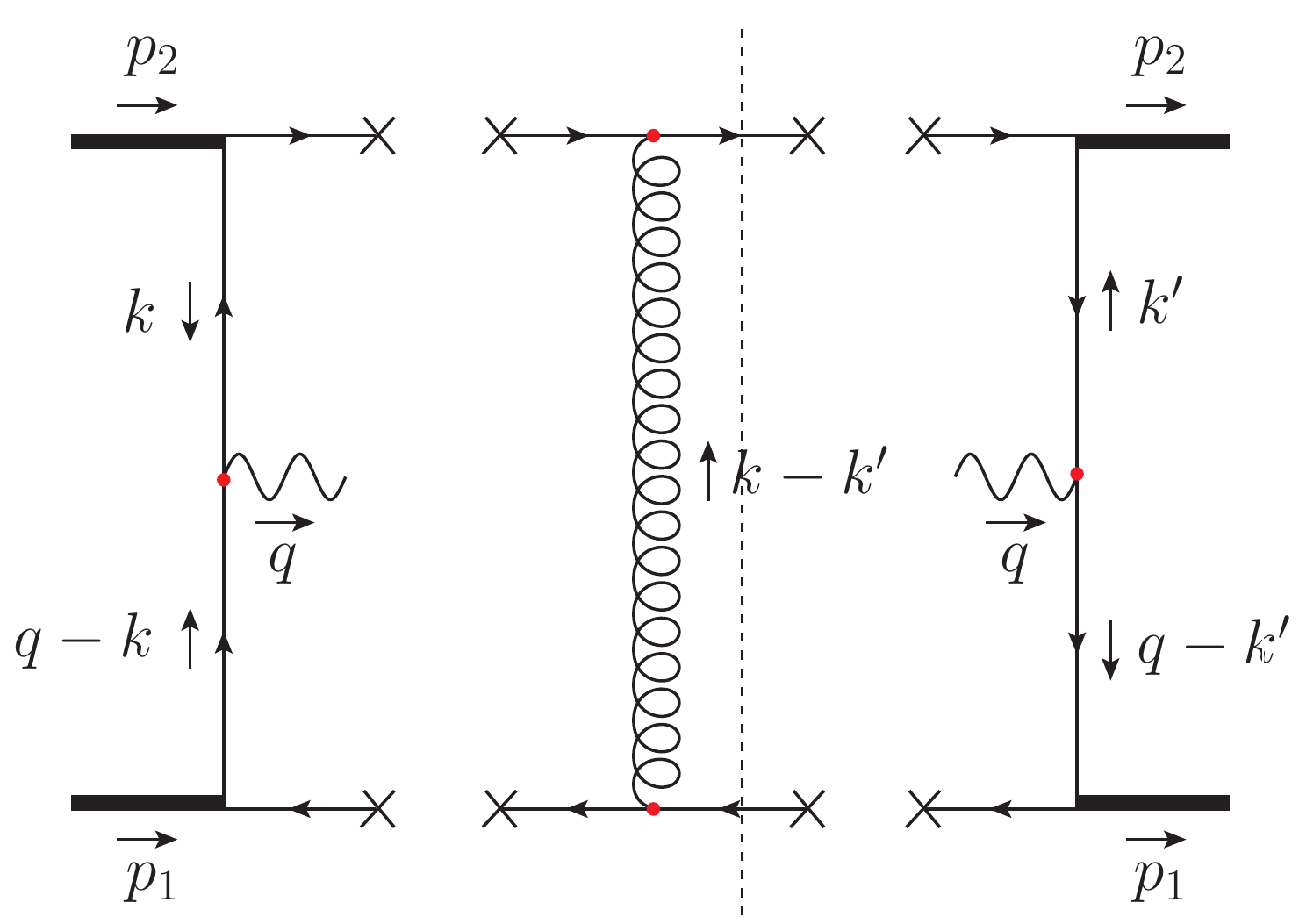}
    \end{flushright}
    \end{subfigure}
    \caption{The single-Glauber exchange graph; only one of the two cut positions is shown. At leading power, the graph on the left can be replaced by the one on the right (the lines with crosses denote on-shell lines).}
\label{f:glauber_right_cut}
\end{figure}

The following discussion is based on reviews in~\cite{Collins:2011zzd,Gaunt:2014ska} and for convenience we consider the same momentum routings. The authors showed that for each cut the spectator propagators that are not already on shell due to the cut can be replaced by on-shell delta functions. Pictorially this means that the graph on the left in figure~\ref{f:glauber_right_cut} can be replaced by the one on the right, and a similar replacement holds for the other cut position. To motivate this replacement, we study the leading-power structure of the diagram on the left in terms of the light-cone components of $k$. Note that in the diagram the momentum $k$ only appears to the left of the cut. The lines $k$ and $q-k$ are left- and right-moving collinear momenta respectively, whereas $q$ is hard. This means that $k^+$ and $q^--k^-$ are negligible compared to the large components $q^+$ and $q^-$ respectively. The precise numerator factors are not relevant in this discussion and will therefore be ignored. Employing Cauchy's residue theorem, we can write
\begin{align}
    &\int \frac{dk^+ dk^-}{(2\pi)^2} \,\frac{i}{(q-k)^2 + i\epsilon} \,\frac{i}{(p_1-q+k)^2 + i\epsilon} \,\frac{i}{k^2 + i\epsilon} \,\frac{i}{(p_2-k)^2 + i\epsilon} \nn \\[3pt]
    = &\left[ \int \frac{dk^-}{2\pi} \,\frac{i}{2q^+(q^--k^-) - (\bm{q}-\bm{k})^2 + i\epsilon} \right. \nn \\
    &\times \left. \!\frac{i}{2(p_1^+-q^+)(p_1^--q^-+k^-) - (\bm{q}-\bm{k})^2 + i\epsilon} \right] \nn \\
    &\times \left[ \int \frac{dk^+}{2\pi} \,\frac{i}{2q^-k^+ - \bm{k}^2 + i\epsilon} \,\frac{i}{2(p_2^--q^-)(p_2^+-k^+) - \bm{k}^2 + i\epsilon} \right] \nn \\[3pt]
    = &\left[ \frac{1}{2(p_1^+-q^+)} \,\frac{i}{2q^+(q^--k_a^-) - (\bm{q}-\bm{k})^2 + i\epsilon} \right] \left[ \frac{1}{2(p_2^--q^-)} \,\frac{i}{2q^-k_a^+ - \bm{k}^2 + i\epsilon} \right] \nn \\[3pt]
    = &\left[ \int \frac{dk^-}{2\pi} \,\frac{i}{2q^+(q^--k^-) - (\bm{q}-\bm{k})^2 + i\epsilon} \right. \nn \\
    &\times \left. 2\pi \,\delta\left( 2(p_1^+-q^+)(p_1^--q^-+k^-) - (\bm{q}-\bm{k})^2 \right) \vphantom{\int \frac{dk^-}{2\pi}} \right] \nn \\
    &\times \left[ \int \frac{dk^+}{2\pi} \,\frac{i}{2q^-k^+ - \bm{k}^2 + i\epsilon} \;2\pi \,\delta\left( 2(p_2^--q^-)(p_2^+-k^+) - \bm{k}^2 \right) \right] \nn \\[3pt]
    = &\int \frac{dk^+ dk^-}{(2\pi)^2} \,\frac{i}{(q-k)^2 + i\epsilon} \;2\pi \,\delta\left( (p_1-q+k)^2 \right) \,\frac{i}{k^2 + i\epsilon} \;2\pi \,\delta\left( (p_2-k)^2 \right) ,
    \label{e:spectators_on-shell}
\end{align}
where the poles $k_a^\pm$ appearing in the third step are given by
\begin{equation}
    k_a^- = q^--p_1^- + \frac{(\bm{q}-\bm{k})^2}{2(p_1^+-q^+)} - i\epsilon , \qquad k_a^+ = p_2^+ - \frac{\bm{k}^2}{2(p_2^--q^-)} + i\epsilon .
\end{equation}
Based on eq.~\eqref{e:spectators_on-shell} we conclude that the effect of the $k^-$ and $k^+$ integrations is to set the lines $p_1-q+k$ and $p_2-k$ on shell, respectively. Hence, we can replace the left graph in figure~\ref{f:glauber_right_cut} by the one on the right. In the same way one can show that for the other cut position (i.e.\ the one directly to the left of the Glauber gluon), the effect of the $k'^-$ and $k'^+$ integrations is to set the lines $p_1-q+k'$ and $p_2-k'$ on shell, respectively.

For both cut positions the left- and rightmost factors are the same -- together those factors constitute the tree-level DY graph (albeit with slightly different momenta to the left and right of the cut). Hence, upon summing over the two cuts, we have a tree-level contribution multiplied by a sum over the two external cuts for the $t$-channel gluon exchange subgraph. The latter sum vanishes due to the Cutkosky rules~\cite{Cutkosky:1960sp}:
\begin{equation} 
    \adjustbox{raise=-0.9cm}{\includegraphics[height=2.0cm]{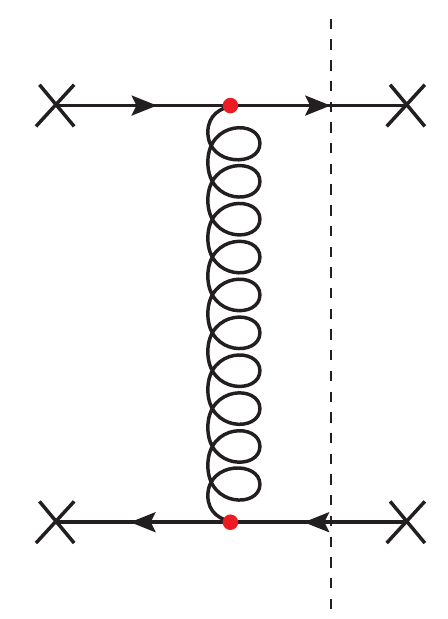}} \;\,+ \adjustbox{raise=-0.9cm}{\includegraphics[height=2.0cm]{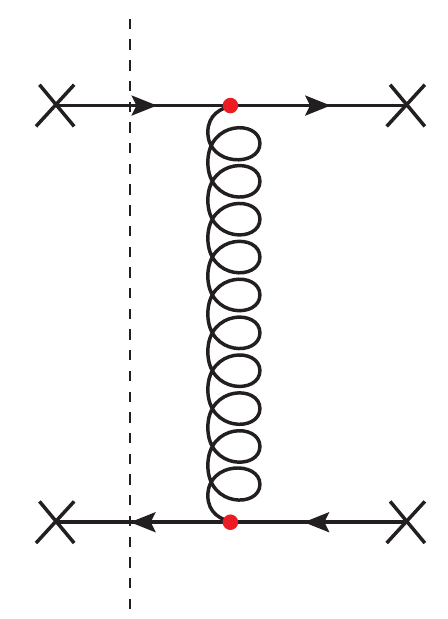}} = 0 .
    \label{e:sum_over_cuts}
\end{equation}
In the following we elaborate on this result and argue that it follows from unitarity of the theory, which in turn relies on conservation of probability.

The probability $P$ that a (normalized) initial state $\ket{i}$ transforms into some (normalized) final state $\ket{f}$ is given by
\begin{equation}
    P(i \to f) = |\bra{f} S \ket{i}|^2 ,
    \label{e:probability}
\end{equation}
where $S$ is the scattering operator. Probability conservation dictates that for a given initial state $\ket{i}$, the sum over all possible final states $\ket{f}$ gives unity. Making use of eq.~\eqref{e:probability}, this condition translates into
\begin{equation}
    1 = \sum_f P(i \to f) = \bra{i} S^\dagger S \ket{i} .
\end{equation}
Since the expectation value on the right-hand side must hold for any $\ket{i}$, it follows that $S^\dagger S = 1$. Similarly, any given final state must arise (i.e.\ with unit probability) from some initial state; this leads to the condition $S S^\dagger = 1$. Hence, $S$ is a unitary operator. Conventionally, we decompose $S$ into the identity operator $I$ and the Hermitian transition operator $T$ as $S = I + iT$. From the unitarity of $S$, it follows that
\begin{equation}
    T-T^\dagger = iT^\dagger T .
    \label{e:T}
\end{equation}
The transition operator $T$, sandwiched between the states $\bra{f}$ and $\ket{i}$, is related to the scattering amplitude $\mathcal{M}$:
\begin{equation}
    \bra{f} T \ket{i} = (2\pi)^4 \,\delta^{(4)}(p_i-p_f) \,i\mathcal{M}(i \to f) .
    \label{e:T_intermsof_M}
\end{equation}
From eqs.~\eqref{e:T} and~\eqref{e:T_intermsof_M} we obtain the so-called generalized optical theorem:
\begin{equation}
    i\left[ \mathcal{M}(i \to f) - \mathcal{M}^*(f \to i) \right] = - \sum_X \int d\Phi_X \;\mathcal{M}(i \to X) \mathcal{M}^*(f \to X) ,
    \label{e:gen_opt_theorem}
\end{equation}
where $X$ is any possible intermediate state and $d\Phi_X$ denotes the corresponding phase space element. The Cutkosky rules are a generalization of eq.~\eqref{e:gen_opt_theorem} -- they hold for \emph{individual} Feynman diagrams with the sum over $X$ on the right-hand side replaced by a sum over internal cuts~\cite{Cutkosky:1960sp,Veltman:1994wz} (see also e.g.~\cite{Gaunt:2014ska,Rietkerk:2016elf}).

The left-hand side of eq.~\eqref{e:sum_over_cuts} is then precisely described by the left-hand side of eq.~\eqref{e:gen_opt_theorem} for a given graph. After all, the Glauber exchange subgraph with the external cut to the right is just the graph itself, $i\mathcal{M}(i \to f)$, while the one with the external cut to the left is given by the complex conjugate of the time-reversed process, $-i\mathcal{M}^*(f \to i)$. As there are no possible internal cuts for our Glauber exchange subgraph, the sum over external cuts evaluates to zero according to the Cutkosky rules. Hence, for the spectator-spectator gluon exchange diagram of this example, the contribution from the Glauber region completely cancels out.\footnote{Strictly speaking, as the Glauber subgraphs are actually imaginary~\cite{Angeles-Martinez:2015bla} (and the tree-level skeleton is real), this particular spectator-spectator exchange graph must always cancel to end up with a real cross section, regardless of the Cutkosky rules.} The general principle illustrated here also applies to more complicated diagrams featuring multiple Glauber exchanges, see e.g.~\cite{Gaunt:2014ska}.

\section{Final steps in obtaining factorization} \label{s:final_steps_factorization}

In the previous section we discussed the complications that arise from Glauber modes regarding the factorization procedure, and how they can be overcome. CSS proved to all orders in perturbation theory that upon summing over all final-state cuts, the final-state poles cancel out due to unitarity of the theory. In some cases, though, this cancellation goes even further -- in the previous section, namely, we showed that for e.g.\ the single-gluon exchange diagram in figure~\ref{f:glauber_right_cut} the Glauber contribution completely vanishes. For the graphs for which this is not the case, the unitarity cancellation of final-state poles allows for contour deformations out of the Glauber region, away from initial-state poles, into soft or collinear regions where the Grammer-Yennie approximations can be applied. After application of those approximations, the contours can be deformed back to the real axes again. The $i\epsilon$ prescriptions in eqs.~\eqref{e:GY-collinear} and~\eqref{e:GY-soft} are chosen such that this is always possible and are consistent with the formation of initial-state Wilson lines, as appropriate for DY.

\begin{figure}[htb]
\centering
    \includegraphics[height=5.03cm]{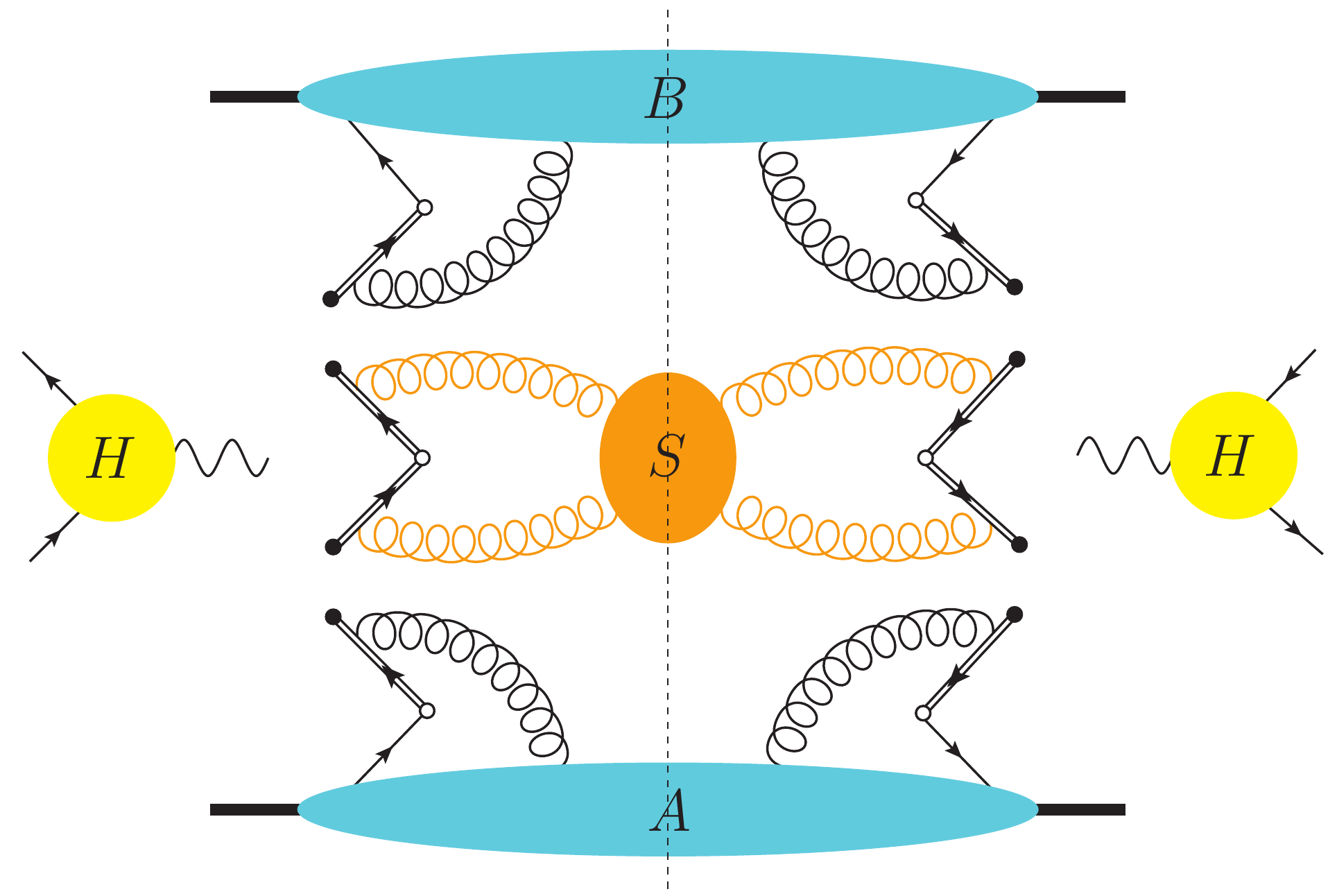}
    \caption{Factorized soft, collinear, and hard subgraphs.}
\label{f:subgraphs2}
\end{figure}

Effectively what happens is that part of the Glauber contributions is cancelled while the other part can be absorbed in the soft or collinear subgraphs. Upon summing over all collinear and soft gluon attachments, Ward identities ensure that the various subgraphs ($S$, $A$, $B$, and $H$) factorize -- this is ultimately linked to gauge invariance of the theory. The result of the factorization procedure is shown schematically in figure~\ref{f:subgraphs2}. The collinear subgraphs $A$ and $B$ each contain a past-pointing Wilson line, and the soft subgraph $S$ is given by the vacuum expectation value of a product of Wilson lines.

\begin{figure}[htb]
\centering
    \includegraphics[height=5.03cm]{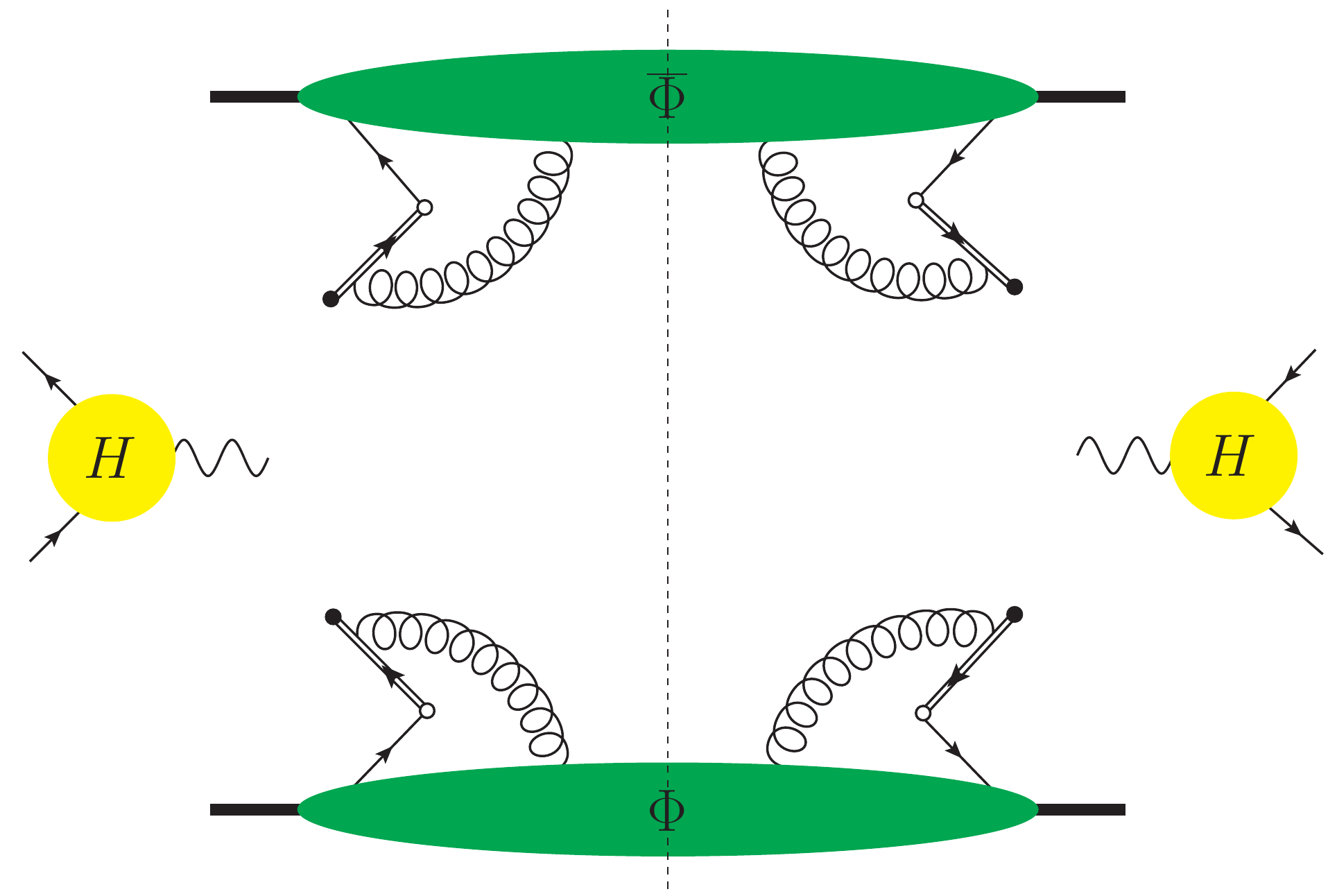}
    \caption{Final factorized form of the DY process. The soft factor has been absorbed in the quark and antiquark TMD correlators.}
\label{f:subgraphs3}
\end{figure}

In contrast to $H$, each of $S$, $A$, and $B$ contains nonperturbative contributions that are not predicted by the theory; those contributions need to be fitted to data. However, since $S$ appears inside a product with two collinear factors, it could never be measured independently.\footnote{Although the soft factor is universal~\cite{Collins:2004nx}, it could not be independently extracted from other processes either (for which TMD factorization has been proven), as it is in fact always accompanied by two collinear factors.} Hence, it was proposed in~\cite{Collins:2011zzd} to combine each of the subgraphs $A$ and $B$ with a square root of $S$. Today, those combinations are considered the proper definitions of TMD parton densities. Another advantage of this definition is that, as it turns out, rapidity divergences precisely cancel in the TMDs, while $S$, $A$, and $B$ separately are divergent. The cancellation of rapidity divergences was recently proven to hold to all orders in perturbation theory~\cite{Vladimirov:2017ksc}. Thus, the result of the CSS procedure is a factorized cross section with two TMDs and a hard function, see figure~\ref{f:subgraphs3}.

\section[\stackunder{Possible issues}{}]{Possible issues} \label{s:possible problems}

In this chapter we have discussed various key elements of the CSS factorization proof for DY scattering. We have argued that upon summing over all graphs, leading momentum regions, and final-state cuts, the DY cross section factorizes into two TMDs and a hard function, see eq.~\eqref{e:fac} and figure~\ref{f:subgraphs3}. As discussed in subsection~\ref{s:process_dependence}, the necessary gauge links in the TMDs are not unique, but depend on the process under consideration. The gauge links are understood to track the color flow in the hard process, which is e.g.\ entirely incoming in DY and entirely outgoing in SIDIS. Consequently, in DY the two TMDs come with past-pointing gauge links, whereas in SIDIS the links are future-pointing. Whenever the color flow is both incoming and outgoing, such as in back-to-back hadron pair production in $pp$ collisions, it has been argued that the gauge links can no longer be disentangled, preventing factorization in terms of separately color gauge invariant factors containing TMDs~\cite{Collins:2007nk,Collins:2007jp,Rogers:2010dm}. This entanglement of color structures hampers the theoretical prediction of the associated observables.

\begin{figure}[htb]
\centering
    \includegraphics[height=4.1cm]{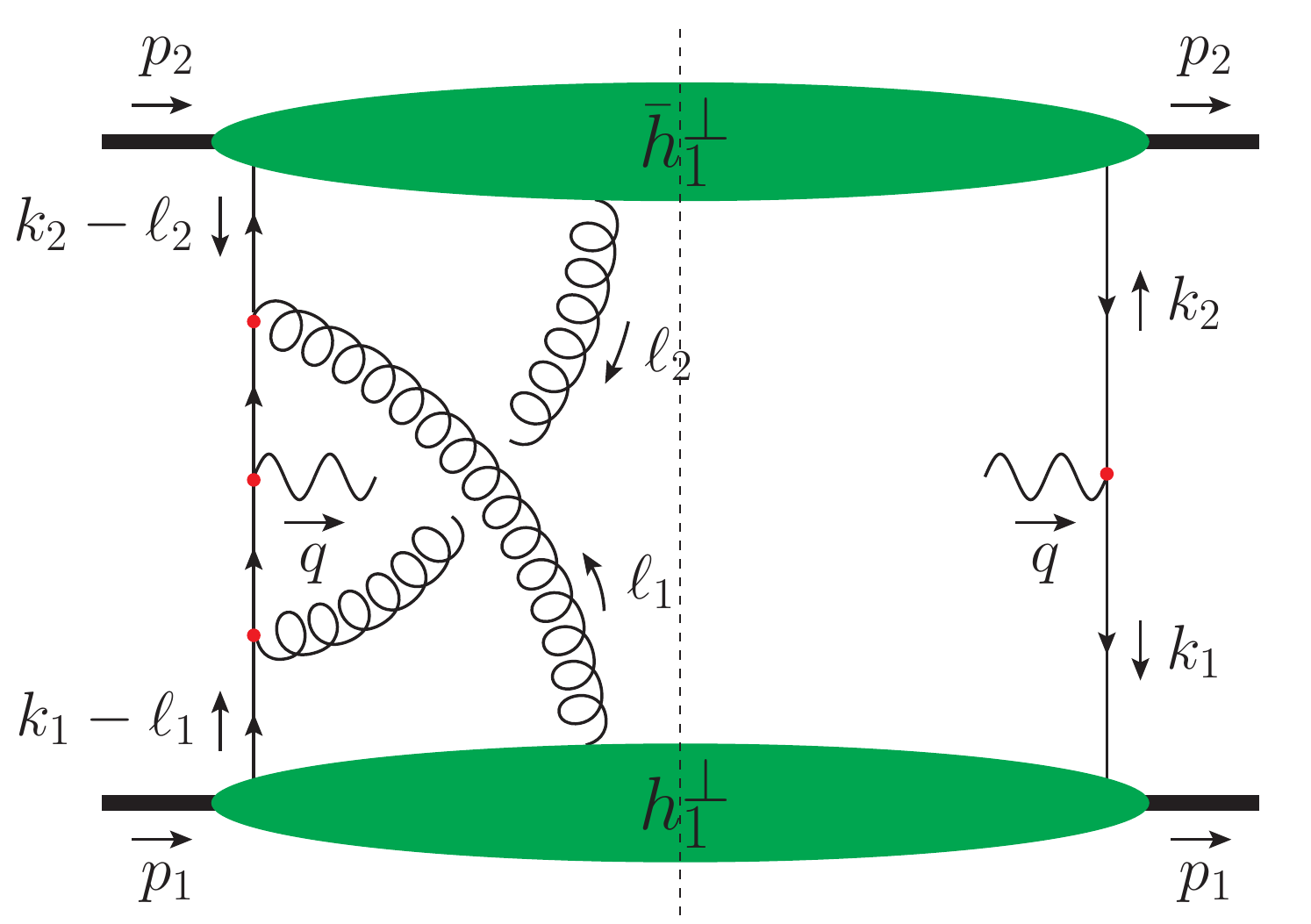}
\caption{An example of a lowest-order graph with a color-entangled structure that gives a nonzero contribution to the dBM term in the DY cross section.}
\label{f:DY2}
\end{figure}

In~\cite{Buffing:2013dxa} it was suggested that color-entangled contributions can also arise in DY scattering. Those contributions, at best, would come in a factorized form with a color factor different from $1/N_c$ (as predicted by the DY factorization theorem). The affected contributions involve only T-odd TMDs, such as the BM or Sivers functions. As explained in subsection~\ref{s:process_dependence}, those functions are special in the sense that their existence completely relies on the presence of gauge links, which causes them to change sign between DY and SIDIS. In~\cite{Buffing:2013dxa}, however, it was claimed that the process dependence of double T-odd contributions goes beyond this sign flip: as a result of color entanglement, those contributions are suppressed and change sign in DY due to an additional color factor of $-1/(N_c^2-1)$. This color-entanglement effect would signal a loophole in the TMD factorization proof for double T-odd contributions that involve polarization. In~\cite{Buffing:2013dxa} factorization was neither proven nor disproven. Since it is unclear how color-entanglement effects would manifest themselves beyond the lowest order considered, it means that factorization could potentially be violated. At lowest order in $\alpha_s$, an entangled color structure contributing to the dBM term for example arises in the graph in figure~\ref{f:DY2} with a single-gluon exchange between each correlator and the active parton coming from the other side. 

The question that is addressed in detail in the next chapter is whether the color-entanglement effect described in~\cite{Buffing:2013dxa} survives after summing over all relevant graphs, leading momentum regions, and final-state cuts, as required according to CSS to obtain factorization. To this end, we perform an explicit calculation of the dBM contribution to DY scattering using a spectator model. The calculation is performed up to the first order at which color entanglement is supposed to appear according to~\cite{Buffing:2013dxa}, i.e.\ up to $\mathcal{O}(\alpha_s^2)$. Note that the $\mathcal{O}(\alpha_s^2)$ level corresponds to two gluon exchanges and thus includes for example the diagram in figure~\ref{f:DY2}.

\chapter{Disentangling colors for azimuthal asymmetries} \label{c:Disentanglement}
\addthumb{\thechapter}{\Large{\thechapter}}{\thumbfg}{\thumbbg}

In this chapter we will investigate whether double T-odd contributions to the DY cross section factorize in the way predicted by CSS, or instead suffer from color-entanglement effects as described in~\cite{Buffing:2013dxa}. As explained in section~\ref{s:possible problems}, those effects would suppress the double T-odd contributions by an additional color factor and could even lead to a breakdown of factorization. Since the affected contributions are measured in experiments, the factorization property of the cross section is highly essential. According to~\cite{Buffing:2013dxa}, the different color structures should be visible already at the lowest nontrivial order, i.e.\ at the level of two gluon exchanges. At this level, we perform an explicit model calculation of the dBM contribution to the DY cross section to investigate its color structure (in section~\ref{s:generalizations} we also briefly discuss the results for the Sivers function). The computation is based on our work in~\cite{Boer:2017hqr} and proceeds largely along the lines of the CSS factorization proof as outlined in the previous chapter. The main differences with the CSS approach are that we will treat the Glauber region distinctly from the soft region in the Collins subtraction procedure (i.e.\ with its own approximator) and that we will not perform any contour deformations.

\section{Status of the Boer-Mulders effect} \label{s:BM_effect}

The BM function $h_1^\perp$ encodes correlations between the transverse polarization and the transverse momentum of a quark inside an unpolarized hadron. As can be seen from eq.~\eqref{e:fac}, this TMD contributes to a $\cos(2\phi)$ azimuthal asymmetry in the DY cross section. Experimentally, the dBM term can be extracted as follows:
\begin{align}
    \mathcal{F}\left[ w(\bm{k}_1,\bm{k}_2) \,h_1^\perp \bar{h}_1^\perp \right] \propto \int_0^{2\pi} d\phi \,\cos(2\phi) \left( \frac{d\sigma}{d\Omega \,dx_1 dx_2 \,d^2\bm{q}} \right) .
\end{align}
Thus, by measuring the $\cos(2\phi)$ angular dependence, DY provides an opportunity to probe the transverse polarization of quarks inside unpolarized hadrons.

There is yet another process that can be used to extract the BM function, namely SIDIS. In the unpolarized SIDIS process, the BM function appears in a convolution with the chiral-odd Collins FF which describes the fragmentation of a transversely polarized quark into an unpolarized hadron~\cite{Collins:1992kk}. This contribution gives rise to a $\cos(2\phi)$ angular dependence as well, where now $\phi$ is the azimuthal angle of the produced hadron with respect to the lepton plane. In both DY and SIDIS, the $\cos(2\phi)$ asymmetries are referred to as BM effects as they can be ascribed to a nonzero BM function.

The fact that intrinsic partonic transverse momenta can give rise to these asymmetries, even at leading twist, was noted in the 1990s by Boer in~\cite{Boer:1999mm} for DY and by Boer and Mulders in~\cite{Boer:1997nt} for SIDIS. Those observations sparked a lot of interest in the field of spin physics. Besides lattice QCD calculations of the BM function~\cite{Musch:2011er,Engelhardt:2015xja}, there have been many studies of the BM function in the context of QCD-inspired models, see e.g.~\cite{Boer:2002ju,Gamberg:2003ey,Lu:2004hu,Gamberg:2005ip,Lu:2005rq,Lu:2006ew,Gamberg:2007wm,Bacchetta:2008af,Courtoy:2009pc,Pasquini:2010af,Pasquini:2014ppa,Wang:2017onm,Wang:2018naw}. Since the BM function cannot be calculated using perturbative QCD methods, model calculations are used to obtain more insight into its nonperturbative properties, such as its size and sign. Moreover, they can be instructive as to what functional form should be used for the experimental extractions.

Perhaps the most studied model is the so-called spectator model in which the hadron is composed of an active quark that participates in the hard scattering, as well as a spectator that groups together the remaining partonic degrees of freedom. In the case of a proton, the spectator is also referred to as a diquark as it serves to represent a combination of two valence quarks. Naturally, the diquark can be of spin $0$ (a scalar) or spin $1$ (an axial vector) -- both cases are studied in the literature, with the former being the simplest. The factorization calculation in this chapter is performed using the scalar spectator model (see the next section for more details).

In unpolarized hadron-hadron collisions, the $\cos(2\phi)$ asymmetry has been measured by the NA10 experiment at CERN~\cite{Falciano:1986wk,Guanziroli:1987rp} and the E615~\cite{Conway:1989fs} and E866/NuSea experiments at Fermilab~\cite{Zhu:2006gx,Zhu:2008sj}, and is currently under active investigation by the COMPASS experiment at CERN~\cite{Aghasyan:2017jop} and the E906/SeaQuest experiment at Fermilab~\cite{Peng:2008sp,Nakano:2016jdm}. Using the available data on $pd$ and $pp$ collisions in~\cite{Zhu:2006gx,Zhu:2008sj}, both the quark~\cite{Zhang:2008nu,Lu:2009ip} and antiquark~\cite{Barone:2010gk} BM functions for the proton have been extracted. For the fits, the $x$- and $\bm{k}$-dependent parts were assumed to factorize, and for the latter a Gaussian parametrization was considered. The results in~\cite{Lu:2009ip} agree with the positive signs for both the up and down quark BM functions as predicted by almost all model calculations. Furthermore, they satisfy the appropriate positivity bounds given in~\cite{Bacchetta:1999kz}. However, as was recently argued in~\cite{Lambertsen:2016wgj}, for proper extractions of BM functions also next-to-leading order perturbative QCD corrections should be incorporated.

For unpolarized SIDIS, two different extractions of the proton BM function have been performed~\cite{Barone:2009hw,Barone:2015ksa}. The latest extraction~\cite{Barone:2015ksa} is based on charged hadron production through $e^\pm p$ and $e^\pm d$ collisions measured by the HERMES experiment at DESY~\cite{Airapetian:2012ki,Airapetian:2012yg}, as well as on charged hadron production through $\mu^\pm d$ collisions measured by the COMPASS experiment at CERN~\cite{Adolph:2013stb,Adolph:2014pwc}. Similar to the extraction from DY, the $x$- and $\bm{k}$-dependent parts of the BM function were assumed to factorize, with a Gaussian profile for the latter. As it turns out, data at higher values of $Q^2$ is needed to properly separate the BM effect from higher-twist contributions~\cite{Barone:2009hw}.

Unfortunately, more data is needed to claim the observation of the expected sign flip of the BM function between DY and SIDIS. Many precision measurements of the $\cos(2\phi)$ asymmetries are planned at existing or future experimental facilities (see e.g.\ the reviews~\cite{Barone:2010zz,Lu:2016pdp}), enabling more precise extractions of the BM functions in both processes.

Since the color-entanglement result for DY as described in~\cite{Buffing:2013dxa} is at odds with the CSS factorization theorem, it is important to check this result in an explicit calculation. After all, the various ongoing and planned experimental extractions of the BM function crucially rely on factorization. To this end, we perform in this chapter an explicit model calculation of the dBM term in the DY cross section, following closely the steps of the CSS factorization proof as reviewed in the previous chapter. In the next section we discuss the details of the spectator model that we adopt for the calculation.

\section{Model setup}

For the factorization study in this chapter we employ a spectator model in which the colorless spin-$\tfrac{1}{2}$ proton couples to a spin-$\tfrac{1}{2}$ quark and a scalar spectator (see e.g.~\cite{Jakob:1997wg,Brodsky:2002cx,Bacchetta:2003rz,Gamberg:2007wm,Meissner:2007rx}). The quark is in the color triplet representation with electrical charge $e_q = 1$, and the scalar is in the color antitriplet representation and is electrically neutral. We take the proton-quark-scalar coupling to be a constant for simplicity (as one would obtain for a fundamental Yukawa-type fermion-fermion-scalar coupling); for convenience this vertex factor will be set to unity in the calculations. The proton and scalar are taken massive with masses $M$ and $m_s$ respectively, and the quark is taken massless.\footnote{To avoid issues related to proton decay, we take $m_s > M$ in our calculations.} The antiproton is treated using the same spectator model as the proton, albeit with quantum numbers appropriately conjugated.

In this chapter we adopt the momenta conventions as specified in sections~\ref{s:DY_scattering} and~\ref{s:momentum_regions_analysis}. This means that in the cross section calculation we consider a proton colliding with an antiproton (in the CM frame of the collision), with right- and left-moving collinear momenta $p_1$ and $p_2$ respectively, see~\eqref{e:scaling_p12}. In this collision, we consider the DY production of an off-shell photon with a hard momentum $q$, which occurs via $q\bar{q}$ fusion. The scalars coupling to either hadron are spectators. To enable the hard scattering, the active quarks must carry right- and left-moving collinear momenta $k_1$ and $k_2$ respectively, see~\eqref{e:scaling_k12}.

In the model calculation we consider QCD corrections to tree-level DY production. The coupling of gluons to quarks, antiquarks, and scalars proceeds via the standard (fermionic or scalar) QCD Feynman rules, see appendix~\ref{a:feynman_rules}. By using the standard couplings, we ensure that the model obeys important physical principles such as unitarity.\footnote{In general, vertices respect unitarity if the corresponding interaction Lagrangian is Hermitian (and this is indeed the case for our model).} The unitarity property of the model will play an important role in the calculations of this chapter. In this part of the thesis we adopt the Feynman gauge.

Since we are interested in the dBM contribution to the cross section, both the quark and antiquark need to be transversely polarized, see eq.~\eqref{e:h_1}. We will for each diagram encounter a Dirac trace of the form $\tr(\Phi H_1 \overline{\Phi} H_2)$, where $H_1$ and $H_2$ represent hard scattering matrices, and $\Phi$ and $\overline{\Phi}$ are matrices for the proton and antiproton pieces, respectively. We can decompose $\Phi$ on a basis of Dirac structures as follows~\cite{Diehl:2011yj,Kasemets:2013nma}:
\begin{align}
    \Phi_{\alpha\beta} &= \tr \left( \tfrac{1}{2} \Phi \right) \tfrac{1}{2} \delta_{\alpha\beta} + \tr \left( \tfrac{1}{2} \gamma^5 \Phi \right) \tfrac{1}{2} (\gamma^5)_{\alpha\beta} + \tr \left( \tfrac{1}{2} \gamma_\mu \Phi \right) \tfrac{1}{2} (\gamma^\mu)_{\alpha\beta} \nn \\
    &\quad\, + \tr \left( \tfrac{1}{2} \gamma^5 \gamma_\mu \Phi \right) \tfrac{1}{2} (\gamma^\mu \gamma^5)_{\alpha\beta} + \tr \left( \tfrac{1}{2}i \sigma_{\nu\mu} \gamma^5 \Phi \right) \tfrac{1}{4}i (\sigma^{\mu\nu} \gamma^5)_{\alpha\beta} \,.
\end{align}
At leading power, the trace with the hard scattering matrix $H_i$ is given by
\begin{align}
    \tr \left( \Phi H_i \right) &= \tr \left( \tfrac{1}{2}\gamma^+ \Phi \right) \tr \left( \tfrac{1}{2}\gamma^- H_i \right) + \tr \left( \tfrac{1}{2}\gamma^+ \gamma^5 \Phi \right) \tr \left( \tfrac{1}{2} \gamma^5 \gamma^- H_i \right) \nn \\
    &\quad\, + \tr \left( \tfrac{1}{2}i \sigma^{j+} \gamma^5 \Phi \right) \tr \left( \tfrac{1}{2}i \sigma^{j-} \gamma^5 H_i \right) .
\end{align}
Doing the same for $\overline{\Phi}$, we can decompose the Dirac trace $\tr(\Phi H_1 \overline{\Phi} H_2)$ as follows:
\begin{align}
    \tr \big( \Phi H_1 \overline{\Phi} H_2 \big) &= \tr \big( \Gamma_U \Phi \big) \tr \big( \overline{\Gamma}_U \overline{\Phi} \big) \tr \big( \overline{\Gamma}_U H_1 \Gamma_U H_2 \big) \nn \\
    &\quad\, + \tr \big( \Gamma_T^j \Phi \big) \tr \big( \overline{\Gamma}_T^k \overline{\Phi} \big) \tr \big( {\overline{\Gamma}_T}_j H_1 {\Gamma_T}_k H_2 \big) + \,\ldots \,,
    \label{e:fierz}
\end{align}
where the Dirac projectors $\Gamma_U$ and $\Gamma_T^j$ select unpolarized and transversely polarized quarks respectively, see eq.~\eqref{e:projector} (a bar on the projector indicates that plus components are replaced by minus components). The ellipsis in eq.~\eqref{e:fierz} represents the double longitudinally polarized contribution as well as mixed terms. For the calculation of the dBM contribution, we only need to consider the second term in this sum.

\section{Graphs and momentum regions} \label{s:graphs_regions}

In this section we discuss which graphs and momentum regions need to be considered for the calculation of the dBM contribution to the DY cross section. We are interested in its precise color structure to check the color-entanglement result of~\cite{Buffing:2013dxa}. We consider the employed spectator model rich enough in structure to settle the issue -- in particular, the color factors involved are the same as those appearing in a full QCD calculation. The study is performed at leading power in $\Lambda/Q$ and at the $\mathcal{O}(\alpha_s^2)$ level, being the lowest order at which color-entanglement effects are supposed to appear according to~\cite{Buffing:2013dxa}.

\subsection{Relevant graphs} \label{s:relevant_graphs}
As will be argued in section~\ref{s:BM_function} based on the factorization theorem, at least one gluon attachment to the spectator is needed for a nonzero BM function. At $\mathcal{O}(\alpha_s)$, we find that the only graph contributing to a nonzero dBM effect is the one that features a (central) Glauber gluon exchange between the two scalar spectators. However, as we illustrated in subsection~\ref{s:unitarity_cancellation}, this graph vanishes upon summing over the two possible final-state cuts. Hence, there are no graphs expected to contribute at the $\mathcal{O}(\alpha_s)$ level.

The lowest order at which graphs contribute to a nonzero dBM term is the $\mathcal{O}(\alpha_s^2)$ level, which corresponds to the exchange of two gluons. In our analysis, we distinguish between graphs that have a `color-entangled' structure and those that do not. By definition, the latter type of graphs, the `unentangled' ones, come with the color factor
\begin{equation}
    \tr(t^a t^a t^b t^b) = \frac{1}{N_c} \,C_\Phi^2 \,,
    \label{e:non-entangled_color}
\end{equation}
where the two factors of $C_\Phi \equiv \tr(t^a t^a)$ are absorbed in the BM functions (see section~\ref{s:BM_function}), and the $1/N_c$ color factor is precisely the overall color factor that is expected from the CSS factorization theorem; it also appears in the leading-order result in eq.~\eqref{e:fac}. The deviating color factor that was found in~\cite{Buffing:2013dxa} for double T-odd contributions originates from diagrams that have a color structure different from that in eq.~\eqref{e:non-entangled_color}.

\begin{figure} 
\centering
\begin{subfigure}{.48\textwidth}
    \begin{flushleft}
        \includegraphics[height=4.0cm]{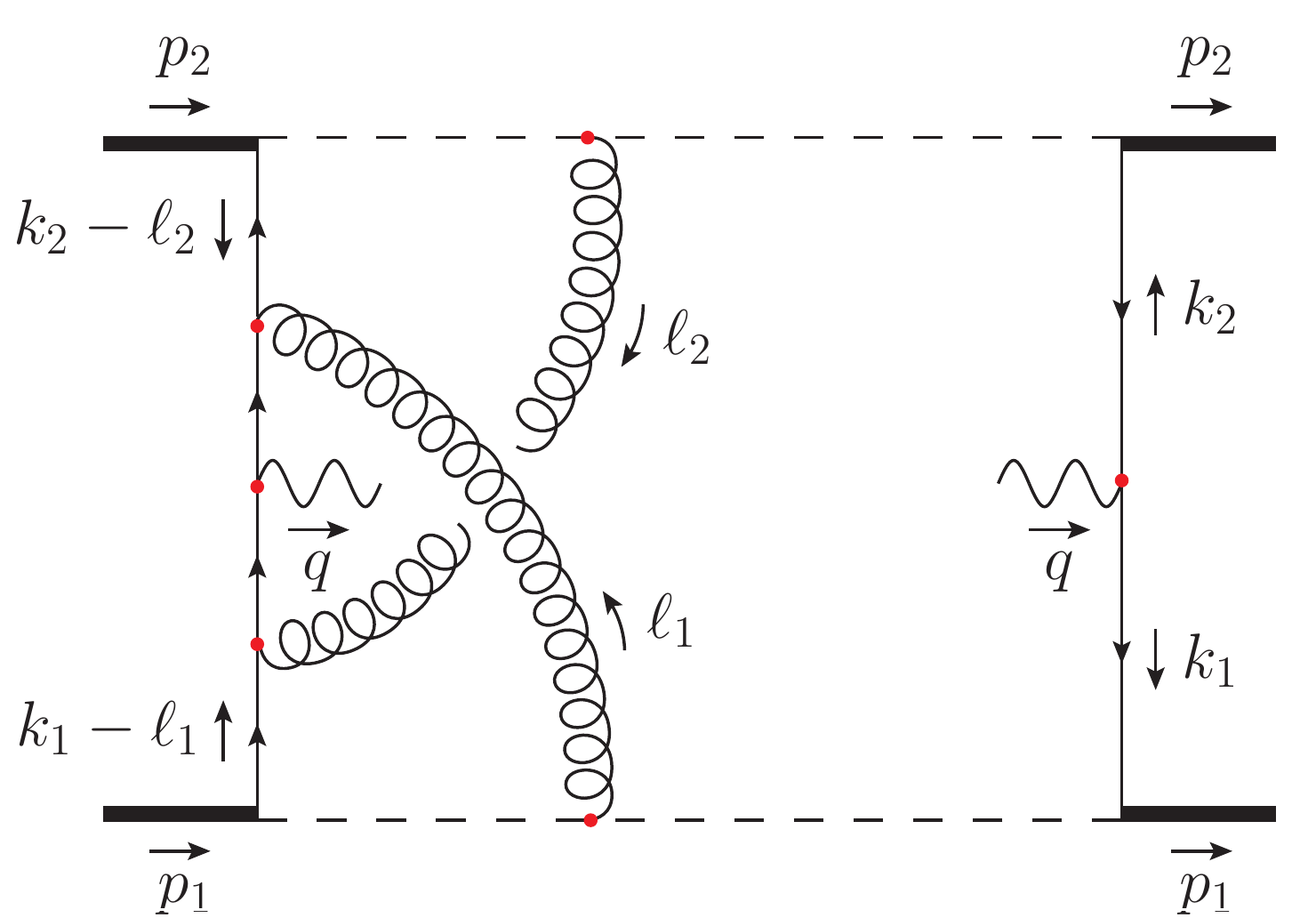}
        \caption*{\hspace{-0.20cm}(a)}
    \end{flushleft}
\end{subfigure}
\begin{subfigure}{.48\textwidth}
    \begin{flushright}
        \includegraphics[height=4.0cm]{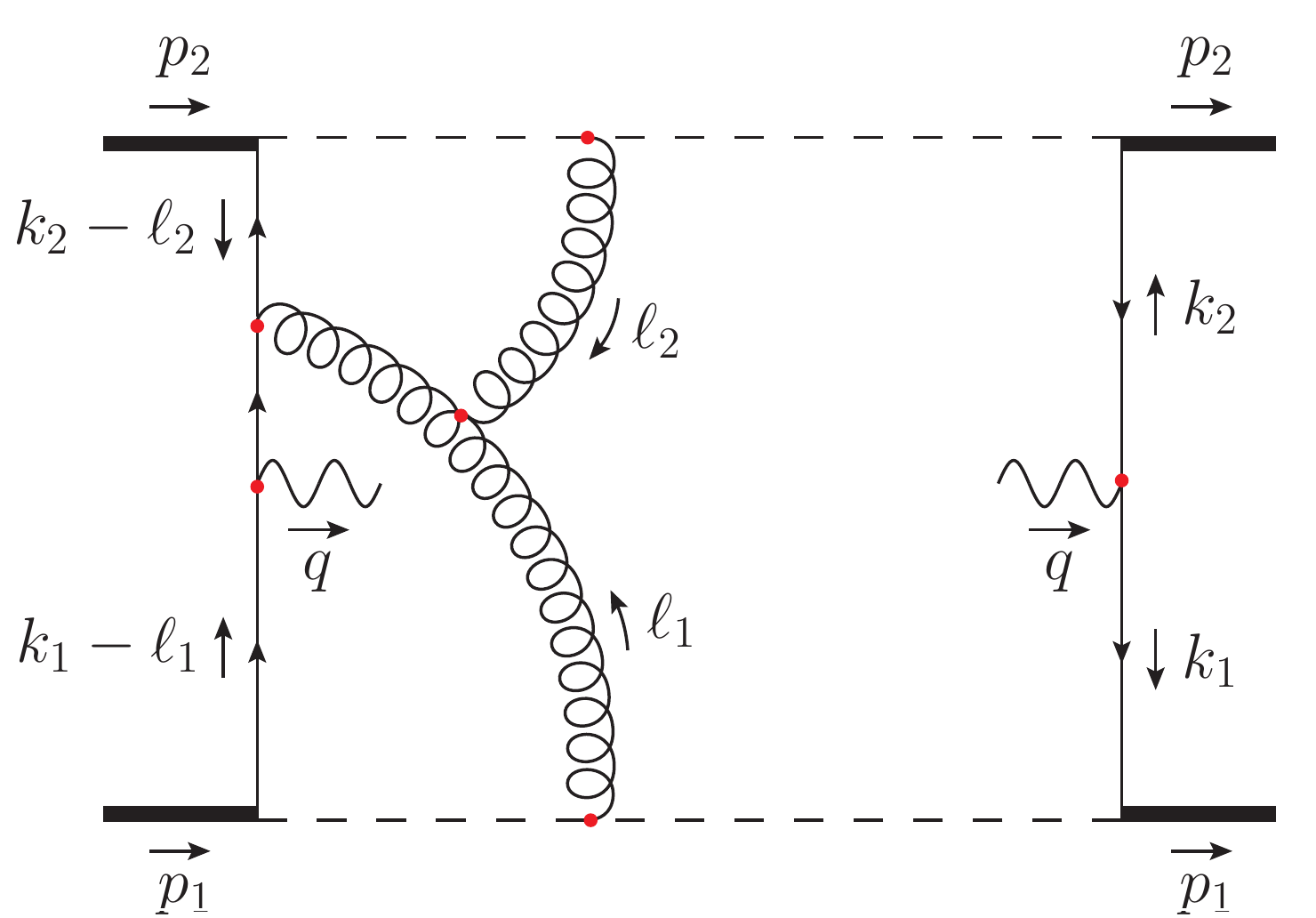}
        \caption*{\hspace{0.85cm}(b)}
    \end{flushright}
\end{subfigure}
\begin{subfigure}{.48\textwidth}
    \begin{flushleft}
        \includegraphics[height=4.0cm]{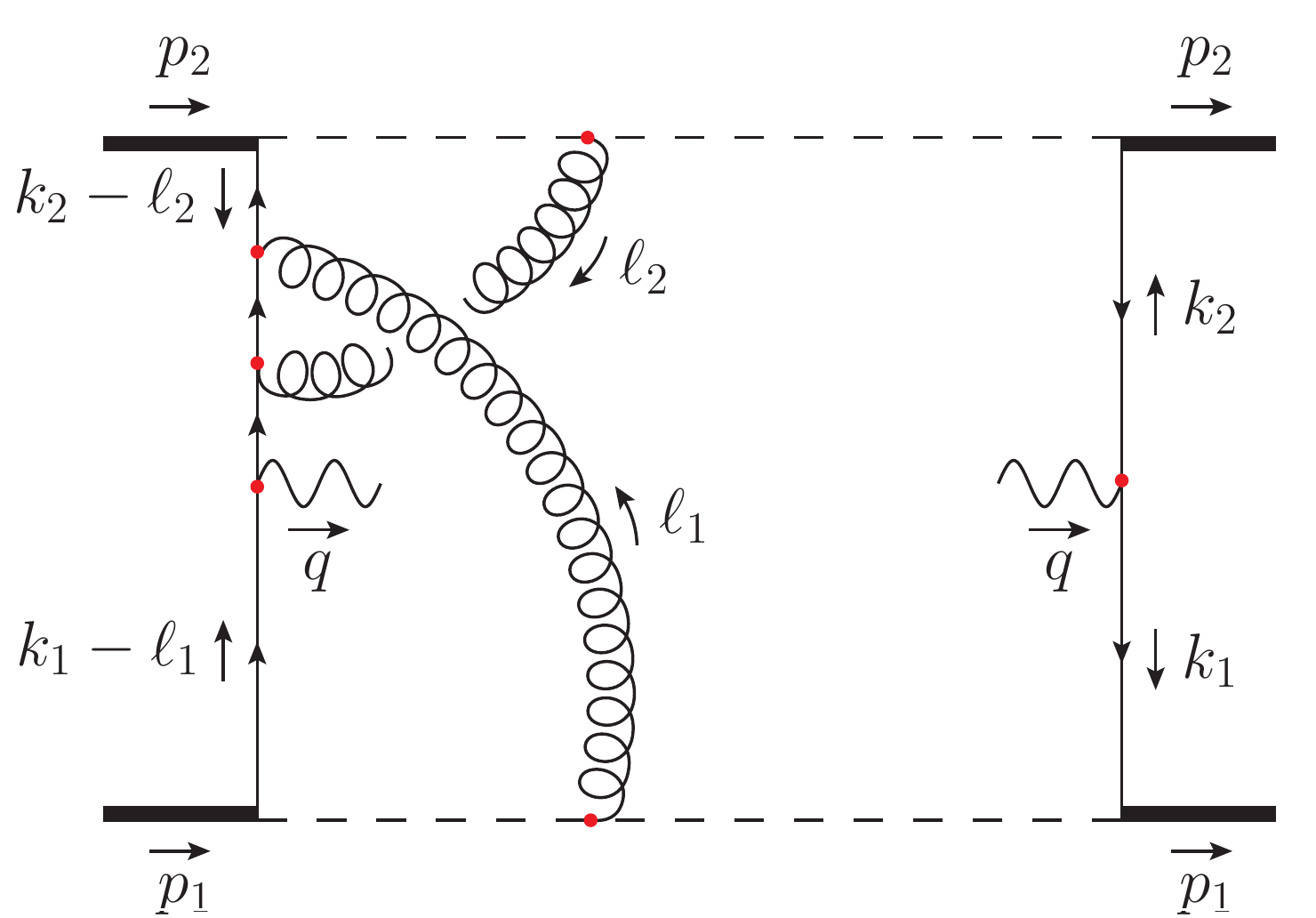}
        \caption*{\hspace{-0.20cm}(c)}
    \end{flushleft}
\end{subfigure}
\begin{subfigure}{.48\textwidth}
    \begin{flushright}
        \includegraphics[height=4.0cm]{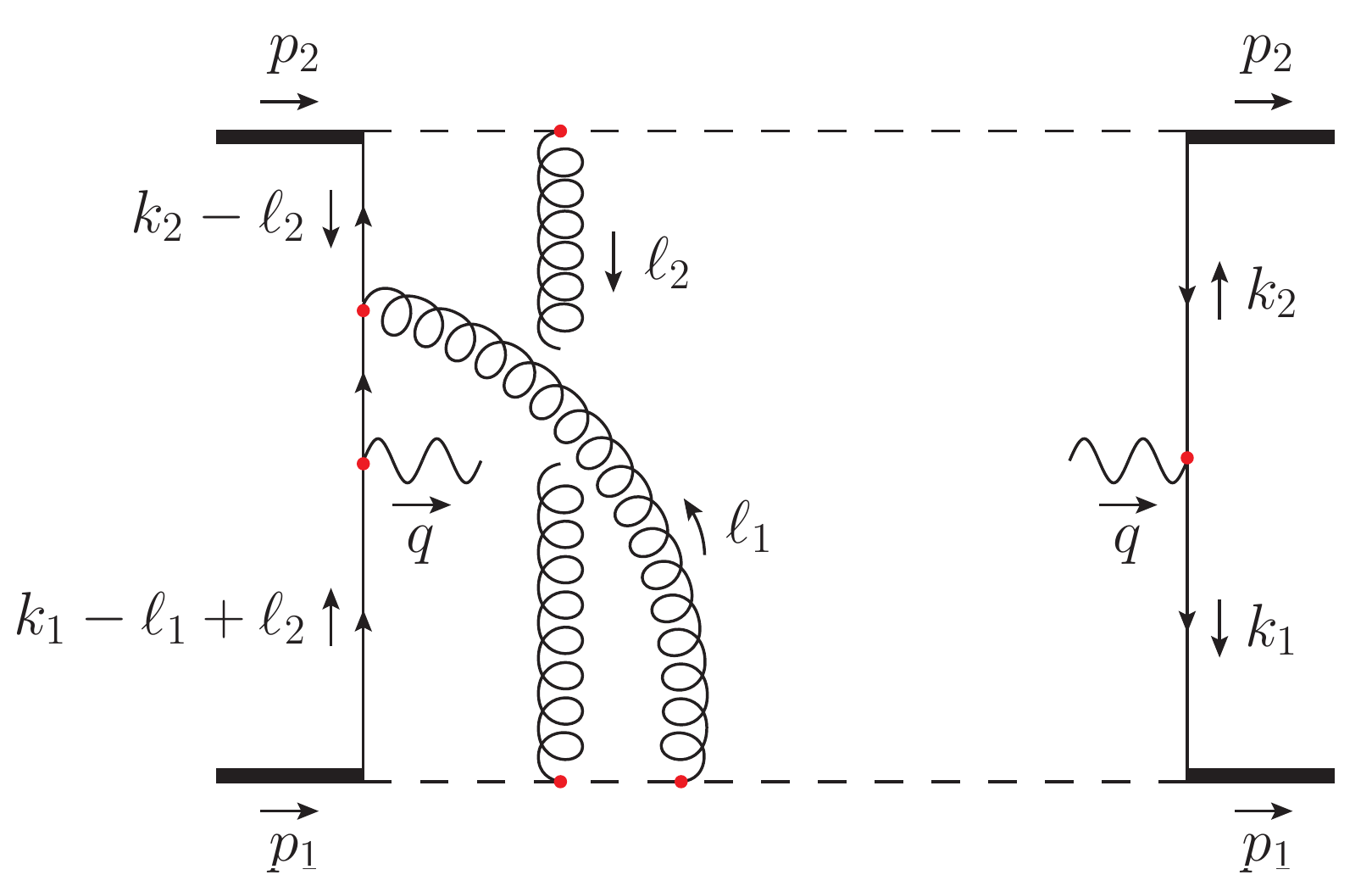}
        \caption*{\hspace{0.87cm}(d)}
    \end{flushright}
\end{subfigure}
\begin{subfigure}{.48\textwidth}
    \begin{center}
    \hspace{-1.0cm}
        \includegraphics[height=4.0cm]{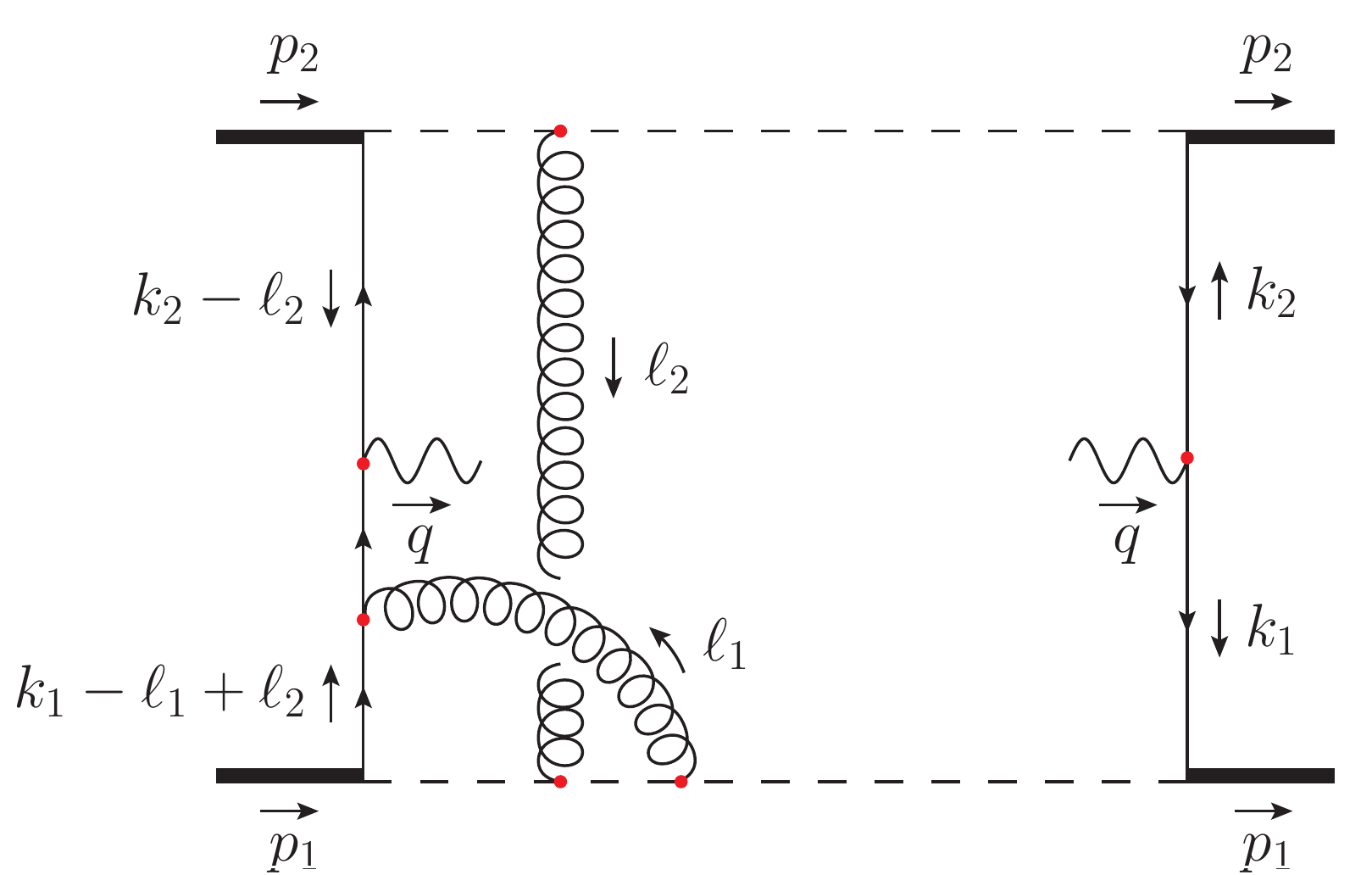}
        \caption*{(e)}
    \end{center}
\end{subfigure}
\caption{Color-entangled diagrams contributing to the dBM part of the DY cross section in the model at $\mathcal{O}(\alpha_s^2)$. This set is supplemented by graphs that can be obtained by $p\leftrightarrow\bar{p}$ or Hermitian conjugation, and for diagrams (d) and (e) there are also seagull versions.}
\label{f:entangled_diagrams}
\end{figure}

\begin{figure}[htb]
\centering
\begin{subfigure}{.48\textwidth}
    \begin{center}
        \includegraphics[height=4.0cm]{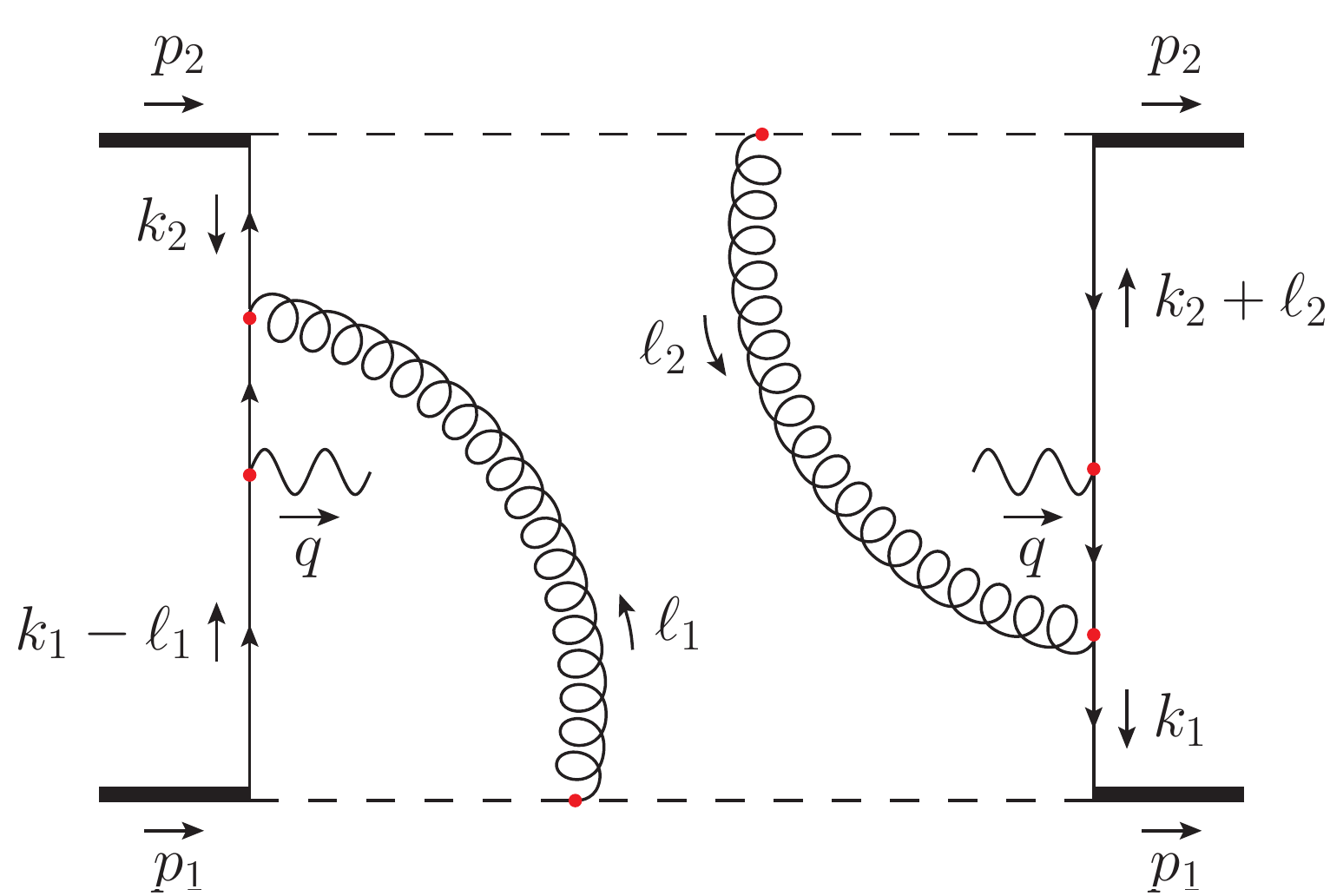}
        \caption*{\hspace{0.10cm}(f)}
    \end{center}
\end{subfigure}
\caption{An example of an unentangled diagram contributing to the dBM part of the DY cross section in the model at $\mathcal{O}(\alpha_s^2)$.}
\label{f:(f)}
\end{figure}

We find by explicit calculation that the (relevant) color-entangled graphs to consider are diagrams (a)--(e) in figure~\ref{f:entangled_diagrams}, plus graphs related to these by Hermitian conjugation or a vertical proton-antiproton flip (denoted by $p\leftrightarrow\bar{p}$).\footnote{For the $p\leftrightarrow\bar{p}$ versions we also flip the particle labels, i.e.\ $1 \leftrightarrow 2$.} Diagrams (d) and (e) also have `seagull' versions where the two gluon attachments to the lower scalar spectator line are merged into one. There are also diagrams that only involve spectator-spectator exchanges. However, the leading-power contribution from this class of diagrams cancels after the sum over possible final-state cuts, analogous to the cancellation of the single-gluon spectator-spectator exchange graph. Besides the color-entangled graphs, there are graphs that already have the color structure anticipated by the factorization theorem. An example of one of those unentangled graphs is diagram (f) in figure~\ref{f:(f)}. Note that for none of the graphs in figures~\ref{f:entangled_diagrams} and~\ref{f:(f)} we have specified any final-state cuts. The reason for this is that the possible cut positions depend on the considered scalings of the gluon momenta; the latter is discussed in the next subsection.

As it turns out, the contributions from diagrams (d) and (e) and their seagull versions cancel after the sum over cuts, along with all unentangled diagrams (except diagram (f) and its Hermitian conjugate); we refer to appendix~A of~\cite{Boer:2017hqr} for a review of these cancellations. Hence, for the calculation of the dBM term we are left with diagrams (a)--(c), and (f). Since diagram (f) already comes with the expected $1/N_c$ color factor, we do not discuss this graph explicitly in the remainder of this chapter.

\subsection{Leading regions} \label{s:leading_regions_model}
As discussed in section~\ref{s:momentum_regions_analysis}, we can identify leading momentum regions for each graph to simplify the calculation. The individual region contributions can be summed together using the Collins subtraction procedure to reconstruct the full graph at leading power. From the power counting analysis applied to the active-spectator gluon exchange graph in figure~\ref{f:power_counting}, it is easy to understand which gluon scalings give a leading-power contribution to diagram (a), which features two active-spectator exchanges. Recall that for the single-gluon exchange graph the leading regions are $U$, $G$, $G_1$, and $C_1$, in the \emph{spin-independent} case. However, now we are interested in contributions to the dBM term; selecting only transversely polarized quarks and antiquarks according to eq.~\eqref{e:fierz}, it turns out that the $U$ region is power suppressed due to the numerator structure. Hence, in diagram (a) the leading regions for the gluon with momentum $\ell_1$ are $G$, $G_1$, and $C_1$, while for the gluon with momentum $\ell_2$ they are $G$, $G_2$, and $C_2$. Even though we know from subsection~\ref{s:trapped_Glauber} that for diagram (a) the integration contours are not trapped in the Glauber regions, we will consider Glauber scalings explicitly in our analysis. As will become clear later, rather than summing each diagram individually over all leading regions and subsequently summing the full graph contributions, it turns out to be convenient to first sum all diagrams for a given region and then combine the various region contributions. Since trappings in the Glauber region occur for other relevant graphs, it is important to consider the Glauber scalings explicitly for all graphs.

For diagrams (a)--(c) and (f) we identify four common momentum regions for the gluon loop momenta that give leading-power contributions. We use the notation $AB$ to describe the various regions, where $A$ denotes the momentum scaling of the gluon with momentum $\ell_1$ and $B$ that of the gluon with momentum $\ell_2$. The four leading regions are $G_1G_2$, $C_1G$, $GC_2$, and $C_1C_2$. For particular graphs one can identify other regions that give a leading-power contribution. However, the contributions from those regions can be absorbed into the contributions from the regions considered (this is analogous to what happens to the $U$ region in eq.~\eqref{e:subtractions_example}). We find that diagram (c) only gives a leading contribution in the $GC_2$ region, so we will not consider this graph explicitly in the other regions (and likewise for the $p\leftrightarrow\bar{p}$ version of diagram (c), which only receives a leading contribution from the $C_1G$ region).

As prescribed by the Collins subtraction procedure (see subsection~\ref{s:subtraction_method}), we consider the contributions from each region with subtractions from the smaller regions. The region $G_1G_2$ is the smallest one, in the sense that the pinch surface it corresponds to is the single point $\ell_1=\ell_2=0$ in the eight-dimensional space spanned by $\ell_1$ and $\ell_2$. The regions $C_1G$ and $GC_2$ are larger than this -- their pinch surfaces are lines that intersect at $\ell_1=\ell_2=0$. Finally, $C_1C_2$ is the largest region with a pinch surface that is a plane. For each momentum region $AB$ we apply an appropriate approximator $T_{AB}$ to the graph $\Gamma$ that reduces to the unit operator (up to power corrections) in the design region $AB$. From eq.~\eqref{e:subtr-def} it follows that the contributions from each leading region are computed as
\begin{align}
    C_{G_1G_2} \Gamma &= T_{G_1G_2} \Gamma, \label{e:C_G1G2} \\
    C_{C_1G} \Gamma &= T_{C_1G} \left( 1 - T_{G_1G_2} \right) \Gamma, \\
    C_{GC_2} \Gamma &= T_{GC_2} \left( 1 - T_{G_1G_2} \right) \Gamma, \\
    C_{C_1C_2} \Gamma &= T_{C_1C_2} \left( 1 - T_{C_1G} - T_{GC_2} \right) \left( 1 - T_{G_1G_2} \right) \Gamma \label{e:C_C1C2} .
\end{align}

The regions $C_1G$, $G_1G_2$, and $GC_2$ are all of the same virtuality (the $C_1C_2$ region is of higher virtuality), meaning that they have the same number of powers of $\lambda$ in their phase space $\int d^4 \ell_1 \,d^4 \ell_2$. In a sense, they are just separated by rapidity. What we mean by this is easiest understood in the context of diagram (b), where we have a gluon with momentum $\ell_1 + \ell_2$ produced by the three-gluon vertex. This gluon has the same virtuality in the $C_1G$, $G_1G_2$, and $GC_2$ regions, but moves in rapidity space from being $C_1$ in the $C_1G$ region, $S$ in the $G_1G_2$ region, and finally $C_2$ in the $GC_2$ region. We know from subsection~\ref{s:rapidity_divergences} that whenever regions are only separated by rapidity, we need to include a rapidity regulator to avoid ill-defined results. For all regions we use regulators of the type~\eqref{e:regulatorbasic}, and we choose slightly different regulators for the different graphs:
\begin{align}
    \text{diagram (a):} &\qquad \frac{1}{2} \left( \left|\frac{\ell_1^+}{\nu}\right|^{-\eta_1} \left|\frac{\ell_2^-}{\nu}\right|^{-\eta_2} + \left|\frac{\ell_1^+}{\nu}\right|^{-\eta_{\bar{1}}} \left|\frac{\ell_2^-}{\nu}\right|^{-\eta_{\bar{2}}} \right) , \label{e:regulator_(a)} \\[5pt]
    \text{diagram (b):} &\qquad \left|\frac{\ell_1^+}{\nu}\right|^{-\eta_1} \left|\frac{\ell_2^-}{\nu}\right|^{-\eta_2} \; \text{with} \;\; \eta_1 \gg \eta_2 \,, \label{e:regulator_(b)} \\[5pt]
    \text{diagram (b) with $p\leftrightarrow\bar{p}$:} &\qquad \left|\frac{\ell_1^+}{\nu}\right|^{-\eta_{\bar{1}}} \left|\frac{\ell_2^-}{\nu}\right|^{-\eta_{\bar{2}}} \; \text{with} \;\; \eta_{\bar{1}} \ll \eta_{\bar{2}} \,. \label{e:regulator_(b)pp}
\end{align}
Choosing regulators on a graph-by-graph basis is allowed as a full graph $\Gamma$ does not suffer from rapidity divergences and thus does not require a rapidity regulator. At the end of the calculation we can take the limits $\eta_i \to 0$. Note however that for the three-gluon vertex graphs we have to take these limits in a specific order to obtain a well-defined result. Once we sum all graphs, diagram (b) needs to be combined with the first term in~\eqref{e:regulator_(a)}, while its $p\leftrightarrow\bar{p}$ version needs to be combined with the second term. Alternative rapidity regulators are discussed in appendix~B of~\cite{Boer:2017hqr}.

In section~\ref{s:diagram_calculations} we will show that upon summing over all graphs, leading regions, and possible final-state cuts we obtain a disentangled result with the standard $1/N_c$ color factor, in agreement with the TMD factorization theorem. As it turns out, the \emph{full} contribution to the dBM term comes from the $G_1G_2$ region -- the contributions from the $C_1G$, $GC_2$, and $C_1C_2$ regions all vanish at leading power. This, however, is a regulator-dependent statement -- it is possible to shift contributions between regions by choosing different rapidity regulators. The fact that the dBM contribution at the considered order can be entirely ascribed to the $G_1G_2$ region implies that these Glauber effects can be absorbed in the TMDs -- after all, there is no explicit Glauber function in the factorization formula. Indeed, we will find that the Glauber gluons can be absorbed in the collinear subgraphs by deformations of the loop momentum contours, fully consistent with the CSS approach; this is discussed in subsection~\ref{s:final_result}.

\section{The Boer-Mulders function} \label{s:BM_function}

\begin{figure}[htb]
\centering
    \includegraphics[height=2.0cm]{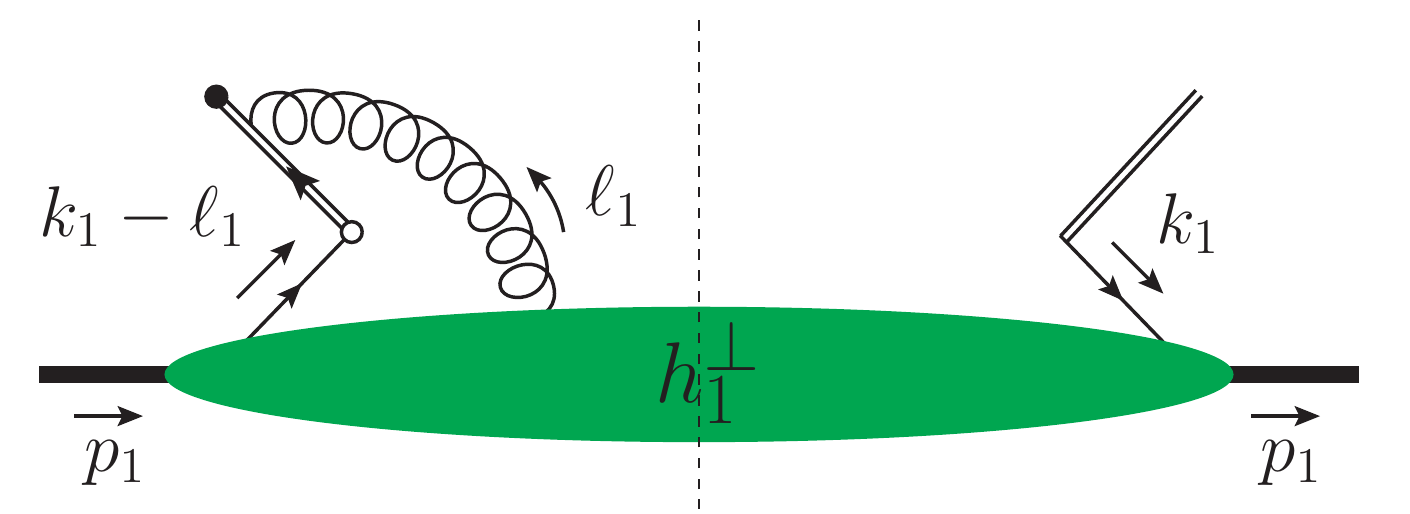}
    \caption{The first-order contribution to the quark BM function $h_1^\perp$ (also the Hermitian conjugate graph is needed).}
\label{f:BM_QCD}
\end{figure}

In our $\mathcal{O}(\alpha_s^2)$ calculation of the dBM contribution in the next section, we will not assume but rather \emph{derive} factorization. To be able to later identify the pieces that represent the quark and antiquark BM functions, we calculate $h_1^\perp$ and $\bar{h}_1^\perp$ to leading power in the model. However, before we resort to the model setting, let us first look at the quark BM function in QCD. The operator definition of $h_1^\perp$ is given in eq.~\eqref{e:h_1}, where the past-pointing gauge link $U_{[0,z]}^{[-]}$ reflects a sum over infinitely many gluon emissions. The $\mathcal{O}(g)$ contribution in the expansion of the gauge link (we will refer to this as the first order), which corresponds to the emission of a single gluon, is given by the diagram in figure~\ref{f:BM_QCD} (plus its Hermitian conjugate).

Whether one ends up with a contribution to either a future- or a past-pointing gauge link is determined by the sign of the $i\epsilon$ term in the denominator of the eikonal propagator. This sign is different for DY and SIDIS, as it depends on the direction of color flow. In the following we will argue that a plus sign results in a past-pointing gauge link. Using the Feynman rules for eikonal lines and vertices given in appendix~\ref{a:feynman_rules}, the BM function in DY is given by\footnote{For convenience we suppress in this section any reference to quark flavors.}
\begin{align}
    \frac{\widetilde{k}_{1\sst}^j}{M} \,h_1^\perp(x_1,\bm{k}_1^2) &= \int \left. \frac{dz^- d^2\bm{z}}{(2\pi)^3} \;e^{ik_1 \cdot z} \bra{p_1} \overline{\psi}(0) \,\Gamma_T^j \,\psi(z) \ket{p_1} \vphantom{\int} \right|_{z^+=0} \nn \\
    &\quad\, + \left\{ \left( -ig t^a n^\mu \right) \int d\ell_1^+ d^2\bm{\ell}_1 \left( - \frac{i}{\ell_1 \cd n + i\epsilon} \right) \int \frac{dz^- d^2\bm{z}}{(2\pi)^3} \int \frac{dz'^- d^2\bm{z}'}{(2\pi)^3} \right. \nn \\
    &\quad\, \times \left. e^{ik_1 \cdot z} \,e^{i\ell_1 \cdot (z'-z)} \bra{p_1} \overline{\psi}(0) \,A_\mu^a(z') \,\Gamma_T^j \,\psi(z) \ket{p_1} \vphantom{\frac{dz^- d^2\bm{z}}{(2\pi)^3}} \right|_{z^+=z'^+=0}  \nn \\
    &\quad\, \left. + \:\text{h.c.} \vphantom{\left( - \frac{i}{\ell_1 \cd n \mp i\epsilon} \right)} \right\} + \mathcal{O}(g^2) \nn \\[3pt]
    &= \int \left. \frac{dz^- d^2\bm{z}}{(2\pi)^3} \;e^{ik_1 \cdot z} \bra{p_1} \overline{\psi}(0) \,\Gamma_T^j \,\psi(z) \ket{p_1} \vphantom{\int} \right|_{z^+=0} \nn \\
    &\quad\, + \left\{ \int \frac{dz^- d^2\bm{z}}{(2\pi)^3} \;e^{ik_1 \cdot z} \bra{p_1} \overline{\psi}(0) \,\Gamma_T^j \left[ ig \int \frac{dz'^- d^2\bm{z}'}{(2\pi)^3} \right.\right. \nn \\
    &\quad\, \times \left.\left. \!\int d\ell_1^+ d^2\bm{\ell}_1 \;\frac{i}{\ell_1 \cd n + i\epsilon} \;e^{i\ell_1 \cdot (z'-z)} \,n \cd A(z') \right] \psi(z) \ket{p_1} \right|_{z^+=z'^+=0} \nn \\
    &\quad\, \left. + \:\text{h.c.} \vphantom{\left( - \frac{i}{\ell_1 \cd n \mp i\epsilon} \right)} \right\} + \mathcal{O}(g^2) .
    \label{e:BM_function_QCD}
\end{align}
The term on the first line represents the $\mathcal{O}(g^0)$ contribution, corresponding to no gluon emissions. Since this term does not depend on an eikonal propagator (in other words, it is \emph{independent} of the direction of color flow), it is the same for DY and SIDIS and thus it must vanish. However, since we will identify this term with the $\mathcal{O}(g^0)$ contribution to the gauge link, it is better to keep it here. The terms in between curly brackets make up the $\mathcal{O}(g)$ contribution. They contain the Fourier transform of the trilocal quark-gluon-quark matrix element, evaluated at the light-front $z^+ = z'^+ = 0$. Making use of the integral representation
\begin{equation}
    \frac{i}{x \pm i\epsilon} = \pm \int_0^\infty d\zeta \;e^{i\zeta(\pm x + i\epsilon)} ,
    \label{e:int_rep}
\end{equation}
the quantity in between square brackets in eq.~\eqref{e:BM_function_QCD} can be rewritten as
\begin{align}
    & \left. ig \int \frac{dz'^- d^2\bm{z}'}{(2\pi)^3} \int d\ell_1^+ d^2\bm{\ell}_1 \;\frac{i}{\ell_1 \cd n + i\epsilon} \;e^{i\ell_1 \cdot (z'-z)} \,n \cd A(z') \right|_{z^+=z'^+=0} \nn \\[3pt]
    =& \left. ig \int_0^\infty d\zeta \int \frac{dz'^- d^2\bm{z}'}{(2\pi)^3} \int d\ell_1^+ d^2\bm{\ell}_1 \;e^{i(\zeta n + z'-z) \cdot \ell_1} \,n \cd A(z') \right|_{z^+=z'^+=0} \nn \\[3pt]
    =& \left. ig \int_0^\infty d\zeta \;n \cd A(z - \zeta n) \right|_{z^+=0} \nn \\[3pt]
    =& \:ig \int_{-\infty}^{z^-} d\eta^- A^+(\eta^+=0,\eta^-,\bm{\eta}=\bm{z}) ,
\end{align}
where in the last step we defined $\eta \equiv z - \zeta n$. Once we also include the graph with one gluon emission on either side of the final-state cut, eq.~\eqref{e:BM_function_QCD} becomes
\begin{align}
    \frac{\widetilde{k}_{1\sst}^j}{M} \,h_1^\perp(x_1,\bm{k}_1^2) &= \int \frac{dz^- d^2\bm{z}}{(2\pi)^3} \;e^{ik_1 \cdot z} \bra{p_1} \overline{\psi}(0) \,\Gamma_T^j \nn \\
    &\quad\, \times \left[ 1 + ig \int_{0}^{-\infty} d\eta^- A^+(\eta^+=0,\eta^-,\bm{\eta}=\bm{0}) \right] \nn \\
    &\quad\, \times \left. \!\left[ 1 + ig \int_{-\infty}^{z^-} d\eta^- A^+(\eta^+=0,\eta^-,\bm{\eta}=\bm{z}) \right] \psi(z) \ket{p_1} \right|_{z^+=0} \nn \\
    &\quad\, + \,\mathcal{O}(g^2) .
\end{align}
The two factors in between square brackets correspond to the expansions up to $\mathcal{O}(g)$ of the Wilson lines $U_{[0,-\infty;\bm{0}]}^n$ and $U_{[-\infty,z^-;\bm{z}]}^n$ defined in eq.~\eqref{e:U^n}. Since we have adopted the Feynman gauge, there is no contribution from the transverse link $U_{[-\infty;\bm{0},\bm{z}]}^T$ defined in eq.~\eqref{e:U^T}. Note that up to the considered order, the path ordering of the Wilson lines does not play a role (this starts at $\mathcal{O}(g^2)$). Summing over all gluon emissions (see e.g.~\cite{Buffing:2015tja}), one indeed precisely recovers eq.~\eqref{e:h_1}, featuring the past-pointing gauge link $U_{[0,z]}^{[-]}$.

If we had taken the $i\epsilon$ term in the eikonal propagator with a minus sign instead, corresponding to the SIDIS case, we would have ended up with the future-pointing gauge link $U_{[0,z]}^{[+]}$. Hence, the final gauge link structure is indeed connected to the $i\epsilon$ term in the eikonal propagator denominator $\ell_1^+ \pm i\epsilon$. Since the BM function is process dependent (it changes sign between DY and SIDIS), it requires sensitivity to the imaginary part of this propagator, which is only the case when $\ell_1^+ \to 0$. As will be shown below, this is consistent with the fact that the first-order contribution to the BM function in the model turns out to originate from the region where the gluon has Glauber scaling.

\begin{figure}[htb]
\centering
    \begin{subfigure}{.445\textwidth}
    \begin{flushleft}
        \includegraphics[height=2.0cm]{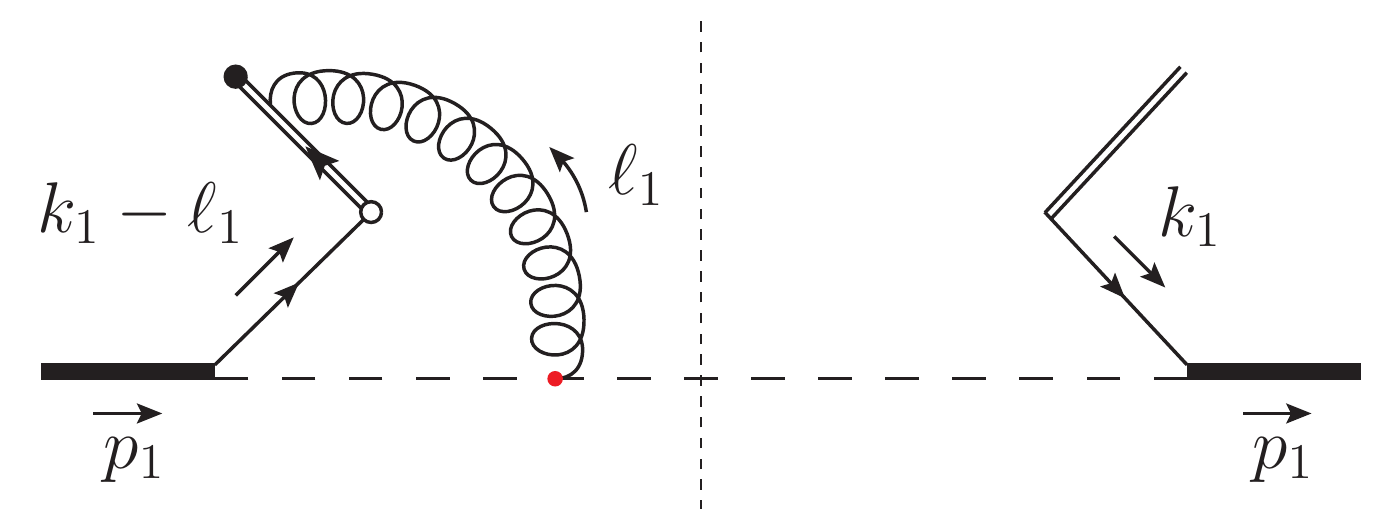}
    \end{flushleft}
    \end{subfigure}
    \begin{subfigure}{.445\textwidth}
    \begin{flushright}
        \includegraphics[height=2.0cm]{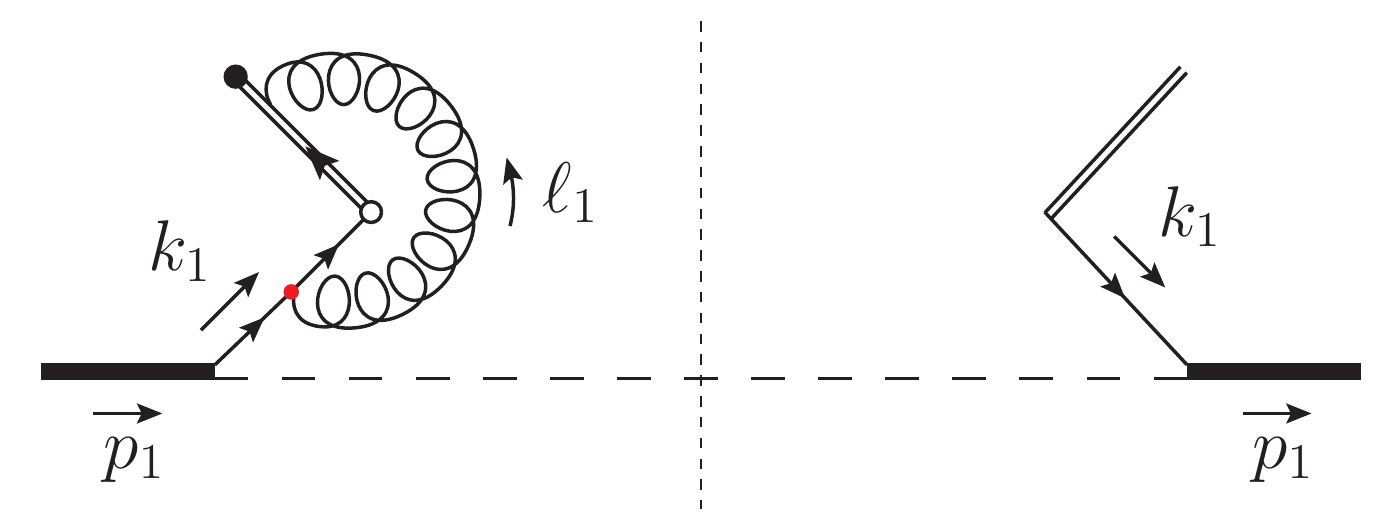}
    \end{flushright}
    \end{subfigure}
    \caption{The two $\mathcal{O}(\alpha_s)$ contributions to the quark BM function $h_1^\perp$ in the scalar spectator model (also the Hermitian conjugate graphs are needed).}
\label{f:BM}
\end{figure}

Now let us switch to the spectator model. Naively, one would need to compute the BM functions up to the order at which we work, namely $\mathcal{O}(\alpha_s^2)$. However, since the BM functions are T-odd they each require a gluon attachment to the eikonal line -- thus, there is no tree-level contribution and it suffices to compute each function only up to $\mathcal{O}(\alpha_s)$. At the $\mathcal{O}(\alpha_s)$ level, there are two contributions to $h_1^\perp$, see figure~\ref{f:BM} (also the Hermitian conjugate graphs are needed). We note that there is no contribution coming from the diagram on the right where the gluon attaches to the active quark line -- its numerator vanishes:
\begin{equation}
    \tr \left( \Gamma_T^j \,\slashed{k}_1 \,(\slashed{p}_1 + M) \,\slashed{k}_1 \,(\slashed{k}_1 - \slashed{\ell}_1) \,\slashed{n} \right) = 0 .
\end{equation}
By contrast, the graph on the left does not vanish and gives the first-order contribution to the past-pointing Wilson line.

To calculate the BM function $h_1^\perp$ in the model, i.e.\ the diagram on the left in figure~\ref{f:BM} plus its conjugate, we first identify two (nontrivial) momentum regions for the gluon momentum $\ell_1$ that could give a leading-power contribution, namely $G_1$ and $C_1$ (the power counting is similar to that for the single-gluon exchange graph in figure~\ref{f:power_counting}). The region $G$ can be absorbed in $G_1$, and $U$ gives a power-suppressed contribution for the same reason as discussed earlier in subsection~\ref{s:leading_regions_model}. From eqs.~\eqref{e:subtr-def} and~\eqref{e:subtr-sum} it follows that the sum over the $G_1$ and $C_1$ region contributions is given by
\begin{equation}
    C_{G_1} \Gamma + C_{C_1} \Gamma = T_{G_1} \Gamma + T_{C_1} \left( 1 - T_{G_1} \right) \Gamma .
    \label{e:bmregions}
\end{equation}
As mentioned already in subsection~\ref{s:leading_regions_model} and as we will show explicitly in section~\ref{s:diagram_calculations}, for our choice of rapidity regulators the full contribution to the dBM term turns out to come from the $G_1G_2$ region. Hence, we expect only the first term in eq.~\eqref{e:bmregions} to be nonzero.

Before applying any momentum approximations, the quark BM function in the spectator model is given by
\begingroup 
\thinmuskip=1.2mu
\medmuskip=1.6mu
\begin{equation}
    \frac{\widetilde{k}_{1\sst}^j}{M} \,h_1^\perp(x_1,\bm{k}_1^2) = - i \,C_\Phi \int \frac{d\ell_1^+}{2\pi} \;(2p_1-2k_1+\ell_1) \cd n \;\chi^j(x_1,\bm{k}_1) \,\frac{\nu^{\eta_1} |\ell_1 \cd n|^{-\eta_1}}{\ell_1 \cd n + i\epsilon} + \text{h.c.} \,,
    \label{e:phi}
\end{equation}
\endgroup
where we have included the rapidity regulator $\eta_1$ (which can be sent to zero at the end of this calculation), as well as the rapidity scale $\nu$. All relevant Feynman rules are listed in appendix~\ref{a:feynman_rules}. Note that $h_1^\perp$ is manifestly real due to the presence of the Hermitian conjugate term (denoted by `h.c.'). The color factor $C_\Phi$ is defined in eq.~\eqref{e:non-entangled_color}, and we have defined
\begin{align}
    \chi^j(x_1,\bm{k}_1) &\equiv \pi g^2 \int \frac{dk_1^-}{(2\pi)^4} \;\theta\big((p_1-k_1)^0\big) \,\delta\big((p_1-k_1)^2 - m_s^2\big) \int \frac{d\ell_1^-}{2\pi} \int \frac{d^2\bm{\ell}_1}{(2\pi)^2} \nn \\
    & \,\quad \times \frac{D_1^j}{[(k_1-\ell_1)^2 + i\epsilon] \,[(p_1-k_1+\ell_1)^2 - m_s^2 + i\epsilon] \,[\ell_1^2 + i\epsilon] \,[k_1^2 - i\epsilon]} \,.
    \label{e:chi}
\end{align}
The Dirac trace $D_1^j$ is given by
\begin{equation}
    D_1^j \equiv \tr \left( \Gamma_T^j \,(\slashed{k}_1 - \slashed{\ell}_1) \,(\slashed{p}_1 + M) \,\slashed{k}_1 \right) = 2iM \left( x_1 p_1^+ \widetilde{\ell}_{1\sst}^{\,j} - \ell_1^+ \widetilde{k}_{1\sst}^{\,j} \right) .
    \label{e:D_1}
\end{equation}

Let us first calculate the contribution from the $G_1$ region. We expand $h_1^\perp$ up to leading power in $\lambda$ and subsequently perform the integrals over $k_1^-$ and $\ell_1^-$. The delta function $\delta\big((p_1-k_1)^2-m_s^2\big)$ is used for the integration over $k_1^-$ and for the integration over $\ell_1^-$ we invoke Cauchy's residue theorem. To leading power, the BM function is given by
\begin{align}
    \frac{\widetilde{k}_{1\sst}^j}{M} \,h_1^\perp(x_1,\bm{k}_1^2) &= - 2i \,C_\Phi \,(1-x_1) p_1^+ \,\chi^j(x_1,\bm{k}_1) \int \frac{d\ell_1^+}{2\pi} \,\frac{\nu^{\eta_1} |\ell_1^+|^{-\eta_1}}{\ell_1^+ + i\epsilon} + \text{h.c.} \,,
    \label{e:phi_final}
\end{align}
where, using the shorthand notation $\Lambda_1^2 \equiv x_1 m_s^2 - x_1(1-x_1) M^2 $,
\begin{align}
    \chi^j(x_1,\bm{k}_1) &= \frac{ig^2}{64\pi^3} \int \frac{d^2\bm{\ell}_1}{(2\pi)^2} \,\frac{\theta(x_1) \,\theta(1-x_1) \,D_1^j}{(p_1^+)^2 \,[(\bm{k}_1-\bm{\ell}_1)^2 + \Lambda_1^2] \,(\bm{k}_1^2 + \Lambda_1^2) \,\bm{\ell}_1^2} \,,
    \label{e:chi_final}
\end{align}
and
\begin{equation}
    D_1^j = 2iM \,x_1 p_1^+ \widetilde{\ell}_{1\sst}^{\,j} \,.
    \label{e:D_1simp}
\end{equation}
Inasmuch as the quantity $\chi$ is real, only the imaginary part of the $\ell_1^+$ integral contributes to $h_1^\perp$ -- its real part is canceled by the Hermitian conjugate term. This imaginary part comes from the region where $\ell_1^+$ is sensitive to the $i\epsilon$ term in the denominator, which is the case for $\ell_1^+ \to 0$, i.e.\ when $\ell_1$ has Glauber scaling. Note that this is consistent with our earlier observation -- the sensitivity to the $i\epsilon$ term is connected to the required process dependence of this T-odd function.

Now let us consider the $C_1$ momentum region. Here, it is sufficient to consider what happens at a fixed nonzero value of $\ell_1^+$, as the small $\ell_1^+$ region is suppressed due to the subtraction (this is explained in subsection~\ref{s:subtraction_method}). This means that we can drop the $i\epsilon$ term in the eikonal propagator, making it purely real. We consider the subtraction and naive graph terms ($T_{C_1}T_{G_1} \Gamma$ and $T_{C_1} \Gamma$, respectively) separately at this nonzero $\ell_1^+$. The contribution to $h_1^\perp$ from the subtraction term is also given by eq.~\eqref{e:phi_final}, but vanishes at finite $\ell_1^+$ due to the cancellation between the amplitude and its conjugate. At leading power, the contribution from the naive graph term is given by
\begingroup 
\thinmuskip=1.8mu
\medmuskip=2.4mu
\begin{equation}
    \frac{\widetilde{k}_{1\sst}^j}{M} \,h_1^\perp(x_1,\bm{k}_1^2) = - i \,C_\Phi \int \frac{d\ell_1^+}{2\pi} \left[ 1 + \frac{2(1-x_1)p_1^+}{\ell_1^+} \right] \left|\frac{\ell_1^+}{\nu}\right|^{-\eta_1} \!\chi^j(x_1,\bm{k}_1,\ell_1^+) + \text{h.c.}
    \label{e:phi_final_C_1}
\end{equation}
\endgroup
The precise form for $\chi$ is not relevant here -- what matters is that it is again real-valued for nonzero $\ell_1^+$ such that the amplitude and conjugate contributions cancel. Hence, the $C_1$ region does not contribute to the BM function and the full contribution comes from the $G_1$ region only.

The $G_1$ region result in eq.~\eqref{e:phi_final} can be simplified further. The integral over $\bm{\ell}_1$ can be performed using standard techniques, including Passarino-Veltman reduction~\cite{Passarino:1978jh} and the use of Feynman parameters. Performing also the integral over $\ell_1^+$ and subsequently taking the limit $\eta_1 \to 0$, we find that the BM function in our spectator model is given by
\begin{equation}
    h_1^\perp(x_1,\bm{k}_1^2) = \frac{g^2}{16\pi^2} \;C_\Phi \,\frac{x_1(1-x_1) M^2}{\bm{k}^2 (\bm{k}^2 + \Lambda_1^2)} \,\ln \left( \frac{\bm{k}^2 + \Lambda_1^2}{\Lambda_1^2} \right) ,
    \label{e:BM_integrated}
\end{equation}
which is consistent with the results in~\cite{Goldstein:2002vv,Bacchetta:2008af}. The fully integrated result for $h_1^\perp$ in eq.~\eqref{e:BM_integrated} is not needed for our calculations in the next section; the form in eq.~\eqref{e:phi_final} turns out to be sufficient for our purposes.

In the same way we can obtain the BM function for the antiquark, where the full contribution comes from the $G_2$ region. Since the kinematical setup is invariant under the simultaneous interchange of plus and minus indices and the particle labels $1$ and $2$, $\bar{h}_1^\perp$ is simply obtained from $h_1^\perp$ by the two substitutions $+ \to -$ and $1 \to 2$.

\section{Calculation of the cross section} \label{s:diagram_calculations}

At the $\mathcal{O}(\alpha_s^2)$ level there are various graphs that potentially contribute to the dBM term in the DY cross section, see figures~\ref{f:entangled_diagrams} and~\ref{f:(f)}. In subsection~\ref{s:relevant_graphs} we argued that some of these graphs vanish upon summing over cuts, and in subsection~\ref{s:leading_regions_model} we already mentioned that in all regions the sum over graphs and cuts turns out to give zero except in the $G_1G_2$ region. In this section we present the leading-power calculation of all nonvanishing color-entangled graphs, considering all possible cut positions. To this end, we use the decomposition in eq.~\eqref{e:fierz} to select transversely polarized quarks and antiquarks. Flavor labels, as well as the sum over different flavors, will be implicit (so the quantity $e^2$ should be thought of as $\sum_q e_q^2$). We discuss the $G_1G_2$ region computation in detail and then comment briefly on the other region contributions. First, let us rearrange the various integrations that appear in the differential cross section contributions and discuss some overall factors.

Each diagram contains integrations over the internal four-momenta $\ell_1$, $\ell_2$, $k_1$, $k_2$, and $q$. The integrations over $\ell_1$ and $\ell_2$, as well as over the components $k_1^-$ and $k_2^+$, are contained in the quark and antiquark BM functions, as discussed in the previous section. The remaining integrals, including the two-particle phase space element, can be cast into the following form at leading power:
\begin{align}
    d\sigma_{\scriptscriptstyle{\text{dBM}}} &\propto \int dk_1^+ d^2\bm{k}_1 \int dk_2^- d^2\bm{k}_2 \;(2\pi)^4 \,\delta^{(4)}(k_1+k_2-q) \int \frac{d^4q}{(2\pi)^4} \nn \\
    &\quad\, \times (2\pi)^4 \,\delta^{(4)}(\ell+\ell'-q) \,\frac{d^3\vec{l}}{(2\pi)^3 \,2l^0} \frac{d^3\vec{l}'}{(2\pi)^3 \,2l'^0} \nn \\
    &= \frac{1}{8} \int dk_1^+ d^2\bm{k}_1 \int dk_2^- d^2\bm{k}_2 \;(2\pi)^4 \,\delta^{(4)}(k_1+k_2-q) \,\frac{d^4q}{(2\pi)^4} \frac{d\Omega}{(2\pi)^2} \nn \\
    &= \frac{p_1^+ p_2^-}{32\pi^2} \int d^2\bm{k}_1 \int d^2\bm{k}_2 \;\delta^{(2)}(\bm{k}_1+\bm{k}_2-\bm{q}) \,d\Omega \,dx_1 dx_2 \,d^2\bm{q} \,,
\end{align}
where $l^0 = \sqrt{\vec{l}^{\,2}}$, $l'^0 = \sqrt{\vec{l}'^{\,2}}$, and $d\Omega \equiv \sin\theta \,d\theta \,d\phi$, in terms of the CS angles $\theta$ and $\phi$ (see figure~\ref{f:CS}).

For our cross section calculation we include the usual flux factor, which to leading power is given by
\begin{equation}
    \frac{1}{4\sqrt{(p_1 \cd p_2)^2 - M^4}} = \frac{1}{4p_1^+ p_2^-} \,.
\end{equation}
The averaging over initial-state polarizations leads to a factor of $\tfrac{1}{4}$ (each hadron can have two spin polarizations); this factor has been absorbed in the BM functions.

\subsection{Contribution from the $G_1G_2$ region} \label{s:G1G2_region_contribution}

\begin{figure}[htb]
\centering
\begin{subfigure}{.48\textwidth}
    \begin{flushleft}
        \includegraphics[height=4.1cm]{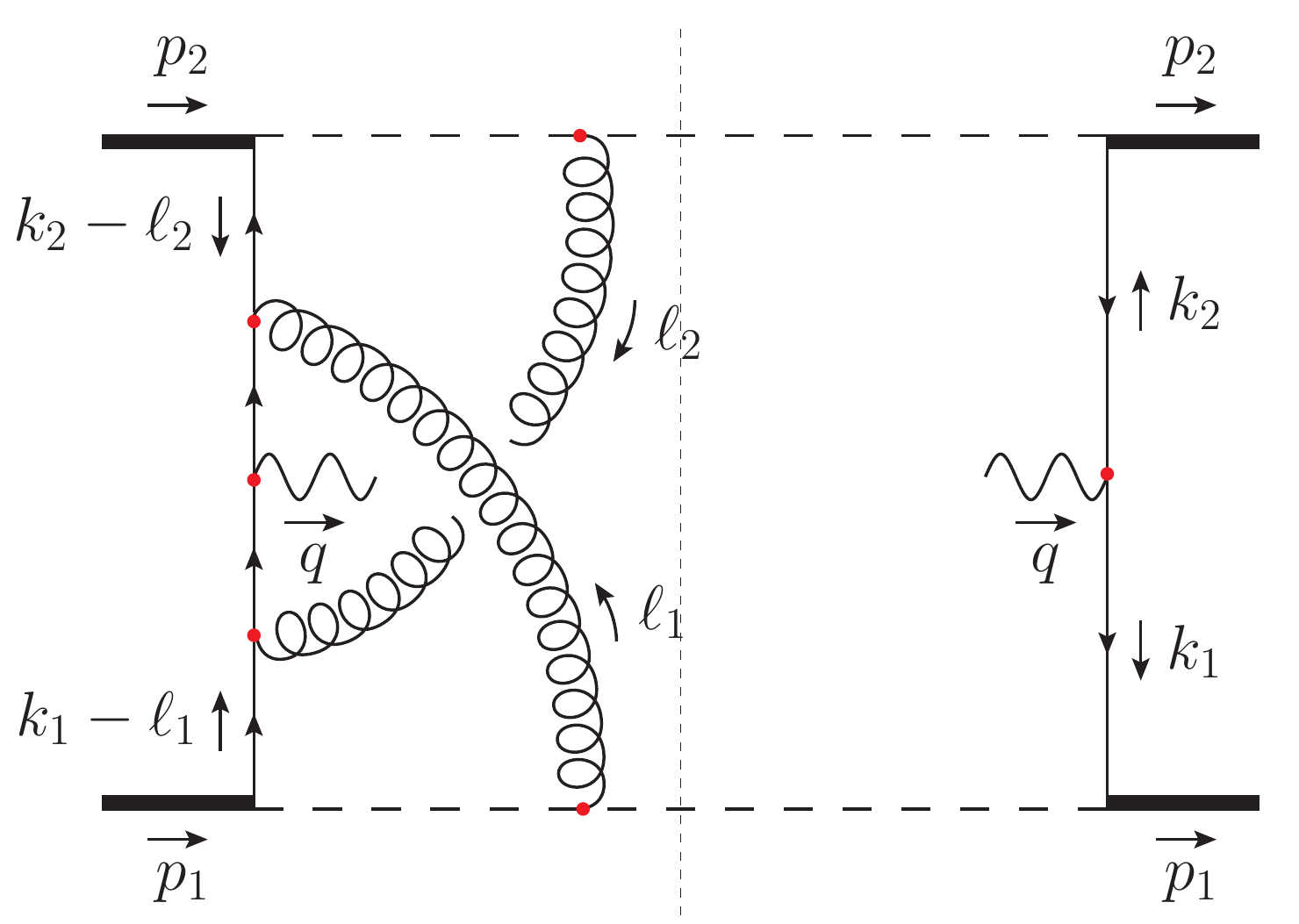}
        \caption*{\hspace{-0.18cm}(a,i)}
    \end{flushleft}
\end{subfigure}
\begin{subfigure}{.48\textwidth}
    \begin{flushright}
        \includegraphics[height=4.1cm]{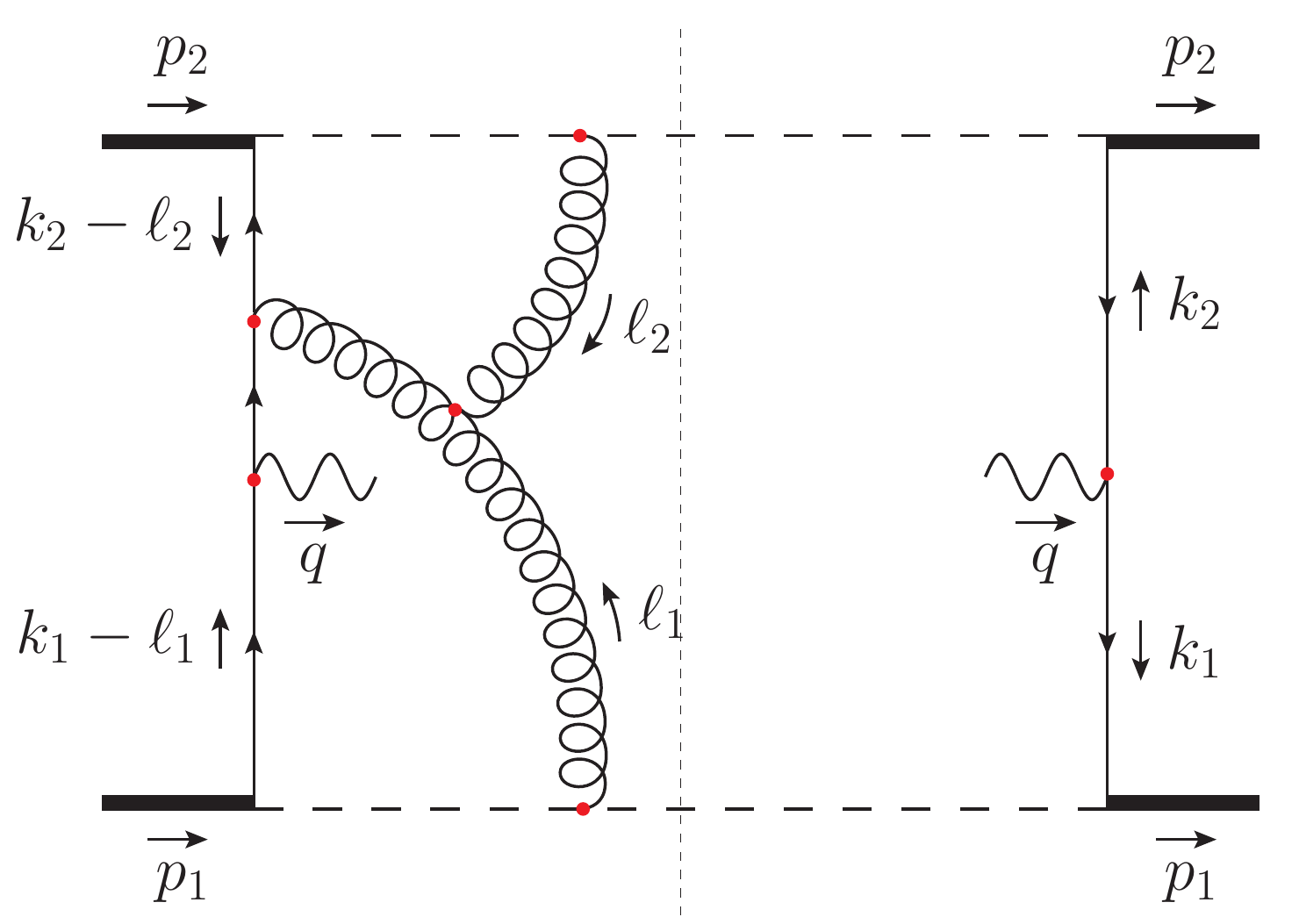}
        \caption*{\hspace{0.78cm}(b,i)}
    \end{flushright}
\end{subfigure}
\begin{subfigure}{.48\textwidth}
    \begin{center}
    \hspace{-0.45cm}
        \includegraphics[height=4.1cm]{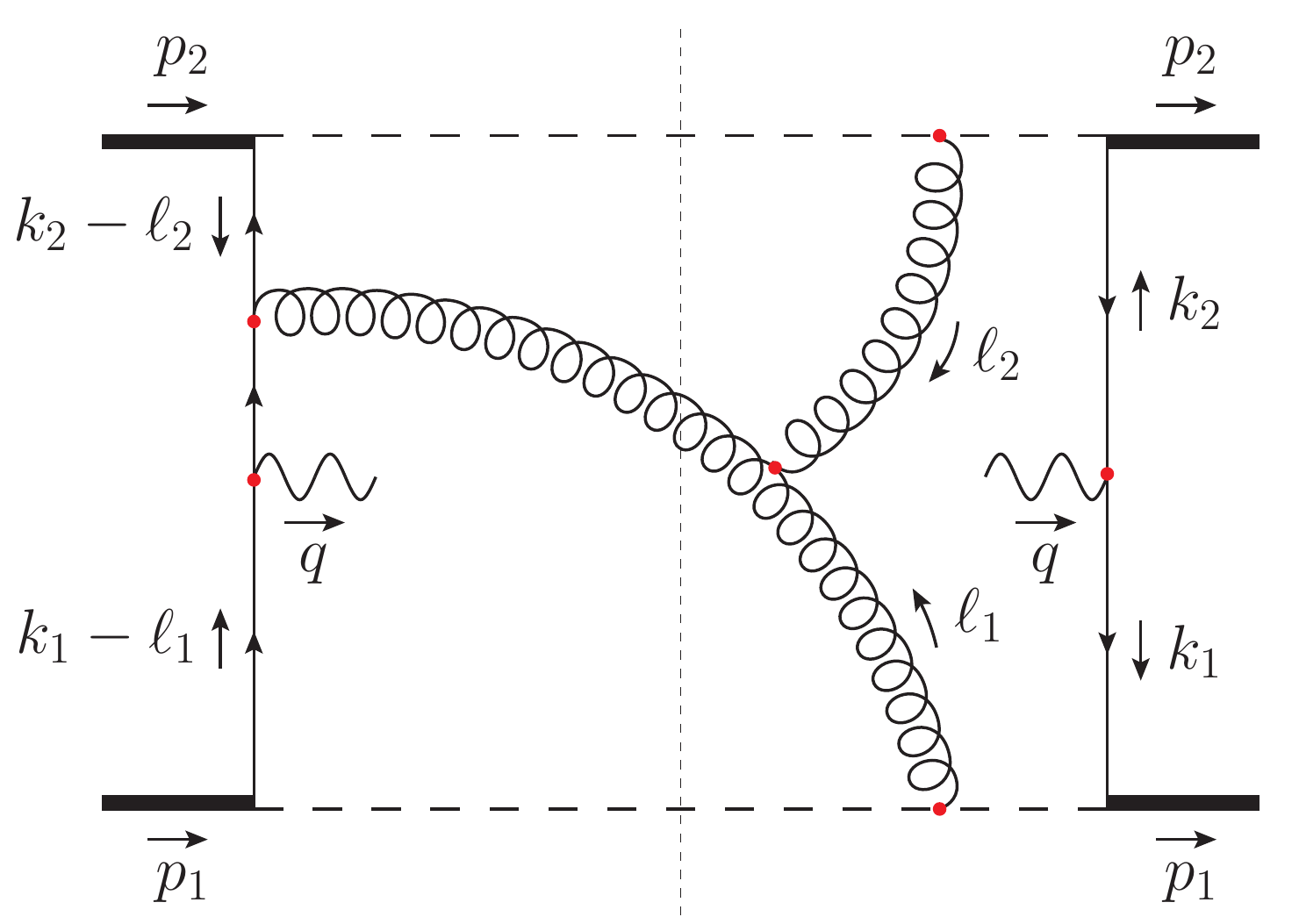}
        \caption*{\hspace{-0.06cm}(b,v)}
    \end{center}
\end{subfigure}
\caption{The color-entangled two-gluon exchange graphs that contribute to the dBM part of the DY cross section in the $G_1G_2$ region. This set is supplemented by graphs that can be obtained by $p\leftrightarrow\bar{p}$ or Hermitian conjugation.}
\label{f:diagrams}
\end{figure}

\noindent For the $G_1G_2$ region, we need to consider all possible final-state cuts for graphs (a) and (b) given in figure~\ref{f:entangled_diagrams}. Since cuts through Glauber modes are not permitted, there is only one possible cut position for graph (a), whereas there are two for graph (b). Denoting the cut positions with roman numerals, the three diagrams that need to be computed are (a,i), (b,i), and (b,v), see figure~\ref{f:diagrams} (in subsection~\ref{s:other_region_contributions} it becomes clear why we refer to the cut through the soft gluon line in graph (b) as the fifth cut). In addition to these diagrams, also the ones that can be obtained through $p\leftrightarrow\bar{p}$ or Hermitian conjugation are needed.

The differential cross section can be computed by summing over the relevant cut graphs. For the $G_1G_2$ region, the sum can be written as follows:
\begin{align}
    \hspace{-0.1cm} \frac{d\sigma_{\scriptscriptstyle{\text{dBM}}}}{d\Omega \,dx_1 dx_2 \,d^2\bm{q}} &= 2 \left[ \frac{1}{2} \left( \frac{d\sigma_{\scriptscriptstyle{\text{dBM}}}}{d\Omega \,dx_1 dx_2 \,d^2\bm{q}} \right)_{\text{(a,i)}} + \left( \frac{d\sigma_{\scriptscriptstyle{\text{dBM}}}}{d\Omega \,dx_1 dx_2 \,d^2\bm{q}} \right)_{\text{(b,i)}} \right. \nn \\
    & \quad\, \left. + \left( \frac{d\sigma_{\scriptscriptstyle{\text{dBM}}}}{d\Omega \,dx_1 dx_2 \,d^2\bm{q}} \right)_{\text{(b,v)}} + \frac{1}{2} \left( \frac{d\sigma_{\scriptscriptstyle{\text{dBM}}}}{d\Omega \,dx_1 dx_2 \,d^2\bm{q}} \right)_{\text{(f,i)}} + \text{h.c.} \right] .
    \label{e:sum_diagrams}
\end{align}
The factor of $2$ in front arises from taking into account the graphs that can be obtained from (b,i) and (b,v) (and their Hermitian conjugates) by $p\leftrightarrow\bar{p}$. Diagram (f,i) is given in figure~\ref{f:(f,i)}. Although this graph already comes with the expected $1/N_c$ color factor and will not play any role in disentangling the color structures of diagrams (a,i), (b,i), and (b,v), it is needed to obtain the full contribution from the $G_1G_2$ region. This graph is not considered here explicitly.

\begin{figure}[htb]
\centering
\begin{subfigure}{.48\textwidth}
    \begin{center}
        \includegraphics[height=4.0cm]{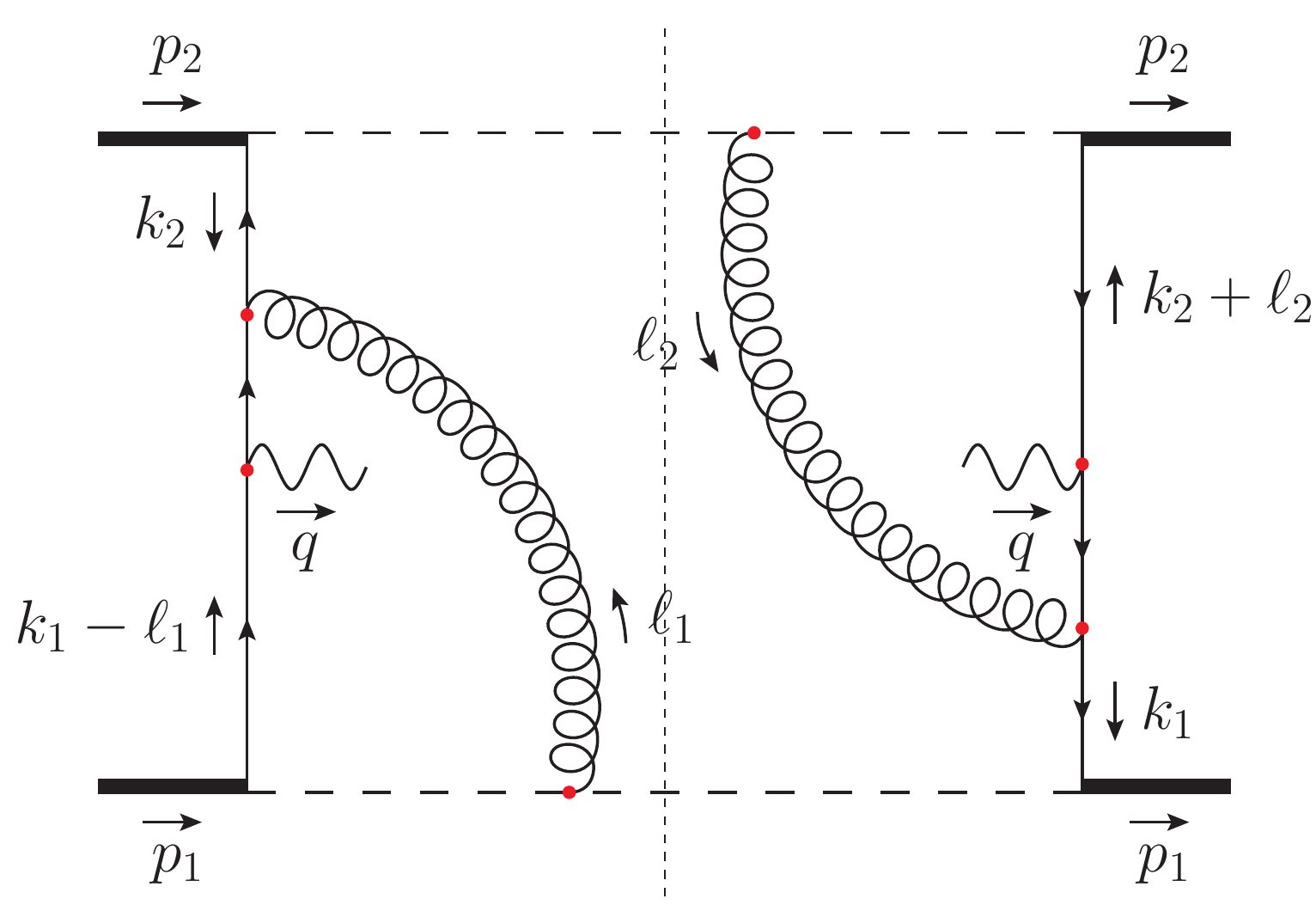}
        \caption*{\hspace{0.08cm}(f,i)}
    \end{center}
\end{subfigure}
\caption{The only unentangled diagram that contributes to the dBM part of the DY cross section in the $G_1G_2$ region (also its Hermitian conjugate is needed).}
\label{f:(f,i)}
\end{figure}

We now proceed with the leading-power calculation of the dBM contributions from diagrams (a,i), (b,i), and (b,v) in the $G_1G_2$ region. In the following, all diagrams are expressed in terms of the functions $\chi$ (defined in eq.~\eqref{e:chi} and simplified in eq.~\eqref{e:chi_final}) and $\overline{\chi}$ (the antiquark analogue of $\chi$). \\

\noindent \textbf{Diagram (a,i).} The dBM contribution from diagram (a,i) is given by
\begin{align}
    \frac{1}{2} \left( \frac{d\sigma_{\scriptscriptstyle{\text{dBM}}}}{d\Omega \,dx_1 dx_2 \,d^2\bm{q}} \right)_{\text{(a,i)}} &= - \frac{1}{16} \frac{\alpha^2}{q^4} \,e^2 \,C_{\text{(a)}} \int d^2\bm{k}_1 \int \frac{d\ell_1^+}{2\pi} \;\chi^j(x_1,\bm{k}_1) \int d^2\bm{k}_2 \int \frac{d\ell_2^-}{2\pi} \nn \\
    &\quad\, \times \overline{\chi}^k(x_2,\bm{k}_2) \,\frac{\nu^{\eta_1 + \eta_2} |\ell_1^+|^{-\eta_1} |\ell_2^-|^{-\eta_2} \,{R_{\text{(a)}}}_{jk;\mu\nu} L^{\mu\nu}}{[(k_2-\ell_2+\ell_1)^2 + i\epsilon] \,[(k_1-\ell_1+\ell_2)^2 + i\epsilon]} \nn \\
    &\quad\, \times \delta^{(2)}(\bm{k}_1+\bm{k}_2-\bm{q}) ,
    \label{e:(a,i)}
\end{align}
where only the first term of the regulator defined in~\eqref{e:regulator_(a)} appears. As explained in subsection~\ref{s:leading_regions_model}, the other `half' of diagram (a,i) comes with the second term in~\eqref{e:regulator_(a)} and is ultimately combined with the $p\leftrightarrow\bar{p}$ versions of diagrams (b,i) and (b,v). The color factor $C_{\text{(a)}}$ is given by\footnote{Whenever a quantity is independent of the cut position, the roman numeral is simply omitted in the subscript.}
\begin{equation}
    C_{\text{(a)}} \equiv \tr \left( t^a t^b t^a t^b \right) = - \frac{1}{N_c^2-1} \frac{1}{N_c} \,C_\Phi^2 \,.
    \label{e:Ca}
\end{equation}
Note that this is precisely the color factor that has been predicted to appear in the final factorized result in~\cite{Buffing:2013dxa}. The Dirac trace ${R_{\text{(a)}}}_{jk;\mu\nu}$ in eq.~\eqref{e:(a,i)} is defined as
\begin{align}
    {R_{\text{(a)}}}_{jk;\mu\nu} &\equiv \tr \left( {\overline{\Gamma}_T}_j \,\gamma_\mu \,{\Gamma_T}_k \,(2 \slashed{p}_1 - 2 \slashed{k}_1 + \slashed{\ell}_1) \,(\slashed{k}_2 - \slashed{\ell}_2 + \slashed{\ell}_1) \right. \nn \\
    &\quad\, \times \left. \!\gamma_\nu \,(\slashed{k}_1 - \slashed{\ell}_1 + \slashed{\ell}_2) \,(2 \slashed{p}_2 - 2 \slashed{k}_2 + \slashed{\ell}_2) \vphantom{{\overline{\Gamma}_T}_j} \right) ,
	\label{e:Ra}
\end{align}
and the spin-averaged leptonic tensor $L^{\mu\nu}$ is given by
\begin{equation}
    L^{\mu\nu} \equiv \tr \left( \slashed{l} \gamma^\mu \slashed{l'} \gamma^\nu \right) = 4 \left( l^\mu l'^\nu + l^\nu l'^\mu - l \cd l' g^{\mu\nu} \right) .
\end{equation}
Up to leading power, its contraction with ${R_{\text{(a)}}}_{jk;\mu\nu}$ reads
\begin{equation}
    {R_{\text{(a)}}}_{jk;\mu\nu} L^{\mu\nu} = -128 \,H_{jk} \,x_1 (1-x_1) x_2 (1-x_2) (p_1^+ p_2^-)^2 ,
\end{equation}
where
\begin{equation}
    H_{jk} \equiv {l_\sst}_j {l'_\sst}_k + {l_\sst}_k {l'_\sst}_j + \bm{l} \cd \bm{l'} {g_\sst}_{jk} \,.
\end{equation}
Hence, to leading power, eq.~\eqref{e:(a,i)} becomes
\begin{align}
    \frac{1}{2} \left( \frac{d\sigma_{\scriptscriptstyle{\text{dBM}}}}{d\Omega \,dx_1 dx_2 \,d^2\bm{q}} \right)_{\text{(a,i)}} &= 2 \,\frac{\alpha^2}{q^4} \,e^2 \,C_{\text{(a)}} \,(1-x_1) (1-x_2) p_1^+ p_2^- \int d^2\bm{k}_1 \;\chi^j(x_1,\bm{k}_1) \nn \\
    &\quad\, \times \int d^2\bm{k}_2 \;\overline{\chi}^k(x_2,\bm{k}_2) \,I_{\text{(a,i)}} \,H_{jk} \;\delta^{(2)}(\bm{k}_1+\bm{k}_2-\bm{q}) ,
    \label{e:(a,i)_final}
\end{align}
where $I_{\text{(a,i)}}$ is an integral over $\ell_1^+$ and $\ell_2^-$, given by
\begin{equation}
    I_{\text{(a,i)}} \equiv \int \frac{d\ell_1^+}{2\pi} \,\frac{\nu^{\eta_1} |\ell_1^+|^{-\eta_1}}{\ell_1^+ + i\epsilon} \int \frac{d\ell_2^-}{2\pi} \,\frac{\nu^{\eta_2} |\ell_2^-|^{-\eta_2}}{\ell_2^- + i\epsilon} \,.
    \label{e:I_(a,i)}
\end{equation}
Note that the integrand has initial-state poles in both $\ell_1^+$ and $\ell_2^-$. \\

\noindent \textbf{Diagram (b,i).} The dBM contribution from diagram (b,i) is given by
\begin{align}
    \left( \frac{d\sigma_{\scriptscriptstyle{\text{dBM}}}}{d\Omega \,dx_1 dx_2 \,d^2\bm{q}} \right)_{\text{(b,i)}} &= - \frac{1}{8} \frac{\alpha^2}{q^4} \,e^2 \,C_{\text{(b)}} \int d^2\bm{k}_1 \int \frac{d\ell_1^+}{2\pi} \;\chi^j(x_1,\bm{k}_1) \int d^2\bm{k}_2 \int \frac{d\ell_2^-}{2\pi} \nn \\
    &\quad\, \times \overline{\chi}^k(x_2,\bm{k}_2) \,\frac{\nu^{\eta_1 + \eta_2} |\ell_1^+|^{-\eta_1} |\ell_2^-|^{-\eta_2} \,{R_{\text{(b)}}}_{jk;\mu\nu} L^{\mu\nu}}{[(k_2+\ell_1)^2 + i\epsilon] \,[(\ell_1+\ell_2)^2 + i\epsilon]} \nn \\
    &\quad\, \times \delta^{(2)}(\bm{k}_1+\bm{k}_2-\bm{q}) .
    \label{e:(b,i)}
\end{align}
The color factor $C_{\text{(b)}}$ reads
\begin{equation}
    C_{\text{(b)}} \equiv \tr \left( i f^{abc} t^a t^b t^c \right) = - \frac{N_c^2}{N_c^2-1} \frac{1}{N_c} \,C_\Phi^2 \,,
    \label{e:Cb}
\end{equation}
and the Dirac trace ${R_{\text{(b)}}}_{jk;\mu\nu}$ is defined as
\begin{align}
    {R_{\text{(b)}}}_{jk;\mu\nu} &\equiv \tr \left( {\overline{\Gamma}_T}_j \,\gamma_\mu \,{\Gamma_T}_k \left[ (2p_1 - 2k_1 + \ell_1) \cdot (2p_2 - 2k_2 + \ell_2) \,(\slashed{\ell}_1 - \slashed{\ell}_2) \vphantom{(2\slashed{p}_1 - 2\slashed{k}_1 + \slashed{\ell}_1)} \right.\right. \nn \\
    & \left.\left. \quad\, + \,(2p_1 - 2k_1 + \ell_1) \cdot (\ell_1 + 2\ell_2) \,(2\slashed{p}_2 - 2\slashed{k}_2 + \slashed{\ell}_2) \right.\right. \nn \\
    & \left.\left. \quad\, - \,(2p_2 - 2k_2 + \ell_2) \cdot (2\ell_1 + \ell_2) \,(2\slashed{p}_1 - 2\slashed{k}_1 + \slashed{\ell}_1) \right] (\slashed{k}_2 + \slashed{\ell}_1) \,\gamma_\nu \vphantom{\overline{\Gamma}_1} \vphantom{{\overline{\Gamma}_T}_j} \right) .
	\label{e:Rb}
\end{align}
Up to leading power, its contraction with the leptonic tensor is given by
\begin{equation}
    {R_{\text{(b)}}}_{jk;\mu\nu} L^{\mu\nu} = 64 \,H_{jk} \,(1-x_1) x_2 (1-x_2) p_1^+ (p_2^-)^2 \ell_1^+ .
\end{equation}
Hence, to leading power, eq.~\eqref{e:(b,i)} becomes
\begin{align}
    \left( \frac{d\sigma_{\scriptscriptstyle{\text{dBM}}}}{d\Omega \,dx_1 dx_2 \,d^2\bm{q}} \right)_{\text{(b,i)}} &= -2 \,\frac{\alpha^2}{q^4} \,e^2 \,C_{\text{(b)}} \,(1-x_1) (1-x_2) p_1^+ p_2^- \int d^2\bm{k}_1 \;\chi^j(x_1,\bm{k}_1) \nn \\
    &\quad\, \times \int d^2\bm{k}_2 \;\overline{\chi}^k(x_2,\bm{k}_2) \,I_{\text{(b,i)}} \,H_{jk} \;\delta^{(2)}(\bm{k}_1+\bm{k}_2-\bm{q}) ,
    \label{e:(b,i)_final}
\end{align}
where $I_{\text{(b,i)}}$ is an integral over $\ell_1^+$ and $\ell_2^-$, given by
\begin{equation}
    I_{\text{(b,i)}} \equiv \int \frac{d\ell_1^+}{2\pi} \,\frac{\nu^{\eta_1} |\ell_1^+|^{-\eta_1}}{\ell_1^+ + i\epsilon} \int \frac{d\ell_2^-}{2\pi} \,\frac{2 \ell_1^+ \nu^{\eta_2} |\ell_2^-|^{-\eta_2}}{2 \ell_1^+ \ell_2^- - (\bm{\ell}_1 + \bm{\ell}_2)^2 + i\epsilon} \,.
    \label{e:I_(b,i)}
\end{equation}
Note that the integrand has an initial-state pole in $\ell_1^+$, and, depending on the sign of $\ell_1^+$, the pole in $\ell_2^-$ is either an initial- or a final-state one. \\

\noindent \textbf{Diagram (b,v).} The dBM contribution from diagram (b,v) is given by
\begin{align}
    \left( \frac{d\sigma_{\scriptscriptstyle{\text{dBM}}}}{d\Omega \,dx_1 dx_2 \,d^2\bm{q}} \right)_{\text{(b,v)}} &= - \frac{\pi^2}{8} \frac{\alpha^2}{q^4} \,e^2 g^4 \,C_{\text{(b)}} \int \frac{dk_1^- d^2\bm{k}_1}{(2\pi)^4} \int \frac{d^4\ell_1}{(2\pi)^4} \nn \\
    &\quad\, \times \theta\big((p_1-k_1+\ell_1)^0\big) \,\delta\big((p_1-k_1+\ell_1)^2 - m_s^2\big) \nn \\
    &\quad\, \times \frac{D_1^j}{[(k_1-\ell_1)^2 + i\epsilon] \,[(p_1-k_1)^2 - m_s^2 - i\epsilon]  \,[\ell_1^2 - i\epsilon] \,[k_1^2 - i\epsilon]} \nn \\
    &\quad\, \times \int \frac{dk_2^+ d^2\bm{k}_2}{(2\pi)^4} \int \frac{d^4\ell_2}{(2\pi)^4} \;\theta\big((p_2-k_2+\ell_2)^0\big) \nn \\
    &\quad\, \times \delta\big((p_2-k_2+\ell_2)^2 - m_s^2\big) \nn \\
    &\quad\, \times \frac{D_2^k}{[(k_2-\ell_2)^2 + i\epsilon] \,[(p_2-k_2)^2 - m_s^2 - i\epsilon] \,[\ell_2^2 - i\epsilon] \,[k_2^2 - i\epsilon]} \nn \\
    &\quad\, \times \frac{\nu^{\eta_1} |\ell_1^+|^{-\eta_1} \,{R_{\text{(b)}}}_{jk;\mu\nu} L^{\mu\nu}}{(k_2+\ell_1)^2 + i\epsilon} \cdot 2\pi i \;\theta\big(-(\ell_1+\ell_2)^0\big) \nn \\
    &\quad\, \times \delta\big((\ell_1+\ell_2)^2\big) \,\nu^{\eta_2} |\ell_2^-|^{-\eta_2} \;\delta^{(2)}(\bm{k}_1+\bm{k}_2-\bm{q}) .
    \label{e:(b,v)}
\end{align}
We now expand this expression to leading power in $\lambda$ and perform the integrals over $k_1^-$, $\ell_1^-$, $k_2^+$, and $\ell_2^+$. The two delta functions $\delta\big((p_1-k_1+\ell_1)^2 - m_s^2\big)$ and $\delta\big((p_2-k_2+\ell_2)^2 - m_s^2\big)$ are used for the integrations over $\ell_1^-$ and $\ell_2^+$, and for the integrations over $k_1^-$ and $k_2^+$ we invoke Cauchy's residue theorem. To leading power, eq.~\eqref{e:(b,v)} becomes
\begin{align}
    \left( \frac{d\sigma_{\scriptscriptstyle{\text{dBM}}}}{d\Omega \,dx_1 dx_2 \,d^2\bm{q}} \right)_{\text{(b,v)}} &= -2 \,\frac{\alpha^2}{q^4} \,e^2 \,C_{\text{(b)}} \,(1-x_1) (1-x_2) p_1^+ p_2^- \int d^2\bm{k}_1 \;\chi^j(x_1,\bm{k}_1) \nn \\
    &\quad\, \times \int d^2\bm{k}_2 \;\overline{\chi}^k(x_2,\bm{k}_2) \,I_{\text{(b,v)}} \,H_{jk} \;\delta^{(2)}(\bm{k}_1+\bm{k}_2-\bm{q}) ,
    \label{e:(b,v)_final}
\end{align}
where $I_{\text{(b,v)}}$ is an integral over $\ell_1^+$ and $\ell_2^-$, given by
\begin{align}
    I_{\text{(b,v)}} &\equiv 4\pi i \int \frac{d\ell_1^+}{2\pi} \,\frac{\nu^{\eta_1} |\ell_1^+|^{-\eta_1}}{\ell_1^+ + i\epsilon} \int \frac{d\ell_2^-}{2\pi} \;\theta(-\ell_1^+) \,\ell_1^+ \nn \\
    &\quad\, \times \delta\big(2 \ell_1^+ \ell_2^- - \left( \bm{\ell}_1 + \bm{\ell}_2 \right)^2\big) \,\nu^{\eta_2} |\ell_2^-|^{-\eta_2} .
    \label{e:I_(b,v)}
\end{align}
Note that the integrand has an initial-state pole in $\ell_1^+$, and both an initial- and a final-state pole in $\ell_2^-$. The latter can be seen from the identity
\begin{equation}
    2\pi i \,\delta(x) = \frac{1}{x-i\epsilon} - \frac{1}{x+i\epsilon} \,,
\end{equation}
which is consistent with eq.~\eqref{e:int_rep}. \\

\noindent \textbf{Sum of the diagrams.} We can use eq.~\eqref{e:sum_diagrams} to combine the results from diagrams (a,i), (b,i), and (b,v) to obtain the full $G_1G_2$ region contribution to the DY cross section. Employing eqs.~\eqref{e:(a,i)_final}, \eqref{e:(b,i)_final}, \eqref{e:(b,v)_final}, and that $C_{\text{(b)}} = N_c^2 \,C_{\text{(a)}}$, gives to leading power
\begin{align}
    \frac{d\sigma_{\scriptscriptstyle{\text{dBM}}}}{d\Omega \,dx_1 dx_2 \,d^2\bm{q}} &= 4 \,\frac{\alpha^2}{q^4} \,e^2 \,C_{\text{(a)}} \,(1-x_1) (1-x_2) p_1^+ p_2^- \int d^2\bm{k}_1 \;\chi^j(x_1,\bm{k}_1) \nn \\
    &\quad\, \times \int d^2\bm{k}_2 \;\overline{\chi}^k(x_2,\bm{k}_2) \left[ I_{\text{(a,i)}} - N_c^2 \left( I_{\text{(b,i)}} + I_{\text{(b,v)}} \right) \right] H_{jk}  \nn \\
    &\quad\, \times \delta^{(2)}(\bm{k}_1+\bm{k}_2-\bm{q}) + \ldots + \text{h.c.} \,,
    \label{e:sum}
\end{align}
where the ellipsis refers to the contribution from diagram (f,i) which we have not considered explicitly.

Let us now have a closer look at the integrals $I_{\text{(b,i)}}$ and $I_{\text{(b,v)}}$. Performing the integrations over $\ell_2^-$ and expanding in the regulator $\eta_2$ (following our regulator prescription in~\eqref{e:regulator_(b)}), we obtain
\begin{align}
    I_{\text{(b,i)}} &= \frac{i}{2} \int \frac{d\ell_1^+}{2\pi} \left[ \theta(-\ell_1^+) - \theta(\ell_1^+) \right] \frac{\nu^{\eta_1} |\ell_1^+|^{-\eta_1}}{\ell_1^+ + i\epsilon} + \mathcal{O}(\eta_2), \\
    I_{\text{(b,v)}} &= -i \int \frac{d\ell_1^+}{2\pi} \;\theta(-\ell_1^+) \,\frac{\nu^{\eta_1} |\ell_1^+|^{-\eta_1}}{\ell_1^+ + i\epsilon} + \mathcal{O}(\eta_2) .
\end{align}
Both $I_{\text{(b,i)}}$ and $I_{\text{(b,v)}}$ have poles in $\eta_1$, which disappear once they are summed together:
\begin{equation}
    I_{\text{(b,i)}} + I_{\text{(b,v)}} = -\frac{i}{2} \int \frac{d\ell_1^+}{2\pi} \,\frac{\nu^{\eta_1} |\ell_1^+|^{-\eta_1}}{\ell_1^+ + i\epsilon} + \mathcal{O}(\eta_2) = I_{\text{(a,i)}} \,.
    \label{e:int_id}
\end{equation}
As expected, summing over all allowed cuts of graph (b), i.e.\ adding up diagrams (b,i) and (b,v), cancels out the final-state poles present in $I_{\text{(b,i)}}$ and $I_{\text{(b,v)}}$. We are now left with $I_{\text{(a,i)}}$ that has initial-state poles only, consistent with the formation of initial-state Wilson lines. This cancellation of final-state poles is a consequence of the unitarity property of our model. From the integral identification~\eqref{e:int_id}, it follows that
\begin{equation}
    C_{\text{(a)}} \left[ I_{\text{(a,i)}} - N_c^2 \left( I_{\text{(b,i)}} + I_{\text{(b,v)}} \right) \right] = \frac{1}{N_c} \,C_\Phi^2 \,I_{\text{(a,i)}} \,.
    \label{e:disentanglement_G1G2}
\end{equation}
Hence, we see that in the $G_1G_2$ region the different color structures completely disentangle, yielding the expected $1/N_c$ color factor.

\subsection{Contributions from other regions} \label{s:other_region_contributions}
According to the Collins subtraction procedure, the full leading-power approximation of the dBM term is given by the sum of the various region contributions given in eqs.~\eqref{e:C_G1G2}--\eqref{e:C_C1C2}. In the previous subsection we calculated the first contribution $C_{G_1G_2} \Gamma$, and, as it turns out, this will be the only nonvanishing contribution. In the following we will argue that (for our choice of rapidity regulators) the dBM term does not receive any contributions from the $C_1G$, $GC_2$, and $C_1C_2$ regions.

\begin{figure}
\centering
\begin{subfigure}{.48\textwidth}
    \begin{flushleft}
        \includegraphics[height=4.1cm]{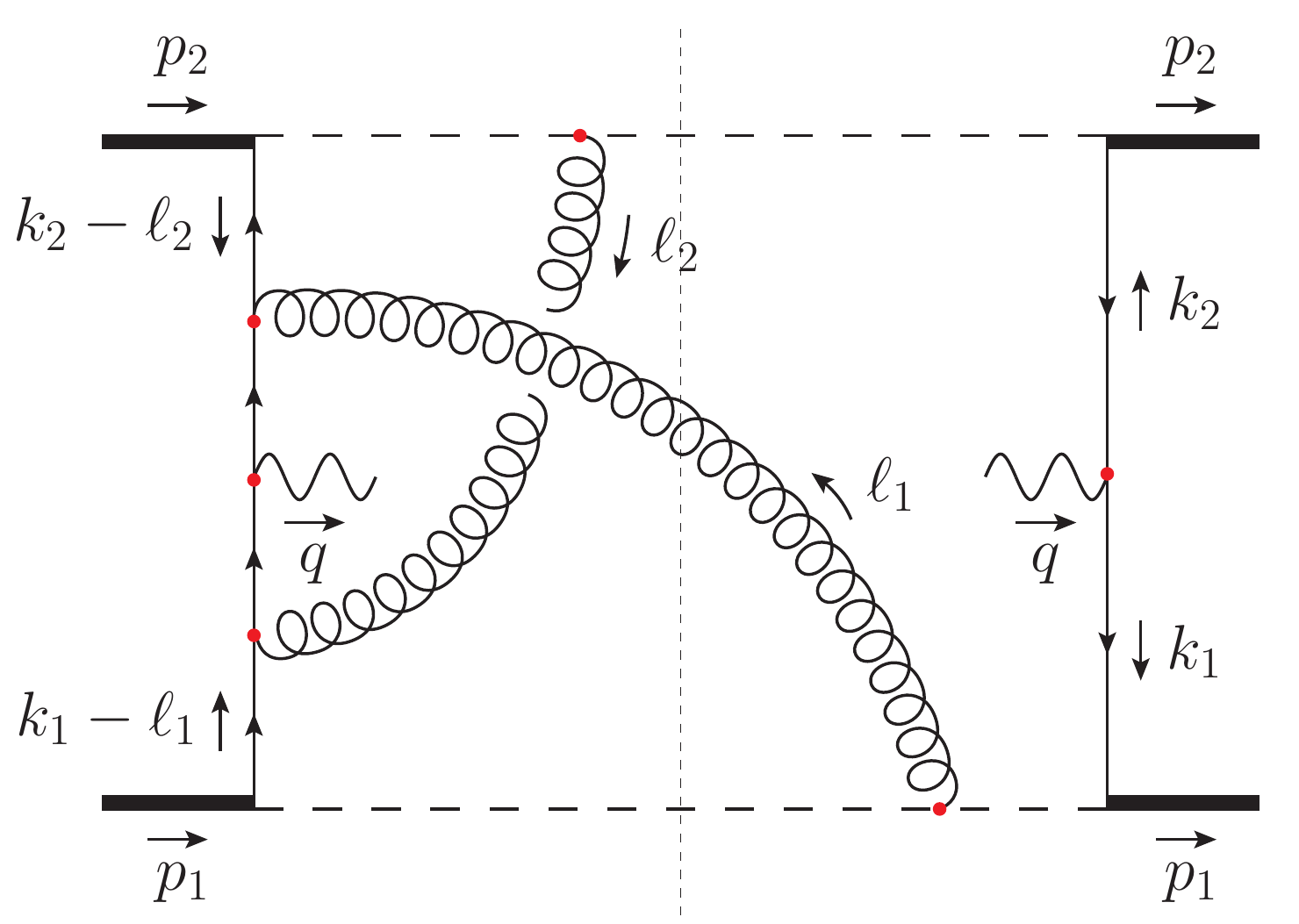}
        \caption*{\hspace{-0.19cm}(a,ii)}
    \end{flushleft}
\end{subfigure}
\begin{subfigure}{.48\textwidth}
    \begin{flushright}
        \includegraphics[height=4.1cm]{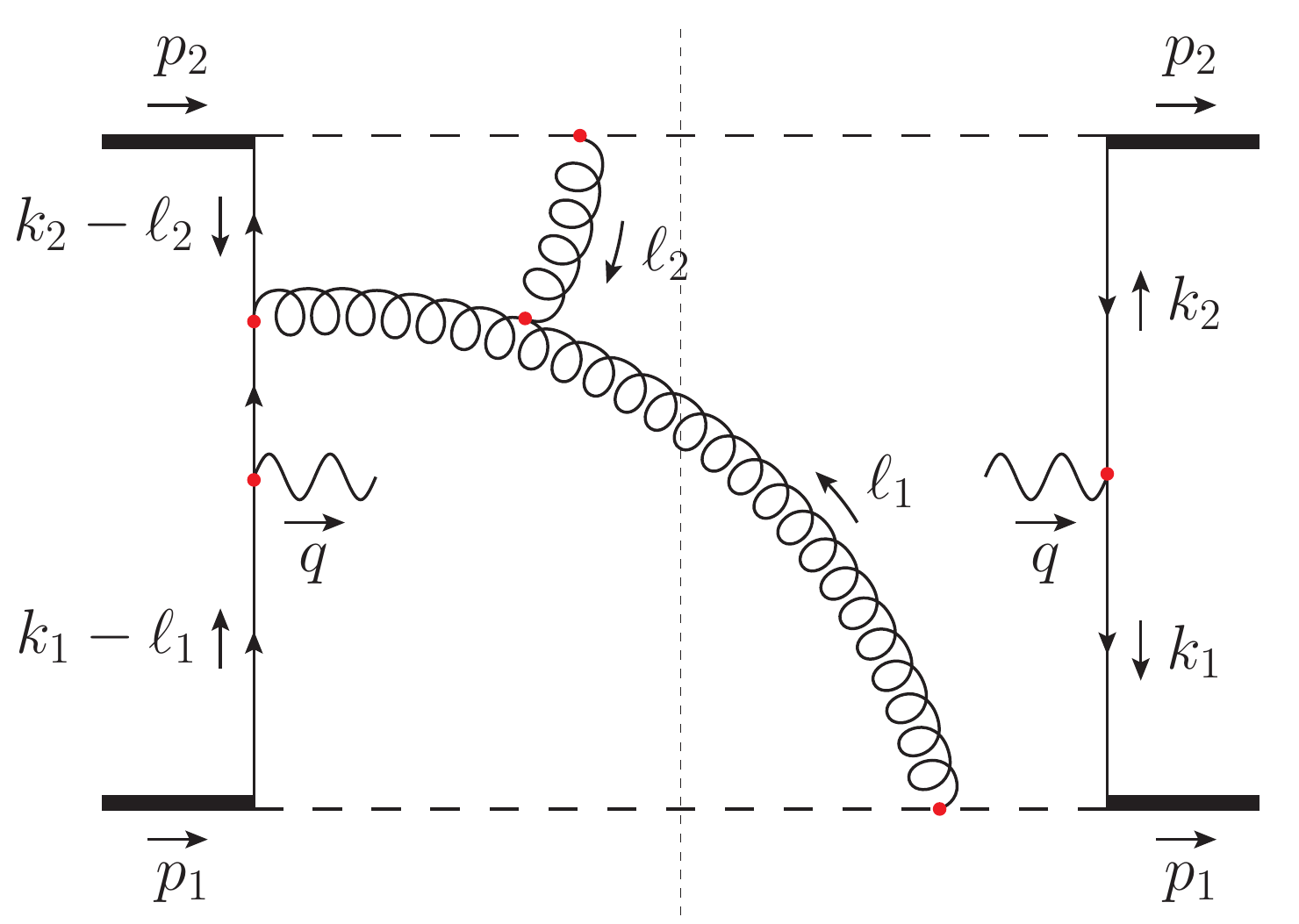}
        \caption*{\hspace{0.77cm}(b,ii)}
    \end{flushright}
\end{subfigure}
\begin{subfigure}{.48\textwidth}
    \begin{flushleft}
        \includegraphics[height=4.1cm]{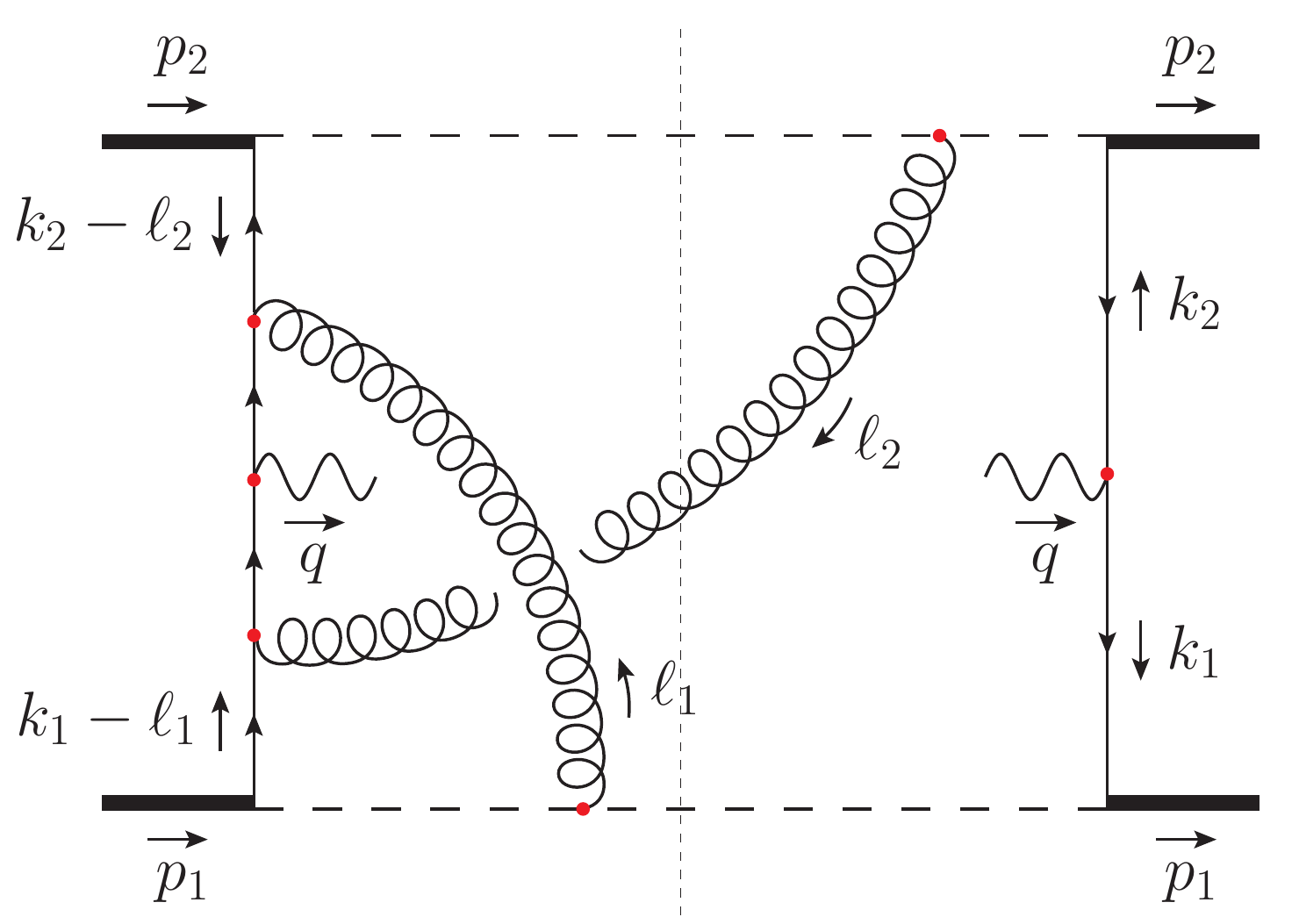}
        \caption*{\hspace{-0.22cm}(a,iii)}
    \end{flushleft}
\end{subfigure}
\begin{subfigure}{.48\textwidth}
    \begin{flushright}
        \includegraphics[height=4.1cm]{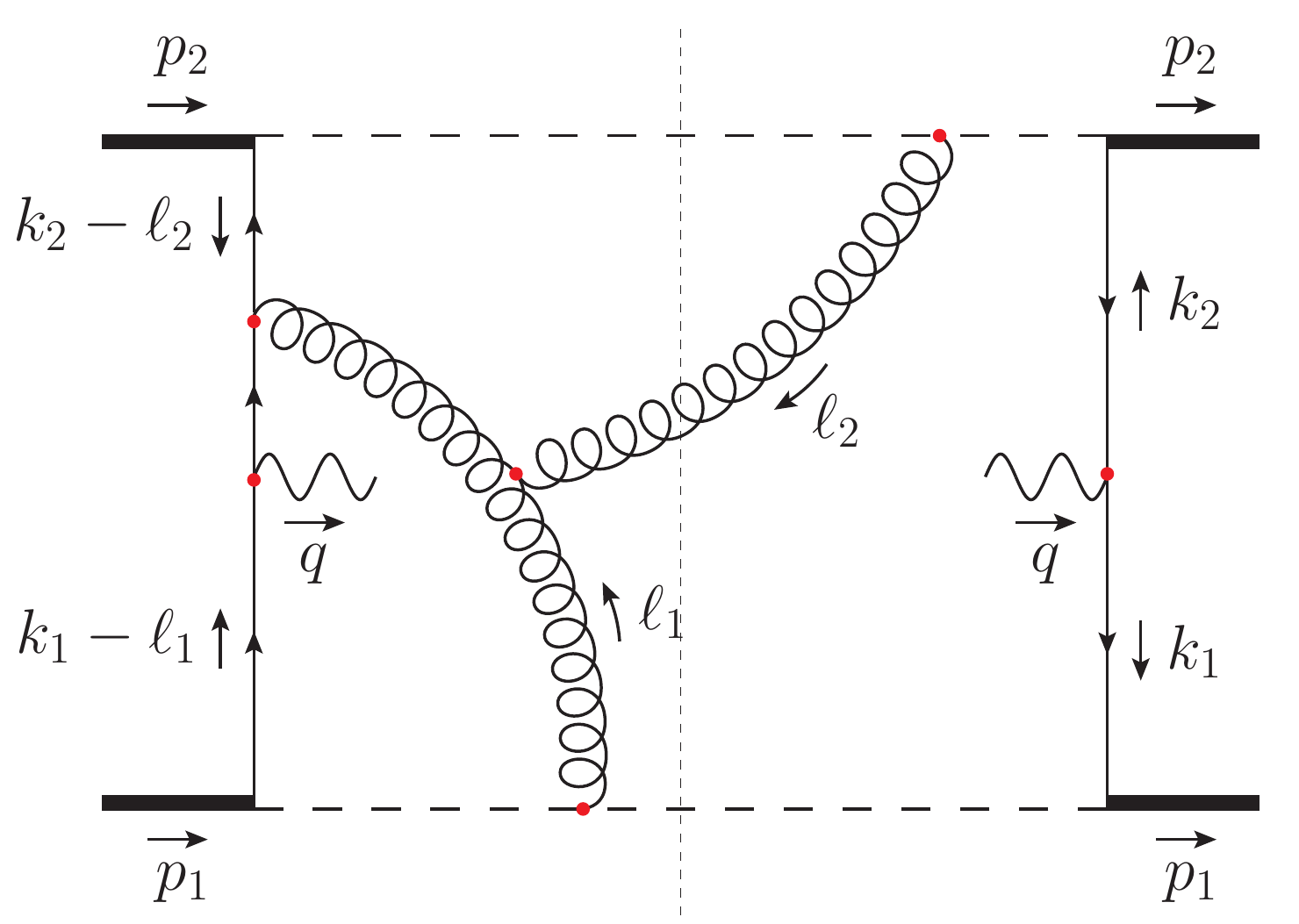}
        \caption*{\hspace{0.74cm}(b,iii)}
    \end{flushright}
\end{subfigure}
\begin{subfigure}{.48\textwidth}
    \begin{flushleft}
        \includegraphics[height=4.1cm]{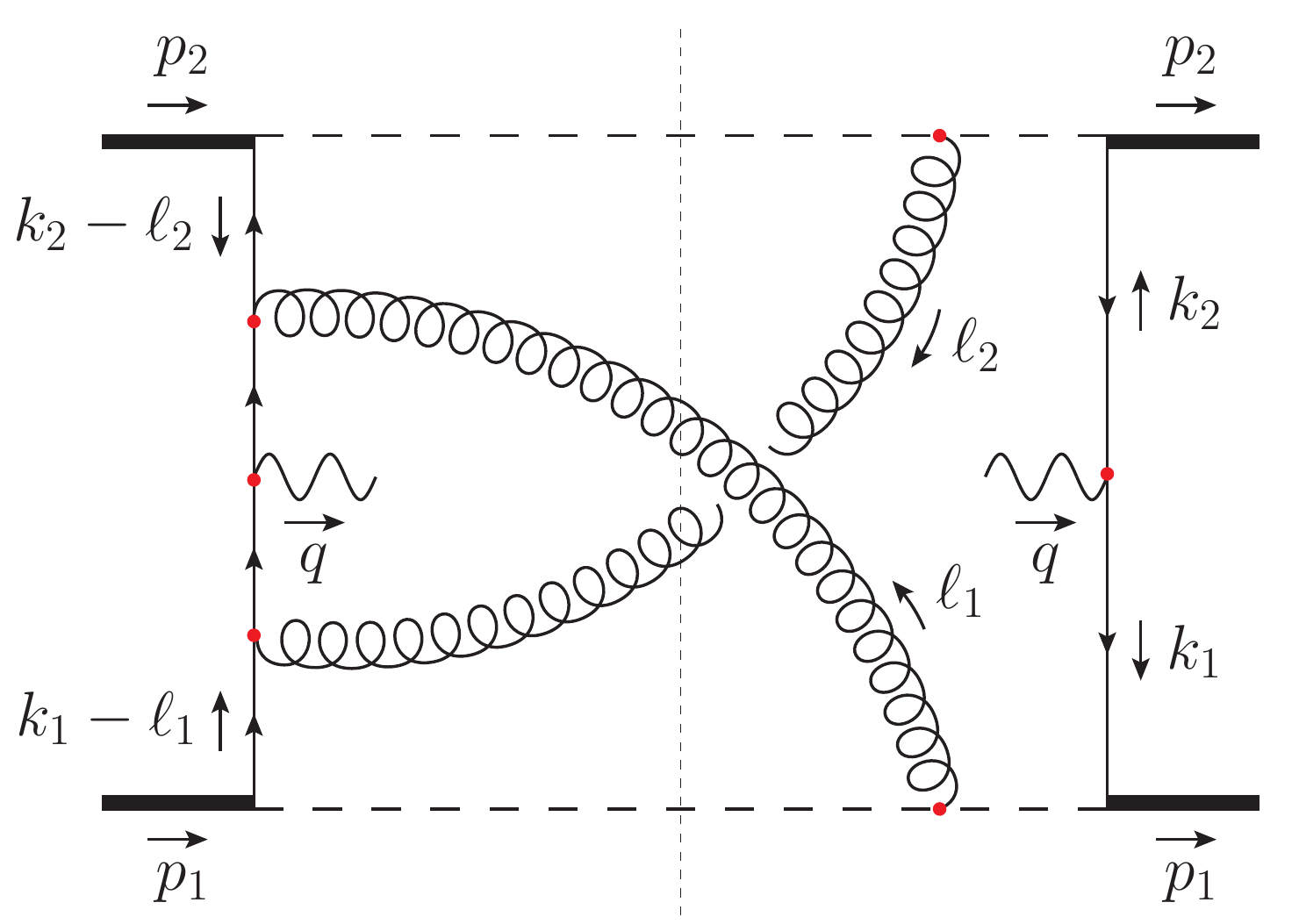}
        \caption*{\hspace{-0.18cm}(a,iv)}
    \end{flushleft}
\end{subfigure}
\begin{subfigure}{.48\textwidth}
    \begin{flushright}
        \includegraphics[height=4.1cm]{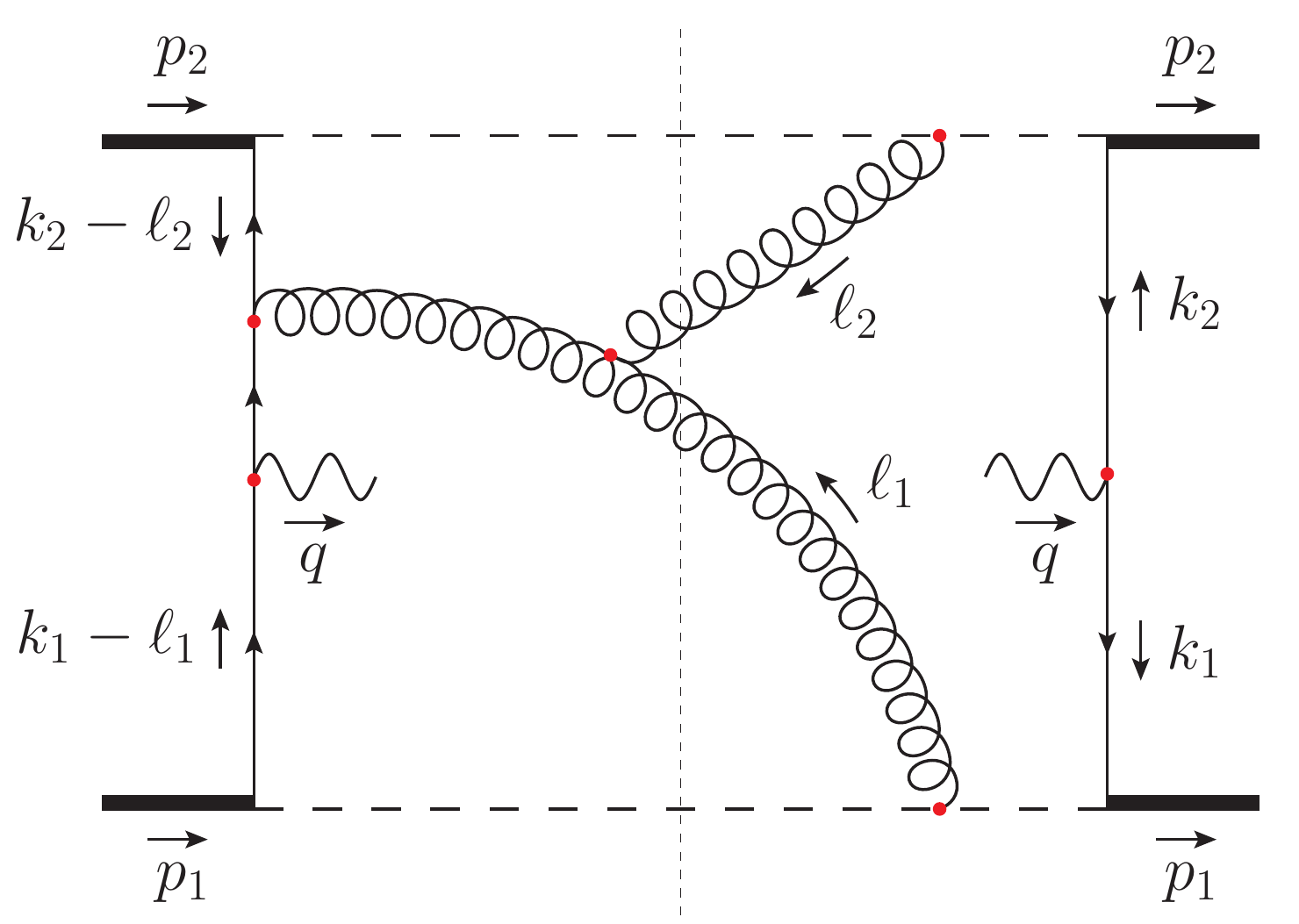}
        \caption*{\hspace{0.78cm}(b,iv)}
    \end{flushright}
\end{subfigure}
\caption{The color-entangled two-gluon exchange graphs that contribute to the dBM part of the DY cross section in the $C_1C_2$ region. This set is supplemented by diagrams (a,i) and (b,i) given in figure~\ref{f:diagrams}, and by graphs that can be obtained by $p\leftrightarrow\bar{p}$ or Hermitian conjugation.}
\label{f:entangled_diagrams_C1C2}
\end{figure}

When the gluons have collinear scalings, more cuts are possible compared to the $G_1G_2$ case. For the $C_1C_2$ region, there are four possible cut positions for both diagrams (a) and (b). Besides (a,i) and (b,i), this region requires consideration of the six cut graphs given in figure~\ref{f:entangled_diagrams_C1C2}, and, as before, we also need to take into account their $p\leftrightarrow\bar{p}$ versions and Hermitian conjugates. Note that graph (b,v) is not allowed this time (the hard gluon line with momentum $\ell_1 + \ell_2$ is off shell).

The cross section contribution from the $C_1C_2$ region is given by eq.~\eqref{e:C_C1C2}. In this region, we can consider the naive graph terms and subtractions separately for fixed nonzero values of $\ell_1^+$ and $\ell_2^-$, as the subtractions remove the regions where $\ell_1^+$ or $\ell_2^-$ are zero. This implies that we can ignore the $i\epsilon$ terms in the eikonal propagators (note that this is analogous to our treatment of the $C_1$ region contribution in section~\ref{s:BM_function}).

Let us consider the naive graph term $T_{C_1C_2} \Gamma$ explicitly here. The starting expressions for diagrams (a,i) and (b,i) are given by eqs.~\eqref{e:(a,i)} and~\eqref{e:(b,i)}. At leading power, the Dirac traces ${R_{\text{(a)}}}_{jk;\mu\nu}$ and ${R_{\text{(b)}}}_{jk;\mu\nu}$, contracted with the leptonic tensor $L^{\mu\nu}$, are now given by
\begin{align}
    {R_{\text{(a)}}}_{jk;\mu\nu} L^{\mu\nu} &= - \frac{2(\ell_1^+ - x_1 p_1^+)(\ell_2^- - x_2 p_2^-)}{\ell_1^+ x_2 p_2^-} \,{R_{\text{(b)}}}_{jk;\mu\nu} L^{\mu\nu} , \\[5pt]
    {R_{\text{(b)}}}_{jk;\mu\nu} L^{\mu\nu} &= 16 \,H_{jk} \,\ell_1^+ x_2 p_2^- \left[ \ell_1^+ + 2(1-x_1)p_1^+ \right] \!\left[ \ell_2^- + 2(1-x_2)p_2^- \right] ,
\end{align}
so that eqs.~\eqref{e:(a,i)} and~\eqref{e:(b,i)} become
\begin{align}
    \frac{1}{2} \left( \frac{d\sigma_{\scriptscriptstyle{\text{dBM}}}}{d\Omega \,dx_1 dx_2 \,d^2\bm{q}} \right)_{\text{(a,i)}} &= \frac{1}{2} \frac{\alpha^2}{q^4} \,e^2 \,C_{\text{(a)}} \int d^2\bm{k}_1 \int \frac{d\ell_1^+}{2\pi} \left[ 1 + \frac{2(1-x_1)p_1^+}{\ell_1^+} \right] \nn \\[3pt]
    &\quad\, \times \chi^j(x_1,\bm{k}_1,\ell_1^+) \int d^2\bm{k}_2 \int \frac{d\ell_2^-}{2\pi} \left[ 1 + \frac{2(1-x_2)p_2^-}{\ell_2^-} \right] \nn \\[3pt]
    &\quad\, \times \overline{\chi}^k(x_2,\bm{k}_2,\ell_2^-) \,H_{jk} \;\delta^{(2)}(\bm{k}_1+\bm{k}_2-\bm{q}) , \\[5pt]
    \left( \frac{d\sigma_{\scriptscriptstyle{\text{dBM}}}}{d\Omega \,dx_1 dx_2 \,d^2\bm{q}} \right)_{\text{(b,i)}} &= -\frac{1}{2} \frac{\alpha^2}{q^4} \,e^2 \,C_{\text{(b)}} \int d^2\bm{k}_1 \int \frac{d\ell_1^+}{2\pi} \left[ 1 + \frac{2(1-x_1)p_1^+}{\ell_1^+} \right] \nn \\[3pt]
    &\quad\, \times \chi^j(x_1,\bm{k}_1,\ell_1^+) \int d^2\bm{k}_2 \int \frac{d\ell_2^-}{2\pi} \left[ 1 + \frac{2(1-x_2)p_2^-}{\ell_2^-} \right] \nn \\[3pt]
    &\quad\, \times \overline{\chi}^k(x_2,\bm{k}_2,\ell_2^-) \,H_{jk} \;\delta^{(2)}(\bm{k}_1+\bm{k}_2-\bm{q}) .
\end{align}
Summing these two contributions and using that $C_{\text{(b)}} = N_c^2 \,C_{\text{(a)}}$, gives
\begin{align}
    &\,\frac{1}{2} \left( \frac{d\sigma_{\scriptscriptstyle{\text{dBM}}}}{d\Omega \,dx_1 dx_2 \,d^2\bm{q}} \right)_{\text{(a,i)}} + \left( \frac{d\sigma_{\scriptscriptstyle{\text{dBM}}}}{d\Omega \,dx_1 dx_2 \,d^2\bm{q}} \right)_{\text{(b,i)}} \nn \\[3pt]
    =& \,\frac{1}{2} \frac{\alpha^2}{q^4} \,e^2 \,\big( 1-N_c^2 \big) \,C_{\text{(a)}} \int d^2\bm{k}_1 \int \frac{d\ell_1^+}{2\pi} \left[ 1 + \frac{2(1-x_1)p_1^+}{\ell_1^+} \right] \chi^j(x_1,\bm{k}_1,\ell_1^+) \nn \\[3pt]
    & \times \int d^2\bm{k}_2 \int \frac{d\ell_2^-}{2\pi} \left[ 1 + \frac{2(1-x_2)p_2^-}{\ell_2^-} \right] \overline{\chi}^k(x_2,\bm{k}_2,\ell_2^-) \,H_{jk} \;\delta^{(2)}(\bm{k}_1+\bm{k}_2-\bm{q}) .
    \label{e:sum_for_C1C2}
\end{align}
Note that since
\begin{equation}
    \big( 1-N_c^2 \big) \,C_{\text{(a)}} = \frac{1}{N_c} \,C_\Phi^2 \,,
\end{equation}
we again obtain the standard $1/N_c$ color factor. This disentanglement of color is consistent with the expectations based on the non-Abelian Ward identity~\cite{Collins:2011zzd}.

In eq.~\eqref{e:sum_for_C1C2} we recognize the first term on the right-hand side of eq.~\eqref{e:phi_final_C_1}. Recall from section~\ref{s:BM_function} that the $C_1$ contribution to the quark BM function vanishes due to the cancellation between the amplitude and its conjugate (and likewise for the antiquark BM function). Hence, it follows that after properly taking into account the contributions from diagram (f,i) and all Hermitian conjugates, also eq.~\eqref{e:sum_for_C1C2} vanishes. Naturally, we arrive at the same conclusion when considering the $p\leftrightarrow\bar{p}$ version of diagram (b,i).

In the preceding discussion on diagrams (a,i) and (b,i), the precise position of the cut was actually unimportant as the Dirac traces and color factors are independent of the cut. This means that the color disentanglement occurs in exactly the same way for the combination of diagrams (a,ii) and (b,ii), and for (a,iii) and (b,iii), etc.\ -- in other words, the disentanglement proceeds in a \emph{cut-by-cut} fashion. Again, these contributions are not only disentangled but in fact completely cancel out.

The subtraction terms can be dealt with in exactly the same way. Also here we find that the different color structures disentangle and ultimately vanish. Hence, the complete region contribution $C_{C_1C_2} \Gamma$ to the dBM term is zero.

For a discussion on the $C_1G$ and $GC_2$ regions we refer to~\cite{Boer:2017hqr}. As it turns out, also for these two regions the color completely disentangles for both the naive graph and subtraction terms separately. Once the Hermitian conjugate diagrams are included, these region contributions precisely cancel out as well. Hence, the dBM term can be completely ascribed to the $G_1G_2$ region.

\subsection{The final result} \label{s:final_result}
We have seen that with the regulators in~\eqref{e:regulator_(a)}--\eqref{e:regulator_(b)pp}, the cancellation of the color entanglement occurs on a region-by-region basis, and the full $\mathcal{O}(\alpha_s^2)$ contribution to the dBM term ends up in the $G_1G_2$ region. With different regulators one can shift contributions between regions; the final result, however, obtained from the sum over all graphs and leading regions, remains the same. We refer to appendix~B of~\cite{Boer:2017hqr} for examples of regulator choices for which the $G_1G_2$ region does not give the full contribution.

We can now compare the result of our model calculation to the leading-order factorization formula~\eqref{e:fac}. The sum of the diagrams in eq.~\eqref{e:sum}, together with eq.~\eqref{e:disentanglement_G1G2}, gives an expression consistent with taking the dBM part of eq.~\eqref{e:fac} and inserting the model expression for the quark BM function in eq.~\eqref{e:phi_final}, as well as the analogous expression for the antiquark BM function. Diagram (f,i) and its conjugate precisely account for the `cross terms', i.e.\ the combinations of the first/second term in eq.~\eqref{e:phi_final} for the quark with the second/first term of the analogous expression for the antiquark. Upon summing all contributions, our factorized result is given by:\footnote{Since $\bar{h}_{1,q}^\perp$ pertains to a left-moving hadron rather than a right-moving one, it comes with a minus sign relative to $h_{1,q}^\perp$ from swapping plus and minus indices in the antisymmetric Levi-Civita symbol (see e.g.~\cite{Kasemets:2012pr}).}
\begin{align}
    \frac{d\sigma_{\scriptscriptstyle{\text{dBM}}}}{d\Omega \,dx_1 dx_2 \,d^2\bm{q}} &= \frac{\alpha^2}{N_c \,q^4} \sum_q e_q^2 \int d^2\bm{k}_1 \,\frac{\widetilde{k}_{1\sst}^j}{M} \,h_{1,q}^\perp(x_1,\bm{k}_1^2) \int d^2\bm{k}_2 \,\frac{\widetilde{k}_{2\sst}^k}{M} \,\bar{h}_{1,q}^\perp(x_2,\bm{k}_2^2) \nn \\
    &\quad\, \times H_{jk} \;\delta^{(2)}(\bm{k}_1+\bm{k}_2-\bm{q}) \nn \\
    &= \frac{\alpha^2}{N_c \,q^2} \sum_q e_q^2 \,B(\theta) \cos(2\phi) \,\mathcal{F}\left[ w(\bm{k}_1,\bm{k}_2) \,h_1^\perp \bar{h}_1^\perp \right] .
    \label{e:sum_final}
\end{align}
Although we have worked in the CM frame of the $p\bar{p}$ collision, we have expressed our final result in terms of the CS angles~\cite{Arnold:2008kf}.

We conclude that the dBM term in the DY cross section precisely factorizes as already anticipated by the CSS works; no loophole in their original proof for this double T-odd contribution is found. In contrast to the findings in~\cite{Buffing:2013dxa}, we do \emph{not} find an additional color factor on top of the standard $1/N_c$ one. We have demonstrated that the entangled color structures are completely disentangled after summing over all relevant diagrams and cuts. The graph containing the three-gluon vertex, diagram (b) in figure~\ref{f:entangled_diagrams}, plays a vital role in this cancellation. This diagram was not taken into account in~\cite{Buffing:2013dxa}.

In our calculation it is possible to identify two mechanisms that underlie the cancellation of the color entanglement. The first of these is the unitarity cancellation of final-state poles after the sum over cuts of a particular diagram. This mechanism gives a finite result with initial-state poles in $\ell_1^+$ and $\ell_2^-$ for diagram (b) after the sum over cuts. The second is the non-Abelian Ward identity. For each region, this identity ensures that when the diagrams are combined we end up with the standard $1/N_c$ color factor. We know from chapter~\ref{c:CSS} that these two principles are also essential in the all-order CSS factorization proof.

As mentioned already in subsection~\ref{s:leading_regions_model}, the fact that our calculation agrees with the factorization formula that contains only TMDs (and a hard function) implies that the Glauber contributions may be absorbed into the TMDs. After the sum over cuts of diagrams (a), (b), and (f), the integrations over the gluon momentum components $\ell_1^+$ and $\ell_2^-$ are no longer pinched in the Glauber region. At the same time, the integrations over $\ell_1^-$ and $\ell_2^+$ are always pinched, even after the sum over cuts. In the case of diagram (b), the numerator structure appears to be responsible for this. Hence, analogous to our example of the active-spectator graph in subsection~\ref{s:trapped_Glauber}, we may deform $\ell_1$ into the $C_1$ region while $\ell_2$ may be deformed into the $C_2$ region. The fact that the Glauber contributions can be absorbed into other region contributions is consistent with the factorization proof. However, the CSS works on the deformation of soft momenta out of the Glauber region are not prescriptive about which momenta can be deformed into the collinear, and which into the central soft regions. With our calculation we have shown that certain Glauber momenta cannot be deformed into the central soft region, but instead must be deformed into a collinear region.

\section{Generalizations} \label{s:generalizations}

In this final section of part~\ref{p:1} we comment on the scope of our analysis. Although, for definiteness, we have only considered collisions between a proton and an antiproton, our calculation of the dBM effect would proceed in exactly the same way for other hadron-hadron collisions (in an appropriately modified spectator model). In particular, the obtained disentanglement result would remain unchanged.

So far we have only discussed the dBM contribution to DY, even though according to~\cite{Buffing:2013dxa} color entanglement is supposed to affect \emph{all} double T-odd contributions. For collisions between spin-$\tfrac{1}{2}$ hadrons there is one other double T-odd contribution, namely the one that involves the quark and antiquark Sivers functions. The Sivers function $f_{1\sst}^\perp$ quantifies the correlation between the transverse momentum of the parton and the transverse spin of the parent hadron, see eq.~\eqref{e:quark_TMD_parametrization_vpolarized}. Hence, to probe the double Sivers effect, the hadron beams need to be transversely polarized~\cite{Boer:1999mm,Boer:2011vq}. It is straightforward to extend the analysis described in this chapter to include this particular contribution as well. Similarly to the dBM part of the cross section, our spectator model also gives a nonzero result for the double Sivers part at $\mathcal{O}(\alpha_s^2)$, coming from the same diagrams as discussed in subsection~\ref{s:relevant_graphs}. Moreover, the BM and Sivers functions have been shown to be identical in this model~\cite{Goldstein:2002vv,Boer:2002ju,Gamberg:2003ey,Bacchetta:2003rz,Bacchetta:2008af}. Via the same steps and mechanisms as for the dBM case, we find here no color entanglement, consistent with the CSS factorization works. Furthermore, by again choosing the regulators given in~\eqref{e:regulator_(a)}--\eqref{e:regulator_(b)pp}, also the double Sivers term completely originates from the $G_1G_2$ momentum region.

As explained in subsection~\ref{s:final_result}, our final results at $\mathcal{O}(\alpha_s^2)$ rely on the unitarity cancellation of final-state poles and the non-Abelian Ward identity. Although we cannot say anything about the possible appearance of color-entangled structures beyond the lowest order considered, we have no reason to doubt that these two mechanisms also drive the cancellation of the color entanglement at higher orders, in line with the arguments of CSS.

Finally, we do not expect the cancellation of the color entanglement to be restricted to the model studied, but rather to hold generally for DY in QCD. After all, the color factors involved in our calculation all appear exactly as in the analogous full QCD calculation. Furthermore, the key mechanisms that lead to our result are rather general field-theoretical principles and also play an essential role in the CSS proof.

\stopthumb
\part{Distributions of gluons inside hadrons} \label{p:2}
\chapter{Gluon TMDs for hadrons of spin $\leq 1$} \label{c:Polarized_gluon_TMDs}
\addthumb{\thechapter}{\Large{\thechapter}}{\thumbfg}{\thumbbg}

This part of the thesis is about multidimensional gluon distributions. More knowledge on gluon distributions could yield new insights into the internal dynamics of hadrons. Fortunately, most of the formalism for quarks as introduced in chapter~\ref{c:PDFs} can be straightforwardly extended to the gluon case. Whereas in chapter~\ref{c:Small_x} we will study gluon distributions at small $x$, here we will study gluon TMDs for general $x$. 

For spin-$\tfrac{1}{2}$ hadrons, the first parametrization of the gluon-gluon TMD correlator in terms of gluon TMDs was presented by Mulders and Rodrigues in 2000 in~\cite{Mulders:2000sh}. Here we extend their work to spin-$1$ hadrons about which little is known. Similarly to the parametrization of the quark-quark TMD correlator in subsection~\ref{s:parametrization_TMD_correlator}, we will parametrize the gluon-gluon TMD correlator for spin-$1$ hadrons in terms of leading-twist gluon TMDs of definite rank; this is based on our work in~\cite{Boer:2016xqr}. Although the TMDs that appear for tensor polarized hadrons are new, their collinear counterparts were already introduced in 1989 in~\cite{Jaffe:1989xy,Artru:1989zv}. Subsequently, in section~\ref{s:positivity_bounds}, we will derive positivity bounds for gluon TMDs for spin-$0$, spin-$\tfrac{1}{2}$, and spin-$1$ hadrons, reflecting our work in~\cite{Cotogno:2017puy}. These bounds are model-independent inequalities that help relating and estimating the magnitude of the gluon TMDs about which very little, or almost nothing, is currently known. In section~\ref{s:process_dependence_gluons} we discuss the process dependence of gluon TMDs and in section~\ref{s:experimental possibilities} we briefly comment on some experimental possibilities to study these functions.

\section{Parametrization of the gluon-gluon correlation function} 

In this section we provide a systematic and efficient parametrization of the gluon-gluon correlation function for spin-$1$ hadrons. We start by giving a parametrization of the unintegrated correlator and then we integrate over $k^-$ to obtain a parametrization of the light-front correlator in terms of TMDs.

\subsection{The unintegrated correlator} 
The unintegrated gluon-gluon correlator for spin-$1$ hadrons is defined as 
\begin{align}
    & \,\Gamma^{[U,U']\,\mu\nu;\rho\sigma}(k,P,S,T,n) \nn \\
    \equiv& \int \frac{d^4z}{(2\pi)^4} \;e^{ik\cdot z} \bra{P;S,T} \tr\left( F^{\mu\nu}(0) \,U_{[0,z]}^{\phantom{'}} \,F^{\rho\sigma}(z) \,U_{[z,0]}' \right) \ket{P;S,T} ,
    \label{e:gamma_unintegrated}
\end{align}
where the trace reflects a summation over color degrees of freedom. Information on hadron spin has been incorporated in a way analogous to eq.~\eqref{e:spin_averaged}. The gauge links $U_{[0,z]}^{\phantom{\prime}}$ and $U_{[z,0]}^\prime$ guarantee color gauge invariance. Similarly to the quark case, these (calculable) gauge links give rise to a process dependence; this will be commented on in section~\ref{s:process_dependence_gluons}. Even though without specifying a process the path integrations could run along arbitrary paths, we have already included a dependence on the four-vector $n$, which enters upon consideration of the staple-like gauge links defined in eq.~\eqref{e:WLs}.

Since the correlator in eq.~\eqref{e:gamma_unintegrated} has open Lorentz indices, it should be parametrized in terms of Lorentz structures. Those structures need to be antisymmetric in $\mu,\nu$ as well as in $\rho,\sigma$ (this follows from the properties of the field strength tensor given in eq.~\eqref{e:fieldstrength}) and can be built from the Levi-Civita symbol $\epsilon$, the metric tensor $g$, and the available vectors and tensors $k$, $P$, $S$, $T$ (they are given in eqs.~\eqref{e:P}, \eqref{e:k}, \eqref{e:spin_vector}, and~\eqref{e:spin_tensor}), and $n$.\footnote{The four-vector $n$ was not included in the original parametrization in~\cite{Mulders:2000sh}. Similarly to the quark case, the inclusion of $n$ does not give rise to any new leading-twist TMDs, which we have checked explicitly for unpolarized hadrons. As we are ultimately only interested in leading-twist functions, we will simply omit terms containing $n$ in the parametrization of the unintegrated correlator.} The different Lorentz structures will be labeled by the coefficient functions $A_i = A_i(k^2,k\cd P,k\cd n,P\cd n)$ with $[A_i] = -4$, and the hadron mass $M$ will be used to give the terms the correct mass dimension, consistent with $[\Gamma] = -2$. As the spin vector and spin tensor appear linearly in the spin density matrix in eq.~\eqref{e:rho}, they can occur at most linearly in the parametrization. Similarly to the quark case, the correlator is constrained by requirements that arise from Hermiticity and parity. For any gauge link structure, these requirements are respectively given by
\begin{align}
    \Gamma^{\mu\nu;\rho\sigma}(k,P,S,T,n) &= \Gamma^{\rho\sigma;\mu\nu\ast}(k,P,S,T,n) , \label{e:constraint_H} \\
    \Gamma^{\mu\nu;\rho\sigma}(k,P,S,T,n) &= \Gamma_{\mu\nu;\rho\sigma}(\bar{k},\bar{P},-\bar{S},\bar{T},\bar{n}) . \label{e:constraint_P}
\end{align}
The Hermiticity constraint requires the coefficient functions $A_i$ to be real and the parity constraint forbids the presence of parity-odd terms (parity-odd structures include $\epsilon$ and $S$, which for example implies that $S$ can only appear together with $\epsilon$). 

As in the case of quarks, time reversal transformations relate correlators with time-reversed gauge link structures and thus do not constrain the actual form of the parametrization. Considering staple-like gauge links only, the time reversal constraints are given by
\begin{align}
    \Gamma^{[\pm,\pm]\,\mu\nu;\rho\sigma\ast}(k,P,S,T,n) &= \Gamma_{\mu\nu;\rho\sigma}^{[\mp,\mp]}(\bar{k},\bar{P},\bar{S},\bar{T},\bar{n}) , \\
    \Gamma^{[\pm,\mp]\,\mu\nu;\rho\sigma\ast}(k,P,S,T,n) &= \Gamma_{\mu\nu;\rho\sigma}^{[\mp,\pm]}(\bar{k},\bar{P},\bar{S},\bar{T},\bar{n}) ,
\end{align}
where Hermitian conjugation is implied for the second gauge link entry. Similarly to eqs.~\eqref{e:T-even} and~\eqref{e:T-odd}, the T-even and T-odd parts of $\Gamma$ are given by~\cite{Bomhof:2007xt}:
\begin{align}
    \Gamma^{(\text{T-even})}(k,P,S,T,n) &= \frac{1}{2} \left[ \Gamma^{[+,+]}(k,P,S,T,n) + \Gamma^{[-,-]}(k,P,S,T,n) \right] , \\
    \Gamma^{\prime(\text{T-even})}(k,P,S,T,n) &= \frac{1}{2} \left[ \Gamma^{[+,-]}(k,P,S,T,n) + \Gamma^{[-,+]}(k,P,S,T,n) \right] , \label{e:T-evenprime} \\
    \Gamma_{(f)}^{(\text{T-odd})}(k,P,S,T,n) &= \frac{1}{2} \left[ \Gamma^{[+,+]}(k,P,S,T,n) - \Gamma^{[-,-]}(k,P,S,T,n) \right] , \\
    \Gamma_{(d)}^{(\text{T-odd})}(k,P,S,T,n) &= \frac{1}{2} \left[ \Gamma^{[+,-]}(k,P,S,T,n) - \Gamma^{[-,+]}(k,P,S,T,n) \right] , \label{e:d-type}
\end{align}
where we suppressed the Lorentz indices to lighten the notation. The two T-odd parts are also sometimes referred to as $f$-type and $d$-type contributions, as they, respectively, are related to the $\text{SU}(3)$ structure constants $f$ and $d$ defined in eq.~\eqref{f_and_d}.

For convenience, let us split the parametrization up according to the type of hadron polarization (unpolarized, vector polarized, or tensor polarized):
\begin{equation}
    \Gamma^{\mu\nu;\rho\sigma}(k,P,S,T,n) = \Gamma^{\mu\nu;\rho\sigma}(k,P,n) + \Delta\Gamma^{\mu\nu;\rho\sigma}(k,P,S,n) + \Delta\Gamma^{\mu\nu;\rho\sigma}(k,P,T,n) .
    \label{e:polarization_splitup}
\end{equation}
A possible parametrization of the first term (for which $S=T=0$) is given by
\begin{align}
    \Gamma^{\mu\nu;\rho\sigma}(k,P,n) &= M^2 A_1 \,\epsilon^{\mu\nu\alpha\beta} \epsilon_{\phantom{\rho\sigma}\alpha\beta}^{\rho\sigma} + A_2 \,P^{[\mu} g^{\nu][\rho} P^{\sigma]} + A_3 \,k^{[\mu} g^{\nu][\rho} k^{\sigma]} \nn \\
    &\quad\, + (A_4 + iA_5) \,P^{[\mu} g^{\nu][\rho} k^{\sigma]} + (A_4 - iA_5) \,k^{[\mu} g^{\nu][\rho} P^{\sigma]} \nn \\
    &\quad\, + \frac{A_6}{M^2} \,P^{[\mu} k^{\nu]} P^{[\rho} k^{\sigma]} ,
    \label{e:gamma_unintegrated_par_up}
\end{align}   
where square brackets denote antisymmetrization of the indices. The term with coefficient $A_5$ is T-odd. 

The second term in eq.~\eqref{e:polarization_splitup} represents the vector polarized case and can be parametrized as follows:
\begin{align}  
    \Delta\Gamma^{\mu\nu;\rho\sigma}(k,P,S,n) &\equiv \frac{1}{2} \left[ \Gamma^{\mu\nu;\rho\sigma}(k,P,S,T,n) - \Gamma^{\mu\nu;\rho\sigma}(k,P,-S,T,n) \right] \nn \\
    &= - 2MA_7 \,\epsilon^{\mu\nu\rho\sigma} \,k \cd S + iMA_8 \left( \epsilon^{\mu\nu P[\rho} S^{\sigma]} - \epsilon^{\rho\sigma P[\mu} S^{\nu]} \right) \nn \\
    &\quad\, + iMA_9 \left( \epsilon^{\mu\nu S[\rho} P^{\sigma]} - \epsilon^{\rho\sigma S[\mu} P^{\nu]} \right) \nn \\
    &\quad\, + iMA_{10} \left( \epsilon^{\mu\nu k[\rho} S^{\sigma]} - \epsilon^{\rho\sigma k[\mu} S^{\nu]} \right) \nn \\
    &\quad\, + iMA_{11} \left( \epsilon^{\mu\nu S[\rho} k^{\sigma]} - \epsilon^{\rho\sigma S[\mu} k^{\nu]} \right) \nn \\
    &\quad\, + \frac{iA_{12}}{M} \left( \epsilon^{\mu\nu P[\rho} P^{\sigma]} - \epsilon^{\rho\sigma P[\mu} P^{\nu]} \right) k \cd S \nn \\
    &\quad\, + \frac{iA_{13}}{M} \left( \epsilon^{\mu\nu k[\rho} k^{\sigma]} - \epsilon^{\rho\sigma k[\mu} k^{\nu]} \right) k \cd S \nn \\
    &\quad\, + \frac{iA_{14}}{M} \left( \epsilon^{\mu\nu P[\rho} k^{\sigma]} - \epsilon^{\rho\sigma P[\mu} k^{\nu]} \right) k \cd S \nn \\
    &\quad\, + \frac{iA_{15}}{M} \left( \epsilon^{\mu\nu k[\rho} P^{\sigma]} - \epsilon^{\rho\sigma k[\mu} P^{\nu]} \right) k \cd S \nn \\
    &\quad\, + \frac{A_{16} + iA_{17}}{M} \,\epsilon^{\mu\nu PS} k^{[\rho} P^{\sigma]} + \frac{A_{16} - iA_{17}}{M} \,\epsilon^{\rho\sigma PS} k^{[\mu} P^{\nu]} \nn \\  
    &\quad\, + \frac{A_{18} + iA_{19}}{M} \,\epsilon^{\mu\nu kS} k^{[\rho} P^{\sigma]} + \frac{A_{18} - iA_{19}}{M} \,\epsilon^{\rho\sigma kS} k^{[\mu} P^{\nu]} \nn \\
    &\quad\, + \frac{A_{20} + iA_{21}}{M} \,\epsilon^{\mu\nu kP} P^{[\rho} S^{\sigma]} + \frac{A_{20} - iA_{21}}{M} \,\epsilon^{\rho\sigma kP} P^{[\mu} S^{\nu]} \nn \\
    &\quad\, + \frac{A_{22} + iA_{23}}{M} \,\epsilon^{\mu\nu kP} k^{[\rho} S^{\sigma]} + \frac{A_{22} - iA_{23}}{M} \,\epsilon^{\rho\sigma kP} k^{[\mu} S^{\nu]} \nn \\
    &\quad\, + \frac{A_{24} + iA_{25}}{M^3} \,\epsilon^{\mu\nu kP} k^{[\rho} P^{\sigma]} \,k \cd S \nn \\
    &\quad\, + \frac{A_{24} - iA_{25}}{M^3} \,\epsilon^{\rho\sigma kP} k^{[\mu} P^{\nu]} \,k \cd S .
    \label{e:gamma_unintegrated_par_vp}
\end{align}    
The terms with coefficients $A_7$, $A_{16}$, $A_{18}$, $A_{20}$, $A_{22}$, and $A_{24}$ are T-odd, and we note that the ones with coefficients $A_8$ up to $A_{15}$ are slightly different from those in the original parametrization in~\cite{Mulders:2000sh}. 

Finally, for the third term in eq.~\eqref{e:polarization_splitup}, representing tensor polarized hadrons, a possible parametrization reads
\begin{align}
    \Delta\Gamma^{\mu\nu;\rho\sigma}(k,P,T,n) &\equiv \frac{1}{2} \left[ \Gamma^{\mu\nu;\rho\sigma}(k,P,S,T,n) - \Gamma^{\mu\nu;\rho\sigma}(k,P,S,-T,n) \right] \nn \\
    &= A_{26} \,k^{[\mu} T^{\nu][\rho} k^{\sigma]} + A_{27} \,P^{[\mu} T^{\nu][\rho} P^{\sigma]} + (A_{28} + iA_{29}) \,k^{[\mu} T^{\nu][\rho} P^{\sigma]} \nn \\
    &\quad\, + (A_{28} - iA_{29}) \,P^{[\mu} T^{\nu][\rho} k^{\sigma]} + \frac{A_{30} + iA_{31}}{M^2} \,k_{\alpha} T^{\alpha[\mu} k^{\nu]} k^{[\rho} P^{\sigma]} \nn \\
    &\quad\, + \frac{A_{30} - iA_{31}}{M^2} \,k_{\alpha} T^{\alpha[\rho} k^{\sigma]} k^{[\mu} P^{\nu]} + \frac{A_{32} + iA_{33}}{M^2} \,k_{\alpha} T^{\alpha[\mu} P^{\nu]} k^{[\rho} P^{\sigma]} \nn \\
    &\quad\, + \frac{A_{32} - iA_{33}}{M^2} \,k_{\alpha} T^{\alpha[\rho} P^{\sigma]} k^{[\mu} P^{\nu]} + M^2 A_{34} \left( g^{\mu[\rho} T^{\sigma]\nu} - g^{\nu[\rho} T^{\sigma]\mu} \right) \nn \\
    &\quad\, + (A_{35} + iA_{36}) \,k_{\alpha} T^{\alpha[\mu} g^{\nu][\rho} k^{\sigma]} + (A_{35} - iA_{36}) \,k_{\alpha} T^{\alpha[\rho} g^{\sigma][\mu} k^{\nu]} \nn \\
    &\quad\, + (A_{37} + iA_{38}) \,k_{\alpha} T^{\alpha[\mu} g^{\nu][\rho} P^{\sigma]} + (A_{37} - iA_{38}) \,k_{\alpha} T^{\alpha[\rho} g^{\sigma][\mu} P^{\nu]} \nn \\
    &\quad\, + A_{39} \,k_{\alpha} k_{\beta} T^{\alpha\beta} \epsilon^{\mu\nu\kappa\lambda} \epsilon_{\phantom{\rho\sigma}\kappa\lambda}^{\rho\sigma} + \frac{A_{40}}{M^2} \,k_{\alpha} k_{\beta} T^{\alpha\beta} P^{[\mu} g^{\nu][\rho} P^{\sigma]} \nn \\
    &\quad\, + \frac{A_{41}}{M^2} \,k_{\alpha} k_{\beta} T^{\alpha\beta} k^{[\mu} g^{\nu][\rho} k^{\sigma]} + \frac{(A_{42} + iA_{43})}{M^2} \,k_{\alpha} k_{\beta} T^{\alpha\beta} P^{[\mu} g^{\nu][\rho} k^{\sigma]} \nn \\
    &\quad\, + \frac{(A_{42} - iA_{43})}{M^2} \,k_{\alpha} k_{\beta} T^{\alpha\beta} k^{[\mu} g^{\nu][\rho} P^{\sigma]} \nn \\
    &\quad\, + \frac{A_{44}}{M^4} \,k_{\alpha} k_{\beta} T^{\alpha\beta} P^{[\mu} k^{\nu]} P^{[\rho} k^{\sigma]} ,
    \label{e:gamma_unintegrated_par_tp}
\end{align}
where the terms with coefficients $A_{29}$, $A_{31}$, $A_{33}$, $A_{36}$, $A_{38}$, and $A_{43}$ are T-odd. 

In the next subsection these results will be used to derive a parametrization of the light-front correlator in terms of gluon TMDs.

\subsection{The TMD correlator} \label{s:parametrization_gluons}
The gluon-gluon TMD correlator can be obtained from the unintegrated correlator in eq.~\eqref{e:gamma_unintegrated} upon integration over $k^-$; it is given by
\begin{align}
    &\,\Gamma^{[U,U']\,\mu\nu;\rho\sigma}(x,\bm{k}) \nn \\
    \equiv& \,\frac{2}{P^+} \int \frac{dz^- d^2\bm{z}}{(2\pi)^3} \;e^{ik\cdot z} \left. \!\bra{P;S,T} \tr\left( F^{\mu\nu}(0) \,U_{[0,z]}^{\phantom{'}} \,F^{\rho\sigma}(z) \,U_{[z,0]}' \right) \ket{P;S,T} \vphantom{\int} \right|_{z^+=0} ,
    \label{e:gamma_TMD}
\end{align}
where we also included a conventional factor of $2/P^+$. Counting powers of the inverse hard scale relevant in the process leads to the definition
of the leading-twist correlator
\begin{equation}
    \Gamma^{ij}(x,\bm{k}) \equiv \Gamma^{+i;+j}(x,\bm{k}) ,
    \label{e:gamma_leadingtwist}
\end{equation}
where $i,j$ are transverse indices. This correlator was first parametrized in terms of gluon TMDs in~\cite{Mulders:2000sh}, covering both unpolarized and vector polarized hadrons. A few years later, a different nomenclature for those TMDs was proposed in~\cite{Meissner:2007rx}, in close analogy to the nomenclature for quarks. We adopt the conventions of the latter work and extend it to spin-$1$ hadrons. As for the quark case, we employ symmetric traceless tensors in $k_\sst$ (see appendix~\ref{a:STTs}) to ensure the TMDs are of definite rank. Integrating the sum of eqs.~\eqref{e:gamma_unintegrated_par_up}, \eqref{e:gamma_unintegrated_par_vp}, and~\eqref{e:gamma_unintegrated_par_tp} over $k^-$, we obtain the following parametrization of the light-front correlator in eq.~\eqref{e:gamma_leadingtwist} in terms of leading-twist, definite-rank gluon TMDs:
\begin{align}
    \Gamma^{ij}(x,\bm{k}) &= \Gamma_U^{ij}(x,\bm{k}) + \Gamma_L^{ij}(x,\bm{k}) + \Gamma_T^{ij}(x,\bm{k}) \nn \\
    &\quad\, + \Gamma_{LL}^{ij}(x,\bm{k}) + \Gamma_{LT}^{ij}(x,\bm{k}) + \Gamma_{TT}^{ij}(x,\bm{k}) ,
    \label{e:gluon_TMD_parametrization}
\end{align}
where
\begin{align}
    \Gamma_U^{ij}(x,\bm{k}) &= x \left[ \delta_\sst^{ij} \,f_1(x,\bm{k}^2) + \frac{k_\sst^{ij}}{M^2} \,h_1^{\perp}(x,\bm{k}^2) \right] \label{e:term1} , \\[5pt]
    \Gamma_L^{ij}(x,\bm{k}) &= x \left[ i \epsilon_\sst^{ij} S_L \,g_1(x,\bm{k}^2) + \frac{{\epsilon_\sst^{\{i}}_\alpha k_\sst^{j\}\alpha} S_L}{2M^2} \,h_{1L}^\perp(x,\bm{k}^2) \right] \label{e:term2} , \\[5pt]
    \Gamma_T^{ij}(x,\bm{k}) &= x \left[ \frac{\delta_\sst^{ij} \,\epsilon_\sst^{S_T k_\sst}}{M} \,f_{1T}^\perp(x,\bm{k}^2) + \frac{i \epsilon_\sst^{ij} \,\bm{k} \cd \bm{S}_T}{M} \,g_{1T}(x,\bm{k}^2) \right. \nn \\[3pt]
    &\quad\, \left. - \,\frac{\epsilon_\sst^{k_\sst\{i} S_T^{j\}} + \epsilon_\sst^{S_T\{i} k_\sst^{j\}}}{4M} \,h_1(x,\bm{k}^2) - \frac{{\epsilon_\sst^{\{i}}_\alpha k_\sst^{j\}\alpha S_T}}{2M^3} \,h_{1T}^\perp(x,\bm{k}^2) \right] \label{e:term3} , \\[5pt]
    \Gamma_{LL}^{ij}(x,\bm{k}) &= x \left[ \delta_\sst^{ij} S_{LL} \,f_{1LL}(x,\bm{k}^2) + \frac{k_\sst^{ij} S_{LL}}{M^2} \,h_{1LL}^{\perp}(x,\bm{k}^2) \right] \label{e:term4} , \\[5pt]
    \Gamma_{LT}^{ij}(x,\bm{k}) &= x \left[ \frac{\delta_\sst^{ij} \,\bm{k} \cd \bm{S}_{LT}}{M} \,f_{1LT}(x,\bm{k}^2) + \frac{i \epsilon_\sst^{ij} \epsilon_\sst^{S_{LT}k_\sst}}{M} \,g_{1LT}(x,\bm{k}^2) \right. \nn \\[3pt]
    &\quad\, \left. + \,\frac{S_{LT}^{\{i} k_\sst^{j\}}}{M} \,h_{1LT}(x,\bm{k}^2) + \frac{k_\sst^{ij\alpha} {S_{LT}}_\alpha}{M^3} \,h_{1LT}^{\perp}(x,\bm{k}^2) \right] \label{e:term5} , \\[5pt]
    \Gamma_{TT}^{ij}(x,\bm{k}) &= x \left[ \frac{\delta_\sst^{ij} \,k_\sst^{\alpha\beta} S_{TT\alpha\beta}}{M^2} \,f_{1TT}(x,\bm{k}^2) + \frac{i \epsilon_\sst^{ij} {\epsilon^{\beta}_\sst}_\gamma k_\sst^{\gamma\alpha} S_{TT\alpha\beta}}{M^2} \,g_{1TT}(x,\bm{k}^2) \right. \nn \\[3pt]
    &\quad\, \left. + \,S_{TT}^{ij} \,h_{1TT}(x,\bm{k}^2) + \frac{{S_{TT}^{\{i}}_\alpha k_\sst^{j\}\alpha}}{M^2} \,h_{1TT}^{\perp}(x,\bm{k}^2) \right. \nn \\[3pt]
    &\quad\, \left. + \,\frac{k_\sst^{ij\alpha\beta} {S_{TT}}_{\alpha\beta}}{M^4} \,h_{1TT}^{\perp\perp}(x,\bm{k}^2) \right] \label{e:term6} .
\end{align}
The expressions of the TMDs in terms of the coefficient functions $A_i$ can be found in appendix~\ref{a:definitions_gluon_TMDs}, and the equivalent parametrization in coordinate space is given in appendix~\ref{a:coordinate_space}. \\

\begin{table*}[!htb]
\begin{center}
{\renewcommand{\arraystretch}{1.4}
\begin{tabular}{|c|c|c|c|}
\hline
& $\delta_\sst^{ij}$ & $i\epsilon_\sst^{ij}$ & $k_\sst^i$, $k_\sst^{ij}$, etc. \\ \hline \hline
U & $\bm{f_1}$ & & $h_1^\perp$ \\ \hline
L & & $\bm{g_1}$ & $\color{red}h_{1L}^\perp$ \\ \hline
T & $\color{red}f_{1T}^\perp$ & $g_{1T}$ & $\color{red}h_1$, $\color{red}h_{1T}^\perp$ \\ \hline
LL & $\bm{f_{1LL}}$ & & $h_{1LL}^\perp$ \\ \hline
LT & $f_{1LT}$ & $\color{red}g_{1LT}$ & $h_{1LT}$, $h_{1LT}^\perp$ \\ \hline
TT & $f_{1TT}$ & $\color{red}g_{1TT}$ & $\bm{h_{1TT}}$, $h_{1TT}^\perp$, $h_{1TT}^{\perp\perp}$ \\
\hline
\end{tabular}} 
\caption{An overview of the leading-twist gluon TMDs for unpolarized (U), vector polarized (L or T), and tensor polarized (LL, LT, or TT) hadrons. The functions indicated in boldface also occur as collinear PDFs, and the ones in red are T-odd. The Lorentz structures $\delta_\sst^{ij}$, $i\epsilon_\sst^{ij}$, and $k_\sst^i$, $k_\sst^{ij}$, etc.\ correspond to unpolarized, circularly polarized, and linearly polarized gluons, respectively.}
\label{t:overview_gluon_tmds}
\end{center}
\end{table*}

In table~\ref{t:overview_gluon_tmds} we have organized the gluon TMDs appearing in eq.~\eqref{e:gluon_TMD_parametrization} by hadron and gluon polarizations and we have indicated which functions are T-odd and which ones have a collinear counterpart (the rank-$0$ TMDs). The gluon polarizations are projected out by the Lorentz structures that come with the functions (analogous to the Dirac structures in the quark case); we use the letters $f$, $g$, and $h$ to indicate unpolarized, circularly polarized, and linearly polarized gluons, respectively. 

With respect to the nomenclature in~\cite{Meissner:2007rx}, we have introduced the definite-rank function\footnote{Despite the notational similarities, this function is quite different from the quark transversity function $h_1$ defined in eq.~\eqref{e:quark_TMD_parametrization_vpolarized}.}
\begin{equation}
    h_1 \equiv h_{1T} + h_{1T}^{\perp(1)} .
\end{equation}
Furthermore, we have defined $g_1 \equiv g_{1L}$, the gluon helicity TMD. Unlike the quark TMDs, most of the gluon TMDs do not go by a special name. The most well-known T-odd gluon function is the gluon Sivers TMD $f_{1T}^\perp$. We note that the collinear PDF $h_{1TT}(x)$ appears in the structure function $\Delta(x)$ defined in~\cite{Jaffe:1989xy} (this structure function is called $\Delta_2G(x)$ in~\cite{Artru:1989zv}).

\section{Positivity bounds} \label{s:positivity_bounds}

In this section we derive bounds for the gluon TMDs that appear in eq.~\eqref{e:gluon_TMD_parametrization}. Positivity bounds for gluon TMDs were derived in~\cite{Mulders:2000sh} for spin-$\tfrac{1}{2}$ hadrons and, by applying the same strategy, we extend their analysis to spin-$1$ hadrons. For completeness, we will also cover the bounds that apply to spin-$\tfrac{1}{2}$ hadrons, completing the study of~\cite{Mulders:2000sh} where T-odd functions were not included. We will also mention the spin-$0$ case. Bounds on collinear gluon PDFs for spin-$1$ hadrons can be found in~\cite{Cotogno:2017mwy,Cotogno:2017puy}. 

Similarly to the quark-quark correlator in eq.~\eqref{e:spin_averaged}, the gluon-gluon correlator in eq.~\eqref{e:gamma_leadingtwist} is defined as
\begin{equation}
    \Gamma^{ij}(x,\bm{k},S,T) \equiv \tr \left( \rho(S,T) \,N^{ij}(x,\bm{k}) \right) = \rho_{s^\prime s}(S,T) \,N^{ij}_{s s^\prime}(x,\bm{k}) ,
    \label{e:spin_averaged_gluon}
\end{equation}
where $s,s^\prime$ label the hadronic polarization states. Our starting point is the idea that the correlator $\Gamma$ can be seen as a $2\times2$ matrix in the two gluon polarizations, and that the quantity $N$ can be regarded as a $6\times6$ matrix in gluon $\otimes$ hadron spin space. Choosing the same basis as in~\cite{Mulders:2000sh}, namely given by the two circular gluon polarizations $\ket{\pm} = \mp \,\tfrac{1}{\sqrt{2}} \left( \ket{1} \pm i\ket{2} \right)$, we find that $N$ is given by
\begin{equation}
    N = x \left( \begin{array}{cc}
    A&B \\
    B^\dagger&C \\
    \end{array} \right) ,
    \label{e:N_matrix}
\end{equation}
where

\vspace{-0.3cm}
\begingroup
\setlength{\arraycolsep}{3pt} 
\medmuskip=0.2mu
{\small
\begin{align}
    A &= \!\left(  \!\begin{array}{ccc}
    {f_{1}}+\frac{{f_{1LL}}}{2}-{g_1} & \frac{|\bm{k}| e^{-i {\phi}} }{\sqrt{2}
    M}\left(\tilde{f}-\tilde{g}+{h_{1LT}}\right) & \frac{\bm{k}^2 e^{-2 i {\phi}}}{M^2}\left({f_{1TT}}+i {g_{1TT}}-{h_{1TT}^\perp}\right) \\[5pt]
    \frac{|\bm{k}| e^{i {\phi}}}{\sqrt{2} M}\left(\tilde{f}^*-\tilde{g}^*+{h_{1LT}}\right) & {f_{1}}-{f_{1LL}} & -\frac{|\bm{k}| e^{-i
    {\phi}}}{\sqrt{2} M} \left(\tilde{f}^*+\tilde{g}^*+{h_{1LT}}\right) \\[5pt]
    \frac{\bm{k}^2 e^{2 i {\phi}} }{M^2} \left({f_{1TT}}-i {g_{1TT}}-{h_{1TT}^\perp}\right)& -\frac{|\bm{k}| e^{i {\phi}}}{\sqrt{2} M}\left(\tilde{f}+\tilde{g}+{h_{1LT}}\right) & {f_{1}}+\frac{{f_{1LL}}}{2}+{g_1} \nn \\
    \end{array} \!\right) , \nn \\[10pt]
    B &= \!\left( \!\begin{array}{ccc}
    -\frac{\bm{k}^2 e^{-2 i {\phi}}}{4 M^2} \left(2 {h_{1}^\perp} + {h_{1LL}^\perp}-2 i {h_{1L}^\perp}\right) & \frac{|\bm{k}|^3 e^{-3 i {\phi}}}{2
    \sqrt{2} M^3}\left({h_{1LT}^\perp}+i {h_{1T}^\perp}\right) & -\frac{\bm{k}^4 e^{-4 i {\phi}}}{2 M^4} \,h_{1TT}^{\perp\perp} \\[5pt]
    -\frac{|\bm{k}| e^{-i {\phi}}}{\sqrt{2} M}\left(2{h_{1LT}} - i{h_{1}}\right) & -\frac{\bm{k}^2 e^{-2 i {\phi}}}{2 M^2}\left({h_{1}^\perp - {h_{1LL}^\perp}}\right) & -\frac{|\bm{k}|^3 e^{-3 i {\phi}}}{2 \sqrt{2} M^3}\left({h_{1LT}^\perp}-i {h_{1T}^\perp}\right) \\[5pt]
    -2 {h_{1TT}} & \frac{|\bm{k}| e^{-i {\phi}} }{\sqrt{2} M}\left(2 {h_{1LT}}+i {h_{1}}\right) & -\frac{\bm{k}^2 e^{-2 i {\phi}}}{4 M^2}\left(2 {h_{1}^\perp + {h_{1LL}^\perp}+2 i{h_{1L}^\perp}}\right) \nn \\
    \end{array} \!\right) , \nn \\[10pt]
    C &= \!\left(  \!\begin{array}{ccc}
    {f_{1}}+\frac{{f_{1LL}}}{2}+{g_1} & \frac{|\bm{k}| e^{-i {\phi}} }{\sqrt{2}
    M}\left(\tilde{f}+\tilde{g}+{h_{1LT}}\right) & \frac{\bm{k}^2 e^{-2 i {\phi}}}{M^2}\left({f_{1TT}}-i {g_{1TT}}-{h_{1TT}^\perp}\right) \\[5pt]
    \frac{|\bm{k}| e^{i {\phi}}}{\sqrt{2} M}\left(\tilde{f}^*+\tilde{g}^*+{h_{1LT}}\right) & {f_{1}}-{f_{1LL}} & -\frac{|\bm{k}| e^{-i
    {\phi}}}{\sqrt{2} M} \left(\tilde{f}^*-\tilde{g}^*+{h_{1LT}}\right) \\[5pt]
    \frac{\bm{k}^2 e^{2 i {\phi}} }{M^2} \left({f_{1TT}}+i {g_{1TT}}-{h_{1TT}^\perp}\right)& -\frac{|\bm{k}| e^{i {\phi}}}{\sqrt{2} M}\left(\tilde{f}-\tilde{g}+{h_{1LT}}\right) & {f_{1}}+\frac{{f_{1LL}}}{2}-{g_1} \nn \\
    \end{array} \!\right) .
\end{align}}
\endgroup

\noindent For convenience we have defined $\tilde{f} \equiv f_{1LT} + if_{1T}^\perp$ and $\tilde{g} \equiv g_{1T} + ig_{1LT}$. Furthermore, we have expressed $\bm{k}$ in terms of its polar coordinates $|\bm{k}|$ and $\phi$. From symmetry considerations it follows that blocks $A$ and $C$ are related by parity and that the off-diagonal blocks are Hermitian conjugates.

To make more apparent the properties of the matrix $N$, we write its elements in the following form:
\begin{align}
    N^{ij}_{s s^\prime}(x,\bm{k}) &\equiv \frac{2}{P^+} \int \left. \frac{dz^- d^2\bm{z}}{(2\pi)^3} \;e^{ik\cdot z} \bra{P;s} \tr \left( F^{+i}(0) \,F^{+j}(z) \right) \ket{P;s^\prime} \right|_{z^+=0} \nn \\
    &= \frac{2}{P^+} \,\sum_m \,\tr \left( \bra{P_m} F^{ni}(0) \ket{P;s}^* \bra{P_m} F^{nj}(0) \ket{P;s^\prime} \right) \nn\\
    & \quad\, \times \,\delta\left(P_m^+ - (1-x)P^+\right) \,\delta^{(2)}\left(\bm{P}_m + \bm{k}\right) ,
    \label{e:positive_definiteness}
\end{align}
where we inserted a complete set of momentum eigenstates $\{\ket{P_m}\}$. From eq.~\eqref{e:positive_definiteness} we infer that, for any basis, the diagonal elements are given by absolute squares. In particular, it follows that the eigenvalues of $N$ in eq.~\eqref{e:N_matrix} must be $\geq 0$, or, equivalently, that $N$ is \emph{positive semidefinite}. This property can be used to set constraints on the TMDs. Given the limited amount of information we have on the gluon functions, we will refrain from diagonalizing the full $6\times6$ matrix and rather restrict ourselves to finding the eigenvalues of its $2\times2$ principal minors. Due to the symmetry properties of $N$, some minors yield the same bounds; we obtain the following nine inequalities for the spin-$1$ case:
\begin{align}
    \frac{\bm{k}^2}{2M^2} \,|h_1^\perp - h_{1LL}^\perp| &\leq f_1 - f_{1LL} \,, \label{e:bound1} \\
    \frac{\bm{k}^4}{16 M^4} \left[ 4 (h_{1L}^\perp)^2 + (2 h_1^\perp + h_{1LL}^\perp)^2 \right] &\leq f^+ f^- , \\
    \frac{\bm{k}^2}{2M^2} \left( h_1^2 + 4 h_{1LT}^2 \right) &\leq (f_1 - f_{1LL}) \,f^+ , \\
    \frac{\bm{k}^6}{8 M^6} \left[ (h_{1T}^\perp)^2 + (h_{1LT}^\perp)^2 \right] &\leq (f_1 - f_{1LL}) \,f^- , \\
    \frac{\bm{k}^2}{2M^2} \left[ (f_{1T}^\perp + g_{1LT})^2 + (f_{1LT} + g_{1T} + h_{1LT})^2 \right] &\leq (f_1 - f_{1LL}) \,f^+ , \\
    \frac{\bm{k}^2}{2M^2} \left[ (f_{1T}^\perp - g_{1LT})^2 + (f_{1LT} - g_{1T} + h_{1LT})^2 \right] &\leq (f_1 - f_{1LL}) \,f^- , \\
    |h_{1TT}| &\leq \frac{f^+}{2} \,, \\
    \frac{\bm{k}^4}{2 M^4} \,|h_{1TT}^{\perp\perp}| &\leq f^- , \\
    \frac{\bm{k}^4}{M^4} \left[ g_{1TT}^2 + (f_{1TT} - h_{1TT}^\perp)^2 \right] &\leq f^+ f^- , \label{e:bound9}
\end{align}
where for convenience we have defined $f^\pm \equiv f_1 + \tfrac{f_{1LL}}{2} \pm g_1$.

Now let us consider spin-$\tfrac{1}{2}$ hadrons. Bounds for this case have been derived already in~\cite{Mulders:2000sh}, however using a different notation and leaving the T-odd TMDs aside. The parametrization of the correlator for a spin-$\tfrac{1}{2}$ hadron is given by the sum of eqs.~\eqref{e:term1}--\eqref{e:term3}, and the spin density matrix can be parametrized in terms of the spin vector only (using the identity matrix and the three Pauli matrices) and is a $2\times2$ matrix in hadron spin space. The quantity $N$ is now a $4\times4$ matrix in gluon $\otimes$ hadron spin space and its explicit form (that \emph{does} contain the T-odd functions) is given in~\cite{Mulders:2000sh}. From that matrix we can extract the following bounds from its $2\times2$ principal minors:
\begin{align}
    |g_1| &\leq f_1 \,, \label{e:boundMR1} \\
    \frac{\bm{k}^4}{4M^4} \left[ (h_{1L}^\perp)^2 + (h_1^\perp)^2 \right] &\leq (f_1 + g_1) (f_1 - g_1) , \\
    \frac{|\bm{k}|}{M} \,|h_1| &\leq f_1 + g_1 \,, \\
    \frac{|\bm{k}|^3}{2M^3} \,|h_{1T}^\perp| &\leq f_1 - g_1 \,, \\
    \frac{\bm{k}^2}{M^2} \left[ (f_{1T}^\perp)^2 + g_{1T}^2 \right] &\leq (f_1 + g_1) (f_1 - g_1) . \label{e:boundMR5}
\end{align}
Upon omitting the functions related to tensor polarization in bounds~\eqref{e:bound1}--\eqref{e:bound9}, which is mathematically equivalent to considering the spin-$\tfrac{1}{2}$ case, one obtains a set of bounds that is less strict than (but consistent with) bounds~\eqref{e:boundMR1}--\eqref{e:boundMR5}. In general, bounds can be sharpened upon considering the eigenvalues of higher-dimensional principal minors.

Finally, let us comment on the spin-$0$ case. For spin-$0$ hadrons the parametrization of the correlator is given by the two functions in eq.~\eqref{e:term1}. Now, as there is no spin density matrix (this means that $\Gamma = N$), we simply need to compute the eigenvalues of the $2\times2$ matrix $\Gamma$; this results in the single bound
\begin{equation}
    \frac{\bm{k}^2}{2M^2} \,|h_1^\perp| \leq f_1 \,.
    \label{e:boundUP}
\end{equation}
This inequality also follows from bounds~\eqref{e:boundMR1}--\eqref{e:boundMR5} upon omitting the functions related to vector polarization.

Note that in eq.~\eqref{e:positive_definiteness} we did not consider the process-dependent gauge link structure explicitly. In fact, the inequalities~\eqref{e:bound1}--\eqref{e:boundUP} do \emph{not} hold generally true for any correlator -- the matrix $N$ is positive semidefinite only for field combinations, including gauge links, that `factorize' into the form $O^\dagger(0) \,O(z)$. The simplest gauge link structures for which this holds are $[+,+]$, $[-,-]$, $[+,-]$, and $[-,+]$ (we will say more about these link structures in the next section). For the same links, one also has the constraint $f_1 \ge 0$ (note that for the spin-$\tfrac{1}{2}$ and spin-$0$ cases this follows already from the provided bounds). 

One might worry about the effects of QCD evolution on the validity of the bounds. In the collinear case, the so-called Soffer bound involving three quark functions~\cite{Soffer:1994ww} has been shown to be preserved up to next-to-leading order accuracy~\cite{Vogelsang:1997ak,Bourrely:1997bx,Martin:1997rz}. However, to our knowledge, there are no studies yet on the stability of bounds under evolution concerning TMDs. The fact that the evolution kernel for TMDs is independent of spin~\cite{Echevarria:2014rua,Echevarria:2015uaa} might suggest that, in the appropriate regime of $\bm{k}$ where TMD factorization is valid, positivity bounds are respected also in this case. 
This topic remains open to further investigation.

Since the bounds follow from the positive semidefiniteness of $N$ and rely primarily on the operator structure of the correlator, they can be considered as tests of the TMD formalism (provided the functions are compared at the same scales). In practical situations, however, the bounds will, rather than serving as tests of the framework, often be more useful as a check in model or lattice calculations, or as a way to obtain an order of magnitude estimate of certain TMDs. The latter is commonly done by saturating the bounds. Those estimates could serve, for instance, as input for an estimate of measurements of particular azimuthal asymmetries.

\section{Process dependence} \label{s:process_dependence_gluons}

\begin{figure}[htb]
\centering
\begin{subfigure}{.42\textwidth}
    \begin{flushleft}
        \includegraphics[height=2.5cm]{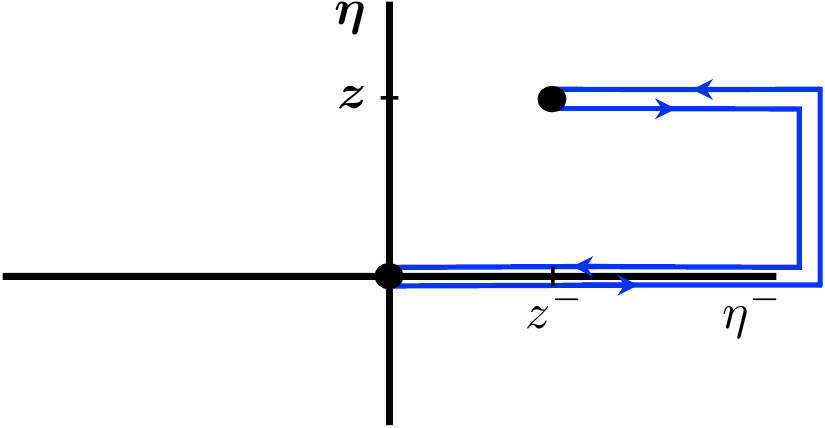}
        \caption*{\hspace{-0.81cm}(a) \vspace{0.40cm}}
    \end{flushleft}
\end{subfigure}
\begin{subfigure}{.42\textwidth}
    \begin{flushright}
        \includegraphics[height=2.5cm]{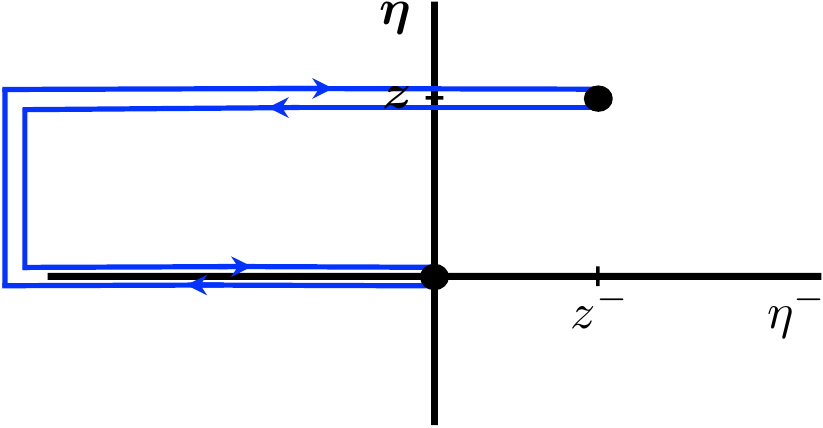}
        \caption*{\hspace{1.02cm}(b) \vspace{0.40cm}}
    \end{flushright}
\end{subfigure} 
\begin{subfigure}{.42\textwidth}
    \begin{flushleft}
        \hspace{-0.09cm} \includegraphics[height=2.5cm]{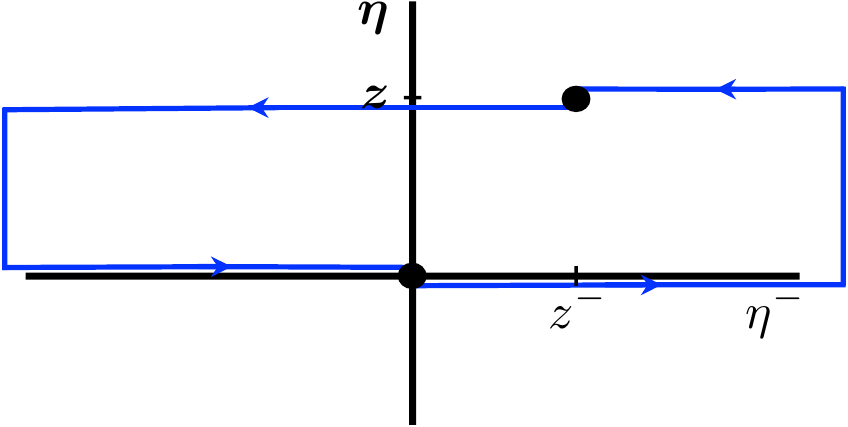}
        \caption*{\hspace{-0.61cm}(c)}
    \end{flushleft}
\end{subfigure}
\hspace{0.21cm} \begin{subfigure}{.42\textwidth}
    \begin{flushright}
        \includegraphics[height=2.5cm]{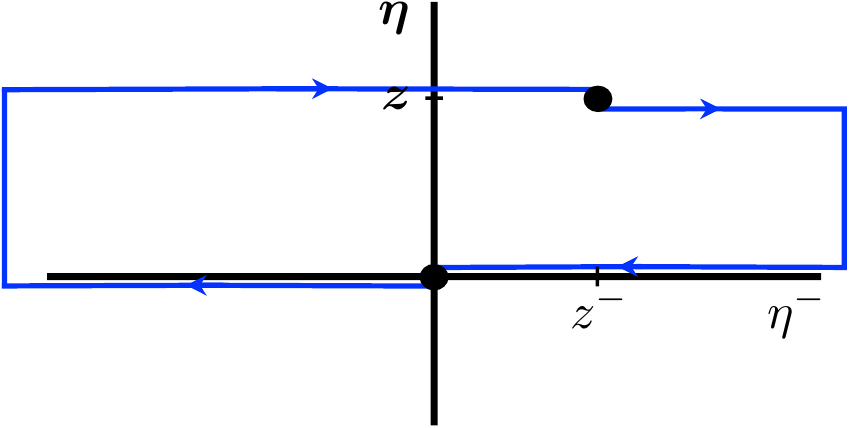}
        \caption*{\hspace{0.60cm}(d)}
    \end{flushright}
\end{subfigure}
\caption{The integration paths of (a) the $[+,+]$ gauge link, (b) the $[-,-]$ gauge link, (c) the $[+,-]$ gauge link, and (d) the $[-,+]$ gauge link. The paths run along the light-front $\eta^+=0$ via plus and/or minus light-cone infinity.}
\label{f:gluon_links}
\end{figure}

In general, gluon TMDs depend on the process under consideration through the gauge links. As explained in subsection~\ref{s:process_dependence}, the integration paths of these links depend on the direction of color flow in the hard scattering process and are calculable. The four simplest and most common gauge links in the gluon case are $[+,+]$, $[-,-]$, $[+,-]$, and $[-,+]$, see figure~\ref{f:gluon_links}. The $[-,-]$ gauge link appears in processes with color flow annihilated within the initial state, such as the (gluonic) DY process or Higgs production through gluon fusion ($gg \to h$)~\cite{Boer:2013fca,Echevarria:2015uaa}. The structure $[+,+]$, on the other hand, is related to color flow into the final state, which is the case for e.g.\ $q\bar{q}$ production in SIDIS~\cite{Pisano:2013cya}. When color flow involves both initial and final states, the gauge links $[+,-]$ and $[-,+]$ appear, which is for instance the case in processes with $qg \to qg$ and $\bar{q}g \to \bar{q}g$ partonic contributions, respectively~\cite{Bomhof:2006dp}. 

The relations between gluon TMDs that feature different gauge link structures have been systematically studied in~\cite{Buffing:2013kca}. For example, for the T-even unpolarized function $f_1$ and the T-odd gluon Sivers function $f_{1T}^\perp$ we have the following relations~\cite{Bomhof:2006dp,Buffing:2013kca,Boer:2016fqd}:
\begin{alignat}{2}
    f_1^{[+,+]} &= f_1^{[-,-]} , \qquad& f_1^{[+,-]} &= f_1^{[-,+]} , \\
    f_{1T}^{\perp[+,+]} &= -f_{1T}^{\perp[-,-]} , \qquad& f_{1T}^{\perp[+,-]} &= -f_{1T}^{\perp[-,+]} . \label{e:gluon_Sivers}
\end{alignat}
As it turns out, $f_1^{[+,+]}$ \emph{cannot} be related to $f_1^{[+,-]}$, and likewise for $f_{1T}^\perp$. Thus, for some processes TMDs are completely independent from each other and encode distinct information. For this reason, different extractions could in fact be \emph{complementary}~\cite{Boer:2015vso}.

In the context of small-$x$ studies, the unpolarized distributions $f_1^{[+,+]}$ and $f_1^{[+,-]}$ are often respectively referred to as the (unintegrated) Weizs\"acker-Williams (WW) and dipole distributions~\cite{Kharzeev:2003wz,Dominguez:2010xd,Dominguez:2011wm}. These distributions are considered as the two fundamental gluon distributions at small $x$. Although in some relatively complicated processes such as dijet production in $pA$ collisions distributions with other gauge link structures can appear, it seems that in the large-$N_c$ limit they can all be related to the WW and dipole distributions~\cite{Dominguez:2011wm,Kotko:2015ura}. We will refer to the $[+,+]$ and $[-,-]$ links as `WW-type' gauge link structures, and to the $[+,-]$ and $[-,+]$ links as `dipole-type` gauge link structures (irrespective of gluon or hadron polarization and for general $x$).

\section[\stackunder{Experimental possibilities}{}]{Experimental possibilities} \label{s:experimental possibilities} 

Over recent years, interest in gluon distributions has increased significantly. However, to this day, the extraction of gluon functions remains very challenging compared to quark functions, for example because they typically require higher-energy scattering processes. We end this chapter with a brief discussion on experimental possibilities to study gluon TMDs.

The gluon TMDs for unpolarized and vector polarized hadrons could be investigated through processes at the Relativistic Heavy Ion Collider (RHIC), at the Large Hadron Collider (LHC), possibly at a future polarized fixed-target experiment at the LHC called AFTER@LHC~\cite{Brodsky:2012vg}, and at a future EIC~\cite{Boer:2011fh}. In fact, recently, using LHCb data on $p p \to J/\Psi \,J/\Psi \,X$ and assuming a Gaussian profile, the transverse momentum dependence of the unpolarized gluon distribution $f_1^{[+,+]}$ was constrained for the first time~\cite{Lansberg:2017dzg}. As yet, there are no fits to data available for the gluon Sivers function $f_{1T}^\perp$. The most promising processes that give access to $f_{1T}^\perp$ are $p^\uparrow p \to \gamma \,\text{jet} \,X$ (dipole-type) at RHIC and at AFTER@LHC~\cite{Schmidt:2005gv,Bacchetta:2007sz}, $p^\uparrow p \to J/\Psi \,\gamma \,X$ (WW-type)~\cite{Dunnen:2014eta} or $p^\uparrow p \to J/\Psi \,J/\Psi \,X$ (WW-type) at AFTER@LHC~\cite{Lansberg:2014myg,Lansberg:2015lva}, and $e^- p^\uparrow \to e^- D\bar{D} X$ (WW-type) at an EIC~\cite{Boer:2015vso}, where in between brackets we indicated for each process the (dominant) gauge link structure that is probed. Today, no all-order TMD factorization proofs are available for these processes, which all involve at least three hadrons. For the same reason that the expected sign change between DY and SIDIS for the quark Sivers function is currently under investigation, it would be interesting from a theoretical point of view to compare extractions of the gluon Sivers function from different processes. The current knowledge on the gluon Sivers function as well as future prospects are reviewed in~\cite{Boer:2015vso}.

The study of gluon TMDs for tensor polarized hadrons would be possible at the experiments proposed to investigate polarized deuterons, e.g.\ at the EIC possibility at Jefferson Lab (JLEIC)~\cite{Boer:2011fh,Abeyratne:2012ah,Abeyratne:2015pma}, or at COMPASS at CERN~\cite{Ball:2006zz}. Some experimental interest has been shown already in the collinear PDF $h_{1TT}(x)$ that represents the distribution of linearly polarized gluons in a transversely tensor polarized hadron. Its extraction has been proposed to occur at Jefferson Lab using nitrogen targets~\cite{DetmoldJLab}, and could also be achieved within the program of the EIC~\cite{Accardi:2012qut}. 

For more information on the phenomenology of gluon TMDs, see e.g.~\cite{Boer:2010zf,Boer:2011fh,Qiu:2011ai,Dominguez:2011wm,Boer:2012bt,Pisano:2013cya,Angeles-Martinez:2015sea,Boer:2015uqa,Boer:2016jnn,Lansberg:2017tlc,Godbole:2018mmh}.
\chapter{Gluon distributions at small $x$} \label{c:Small_x}
\addthumb{\thechapter}{\Large{\thechapter}}{\thumbfg}{\thumbbg}

In this chapter we will discuss dipole-type gluon TMDs, GTMDs, and Wigner distributions in the small-$x$ region where hadrons mainly consist of gluons. The field of research concerned with small-$x$ physics is also referred to as high-energy QCD, for a high CM energy is required in collisions to probe small values of $x$. Most of this chapter is based on our work in~\cite{Boer:2016xqr,Cotogno:2017puy,Boer:2018vdi}.

First, we provide a parametrization of the gluon-gluon GTMD correlator for unpolarized hadrons in terms of four leading-twist gluon GTMDs. Then we show that for the dipole-type gauge link structure the GTMD correlator simplifies in the small-$x$ limit to the Fourier transform of a hadronic matrix element containing a rectangular Wilson loop. This so-called off-forward Wilson loop correlator can be parametrized in terms of a single function (for unpolarized hadrons), which implies that at small $x$ there is only one independent dipole-type GTMD and not four. This Wilson loop GTMD contains both so-called pomeron and odderon contributions. The same strategy can be applied to the TMD case. As it turns out, in this analysis the small-$x$ limit is independent of hadron polarization, which means that the obtained results for the unpolarized case in fact apply to each individual hadronic polarization state. As a result, we obtain a greatly simplified picture for the TMDs that were introduced in subsection~\ref{s:parametrization_gluons} for spin-$1$ hadrons. In the TMD case, the only odderon contribution that we identify originates from transversely polarized hadrons, consistent with the findings in~\cite{Boer:2015pni}. We will also derive positivity bounds for TMDs that apply to the small-$x$ region. 

In the second part of this chapter we will study azimuthal correlations in hadron production from high-energy collisions that originate from dipole Wigner distributions of gluons in the nucleus. In section~\ref{s:MV_model} we introduce the so-called McLerran-Venugopalan (MV) model (and an extension thereof) that will be used to calculate the expectation values of the pomeron and odderon operators in a nuclear environment. Those results will feature in the calculation of pomeron and odderon Wigner distributions that contribute to the cross section of dihadron production in $pA$ collisions. We show that the odderon Wigner distribution can generate odd harmonics in the two-particle azimuthal correlations and we will provide a numerical estimate of the size of the first odd-harmonic contribution, quantified by the so-called directed flow coefficient.

\section{Parametrization of the GTMD correlator}

Analogous to the quark-quark GTMD correlator in eq.~\eqref{e:qq_GTMDcorrelator}, the gluon-gluon GTMD correlator for an unpolarized hadron is defined as~\cite{Lorce:2013pza}:
\begin{align}
    & \,G^{[U,U']\,\mu\nu;\rho\sigma}(x,\bm{k},\xi,\bm{\Delta}) \nn \\
    \equiv& \,\frac{2}{P^+} \int \frac{dz^- \,d^2\bm{z}}{(2\pi)^3} \;e^{ik\cdot z} \left. \!\bra{p'} \tr \left( F^{\mu\nu} \!\left(-\tfrac{z}{2}\right) U_{\left[-\tfrac{z}{2},\tfrac{z}{2}\right]}^{\phantom{\prime}} \,F^{\rho\sigma} \!\left(\tfrac{z}{2}\right) U_{\left[\tfrac{z}{2},-\tfrac{z}{2}\right]}^\prime \right) \ket{p} \vphantom{\int} \right|_{z^+ = 0} .
\end{align}
The off-forwardness $\Delta \equiv p'-p$ is given in eq.~\eqref{e:Delta}. Similarly to eq.~\eqref{e:gamma_leadingtwist}, the leading-twist correlator is given by
\begin{equation}
    G^{ij}(x,\bm{k},\xi,\bm{\Delta}) \equiv G^{+i;+j}(x,\bm{k},\xi,\bm{\Delta}) ,
    \label{e:G_GTMD}
\end{equation}
where $i,j$ are transverse indices. This correlator can be parametrized in terms of GTMDs, in a way analogous to the parametrization of the TMD correlator in terms of TMDs in chapter~\ref{c:Polarized_gluon_TMDs}. This has been done already in~\cite{Lorce:2013pza} based on the light-front formalism (in fact, up to all powers in the inverse hard scale and including vector polarization) and here we present an alternative, but equivalent, parametrization. To ensure that the distribution functions are of definite rank, we use symmetric traceless tensors in both $k_\sst$ and $\Delta_\sst$. As explained in~\cite{Lorce:2013pza}, requirements from Hermiticity and time reversal (unlike parity) do not affect the Lorentz structure of the parametrization -- they rather impose constraints on the GTMDs. A possible parametrization of the correlator in eq.~\eqref{e:G_GTMD} is given by\footnote{In principle one could also have a function that comes with the symmetric and traceless Lorentz structure $k_\sst^{\{i} \Delta_\sst^{j\}} + (\bm{k} \cd \bm{\Delta}) \,g_\sst^{ij}$. However, this function would not be independent from the other ones; more specifically, it could be eliminated by suitable redefinitions of the functions $\mathcal{F}_2$ and $\mathcal{F}_3$ through the (two-dimensional) relation $\bm{k} \cd \bm{\Delta} \left[ k_\sst^{\{i} \Delta_\sst^{j\}} + (\bm{k} \cd \bm{\Delta}) \,g_\sst^{ij} \right] = \bm{\Delta}^2 k_\sst^{ij} + \bm{k}^2 \Delta_\sst^{ij}$.}
\begin{align}
    G^{ij}(x,\bm{k},\xi,\bm{\Delta}) &= x \left[ \delta_\sst^{ij} \,\mathcal{F}_1(x,\bm{k}^2,\xi,\bm{\Delta}^2,\bm{k} \cd \bm{\Delta}) + \frac{k_\sst^{ij}}{M^2} \,\mathcal{F}_2(x,\bm{k}^2,\xi,\bm{\Delta}^2,\bm{k} \cd \bm{\Delta}) \right. \nn \\
    &\quad\, \left. + \,\frac{\Delta_\sst^{ij}}{M^2} \,\mathcal{F}_3(x,\bm{k}^2,\xi,\bm{\Delta}^2,\bm{k} \cd \bm{\Delta}) + \frac{k_\sst^{[i} \Delta_\sst^{j]}}{M^2} \,\mathcal{F}_4(x,\bm{k}^2,\xi,\bm{\Delta}^2,\bm{k} \cd \bm{\Delta}) \right] ,
    \label{e:G_GTMD_par}
\end{align}
where the functions $\mathcal{F}_{1-4}$ are (complex-valued) GTMDs. In contrast to the TMD case, no positivity bounds exist for GTMDs. After all, the GTMD analogue of $N$ in eq.~\eqref{e:positive_definiteness} is \emph{not} positive semidefinite due to $\Delta$ being nonzero.

In the forward limit, the GTMDs in eq.~\eqref{e:G_GTMD_par} are related to the TMDs for unpolarized hadrons in eq.~\eqref{e:term1}; to be specific:
\begin{equation}
    \lim_{\Delta \to 0} \mathcal{F}_1 = f_1 \,, \qquad \lim_{\Delta \to 0} \mathcal{F}_2 = h_1^\perp , \qquad \lim_{\Delta \to 0} \mathcal{F}_3 = \lim_{\Delta \to 0} \mathcal{F}_4 = 0 .
\end{equation}
Upon integration over $\bm{k}$, one recovers from the GTMDs the leading-twist GPDs~\cite{Meissner:2007rx}.

\section{Dipole-type distributions in the small-$x$ limit} \label{s:dipole_smallx}

The parametrization in eq.~\eqref{e:G_GTMD_par} holds for general $x$ and for any gauge link structure. In this section we will study the small-$x$ limit of the correlator for the dipole-type gauge link structure. We will start with the case of GTMDs and discuss the forward limit in subsection~\ref{e:forward_WL}. Hadron polarization will not be considered until subsection~\ref{s:WL_TMDs_pol_hadrons}.

\subsection{The off-forward Wilson loop correlator} \label{e:off-forward_WL}
The dipole-type GTMD correlator is given by
\begingroup
\thinmuskip=2.4mu 
\medmuskip=3.2mu
\begin{align}
    \hspace{-0.1cm} G^{[+,-]\,ij}(x,\bm{k},\xi,\bm{\Delta}) &= \frac{2}{P^+} \int \frac{dz^- d^2\bm{z}}{(2\pi)^3} \;e^{ik\cdot z} \nn \\
    &\quad\, \times \left. \!\bra{p'} \tr \left( F^{+j} \!\left(\tfrac{z}{2}\right) U_{\left[-\tfrac{z}{2},\tfrac{z}{2}\right]}^{[-]\dag} \,F^{+i} \!\left(-\tfrac{z}{2}\right) U_{\left[-\tfrac{z}{2},\tfrac{z}{2}\right]}^{[+]} \right) \ket{p} \vphantom{\int} \right|_{z^+=0} \nn \\[3pt]
    &= \frac{2}{P^+} \int \frac{dz^- d^2\bm{z}}{(2\pi)^3} \;e^{ik\cdot z- i\Delta\cdot b} \nn \\
    &\quad\, \times \left. \!\bra{p'} \tr \left( F^{+j}(x) \,U_{\left[y,x\right]}^{[-]\dag} \,F^{+i}(y) \,U_{\left[y,x\right]}^{[+]} \right) \ket{p} \vphantom{\int} \right|_{x^+=y^+=b^+=z^+=0} ,
\end{align}
\endgroup
where $x \equiv b + \tfrac{z}{2}$ and $y \equiv b - \tfrac{z}{2}$. Using that the limit $\Delta \to 0$ of 
\begin{equation}
    \braket{p'|p} = (2\pi)^3 \,2p^+ \delta(\Delta^+) \,\delta^{(2)}(\bm{\Delta})
    \label{e:normalization}
\end{equation}
is given by $\braket{P|P} = 2P^+ \!\int db^- d^2\bm{b}$, we can write
\begin{align}
    G^{[+,-]\,ij}(x,\bm{k},\xi,\bm{\Delta}) &= \frac{4}{\braket{P|P}} \int \frac{dx^- d^2\bm{x} \;dy^- d^2\bm{y}}{(2\pi)^3} \;e^{ik\cdot (x - y) - i\Delta \cdot \frac{x+y}{2}} \nn \\
    &\quad\, \times \left. \!\bra{p'} \tr \left( F^{+j}(x) \,U_{[x,y]}^{[-]} \,F^{+i}(y) \,U_{[y,x]}^{[+]} \right) \ket{p} \vphantom{\int} \right|_{x^+=y^+=0} .
\end{align}

\begin{figure}[htb]
\centering
    \includegraphics[height=2.5cm]{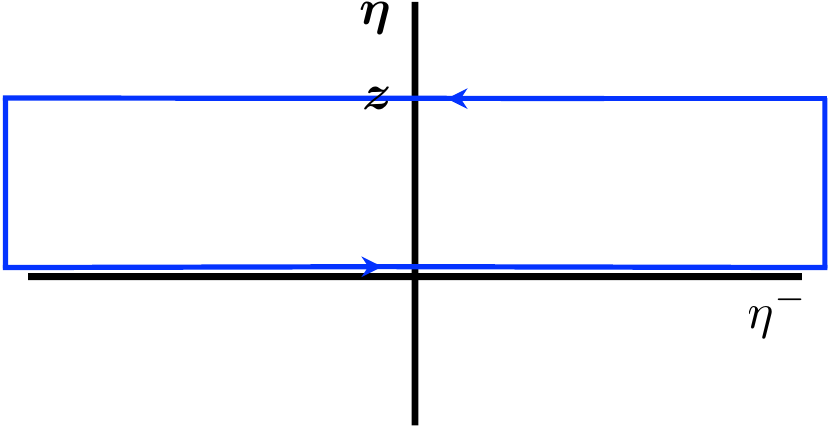}
    \caption{The integration path of the Wilson loop $U_{[\bm{0},\bm{z}]}^{[\Box]}$ with transverse extent $|\bm{z}|$. The closed path runs along the light-front $\eta^+=0$ via plus and minus light-cone infinity.}
\label{f:Wilson_loop}
\end{figure}

Now, let us consider the region where $x \sim \xi$ is small. To that end, we perform the limit
\begin{equation}
    G^{[+,-]\,ij}(\bm{k},\bm{\Delta}) \equiv \lim_{x,\xi \to 0} G^{[+,-]\,ij}(x,\bm{k},\xi,\bm{\Delta}) ,
    \label{e:correlator_atsmallx}
\end{equation}
in which case the $x^-$ and $y^-$ integrations can be performed. Those integrations can be used to express the result in terms of the operators
\begin{equation}
    G_\sst^k(z) \equiv \frac{1}{2} \int_{-\infty}^\infty d\eta^- \,U_{[z^-,\eta^-;\bm{z}]}^n \,F^{+k}(z^+,\eta^-,\bm{z}) \,U_{[\eta^-,z^-;\bm{z}]}^n \,.
\end{equation} 
Upon using the relation~\cite{Buffing:2013kca}
\begin{equation}
    \left[ i \partial_{\bm{z}}^k, U_{[a,z]}^{[\pm]} \right] = \mp \,g U_{[a,z]}^{[\pm]} \,G_\sst^k(z) ,
\end{equation}
the dipole-type GTMD correlator can be written as the correlator of a Wilson loop:
\begin{align}
    G^{[+,-]\,ij}(\bm{k},\bm{\Delta}) &= \frac{16}{\braket{P|P}} \int \frac{d^2\bm{x} \,d^2\bm{y}}{(2\pi)^3} \;e^{- i\bm{k} \cd (\bm{x} - \bm{y}) + i\bm{\Delta} \cd \frac{\bm{x}+\bm{y}}{2}} \nn \\
    &\quad\, \times \left. \!\bra{p'} \tr \left( G_\sst^j(x) \,U_{[x,y]}^{[-]} \,G_\sst^i(y) \,U_{[y,x]}^{[+]} \right) \ket{p} \vphantom{\int} \right|_{x^+=y^+=0} \nn \\
    &= \frac{4}{g^2 \braket{P|P}} \int \frac{d^2\bm{x} \,d^2\bm{y}}{(2\pi)^3} \;e^{- i\bm{k} \cd (\bm{x} - \bm{y}) + i\bm{\Delta} \cd \frac{\bm{x}+\bm{y}}{2}} \nn \\
    &\quad\, \times \left. \!\bra{p'} \partial_{\bm{x}}^j \partial_{\bm{y}}^i \tr \left( U_{[\bm{y},\bm{x}]}^{[\Box]} \right) \ket{p} \vphantom{\int} \right|_{x^+=y^+=0} ,
    \label{e:G_GTMD_dipole}
\end{align}
where now the off-forwardness is only transverse ($\bm{\Delta}$) and where
\begin{equation}
    U_{[\bm{y},\bm{x}]}^{[\Box]} \equiv U_{[y,x]}^{[+]} \,U_{[x,y]}^{[-]}
\end{equation}
is a rectangular Wilson loop with transverse distance $\bm{x}-\bm{y}$, stretched to infinity in the minus direction (see figure~\ref{f:Wilson_loop}). 

Next, we can apply partial integration twice and write eq.~\eqref{e:G_GTMD_dipole} as\footnote{Eq.~\eqref{e:G_GTMD_S} is consistent with the result in~\cite{Hatta:2016dxp} where only the term with $\delta_\sst^{ij}$ was considered.}
\begin{equation}
    G^{[+,-]\,ij}(\bm{k},\bm{\Delta}) = \frac{1}{2\pi g^2} \left[ \frac{1}{2} \left( \bm{k}^2 - \frac{\bm{\Delta}^2}{4} \right) \delta_\sst^{ij} + k_\sst^{ij} - \frac{\Delta_\sst^{ij}}{4} - \frac{k_\sst^{[i} \Delta_\sst^{j]}}{2} \right] G^{[\Box]}(\bm{k},\bm{\Delta}) ,
    \label{e:G_GTMD_S}
\end{equation}
where
\begin{equation}
    G^{[\Box]}(\bm{k},\bm{\Delta}) \equiv \frac{4 N_c}{\braket{P|P}} \int \frac{d^2\bm{x} \,d^2\bm{y}}{(2\pi)^2} \;e^{- i\bm{k} \cd (\bm{x} - \bm{y}) + i\bm{\Delta} \cd \frac{\bm{x}+\bm{y}}{2}} \left. \!\bra{p'} S^{[\Box]}(\bm{x},\bm{y}) \ket{p} \vphantom{\int} \right|_{x^+=y^+=0} ,
    \label{e:G_at_small_x}
\end{equation}
is the Fourier transform of the Wilson loop operator
\begin{equation}
    S^{[\Box]}(\bm{x},\bm{y}) \equiv \frac{1}{N_c} \tr \left( U_{[\bm{y},\bm{x}]}^{[\Box]} \right) .
    \label{e:dipole_operator}
\end{equation}
We will refer to $G^{[\Box]}$ as the off-forward Wilson loop correlator. Naturally, also this correlator can be parametrized in terms of GTMDs, analogous to the parametrization in eq.~\eqref{e:G_GTMD_par}; we choose the following parametrization:
\begin{equation}
    G^{[\Box]}(\bm{k},\bm{\Delta}) = \frac{2\pi g^2}{M^2} \;\mathcal{E}(\bm{k}^2,\bm{\Delta}^2,\bm{k} \cd \bm{\Delta}) ,
    \label{e:par_of_G}
\end{equation}
where $\mathcal{E}$ is a (complex-valued) Wilson loop GTMD. 

From eq.~\eqref{e:par_of_G} we infer something important, namely that at vanishing $x$ and $\xi$ the picture is very simple -- there is only \emph{one} independent GTMD. From eqs.~\eqref{e:G_GTMD_S} and~\eqref{e:par_of_G} it follows that in the limit of small $x$ and $\xi$ the GTMDs defined in eq.~\eqref{e:G_GTMD_par} are related as follows:
\begin{equation}
    \lim_{x,\xi \to 0} x\mathcal{F}_1 = \lim_{x,\xi \to 0} x\mathcal{F}_2^{\{1\}} = -4 \lim_{x,\xi \to 0} x\mathcal{F}_3^{\{1\}} = -2 \lim_{x,\xi \to 0} x\mathcal{F}_4^{\{1\}} = \mathcal{E}^{\{1\}} ,
    \label{e:relbetweenfcts}
\end{equation}
where we used the shorthand notation $\mathcal{F}_i^{\{n\}} \equiv [(\bm{k}^2 - \bm{\Delta}^2/4)/(2M^2)]^n \,\mathcal{F}_i$, which reduces to eq.~\eqref{e:moment_notation} in the forward limit. 

Let us now have a closer look at the Wilson loop operator $S^{[\Box]}$. This operator can be written in terms of its real and imaginary parts as follows:
\begin{equation}
    S^{[\Box]}(\bm{x},\bm{y}) = \mathcal{P}(\bm{x},\bm{y}) + i\mathcal{O}(\bm{x},\bm{y}) ,
    \label{e:S_real_im}
\end{equation}
with
\begin{align}
    \mathcal{P}(\bm{x},\bm{y}) &\equiv \frac{1}{2N_c} \tr \left( U_{[\bm{y},\bm{x}]}^{[\Box]} + U_{[\bm{y},\bm{x}]}^{[\Box]\dag} \right) , \label{e:pomeron_operator} \\[2pt]
    \mathcal{O}(\bm{x},\bm{y}) &\equiv \frac{1}{2iN_c} \tr \left( U_{[\bm{y},\bm{x}]}^{[\Box]} - U_{[\bm{y},\bm{x}]}^{[\Box]\dag} \right) . \label{e:odderon_operator}
\end{align}
As it turns out, these parts behave differently under charge conjugation (C): the real part $\mathcal{P}$ is C-even and called the pomeron, while the imaginary part $\mathcal{O}$ is C-odd and called the odderon~\cite{Hatta:2005as}. These charge conjugation properties can be easily understood starting from the transformation of the gauge field:
\begin{equation}
    C A_\mu C^{-1} = -A_\mu \,.
\end{equation}
It then follows that the gauge links transform as
\begin{equation}
    C \,U_{[x,y]}^{[\pm]} \,C^{-1} = U_{[x,y]}^{[\pm]\dag} \,, \qquad C \,U_{[\bm{y},\bm{x}]}^{[\Box]} \,C^{-1} = U_{[\bm{y},\bm{x}]}^{[\Box]\dag} \,,
\end{equation}
so that, indeed,
\begin{equation}
    C \,\mathcal{P}(\bm{x},\bm{y}) \,C^{-1} = \mathcal{P}(\bm{x},\bm{y}) , \qquad C \,\mathcal{O}(\bm{x},\bm{y}) \,C^{-1} = -\mathcal{O}(\bm{x},\bm{y}) .
\end{equation}
In this chapter special attention will be given to the odderon, which was introduced already in 1973 in~\cite{Lukaszuk:1973nt}. At lowest order in perturbation theory, an odderon corresponds to three gluons in a symmetric color state~\cite{Levin:1990gg}. First experimental indications for the existence of odderon effects have come from the recent results on $pp$ collisions by the TOTEM experiment at CERN~\cite{Martynov:2017zjz,Antchev:2017yns,Csorgo:2018uyp}, combined with data on $p\bar{p}$ collisions by the D{\O} experiment at Fermilab~\cite{Martynov:2018sga}.

Similar to eq.~\eqref{e:d-type}, the T-odd contribution to the dipole-type GTMD correlator in eq.~\eqref{e:G_GTMD_S} is given by:
\begin{align}
    G_{(d)}^{(\text{T-odd})\,ij}(\bm{k},\bm{\Delta}) &\equiv \frac{1}{2} \left[ G^{[+,-]\,ij}(\bm{k},\bm{\Delta}) - G^{[-,+]\,ij}(\bm{k},\bm{\Delta}) \right] \nn \\[5pt]
    &= \frac{1}{4\pi g^2} \left[ \frac{1}{2} \left( \bm{k}^2 - \frac{\bm{\Delta}^2}{4} \right) \delta_\sst^{ij} + k_\sst^{ij} - \frac{\Delta_\sst^{ij}}{4} - \frac{k_\sst^{[i} \Delta_\sst^{j]}}{2} \right] \nn \\[3pt]
    &\quad\, \times \left[ G^{[\Box]}(\bm{k},\bm{\Delta}) - G^{[\Box^\dag]}(\bm{k},\bm{\Delta}) \right] \nn \\[5pt]
    &= \frac{2iN_c}{\pi g^2 \braket{P|P}} \left[ \frac{1}{2} \left( \bm{k}^2 - \frac{\bm{\Delta}^2}{4} \right) \delta_\sst^{ij} + k_\sst^{ij} - \frac{\Delta_\sst^{ij}}{4} - \frac{k_\sst^{[i} \Delta_\sst^{j]}}{2} \right] \nn \\[3pt]
    &\quad\, \times \left. \!\int \frac{d^2\bm{x} \,d^2\bm{y}}{(2\pi)^2} \;e^{- i\bm{k} \cd (\bm{x} - \bm{y}) + i\bm{\Delta} \cd \frac{\bm{x}+\bm{y}}{2}} \bra{p'} \mathcal{O}(\bm{x},\bm{y}) \ket{p} \vphantom{\int} \right|_{x^+=y^+=0} ,
    \label{e:odderon}
\end{align}
which is proportional to a matrix element of the odderon operator. Thus, the odderon contribution to the dipole-type GTMD correlator is not only C-odd but also T-odd. In the same way it follows that the T-even contribution is proportional to a matrix element of the pomeron operator. From the Hermiticity and time reversal constraints, respectively given by
\begin{equation}
    G^{[\Box]*}(\bm{k},\bm{\Delta}) = G^{[\Box]}(\bm{k},-\bm{\Delta}) , \qquad G^{[\Box]*}(\bm{k},\bm{\Delta}) = G^{[\Box^\dag]}(-\bm{k},-\bm{\Delta}) ,
    \label{e:constraints_offforwardWL}
\end{equation}
it follows that the combination $G^{[\Box]}(\bm{k},\bm{\Delta}) - G^{[\Box^\dag]}(\bm{k},\bm{\Delta})$ appearing on the second line of eq.~\eqref{e:odderon} only contains \emph{odd} powers of $\bm{k} \cd \bm{\Delta}$. Likewise, the T-even contribution only contains \emph{even} powers of $\bm{k} \cd \bm{\Delta}$. Hence, if we were to expand $\mathcal{E}$ in powers of $\bm{k} \cd \bm{\Delta}$ and extract the angle $\phi \equiv \phi_{\bm{k}} - \phi_{\bm{\Delta}}$ (we will do something similar for the gluon Wigner distributions in subsection~\ref{s:directed_flow}), then the pomeron and odderon contributions would, respectively, be constituted by all even harmonics $\cos\left( 2n\phi \right)$ and all odd harmonics $\cos\left( (2n+1)\phi \right)$, where $n \geq 0$.

\subsection{The forward Wilson loop correlator} \label{e:forward_WL}
In the remainder of this section we will consider the TMD case. In the forward limit, the dipole-type GTMD correlator reduces to the dipole-type TMD correlator:
\begin{equation}
    \lim_{\Delta \to 0} G^{[+,-]\,ij}(x,\bm{k},\xi,\bm{\Delta}) = \Gamma^{[+,-]\,ij}(x,\bm{k}) .
    \label{e:deltalimit1}
\end{equation}
Employing eqs.~\eqref{e:correlator_atsmallx}, \eqref{e:G_GTMD_S}, and the $\Delta \to 0$ limit of~\eqref{e:normalization}, it follows that in the small-$x$ limit the TMD correlator is given by
\begin{align}
    \lim_{x\to0} \,\Gamma^{[+,-]\,ij}(x,\bm{k}) &= \lim_{\bm{\Delta} \to 0} G^{[+,-]\,ij}(\bm{k},\bm{\Delta}) \nn \\
    &= \frac{1}{2\pi g^2 L} \left( \frac{1}{2} \bm{k}^2 \delta_\sst^{ij} + k_\sst^{ij} \right) \Gamma^{[\Box]}(\bm{k}) ,
    \label{e:deltalimit2}
\end{align}
where $L \equiv 2\pi \,\delta(\Delta^+=0)$ and
\begin{align}
    \Gamma^{[\Box]}(\bm{k}) &\equiv \lim_{\bm{\Delta} \to 0} \,L \,G^{[\Box]}(\bm{k},\bm{\Delta}) \nn \\
    &= \frac{2N_c}{P^+} \left. \!\int \frac{d^2\bm{z}}{(2\pi)^2} \;e^{- i\bm{k} \cd \bm{z}} \bra{P} S^{[\Box]}(\bm{z},\bm{0}) \ket{P} \right|_{z^+=0} 
    \label{e:relation_WL_correlators}
\end{align}
is the \emph{forward} Wilson loop correlator. We choose to parametrize this correlator as follows (the usual Hermiticity, parity, and time reversal constraints will be given in the next subsection where we will include hadron spin):
\begin{equation}
    \Gamma^{[\Box]}(\bm{k}) = \frac{2\pi g^2 L}{M^2} \;e(\bm{k}^2) ,
    \label{e:forwardWL_par}
\end{equation}
where $e$ is a (real) T-even Wilson loop TMD.

Similar to the GTMD case in eq.~\eqref{e:par_of_G}, there is only \emph{one} independent TMD in the small-$x$ limit. From eqs.~\eqref{e:deltalimit2} and~\eqref{e:forwardWL_par} it follows that the two TMDs appearing in eq.~\eqref{e:term1} are related as
\begin{equation}
    \lim_{x\to0} xf_1 = \lim_{x\to0} xh_1^{\perp(1)} = e^{(1)} .
    \label{e:upol_result}
\end{equation} 
This result implies that bound~\eqref{e:boundUP} is saturated. It also indicates that the unpolarized dipole gluon distribution grows as $1/x$ towards small $x$, apart from subdominant modifications arising from resummation of large logarithms in $1/x$ (see e.g.~\cite{Ball:2017otu}) and higher-twist effects. From a recent global fit called NNPDF3.0 it follows that the small-$x$ asymptotic behavior of the collinear gluon PDF for the proton at $x=10^{-4}$ is $1/x^{1.20\pm0.46}$ at $Q^2 = 10$ GeV$^2$ and $1/x^{1.382\pm0.011}$ at $Q^2 = 10^4$ GeV$^2$~\cite{Ball:2016spl}.

\begin{figure}[htb]
\centering
    \hspace{0.21cm} \includegraphics[height=3.45cm]{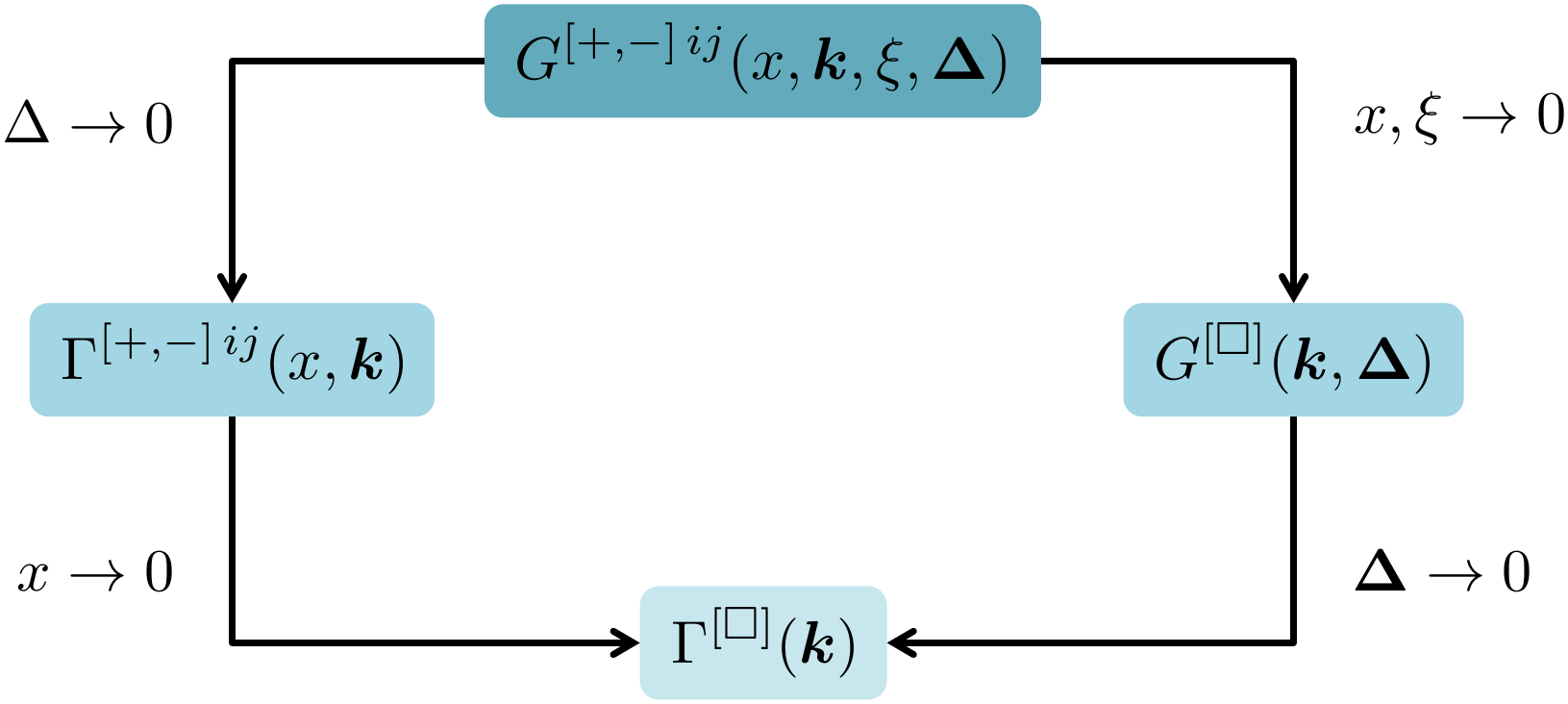}
    \caption{Relations (schematically) between the dipole-type TMD and GTMD correlators and the forward and off-forward Wilson loop correlators. The precise relations are given in the main text.}
\label{f:scheme_correlators}
\end{figure}

\begin{figure}[htb]
\centering
    \includegraphics[height=3.45cm]{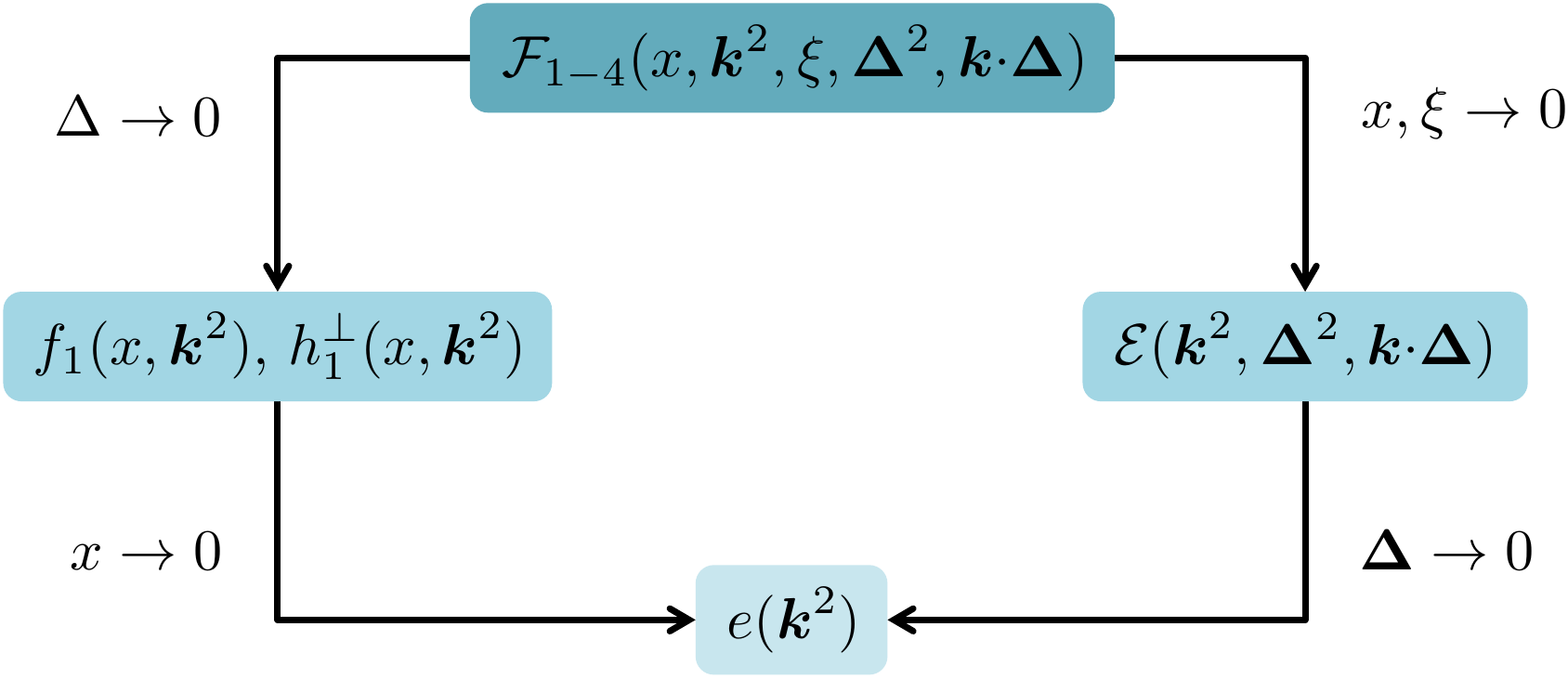}
    \caption{Relations (schematically) between the gluon and Wilson loop TMDs and GTMDs for unpolarized hadrons. The precise relations are given in the main text.}
\label{f:scheme_functions}
\end{figure}

From eqs.~\eqref{e:par_of_G}, \eqref{e:relation_WL_correlators}, and~\eqref{e:forwardWL_par} we infer that the Wilson loop function $\mathcal{E}$ reduces to $e$ in the forward limit:
\begin{equation}
    \lim_{\bm{\Delta} \to 0} \mathcal{E}(\bm{k}^2,\bm{\Delta}^2,\bm{k} \cd \bm{\Delta}) = e(\bm{k}^2) .
\end{equation}
In the previous subsection we argued that $\mathcal{E}$ contains both pomeron and odderon contributions. In the forward limit, only the angular-independent harmonic ($n=0$) contained in $\mathcal{E}$ survives -- this is a pomeron contribution and precisely corresponds with the T-even function $e$. Thus, at least for unpolarized hadrons and at leading twist, there are \emph{no} odderon contributions in the TMD case. 
Figures~\ref{f:scheme_correlators} and~\ref{f:scheme_functions} summarize the main results of this section so far. They provide the relations between the various correlators and distribution functions that we have discussed. We stress that the main results, eqs.~\eqref{e:relbetweenfcts} and~\eqref{e:upol_result}, only hold at leading twist and for the dipole-type gauge link structure. In different contexts, the small-$x$ limit has also been considered for the WW-type gauge link structure, see e.g.~\cite{Marquet:2016cgx}. However, it seems in that case impossible to relate the result to a Wilson loop or any other `simple structure' that would allow for partial integration as in eq.~\eqref{e:G_GTMD_S}. Hence, we think that a similar analysis for the WW case is not possible.

\subsection{Wilson loop TMDs for polarized hadrons} \label{s:WL_TMDs_pol_hadrons}
In subsection~\ref{e:off-forward_WL} we studied the limit of small $x,\xi$ of the GTMD correlator for \emph{unpolarized} hadrons, and in subsection~\ref{e:forward_WL} we considered the forward limit of those results. However, it is easy to see that all these limits are actually completely independent of hadron polarization. In particular, eq.~\eqref{e:deltalimit2} \emph{also} holds for vector and tensor polarized hadrons (this result was already obtained for transversely polarized hadrons in~\cite{Boer:2015pni}). The left-hand side of that equation has been parametrized in terms of TMDs in eq.~\eqref{e:gluon_TMD_parametrization} for spin-$1$ hadrons. Naturally, we can do the same thing for the right-hand side, i.e.\ we can extend the parametrization of the forward Wilson loop correlator for unpolarized hadrons in eq.~\eqref{e:forwardWL_par} to the case of spin-$1$ hadrons; this will be done below.
 
For spin-$1$ hadrons, the forward Wilson loop correlator is defined as
\begin{equation}
    \Gamma^{[\Box]}(\bm{k}) \equiv \frac{2N_c}{P^+} \left. \!\int \frac{d^2\bm{z}}{(2\pi)^2} \;e^{- i\bm{k} \cd \bm{z}} \bra{P;S,T} S^{[\Box]}(\bm{z},\bm{0}) \ket{P;S,T} \right|_{z^+=0} .
    \label{e:relation_WL_correlators_spin1}
\end{equation}
Since this correlator has no open indices, a parametrization in terms of independent Wilson loop TMDs is rather straightforward. The relevant Hermiticity and parity constrains follow directly from eqs.~\eqref{e:constraint_H} and~\eqref{e:constraint_P} and are respectively given by
\begin{align}
    \Gamma^{[\Box]}(\bm{k},P,S,T,n) &= \Gamma^{[\Box]\ast}(\bm{k},P,S,T,n) , \label{e:constraint_H_WL} \\
    \Gamma^{[\Box]}(\bm{k},P,S,T,n) &= \Gamma^{[\Box]}(-\bm{k},\bar{P},-\bar{S},\bar{T},\bar{n}) . \label{e:constraint_P_WL}
\end{align}
We propose the following parametrization that respects these constraints:
\begin{align}
    \Gamma^{[\Box]}(\bm{k}) &= \frac{2\pi g^2 L}{M^2} \left[ e(\bm{k}^2) + \frac{\epsilon_\sst^{S_T k_\sst}}{M} \,e_T(\bm{k}^2) + S_{LL} \,e_{LL}(\bm{k}^2) \vphantom{\frac{k_\sst^{\alpha\beta} {S_{TT}}_{\alpha\beta}}{M^2}} \right. \nn \\
    &\quad\, \left. + \,\frac{\bm{k} \cd \bm{S}_{LT}}{M} \,e_{LT}(\bm{k}^2) + \frac{k_\sst^{\alpha\beta} {S_{TT}}_{\alpha\beta}}{M^2} \,e_{TT}(\bm{k}^2) \right] .
    \label{e:forwardWL_par_pol}
\end{align}

According to the time reversal constraint 
\begin{align}
    \Gamma^{[\Box]\ast}(\bm{k},P,S,T,n) = \Gamma^{[\Box^\dag]}(-\bm{k},\bar{P},\bar{S},\bar{T},\bar{n}) ,
\end{align}
the function $e_T$ is T-odd while the other functions are T-even. From eqs.~\eqref{e:d-type}, \eqref{e:odderon_operator}, \eqref{e:deltalimit2}, and~\eqref{e:relation_WL_correlators_spin1} it follows that in the small-$x$ limit the T-odd contribution to the dipole-type TMD correlator is given by
\begin{align}
    \lim_{x\to0} \,\Gamma_{(d)}^{(\text{T-odd})\,ij}(x,\bm{k}) &\equiv \frac{1}{2} \lim_{x\to0} \left[ \Gamma^{[+,-]\,ij}(x,\bm{k}) - \Gamma^{[-,+]\,ij}(x,\bm{k}) \right] \nn \\[5pt]
    &= \frac{1}{4\pi g^2L} \left( \frac{1}{2} \bm{k}^2 \delta_\sst^{ij} + k_\sst^{ij} \right) \left[ \Gamma^{[\Box]}(\bm{k}) - \Gamma^{[\Box^\dag]}(\bm{k}) \right] \nn \\[5pt]
    &= \frac{iN_c}{\pi g^2 P^+ L} \left( \frac{1}{2} \bm{k}^2 \delta_\sst^{ij} + k_\sst^{ij} \right) \nn \\
    &\quad\, \times \left. \int \frac{d^2\bm{z}}{(2\pi)^2} \;e^{- i\bm{k} \cd \bm{z}} \bra{P;S,T} \mathcal{O}(\bm{z},\bm{0}) \ket{P;S,T} \right|_{z^+=0} .
    \label{e:odderonTMD}
\end{align}
Hence, analogous to eq.~\eqref{e:odderon} for the GTMD case, the T-odd contribution features the odderon operator. Since there is only one T-odd Wilson loop TMD, this contribution is completely captured by the function $e_T$. 

Note that in eq.~\eqref{e:forwardWL_par_pol} there is no Wilson loop TMD corresponding to longitudinally polarized hadrons. Thus, it follows from eqs.~\eqref{e:term2} and~\eqref{e:deltalimit2} that $g_1$ and $h_{1L}^\perp$ vanish in the small-$x$ limit, or rather that they are less divergent than $1/x$ in the limit of small $x$. The latter is in agreement with the results in~\cite{Kovchegov:2017lsr,Kovchegov:2018znm} where large logarithms in $1/x$ were resummed for the gluon helicity function $g_1$. For the other hadron polarizations we can derive relations similar to the one for unpolarized hadrons, eq.~\eqref{e:upol_result}. Notice that as the right-hand side of eq.~\eqref{e:deltalimit2} is symmetric in $i,j$, the $g$-type functions (that come with the antisymmetric structure $i\epsilon_\sst^{ij}$) generally vanish in the small-$x$ limit.

Now let us consider transversely polarized hadrons. From eqs.~\eqref{e:deltalimit2} and~\eqref{e:forwardWL_par_pol} it follows that the TMDs appearing in eq.~\eqref{e:term3} are related as
\begin{equation}
    \lim_{x\to0} x f_{1T}^\perp = \lim_{x\to0} x h_1 = - \lim_{x\to0} x h_{1T}^{\perp(1)} = e_T^{(1)} .
    \label{e:tpol_result}
\end{equation}
This is in agreement with the results in~\cite{Boer:2015pni}. The fact that a transversely polarized hadron can give rise to an odderon contribution was already noted in~\cite{Ryskin:1987ya,Buttimore:1998rj,Leader:1999ua}, but without discussion of its operator structure. The precise operator structure was first given in~\cite{Zhou:2013gsa} and was referred to as the `spin-dependent' odderon. As discussed in the previous subsection, we have not identified a spin-\emph{independent} odderon. However, as we argued below eq.~\eqref{e:constraints_offforwardWL}, odderon contributions do appear for unpolarized hadrons in the GTMD case.

Finally, let us consider tensor polarized hadrons. In the small-$x$ limit, the TMDs for longitudinal-longitudinal (LL) polarization appearing in eq.~\eqref{e:term4} are related as
\begin{equation}
    \lim_{x\to0} x f_{1LL} = \lim_{x\to0} x h_{1LL}^{\perp(1)} = e_{LL}^{(1)} .
    \label{e:LLpol_result}
\end{equation}
For longitudinal-transverse (LT) polarization, see eq.~\eqref{e:term5}, we obtain
\begin{equation}
    \lim_{x\to0} x f_{1LT} = \lim_{x\to0} x h_{1LT} = -\frac{1}{2} \lim_{x\to0} x h_{1LT}^{\perp(1)} = \frac{1}{2} \,e_{LT}^{(1)} ,
\end{equation}
and for the TMDs corresponding to transverse-transverse (TT) polarization in eq.~\eqref{e:term6} we find
\begin{equation}
    \lim_{x\to0} x f_{1TT}^{(1)} = \frac{1}{3} \lim_{x\to0} x h_{1TT} = - \frac{1}{2} \lim_{x\to0} x h_{1TT}^{\perp(1)} = \frac{1}{3} \lim_{x\to0} x h_{1TT}^{\perp\perp(2)} = \frac{1}{3} \,e_{TT}^{(2)} \,.
\end{equation}
Note that for all hadron polarizations the time reversal properties of the dipole-type TMDs and the Wilson loop TMDs correctly match.

Similarly to the gluon TMDs in section~\ref{s:positivity_bounds}, we can also derive bounds for the Wilson loop TMDs that appear in the parametrization of the forward Wilson loop correlator given by
\begin{equation}
    \Gamma^{[\Box]}(\bm{k},S,T) \equiv \tr \left( \rho(S,T) \,N^{[\Box]}(\bm{k}) \right) = \rho_{s^\prime s}(S,T) \,N_{s s^\prime}^{[\Box]}(\bm{k}) .
\end{equation}
Upon inverting this relation, we find that 
\begin{equation}
\setlength{\arraycolsep}{3pt}
\medmuskip=1.5mu
    N^{[\Box]} = \frac{2\pi g^2 L}{M^2} \left( \begin{array}{ccc}
    e + \frac{e_{LL}}{2} & \frac{|\bm{k}|e^{-i\phi}}{\sqrt{2} M} \left(e_{LT} + ie_T\right) & \frac{\bm{k}^2 e^{-2i\phi}}{M^2} \,e_{TT} \\[5 pt]
    \frac{|\bm{k}| e^{i\phi}}{\sqrt{2} M} \left(e_{LT} - ie_T\right) & e - e_{LL} & -\frac{|\bm{k}| e^{-i\phi}}{\sqrt{2} M} \left(e_{LT} - ie_T\right) \\[5 pt]
    \frac{\bm{k}^2 e^{2i\phi}}{M^2} \,e_{TT}& -\frac{|\bm{k}| e^{i\phi}}{\sqrt{2} M}\left(e_{LT} + ie_T\right) & e + \frac{e_{LL}}{2} \\
    \end{array} \right) .
    \label{e:spinmatrix_loop}
\end{equation}
In analogy to eq.~\eqref{e:positive_definiteness}, we can write the elements of $N^{[\Box]}$ in the following form:
\begin{align}
    N_{s s^\prime}^{[\Box]}(\bm{k}) &\equiv \frac{2N_c}{P^+} \left. \!\int \frac{d^2\bm{z}}{(2\pi)^2} \;e^{- i\bm{k} \cd \bm{z}} \bra{P;s} S^{[\Box]}(\bm{z},\bm{0}) \ket{P;s^\prime} \right|_{z^+=0} \nn \\
    &= \frac{2}{P^+} \,\sum_m \,\tr \left( \bra{P_m} U^T_{[\infty;\bm{\infty},\bm{0}]} \,U^n_{[\infty,-\infty;\bm{0}]} \,U^T_{[-\infty;\bm{0},\bm{\infty}]} \ket{P;s}^* \right. \nn \\
    &\quad\, \left.\left. \times \bra{P_m} U^T_{[\infty;\bm{\infty},\bm{0}]} \,U^n_{[\infty,-\infty;\bm{0}]} \,U^T_{[-\infty;\bm{0},\bm{\infty}]} \ket{P;s^\prime} \right) \right|_{z^+=0} \delta^{(2)}({\bm{P}_m} + \bm{k}) ,
    \label{e:posdef_loop}
\end{align}
where we inserted a complete set of momentum eigenstates $\{\ket{P_m}\}$ and we used that 
\begin{align}
    U_{[\bm{0},\bm{z}]}^{[\Box]} &= U^n_{[-\infty,\infty;\bm{0}]} \,U^T_{[\infty;\bm{0},\bm{z}]} \,U^n_{[\infty,-\infty;\bm{z}]} \,U^T_{[-\infty;\bm{z},\bm{0}]} \nn \\
    &= \left( U^T_{[-\infty;\bm{\infty},\bm{0}]} \,U^n_{[-\infty,\infty;\bm{0}]} \,U^T_{[\infty;\bm{0},\bm{\infty}]} \right) \left( U^T_{[-\infty;\bm{\infty},\bm{z}]} \,U^n_{[-\infty,\infty;\bm{z}]} \,U^T_{[\infty;\bm{z},\bm{\infty}]} \right)^\dagger .
    \label{e:ubox}
\end{align}
From eq.~\eqref{e:posdef_loop} it follows that $N^{[\Box]}$ is positive semidefinite -- in other words, its eigenvalues must be $\geq 0$. To establish bounds for the Wilson loop TMDs, we restrict ourselves to two-dimensional principal minors and find
\begin{align}
    \frac{\bm{k}^2}{2M^2} \left( e_T^2 + e_{LT}^2 \right) &\leq (e - e_{LL}) \left( e + \frac{e_{LL}}{2} \right) , \label{e:ebound1} \\
    \frac{\bm{k}^2}{M^2} \,|e_{TT}| &\leq e + \frac{e_{LL}}{2} \,. 
    \label{e:ebound2}
\end{align}
Applying the small-$x$ limit to bounds~\eqref{e:bound1}--\eqref{e:bound9}, one indeed recovers bounds~\eqref{e:ebound1} and~\eqref{e:ebound2}. Besides these two bounds, we also have $e \geq 0$ (this follows directly from eq.~\eqref{e:ubox}).

Let us finally also comment on the case of a spin-$\tfrac{1}{2}$ hadron. The parametrization of the forward Wilson loop correlator for spin-$\tfrac{1}{2}$ hadrons is given in terms of the two functions $e$ and $e_T$. For $N^{[\Box]}$ we now obtain
\begin{equation}
    N^{[\Box]} = \frac{2\pi g^2 L}{M^2} \left( \begin{array}{cc}
    e & \frac{i|\bm{k}|e^{-i\phi}}{M} \,e_T \\[5 pt]
    -\frac{i|\bm{k}|e^{i\phi}}{M} \,e_T & e \\
    \end{array} \right) ,
    \label{e:spin1/2}
\end{equation}
from which we can derive the following upper bound for the magnitude of the spin-dependent odderon $e_T$:
\begin{equation}
    \frac{|\bm{k}|}{M} \,|e_T| \leq e .
    \label{e:spinhalfsmallx}
\end{equation}
Note that upon omitting the functions related to tensor polarization, the bounds~\eqref{e:ebound1} and~\eqref{e:ebound2} reduce to a bound that is consistent with but less strict than~\eqref{e:spinhalfsmallx}. Aside from diagonalizing higher-dimensional minors to sharpen bounds~\eqref{e:ebound1} and~\eqref{e:ebound2}, we could also obtain~\eqref{e:spinhalfsmallx} by applying the small-$x$ limit to bounds~\eqref{e:boundMR1}--\eqref{e:boundMR5} for spin-$\tfrac{1}{2}$ hadrons.

\section{The McLerran-Venugopalan model} \label{s:MV_model}

Probed at increasingly smaller values of $x$, the gluon density of hadrons and nuclei rises. This implies that the description of an incoming parton scattering off a single gluon starts to fall short -- due to the high density of gluons, scattering off many gluons needs to be taken into account. Rather than describing gluons inside nuclei by PDFs, we could model the internal structure of nuclei by an effective color field called the color glass condensate (CGC). The term ‘glass’ refers to the state of ‘frozen disorder' as seen by a probe (due to time dilation), and the term ‘condensate’ refers to the very high occupation numbers of the gluon field. The CGC effective theory is the most widely accepted description of the gluon content of ultrarelativistic nuclei. This theory will be employed in the remainder of this chapter, in particular to calculate the expectation values of the pomeron and odderon operators in a nuclear environment in section~\ref{s:dipole_operator_CGC}. Reviews on the CGC formalism can be found in e.g.~\cite{Iancu:2002xk,Iancu:2003xm,Wessels:2009ona,Gelis:2010nm,Dominguez:2011saa,Kovchegov:2012mbw,Taels:2017shj}.

The starting point of the CGC picture is provided by the MV model~\cite{McLerran:1993ka,McLerran:1993ni,McLerran:1994vd} in which the large-$x$ partons (the valence quarks) are considered as static, acting as color charge sources for the smaller-$x$ gluons that constitute the color field. The typical size of fluctuations in the color charge density is characterized by the so-called saturation scale $Q_s$. As it turns out, this scale grows with increasing atomic number $A$ and with decreasing $x$; more specifically $Q_s^2 \propto A^{1/3} / x^\lambda$, with $\lambda \sim 0.2$-$0.3$~\cite{Kovchegov:2012mbw,Kovchegov:2014kua}. Hence, for sufficiently large nuclei (i.e.\ $A \gg 1$) and small values of $x$ (typically one requires $x \LessSim 0.01$) we have $Q_s \gg \Lambda_\text{QCD}$ and $\alpha_s(Q_s) \ll 1$. This implies that the small-$x$ gluon field of a large nucleus can be described by a \emph{classical} color field -- quantum corrections are suppressed by extra powers of $\alpha_s$~\cite{McLerran:1993ka,McLerran:1993ni}. This color field is given by the classical Yang-Mills equations of motion~\cite{Yang:1954ek}:\footnote{In the following discussion, several quantities ‘live' in the adjoint representation of $\text{SU}(3)$, e.g.\ we have $J^\nu \equiv J^{a\nu} T^a$.}
\begin{equation}
    D_\mu F^{\mu\nu} = J^\nu .
    \label{e:YM}
\end{equation}
For an ultrarelativistic nucleus moving along the plus direction, the source current $J$ only has a nonzero plus component:
\begin{equation}
    J^\nu(x^-,\bm{x}) = \delta^{\nu+} \rho(x^-,\bm{x}) ,
\end{equation}
where $\rho$ is the color charge density of the sources (due to current conservation this density is independent of $x^+$). 

Often CGC calculations are performed in a covariant gauge with the only nonzero field component given by $\alpha \equiv A^+$ (the minus component can be neglected in our field configuration); we will stick to this gauge in the remainder of this chapter. In this particular gauge, the solution to eq.~\eqref{e:YM} is given by~\cite{Iancu:2000hn}
\begin{equation}
    \nabla_{\!\bm{x}}^2 \,\alpha(x^-,\bm{x}) = - \rho(x^-,\bm{x}) .
\end{equation}
Upon inverting this relation, we obtain the color field in terms of the color sources:
\begin{equation}
    \alpha^a(x^-,\bm{x}) = \int d^2\bm{z} \;G(\bm{x}-\bm{z}) \,\rho^a(x^-,\bm{z}) ,
    \label{e:alpha_in_rho}
\end{equation}
where the Green's function associated to the two-dimensional Laplacian is given by
\begin{equation}
    G(\bm{x}-\bm{z}) = \frac{1}{4\pi} \ln\left( \frac{1}{(\bm{x}-\bm{z})^2 \Lambda^2} \right) .
    \label{e:Green's_fct}
\end{equation}
Here $\Lambda$ is an infrared cut-off scale, commonly taken to be $\Lambda_\text{QCD}$. 

The result in eq.~\eqref{e:alpha_in_rho} provides the classical color field that arises from a single configuration of color sources described by the density $\rho$. Even though the sources (three valence quarks per nucleon) are considered static in this model, their precise distribution can vary between collisions -- the density $\rho$ is \emph{arbitrary}. Hence, to calculate an observable quantity, we need to \emph{average} the corresponding operator over all possible color source configurations. This can be achieved by a functional average that is weighted by a suitable weight functional. In the (original) MV model, this weight functional is taken to be Gaussian in the color sources. The expectation value of an operator $O = O[\rho]$ (also referred to as a CGC average) can be calculated as
\begin{equation}
    \left\langle O \right\rangle_G = \int \mathcal{D}\rho \;w_G[\rho] \,O[\rho] ,
    \label{e:MV_model}
\end{equation}
provided the weight functional $w_G$ is properly normalized, i.e.\ $\int \mathcal{D}\rho \;w_G[\rho] = 1$. The Gaussian weight functional is given by
\begin{equation}
    w_G[\rho] = \exp\left( -\int d^2\bm{x} \;\frac{\delta^{ab} \rho^a(\bm{x}) \,\rho^b(\bm{x})}{2\mu_{\!A}} \right) , 
    \label{e:Gaussian_weight}
\end{equation}
with $\rho^a(\bm{x}) \equiv \int dx^- \,\rho^a(x^-,\bm{x})$, and $\mu_{\!A} = g^2 A/(2\pi R_A^2)$ represents the average color charge squared per unit area ($R_A$ denotes the nuclear radius). 

In general, the calculation of a functional average of an arbitrary product of sources can be expressed in terms of the average of a product of two sources,
\begin{equation}
    \left\langle \rho^a(\bm{x}) \,\rho^b(\bm{y}) \right\rangle_G = \mu_{\!A} \,\delta^{ab} \delta^{(2)}(\bm{x}-\bm{y}) .
\end{equation}
While the average of a product of an even number of sources can be obtained by summing over all possible pairings of two sources, the average of a product of an odd number of sources vanishes~\cite{Gelis:2001da}. As was first argued in~\cite{Hatta:2005as}, the latter implies that the expectation value of the odderon operator in the (original) MV model vanishes in a trivial way. In the next section we will explain how the functional averaging can be modified to properly account for odderon excitations in the CGC theory.

As mentioned earlier, the MV model is based on the separation between large-$x$ partons, the sources, and small-$x$ partons, the gluons. Naturally, an observable should not depend on what values of $x$ are called small or large with respect to some arbitrary reference scale. This requirement gives rise to the so-called Jalilian-Marian-Iancu-McLerran-Weigert-Leonidov-Kovner (JIMWLK) evolution equation~\cite{JalilianMarian:1997jx,JalilianMarian:1997gr,JalilianMarian:1997dw,Weigert:2000gi,Iancu:2000hn,Iancu:2001ad,Ferreiro:2001qy}. This is a complicated differential equation in $x$ for the weight functional, with the Gaussian weight $w_G$ serving as its initial condition. Evolution in $x$ does not preserve the Gaussian form and goes beyond the MV model (and beyond the scope of this thesis). In the remainder of this chapter, all CGC averages are (implicitly) considered at some small (but nonzero) longitudinal momentum fraction $x$.

\section{The dipole operator in a nuclear color glass condensate} \label{s:dipole_operator_CGC}

In section~\ref{s:pA_collisions} we will study dihadron production in $pA$ collisions through two incoming quarks from the proton that scatter off the small-$x$ gluon field of the nucleus, modeled by a CGC. As it turns out, the cross section of that process features a product of two dipole operators. The dipole operator corresponds to the $S$-matrix operator for a $q\bar{q}$ pair (a `color dipole') that scatters off a nucleus through infinitely many gluon exchanges; with the antiquark at position $\bm{x}$ and the quark at position $\bm{y}$, the dipole operator is given by 
\begin{equation}
    S(\bm{x},\bm{y}) \equiv \frac{1}{N_c} \tr \left( U_{\bm{x}}^\dag U_{\bm{y}}^{\vphantom{\dag}} \right) ,
    \label{e:dipole_operator_CGC}
\end{equation}
where
\begin{equation}
    U_{\bm{z}} \equiv U_{[-\infty,\infty;\bm{z}]}^n = \mathcal{P} \exp\left( ig \int_{-\infty}^{\infty} d\eta^- \alpha^a(\eta^-,\bm{z}) \,t^a \right) .
    \label{e:U_z} 
\end{equation}
The color gauge invariant version of the dipole operator is the Wilson loop operator $S^{[\Box]}$ given in eq.~\eqref{e:dipole_operator}. To facilitate the study of $pA$ collisions in the CGC theory, we will calculate in this section the CGC averages of the real (C-even) and imaginary (C-odd) parts of the dipole operator. In our covariant gauge, those pomeron and odderon contributions are respectively given by
\begin{align}
    \mathcal{P}(\bm{x},\bm{y}) &\equiv \frac{1}{2N_c} \tr \left( U_{\bm{x}}^\dag U_{\bm{y}}^{\vphantom{\dag}} + U_{\bm{y}}^\dag U_{\bm{x}}^{\vphantom{\dag}} \right) , \label{e:pomeron_operator_cg} \\[2pt]
    \mathcal{O}(\bm{x},\bm{y}) &\equiv \frac{1}{2iN_c} \tr \left( U_{\bm{x}}^\dag U_{\bm{y}}^{\vphantom{\dag}} - U_{\bm{y}}^\dag U_{\bm{x}}^{\vphantom{\dag}} \right) . \label{e:odderon_operator_cg}
\end{align}    

\begin{figure}[htb]
\centering
    \includegraphics[height=4.0cm]{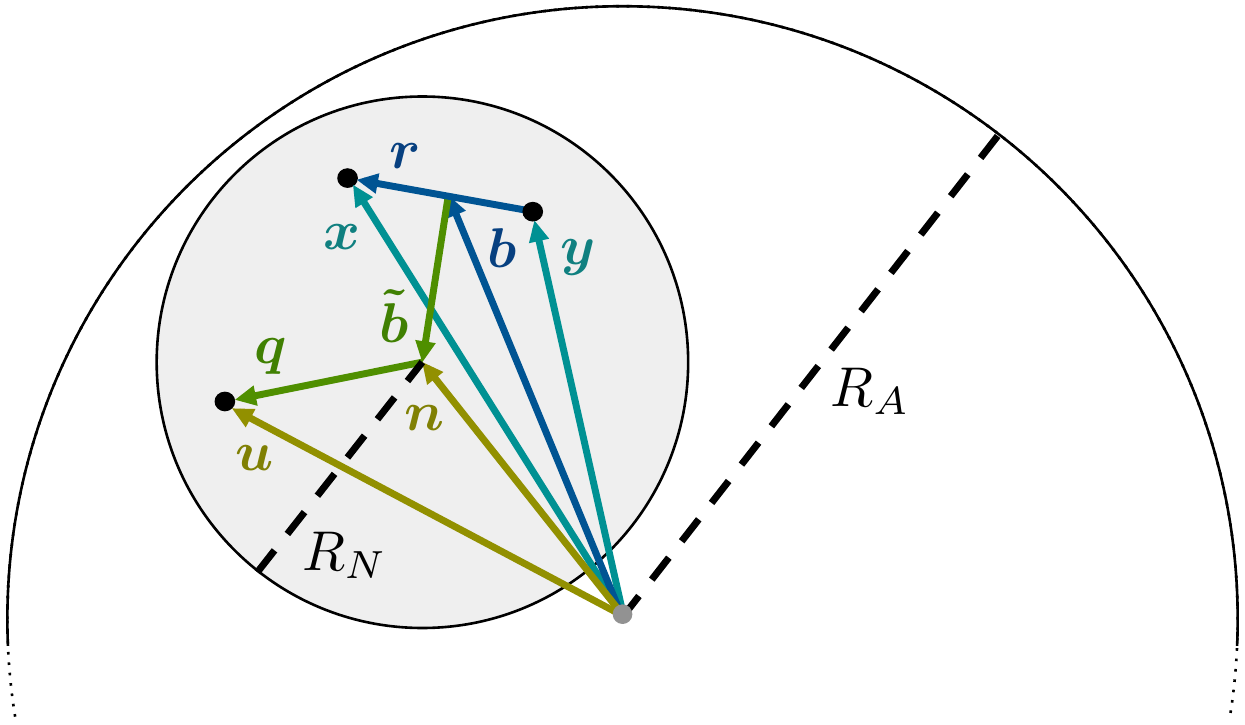}
    \caption{Geometrical picture of a dipole hitting a nucleon (that contains one valence quark) inside a nucleus.}
\label{f:nucleus}
\end{figure}

In the following calculations, we will deviate from the original MV model described in section~\ref{s:MV_model} in a number of ways. First of all, we will calculate the expectation values of the pomeron and odderon operators for a nuclear environment where the different nucleons are distributed in transverse space according to some profile function. The latter is in contrast with the original MV model where the nuclear profile is taken to be constant -- that assumption is reflected in the constantness of $\mu_A$ in eq.~\eqref{e:Gaussian_weight}. Since for our analysis in section~\ref{s:pA_collisions} it is important to allow for a nontrivial transverse structure of the nucleus, we perform the replacement $\mu_A \to \mu(\bm{u})$, with $\bm{u}$ the position of a color source with respect to the center of the nucleus. Another difference with the original MV model lies in the modeling of nucleons: for simplicity, each nucleon is modeled to contain one valence quark only. Finally, as already mentioned in the previous section, the odderon operator requires a modified weight functional. As it turns out, this modification amounts to adding an extra term to the exponent that is \emph{cubic} in the color charge density~\cite{Jeon:2005cf}. This extra term, that goes beyond the original MV model formulation, was derived in~\cite{Jeon:2004rk} and was shown to be subleading in the large-$A$ limit as compared to the Gaussian term. With all these modifications, the CGC average for a given operator $O = O[\rho]$ is given by
\begin{equation}
    \left\langle O \right\rangle = \int \mathcal{D}\rho \;w[\rho] \,O[\rho] .
    \label{e:MV_model_extended}
\end{equation}
The weight functional now reads
\begin{equation}
    w[\rho] = w_G[\rho] \,w_c[\rho] ,
\end{equation}
with the Gaussian and cubic weights respectively given by
\begin{align}
    w_G[\rho] &= \exp\left( -\int d^2\bm{n} \int \frac{d^2\bm{q}}{\pi R_N^2} \;\frac{\delta^{ab} \rho^a(\bm{u}) \,\rho^b(\bm{u})}{2\mu(\bm{u})} \right) , \\[3pt]
    w_c[\rho] &= \exp\left( \int d^2\bm{n} \int \frac{d^2\bm{q}}{\pi R_N^2} \;\frac{d^{abc} \rho^a(\bm{u}) \,\rho^b(\bm{u}) \,\rho^c(\bm{u})}{\kappa(\bm{u})} \right) ,
\end{align}
where, again, the weights are properly normalized. The quantities $\mu$ and $\kappa$ can be related to the nuclear profile function $T$ as follows:
\begin{equation}
    \mu(\bm{u}) = \frac{g^2}{2N_c} \,T(\bm{u}), \qquad \kappa(\bm{u}) = \frac{g^3}{N_c} \,T^2(\bm{u}) .
    \label{e:mu_kappa}
\end{equation}
If the nuclear profile were taken to be constant within the nucleus, the function $T$ would simply equal the average number of nucleons per unit of area, $A/(\pi R_A^2)$. Rather than integrating over $\bm{u} = \bm{n} + \bm{q}$ in the CGC weights, we instead average over the position $\bm{q}$ of the valence quark with respect to the center of the nucleon (we denote the radius of the nucleon by $R_N$), as well as integrate over all positions $\bm{n}$ of nucleons with respect to the center of the nucleus (note that at this point the two ways of writing the integrations are completely equivalent as the integrands depend on $\bm{u}$ only; the latter form will, however, prove more convenient in what follows). The geometrical picture is schematically illustrated in figure~\ref{f:nucleus}.

\subsection{The pomeron operator}   
In the following we will calculate the CGC averages of the real and imaginary parts of the dipole operator according to eq.~\eqref{e:MV_model_extended}. We start by providing the expectation value of the pomeron operator given in eq.~\eqref{e:pomeron_operator_cg}. The CGC average of this operator can be obtained by considering the dipole operator with a Gaussian weight (the cubic weight does not affect the real part). The expectation value of the pomeron operator is given by the well-known result~\cite{Mueller:1989st,Gelis:2001da,Kovchegov:2012ga}
\begin{align}
    \left\langle \mathcal{P}(\bm{x},\bm{y}) \right\rangle &= \left\langle S(\bm{x},\bm{y}) \right\rangle_G \nn \\
    &= \frac{1}{N_c} \int \mathcal{D}\rho \;w_G[\rho] \,\tr \left( U_{\bm{x}}^\dag U_{\bm{y}}^{\vphantom{\dag}} \right) \nn \\
    &= \exp \left( -\frac{1}{4} |\bm{r}|^2 Q_s^2(|\bm{b}|) \ln \left( \frac{1}{|\bm{r}| \Lambda} \right) \right) ,
    \label{e:dipole_CGC}
\end{align}    
where $\bm{b} \equiv (\bm{x}+\bm{y})/2$ represents the impact parameter of the dipole with respect to the center of the nucleus and $|\bm{r}| \equiv |\bm{x}-\bm{y}|$ is the dipole size (see also figure~\ref{f:nucleus}). The saturation scale $Q_s$ is defined in terms of the nuclear profile function $T$ as
\begin{equation}
    Q_s^2(|\bm{b}|) \equiv 2\pi \alpha_s^2 \,\frac{N_c^2-1}{N_c^2} \,T(|\bm{b}|) .
\end{equation}

\subsection{The odderon operator}
Next, we will calculate the expectation value of the odderon operator given in eq.~\eqref{e:odderon_operator_cg}. In the following we generalize the derivation of~\cite{Jeon:2005cf} by taking into account inhomogeneities in the transverse structure of the nucleus. Our final result matches with the result given in~\cite{Kovchegov:2012ga}, which was obtained in~\cite{Kovchegov:2003dm} following a diagrammatic approach. 

The CGC average of the odderon operator is given by
\begin{align}
    \left\langle \mathcal{O}(\bm{x},\bm{y}) \right\rangle &= \frac{1}{2iN_c} \int \mathcal{D}\rho \;w_c[\rho] \,w_G[\rho] \,\tr \left( U_{\bm{x}}^\dag U_{\bm{y}}^{\vphantom{\dag}} - U_{\bm{y}}^\dag U_{\bm{x}}^{\vphantom{\dag}} \right) \nn \\[3pt]
    &\approx \frac{d^{abc}}{2iN_c} \int d^2\bm{n} \,\frac{1}{\kappa(|\bm{n}|)} \int \frac{d^2\bm{q}}{\pi R_N^2} \int \mathcal{D}\rho \;\rho^a(\bm{u}) \,\rho^b(\bm{u}) \,\rho^c(\bm{u}) \,w_G[\rho] \nn \\
    &\quad\, \times \tr \left( U_{\bm{x}}^\dag U_{\bm{y}}^{\vphantom{\dag}} - U_{\bm{y}}^\dag U_{\bm{x}}^{\vphantom{\dag}} \right) ,
\end{align}  
where we have expanded the cubic weight up to first nontrivial order. Furthermore, we have used that $T(\bm{u}) \approx T(|\bm{n}|)$, i.e.\ we have neglected the difference between the location of a given nucleon and the location of a valence quark inside that nucleon and we have assumed rotational symmetry. From the functional derivative acting on the Gaussian weight,
\begin{equation}
    \delta^a w_G[\rho] \equiv \frac{\delta}{\delta \rho^a(u^-,\bm{u})} \;w_G[\rho] = -\frac{\rho^a(\bm{u})}{\mu(\bm{n})} \;w_G[\rho] ,
\end{equation}
it follows that
\begingroup 
\thinmuskip=2.4mu 
\medmuskip=3.2mu
\begin{align}
    \hspace{-0.2cm}\left\langle \mathcal{O}(\bm{x},\bm{y}) \right\rangle &= -\frac{d^{abc}}{2iN_c} \int d^2\bm{n} \;\frac{\mu^3(|\bm{n}|)}{\kappa(|\bm{n}|)} \int \frac{d^2\bm{q}}{\pi R_N^2} \int \mathcal{D}\rho \left( \delta^a \delta^b \delta^c \,w_G[\rho] \right) \tr \left( U_{\bm{x}}^\dag U_{\bm{y}}^{\vphantom{\dag}} - U_{\bm{y}}^\dag U_{\bm{x}}^{\vphantom{\dag}} \right) \nn \\[3pt]
    &= \frac{g^3}{16iN_c^3} \int d^2\bm{n} \;T(|\bm{n}|) \int \frac{d^2\bm{q}}{\pi R_N^2} \,\left\langle d^{abc} \delta^a \delta^b \delta^c \tr \left( U_{\bm{x}}^\dag U_{\bm{y}}^{\vphantom{\dag}} - U_{\bm{y}}^\dag U_{\bm{x}}^{\vphantom{\dag}} \right) \right\rangle_G ,
    \label{e:odderon_firststep}
\end{align}
\endgroup
where in the second step we used eq.~\eqref{e:mu_kappa} and integration by parts.
    
For convenience, we will now only consider the first term of eq.~\eqref{e:odderon_firststep} explicitly. At the end, the second term can be obtained from the first term by changing the overall sign and interchanging $\bm{x}$ and $\bm{y}$. Using the symmetry properties of $d$, we obtain
\begin{align}
    d^{abc} \delta^a \delta^b \delta^c \tr \left( U_{\bm{x}}^\dag U_{\bm{y}}^{\vphantom{\dag}} \right) &= d^{abc} \tr \left( (\delta^a \delta^b \delta^c U_{\bm{x}}^\dag) U_{\bm{y}}^{\vphantom{\dag}} + U_{\bm{x}}^\dag (\delta^a \delta^b \delta^c U_{\bm{y}}^{\vphantom{\dag}}) \right. \nn \\
    &\quad\, \left. + \,3 (\delta^a \delta^b U_{\bm{x}}^\dag) (\delta^c U_{\bm{y}}^{\vphantom{\dag}}) + 3 (\delta^a U_{\bm{x}}^\dag) (\delta^b \delta^c U_{\bm{y}}^{\vphantom{\dag}}) \right) .
    \label{e:fourterms}
\end{align}    
A single functional derivative acting on a Wilson line gives
\begin{equation}
    \delta^c U_{[a^-,b^-;\bm{z}]}^n = ig \int_{a^-}^{b^-} d\eta^- \,U_{[a^-,\eta^-;\bm{z}]}^n \left[ \delta^c \alpha(\eta^-,\bm{z}) \right] U_{[\eta^-,b^-;\bm{z}]}^n \,,
\end{equation} 
where, using eq.~\eqref{e:alpha_in_rho}, 
\begin{equation}
    \delta^c \alpha(\eta^-,\bm{z}) = \delta(\eta^- - u^-) \,G(\bm{z}-\bm{u}) \,t^c .
\end{equation}
Keeping in mind that
\begin{equation}
    \int_{-\infty}^{u^-} d\eta^- \,\delta(\eta^- - u^-) = \frac{1}{2} \,,
\end{equation}
we can now compute the first, second, and third order functional derivatives of a Wilson line:
\begingroup 
\thinmuskip=1.05mu 
\medmuskip=1.40mu
\begin{align}
    \hspace{-0.10cm}\delta^c U_{\bm{z}} &= ig \,G(\bm{z} - \bm{u}) \,U_{[-\infty,u^-;\bm{z}]}^n \,t^c \,U_{[u^-,\infty;\bm{z}]}^n \,, \\
    \hspace{-0.10cm}\delta^b \delta^c U_{\bm{z}} &= -\frac{g^2}{2} \,G^2(\bm{z} - \bm{u}) \,U_{[-\infty,u^-;\bm{z}]}^n \left( t^b t^c + t^c t^b \right) U_{[u^-,\infty;\bm{z}]}^n \,, \\
    \hspace{-0.10cm}\delta^a \delta^b \delta^c U_{\bm{z}} &= -\frac{ig^3}{4} \,G^3(\bm{z} - \bm{u}) \,U_{[-\infty,u^-;\bm{z}]}^n \left( t^a t^b t^c + t^a t^c t^b + t^b t^c t^a + t^c t^b t^a \right) U_{[u^-,\infty;\bm{z}]}^n \,.
\end{align}
\endgroup
With these results we proceed with the computation of the (Gaussian) expectation values of the four terms in eq.~\eqref{e:fourterms}:
\begin{align}
    d^{abc} \left\langle \tr \left( (\delta^a \delta^b \delta^c U_{\bm{x}}^\dag) U_{\bm{y}}^{\vphantom{\dag}} \right) \right\rangle_G &= \frac{g^3 (N_c^2-4)}{N_c} \frac{i(N_c^2-1)}{4N_c} \,G^3(\bm{x} - \bm{u}) \nn \label{e:gaussian_expvalue_-1} \\
    &\quad\, \times \left\langle \tr\left( U_{\bm{x}}^\dag U_{\bm{y}}^{\vphantom{\dag}} \right) \right\rangle_G , \\[5pt]
    d^{abc} \left\langle \tr \left( U_{\bm{x}}^\dag (\delta^a \delta^b \delta^c U_{\bm{y}}^{\vphantom{\dag}}) \right) \right\rangle_G &= -\frac{g^3 (N_c^2-4)}{N_c} \frac{i(N_c^2-1)}{4N_c} \,G^3(\bm{y} - \bm{u}) \nn \\
    &\quad\, \times \left\langle \tr\left( U_{\bm{x}}^\dag U_{\bm{y}}^{\vphantom{\dag}} \right) \right\rangle_G , \\[5pt]
    d^{abc} \left\langle \tr \left( 3 (\delta^a \delta^b U_{\bm{x}}^\dag) (\delta^c U_{\bm{y}}^{\vphantom{\dag}}) \right) \right\rangle_G &= -\frac{g^3 (N_c^2-4)}{N_c} \frac{3i}{2} \,G^2(\bm{x} - \bm{u}) \,G(\bm{y} - \bm{u}) \nn \\
    &\quad\, \times \left\langle \tr\left( t^a \,U_{[u^-,\infty;\bm{y}]}^n \,U_{[\infty,u^-;\bm{x}]}^n \right.\right. \nn \\[2pt]
    &\quad\, \times \left.\left. \!t^a \,U_{[u^-,-\infty;\bm{x}]}^n \,U_{[-\infty,u^-;\bm{y}]}^n \right) \right\rangle_G , \label{e:gaussian_expvalue_1} \\[5pt]
    d^{abc} \left\langle \tr \left( 3 (\delta^a U_{\bm{x}}^\dag) (\delta^b \delta^c U_{\bm{y}}^{\vphantom{\dag}}) \right) \right\rangle_G &= \frac{g^3 (N_c^2-4)}{N_c} \frac{3i}{2} \,G(\bm{x} - \bm{u}) \,G^2(\bm{y} - \bm{u}) \nn \\
    &\quad\, \times \left\langle \tr\left( t^a \,U_{[u^-,\infty;\bm{y}]}^n \,U_{[\infty,u^-;\bm{x}]}^n \right.\right. \nn \\[2pt]
    &\quad\, \times \left.\left. \!t^a \,U_{[u^-,-\infty;\bm{x}]}^n \,U_{[-\infty,u^-;\bm{y}]}^n \right) \right\rangle_G , \label{e:gaussian_expvalue_2}
\end{align}
where we used that
\begin{equation}
    d^{abc} t^a t^b t^c = \frac{(N_c^2-4)(N_c^2-1)}{4N_c^2} \,I , \qquad d^{abc} t^a t^b = \frac{N_c^2-4}{2N_c} \,t^c .
\end{equation}
The Gaussian CGC average appearing in eqs.~\eqref{e:gaussian_expvalue_1} and~\eqref{e:gaussian_expvalue_2} can be simplified to~\cite{Gelis:2001da}
\begin{align}
    &\left\langle t^a \,U_{[u^-,\infty;\bm{y}]}^n \,U_{[\infty,u^-;\bm{x}]}^n \,t^a \,U_{[u^-,-\infty;\bm{x}]}^n \,U_{[-\infty,u^-;\bm{y}]}^n \right\rangle_G \nn \\
    =& \,t^a \left\langle U_{[u^-,\infty;\bm{y}]}^n \,U_{[\infty,u^-;\bm{x}]}^n \right\rangle_G t^a \left\langle U_{[u^-,-\infty;\bm{x}]}^n \,U_{[-\infty,u^-;\bm{y}]}^n \right\rangle_G \nn \\
    =& \,t^a t^a \left\langle U_{\bm{x}}^\dag U_{\bm{y}}^{\vphantom{\dag}} \right\rangle_G ,
    \label{e:I2_simplified}
\end{align}
with $t^a t^a = (N_c^2-1)/(2N_c) \,I$. In the last step we used that both Gaussian averages contain $t^a t^a \propto I$~\cite{Gelis:2001da} and thus commute with all $t^a$. 

Employing eqs.~\eqref{e:Green's_fct}, \eqref{e:dipole_CGC}, \eqref{e:gaussian_expvalue_-1}--\eqref{e:gaussian_expvalue_2}, and~\eqref{e:I2_simplified}, the expectation value of the odderon operator in eq.~\eqref{e:odderon_firststep} becomes
\begin{align}
    \left\langle \mathcal{O}(\bm{x},\bm{y}) \right\rangle &= \frac{g^6 (N_c^2-4)}{8iN_c^4} \int d^2\bm{n} \;T(|\bm{n}|) \int \frac{d^2\bm{q}}{\pi R_N^2} \nn \\
    &\quad\, \times \tr \left( \frac{i(N_c^2-1)}{4N_c} \left[ G^3(\bm{x} - \bm{u}) - G^3(\bm{y} - \bm{u}) \right] \left\langle U_{\bm{x}}^\dag U_{\bm{y}}^{\vphantom{\dag}} \right\rangle_G \right. \nn \\
    &\quad\, \left. - \,\frac{3i}{2} \left[ G^2(\bm{x} - \bm{u}) \,G(\bm{y} - \bm{u}) - G(\bm{x} - \bm{u}) \,G^2(\bm{y} - \bm{u}) \right] t^a t^a \left\langle U_{\bm{x}}^\dag U_{\bm{y}}^{\vphantom{\dag}} \right\rangle_G \right) \nn \\[5pt]
    &= \frac{g^6 (N_c^2-4) (N_c^2-1)}{32N_c^5} \,\left\langle \tr\left( U_{\bm{x}}^\dag U_{\bm{y}}^{\vphantom{\dag}} \right) \right\rangle_G \,\int d^2\bm{n} \;T(|\bm{n}|) \nn \\
    &\quad\, \times \int \frac{d^2\bm{q}}{\pi R_N^2} \,\left[ G(\bm{x} - \bm{u}) - G(\bm{y} - \bm{u}) \right]^3 \nn \\[5pt]
    &= c_0 \alpha_s^3 \,\exp \left( -\frac{1}{4} |\bm{r}|^2 Q_s^2(|\bm{b}|) \ln \left( \frac{1}{|\bm{r}| \Lambda} \right) \right) \int d^2\bm{n} \;T(|\bm{n}|) \nn \\
    &\quad\, \times \int \frac{d^2\bm{q}}{\pi R_N^2} \,\ln^3 \left( \frac{|\bm{x}-\bm{n}-\bm{q}|}{|\bm{y}-\bm{n}-\bm{q}|} \right) ,
    \label{e:odderon_almostfinal}
\end{align}
where the color factor $c_0$ is defined as\footnote{This is the color factor given in~\cite{Kovchegov:2012ga}. Our calculation actually gives the slightly modified color factor $c_0' = 3c_0/N_c$. It is unclear to us where this difference comes from (for $N_c = 3$ the results are the same, of course).} 
\begin{equation}
    c_0 \equiv -\frac{(N_c^2-4)(N_c^2-1)}{12N_c^3} \,.
\end{equation}

Following the same steps as in~\cite{Kovchegov:2012ga}, we can perform the integration over $\bm{q}$ in eq.~\eqref{e:odderon_almostfinal}, which is constrained to the area of the nucleon. It was already noted in~\cite{Kovchegov:2003dm} that
\begin{equation}
    \int d^2\bm{q} \,\ln^3 \left( \frac{|\bm{x}-\bm{n}-\bm{q}|}{|\bm{y}-\bm{n}-\bm{q}|} \right) = 0 
\end{equation}
in case the integration would carry over the entire transverse plane. This result can be used to write
\begingroup 
\thinmuskip=1.2mu 
\medmuskip=1.6mu
\begin{equation}
    \int d^2\bm{q} \;\ln^3 \left( \frac{|\bm{x}-\bm{n}-\bm{q}|}{|\bm{y}-\bm{n}-\bm{q}|} \right) \,\theta(R_N-|\bm{q}|) = -\int d^2\bm{q} \;\ln^3 \left( \frac{|\bm{x}-\bm{n}-\bm{q}|}{|\bm{y}-\bm{n}-\bm{q}|} \right) \,\theta(|\bm{q}|-R_N) .
\end{equation}
\endgroup
The integral on the right-hand side can be performed if we expand the integrand in powers of $|\bm{x}-\bm{n}|/|\bm{q}|$ and $|\bm{y}-\bm{n}|/|\bm{q}|$. A Taylor expansion up to the second derivative of the logarithm with respect to $|\bm{q}|$ (this turns out to correspond to the first nontrivial order after integration), gives
\begin{equation}
    \ln\left( |\bm{q}-(\bm{x}-\bm{n})| \right) = \ln(|\bm{q}|) + \frac{1}{\bm{q}^2} \left[ \frac{(\bm{x}-\bm{n})^2}{2} - \bm{q} \cdot (\bm{x}-\bm{n}) \right] - \frac{\left( \bm{q} \cdot (\bm{x}-\bm{n}) \right)^2}{\bm{q}^4} \,,
\end{equation}
and likewise for the logarithm containing $\bm{y}$. At this order, the integral evaluates to
\begin{equation}
    \int d^2\bm{q} \,\ln^3 \left( \frac{|\bm{x}-\bm{n}-\bm{q}|}{|\bm{y}-\bm{n}-\bm{q}|} \right) \theta(R_N-|\bm{q}|) \approx -\frac{3\pi}{4R_N^2} \,|\bm{r}|^2 \,\bm{r} \cd \bm{\tilde{b}} ,
    \label{e:int_over_q}
\end{equation}
where we used that $\phi_{\bm{x}} \approx \phi_{\bm{y}} \approx \phi_{\bm{b}}$, as appropriate for a large nucleus. Furthermore, we have introduced the quantity $-\bm{\tilde{b}} \equiv \bm{b}-\bm{n}$, which represents the impact parameter of the dipole with respect to the center of the struck nucleon (see also figure~\ref{f:nucleus}). The result in eq.~\eqref{e:int_over_q} is a good approximation for $|\bm{r}| < R_N$ (which means that the dipole is `perturbatively small') and for $|\bm{\tilde{b}}| \ll R_N$. Substituting the result of the integration in eq.~\eqref{e:odderon_almostfinal}, we obtain
\begin{align}
    \left\langle \mathcal{O}(\bm{x},\bm{y}) \right\rangle &= - \frac{3}{4R_N^4} \,c_0 \alpha_s^3 \,|\bm{r}|^2 \exp \left( -\frac{1}{4} |\bm{r}|^2 Q_s^2(|\bm{b}|) \ln \left( \frac{1}{|\bm{r}| \Lambda} \right) \right) \nn \\[2pt]
    &\quad\, \times \bm{r} \cdot \!\int d^2\bm{\tilde{b}} \;\bm{\tilde{b}} \,T\left( |\bm{b}+\bm{\tilde{b}}| \right) .
    \label{e:odderon_final}
\end{align}
Without any further knowledge or assumptions on the nuclear profile function, this expression cannot be simplified further.

\section{Angular correlations in proton-nucleus collisions} \label{s:pA_collisions}

This section provides a possible explanation for certain angular correlations that have been observed in $pA$ collisions referred to as collectivity. After a brief introduction to collectivity, we will calculate a particular flow coefficient for dihadron production in $pA$ collisions that originates from odderon effects. To that end, we will employ the CGC results of the previous section.

\subsection{Collective behavior} \label{s:collective_flow}
Collective phenomena are common in physics and are characterized by collective behavior of a group of physical entities. A goal for theorists is to explain the observed behavior in terms of the fundamental degrees of freedom of the system. A well-known example of a collective phenomenon is collectivity in ultrarelativistic $AA$ collisions; it refers to the observed anisotropies in the angular distributions of final-state hadrons. This phenomenon is often studied through correlations between transverse momenta of produced particle pairs. A characteristic feature of collectivity is that for relative azimuthal angles $\Delta\phi \sim 0$ and $\pi$ the correlations extend over a long range in the relative pseudorapidity $\Delta\eta$. These `ridge-like' two-particle correlations can be characterized by Fourier harmonics in the relative azimuthal angle $\Delta\phi$ and are quantified by so-called flow coefficients. The term `flow' is used here as collectivity in $AA$ collisions can be explained by hydrodynamic flow models of strong interactions between final-state particles; for that reason this phenomenon is also referred to as collective flow.

\begin{figure}[htb]
\centering
    \includegraphics[height=4.7cm]{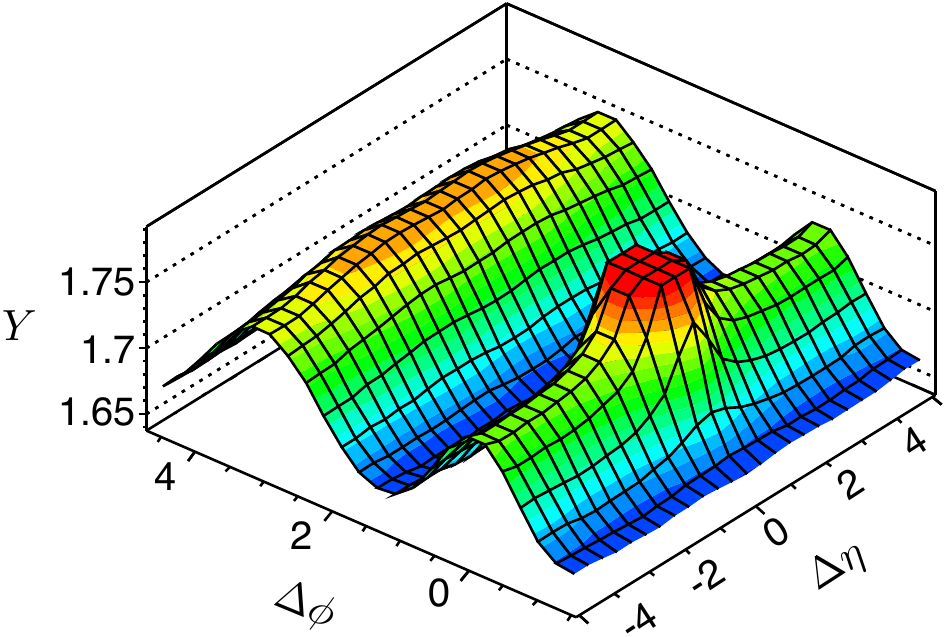}
    \caption{An illustration of ridge-like structures as measured by the ATLAS experiment at CERN. The two-particle correlation function $Y(\Delta\phi,\Delta\eta)$ is shown for proton-lead collisions with a CM energy of $\sqrt{s} = 5.02$ TeV~\cite{Aad:2014lta}.}
\label{f:ridges}
\end{figure}

Over the last decade, it has come as a surprise that the same azimuthal correlations as in $AA$ collisions (of the same order of magnitude) have also been observed in hadron production from $pp$ and $pA$ collisions at various experiments at the LHC~\cite{Abelev:2012ola,ABELEV:2013wsa,Abelev:2014mda,Adam:2015bka,Acharya:2018dxy,Aad:2012gla,Aad:2013fja,Aad:2014lta,Aad:2015gqa,Aaboud:2016yar,Aaboud:2017acw,Aaboud:2017blb,Khachatryan:2010gv,CMS:2012qk,Chatrchyan:2013nka,Khachatryan:2015lva,Khachatryan:2015oea,Khachatryan:2015waa,Khachatryan:2016txc,Sirunyan:2017igb,Sirunyan:2017uyl,Aaij:2015qcq} and RHIC~\cite{Adare:2013piz,Adare:2014keg,Belmont:2017lse,Adamczyk:2014fcx,Adamczyk:2015xjc}. An example of measured ridge-like correlations in $pA$ collisions is shown in figure~\ref{f:ridges}. Whether these observations, which concern much smaller colliding systems, can be correctly described by hydrodynamic flow models remains an issue of debate. Although they can provide reasonably accurate descriptions of the data~\cite{dEnterria:2010xip,Bozek:2010pb,Bozek:2012gr,Qin:2013bha,Werner:2013ipa,Bozek:2013ska,Bozek:2013uha,Bzdak:2014dia,Weller:2017tsr}, hydrodynamics might not be appropriate for such small systems~\cite{Dusling:2015gta}. For this reason, theorists have mainly focused on other ideas, particularly explanations offered by so-called initial-state models. These models do not rely on hydrodynamic approaches, but assume that the asymmetry is created at the moment of particle production due to correlations between gluons that exist within the nuclear CGC. This category of models has received a lot of support over recent years, see e.g.~\cite{Avsar:2010rf,Dumitru:2010iy,Kovner:2010xk,Levin:2011fb,Iancu:2011ns,Kovner:2011pe,Dusling:2012iga,Kovner:2012jm,Dusling:2013qoz,Kovner:2014qea,Dumitru:2014yza,Schenke:2015aqa,Altinoluk:2015dpi,Dusling:2015rja,Kovner:2015rna,Schenke:2016lrs,Rezaeian:2016szi,Dusling:2017aot,Dusling:2017dqg,Fukushima:2017mko,Altinoluk:2018hcu,Davy:2018hsl}. Perhaps collectivity in high-energy $pp$ and $pA$ collisions should be attributed to both initial- and final-state effects~\cite{Bzdak:2013zma}. The experimental and theoretical status of collectivity in small colliding systems is reviewed in~\cite{Dusling:2015gta}.

In the next subsection we will study azimuthal anisotropies in the production of two hadrons in $pA$ collisions. We will focus on generating nonzero odd-harmonic two-particle correlations entirely from initial-state effects, which has been studied less than generating even-harmonic correlations. Odd harmonics from initial-state effects have been studied in different settings in~\cite{Dumitru:2014dra,Dumitru:2014vka,Lappi:2015vha,Lappi:2015vta,Kovner:2016jfp,Kovchegov:2018jun}. The cross section for hadron production from an incoming quark that scatters off a nucleus is proportional to the Fourier transform of the expectation value of the dipole operator $S$. The real part (the pomeron) of $S$ produces even harmonics, while the imaginary part (the odderon) generates odd harmonics. We extend the work of~\cite{Hagiwara:2017ofm} and show that the odderon Wigner distribution at small $x$ can give rise to a nonzero $v_1$ flow coefficient that characterizes the so-called directed flow ($v_1$ is associated to the first odd-harmonic contribution to the cross section). Compared to previous literature, the source of odd azimuthal correlations in our model is the relative orientation between the impact parameter and transverse momentum of the produced particle for a nuclear CGC that is inhomogeneous in the transverse plane. The same mechanism was discussed in~\cite{Iancu:2017fzn} for the case of even harmonics and for single-inclusive hadron production. Although $v_1$ can arise from C-odd fluctuations in a rotationally noninvariant nucleus~\cite{Dumitru:2014dra}, we show that it can also arise from C-odd correlations without breaking of rotational symmetry.

\subsection[\stackunder{Directed flow in dihadron production}{}]{Directed flow in dihadron production} \label{s:directed_flow}
Let us consider dihadron production from high-energy $pA$ collisions, $pA \to H_1 H_2 X$, where the proton and nucleus are respectively moving along the minus and plus directions. At forward rapidities (in the direction of the proton), the produced hadrons probe the proton at large longitudinal momentum fractions, while the nucleus is probed at small longitudinal momentum fractions $x \ll 1$, see e.g.~\cite{Dominguez:2011wm,Marquet:2016cgx,Petreska:2018cbf}. This means that in the forward-rapidity region the proton is `dilute' and mainly consists of valence quarks, whereas the nucleus is characterized by a high gluon density that can be modeled by a CGC. Furthermore, it has been shown that at forward rapidities the cross section may be dominated by contributions from \emph{double parton scattering} rather than from single parton scattering~\cite{Strikman:2010bg,Stasto:2011ru,Lappi:2012nh}. For that reason, we will study contributions from the former and consider two quarks coming from the proton that scatter (independently) off the CGC system, producing two final-state hadrons. The distribution of a pair of quarks inside the proton can be described by the so-called (collinear) double parton distribution (DPD) $F_p(x_1,x_2,\bm{b}_1-\bm{b}_2)$, where $x_1$ and $x_2$ denote the longitudinal momentum fractions of the quarks and $\bm{b}_1$ and $\bm{b}_2$ denote their scattering positions with respect to the center of the nucleus (for more information on DPDs, see e.g.~\cite{Diehl:2017wew}). The description of `dilute-dense' collisions by a collinear PDF on the dilute proton side and a CGC (that includes transverse momentum effects) on the dense nucleus side, is called the hybrid approach~\cite{Dumitru:2005gt}. 

The transverse momenta of the final-state quarks are acquired through multiple rescatterings off the gluon field of the nucleus which are resummed into Wilson lines. The operator that describes the scattering of two quarks off the CGC at the level of the cross section is given by the product of two dipole operators, $S(\bm{x}_1,\bm{y}_1) \,S(\bm{x}_2,\bm{y}_2)$. From the physical interpretation of the dipole operator given at the beginning of section~\ref{s:dipole_operator_CGC}, it follows that the picture of two quarks scattering off a nucleus is mathematically equivalent to the scattering of two color dipoles off a nucleus, see e.g.~\cite{Kovchegov:2012mbw} for more details.

Assuming factorization, one can write the cross section for the production of two quarks with momenta $\bm{k}_1$ and $\bm{k}_2$ and rapidities $y_1$ and $y_2$ as (see e.g.~\cite{Lappi:2015vta}):\footnote{For simplicity we shall not consider explicitly the fragmentation of the produced partons into hadrons.}
\begingroup 
\thinmuskip=1.5mu 
\medmuskip=2.0mu
\begin{align}
    \frac{d\sigma^{pA}}{dy_1 dy_2 \,d^2\bm{k}_1 d^2 \bm{k}_2} &\propto \int d^2\bm{b}_1 \int d^2 \bm{b}_2 \;F_p(x_1,x_2,\bm{b}_1-\bm{b}_2) \int \frac{d^2\bm{r}_1}{(2\pi)^2} \;e^{- i\bm{k}_1 \cdot \bm{r}_1} \int \frac{d^2 \bm{r}_2}{(2\pi)^2} \;e^{- i\bm{k}_2 \cdot \bm{r}_2} \nn \\[2pt]
    &\quad\, \times \left\langle S\left(\bm{b}_1+\tfrac{\bm{r}_1}{2},\bm{b}_1-\tfrac{\bm{r}_1}{2}\right) \,S\left(\bm{b}_2+\tfrac{\bm{r}_2}{2},\bm{b}_2-\tfrac{\bm{r}_2}{2}\right) \right\rangle ,
    \label{e:crosssection_step1}
\end{align}
\endgroup
where $\bm{r}_1 \equiv \bm{x}_1-\bm{y}_1$ and $\bm{r}_2 \equiv \bm{x}_2-\bm{y}_2$ are the dipole orientations, and $\bm{b}_1 \equiv (\bm{x}_1+\bm{y}_1)/2$ and $\bm{b}_2 \equiv (\bm{x}_2+\bm{y}_2)/2$ are the corresponding impact parameters. We will work in the large-$N_c$ limit and consider only the leading contributions in $1/N_c$ to the azimuthal correlations. At large $N_c$, the expectation value of the product of traces factorizes (see e.g.~\cite{Lappi:2015vta}):
\begin{align}
    &\left\langle S\left(\bm{b}_1+\tfrac{\bm{r}_1}{2},\bm{b}_1-\tfrac{\bm{r}_1}{2}\right) \,S\left(\bm{b}_2+\tfrac{\bm{r}_2}{2},\bm{b}_2-\tfrac{\bm{r}_2}{2}\right) \right\rangle \nn \\[3pt]
    \approx &\left\langle S\left(\bm{b}_1+\tfrac{\bm{r}_1}{2},\bm{b}_1-\tfrac{\bm{r}_1}{2}\right) \right\rangle \left\langle S\left(\bm{b}_2+\tfrac{\bm{r}_2}{2},\bm{b}_2-\tfrac{\bm{r}_2}{2}\right) \right\rangle .
    \label{e:approx_for_S^2}
\end{align}
The neglected corrections have been shown to give rise to azimuthal asymmetries in~\cite{Lappi:2015vta,Dusling:2017aot,Dusling:2017dqg}; here we demonstrate that odd azimuthal asymmetries can arise already at leading power in $1/N_c$. In the large-$N_c$ limit, the cross section in eq.~\eqref{e:crosssection_step1} can be written as a convolution of two dipole Wigner distributions:
\begin{align}
    \frac{d\sigma^{pA}}{dy_1 dy_2 \,d^2\bm{k}_1 d^2 \bm{k}_2} &\propto \int d^2\bm{b}_1 \int d^2 \bm{b}_2 \;F_p(x_1,x_2,\bm{b}_1-\bm{b}_2) \nn \\
    &\quad\, \times xW(x,\bm{b}_1,\bm{k}_1) \,xW(x,\bm{b}_2,\bm{k}_2) ,
\label{e:sigma_DPS}
\end{align}
where the dipole Wigner distribution $W$ is defined in the CGC theory as
\begin{equation}
    xW(x,\bm{b},\bm{k}) \equiv \int \frac{d^2\bm{r}}{(2\pi)^2} \;e^{-i\bm{k} \cdot \bm{r}} \left\langle S\left(\bm{b}+\tfrac{\bm{r}}{2},\bm{b}-\tfrac{\bm{r}}{2}\right) \right\rangle .
    \label{e:Wigner}
\end{equation}
We refer to~\cite{Boer:2018vdi} for a motivation of this definition.
    
The source for azimuthal angular correlations (both even and odd) in the cross section of this process is the relative orientation between the transverse momenta of the produced particles and their impact parameters, which is encoded in the dipole Wigner distributions. To extract the angular correlations, we parametrize the Wigner distribution of eq.~\eqref{e:Wigner} in terms of the different harmonic contributions~\cite{Hatta:2016dxp}:
\begin{align}
    xW(x,\bm{b},\bm{k}) &= x\mathcal{W}_0(x,\bm{b}^2,\bm{k}^2) + 2\cos(\phi_{\bm{b}}-\phi_{\bm{k}}) \,x\mathcal{W}_1(x,\bm{b}^2,\bm{k}^2) \nn \\[1pt]
    &\quad + 2 \cos(2(\phi_{\bm{b}}-\phi_{\bm{k}})) \,x\mathcal{W}_2(x,\bm{b}^2,\bm{k}^2) + \,\ldots
    \label{e:WignerParametrization}
\end{align}
The so-called elliptic flow resulting from the elliptic Wigner distribution $\mathcal{W}_2$ has been studied in e.g.~\cite{Hatta:2016dxp,Hagiwara:2016kam,Zhou:2016rnt,Hatta:2017cte,Hagiwara:2017fye,Iancu:2017fzn,Hagiwara:2017ofm}. As explained in subsection~\ref{e:off-forward_WL} for the case of dipole GTMDs, the even and odd harmonics in eq.~\eqref{e:WignerParametrization} constitute, respectively, the pomeron and odderon contributions to the dipole Wigner distribution $W$. In this analysis we will focus on the odd-harmonic correlations that are generated by the odderon operator. We will only consider the first odd contribution explicitly, i.e.\ the one associated to the function $\mathcal{W}_1$, which we will refer to as the `odderon Wigner distribution'.

For simplicity, we assume that the dependence on $\bm{b}_1-\bm{b}_2$ in the DPD factorizes and we take a Gaussian ansatz for the transverse density profile (see e.g.~\cite{Blok:2010ge,Diehl:2011yj}):
\begin{equation}
    F_p(x_1,x_2,\bm{b}_1-\bm{b}_2) =  \frac{1}{4\pi R_N^2} \,f_p(x_1,x_2) \;e^{-\frac{(\bm{b}_1-\bm{b}_2)^2}{4 R_N^2}} .
\label{e:Proton_DPD}
\end{equation}
Using eqs.~\eqref{e:WignerParametrization} and~\eqref{e:Proton_DPD} in the cross section~\eqref{e:sigma_DPS}, and integrating over the angles $\phi_{\bm{b}_1}$ and $\phi_{\bm{b}_2}$, we obtain
\begin{align}
    \frac{d\sigma^{pA}}{dy_1 dy_2 \,d^2\bm{k}_1 d^2 \bm{k}_2} &\propto \frac{\pi}{8R_N^2} \,f_p(x_1,x_2) \int_0^\infty db_1^2 \int_0^\infty db_2^2 \;\,e^{-\frac{b_1^2+b_2^2}{4R_N^2}} \nn \\[2pt]
    &\quad\, \times \left[ 2 I_0\left(\frac{b_1b_2}{2R_N^2}\right) x\mathcal{W}_0(x,\bm{b}_1^2,\bm{k}_1^2) \,x\mathcal{W}_0(x,\bm{b}_2^2,\bm{k}_2^2) \right. \nn \\[2pt]
    &\quad\, \left. + \;4\cos(\phi_{\bm{k}_1}-\phi_{\bm{k}_2}) \,I_1\left(\frac{b_1b_2}{2R_N^2}\right) x\mathcal{W}_1(x,\bm{b}_1^2,\bm{k}_1^2) \,x\mathcal{W}_1(x,\bm{b}_2^2,\bm{k}_2^2) \right. \nn \\[2pt]
    &\quad\, \left. + \;4\cos(2(\phi_{\bm{k}_1}-\phi_{\bm{k}_2})) \,I_2\left(\frac{b_1b_2}{2R_N^2}\right) x\mathcal{W}_2(x,\bm{b}_1^2,\bm{k}_1^2) \,x\mathcal{W}_2(x,\bm{b}_2^2,\bm{k}_2^2) \right] \nn \\[2pt]
    &\quad\: + \,\ldots \,,
    \label{e:cross-section_final}
\end{align}
where $b_{1,2} \equiv |\bm{b}_{1,2}|$ and $I_n$ is the $n$th modified Bessel function of the first kind. We note that the odderon Wigner distribution $\mathcal{W}_1$ of the nucleus gives rise to an odd azimuthal correlation between the transverse momenta of the produced particles. 

The standard way of quantifying azimuthal correlations in particle production is through the flow coefficients $v_n$~\cite{Voloshin:1994mz}. For two-particle correlations, we have
\begin{equation}
    v_n(\bm{k},\bm{k}^\text{ref}) \equiv \frac{V_n(\bm{k},\bm{k}^\text{ref})}{\sqrt{V_n(\bm{k}^\text{ref},\bm{k}^\text{ref})}} \,,
    \label{e:vn}
\end{equation}
where the coefficients $V_n$ are obtained from the decomposition of the cross section into Fourier modes in the relative azimuthal angle between the produced particles, 
\begin{equation}
    \frac{d\sigma^{pA}}{dy_1 dy_2 \,d^2\bm{k}_1 d^2 \bm{k}_2} \propto 1 + \sum_{n=1}^\infty 2V_n(\bm{k}_1,\bm{k}_2) \cos(n(\phi_{\bm{k}_1}-\phi_{\bm{k}_2})) ,
    \label{e:FD}
\end{equation}
and $\bm{k}^\text{ref}$ is a reference momentum corresponding to an experimental reference bin. From eqs.~\eqref{e:cross-section_final} and~\eqref{e:FD} it follows that the first odd coefficient, which characterizes the directed flow, is given by
\begingroup 
\thinmuskip=2.1mu 
\medmuskip=2.8mu
\begin{equation}
    V_1(\bm{k}_1,\bm{k}_2) = \frac{\int_0^\infty db_1^2 \int_0^\infty db_2^2 \;e^{-\frac{b_1^2+b_2^2}{4R_N^2}} \,I_1\left(\frac{b_1b_2}{2R_N^2}\right) x\mathcal{W}_1(x,\bm{b}_1^2,\bm{k}_1^2) \,x\mathcal{W}_1(x,\bm{b}_2^2, \bm{k}_2^2)}{\int_0^\infty db_1^2 \int_0^\infty db_2^2 \;e^{-\frac{b_1^2+b_2^2}{4R_N^2}} \,I_0\left(\frac{b_1b_2}{2R_N^2}\right) x\mathcal{W}_0(x,\bm{b}_1^2,\bm{k}_1^2) \,x\mathcal{W}_0(x,\bm{b}_2^2,\bm{k}_2^2)} \,.
    \label{e:General_V1}
\end{equation}
\endgroup

In the following we will derive an explicit expression for $V_1$ in the CGC theory for small (but nonzero) $x$. We will make use of the CGC averages that we calculated in section~\ref{s:dipole_operator_CGC}. The expression for the angular-independent contribution, $\mathcal{W}_0$, can be obtained from the CGC average of the pomeron operator given in eq.~\eqref{e:dipole_CGC}:
\begin{align}
    x\mathcal{W}_0(x,\bm{b}^2,\bm{k}^2) &= \int \frac{d^2\bm{r}}{(2\pi)^2} \;e^{-i \bm{k} \cdot \bm{r}} \left\langle \mathcal{P}\left(\bm{b}+\tfrac{\bm{r}}{2},\bm{b}-\tfrac{\bm{r}}{2}\right) \right\rangle \nn \\[3pt]
    &= \frac{1}{2\pi} \int_0^\infty dr \,r \,J_0(k\,r) \,\exp\left(-\frac{1}{4} r^2 Q_s^2(b) \ln\left(\frac{1}{r \Lambda}\right) \right) ,
    \label{e:W0}
\end{align}
where $k \equiv |\bm{k}|$, $r \equiv |\bm{r}|$, and $J_n$ is the $n$th Bessel function of the first kind.

Next, we turn to the CGC calculation of the odderon Wigner function $\mathcal{W}_1$. This function can be obtained from the expectation value of the odderon operator given in eq.~\eqref{e:odderon_final}, which comes with the nuclear profile function $T = T\left(|\bm{b}+\bm{\tilde{b}}|\right)$. Since we consider angular asymmetries generated by the inhomogeneity of the nucleus in the transverse plane, which is naturally larger at the edge of the nucleus than at its center, our main focus is on \emph{peripheral} collisions. As $\tilde{b} \equiv |\bm{\tilde{b}}|$ is confined to the area of the nucleon, those type of collisions are characterized by $\tilde{b} \ll b$. Hence, we can expand the nuclear profile function in powers of $\tilde{b}/b$:
\begin{equation}
    T\left(|\bm{b}+\bm{\tilde{b}}|\right) = \left[1+ \tilde{b}^i\frac{\partial}{\partial b^i} + \frac{1}{2} \,\tilde{b}^i \tilde{b}^j\frac{\partial^2}{\partial b^i \partial b^j} + \frac{1}{3!} \,\tilde{b}^i \tilde{b}^j\tilde{b}^k\frac{\partial^3}{\partial b^i \partial b^j \partial b^k} + \ldots \right] T(b) .
    \label{e:T_expansion}
\end{equation}
Plugging this expansion (up to second order in $\tilde{b}/b$) in eq.~\eqref{e:odderon_final} and performing the integration over $\bm{\tilde{b}}$, we obtain:
\begin{align}
    \left\langle \mathcal{O}\left(\bm{b}+\tfrac{\bm{r}}{2},\bm{b}-\tfrac{\bm{r}}{2}\right) \right\rangle &=  - \frac{3\pi}{16} \,c_0 \alpha_s^3 r^2 \exp\left(-\frac{1}{4} r^2 Q_s^2(b) \ln\left(\frac{1}{r \Lambda}\right) \right) r^i \frac{\partial}{\partial b^i} \,T(b) \nn \\[5pt]
    &= - \frac{3\pi}{16} \,c_0 \alpha_s^3 r^3 \exp\left(-\frac{1}{4} r^2 Q_s^2(b) \ln\left(\frac{1}{r \Lambda}\right) \right) \nn \\[2pt]
    &\quad\, \times \cos(\phi_{\bm{b}} -\phi_{\bm{r}}) \,T'(b) .
    \label{e:odderon_Texp}
\end{align}
Note that only the linear term in the expansion survives the integration; it has given rise to a $\cos(\phi_{\bm{b}} -\phi_{\bm{r}})$ correlation.\footnote{The quadratic term in the expansion gives rise to a $\cos(2(\phi_{\bm{b}} -\phi_{\bm{r}}))$ correlation in the \emph{real} part of the dipole operator and gives rise to elliptic flow~\cite{Iancu:2017fzn}. The third derivative of the profile function, combined with higher-order corrections to eq.~\eqref{e:odderon_final}, can give rise to a nonzero $v_3$ flow coefficient.} From the Fourier transform of eq.~\eqref{e:odderon_Texp},
\begin{align}
    \int \frac{d^2\bm{r}}{(2\pi)^2} \;e^{-i \bm{k} \cdot \bm{r}} \left\langle \mathcal{O}\left(\bm{b}+\tfrac{\bm{r}}{2},\bm{b}-\tfrac{\bm{r}}{2}\right) \right\rangle &= \frac{3}{32} \,i c_0 \alpha_s^3 \cos(\phi_{\bm{b}}-\phi_{\bm{k}}) \,T'(b) \int_0^\infty dr \,r^4 \,J_1(k\,r) \nn \\[2pt]
    &\quad\, \times \exp\left(-\frac{1}{4} r^2 Q_s^2(b) \ln\left(\frac{1}{r \Lambda}\right) \right) ,
\end{align}
combined with eq.~\eqref{e:WignerParametrization}, it follows that the odderon Wigner distribution is given by
\begin{align}
    x\mathcal{W}_1(x,\bm{b}^2,\bm{k}^2) &= -\frac{3}{64} \,c_0 \alpha_s^3 \;T'(b) \int_0^\infty dr \,r^4 \,J_1(k\,r) \nn \\[2pt]
    &\quad\, \times \exp\left(-\frac{1}{4} r^2 Q_s^2(b) \ln\left(\frac{1}{r\Lambda}\right)\right) .
    \label{e:Wodd}
\end{align}
From this expression we infer that the C-odd interaction described by $\mathcal{W}_1$ vanishes for a nucleus that is homogeneous in the transverse plane as then $T'(b)=0$ (this is \emph{not} the case for the C-even exchange described by $\mathcal{W}_0$).  

With the results in eqs.~\eqref{e:W0} and~\eqref{e:Wodd} we can calculate the directed flow coefficient $v_1$, defined in eq.~\eqref{e:vn}. To numerically estimate its size in our CGC model, we need to assume a certain nuclear profile. For a rotationally symmetric nucleus, the nuclear profile function $T$ can be written as follows: 
\begin{equation}
    T(b) = \int_{-\infty}^\infty dz \;\rho_A \!\left( \sqrt{\bm{b}^2 + z^2} \,\right) ,
    \label{e:profile}
\end{equation}
where $\rho_A$ is the density of nucleons in the nucleus. For a large nucleus one typically takes a Woods-Saxon profile~\cite{Woods:1954zz},
\begin{equation}
    \rho_A(r) = \frac{N_A}{1+e^{(r-R_A)/\delta}} \,,
\end{equation}
with $R_A = (1.12 \;\text{fm}) A^{1/3}$ the nuclear radius, $\delta = 0.54$ fm the width of the `nuclear edge'~\cite{Iancu:2017fzn}, and $N_A$ is a normalization factor such that $\int d^3\vec{r} \;\rho_A(\vec{r}) = A$, the number of nucleons inside the nucleus. We will calculate $v_1$ for lead and copper nuclei with, respectively, $A=208$ and $A=63$, and we will use the following approximation for the nuclear profile function:
\begin{equation}
    T(b) \approx \beta \left( 1 - \frac{1}{1 + e^{-(b/b_0 - \gamma)}} \right) ,
\end{equation}
with $\beta = 2.13$ fm$^{-2}$ and $\gamma = 5.4$ for lead, and $\beta = 1.38$ fm$^{-2}$ and $\gamma = 3.5$ for copper, and $b_0 = 1$ fm. This form approximates the Woods-Saxon distribution sufficiently well for our purposes. To save computation time, we take $b_{1,2} \GtrSim R_N \sim 1$ fm and we only consider the leading terms in the asymptotic expansions of the Bessel functions $I_n \Big( \tfrac{b_1b_2}{2R_N^2} \Big)$. 

\begin{figure}[htb]
\centering
    \includegraphics[height=4.7cm]{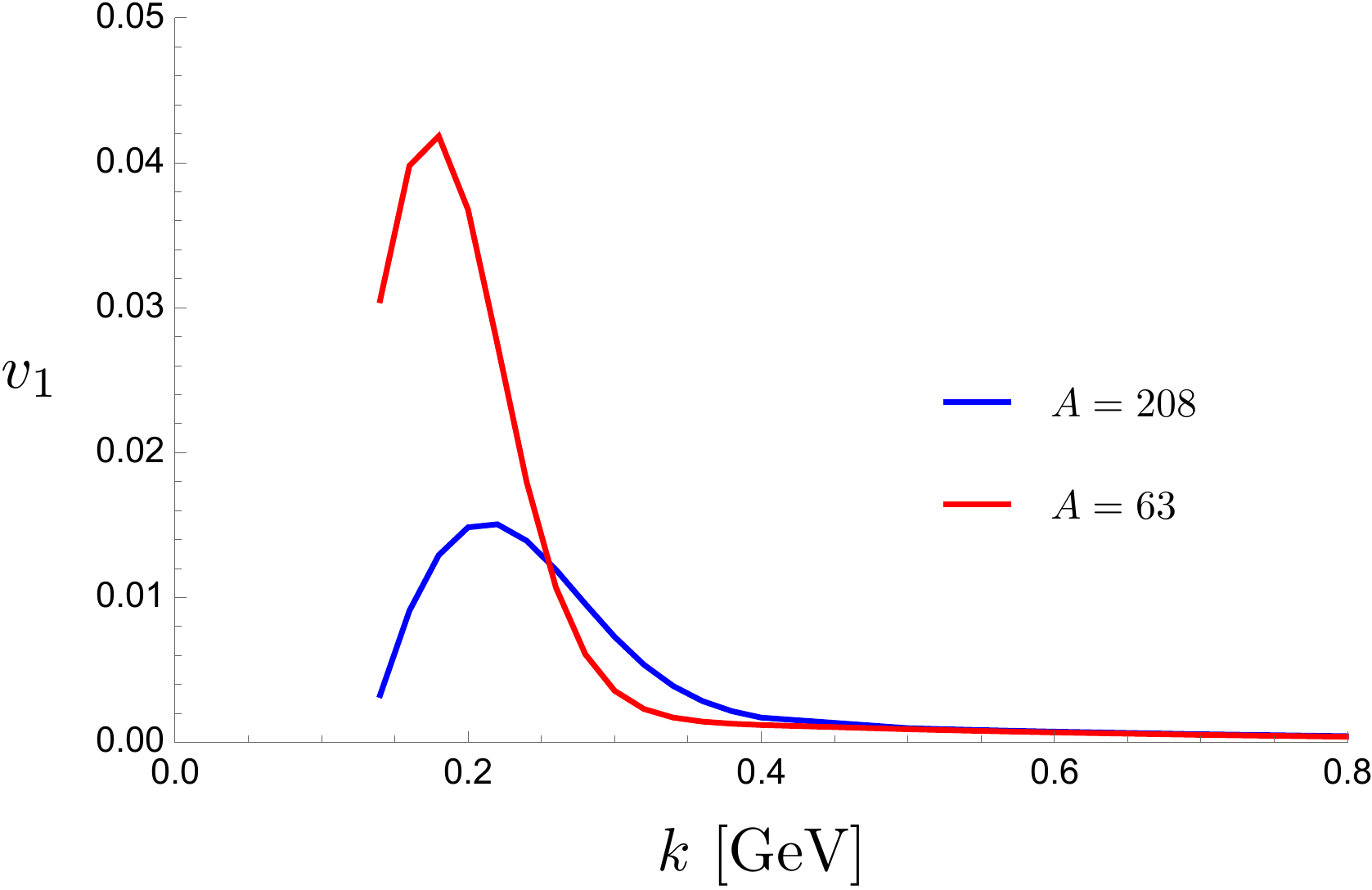}
    \caption{The directed flow $v_1(k)$ for a lead nucleus with $A=208$ (blue curve) and for a copper nucleus with $A=63$ (red curve). For the reference momentum we have taken $k^{\text{ref}} = 0.80$ GeV.}
\label{f:v1}
\end{figure}

In figure~\ref{f:v1} we have plotted $v_1$ as a function of $k$ for $k^{\text{ref}} = 0.80$ GeV (the behavior and magnitude of $v_1$ are similar for other choices of $k^{\text{ref}}$). We have taken $\alpha_s = 0.3$ and $\Lambda = \Lambda_\text{QCD} = 0.24$ GeV~\cite{Patrignani:2016xqp}. The two curves for lead and copper start at $k = 0.14$~GeV (unphysical fluctuations are present at lower values of $k$, where $v_1$ should go to zero as $k \to 0$). A word of caution is required, however, for small values of $k$: we expect the calculation to break down for small momenta $k < \Lambda_\text{QCD}$ as our result only applies to perturbatively small dipoles for which $r<R_N \sim 1/\Lambda_\text{QCD}$. 

Our numerical estimate for lead with $A=208$ can be compared to a $v_1$ measurement by the ATLAS experiment at CERN~\cite{Aad:2014lta} (see also figure~\ref{f:ridges}), keeping in mind that their observable does not correspond exactly to ours. We conclude that the sign change observed in~\cite{Aad:2014lta} is not present in our result and that the behavior at large $k$ is different.\footnote{Note that, as opposed to the analysis in~\cite{Aad:2014lta}, we have \emph{not} included the sign function $\text{sgn}(k^{\text{ref}}-k^0)$ in the definition of $v_1$ (the sign function is defined to be negative for $k^{\text{ref}} < k^0 = 1.5$ GeV and is positive otherwise).} However, our goal here is \emph{not} to describe the data, but rather to illustrate that our simple model can produce a nonzero $v_1$ that is of the right order of magnitude. Indeed, we find that the magnitude of $v_1$ for $k \sim 0.2$ GeV is of the same order as the maximal value observed in the data for $k \LessSim 1$ GeV~\cite{Aad:2014lta}. We emphasize that we have not made use of any fit parameters in our analysis.

With regards to the dependence on $A$, we note that $v_1$ decreases as $A$ increases. This observation is not surprising given that we have computed $\mathcal{W}_1$ in the MV model \emph{extended} by a cubic weight. As already mentioned in section~\ref{s:dipole_operator_CGC}, the cubic weight is subleading in $1/A$ compared to the Gaussian weight (which has been used to calculate $\mathcal{W}_0$) and therefore our observable is suppressed in the large-$A$ limit. Note that the precise dependence of our result on $A$ is not straightforward to extract from our expressions due to the nontrivial dependence on $Q_s$.

Finally, let us comment on particle production at more central rapidities. We have derived the directed flow coefficient for the production of two hadrons by two incoming \emph{quarks} from the proton that scatter off a nuclear CGC. This description is valid at forward rapidity, such that the proton consists of valence quarks only. Moving away from the forward-rapidity region, one also needs to account for the presence of sea quarks and antiquarks inside the proton. The scattering of an antiquark off a nucleus is given by the complex conjugate of the quark's dipole operator, which implies that $\mathcal{W}_1^q = -\mathcal{W}_1^{\bar{q}}$. Hence, if one assumes that away from the valence region the proton consists of an equal number of quarks and antiquarks with exactly the same DPDs for $qq$, $\bar{q}\bar{q}$, and $q\bar{q}$ pairs, then the sum over all parton pairs that could scatter off a nucleus would not give any odd harmonics.\footnote{The scattering of gluons off a nuclear CGC does not contribute to odd harmonics in our model as the \emph{adjoint} dipole operator is purely real.} However, it is known that for values of $x \LessSim 10^{-1}$, the DPDs \emph{differ} for different pairs of partons with an $x$-dependent width in the Gaussian model~\cite{Diehl:2004cx,Diehl:2014vaa}. In general, the proton DPD takes the following form:
\begin{equation}
    F_p(x_1,x_2,\bm{b}_1-\bm{b}_2) = \frac{1}{4\pi h_{ab}(x_1,x_2)} \,f_{ab}(x_1,x_2) \;e^{-\frac{(\bm{b}_1-\bm{b}_2)^2}{4 h_{ab}(x_1,x_2)}} ,
    \label{e:Proton_DPD_xDependent}
\end{equation}
where the labels $a,b$ refer to quarks or antiquarks. The precise form of $h_{ab}$, including numerical values for the relevant parameters, can be found in~\cite{Diehl:2014vaa}. Because the proton DPDs differ for different combinations of quarks and antiquarks, our mechanism would \emph{also} produce nonzero odd azimuthal correlations away from the forward-rapidity region.

\chapter[\stackunder{Summary and conclusions}{}]{Summary and conclusions} \label{c:Conclusions}
\addthumb{\thechapter}{\Large{\thechapter}}{\thumbfg}{\thumbbg}

In this thesis we have studied multidimensional quark and gluon PDFs as probes of the internal structure of hadrons. In high-energetic collisions involving hadrons, multidimensional PDFs are often related to asymmetries in the angular distributions of the produced particles. Thus, measuring a certain angular harmonic contribution to a cross section can imply the presence of certain partonic correlations inside hadrons. More knowledge on the rich dynamics of the hadronic constituents can bring us closer to understanding the origins of for example hadron masses and spin angular momenta. Moreover, it can allow for more accurate descriptions of processes in which QCD effects play an important role. 

In chapter~\ref{c:PDFs} we introduced the quark-quark correlator as the Fourier transform of a bilocal hadronic matrix element containing two quark fields. This quantity forms the starting point for any study on multidimensional PDFs. Correlators can depend on different sets of kinematical variables, depending on which projection or limit is considered. The most general correlator that we have studied is the GTMD correlator that features an off-forward matrix element; it depends on the variables $x$, $\bm{k}$, and $\Delta$. In the forward limit ($\Delta \to 0$), the GTMD correlator reduces to the TMD correlator which depends on $x$ and $\bm{k}$. Upon integration over the partonic transverse momentum $\bm{k}$, the TMD correlator reduces to the collinear correlator that depends only on the longitudinal momentum fraction $x$. 

Since correlators encode nonperturbative physics describing the dynamics of quarks and gluons inside hadrons, they cannot be calculated using perturbation theory. Instead, they are typically parametrized in terms of a priori unknown functions, the PDFs, that can be fitted to data. In general, correlators satisfy certain field theoretical constraints that can limit the number of allowed structures in such parametrizations or impose constraints on the PDFs. Both parton and hadron spin effects give rise to a wealth of different TMDs and play a central role in TMD phenomenology. For example, the quark-quark TMD correlator, constrained by Hermiticity, parity, and time reversal requirements, can be parametrized (at leading twist) in terms of two, six, and ten quark TMDs for unpolarized, vector polarized, and tensor polarized hadrons, respectively.

An important ingredient of any hadronic correlator is the gauge link, a path ordered exponential of the gluon field that ensures color gauge invariance. The presence of a gauge link in correlators has important consequences for TMDs, GTMDs, and Wigner distributions that all depend on the partonic transverse momentum. For those functions, in contrast to for example collinear PDFs, the integration paths of the associated gauge links are not unique but depend on the process under consideration. Consequently, those functions themselves become nonuniversal. The most well-known example of the nonuniversality of TMDs is the expected sign change for T-odd quark TMDs between DY and SIDIS, which results from the fact that these processes come with past-pointing ($[-]$) and future-pointing ($[+]$) gauge links, respectively. Experimental verification of this sign flip is ongoing and forms a vital test of the TMD formalism.


In part~\ref{p:1} we looked in detail at the DY process, which is for example essential for the description of $pp$ collisions at the LHC. We considered the DY cross section differential in the transverse momentum of the produced lepton pair (in general, the less inclusive a process, the more sensitive it is to multidimensional PDFs). For small transverse momenta, the DY cross section satisfies a TMD factorization theorem, which means that it can be factorized in terms of a perturbatively calculable hard scattering factor and two TMDs. Factorization formulas provide the foundation for QCD phenomenology. TMD factorization for DY scattering was established by Collins, Soper, and Sterman (CSS) at leading power in the small parameter $\Lambda/Q$ and to all orders in perturbation theory. In chapter~\ref{c:CSS} we reviewed various elements of their proof.

The CSS proof is based on a momentum regions analysis. Graphs can receive leading-power contributions from loop momenta with hard, collinear, soft, or Glauber scalings. The latter type of scalings represents virtual modes characterized by $|\ell^+\ell^-| \ll \bm{\ell}^2 \ll Q^2$. To reconstruct a full graph from its leading-power contributions, the leading regions need to be summed over and overlap contributions can be accounted for by the Collins subtraction method. The DY graphs can be organized in terms of subgraphs that each characterize a particular momentum scaling. The goal of the factorization procedure is to prove that those subgraphs, that are a priori connected through soft and collinear gluon exchanges, become disconnected (at leading power) upon summing over all contributions. By virtue of Ward identities, the hard, collinear, and soft subgraphs factorize upon summing over all possible soft and collinear gluon attachments; the Grammer-Yennie approximation plays an essential role in this. The latter approximation, however, is not valid for Glauber modes. To establish factorization, it needs to be shown that Glauber effects ultimately cancel out or can be captured by other region contributions. Through suitable contour deformations, some (but not all) Glauber exchanges can be absorbed into soft or collinear regions. As it turns out, upon summing over all possible final-state cuts all Glauber modes become untrapped, allowing for contour deformations into different momentum regions; the cancellation of final-state poles that is responsible for this is related to unitarity of the theory. This means that all Glauber partons can eventually be absorbed into soft or collinear regions where Grammer-Yennie approximations can be applied. Hence, upon summing over all graphs, cuts, and leading regions, the DY cross section factorizes into two TMD parton densities (these arise from combining soft and collinear subgraphs) and a hard scattering factor.


Over recent years, there has been debate about possible color-entanglement effects in DY that arise from double T-odd contributions. These effects would go beyond the expected sign flip between DY and SIDIS and would imply a loophole in the CSS proof for T-odd TMDs that involve polarization. An example of a double T-odd contribution to the DY cross section is the $\cos(2\phi)$ angular modulation associated to two BM functions (the dBM effect). As color-entanglement effects could lead to modified color factors in the (factorized) cross section, or even to a breakdown of factorization, we have reexamined this issue in detail. 

By largely following the steps of the CSS factorization proof, we investigated in chapter~\ref{c:Disentanglement} by an explicit model calculation whether color-entanglement effects persist in the dBM contribution to the cross section after properly summing over all relevant graphs, cuts, and leading regions. This study was performed at the lowest order in perturbation theory for which color-entanglement effects have been argued to be present -- this is $\mathcal{O}(\alpha_s^2)$, corresponding to the level of two gluon exchanges. At this order we identified various graphs that contribute to the dBM term and determined four common leading regions of loop momenta involving collinear and Glauber scalings only. Upon summing over all contributions, we find that color-entanglement effects completely cancel on a region-by-region basis; the final result conforms with the predictions of the CSS factorization theorem. Furthermore, the Glauber region turns out to play an important role: depending on the choice of rapidity regulators, it is possible to assign the full contribution to the dBM term at the considered order to the region in which the two gluons have Glauber scalings.

Similarly to the CSS proof, the two mechanisms that lie at the heart of our findings are the non-Abelian Ward identity and the unitarity cancellation of final-state poles. Furthermore, following the summation over all possible final-state cuts, the Glauber gluons can be absorbed into other region contributions, consistent with the CSS works. More specifically, we find that the Glauber momenta can be deformed into the collinear regions. Although our model calculation was performed at $\mathcal{O}(\alpha_s^2)$, we have no reasons to doubt that the same physical principles also drive the disentanglement of different color structures at higher orders in perturbation theory. Finally, we expect our results to go beyond our model setup and to apply also to DY in QCD. 


In part~\ref{p:2} of this thesis we studied multidimensional gluon distributions. Compared to the quark case, the extraction of gluon functions remains challenging. In this respect the possible construction of the EIC is very promising, as it could lead to unraveling the gluon content of (polarized) hadrons by measuring a wide variety of gluon observables in the small-$x$ region where gluons dominate over quarks. Today, no gluon TMDs other than the unpolarized one have been fitted to data yet. 

To facilitate the study of gluon observables, we provided in chapter~\ref{c:Polarized_gluon_TMDs} an instructive parametrization of the gluon-gluon TMD correlator in terms of leading-twist gluon TMDs that appear in the description of spin-$0$, spin-$\tfrac{1}{2}$, and spin-$1$ hadrons. The ones related to tensor polarization are new. As from the experimental side the gluon functions are largely unknown, we
have also derived model-independent inequalities for combinations of TMDs that follow from the operator structure of the correlator. These bounds may be used to estimate maximal contributions of certain functions to observables.


In chapter~\ref{c:Small_x} we focused on gluon distributions at small $x$. We only looked at one particular gauge link structure, namely the dipole-type structure $[+,-]$. We have shown that for this gauge link structure the gluon-gluon GTMD correlator is related to the correlator of a Wilson loop ($[\Box]$) in the limits of small $x$ and $\xi$. For unpolarized hadrons, the off-forward Wilson loop correlator can be parametrized in terms of a single GTMD. The fact that the gluon-gluon GTMD correlator is parametrized in terms of four GTMDs implies that at vanishing $x$ and $\xi$ these functions all become related to each other. Hence, one expects that when hadrons are probed at very high energies the picture of gluon GTMDs becomes very simple. By taking the forward limit of these results, we arrive at similar conclusions for dipole-type TMDs: for each type of hadron polarization we find that in the small-$x$ limit dipole-type TMDs either vanish or become proportional to each other. The nonvanishing gluon TMDs and GTMDs are expected to behave as $1/x$, or a slightly modified power after resummation of large logarithms in $1/x$ from $\alpha_s$ corrections. 

For transversely polarized hadrons we identified a C-odd TMD contribution ($e_T$), which has been referred to as the spin-dependent odderon. We also derived an upper bound for this function. In contrast to the TMD case, the dipole-type GTMDs also contain odderon contributions in the unpolarized situation; these contributions originate from correlations between the off-forwardness and the gluon momentum in the transverse plane. The same is true for dipole Wigner distributions, with the off-forwardness replaced by its Fourier conjugate, the impact parameter $\bm{b}$. 

Finally, we considered partonic correlations within the nuclear wave function as sources for the observed collective behavior in ultrarelativistic $pA$ collisions. The high density of gluons in the nuclei can be modeled by a CGC, which is an effective color field generated by static color sources (the valence quarks). In the MV model, the expectation value of an operator in the CGC theory can be obtained by performing a functional average over all possible configurations of color sources in the nucleus. For dihadron production through two incoming quarks from the proton that scatter off a nuclear CGC, the cross section is proportional to a convolution of two dipole Wigner distributions. These distributions can give rise to both even and odd harmonics in the two-particle angular correlations, the latter being due to odderon effects. Within a simple CGC model (with a cubic weight functional) and in the large-$N_c$ limit, we have calculated the directed flow coefficient $v_1$ that quantifies the contribution of the first odd harmonic with respect to the angular-independent contribution. For peripheral collisions involving a lead nucleus described by a Woods-Saxon-like profile, we find a $v_1$ at the percent level, which is the same order of magnitude as is observed experimentally at the LHC. The shape of $v_1$, however, is not well-described by our model as it is probably not realistic enough due to many simplifying assumptions.

\begin{appendix}
\chapter[\stackunder{Conventions}{}]{Conventions} \label{a:conventions}
\addthumb{\thechapter}{\Large{\thechapter}}{\thumbfg}{\thumbbg}

Below we list the most important conventions that are used throughout this thesis:

\begin{itemize}
\item We adopt natural units where $\hbar = c = 1$. With this choice, mass is the only dimension that is left: $[\text{mass}] = [\text{energy}] = [\text{length}]^{-1} = [\text{time}]^{-1}$, where the square brackets denote the mass dimension.

\item Spacetime components are labeled by Greek indices and the roman indices $i,j,k,l = 1,2$ refer to transverse components only. For Dirac indices we use the Greek letters $\alpha,\beta = 1,2,3,4$, and the roman indices $a,b,c = 1,2,\ldots,8$ label the generators of the $\text{SU}(3)$ Lie algebra, unless explicitly stated otherwise.

\item For the Minkowski metric $g$ we adopt the `mostly minus' convention, i.e.\ its nonzero components read $g^{00} = - g^{11} = - g^{22} = - g^{33} = 1$.

\item A four-vector $a$ can be represented in terms of its light-cone components as $a = (a^+,a^-,\bm{a})$, where $a^\pm \equiv (a^0 \pm a^3)/\sqrt{2}$ and $\bm{a} \equiv (a^1, a^2)$. A dot product between two four-vectors $a$ and $b$ is given by $a \cd b = a^+b^- + a^-b^+ - \bm{a} \cd \bm{b}$.

\item The lightlike vectors $\bar{n}$ and $n$ are defined as $\bar{n} \equiv (1,0,\bm{0})$ and $n \equiv (0,1,\bm{0})$, such that $n \cd \bar{n} = 1$. Any four-vector $a$ can be decomposed in terms of $\bar{n}$ and $n$ as $a^\mu = a^+ \bar{n}^\mu + a^- n^\mu + a_\sst^\mu$, where we have defined $a_\sst \equiv (0,0,\bm{a})$ (note that $a_\sst^2 = -\bm{a}^2$).

\item Given two four-vectors $a$ and $b$, their symmetrized product is defined as $a^{\{\mu} \,b^{\nu\}} \equiv a^\mu b^\nu + a^\nu b^\mu$ and their antisymmetrized product as $a^{[\mu} \,b^{\nu]} \equiv a^\mu b^\nu - a^\nu b^\mu$.

\item For a four-vector $a$ and a rank-two tensor $b$ we define $\bar{a}^\mu \equiv \delta^{\mu\nu} a_\nu$ and $\bar{b}^{\mu\nu} \equiv \delta^{\mu\rho} \delta^{\nu\sigma} b_{\rho\sigma}$.

\item The Levi-Civita symbol $\epsilon$ is fixed by taking $\epsilon^{0123} = \epsilon^{-+12} = 1$. Furthermore, we employ the notation $\epsilon^{abcd} \equiv \epsilon^{\mu\nu\rho\sigma} a_\mu b_\nu c_\rho d_\sigma$.

\item The metric tensor in transverse space is defined as $g_\sst^{\mu\nu} \equiv g^{\mu\nu} - \bar{n}^{\{\mu} n^{\nu\}}$ and has nonzero components $g_\sst^{11} = g_\sst^{22} = -1$ (note that $g_\sst^{ij} = -\delta_\sst^{ij}$).

\item The Levi-Civita symbol in the transverse plane is given by $\epsilon_\sst^{\mu\nu} \equiv \epsilon^{\mu\nu-+}$ with nonzero components $\epsilon_\sst^{12} = -\epsilon_\sst^{21} = 1$. We also define $\widetilde{a}_\sst^{\nu} \equiv \epsilon_\sst^{\mu\nu} {a_\sst}_\mu$.

\item The Dirac matrices satisfy the anticommutation relation $\{ \gamma^\mu,\gamma^\nu \} = 2g^{\mu\nu} I$ (with $I$ the identity matrix) and the Hermiticity condition $(\gamma^\mu)^\dag = \gamma^0 \gamma^\mu \gamma^0$. We define a fifth matrix $\gamma^5 \equiv i\gamma^0 \gamma^1 \gamma^2 \gamma^3$ and the antisymmetric structure $\sigma^{\mu\nu} \equiv \tfrac{i}{2} [\gamma^\mu,\gamma^\nu]$.
\end{itemize}

\chapter[\stackunder{Symmetric traceless tensors}{}]{Symmetric traceless tensors} \label{a:STTs}

In this appendix we list the completely symmetric and traceless tensors $k_\sst^{i_1 \ldots i_n}$ that are built from the partonic momentum $k_\sst$. Up to rank $n=4$, these tensors are given by
\begin{align}
    k_\sst^{ij} &\equiv k_\sst^i k_\sst^j + \frac{1}{2} \bm{k}^2 g_\sst^{ij} , \\
    k_\sst^{ijk} &\equiv k_\sst^i k_\sst^j k_\sst^k + \frac{1}{4} \bm{k}^2 \left(
    g_\sst^{ij} k_\sst^k + g_\sst^{ik} k_\sst^j + g_\sst^{jk} k_\sst^i \right) ,
    \label{kT3}\\
    k_\sst^{ijkl} &\equiv k_\sst^i k_\sst^j k_\sst^k k_\sst^l + \frac{1}{6} \bm{k}^2 \left( g_\sst^{ij} k_\sst^{kl} + g_\sst^{ik} k_\sst^{jl} + g_\sst^{il} k_\sst^{jk} + g_\sst^{jk} k_\sst^{il} + g_\sst^{jl} k_\sst^{ik} + g_\sst^{kl} k_\sst^{ij} \right) \nn \\
    &\quad\, - \frac{1}{8} \bm{k}^4 \left( g_\sst^{ij} g_\sst^{kl} + g_\sst^{ik} g_\sst^{jl} + g_\sst^{il} g_\sst^{jk} \right) ,
\end{align}
satisfying
\begin{equation}
    {g_\sst}_{ij} \,k_\sst^{ij} = {g_\sst}_{ij} \,k_\sst^{ijk} = {g_\sst}_{ij} \,k_\sst^{ijkl} = 0 .
\end{equation}
Products of symmetric traceless tensors can be decomposed as follows:
\begin{align}
    k_\sst^i k_\sst^j &= k_\sst^{ij} - \frac{1}{2} \bm{k}^2 g_\sst^{ij} , \\
    k_\sst^i k_\sst^{jk} &= k_\sst^{ijk} - \frac{1}{4} \bm{k}^2 \left( g_\sst^{ij} k_\sst^k + g_\sst^{ik} k_\sst^{j} - g_\sst^{jk} k_\sst^i \right) , \\
    k_\sst^{ij} k_\sst^{kl} &= k_\sst^{ijkl} - \frac{1}{6} \bm{k}^2 \left( g_\sst^{ik} k_\sst^{jl} + g_\sst^{il} k_\sst^{jk} + g_\sst^{jk} k_\sst^{il} + g_\sst^{jl} k_\sst^{ik} - 2 g_\sst^{ij} k_\sst^{kl} - 2 g_\sst^{kl} k_\sst^{ij} \right) \nn \\
    &\quad\, + \frac{1}{8} \bm{k}^4 \left( g_\sst^{ik} g_\sst^{jl} + g_\sst^{il} g_\sst^{jk} - g_\sst^{ij} g_\sst^{kl} \right) .
\end{align}
The symmetric traceless tensor $k_\sst^{i_1 \ldots i_n}$ of rank $n \geq 1$ only has two independent components, which allows for the following decomposition:
\begin{equation}
    k_\sst^{i_1 \ldots i_n} \;\to\; \frac{|\bm{k}|^n}{2^{n-1}} \,e^{\pm in\varphi} ,
    \label{e:STT}
\end{equation}
in terms of the two real numbers $|\bm{k}|$ and $\varphi$, the polar coordinates of the transverse vector $k_\sst$.

\newpage
\stopthumb

\chapter[\stackunder{Feynman rules for spectator model}{}]{Feynman rules for spectator model} \label{a:feynman_rules}
\continuethumb

In this appendix we provide the Feynman rules for the scalar spectator model that is employed in chapter~\ref{c:Disentanglement}, including those for eikonal lines and vertices. In table~\ref{t:feynman_rules} we list the relevant rules for the model, adopting the Feynman gauge; they are given for both amplitudes and conjugate amplitudes.

\begin{center}
\renewcommand{\arraystretch}{2.1}
\begin{longtable}{|C L : C L|}
    \hline
    \multicolumn{2}{|c:}{\textbf{Amplitude}} & \multicolumn{2}{c|}{\textbf{Conjugate amplitude}} \\ \hline
    $\;\;\parbox{16.6mm}{\begin{fmfgraph*}(16.6,5)
    \fmfleft{i1}
    \fmfright{o1}
    \marrow{a}{down}{bot}{$p$}{i1,o1}
    \fmf{dashes}{i1,o1}
    \end{fmfgraph*}}$ & \;$\displaystyle \frac{i}{p^2 - m_s^2 + i\epsilon}$ &
    $\;\;\parbox{16.6mm}{\begin{fmfgraph*}(16.6,5)
    \fmfleft{i1}
    \fmfright{o1}
    \marrow{a}{down}{bot}{$p$}{i1,o1}
    \fmf{dashes}{i1,o1}
    \end{fmfgraph*}}$ & \;$\displaystyle -\frac{i}{p^2 - m_s^2 - i\epsilon}$ \\[10pt]

    $\;\parbox{15mm}{\begin{fmfgraph*}(15,5)
    \fmfleft{i1}
    \fmfright{o1}
    \marrow{b}{down}{bot}{$p$}{i1,o1}
    \fmf{fermion}{i1,o1}
    \end{fmfgraph*}}$ & \;$\displaystyle \frac{i(\slashed{p}+m_f)}{p^2 - m_f^2 + i\epsilon}$ &
    $\;\parbox{15mm}{\begin{fmfgraph*}(15,5)
    \fmfleft{i1}
    \fmfright{o1}
    \marrow{b}{down}{bot}{$p$}{i1,o1}
    \fmf{fermion}{i1,o1}
    \end{fmfgraph*}}$ & \;$\displaystyle -\frac{i(\slashed{p}+m_f)}{p^2 - m_f^2 - i\epsilon}$ \\ [10pt]

    $\;\parbox{15mm}{\begin{fmfgraph*}(15,5)
    \fmfleft{i1}
    \fmfright{o1}
    \marrow{c}{down}{bot}{$p$}{i1,o1}
    \fmf{fermion}{o1,i1}
    \end{fmfgraph*}}$ & \;$\displaystyle -\frac{i(\slashed{p}-m_f)}{p^2 - m_f^2 + i\epsilon}$ &
    $\;\parbox{15mm}{\begin{fmfgraph*}(15,5)
    \fmfleft{i1}
    \fmfright{o1}
    \marrow{c}{down}{bot}{$p$}{i1,o1}
    \fmf{fermion}{o1,i1}
    \end{fmfgraph*}}$ & \;$\displaystyle \frac{i(\slashed{p}-m_f)}{p^2 - m_f^2 - i\epsilon}$ \\[10pt]

    $\;\parbox{15mm}{\begin{fmfgraph*}(15,5)
    \fmfleft{i1}
    \fmfright{o1}
    \fmflabel{$\mu$}{i1}
    \fmflabel{$\nu$}{o1}
    \fmf{photon}{i1,o1}
    \Marrow{d}{down}{bot}{$p$}{i1,o1}{8}
    \end{fmfgraph*}}$ & \;$\displaystyle -\frac{i g^{\mu\nu}}{p^2 + i\epsilon}$ &
    $\;\parbox{15mm}{\begin{fmfgraph*}(15,5)
    \fmfleft{i1}
    \fmfright{o1}
    \fmflabel{$\mu$}{i1}
    \fmflabel{$\nu$}{o1}
    \fmf{photon}{i1,o1}
    \Marrow{d}{down}{bot}{$p$}{i1,o1}{8}
    \end{fmfgraph*}}$ & \;$\displaystyle \frac{i g^{\mu\nu}}{p^2 - i\epsilon}$ \\[10pt]

    $\;\parbox{15mm}{\begin{fmfgraph*}(15,5)
    \fmfleft{i1}
    \fmfright{o1}
    \fmflabel{$\mu,a$}{i1}
    \fmflabel{$\nu,b$}{o1}
    \fmf{gluon}{i1,o1}
    \Marrow{d}{down}{bot}{$p$}{i1,o1}{8}
    \end{fmfgraph*}}$ & \;$\displaystyle -\frac{i \delta^{ab} g^{\mu\nu}}{p^2 + i\epsilon}$ &
    $\;\parbox{15mm}{\begin{fmfgraph*}(15,5)
    \fmfleft{i1}
    \fmfright{o1}
    \fmflabel{$\mu,a$}{i1}
    \fmflabel{$\nu,b$}{o1}
    \fmf{gluon}{i1,o1}
    \Marrow{d}{down}{bot}{$p$}{i1,o1}{8}
    \end{fmfgraph*}}$ & \;$\displaystyle \frac{i \delta^{ab} g^{\mu\nu}}{p^2 - i\epsilon}$ \\[10pt]

    \mybox{\hspace{4.97cm} $\parbox{16.6mm}{\begin{fmfgraph*}(16.6,5)
    \fmfleft{i1}
    \fmfright{o1}
    \marrow{a}{down}{bot}{$p$}{i1,o1}
    \fmf{dashes}{i1,o1}
    \end{fmfgraph*}} \qquad\;\;\;\, 2\pi \;\theta(p^0) \,\delta(p^2 - m_s^2)$}& & & \\[6pt]

    \mybox{\hspace{5.28cm} $\parbox{15mm}{\begin{fmfgraph*}(15,5)
    \fmfleft{i1}
    \fmfright{o1}
    \marrow{b}{down}{bot}{$p$}{i1,o1}
    \fmf{fermion}{i1,o1}
    \end{fmfgraph*}} \qquad\;\;\;
    \begin{array}{l}
    2\pi \;\theta(p^0) \,\delta(p^2 - m_f^2) \\[-13pt]
    \quad\! \times (\slashed{p}+m_f)
    \end{array}$}& & & \\[-4pt]

    \mybox{\hspace{5.62cm} $\parbox{15mm}{\begin{fmfgraph*}(15,5)
    \fmfleft{i1}
    \fmfright{o1}
    \marrow{c}{down}{bot}{$p$}{i1,o1}
    \fmf{fermion}{o1,i1}
    \end{fmfgraph*}} \qquad\;\;\;
    \begin{array}{l}
    -\,2\pi \;\theta(p^0) \,\delta(p^2 - m_f^2) \\[-13pt]
    \quad\! \times (\slashed{p}-m_f)
    \end{array}$}& & & \\[1pt]

    \mybox{\hspace{5.14cm} $\parbox{15mm}{\begin{fmfgraph*}(15,5)
    \fmfleft{i1}
    \fmfright{o1}
    \fmflabel{$\mu$}{i1}
    \fmflabel{$\nu$}{o1}
    \fmf{photon}{i1,o1}
    \Marrow{d}{down}{bot}{$p$}{i1,o1}{8}
    \end{fmfgraph*}} \qquad\;\;\;\; -2\pi \;\theta(p^0) \,\delta(p^2) \,g^{\mu\nu}$}& & & \\[10pt]

    \mybox{\hspace{5.66cm} $\parbox{15mm}{\begin{fmfgraph*}(15,5)
    \fmfleft{i1}
    \fmfright{o1}
    \fmflabel{$\mu,a$}{i1}
    \fmflabel{$\nu,b$}{o1}
    \fmf{gluon}{i1,o1}
    \Marrow{d}{down}{bot}{$p$}{i1,o1}{8}
    \end{fmfgraph*}} \qquad\;\;\;\; -2\pi \;\theta(p^0) \,\delta(p^2) \,\delta^{ab} \,g^{\mu\nu}$}& & & \\[20pt]

    $\;\parbox{15mm}{\begin{fmfgraph*}(15,11)
    \fmfdot{v}
    \fmfleft{i1,i2}
    \fmfright{o1}
    \fmf{fermion}{v,i1}
    \fmf{fermion}{i2,v}
    \fmf{photon}{v,o1}
    \fmflabel{$\mu$}{o1}
    \end{fmfgraph*}}$ & \;$-ie_q e \gamma^\mu$ &
    $\;\parbox{15mm}{\begin{fmfgraph*}(15,11)
    \fmfdot{v}
    \fmfleft{i1,i2}
    \fmfright{o1}
    \fmf{fermion}{v,i1}
    \fmf{fermion}{i2,v}
    \fmf{photon}{v,o1}
    \fmflabel{$\mu$}{o1}
    \end{fmfgraph*}}$ & \;$ie_q e \gamma^\mu$ \\[35pt]

    $\;\parbox{15mm}{\begin{fmfgraph*}(15,11)
    \fmfdot{v}
    \fmfleft{i1,i2}
    \fmfright{o1}
    \fmf{scalar,fore=blue}{v,i1}
    \fmf{scalar,fore=blue}{i2,v}
    \Marrow{e}{down}{bot}{$\quad\enspace p_2$}{v,i1}{8}
    \Marrow{f}{up}{top}{$\quad\enspace p_1$}{i2,v}{8}
    \fmf{gluon}{v,o1}
    \fmflabel{$\mu,a$}{o1}
    \end{fmfgraph*}}$ & \;$-ig t^a (p_1+p_2)^\mu$ &
    $\;\parbox{15mm}{\begin{fmfgraph*}(15,11)
    \fmfdot{v}
    \fmfleft{i1,i2}
    \fmfright{o1}
    \fmf{scalar,fore=blue}{v,i1}
    \fmf{scalar,fore=blue}{i2,v}
    \Marrow{e}{down}{bot}{$\quad\enspace p_2$}{v,i1}{8}
    \Marrow{f}{up}{top}{$\quad\enspace p_1$}{i2,v}{8}
    \fmf{gluon}{v,o1}
    \fmflabel{$\mu,a$}{o1}
    \end{fmfgraph*}}$ & \;$ig t^a (p_1+p_2)^\mu$ \\[35pt]

    $\;\parbox{15mm}{\begin{fmfgraph*}(15,11)
    \fmfdot{v}
    \fmfleft{i1,i2}
    \fmfright{o1}
    \fmf{scalar,fore=blue}{i1,v}
    \fmf{scalar,fore=blue}{v,i2}
    \Marrow{g}{down}{bot}{$\quad\enspace p_2$}{v,i1}{8}
    \Marrow{h}{up}{top}{$\quad\enspace p_1$}{i2,v}{8}
    \fmf{gluon}{v,o1}
    \fmflabel{$\mu,a$}{o1}
    \end{fmfgraph*}}$ & \;$ig t^a (p_1+p_2)^\mu$ &
    $\;\parbox{15mm}{\begin{fmfgraph*}(15,11)
    \fmfdot{v}
    \fmfleft{i1,i2}
    \fmfright{o1}
    \fmf{scalar,fore=blue}{i1,v}
    \fmf{scalar,fore=blue}{v,i2}
    \Marrow{g}{down}{bot}{$\quad\enspace p_2$}{v,i1}{8}
    \Marrow{h}{up}{top}{$\quad\enspace p_1$}{i2,v}{8}
    \fmf{gluon}{v,o1}
    \fmflabel{$\mu,a$}{o1}
    \end{fmfgraph*}}$ & \;$-ig t^a (p_1+p_2)^\mu$ \\[22pt] 

    $\;\parbox{15mm}{\begin{fmfgraph*}(15,11)
    \fmfdot{v}
    \fmfleft{i1,i2}
    \fmfright{o1}
    \fmf{fermion}{v,i1}
    \fmf{fermion}{i2,v}
    \fmf{gluon}{v,o1}
    \fmflabel{$\mu,a$}{o1}
    \end{fmfgraph*}}$ & \;$-ig t^a \gamma^\mu$ &
    $\;\parbox{15mm}{\begin{fmfgraph*}(15,11)
    \fmfdot{v}
    \fmfleft{i1,i2}
    \fmfright{o1}
    \fmf{fermion}{v,i1}
    \fmf{fermion}{i2,v}
    \fmf{gluon}{v,o1}
    \fmflabel{$\mu,a$}{o1}
    \end{fmfgraph*}}$ & \;$ig t^a \gamma^\mu$ \\[35pt]

    $\;\parbox{15mm}{\begin{fmfgraph*}(15,11)
    \fmfdot{v}
    \fmfleft{i1,i2}
    \fmfright{o1}
    \fmf{gluon}{i1,v}
    \fmf{gluon}{i2,v}
    \Marrow{i}{down}{bot}{$\quad\enspace p_2$}{i1,v}{12}
    \Marrow{j}{up}{top}{$\quad\enspace p_1$}{i2,v}{9}
    \fmflabel{$\nu,b$}{i1}
    \fmflabel{$\mu,a$}{i2}
    \fmf{gluon}{v,o1}
    \Marrow{k}{up}{top}{$p_3$}{o1,v}{6}
    \fmflabel{$\rho,c$}{o1}
    \end{fmfgraph*}}$ & \;$-g f^{abc} \,V^{\mu\nu\rho}$ &
    $\;\parbox{15mm}{\begin{fmfgraph*}(15,11)
    \fmfdot{v}
    \fmfleft{i1,i2}
    \fmfright{o1}
    \fmf{gluon}{i1,v}
    \fmf{gluon}{i2,v}
    \Marrow{i}{down}{bot}{$\quad\enspace p_2$}{i1,v}{12}
    \Marrow{j}{up}{top}{$\quad\enspace p_1$}{i2,v}{9}
    \fmflabel{$\nu,b$}{i1}
    \fmflabel{$\mu,a$}{i2}
    \fmf{gluon}{v,o1}
    \Marrow{k}{up}{top}{$p_3$}{o1,v}{6}
    \fmflabel{$\rho,c$}{o1}
    \end{fmfgraph*}}$ & \;$g f^{abc} \,V^{\mu\nu\rho}$ \\[35pt]

    $\;\parbox{15mm}{\begin{fmfgraph*}(15,11)
    \fmfdot{v}
    \fmfleft{i1,i2}
    \fmfright{o1,o2}
    \fmf{dashes}{i1,v}
    \fmf{dashes}{i2,v}
    \fmf{gluon}{v,o1}
    \fmf{gluon}{v,o2}
    \fmflabel{$\nu,b$}{o1}
    \fmflabel{$\mu,a$}{o2}
    \end{fmfgraph*}}$ & \;$i g^2 \{ t^a,t^b \} g^{\mu\nu}$ &
    $\;\parbox{15mm}{\begin{fmfgraph*}(15,11)
    \fmfdot{v}
    \fmfleft{i1,i2}
    \fmfright{o1,o2}
    \fmf{dashes}{i1,v}
    \fmf{dashes}{i2,v}
    \fmf{gluon}{v,o1}
    \fmf{gluon}{v,o2}
    \fmflabel{$\nu,b$}{o1}
    \fmflabel{$\mu,a$}{o2}
    \end{fmfgraph*}}$ & \;$-i g^2 \{ t^a,t^b \} g^{\mu\nu}$ \\[-9pt] & & & \\ \hline
\caption{The (relevant) Feynman rules for the scalar spectator model. The dashed line separating the columns represents the final-state cut. In the calculations of chapter~\ref{c:Disentanglement} we set the quark mass to zero ($m_f=0$).}
\label{t:feynman_rules}
\end{longtable}
\end{center} \vspace{-0.6cm}

The rules for conjugate amplitudes are easily obtained from those for amplitudes through complex conjugation. There is one exception to this, though: the three-gluon vertex changes sign when it crosses the final-state cut~\cite{Belitsky:1997ay}.\footnote{This rule in fact applies to all momentum-dependent vertex factors that change sign when all momenta are reversed.} Note that table~\ref{t:feynman_rules} also provides the rules for cutting lines. For cutting a line with mass $m$ and momentum $p$ going to the right, we replace the denominator of its propagator by an on-shell delta function (see e.g.~\cite{Peskin:1995ev}):
\begin{equation}
    \frac{1}{p^2 - m^2 + i\epsilon} \;\to\; -2\pi i \;\theta(p^0) \,\delta(p^2 - m^2) ,
\end{equation}
where the theta function ensures a positive energy flow from left to right through the cut. The numerator remains unchanged.

The sign of the scalar-scalar-gluon coupling depends on the direction of color flow through the scalar lines with respect to the direction of momentum flow. To indicate the direction of color flow, we use blue arrows on the scalar lines. Furthermore, for convenience we define the following object for the three-gluon vertex:
\begin{equation}
    V^{\mu\nu\rho} \equiv g^{\mu\nu} (p_1-p_2)^\rho + g^{\nu\rho} (p_2-p_3)^\mu + g^{\rho\mu} (p_3-p_1)^\nu .
\end{equation}
All listed vertex factors would come with an opposite sign if in the covariant derivatives the interaction terms with the vector boson fields would be defined with a minus rather than with a plus sign~\cite{Romao:2012pq}.

In table~\ref{t:eikonal_rules} we list the relevant eikonal Feynman rules. We make use of an auxiliary vector $v$ that depends on whether the fermion is left-moving or right-moving. In the former case we can take $v = n$ and in the latter case we can choose $v = \bar{n}$. Throughout the thesis we use the following notation for eikonal lines~\cite{Diehl:2011yj,Buffing:2017mqm}: the circles at the ends of an eikonal line indicate the direction of momentum flow (from the full to the empty circle) of the original fermion, and the arrow on the line denotes the direction of color flow (and thus also the direction of fermion number flow).

\begin{center}
\renewcommand{\arraystretch}{2.1}
\begin{longtable}{|C L : C L|}
    \hline
    \multicolumn{2}{|c:}{\textbf{Amplitude}} & \multicolumn{2}{c|}{\textbf{Conjugate amplitude}} \\ \hline
    $\;\parbox{15mm}{\begin{fmfgraph*}(15,5)
    \fmfleft{i1}
    \fmfright{o1}
    \marrow{a}{down}{bot}{$\ell$}{i1,o1}
    \fmf{dbl_plain_arrow}{i1,o1}
    \fmfv{decor.shape=circle,decor.filled=full,decor.size=2.5thick}{i1}
    \fmfv{decor.shape=circle,decor.filled=empty,decor.size=2.5thick}{o1}
    \end{fmfgraph*}}$ & \;$\displaystyle \frac{i}{\ell \cd v + i\epsilon}$ &
    $\;\parbox{15mm}{\begin{fmfgraph*}(15,5)
    \fmfleft{i1}
    \fmfright{o1}
    \marrow{a}{down}{bot}{$\ell$}{i1,o1}
    \fmf{dbl_plain_arrow}{i1,o1}
    \fmfv{decor.shape=circle,decor.filled=full,decor.size=2.5thick}{o1}
    \fmfv{decor.shape=circle,decor.filled=empty,decor.size=2.5thick}{i1}
    \end{fmfgraph*}}$ & \;$\displaystyle -\frac{i}{\ell \cd v - i\epsilon}$ \\[10pt]

    $\;\parbox{15mm}{\begin{fmfgraph*}(15,5)
    \fmfleft{i1}
    \fmfright{o1}
    \marrow{a}{down}{bot}{$\ell$}{i1,o1}
    \fmf{dbl_plain_arrow}{o1,i1}
    \fmfv{decor.shape=circle,decor.filled=full,decor.size=2.5thick}{i1}
    \fmfv{decor.shape=circle,decor.filled=empty,decor.size=2.5thick}{o1}
    \end{fmfgraph*}}$ & \;$\displaystyle - \frac{i}{\ell \cd v + i\epsilon}$ &
    $\;\parbox{15mm}{\begin{fmfgraph*}(15,5)
    \fmfleft{i1}
    \fmfright{o1}
    \marrow{a}{down}{bot}{$\ell$}{i1,o1}
    \fmf{dbl_plain_arrow}{o1,i1}
    \fmfv{decor.shape=circle,decor.filled=full,decor.size=2.5thick}{o1}
    \fmfv{decor.shape=circle,decor.filled=empty,decor.size=2.5thick}{i1}
    \end{fmfgraph*}}$ & \;$\displaystyle \frac{i}{\ell \cd v - i\epsilon}$ \\[20pt]

    $\;\parbox{15mm}{\begin{fmfgraph*}(15,11)
    \fmfdot{v}
    \fmfleft{i1,i2}
    \fmfright{o1}
    \fmf{dbl_plain}{v,i1}
    \fmf{dbl_plain}{i2,v}
    \fmf{gluon}{v,o1}
    \fmflabel{$\mu,a$}{o1}
    \end{fmfgraph*}}$ & \;$-ig t^a v^\mu$ &
    $\;\parbox{15mm}{\begin{fmfgraph*}(15,11)
    \fmfdot{v}
    \fmfleft{i1,i2}
    \fmfright{o1}
    \fmf{dbl_plain}{v,i1}
    \fmf{dbl_plain}{i2,v}
    \fmf{gluon}{v,o1}
    \fmflabel{$\mu,a$}{o1}
    \end{fmfgraph*}}$ & \;$ig t^a v^\mu$ \\[-21pt] & & & \\ \hline
\caption{The (relevant) Feynman rules for eikonal lines and vertices. The dashed line separating the columns represents the final-state cut.}
\label{t:eikonal_rules}
\end{longtable}
\end{center}

\thispagestyle{empty} 
\stopthumb

\chapter[\stackunder{Definitions of gluon TMDs}{}]{Definitions of gluon TMDs} \label{a:definitions_gluon_TMDs}
\continuethumb

In this appendix we provide the definitions of the leading-twist gluon TMDs in eq.~\eqref{e:gluon_TMD_parametrization} in terms of the coefficient functions $A_i$ that appear in eqs.~\eqref{e:gamma_unintegrated_par_up}--\eqref{e:gamma_unintegrated_par_tp}. The TMD correlator in eq.~\eqref{e:gamma_leadingtwist} is related to the unintegrated correlator as
\begin{equation}
    \Gamma^{ij}(x,\bm{k}) \equiv \frac{2}{P^+} \int dk^- \,\Gamma^{+i;+j}(k) = \frac{M^2}{(P^+)^2} \int [d\sigma d\tau] \;\Gamma^{+i;+j}(k) ,
    \label{e:sigma_tau}
\end{equation}
where we have introduced the shorthand notation
\begin{equation}
    [d\sigma d\tau] \equiv d\sigma d\tau \;\delta \left( \tau - x\sigma + x^2 + \frac{\bm{k}^2}{M^2} \right) ,
\end{equation}
with the dimensionless invariants $\sigma$ and $\tau$ given by
\begin{equation}
    \sigma \equiv \frac{2k \cd P}{M^2} \,, \qquad \tau \equiv \frac{k^2}{M^2} \,.
\end{equation}
Making use of eq.~\eqref{e:sigma_tau}, the gluon TMDs are related to the coefficient functions as follows:
\begin{align}
    x f_1(x,\bm{k}^2) &\equiv M^2 \int [d\sigma d\tau] \left( A_2 + 2x A_4 + x^2 A_3 + \frac{\bm{k}^2}{2M^2} \,A_6 \right) , \label{e:D4} \\
    x h_1^\perp(x,\bm{k}^2) &\equiv M^2 \int [d\sigma d\tau] \,A_6 \,, \\
    x g_{1}(x,\bm{k}^2) &\equiv 2M^2 \int [d\sigma d\tau] \left\{ \vphantom{\frac{\bm{k}^2}{M^2}} A_8 + A_9 + x \left( A_{10} + A_{11} \right) \right. \nn \\
    &\quad\, + \left( \frac{\sigma}{2} - x \right) \left[ A_{12} + x \left( A_{14} + A_{15} \right) + x^2 A_{13} \right] \nn \\
    &\quad\, \left. + \,\frac{\bm{k}^2}{2M^2} \left[ A_{19} + A_{23} + \left( \frac{\sigma}{2} - x \right) A_{25} \right] \right\} , \\
    x h_{1L}^\perp(x,\bm{k}^2) &\equiv -2M^2 \int [d\sigma d\tau] \left[ A_{18} + A_{22} + \left( \frac{\sigma}{2} - x \right) A_{24} \right] , \\
    x f_{1T}^\perp(x,\bm{k}^2) &\equiv M^2 \int [d\sigma d\tau] \left[ A_{16} - A_{20} + x \left( A_{18} - A_{22} \right) \right] , \\
    x g_{1T}(x,\bm{k}^2) &\equiv -M^2 \int [d\sigma d\tau] \left[ 2A_{12} + A_{17} + A_{21} + 2x \left( A_{14} + A_{15} \right) \vphantom{\frac{\bm{k}^2}{M^2}} \right. \nn \\
    &\quad\, \left. + \,x \left( A_{19} + A_{23} \right) + 2x^2 A_{13} + \frac{\bm{k}^2}{M^2} \,A_{25} \right] , \\
    x h_{1}(x,\bm{k}^2) &\equiv 2M^2 \int [d\sigma d\tau] \left[ A_{16} + A_{20} + x \left( A_{18} + A_{22} \right) + \frac{\bm{k}^2}{2M^2} \,A_{24} \right] , \\
    x h_{1T}^\perp(x,\bm{k}^2) &\equiv 2M^2 \int [d\sigma d\tau] \,A_{24} \,, \\
    x f_{1LL}(x,\bm{k}^2) &\equiv \frac{M^2}{3} \int [d\sigma d\tau] \left\{ A_{27} - 2A_{34} + 2x A_{28} + x^2 A_{26} \vphantom{\left( \frac{(\sigma-2x)^2}{4} - \frac{\bm{k}^2}{2M^2} \right)} \right. \nn \\
    &\quad\, + 2 (\sigma-2x) \left( A_{37} + x A_{35} \right) + \frac{(\sigma-2x)^2}{2} \left( A_{40} + 2x A_{42} + x^2 A_{41} \right) \nn \\
    &\quad\, - \frac{\bm{k}^2}{M^2} \left[ A_{26} - A_{32} + \,A_{40} + 2x A_{42} + x^2 A_{41} + (\sigma-3x) A_{30} \vphantom{\left( \frac{(\sigma-2x)^2}{4} - \frac{\bm{k}^2}{2M^2} \right)} \right. \nn \\
    &\quad\, \left.\left. \!- \left( \frac{(\sigma-2x)^2}{4} - \frac{\bm{k}^2}{2M^2} \right) A_{44} \right] \right\} , \\
    x h_{1LL}^{\perp}(x,\bm{k}^2) &\equiv - \frac{2M^2}{3} \int [d\sigma d\tau] \left[ A_{26} - A_{32} + (\sigma-3x) A_{30} \vphantom{\left( \frac{(\sigma-2x)^2}{4} - \frac{\bm{k}^2}{2M^2} \right)} \right. \nn \\
    &\quad\, \left. - \,\left( \frac{(\sigma-2x)^2}{4} - \frac{\bm{k}^2}{2M^2} \right) A_{44} \right] , \\
    x f_{1LT}(x,\bm{k}^2) &\equiv -M^2 \int [d\sigma d\tau] \left\{ A_{37} + x A_{35} + \left( \frac{\sigma}{2} - x \right) \left( A_{40} + 2x A_{42} + x^2 A_{41} \right) \vphantom{\frac{\bm{k}^2}{2M^2}} \right. \nn \\
    &\quad\, \left. - \,\frac{\bm{k}^2}{4M^2} \left[ A_{30} + \left( x - \frac{\sigma}{2} \right) A_{44} \right] \right\} , \\
    x g_{1LT}(x,\bm{k}^2) &\equiv -\frac{M^2}{2} \int [d\sigma d\tau] \left[ A_{29} + \left( x - \frac{\sigma}{2} \right) \left( A_{33} + xA_{31} \right) \right] , \\
    x h_{1LT}(x,\bm{k}^2) &\equiv \frac{M^2}{2} \int [d\sigma d\tau] \left\{ A_{28} + x A_{26} + \left( \frac{\sigma}{2} - x \right) \left( A_{32} + x A_{30} \right) \vphantom{\frac{\bm{k}^2}{2M^2}} \right. \nn \\
    &\quad\, \left. + \,\frac{\bm{k}^2}{2M^2} \left[ A_{30} + \left( x - \frac{\sigma}{2} \right) A_{44} \right] \right\} , \\
    x h_{1LT}^{\perp}(x,\bm{k}^2) &\equiv -M^2 \int [d\sigma d\tau] \left[ A_{30} + \left( x - \frac{\sigma}{2} \right) A_{44} \right] , \\
    x f_{1TT}(x,\bm{k}^2) &\equiv \frac{M^2}{2} \int [d\sigma d\tau] \left( A_{40} + 2x A_{42} + x^2 A_{41} + \frac{\bm{k}^2}{6M^2} \,A_{44} \right) , \\
    x g_{1TT}(x,\bm{k}^2) &\equiv \frac{M^2}{2} \int [d\sigma d\tau] \left( A_{33} + x A_{31} \right) , \\
    x h_{1TT}(x,\bm{k}^2) &\equiv -\frac{M^2}{2} \int [d\sigma d\tau] \left[ A_{27} + 2x A_{28} + x^2 A_{26} \vphantom{\frac{\bm{k}^4}{4M^4}} \right. \nn \\
    &\quad\, \left. + \,\frac{\bm{k}^2}{M^2} \left( A_{32} + x A_{30} \right) - \frac{\bm{k}^4}{4M^4} \,A_{44} \right] , \\
    x h_{1TT}^{\perp}(x,\bm{k}^2) &\equiv \frac{M^2}{2} \int [d\sigma d\tau] \left( A_{32} + x A_{30} - \frac{\bm{k}^2}{3M^2} \,A_{44} \right) , \\
    x h_{1TT}^{\perp\perp}(x,\bm{k}^2) &\equiv \frac{M^2}{2} \int [d\sigma d\tau] \,A_{44} \,. \label{e:D22}
\end{align}

\newpage
\stopthumb

\chapter{Correlators in coordinate space} \label{a:coordinate_space}
\continuethumb

In subsections~\ref{s:parametrization_TMD_correlator} and~\ref{s:parametrization_gluons} the quark-quark and gluon-gluon TMD correlators are parametrized in terms of leading-twist quark and gluon TMDs, respectively. Those parametrizations are given in momentum space ($k_\sst$-space). In this appendix we provide parametrizations of those correlators in coordinate space ($z_\sst$-space), based on our work in~\cite{Boer:2016xqr,vanDaal:2016glj}.

To facilitate the translation into $z_\sst$-space, let us denote a generic TMD correlator by $\Omega$ and a generic TMD function by $f$. Their counterparts in $z_\sst$-space are related by a Fourier transformation~\cite{Boer:2011xd}:
\begin{align}
    \tilde{\Omega}(x,\bm{z}) &\equiv \int d^2\bm{k} \;e^{i \bm{k} \cd \bm{z}} \,\Omega(x,\bm{k}) , \label{e:FT} \\
    \tilde{f}(x,\bm{z}^2) &\equiv \int d^2\bm{k} \;e^{i \bm{k} \cd \bm{z}} \,f(x,\bm{k}^2) . \label{e:FT2}
\end{align}

In momentum space, the parametrization of the correlator in terms of TMDs $f_j$ with rank $n = n(j) \geq 0$ takes the form\footnote{For simplicity, possible Lorentz indices on $\Omega$ or $C_j$ are omitted.}
\begin{equation}
    \Omega(x,\bm{k}) = \sum_j C_j \,\frac{k_\sst^{i_1 \ldots i_n}}{M^n} \,f_j(x,\bm{k}^2) ,
    \label{e:mom space}
\end{equation}
where $C_j$ is a coefficient independent of $k$ that contains information on both the hadron and parton polarization. Using eqs.~\eqref{e:STT}, \eqref{e:FT}, and~\eqref{e:mom space}, the correlator in $z_\sst$-space is given by
\begin{align}
    \tilde{\Omega}(x,\bm{z}) &= \sum_{j} C_j \int d^2\bm{k} \;e^{i \bm{k} \cd \bm{z}} \,\frac{k_\sst^{i_1 \ldots i_n}}{M^n} \,f_j(x,\bm{k}^2) \nn \\
    &= \sum_{j} C_j \,\frac{z_\sst^{i_1 \ldots i_n}}{M^n} \int_0^\infty d|\bm{k}| \,|\bm{k}| \left( \frac{|\bm{k}|}{|\bm{z}|} \right)^n (2\pi i^n) \,J_n(|\bm{k}||\bm{z}|) \,f_j(x,\bm{k}^2) \nn \\
    &= \sum_{j} \frac{i^n}{n!} \,C_j \,M^n \,z_\sst^{i_1 \ldots i_n} \,\tilde{f}_j^{(n)}(x,\bm{z}^2) ,
    \label{e:corrinbtspace}
\end{align}
where on the second line the Bessel function of the first kind $J_k(u)$ arose from the integral identity
\begin{equation}
    \int_0^{2\pi} d\alpha \;e^{ik\alpha} e^{iu\cos(\alpha-\beta)} = 2\pi i^k J_k(u) \,e^{ik\beta} .
\end{equation}
Following the conventions in~\cite{Boer:2011xd}, we define the function $\tilde{f}^{(n)}$ as
\begin{align}
    \tilde{f}^{(n)}(x,\bm{z}^2) &\equiv n! \left( - \frac{2}{M^2} \frac{\p}{\p \bm{z}^2} \right)^n \tilde{f}(x,\bm{z}^2) \nn \\[3pt]
    &= \frac{2\pi n!}{M^{2n}} \int_0^\infty d|\bm{k}| \,|\bm{k}| \left( \frac{|\bm{k}|}{|\bm{z}|} \right)^n J_n(|\bm{k}||\bm{z}|) \,f(x,\bm{k}^2) ,
    \label{e:ftilde}
\end{align}
where we used eq.~\eqref{e:FT2}, as well as the recurrence relation
\begin{equation}
    \left( \frac{1}{u} \frac{d}{du} \right)^m \left[ \frac{J_k(u)}{u^k} \right] = (-1)^m \,\frac{J_{k+m}(u)}{z^{k+m}} \,,
\end{equation}
with $u = |\bm{k}||\bm{z}|$ (with $|\bm{k}|$ fixed), $m = n$, and $k = 0$.

From eqs.~\eqref{e:corrinbtspace} and~\eqref{e:ftilde} it follows that in the parametrization of the correlator in $z_\sst$-space the $n$th derivative $\tilde{f}^{(n)}$ with respect to $\bm{z}^2$ appears rather than the function $\tilde{f} = \tilde{f}^{(0)}$ itself. Furthermore, we infer from eq.~\eqref{e:ftilde} that for definite-rank TMDs there is a one-to-one correspondence between the functions in momentum and coordinate space. The motivation for this particular definition of $\tilde{f}^{(n)}$ in eq.~\eqref{e:ftilde} becomes obvious once we set $\bm{z} = \bm{0}$, which is equivalent to integration over transverse momentum in $k_\sst$-space. Using the limit
\begin{equation}
    \lim_{u\to0} \frac{J_k(u)}{u^k} = \frac{1}{2^k k!} \,,
\end{equation}
we see that
\begin{equation}
    \lim_{|\bm{z}| \to 0} \tilde{f}^{(n)}(x,\bm{z}^2) = \int d^2\bm{k} \,\left( \frac{\bm{k}^2}{2M^2} \right)^n f(x,\bm{k}^2) ,
\end{equation}
which is precisely the conventional $n$th moment $f^{(n)}(x)$ of the TMD. Hence, by construction, the derivatives in $z_\sst$-space are directly related to moments in $k_\sst$-space.

\section{The quark-quark TMD correlator}
We can use eq.~\eqref{e:corrinbtspace} to translate the quark-quark TMD correlator in eq.~\eqref{e:quark_TMD_parametrization} to $z_\sst$-space. In~\cite{Boer:2011xd} this was already done for spin-$\tfrac{1}{2}$ hadrons. For the spin-$1$ case we have
\begin{align}
    \tilde{\Phi}(x,\bm{z}) &= \tilde{\Phi}_U(x,\bm{z}) + \tilde{\Phi}_L(x,\bm{z}) + \tilde{\Phi}_T(x,\bm{z}) \nn \\
    &\quad\, + \tilde{\Phi}_{LL}(x,\bm{z}) + \tilde{\Phi}_{LT}(x,\bm{z}) + \tilde{\Phi}_{TT}(x,\bm{z}) ,
\end{align}
where
\begin{align}
    \tilde{\Phi}_U(x,\bm{z}) &= \frac{1}{2} \left[ \slashed{\bar{n}} \,\tilde{f}_1(x,\bm{z}^2) + iM \,\sigma_{\mu\nu} z_\sst^\mu \bar{n}^\nu \,\tilde{h}_1^{\perp(1)}(x,\bm{z}^2) \right] , \\[5pt]
    \tilde{\Phi}_L(x,\bm{z}) &= \frac{1}{2} \left[ \gamma^5 \slashed{\bar{n}} \,S_L \,\tilde{g}_1(x,\bm{z}^2) - M \,\sigma_{\mu\nu} \gamma^5 \bar{n}^\mu z_\sst^\nu \,S_L \,\tilde{h}_{1L}^{\perp(1)}(x,\bm{z}^2) \right] , \\[5pt]
    \tilde{\Phi}_T(x,\bm{z}) &= \frac{1}{2} \left[ iM \,\slashed{\bar{n}} \,\epsilon_\sst^{S_T z_\sst} \tilde{f}_{1T}^{\perp(1)}(x,\bm{z}^2) + iM \gamma^5 \slashed{\bar{n}} \,\bm{z} \cd \bm{S}_T \,\tilde{g}_{1T}^{(1)}(x,\bm{z}^2) \vphantom{\frac{iM^2 \sigma_{\mu\nu} \gamma^5 \bar{n}^\mu z_\sst^{\nu\rho} {S_T}_\rho}{2}} \right. \nn \\
    &\quad\, \left. + \,i\sigma_{\mu\nu} \gamma^5 \bar{n}^\mu S_T^\nu \,\tilde{h}_1(x,\bm{z}^2) + \frac{iM^2 \sigma_{\mu\nu} \gamma^5 \bar{n}^\mu z_\sst^{\nu\rho} {S_T}_\rho}{2} \,\tilde{h}_{1T}^{\perp(2)}(x,\bm{z}^2) \right] , \\[5pt]
    \tilde{\Phi}_{LL}(x,\bm{z}) &= \frac{1}{2} \left[ \slashed{\bar{n}} \,S_{LL} \,\tilde{f}_{1LL}(x,\bm{z}^2) + iM \,\sigma_{\mu\nu} z_\sst^\mu \bar{n}^\nu S_{LL} \,\tilde{h}_{1LL}^{\perp(1)}(x,\bm{z}^2) \right] , \\[5pt]
    \tilde{\Phi}_{LT}(x,\bm{z}) &= \frac{1}{2} \left[ iM \,\slashed{\bar{n}} \,\bm{z} \cd \bm{S}_{LT} \,\tilde{f}_{1LT}^{(1)}(x,\bm{z}^2) + iM \gamma^5 \slashed{\bar{n}} \,\epsilon_\sst^{S_{LT} z_\sst} \,\tilde{g}_{1LT}^{(1)}(x,\bm{z}^2) \vphantom{\frac{M^2 \sigma_{\mu\nu} \bar{n}^\nu b_\sst^{\mu\rho} {S_{LT}}_\rho}{2}} \right. \nn \\
    &\quad\, \left. + \,\sigma_{\mu\nu} \bar{n}^\nu S_{LT}^\mu \,\tilde{h}_{1LT}(x,\bm{z}^2) + \frac{M^2 \sigma_{\mu\nu} \bar{n}^\nu z_\sst^{\mu\rho} {S_{LT}}_\rho}{2} \,\tilde{h}_{1LT}^{\perp(2)}(x,\bm{z}^2) \right] , \\[5pt]
    \tilde{\Phi}_{TT}(x,\bm{z}) &= \frac{1}{2} \left[ - \,\frac{M^2 \slashed{\bar{n}} \,z_\sst^{\mu\nu} {S_{TT}}_{\mu\nu}}{2} \,\tilde{f}_{1TT}^{(2)}(x,\bm{z}^2) + \frac{M^2 \gamma^5 \slashed{\bar{n}} \,{\epsilon_\sst}_{\mu\nu} z_\sst^{\mu\rho} {S_{TT}^\nu}_\rho}{2} \,\tilde{g}_{1TT}^{(2)}(x,\bm{z}^2) \right. \nn \\[3pt]
    &\quad\, - iM \,\sigma_{\mu\nu} \bar{n}^\nu z_\sst^\rho {S_{TT}^\mu}_\rho \,\tilde{h}_{1TT}^{(1)}(x,\bm{z}^2) \nn \\[3pt]
    &\quad\, \left. - \,\frac{iM^3 \sigma_{\mu\nu} \bar{n}^\nu z_\sst^{\mu\rho\sigma} {S_{TT}}_{\rho\sigma}}{6} \,\tilde{h}_{1TT}^{\perp(3)}(x,\bm{z}^2) \vphantom{\frac{M^2 \gamma^5 \slashed{\bar{n}} \,{\epsilon_\sst}_{\mu\nu} z_\sst^{\rho\mu} {S_{TT}^\nu}_\rho}{2}} \right] .
\end{align}
The quark TMDs in $z_\sst$-space are one-to-one related to their $k_\sst$-space counterparts through eq.~\eqref{e:ftilde}.

\section[\stackunder{The gluon-gluon TMD correlator}{}]{The gluon-gluon TMD correlator}
We can use eq.~\eqref{e:corrinbtspace} to translate the gluon-gluon TMD correlator in eq.~\eqref{e:gluon_TMD_parametrization} to $z_\sst$-space. It is given by
\begin{align}
    \tilde{\Gamma}^{ij}(x,\bm{z}) &= \tilde{\Gamma}_U^{ij}(x,\bm{z}) + \tilde{\Gamma}_L^{ij}(x,\bm{z}) + \tilde{\Gamma}_T^{ij}(x,\bm{z}) \nn \\
    &\quad\, + \tilde{\Gamma}_{LL}^{ij}(x,\bm{z}) + \tilde{\Gamma}_{LT}^{ij}(x,\bm{z}) + \tilde{\Gamma}_{TT}^{ij}(x,\bm{z}) ,
\end{align}
where
\begin{align}
    \tilde{\Gamma}_U^{ij}(x,\bm{z}) &= \frac{xP^+}{2} \left[ \delta_\sst^{ij} \,\tilde{f}_1(x,\bm{z}^2) - \frac{M^2 z_\sst^{ij}}{2} \,\tilde{h}_1^{\perp(2)}(x,\bm{z}^2) \right] , \\[5pt]
    \tilde{\Gamma}_L^{ij}(x,\bm{z}) &= \frac{xP^+}{2} \left[ i \epsilon_\sst^{ij} S_L \,\tilde{g}_1(x,\bm{z}^2) - \frac{M^2 {\epsilon_\sst^{\{i}}_\alpha z_\sst^{j\}\alpha} S_L}{4} \,\tilde{h}_{1L}^{\perp(2)}(x,\bm{z}^2) \right] , \\[5pt]
    \tilde{\Gamma}_T^{ij}(x,\bm{z}) &= \frac{xP^+}{2} \left[ iM \,\delta_\sst^{ij} \,\epsilon_\sst^{S_T z_\sst} \tilde{f}_{1T}^{\perp(1)}(x,\bm{z}^2) - M \,\epsilon_\sst^{ij} \,\bm{z} \cd \bm{S}_T \,\tilde{g}_{1T}^{(1)}(x,\bm{z}^2) \vphantom{\frac{iM^3 {\epsilon_\sst^{\{i}}_\alpha z_\sst^{j\}\alpha S_T}}{12}} \right. \nn \\[3pt]
    &\quad\, - \frac{iM \Big( \epsilon_\sst^{z_\sst\{i} S_T^{j\}} + \epsilon_\sst^{S_T\{i} z_\sst^{j\}} \Big)}{4} \,\tilde{h}_1^{(1)}(x,\bm{z}^2) \nn \\[3pt]
    &\quad\, \left. + \,\frac{iM^3 {\epsilon_\sst^{\{i}}_\alpha z_\sst^{j\}\alpha S_T}}{12} \,\tilde{h}_{1T}^{\perp(3)}(x,\bm{z}^2) \right] , \\[5pt]
    \tilde{\Gamma}_{LL}^{ij}(x,\bm{z}) &= \frac{xP^+}{2} \left[ \delta_\sst^{ij} S_{LL} \,\tilde{f}_{1LL}(x,\bm{z}^2) - \frac{M^2 z_\sst^{ij} S_{LL}}{2} \,\tilde{h}_{1LL}^{\perp(2)}(x,\bm{z}^2) \right] , \\[5pt]
    \tilde{\Gamma}_{LT}^{ij}(x,\bm{z}) &= \frac{xP^+}{2} \left[ iM \,\delta_\sst^{ij} \,\bm{z} \cd \bm{S}_{LT} \,\tilde{f}_{1LT}^{(1)}(x,\bm{z}^2) - M \,\epsilon_\sst^{ij} \epsilon_\sst^{S_{LT} z_\sst} \,\tilde{g}_{1LT}^{(1)}(x,\bm{z}^2) \vphantom{\frac{iM^3 z_\sst^{ij\alpha} {S_{LT}}_\alpha}{6}} \right. \nn \\
    &\quad\, \left. + \,iM \,S_{LT}^{\{i} z_\sst^{j\}} \,\tilde{h}_{1LT}^{(1)}(x,\bm{z}^2) - \frac{iM^3 z_\sst^{ij\alpha} {S_{LT}}_\alpha}{6} \,\tilde{h}_{1LT}^{\perp(3)}(x,\bm{z}^2) \right] , \\[5pt]
    \tilde{\Gamma}_{TT}^{ij}(x,\bm{z}) &= \frac{xP^+}{2} \left[ - \frac{M^2 \delta_\sst^{ij} z_\sst^{\alpha\beta} {S_{TT}}_{\alpha\beta}}{2} \,\tilde{f}_{1TT}^{(2)}(x,\bm{z}^2) - \frac{i M^2 \epsilon_\sst^{ij} {\epsilon^{\beta}_\sst}_\gamma z_\sst^{\gamma\alpha} {S_{TT}}_{\alpha\beta}}{2} \,\tilde{g}_{1TT}^{(2)}(x,\bm{z}^2) \right. \nn \\[3pt]
    &\quad\, \left. + \,S_{TT}^{ij} \,\tilde{h}_{1TT}(x,\bm{z}^2) - \frac{M^2 {S_{TT}^{\{i}}_\alpha z_\sst^{j\}\alpha}}{2} \,\tilde{h}_{1TT}^{\perp(2)}(x,\bm{z}^2) \right. \nn \\[3pt]
    &\quad\, \left. + \,\frac{M^4 z_\sst^{ij\alpha\beta} {S_{TT}}_{\alpha\beta}}{24} \,\tilde{h}_{1TT}^{\perp\perp(4)}(x,\bm{z}^2) \right] .
\end{align}
The gluon TMDs in $z_\sst$-space are one-to-one related to their $k_\sst$-space counterparts through eq.~\eqref{e:ftilde}.

\end{appendix}



\stopthumb
\addchap[\stackunder{Lay summary}{}]{Lay summary}
\renewcommand{\thefigure}{\arabic{figure}}

\begin{figure}[htb]
\centering
    \includegraphics[height=5.0cm]{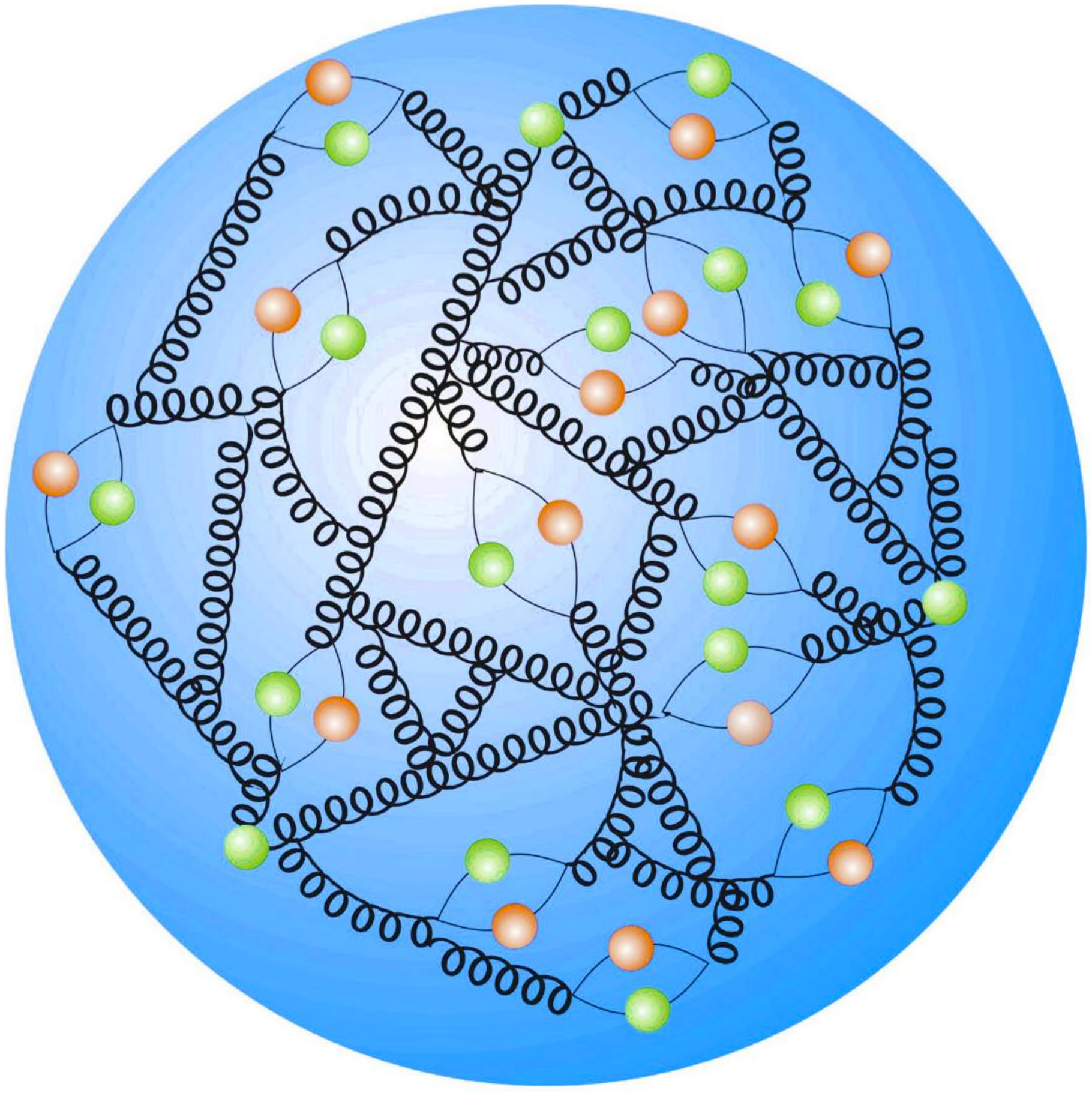}
    \caption{A schematic representation of the internal structure of the proton seen at high energy. In addition to three quarks, one can find many antiquarks and gluons inside the proton. (Source: the DESY website.)}
\label{f:proton_eng}
\end{figure}

In high school one is taught that a proton consists of two up quarks and one down quark. However, the veracity of that claim entirely depends on the resolution that is used to look at the proton. The internal structure of the proton can only be unraveled by studying interactions of the proton with other particles. The higher the energy of the particle that bumps into the proton (a higher energy implies a smaller wavelength), the smaller the length scale that can be studied. In collisions at low energies we see the proton as a whole; it behaves the same as a point particle. As the collision energy increases, the size of the proton plays an increasingly important role. The substructure then appears to be described by three point particles, the quarks. This was first observed in 1968. However, when the proton is involved in a very high-energetic collision, it appears to consist of a `soup' of innumerable quarks, antiquarks, and gluons, see figure~\ref{f:proton_eng}. At high resolution, the proton is an immensely chaotic collection of elementary particles that move around with almost the speed of light and in which quark-antiquark pairs are constantly formed and disappear according to Einstein's well-known formula $E=mc^2$. Contrary to what the picture suggests, there is no physical edge that limits the proton -- the density of the particle soup gradually decreases as the distance to the center increases.

The particles in the proton are called \emph{partons} and their distribution is described by \emph{parton distribution functions} (PDFs). The so-called \emph{collinear} PDFs $f_i(x,Q)$ say something about the probability to find a parton of type $i$ (an up quark, gluon, etc.) with a fraction $x$ of the total proton momentum (in the direction in which the proton moves) at a resolution scale $Q$ inside the proton. Since PDFs depend on a resolution scale, the inner structure of the proton is not a static but rather a dynamic property. You could compare this with how you look when you are lit (and therefore interact) with photons of different wavelengths. When visible light is used, the image is clearly different from an X-ray image in which electromagnetic radiation with a much smaller wavelength is used. That what is `constant' in the proton, which defines the proton, is that the number of up quarks minus the number of anti-up quarks equals $2$ and that the number of down quarks minus the number of antidown quarks equals $1$.

The internal structure of the proton is a consequence of the \emph{strong interaction}. The strong interaction is responsible for the strong nuclear force that holds atomic nuclei together and is one of the four fundamental interactions in nature (the other three are the electromagnetic and weak interactions, and gravity). At the level of a proton, the strength of the strong interaction is respectively $10^2$ and $10^6$ times greater than that of the electromagnetic and weak interactions, hence the indication `strong'. The strong interaction is described by a theory called \emph{quantum chromodynamics} (QCD). This name refers to the Greek word `chroma' that means color. QCD, namely, describes the dynamics of elementary particles that have a \emph{color charge}. Color charge is somewhat analogous to electric charge. Electrically charged particles, such as electrons, interact with each other through the exchange of photons. In addition to an electric charge, quarks also have a color charge, allowing them to interact through the strong interaction. The transmitter of the strong interaction is the gluon, the QCD version of the photon.

An important difference between photons and gluons is that the photon is electrically neutral while the gluon has a color charge. The latter implies that gluons can also engage in the strong interaction with each other. A consequence of this property is that in contrast to the other fundamental interactions the strength of the strong interaction, which is characterized by the coupling parameter $\alpha_s$, increases as the distance between two color charges increases. Since at some point it becomes energetically more favorable to create a quark-antiquark pair than to further separate the color charges, partons only occur in bound states called \emph{hadrons} (protons and neutrons are examples of hadrons). This phenomenon is called \emph{confinement}. The counterpart of confinement is \emph{asymptotic freedom}: at very small length scales, smaller than the proton radius of about $1$ fm $= 10^{-15}$ m, the strong interaction is very weak; so weak that the partons can move almost freely (in that case $\alpha_s$ is very close to zero).

The partons form the (fundamental) degrees of freedom of QCD at high energies, whereas at low energies the (effective) degrees of freedom are formed by the hadrons. We still do not understand how certain features of hadrons can be explained in terms of the partons. For example, it is unclear where exactly the proton mass of $938$ MeV $= 1.67 \times 10^{-27}$ kg comes from. The Higgs mechanism only explains the masses of the elementary quarks that are at most a few MeVs. The remainder of the proton mass, which amounts to about $98$\%, comes from the binding energy in the proton. This energy corresponds to the kinetic and potential energy stored in the `sea' of quarks, antiquarks, and gluons and its size is mainly determined by the strong interaction. When you are standing on a scale, the number before the decimal point roughly corresponds to the total binding energy of all protons and neutrons in your body ($m = E/c^2$); the number after the decimal point comes from your interaction with the Higgs field (and the fact that you do not sink through the scale is the result of the electromagnetic interaction).

Why can we not just calculate the properties of hadrons? After all, we have a theory that describes the strong interaction, namely QCD. It turns out that what can and cannot be calculated with QCD depends on the energy scale involved in the process under consideration. Thanks to asymptotic freedom, high-energy interactions between partons (these interactions are weak) can be calculated using \emph{perturbation theory}. This means that the probability that a certain process occurs, which is characterized by the so-called \emph{scattering cross section}, can be expanded in powers of $\alpha_s$. Since for sufficiently high energies the (dimensionless) parameter $\alpha_s$ is much smaller than $1$, it is usually sufficient to calculate only a few terms in this expansion. Contributions to these terms are graphically represented by the well-known Feynman diagrams. However, at low energies $\alpha_s$ can become greater than $1$. In that case, higher-order terms yield larger contributions than the lowest-order terms -- the expansion does not converge and has thus become meaningless. Since the internal structure of hadrons is shaped by low-energy interactions between partons (the partons are strongly bound) it cannot be calculated with perturbative techniques. This means that we for example cannot calculate the mass of the proton using Feynman diagrams.

As noted earlier, the internal structure of hadrons is described by PDFs. These functions contain information about the dynamics of quarks and gluons inside hadrons and can depend on various kinematic variables. Since PDFs cannot be calculated using perturbative QCD, they must be extracted from collision experiments. In this thesis we study measurable quantities, also called \emph{observables}, that depend on different types of PDFs. By measuring these observables we can learn more about what exactly is happening inside hadrons. Moreover, a greater understanding of the internal structure of the proton for example allows for more accurate descriptions of high-energy collisions between protons in the Large Hadron Collider. The latter can be of great significance for the search for new elementary particles.

\begin{figure}[htb]
\centering
    \includegraphics[height=5.0cm]{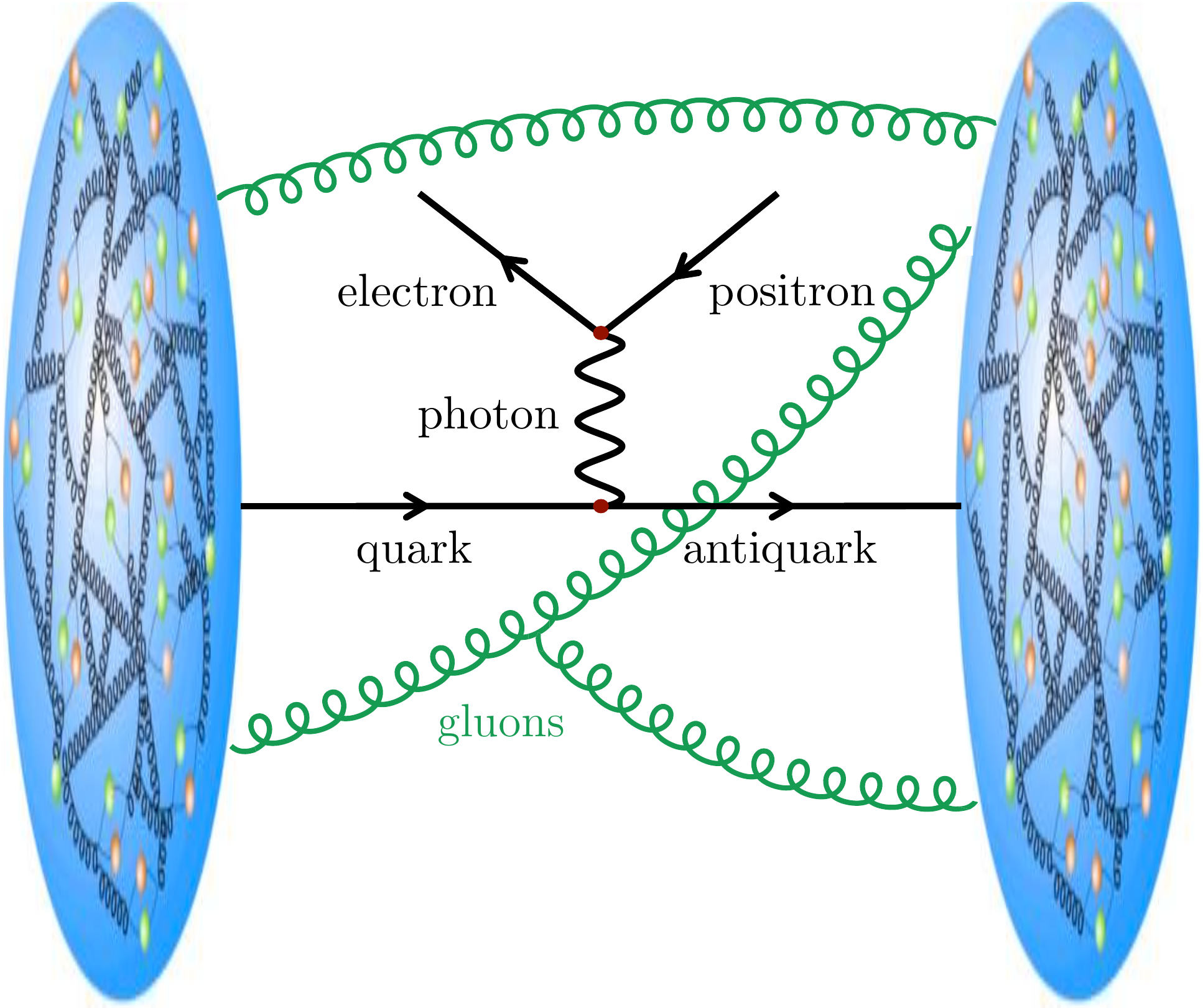}
    \caption{An example of a DY process: a quark and an antiquark from two colliding protons produce a photon that subsequently decays into an electron-positron pair. An arbitrary number of gluons is exchanged between the strongly Lorentz-contracted protons.}
\label{f:DY_ENG}
\end{figure}

An example of an observable that `feels' the internal structure of hadrons is the cross section of the so-called \emph{Drell-Yan} (DY) process. DY scattering is a common scenario in high-energy collisions between hadrons; during such a collision, a quark from one hadron and an antiquark from the other interact with each other, producing an electrically charged lepton-antilepton pair (an example of which is an electron-positron pair), see figure~\ref{f:DY_ENG}. Since the lepton momenta can be measured in particle detectors, the collision can be reconstructed. For the reconstruction it is important to know the probabilities for finding quarks and antiquarks with certain momenta inside hadrons -- this is exactly the information contained in the PDFs. As the interaction between the quark and antiquark takes place at high energy (this is also called the `hard' process) it can be calculated perturbatively. Ultimately, the total cross section of DY scattering is given by the product of the hard scattering cross section with two PDFs (one for the quark and one for the antiquark). The possibility to write a cross section as a product means that the process can be divided into separate contributions that do not `talk' to each other. This is called \emph{factorization}. Factorization is essential for the experimental determination of PDFs and forms the foundation of hadron physics.

However, the property of factorization should not be taken for granted. In order to calculate the total cross section for the DY process, the superposition principle of quantum mechanics requires summing over all possible interactions that could possibly take place, including those between the hadrons. Since for example arbitrarily many low-energy (`soft') gluons can be exchanged between the hadrons, it is a priori unclear that the contributions from the two hadrons can be completely disentangled. In other words, it is unclear whether factorization is permitted. That factorization does hold was proven in the 1980s by Collins, Soper, and Sterman (CSS) for collinear PDFs. The fact that interference effects between the two hadrons are indeed negligible is intuitively easy to understand. After all, as the hadrons move towards each other at almost the speed of light (hence they are relativistic objects), there is length contraction along the direction of movement as well as time dilation of the interactions inside the hadrons. As a result, the hadrons can be considered as two `frozen pancakes' that hardly `feel' each other prior to the collision. Factorization implies that collinear PDFs are \emph{universal}; they pertain to individual hadrons and are completely independent of any hard scattering processes. Hence, collinear PDFs can be measured in one process and then used for the description of hadrons in other processes.

As it turns out, not all types of PDFs are universal. In this thesis we look at \emph{multidimensional} PDFs that in addition to the longitudinal component $x$ also depend on the transverse parton momentum $\bm{k}$. These PDFs contain more information about the internal structure of hadrons than the one-dimensional collinear PDFs and may differ from process to process. The process dependence of multidimensional PDFs can be calculated and isolated in a so-called \emph{gauge link} that depends on the direction in which color charges move in the hard process. An example of such a nonuniversal multidimensional PDF is the so-called \emph{Boer-Mulders} (BM) function $h_{1,i}^{\perp q}(x,\bm{k},Q)$. This (three-dimensional) function quantifies the net transverse spin polarization of a quark or antiquark of type $i$ in an unpolarized hadron. In the DY process, the BM effect generates a very specific asymmetry in the directions of the lepton momenta in the transverse plane. Although this asymmetry in the DY cross section has been measured by various experiments, more data (among other things) is needed for an accurate extraction of the BM function.

In part~\ref{p:1} of this thesis we look at color effects that may influence the factorization property of the BM asymmetry in the DY process. These effects originate from `color-entangled' gluons that are exchanged between the hadrons and can, at worst, give rise to a violation of factorization. Since the BM effect is actively being investigated by various experimental collaborations, it is essential to know how exactly it manifests itself in the cross section. By largely following the steps of the CSS proof, we find that the contribution of the BM effect to the DY cross section factorizes in the `standard way'. This means that the BM asymmetry can be written as a product of a hard scattering cross section with quark and antiquark BM functions, without any complications.

Part~\ref{p:2} deals with multidimensional distributions of gluons inside hadrons. At very high collision energies, the internal structure of hadrons is dominated by gluons; this corresponds to the regime of small values of the momentum component $x$. We introduce several new multidimensional gluon distributions for spin-$1$ hadrons that contain interesting information about correlations between transverse gluon momenta and spin components. Because gluon PDFs are still largely unknown, we also derive limits for maximal values that certain functions may take. For very small values of $x$, different gluon PDFs appear to be related to each other. This results in a simpler description of the distributions of gluons inside hadrons at high energies. Finally, we illustrate with a model calculation that certain asymmetries observed in the angular distributions of two produced hadrons in collisions between protons and lead nuclei may originate from quantum correlations that exist between gluons inside the lead nuclei.

Nowadays we know quite well what the proton looks like from the inside wearing `one-dimensional glasses'. Through a combination of experimental and theoretical efforts since the 1970s, the one-dimensional collinear PDFs have been determined fairly precisely (for not too small values of $x$ that is). Since the last turn of the century, the research field has expanded to include measurements on the multidimensional inner structure of hadrons; a few multidimensional PDFs have already been extracted. In the coming years we expect to learn more about interesting quantum effects in hadrons, an example of which is the BM effect that we investigate in part~\ref{p:1} of this thesis. Moreover, as measurements involve various different processes, we could also test some QCD predictions with regards to the process-dependent gauge links. For the research at small values of $x$, the future Electron-Ion Collider (EIC) could be of great importance. This particle accelerator could collide electrons with polarized protons and light atomic nuclei at very high energy and thus shed new light on the gluonic structure of hadrons. With the EIC, several theoretical predictions from part~\ref{p:2} of this thesis could be experimentally tested. In short, the research into the properties of the particle soup inside hadrons still has a long road ahead. However, we find ourselves in interesting times, gradually unraveling the rich multidimensional inner structure of hadrons through which we learn more about the most important building blocks of our (visible) universe.

\selectlanguage{dutch}
\addchap[\stackunder{Lekensamenvatting}{}]{Lekensamenvatting}

\begin{figure}[htb]
\centering
    \includegraphics[height=5.0cm]{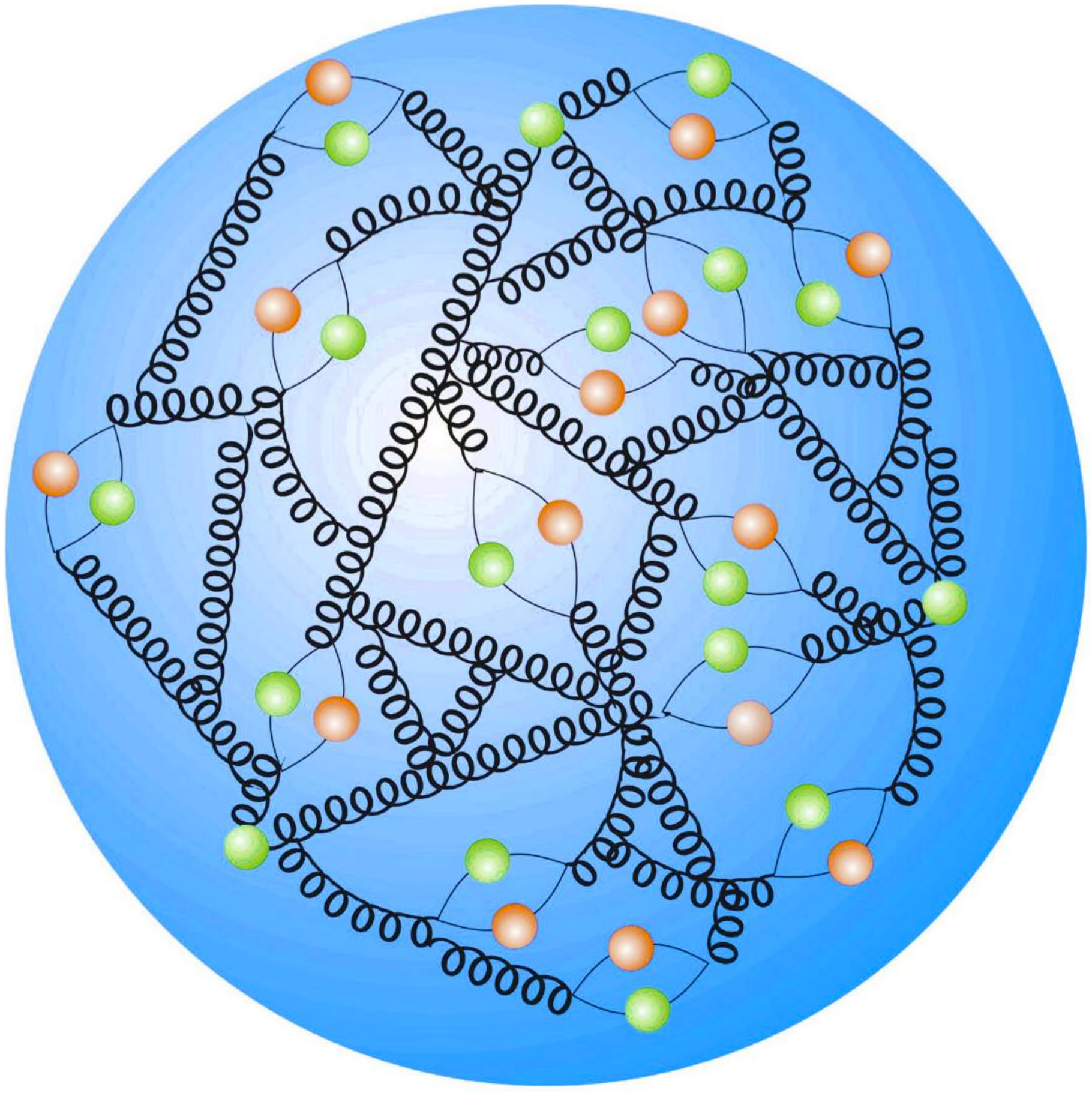}
    \caption{Een schematische weergave van de interne structuur van het proton gezien bij hoge energie. Naast drie quarks kun je in het proton vele antiquarks en gluonen aantreffen. (Bron: de DESY website.)}
\label{f:proton}
\end{figure}

Op de middelbare school leer je dat een proton bestaat uit twee upquarks en \'e\'en downquark. Het waarheidsgehalte van die bewering hangt echter helemaal af van de resolutie waarmee je naar het proton kijkt. De interne structuur van het proton kan alleen worden ontrafeld door interacties te bestuderen van het proton met andere deeltjes. Hoe hoger de energie van het deeltje dat op het proton invalt (een hogere energie betekent een kleinere golflengte), des te kleiner is de lengteschaal die kan worden bestudeerd. In een botsing bij lage energie zien we het proton in z'n geheel; het gedraagt zich hetzelfde als een puntdeeltje. Naarmate de botsingsenergie wordt opgevoerd, gaat de afmeting van het proton een steeds grotere rol spelen. De substructuur blijkt dan te kunnen worden beschreven door drie puntdeeltjes, de quarks. Dit werd voor het eerst waargenomen in 1968. Echter, wanneer het proton betrokken is bij een zeer hoog-energetische botsing, dan blijkt het te bestaan uit een `soep' van ontelbaar veel quarks, antiquarks en gluonen, zie figuur~\ref{f:proton}. Bij hoge resolutie is het proton een immens chaotische collectie van elementaire deeltjes die rondrazen met bijna de lichtsnelheid en waarin voortdurend quark-antiquark paren worden gevormd en weer verdwijnen volgens Einsteins bekende formule $E=mc^2$. In tegenstelling tot wat het plaatje suggereert is er geen fysieke rand die het proton begrenst -- de dichtheid van de deeltjessoep neemt geleidelijk af naarmate de afstand tot het centrum groter wordt.

De deeltjes in het proton worden \emph{partonen} genoemd en hun verdeling wordt beschreven door \emph{partondistributiefuncties} (PDFs). De zogenaamde \emph{collineaire} PDFs $f_i(x,Q)$ zeggen iets over de waarschijnlijkheid om bij een resolutie $Q$ een parton van het type $i$ (een upquark, gluon etc.) met een fractie $x$ van de totale protonimpuls (in de richting waarin het proton beweegt) in het proton aan te treffen. Aangezien PDFs afhangen van een resolutieschaal, is de inwendige structuur van het proton niet een statische maar een dynamische eigenschap. Je zou dit kunnen vergelijken met hoe jijzelf eruit ziet als je wordt beschenen (en dus interacties aangaat) met fotonen van verschillende golflengtes. Bij gebruik van zichtbaar licht is het plaatje duidelijk anders dan op een r\"ontgenfoto waarbij gebruik is gemaakt van elektromagnetische straling met een veel kleinere golflengte. De `constante' in het proton, datgene wat het proton definieert, is dat het aantal upquarks minus het aantal anti-upquarks gelijk is aan $2$ en dat het aantal downquarks minus het aantal antidownquarks gelijk is aan $1$.

De interne structuur van het proton is een gevolg van de \emph{sterke interactie}. De sterke interactie is verantwoordelijk voor de sterke kernkracht die atoomkernen bijeenhoudt en is een van de vier fundamentale interacties in de natuur (de andere drie zijn de elektromagnetische en zwakke interactie en gravitatie). Op het niveau van een proton is de kracht van de sterke interactie respectievelijk $10^2$ en $10^6$ keer groter dan die van de elektromagnetische en zwakke interactie, vandaar de aanduiding `sterk'. De sterke interactie wordt beschreven door een theorie die \emph{kwantumchromodynamica} (QCD) heet. Deze naam refereert aan het Griekse woord `chroma' dat kleur betekent. QCD beschrijft namelijk de dynamica van elementaire deeltjes die een \emph{kleurlading} hebben. Kleurlading is enigszins analoog aan elektrische lading. Elektrisch geladen deeltjes, zoals elektronen, interageren met elkaar door de uitwisseling van fotonen. Quarks hebben naast een elektrische lading ook een kleurlading. Hierdoor kunnen ze ook interageren via de sterke interactie. De overbrenger van de sterke interactie is het gluon, de QCD-versie van het foton.

Een belangrijk verschil tussen fotonen en gluonen is dat het foton elektrisch neutraal is terwijl het gluon een kleurlading heeft. Dit laatste impliceert dat gluonen ook met zichzelf de sterke interactie kunnen aangaan. Deze eigenschap heeft tot gevolg dat -- in tegenstelling tot de andere fundamentele interacties -- de sterkte van de sterke interactie, die gekarakteriseerd wordt door de koppelingsparameter $\alpha_s$, toeneemt naarmate de afstand tussen twee kleurladingen groter wordt. Omdat het op een gegeven moment energetisch gunstiger is om een quark-antiquark paar te cre\"eren dan om de kleurladingen verder uit elkaar te brengen, komen partonen alleen voor in gebonden toestanden genaamd \emph{hadronen} (protonen en neutronen zijn voorbeelden van hadronen). Dit fenomeen wordt \emph{confinement} genoemd. De tegenhanger van confinement is \emph{asymptotische vrijheid}: op zeer kleine lengteschalen, kleiner dan de protonstraal van ca.\ $1$ fm $= 10^{-15}$ m, is de sterke interactie erg zwak; zo zwak zelfs dat de partonen zich nagenoeg vrij kunnen bewegen ($\alpha_s$ is dan bijna gelijk aan nul).

De partonen vormen de (fundamentele) vrijheidsgraden van QCD bij hoge energie\"en terwijl bij lage energie\"en de (effectieve) vrijheidsgraden worden gevormd door de hadronen. We begrijpen nog altijd niet goed hoe bepaalde eigenschappen van hadronen kunnen worden verklaard in termen van de partonen. Zo is het bijvoorbeeld onduidelijk waar de protonmassa van $938$ MeV $= 1,67 \times 10^{-27}$ kg precies vandaan komt. Het Higgsmechanisme verklaart enkel de massa's van de elementaire quarks die hooguit een paar MeV bedragen. De rest van de protonmassa, ca.\ $98$\%, komt van de bindingsenergie in het proton. Deze energie correspondeert met de kinetische en potenti\"ele energie die zit opgeslagen in de `zee' van quarks, antiquarks en gluonen en de grootte ervan wordt voornamelijk bepaald door de sterke interactie. Wanneer je op de weegschaal staat dan komt het getal v\'o\'or de komma ruwweg overeen met de totale bindingsenergie van alle protonen en neutronen in je lichaam ($m = E/c^2$); het getal achter de komma komt van je interactie met het Higgsveld (en het feit dat je niet door de weegschaal zakt is het gevolg van de elektromagnetische wisselwerking).

Waarom kunnen we de eigenschappen van hadronen niet gewoon uitrekenen? We hebben immers een theorie die de sterke interactie beschrijft, namelijk QCD. Het blijkt dat wat wel en wat niet met QCD kan worden uitgerekend afhangt van de energieschaal die gemoeid gaat met het proces dat we bekijken. Dankzij asymptotische vrijheid kunnen hoog-energetische interacties tussen partonen (deze interacties zijn zwak) worden uitgerekend met behulp van \emph{perturbatietheorie}. Dit houdt in dat de waarschijnlijkheid dat een bepaald proces zich voordoet, gekarakteriseerd door de zogenaamde \emph{botsingsdoorsnede}, kan worden ge\"expandeerd in machten van $\alpha_s$. Omdat bij voldoende hoge energie\"en de (dimensieloze) parameter $\alpha_s$ veel kleiner is dan $1$, volstaat het meestal om slechts enkele termen in deze expansie uit te rekenen. Bijdrages aan deze termen worden grafisch weergegeven met de bekende Feynmandiagrammen. Echter, bij lage energie\"en kan $\alpha_s$ groter worden dan $1$. In dat geval leveren hogere-orde termen grotere bijdrages dan de laagste-orde termen -- de expansie convergeert niet en is daarom betekenisloos geworden. Aangezien de interne structuur van hadronen wordt vormgegeven door laag-energetische interacties tussen partonen (de partonen zijn immers sterk gebonden) kan deze niet worden doorgerekend met perturbatieve technieken. Dit betekent dat we bijvoorbeeld de massa van het proton niet met behulp van Feynmandiagrammen kunnen uitrekenen.

Zoals al eerder opgemerkt wordt de interne structuur van hadronen beschreven door PDFs. Deze functies bevatten informatie over de dynamica van quarks en gluonen in hadronen en kunnen van verschillende kinematische variabelen afhangen. Aangezien PDFs niet kunnen worden uitgerekend met behulp van perturbatieve QCD, moeten ze uit metingen aan botsingsexperimenten worden gehaald. In dit proefschrift bestuderen we meetbare grootheden, ook wel \emph{observabelen} genoemd, die afhangen van verschillende typen PDFs. Door het meten van die observabelen kunnen we meer te weten komen over wat er zich precies afspeelt in hadronen. Bovendien stelt een groter begrip van de interne structuur van het proton ons in staat om bijvoorbeeld hoog-energetische botsingen tussen protonen in de Large Hadron Collider nauwkeuriger te kunnen beschrijven. Dit laatste kan van grote betekenis zijn voor de zoektocht naar nieuwe elementaire deeltjes.

\begin{figure}[htb]
\centering
    \includegraphics[height=5.0cm]{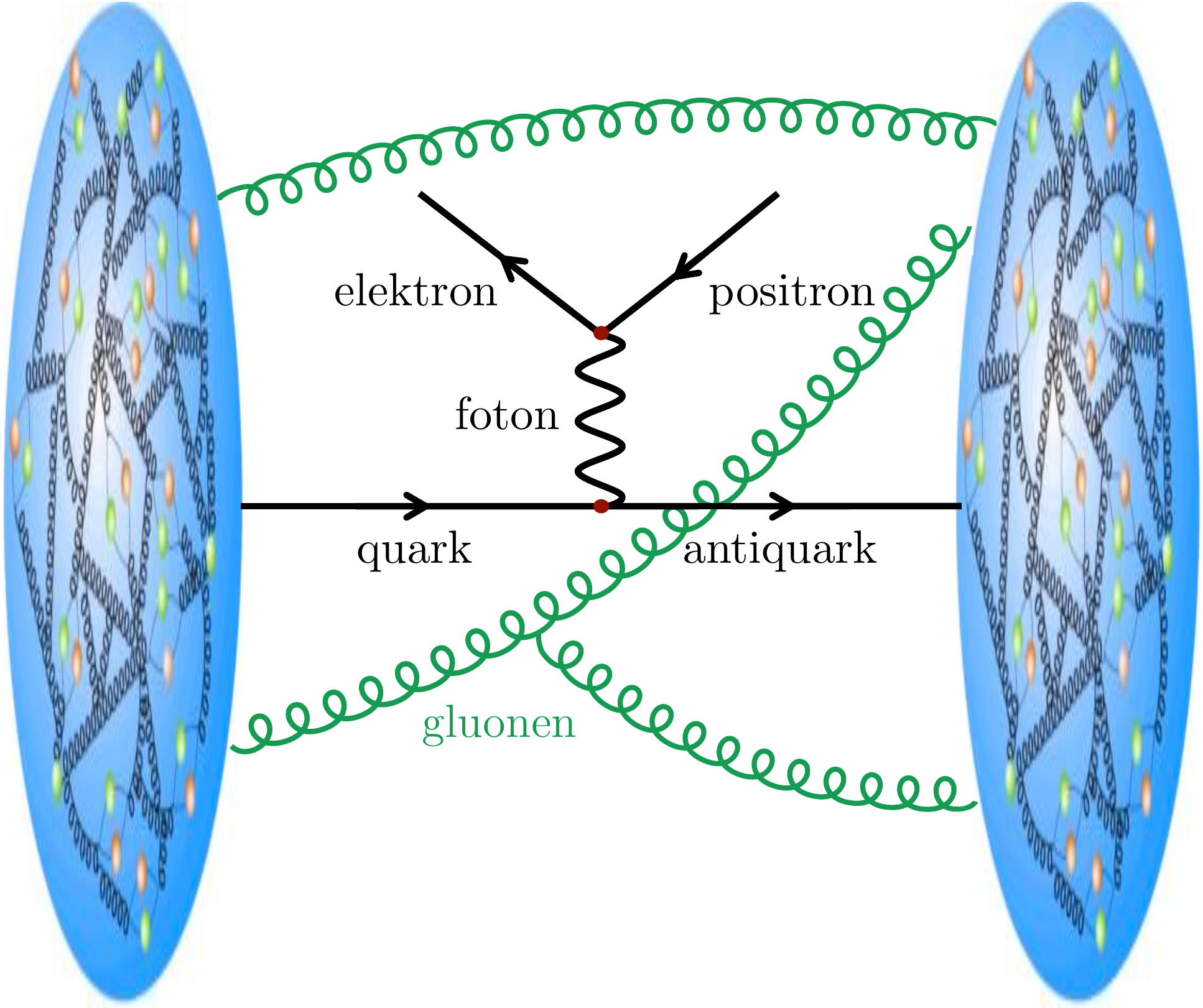}
    \caption{Een voorbeeld van een DY proces: een quark en antiquark uit twee botsende protonen produceren een foton dat vervolgens vervalt in een elektron-positron paar. De sterk Lorentz-gecontraheerde protonen wisselen een willekeurig aantal gluonen met elkaar uit.}
\label{f:DY_NL}
\end{figure}

Een voorbeeld van een observabele die de interne structuur van hadronen `voelt' is de botsingsdoorsnede van het zogenaamde \emph{Drell-Yan} (DY) proces. DY verstrooiing is een veelvoorkomend scenario in hoog-energetische botsingen tussen hadronen; tijdens zo'n botsing gaan een quark uit het ene en een antiquark uit het andere hadron een interactie met elkaar aan en produceren vervolgens een elektrisch geladen lepton-antilepton paar (een voorbeeld hiervan is een elektron-positron paar), zie figuur~\ref{f:DY_NL}. Doordat de impulsen van de leptonen gemeten kunnen worden in deeltjesdetectoren kan de botsing worden gereconstrueerd. Bij de reconstructie is het van belang met welke kans het quark en het antiquark met een bepaalde impuls in de hadronen voorkomen -- dit is precies de informatie die bevat is in de PDFs. Aangezien de interactie tussen het quark en antiquark plaatsvindt bij hoge energie (dit wordt ook wel het `harde' proces genoemd) kan deze perturbatief worden uitgerekend. De totale botsingsdoorsnede van DY verstrooiing wordt uiteindelijk gegeven door het product van de harde botsingsdoorsnede met twee PDFs (eentje voor het quark en eentje voor het antiquark). De mogelijkheid om een botsingsdoorsnede te schrijven als een product betekent dat het proces kan worden opgesplitst in afzonderlijke bijdrages die niet met elkaar `praten'. Dit wordt \emph{factorisatie} genoemd. Factorisatie is essentieel voor de experimentele bepaling van PDFs en vormt het fundament van de hadronfysica.

De eigenschap van factorisatie is echter niet vanzelfsprekend. Om de totale botsingsdoorsnede van het DY proces uit te rekenen, moet er volgens het superpositieprincipe in de kwantummechanica gesommeerd worden over alle mogelijke interacties die zouden kunnen plaatsvinden, ook die tussen de hadronen. Aangezien er bijvoorbeeld willekeurig veel laag-energetische (`zachte') gluonen tussen de hadronen kunnen worden uitgewisseld, is het a priori onduidelijk dat de bijdrages van de twee hadronen volledig opgesplitst kunnen worden. Met andere woorden, het is maar de vraag of dat factorisatie is toegestaan. Dat dit wel zo is, werd in de jaren 1980 bewezen door Collins, Soper en Sterman (CSS) voor collineaire PDFs. Het feit dat interferentie-effecten tussen de twee hadronen inderdaad verwaarloosbaar zijn is intuitief goed te begrijpen. Immers, omdat de hadronen met bijna de lichtsnelheid op elkaar afbewegen (het zijn dus relativistische objecten), vindt er lengtecontractie plaats in de bewegingsrichting alsook tijddilatatie van de interacties in de hadronen. Hierdoor kunnen de hadronen worden beschouwd als twee `bevroren pannenkoeken' die elkaar nauwelijks `voelen' voorafgaand aan de botsing. Factorisatie impliceert dat collineaire PDFs \emph{universeel} zijn; ze horen bij individuele hadronen en zijn volledig onafhankelijk van harde processen. Collineaire PDFs kunnen dus in een bepaald proces worden gemeten en vervolgens gebruikt worden voor de beschrijving van hadronen in andere processen.

Het blijkt dat universaliteit geen eigenschap is van alle soorten PDFs. In dit proefschrift kijken we naar \emph{multidimensionale} PDFs die naast de longitudinale component $x$ ook afhangen van de transversale partonimpuls $\bm{k}$. Deze PDFs bevatten meer informatie over de interne structuur van hadronen dan de eendimensionale collineaire PDFs en kunnen verschillen van proces tot proces. De procesafhankelijkheid van multidimensionale PDFs kan worden uitgerekend en ge\"isoleerd in een zogenaamde \emph{ijklink} die afhangt van de richting waarin kleurladingen zich verplaatsen in het harde proces. Een voorbeeld van zo'n niet-universele, multidimensionale PDF is de zogenaamde \emph{Boer-Mulders} (BM) functie $h_{1,i}^{\perp q}(x,\bm{k},Q)$. Deze (driedimensionale) functie kwantificeert de netto transversale spinpolarisatie van een quark of antiquark van het type $i$ in een ongepolariseerd hadron. Het BM effect genereert in het DY proces een heel specifieke asymmetrie in de richtingen van de leptonimpulsen in het transversale vlak. Hoewel deze asymmetrie in de DY botsingsdoorsnede door verschillende experimenten is gemeten, is voor een nauwkeurige extractie van de BM functie (onder andere) meer data nodig.

In deel~\ref{p:1} van dit proefschrift kijken we naar kleureffecten die mogelijk de factorisatie-eigenschap van de BM asymmetrie in het DY proces be\"invloeden. Deze effecten zijn afkomstig van `kleurverstrengelde' gluonen die tussen de hadronen worden uitgewisseld en kunnen in het ergste geval aanleiding geven tot een schending van factorisatie. Aangezien het BM effect door verschillende experimentele collaboraties actief wordt onderzocht, is het belangrijk om te weten hoe het zich precies manisfesteert in de botsingsdoorsnede. Door grotendeels de stappen in het CSS bewijs te volgen, vinden we dat de bijdrage van het BM effect aan de DY botsingsdoorsnede op de `standaardwijze' factoriseert. Dit houdt in dat de BM asymmetrie geschreven kan worden als een product van een harde botsingsdoorsnede met quark en antiquark BM functies, zonder extra complicaties.

Deel~\ref{p:2} gaat over multidimensionale distributies van gluonen in hadronen. Bij heel hoge botsingsenergie\"en wordt de interne structuur van hadronen gedomineerd door gluonen; dit correspondeert met het regime van kleine waardes van de impulscomponent $x$. We introduceren verschillende nieuwe multidimensionale gluondistributies voor spin-$1$ hadronen die interessante informatie bevatten over correlaties tussen transversale gluonimpulsen en spincomponenten. Omdat gluon PDFs nog grotendeels onbekend zijn, leiden we ook limieten af voor maximale waardes die bepaalde functies mogen aannemen. Voor heel kleine waardes van $x$ blijken verschillende gluon PDFs aan elkaar gerelateerd te zijn. Dit resulteert in een simpelere beschrijving van de distributies van gluonen in hadronen bij hoge energie\"en. Tot slot illustreren we met een modelberekening dat bepaalde asymmetrie\"en die zijn waargenomen in de hoekverdeling van twee geproduceerde hadronen in botsingen tussen protonen en loodkernen mogelijk hun oorsprong vinden in kwantumcorrelaties die bestaan tussen gluonen in de loodkernen.

Tegenwoordig weten we redelijk goed hoe het proton er met een `eendimensionale bril' op vanbinnen uitziet. Door een combinatie van experimenteel en theoretisch onderzoek sinds de jaren 1970 zijn namelijk de eendimensionale collineaire PDFs -- voor niet al te kleine waardes van $x$ -- vrij precies bekend. Sinds de laatste eeuwwisseling heeft het onderzoeksveld zich uitgebreid naar metingen aan de multidimensionale interne structuur van hadronen en inmiddels zijn er ook een paar multidimensionale PDFs bekend. We verwachten de komende jaren meer te zullen leren over interessante kwantumeffecten in hadronen zoals het BM effect dat we onderzoeken in deel~\ref{p:1} van dit proefschrift. Aangezien er metingen worden verricht aan verschillende botsingsprocessen, zouden we tevens enkele QCD voorspellingen met betrekking tot de procesafhankelijke ijklinks kunnen testen. Voor het onderzoek bij kleine waardes van $x$ zou de toekomstige Electron-Ion Collider (EIC) van groot belang kunnen zijn. Deze deeltjesversneller zou bij heel hoge energie elektronen op gepolariseerde protonen en lichte atoomkernen kunnen laten botsen en daarmee nieuw licht kunnen werpen op de gluonstructuur van hadronen. Met de EIC zouden verschillende theoretische voorspellingen uit deel~\ref{p:2} van dit proefschrift experimenteel getoetst kunnen worden. Kortom, het onderzoek naar de eigenschappen van de deeltjessoep in hadronen heeft nog een lange weg te gaan. Echter, we bevinden ons in een interessante tijd waarin we stapje voor stapje de rijke multidimensionale interne structuur van hadronen ontrafelen en daarmee meer leren over de belangrijkste bouwstenen van ons (zichtbare) universum.

\selectlanguage{english}
\addchap[\stackunder{Academic biography}{}]{Academic biography}

In 2008, Tom enrolled in the Bachelor's program of Physics \& Astronomy at Radboud University Nijmegen. After obtaining his B.Sc.\ degree in 2011, he continued with a Master's specialization in Theoretical Physics at the same university. In 2012, he spent a summer at CERN in Geneva, Switzerland and contributed to the search for dark matter particles at the ATLAS experiment. A year later, he participated in the International Student Exchange Program and spent one semester at New Mexico State University in Las Cruces, United States. During the final year of his Master's program in Nijmegen he conducted research at the department of Theoretical High-Energy Physics on renormalization group invariant quantities in supersymmetric models under the supervision of Prof.\ dr.\ W.J.P.\ Beenakker. Finally, in 2014, he obtained his M.Sc.\ degree cum laude.

After his graduation, Tom decided to pursue a Ph.D.\ degree at the National Institute for Subatomic Physics in Amsterdam under the supervision of Prof.\ dr.\ D.\ Boer and Prof.\ dr.\ P.J.G.\ Mulders. This thesis is the result of four years of theoretical research on the multidimensional inner structure of hadrons, which is shaped by the strong interaction. Besides research, Tom was involved as a teaching assistant for various courses at VU University Amsterdam and he was a member of the PhD National Student Council.

\addchap[\stackunder{Acknowledgments}{}]{Acknowledgments}
\pagestyle{plain}

This work would not have been possible without the help and support of various people, both at a personal and a professional level.

First and foremost, I would like to thank my supervisors Dani\"el Boer and Piet Mulders. Dani\"el and Piet, it was a great pleasure to be part of your groups. You gave me the opportunity to work on interesting projects and encouraged me to attend multiple national and international schools, conferences, and workshops, which helped me greatly to develop myself as a scientist. I also truly appreciate all the trust and freedom that I was granted right from the start and the useful advice you gave me throughout the years. Dani\"el, I always very much enjoyed my visits to Groningen where I would learn a lot from your intuitive explanations of complicated concepts. Your structured way of working was a delight for me and your detailed reading of this thesis was very helpful. Piet, your enthusiasm is contagious and continues to inspire me. Your door was always open and no matter how busy you were, you would always make time for me. And for questions on the Boer-Mulders effect, who better to ask than either of you!

I also highly benefited from three `mentors' during my PhD: Tomas Kasemets, Jonathan Gaunt, and Elena Petreska. Tomas, Jo, and Elena, it was a great pleasure to work together with you and this thesis would not have been remotely the same if it weren't for you. Tomas, as my longest-overlapping office mate you assisted me in various ways. I appreciate your help in getting me started in the field of TMDs and your guidance in the DY calculations. Jo, although Glaubers are ``a pain in the~\ldots'', I believe they do give that extra spark to the results in our paper. I learned a lot from discussions with you about factorization and your feedback on part~\ref{p:1} of this thesis was very useful. Elena, you introduced me to the wonderful world of small-$x$ physics for which I am very grateful. Your expertise in the CGC project was vital, as well as your feedback on part~\ref{p:2} of this thesis. All three of you are great scientists and I wish you the best of luck in your future careers.

Furthermore, I enjoyed very interesting and fruitful collaborations with Andrea Signori, Sabrina Cotogno, and Ya-Jin Zhou, and further benefited from discussions with Asmita Mukherjee, Cristian Pisano, Florent Scarpa, Jian Zhou, Maarten Buffing, and Miguel Garc\'ia Echevarr\'ia. I am indebted to Michael Engelhardt and Matthias Burkardt for their hospitality at New Mexico State University.

I also would like to thank Andreas Sch\"afer, Ignazio Scimemi, and Rob Timmermans for their careful examination of this manuscript as members of the assessment committee.

The Nikhef theory group provided a very pleasant and stimulating work environment. Thank you Eric, Robert, Giulio, Ruben, Jordy, Bert, Gillian, Pedro, Jake, Eleftheria, Ben, Kristof, Jorinde, Lorenzo, Juan, Elisa, Michael W., Mathias, Evangelos, Adolfo, Darren, Emanuele, Wouter, Kalliopi, Melissa, Jort, Jan-Willem, Lisa, Nathan, Giuseppe, Marieke, Robbert, Eleni, Sonia, Wim, Satish, Jos, Gilberto, Leonardo, Daniela, Takahiro, Marco, Valerio, Franz, Andreas, Rabah, Michael B., Jacopo, and Domenico for the wonderful time I have had at Nikhef over the past four years. Among other things, the brewery meet-ups and the theory days out were highly enjoyable.

On a more personal level, I would like to thank all my friends from my times in Boxmeer, Nijmegen, Geneva, Las Cruces, and Amsterdam. Without you, life would be substantially less interesting. Roy, thanks a lot for your help with the design of the front cover, and Adriaan thank you for the nice `weekly' work-outs and your support throughout the years.

I am very grateful for the love and support from my family spread across the Netherlands, Mexico, and the United States. Muchas gracias Laura, Sheny, Leia, y Luc\'ia por todos los lindos recuerdos y por siempre darme buenas vibras. Lotte, Guus en Elke, bedankt voor alle mooie momenten door de jaren heen. Pap en mam, bedankt voor alles wat jullie voor me hebben gedaan. Without you in my life I would have never made it to where I am today.

Finally, I would like to thank Karen for adding so much color to my life. You moved across the Atlantic for me and your unconditional love and support have benefited me greatly.

\bibliographystyle{JHEP}
\bibliography{references.bib}

\providecommand{\href}[2]{#2}\begingroup\raggedright\begin{thebibliography}{100}

\bibitem{Boer:2016xqr}
D.~Boer, S.~Cotogno, T.~van Daal, P.~J. Mulders, A.~Signori and Y.-J. Zhou,
  \emph{{Gluon and Wilson loop TMDs for hadrons of spin $\leq 1$}},
  \href{http://dx.doi.org/10.1007/JHEP10(2016)013}{\emph{JHEP} {\bf 10} (2016)
  013}, [\href{http://arxiv.org/abs/1607.01654}{{\tt 1607.01654}}].

\bibitem{Boer:2017hqr}
D.~Boer, T.~van Daal, J.~R. Gaunt, T.~Kasemets and P.~J. Mulders, \emph{{Colour
  unwound -- disentangling colours for azimuthal asymmetries in Drell-Yan
  scattering}},
  \href{http://dx.doi.org/10.21468/SciPostPhys.3.6.040}{\emph{SciPost Phys.}
  {\bf 3} (2017) 040}, [\href{http://arxiv.org/abs/1709.04935}{{\tt
  1709.04935}}].

\bibitem{Cotogno:2017puy}
S.~Cotogno, T.~van Daal and P.~J. Mulders, \emph{{Positivity bounds on gluon
  TMDs for hadrons of spin $\le 1$}},
  \href{http://dx.doi.org/10.1007/JHEP11(2017)185}{\emph{JHEP} {\bf 11} (2017)
  185}, [\href{http://arxiv.org/abs/1709.07827}{{\tt 1709.07827}}].

\bibitem{Boer:2018vdi}
D.~Boer, T.~van Daal, P.~J. Mulders and E.~Petreska, \emph{{Directed flow from
  C-odd gluon correlations at small $x$}},
  \href{http://dx.doi.org/10.1007/JHEP07(2018)140}{\emph{JHEP} {\bf 07} (2018)
  140}, [\href{http://arxiv.org/abs/1805.05219}{{\tt 1805.05219}}].

\bibitem{GellMann:1964nj}
M.~Gell-Mann, \emph{{A Schematic Model of Baryons and Mesons}},
  \href{http://dx.doi.org/10.1016/S0031-9163(64)92001-3}{\emph{Phys. Lett.}
  {\bf 8} (1964) 214--215}.

\bibitem{Zweig:1981pd}
G.~Zweig, \emph{{An $\text{SU}(3)$ model for strong interaction symmetry and
  its breaking. Version 1}},
\newblock 1964.

\bibitem{Zweig:1964jf}
G.~Zweig, \emph{{An $\text{SU}(3)$ model for strong interaction symmetry and
  its breaking. Version 2}},  in \emph{Developments in the Quark Theory of
  Hadrons, Volume 1. 1964 - 1978} (D.~Lichtenberg and S.~P. Rosen, eds.),
  pp.~22--101.
\newblock 1964.

\bibitem{Bloom:1969kc}
E.~D. Bloom et~al., \emph{{High-Energy Inelastic $e p$ Scattering at
  $6$-Degrees and $10$-Degrees}},
  \href{http://dx.doi.org/10.1103/PhysRevLett.23.930}{\emph{Phys. Rev. Lett.}
  {\bf 23} (1969) 930--934}.

\bibitem{Breidenbach:1969kd}
M.~Breidenbach, J.~I. Friedman, H.~W. Kendall, E.~D. Bloom, D.~H. Coward, H.~C.
  DeStaebler et~al., \emph{{Observed Behavior of Highly Inelastic
  Electron-Proton Scattering}},
  \href{http://dx.doi.org/10.1103/PhysRevLett.23.935}{\emph{Phys. Rev. Lett.}
  {\bf 23} (1969) 935--939}.

\bibitem{Kobayashi:1973fv}
M.~Kobayashi and T.~Maskawa, \emph{{CP Violation in the Renormalizable Theory
  of Weak Interaction}},
  \href{http://dx.doi.org/10.1143/PTP.49.652}{\emph{Prog. Theor. Phys.} {\bf
  49} (1973) 652--657}.

\bibitem{Abe:1995hr}
{\scshape CDF} collaboration, F.~Abe et~al., \emph{{Observation of top quark
  production in $p\bar{p}$ collisions}},
  \href{http://dx.doi.org/10.1103/PhysRevLett.74.2626}{\emph{Phys. Rev. Lett.}
  {\bf 74} (1995) 2626--2631}, [\href{http://arxiv.org/abs/hep-ex/9503002}{{\tt
  hep-ex/9503002}}].

\bibitem{Brandelik:1979bd}
{\scshape TASSO} collaboration, R.~Brandelik et~al., \emph{{Evidence for Planar
  Events in $e^+ e^-$ Annihilation at High-Energies}},
  \href{http://dx.doi.org/10.1016/0370-2693(79)90830-X}{\emph{Phys. Lett.} {\bf
  86B} (1979) 243--249}.

\bibitem{Barber:1979yr}
D.~P. Barber et~al., \emph{{Discovery of Three Jet Events and a Test of Quantum
  Chromodynamics at PETRA Energies}},
  \href{http://dx.doi.org/10.1103/PhysRevLett.43.830}{\emph{Phys. Rev. Lett.}
  {\bf 43} (1979) 830}.

\bibitem{Berger:1979cj}
{\scshape PLUTO} collaboration, C.~Berger et~al., \emph{{Evidence for Gluon
  Bremsstrahlung in $e^+ e^-$ Annihilations at High-Energies}},
  \href{http://dx.doi.org/10.1016/0370-2693(79)90869-4}{\emph{Phys. Lett.} {\bf
  86B} (1979) 418--425}.

\bibitem{Bartel:1979ut}
{\scshape JADE} collaboration, W.~Bartel et~al., \emph{{Observation of Planar
  Three Jet Events in $e^+ e^-$ Annihilation and Evidence for Gluon
  Bremsstrahlung}},
  \href{http://dx.doi.org/10.1016/0370-2693(80)90680-2}{\emph{Phys. Lett.} {\bf
  91B} (1980) 142--147}.

\bibitem{Gross:1973id}
D.~J. Gross and F.~Wilczek, \emph{{Ultraviolet Behavior of Nonabelian Gauge
  Theories}}, \href{http://dx.doi.org/10.1103/PhysRevLett.30.1343}{\emph{Phys.
  Rev. Lett.} {\bf 30} (1973) 1343--1346}.

\bibitem{Politzer:1973fx}
H.~D. Politzer, \emph{{Reliable Perturbative Results for Strong
  Interactions?}},
  \href{http://dx.doi.org/10.1103/PhysRevLett.30.1346}{\emph{Phys. Rev. Lett.}
  {\bf 30} (1973) 1346--1349}.

\bibitem{Feynman:1969ej}
R.~P. Feynman, \emph{{Very high-energy collisions of hadrons}},
  \href{http://dx.doi.org/10.1103/PhysRevLett.23.1415}{\emph{Phys. Rev. Lett.}
  {\bf 23} (1969) 1415--1417}.

\bibitem{Fritzsch:1973pi}
H.~Fritzsch, M.~Gell-Mann and H.~Leutwyler, \emph{{Advantages of the Color
  Octet Gluon Picture}},
  \href{http://dx.doi.org/10.1016/0370-2693(73)90625-4}{\emph{Phys. Lett.} {\bf
  47B} (1973) 365--368}.

\bibitem{GellMann:1962xb}
M.~Gell-Mann, \emph{{Symmetries of baryons and mesons}},
  \href{http://dx.doi.org/10.1103/PhysRev.125.1067}{\emph{Phys. Rev.} {\bf 125}
  (1962) 1067--1084}.

\bibitem{Baikov:2016tgj}
P.~A. Baikov, K.~G. Chetyrkin and J.~H. Kühn, \emph{{Five-Loop Running of the
  QCD coupling constant}},
  \href{http://dx.doi.org/10.1103/PhysRevLett.118.082002}{\emph{Phys. Rev.
  Lett.} {\bf 118} (2017) 082002}, [\href{http://arxiv.org/abs/1606.08659}{{\tt
  1606.08659}}].

\bibitem{Herzog:2017ohr}
F.~Herzog, B.~Ruijl, T.~Ueda, J.~A.~M. Vermaseren and A.~Vogt, \emph{{The
  five-loop beta function of Yang-Mills theory with fermions}},
  \href{http://dx.doi.org/10.1007/JHEP02(2017)090}{\emph{JHEP} {\bf 02} (2017)
  090}, [\href{http://arxiv.org/abs/1701.01404}{{\tt 1701.01404}}].

\bibitem{Patrignani:2016xqp}
{\scshape Particle Data Group} collaboration, C.~Patrignani et~al.,
  \emph{{Review of Particle Physics}},
  \href{http://dx.doi.org/10.1088/1674-1137/40/10/100001}{\emph{Chin. Phys.}
  {\bf C40} (2016) 100001}.

\bibitem{Zee:1973gn}
A.~Zee, \emph{{Study of the renormalization group for small coupling
  constants}}, \href{http://dx.doi.org/10.1103/PhysRevD.7.3630}{\emph{Phys.
  Rev.} {\bf D7} (1973) 3630--3636}.

\bibitem{Coleman:1973jx}
S.~R. Coleman and E.~J. Weinberg, \emph{{Radiative Corrections as the Origin of
  Spontaneous Symmetry Breaking}},
  \href{http://dx.doi.org/10.1103/PhysRevD.7.1888}{\emph{Phys. Rev.} {\bf D7}
  (1973) 1888--1910}.

\bibitem{Signori:2016lvd}
A.~Signori, \emph{{Flavor and Evolution Effects in TMD Phenomenology}}.
\newblock PhD thesis, Vrije U., Amsterdam, 2016.

\bibitem{Gribov:1972ri}
V.~N. Gribov and L.~N. Lipatov, \emph{{Deep inelastic $e p$ scattering in
  perturbation theory}}, {\emph{Sov. J. Nucl. Phys.} {\bf 15} (1972) 438--450}.

\bibitem{Dokshitzer:1977sg}
Y.~L. Dokshitzer, \emph{{Calculation of the Structure Functions for Deep
  Inelastic Scattering and $e^+ e^-$ Annihilation by Perturbation Theory in
  Quantum Chromodynamics.}}, {\emph{Sov. Phys. JETP} {\bf 46} (1977) 641--653}.

\bibitem{Altarelli:1977zs}
G.~Altarelli and G.~Parisi, \emph{{Asymptotic Freedom in Parton Language}},
  \href{http://dx.doi.org/10.1016/0550-3213(77)90384-4}{\emph{Nucl. Phys.} {\bf
  B126} (1977) 298--318}.

\bibitem{Collins:2011zzd}
J.~Collins, \emph{{Foundations of perturbative QCD}}.
\newblock Cambridge University Press, 2013.

\bibitem{Wilson:1974sk}
K.~G. Wilson, \emph{{Confinement of Quarks}},
  \href{http://dx.doi.org/10.1103/PhysRevD.10.2445}{\emph{Phys. Rev.} {\bf D10}
  (1974) 2445--2459}.

\bibitem{Efremov:1978xm}
A.~V. Efremov and A.~V. Radyushkin, \emph{{Field Theoretic Treatment of High
  Momentum Transfer Processes. 3. Gauge Theories}},
  \href{http://dx.doi.org/10.1007/BF01029042}{\emph{Theor. Math. Phys.} {\bf
  44} (1981) 774}.

\bibitem{Buffing:2015tja}
M.~G.~A. Buffing, \emph{{Color and TMD Universality in Hadronic Interactions}}.
\newblock PhD thesis, NIKHEF, Amsterdam, 2015-09-02.

\bibitem{Collins:1981uw}
J.~C. Collins and D.~E. Soper, \emph{{Parton Distribution and Decay
  Functions}},
  \href{http://dx.doi.org/10.1016/0550-3213(82)90021-9}{\emph{Nucl. Phys.} {\bf
  B194} (1982) 445--492}.

\bibitem{Tangerman:1994eh}
R.~D. Tangerman and P.~J. Mulders, \emph{{Intrinsic transverse momentum and the
  polarized Drell-Yan process}},
  \href{http://dx.doi.org/10.1103/PhysRevD.51.3357}{\emph{Phys. Rev.} {\bf D51}
  (1995) 3357--3372}, [\href{http://arxiv.org/abs/hep-ph/9403227}{{\tt
  hep-ph/9403227}}].

\bibitem{Mulders:1995dh}
P.~J. Mulders and R.~D. Tangerman, \emph{{The Complete tree level result up to
  order $1/Q$ for polarized deep inelastic leptoproduction}},
  \href{http://dx.doi.org/10.1016/S0550-3213(96)00648-7,
  10.1016/0550-3213(95)00632-X}{\emph{Nucl. Phys.} {\bf B461} (1996) 197--237},
  [\href{http://arxiv.org/abs/hep-ph/9510301}{{\tt hep-ph/9510301}}].

\bibitem{Ball:2017nwa}
{\scshape NNPDF} collaboration, R.~D. Ball et~al., \emph{{Parton distributions
  from high-precision collider data}},
  \href{http://dx.doi.org/10.1140/epjc/s10052-017-5199-5}{\emph{Eur. Phys. J.}
  {\bf C77} (2017) 663}, [\href{http://arxiv.org/abs/1706.00428}{{\tt
  1706.00428}}].

\bibitem{Burkardt:2015qoa}
M.~Burkardt and B.~Pasquini, \emph{{Modelling the nucleon structure}},
  \href{http://dx.doi.org/10.1140/epja/i2016-16161-7}{\emph{Eur. Phys. J.} {\bf
  A52} (2016) 161}, [\href{http://arxiv.org/abs/1510.02567}{{\tt 1510.02567}}].

\bibitem{Diehl:2015uka}
M.~Diehl, \emph{{Introduction to GPDs and TMDs}},
  \href{http://dx.doi.org/10.1140/epja/i2016-16149-3}{\emph{Eur. Phys. J.} {\bf
  A52} (2016) 149}, [\href{http://arxiv.org/abs/1512.01328}{{\tt 1512.01328}}].

\bibitem{Bacchetta:2016ccz}
A.~Bacchetta, \emph{{Where do we stand with a $3$-D picture of the proton?}},
  \href{http://dx.doi.org/10.1140/epja/i2016-16163-5}{\emph{Eur. Phys. J.} {\bf
  A52} (2016) 163}.

\bibitem{Echevarria:2016mrc}
M.~G. Echevarría, A.~Idilbi, K.~Kanazawa, C.~Lorcé, A.~Metz, B.~Pasquini
  et~al., \emph{{Proper definition and evolution of generalized transverse
  momentum dependent distributions}},
  \href{http://dx.doi.org/10.1016/j.physletb.2016.05.086}{\emph{Phys. Lett.}
  {\bf B759} (2016) 336--341}, [\href{http://arxiv.org/abs/1602.06953}{{\tt
  1602.06953}}].

\bibitem{Diehl:2003ny}
M.~Diehl, \emph{{Generalized parton distributions}},
  \href{http://dx.doi.org/10.1016/j.physrep.2003.08.002,
  10.3204/DESY-THESIS-2003-018}{\emph{Phys. Rept.} {\bf 388} (2003) 41--277},
  [\href{http://arxiv.org/abs/hep-ph/0307382}{{\tt hep-ph/0307382}}].

\bibitem{Meissner:2009ww}
S.~Mei{\ss}ner, A.~Metz and M.~Schlegel, \emph{{Generalized parton correlation
  functions for a spin-$\tfrac{1}{2}$ hadron}},
  \href{http://dx.doi.org/10.1088/1126-6708/2009/08/056}{\emph{JHEP} {\bf 08}
  (2009) 056}, [\href{http://arxiv.org/abs/0906.5323}{{\tt 0906.5323}}].

\bibitem{Bhattacharya:2017bvs}
S.~Bhattacharya, A.~Metz and J.~Zhou, \emph{{Generalized TMDs and the exclusive
  double Drell-Yan process}},
  \href{http://dx.doi.org/10.1016/j.physletb.2017.05.081}{\emph{Phys. Lett.}
  {\bf B771} (2017) 396--400}, [\href{http://arxiv.org/abs/1702.04387}{{\tt
  1702.04387}}].

\bibitem{Ji:2003ak}
X.~Ji, \emph{{Viewing the proton through `color' filters}},
  \href{http://dx.doi.org/10.1103/PhysRevLett.91.062001}{\emph{Phys. Rev.
  Lett.} {\bf 91} (2003) 062001},
  [\href{http://arxiv.org/abs/hep-ph/0304037}{{\tt hep-ph/0304037}}].

\bibitem{Belitsky:2003nz}
A.~V. Belitsky, X.~Ji and F.~Yuan, \emph{{Quark imaging in the proton via
  quantum phase space distributions}},
  \href{http://dx.doi.org/10.1103/PhysRevD.69.074014}{\emph{Phys. Rev.} {\bf
  D69} (2004) 074014}, [\href{http://arxiv.org/abs/hep-ph/0307383}{{\tt
  hep-ph/0307383}}].

\bibitem{Lorce:2011dv}
C.~Lorcé, B.~Pasquini and M.~Vanderhaeghen, \emph{{Unified framework for
  generalized and transverse-momentum dependent parton distributions within a
  $3$Q light-cone picture of the nucleon}},
  \href{http://dx.doi.org/10.1007/JHEP05(2011)041}{\emph{JHEP} {\bf 05} (2011)
  041}, [\href{http://arxiv.org/abs/1102.4704}{{\tt 1102.4704}}].

\bibitem{Lorce:2011ni}
C.~Lorcé, B.~Pasquini, X.~Xiong and F.~Yuan, \emph{{The quark orbital angular
  momentum from Wigner distributions and light-cone wave functions}},
  \href{http://dx.doi.org/10.1103/PhysRevD.85.114006}{\emph{Phys. Rev.} {\bf
  D85} (2012) 114006}, [\href{http://arxiv.org/abs/1111.4827}{{\tt
  1111.4827}}].

\bibitem{Mukherjee:2014nya}
A.~Mukherjee, S.~Nair and V.~K. Ojha, \emph{{Quark Wigner Distributions and
  Orbital Angular Momentum in Light-front Dressed Quark Model}},
  \href{http://dx.doi.org/10.1103/PhysRevD.90.014024}{\emph{Phys. Rev.} {\bf
  D90} (2014) 014024}, [\href{http://arxiv.org/abs/1403.6233}{{\tt
  1403.6233}}].

\bibitem{Liu:2014vwa}
T.~Liu, \emph{{Quark orbital motions from Wigner distributions}},
  \href{http://arxiv.org/abs/1406.7709}{{\tt 1406.7709}}.

\bibitem{Liu:2015eqa}
T.~Liu and B.-Q. Ma, \emph{{Quark Wigner distributions in a light-cone
  spectator model}},
  \href{http://dx.doi.org/10.1103/PhysRevD.91.034019}{\emph{Phys. Rev.} {\bf
  D91} (2015) 034019}, [\href{http://arxiv.org/abs/1501.07690}{{\tt
  1501.07690}}].

\bibitem{Lorce:2015sqe}
C.~Lorcé and B.~Pasquini, \emph{{Multipole decomposition of the nucleon
  transverse phase space}},
  \href{http://dx.doi.org/10.1103/PhysRevD.93.034040}{\emph{Phys. Rev.} {\bf
  D93} (2016) 034040}, [\href{http://arxiv.org/abs/1512.06744}{{\tt
  1512.06744}}].

\bibitem{More:2017zqq}
J.~More, A.~Mukherjee and S.~Nair, \emph{{Quark Wigner Distributions Using
  Light-Front Wave Functions}},
  \href{http://dx.doi.org/10.1103/PhysRevD.95.074039}{\emph{Phys. Rev.} {\bf
  D95} (2017) 074039}, [\href{http://arxiv.org/abs/1701.00339}{{\tt
  1701.00339}}].

\bibitem{Chakrabarti:2017teq}
D.~Chakrabarti, T.~Maji, C.~Mondal and A.~Mukherjee, \emph{{Quark Wigner
  distributions and spin-spin correlations}},
  \href{http://dx.doi.org/10.1103/PhysRevD.95.074028}{\emph{Phys. Rev.} {\bf
  D95} (2017) 074028}, [\href{http://arxiv.org/abs/1701.08551}{{\tt
  1701.08551}}].

\bibitem{Kaur:2018dns}
S.~Kaur and H.~Dahiya, \emph{{Wigner distributions and GTMDs in a proton using
  light-front quark-diquark model}},
  \href{http://arxiv.org/abs/1810.09099}{{\tt 1810.09099}}.

\bibitem{Diehl:2001pm}
M.~Diehl, \emph{{Generalized parton distributions with helicity flip}},
  \href{http://dx.doi.org/10.1007/s100520100635}{\emph{Eur. Phys. J.} {\bf C19}
  (2001) 485--492}, [\href{http://arxiv.org/abs/hep-ph/0101335}{{\tt
  hep-ph/0101335}}].

\bibitem{Ji:1996ek}
X.~Ji, \emph{{Gauge-Invariant Decomposition of Nucleon Spin}},
  \href{http://dx.doi.org/10.1103/PhysRevLett.78.610}{\emph{Phys. Rev. Lett.}
  {\bf 78} (1997) 610--613}, [\href{http://arxiv.org/abs/hep-ph/9603249}{{\tt
  hep-ph/9603249}}].

\bibitem{Ralston:1979ys}
J.~P. Ralston and D.~E. Soper, \emph{{Production of Dimuons from High-Energy
  Polarized Proton-Proton Collisions}},
  \href{http://dx.doi.org/10.1016/0550-3213(79)90082-8}{\emph{Nucl. Phys.} {\bf
  B152} (1979) 109}.

\bibitem{Barone:2010zz}
V.~Barone, F.~Bradamante and A.~Martin, \emph{{Transverse-spin and
  transverse-momentum effects in high-energy processes}},
  \href{http://dx.doi.org/10.1016/j.ppnp.2010.07.003}{\emph{Prog. Part. Nucl.
  Phys.} {\bf 65} (2010) 267--333}, [\href{http://arxiv.org/abs/1011.0909}{{\tt
  1011.0909}}].

\bibitem{Bacchetta:2000jk}
A.~Bacchetta and P.~J. Mulders, \emph{{Deep inelastic leptoproduction of
  spin-one hadrons}},
  \href{http://dx.doi.org/10.1103/PhysRevD.62.114004}{\emph{Phys. Rev.} {\bf
  D62} (2000) 114004}, [\href{http://arxiv.org/abs/hep-ph/0007120}{{\tt
  hep-ph/0007120}}].

\bibitem{Bacchetta:2002}
A.~Bacchetta, \emph{Probing the Transverse Spin of Quarks in Deep Inelastic
  Scattering}.
\newblock PhD thesis, {P}h.D. thesis at VU University (Amsterdam), 2002.

\bibitem{GarciaEchevarria:2011rb}
M.~G. Echevarría, A.~Idilbi and I.~Scimemi, \emph{{Factorization Theorem For
  Drell-Yan At Low $q_T$ And Transverse Momentum Distributions
  On-The-Light-Cone}},
  \href{http://dx.doi.org/10.1007/JHEP07(2012)002}{\emph{JHEP} {\bf 07} (2012)
  002}, [\href{http://arxiv.org/abs/1111.4996}{{\tt 1111.4996}}].

\bibitem{Boer:1997nt}
D.~Boer and P.~J. Mulders, \emph{{Time reversal odd distribution functions in
  leptoproduction}},
  \href{http://dx.doi.org/10.1103/PhysRevD.57.5780}{\emph{Phys. Rev.} {\bf D57}
  (1998) 5780--5786}, [\href{http://arxiv.org/abs/hep-ph/9711485}{{\tt
  hep-ph/9711485}}].

\bibitem{Hoodbhoy:1988am}
P.~Hoodbhoy, R.~L. Jaffe and A.~Manohar, \emph{{Novel Effects in Deep Inelastic
  Scattering from Spin $1$ Hadrons}},
  \href{http://dx.doi.org/10.1016/0550-3213(89)90572-5}{\emph{Nucl. Phys.} {\bf
  B312} (1989) 571--588}.

\bibitem{Goeke:2003az}
K.~Goeke, A.~Metz, P.~V. Pobylitsa and M.~V. Polyakov, \emph{{Lorentz
  invariance relations among parton distributions revisited}},
  \href{http://dx.doi.org/10.1016/S0370-2693(03)00870-0}{\emph{Phys. Lett.}
  {\bf B567} (2003) 27--30}, [\href{http://arxiv.org/abs/hep-ph/0302028}{{\tt
  hep-ph/0302028}}].

\bibitem{Goeke:2005hb}
K.~Goeke, A.~Metz and M.~Schlegel, \emph{{Parameterization of the quark-quark
  correlator of a spin-$\tfrac{1}{2}$ hadron}},
  \href{http://dx.doi.org/10.1016/j.physletb.2005.05.037}{\emph{Phys. Lett.}
  {\bf B618} (2005) 90--96}, [\href{http://arxiv.org/abs/hep-ph/0504130}{{\tt
  hep-ph/0504130}}].

\bibitem{vanDaal:2016glj}
T.~van Daal, \emph{{Quark and gluon TMD correlators in momentum and coordinate
  space}},  in \emph{{22nd International Symposium on Spin Physics (SPIN 2016)
  Urbana, IL, USA, September 25-30, 2016}}, 2016.
\newblock \href{http://arxiv.org/abs/1612.06585}{{\tt 1612.06585}}.

\bibitem{Bacchetta:1999kz}
A.~Bacchetta, M.~Boglione, A.~Henneman and P.~J. Mulders, \emph{{Bounds on
  transverse momentum dependent distribution and fragmentation functions}},
  \href{http://dx.doi.org/10.1103/PhysRevLett.85.712}{\emph{Phys. Rev. Lett.}
  {\bf 85} (2000) 712--715}, [\href{http://arxiv.org/abs/hep-ph/9912490}{{\tt
  hep-ph/9912490}}].

\bibitem{Bacchetta:2001rb}
A.~Bacchetta and P.~J. Mulders, \emph{{Positivity bounds on spin one
  distribution and fragmentation functions}},
  \href{http://dx.doi.org/10.1016/S0370-2693(01)01051-6}{\emph{Phys. Lett.}
  {\bf B518} (2001) 85--93}, [\href{http://arxiv.org/abs/hep-ph/0104176}{{\tt
  hep-ph/0104176}}].

\bibitem{Belitsky:2002sm}
A.~V. Belitsky, X.~Ji and F.~Yuan, \emph{{Final state interactions and gauge
  invariant parton distributions}},
  \href{http://dx.doi.org/10.1016/S0550-3213(03)00121-4}{\emph{Nucl. Phys.}
  {\bf B656} (2003) 165--198}, [\href{http://arxiv.org/abs/hep-ph/0208038}{{\tt
  hep-ph/0208038}}].

\bibitem{Boer:2003cm}
D.~Boer, P.~J. Mulders and F.~Pijlman, \emph{{Universality of T odd effects in
  single spin and azimuthal asymmetries}},
  \href{http://dx.doi.org/10.1016/S0550-3213(03)00527-3}{\emph{Nucl. Phys.}
  {\bf B667} (2003) 201--241}, [\href{http://arxiv.org/abs/hep-ph/0303034}{{\tt
  hep-ph/0303034}}].

\bibitem{Bomhof:2006dp}
C.~J. Bomhof, P.~J. Mulders and F.~Pijlman, \emph{{The Construction of
  gauge-links in arbitrary hard processes}},
  \href{http://dx.doi.org/10.1140/epjc/s2006-02554-2}{\emph{Eur. Phys. J.} {\bf
  C47} (2006) 147--162}, [\href{http://arxiv.org/abs/hep-ph/0601171}{{\tt
  hep-ph/0601171}}].

\bibitem{Sivers:1989cc}
D.~W. Sivers, \emph{{Single Spin Production Asymmetries from the Hard
  Scattering of Point-Like Constituents}},
  \href{http://dx.doi.org/10.1103/PhysRevD.41.83}{\emph{Phys. Rev.} {\bf D41}
  (1990) 83}.

\bibitem{Sivers:1990fh}
D.~W. Sivers, \emph{{Hard scattering scaling laws for single spin production
  asymmetries}}, \href{http://dx.doi.org/10.1103/PhysRevD.43.261}{\emph{Phys.
  Rev.} {\bf D43} (1991) 261--263}.

\bibitem{Brodsky:2002cx}
S.~J. Brodsky, D.~S. Hwang and I.~Schmidt, \emph{{Final state interactions and
  single spin asymmetries in semi-inclusive deep inelastic scattering}},
  \href{http://dx.doi.org/10.1016/S0370-2693(02)01320-5}{\emph{Phys. Lett.}
  {\bf B530} (2002) 99--107}, [\href{http://arxiv.org/abs/hep-ph/0201296}{{\tt
  hep-ph/0201296}}].

\bibitem{Airapetian:2004tw}
{\scshape HERMES} collaboration, A.~Airapetian et~al., \emph{{Single-spin
  asymmetries in semi-inclusive deep-inelastic scattering on a transversely
  polarized hydrogen target}},
  \href{http://dx.doi.org/10.1103/PhysRevLett.94.012002}{\emph{Phys. Rev.
  Lett.} {\bf 94} (2005) 012002},
  [\href{http://arxiv.org/abs/hep-ex/0408013}{{\tt hep-ex/0408013}}].

\bibitem{Collins:2002kn}
J.~C. Collins, \emph{{Leading twist single transverse-spin asymmetries:
  Drell-Yan and deep inelastic scattering}},
  \href{http://dx.doi.org/10.1016/S0370-2693(02)01819-1}{\emph{Phys. Lett.}
  {\bf B536} (2002) 43--48}, [\href{http://arxiv.org/abs/hep-ph/0204004}{{\tt
  hep-ph/0204004}}].

\bibitem{Buffing:2012sz}
M.~G.~A. Buffing, A.~Mukherjee and P.~J. Mulders, \emph{{Generalized
  Universality of Higher Transverse Moments of Quark TMD Correlators}},
  \href{http://dx.doi.org/10.1103/PhysRevD.86.074030}{\emph{Phys. Rev.} {\bf
  D86} (2012) 074030}, [\href{http://arxiv.org/abs/1207.3221}{{\tt
  1207.3221}}].

\bibitem{Buffing:2013kca}
M.~G.~A. Buffing, A.~Mukherjee and P.~J. Mulders, \emph{{Generalized
  Universality of Definite Rank Gluon Transverse Momentum Dependent
  Correlators}},
  \href{http://dx.doi.org/10.1103/PhysRevD.88.054027}{\emph{Phys. Rev.} {\bf
  D88} (2013) 054027}, [\href{http://arxiv.org/abs/1306.5897}{{\tt
  1306.5897}}].

\bibitem{Boer:2015kxa}
D.~Boer, M.~G.~A. Buffing and P.~J. Mulders, \emph{{Operator analysis of
  $p_T$-widths of TMDs}},
  \href{http://dx.doi.org/10.1007/JHEP08(2015)053}{\emph{JHEP} {\bf 08} (2015)
  053}, [\href{http://arxiv.org/abs/1503.03760}{{\tt 1503.03760}}].

\bibitem{Airapetian:2009ae}
{\scshape HERMES} collaboration, A.~Airapetian et~al., \emph{{Observation of
  the Naive-T-odd Sivers Effect in Deep-Inelastic Scattering}},
  \href{http://dx.doi.org/10.1103/PhysRevLett.103.152002}{\emph{Phys. Rev.
  Lett.} {\bf 103} (2009) 152002}, [\href{http://arxiv.org/abs/0906.3918}{{\tt
  0906.3918}}].

\bibitem{Ageev:2006da}
{\scshape COMPASS} collaboration, E.~S. Ageev et~al., \emph{{A New measurement
  of the Collins and Sivers asymmetries on a transversely polarised deuteron
  target}},
  \href{http://dx.doi.org/10.1016/j.nuclphysb.2006.10.027}{\emph{Nucl. Phys.}
  {\bf B765} (2007) 31--70}, [\href{http://arxiv.org/abs/hep-ex/0610068}{{\tt
  hep-ex/0610068}}].

\bibitem{Alekseev:2008aa}
{\scshape COMPASS} collaboration, M.~Alekseev et~al., \emph{{Collins and Sivers
  asymmetries for pions and kaons in muon-deuteron DIS}},
  \href{http://dx.doi.org/10.1016/j.physletb.2009.01.060}{\emph{Phys. Lett.}
  {\bf B673} (2009) 127--135}, [\href{http://arxiv.org/abs/0802.2160}{{\tt
  0802.2160}}].

\bibitem{Adolph:2012sp}
{\scshape COMPASS} collaboration, C.~Adolph et~al., \emph{{II -- Experimental
  investigation of transverse spin asymmetries in $\mu$-$p$ SIDIS processes:
  Sivers asymmetries}},
  \href{http://dx.doi.org/10.1016/j.physletb.2012.09.056}{\emph{Phys. Lett.}
  {\bf B717} (2012) 383--389}, [\href{http://arxiv.org/abs/1205.5122}{{\tt
  1205.5122}}].

\bibitem{Adolph:2014zba}
{\scshape COMPASS} collaboration, C.~Adolph et~al., \emph{{Collins and Sivers
  asymmetries in muonproduction of pions and kaons off transversely polarised
  protons}},
  \href{http://dx.doi.org/10.1016/j.physletb.2015.03.056}{\emph{Phys. Lett.}
  {\bf B744} (2015) 250--259}, [\href{http://arxiv.org/abs/1408.4405}{{\tt
  1408.4405}}].

\bibitem{Adolph:2016dvl}
{\scshape COMPASS} collaboration, C.~Adolph et~al., \emph{{Sivers asymmetry
  extracted in SIDIS at the hard scales of the Drell-Yan process at COMPASS}},
  \href{http://dx.doi.org/10.1016/j.physletb.2017.04.042}{\emph{Phys. Lett.}
  {\bf B770} (2017) 138--145}, [\href{http://arxiv.org/abs/1609.07374}{{\tt
  1609.07374}}].

\bibitem{Qian:2011py}
{\scshape Jefferson Lab Hall A} collaboration, X.~Qian et~al., \emph{{Single
  Spin Asymmetries in Charged Pion Production from Semi-Inclusive Deep
  Inelastic Scattering on a Transversely Polarized $^3$He Target}},
  \href{http://dx.doi.org/10.1103/PhysRevLett.107.072003}{\emph{Phys. Rev.
  Lett.} {\bf 107} (2011) 072003}, [\href{http://arxiv.org/abs/1106.0363}{{\tt
  1106.0363}}].

\bibitem{Adamczyk:2015gyk}
{\scshape STAR} collaboration, L.~Adamczyk et~al., \emph{{Measurement of the
  transverse single-spin asymmetry in $p^\uparrow p \to W^{\pm}/Z^0$ at RHIC}},
  \href{http://dx.doi.org/10.1103/PhysRevLett.116.132301}{\emph{Phys. Rev.
  Lett.} {\bf 116} (2016) 132301}, [\href{http://arxiv.org/abs/1511.06003}{{\tt
  1511.06003}}].

\bibitem{Aghasyan:2017jop}
{\scshape COMPASS} collaboration, M.~Aghasyan et~al., \emph{{First measurement
  of transverse-spin-dependent azimuthal asymmetries in the Drell-Yan
  process}},
  \href{http://dx.doi.org/10.1103/PhysRevLett.119.112002}{\emph{Phys. Rev.
  Lett.} {\bf 119} (2017) 112002}, [\href{http://arxiv.org/abs/1704.00488}{{\tt
  1704.00488}}].

\bibitem{Bacchetta:2017gcc}
A.~Bacchetta, F.~Delcarro, C.~Pisano, M.~Radici and A.~Signori,
  \emph{{Extraction of partonic transverse momentum distributions from
  semi-inclusive deep-inelastic scattering, Drell-Yan and Z-boson production}},
  \href{http://dx.doi.org/10.1007/JHEP06(2017)081}{\emph{JHEP} {\bf 06} (2017)
  081}, [\href{http://arxiv.org/abs/1703.10157}{{\tt 1703.10157}}].

\bibitem{Accardi:2012qut}
A.~Accardi et~al., \emph{{Electron Ion Collider: The Next QCD Frontier}},
  \href{http://dx.doi.org/10.1140/epja/i2016-16268-9}{\emph{Eur. Phys. J.} {\bf
  A52} (2016) 268}, [\href{http://arxiv.org/abs/1212.1701}{{\tt 1212.1701}}].

\bibitem{Ent:2016lod}
R.~Ent, \emph{{TMDs and GPDs at a future Electron-Ion Collider}},
  \href{http://dx.doi.org/10.1140/epja/i2016-16162-6}{\emph{Eur. Phys. J.} {\bf
  A52} (2016) 162}.

\bibitem{Boer:2011fh}
D.~Boer et~al., \emph{{Gluons and the quark sea at high energies:
  Distributions, polarization, tomography}},
  \href{http://arxiv.org/abs/1108.1713}{{\tt 1108.1713}}.

\bibitem{Aidala:2012mv}
C.~A. Aidala, S.~D. Bass, D.~Hasch and G.~K. Mallot, \emph{{The Spin Structure
  of the Nucleon}},
  \href{http://dx.doi.org/10.1103/RevModPhys.85.655}{\emph{Rev. Mod. Phys.}
  {\bf 85} (2013) 655--691}, [\href{http://arxiv.org/abs/1209.2803}{{\tt
  1209.2803}}].

\bibitem{Angeles-Martinez:2015sea}
R.~Ángeles Martínez et~al., \emph{{Transverse Momentum Dependent (TMD) parton
  distribution functions: status and prospects}},
  \href{http://dx.doi.org/10.5506/APhysPolB.46.2501}{\emph{Acta Phys. Polon.}
  {\bf B46} (2015) 2501--2534}, [\href{http://arxiv.org/abs/1507.05267}{{\tt
  1507.05267}}].

\bibitem{Boglione:2015zyc}
M.~Boglione and A.~Prokudin, \emph{{Phenomenology of transverse spin: past,
  present and future}},
  \href{http://dx.doi.org/10.1140/epja/i2016-16154-6}{\emph{Eur. Phys. J.} {\bf
  A52} (2016) 154}, [\href{http://arxiv.org/abs/1511.06924}{{\tt 1511.06924}}].

\bibitem{Collins:1985ue}
J.~C. Collins, D.~E. Soper and G.~F. Sterman, \emph{{Factorization for Short
  Distance Hadron-Hadron Scattering}},
  \href{http://dx.doi.org/10.1016/0550-3213(85)90565-6}{\emph{Nucl. Phys.} {\bf
  B261} (1985) 104--142}.

\bibitem{Collins:1988ig}
J.~C. Collins, D.~E. Soper and G.~F. Sterman, \emph{{Soft Gluons and
  Factorization}},
  \href{http://dx.doi.org/10.1016/0550-3213(88)90130-7}{\emph{Nucl. Phys.} {\bf
  B308} (1988) 833--856}.

\bibitem{Becher:2010tm}
T.~Becher and M.~Neubert, \emph{{{Drell-Yan} Production at Small $q_T$,
  Transverse Parton Distributions and the Collinear Anomaly}},
  \href{http://dx.doi.org/10.1140/epjc/s10052-011-1665-7}{\emph{Eur. Phys. J.}
  {\bf C71} (2011) 1665}, [\href{http://arxiv.org/abs/1007.4005}{{\tt
  1007.4005}}].

\bibitem{Echevarria:2012js}
M.~G. Echevarría, A.~Idilbi and I.~Scimemi, \emph{{Soft and Collinear
  Factorization and Transverse Momentum Dependent Parton Distribution
  Functions}},
  \href{http://dx.doi.org/10.1016/j.physletb.2013.09.003}{\emph{Phys. Lett.}
  {\bf B726} (2013) 795--801}, [\href{http://arxiv.org/abs/1211.1947}{{\tt
  1211.1947}}].

\bibitem{Buffing:2013dxa}
M.~G.~A. Buffing and P.~J. Mulders, \emph{{Color entanglement for azimuthal
  asymmetries in the Drell-Yan process}},
  \href{http://dx.doi.org/10.1103/PhysRevLett.112.092002}{\emph{Phys. Rev.
  Lett.} {\bf 112} (2014) 092002}, [\href{http://arxiv.org/abs/1309.4681}{{\tt
  1309.4681}}].

\bibitem{Diehl:2011yj}
M.~Diehl, D.~Ostermeier and A.~Sch{\"a}fer, \emph{{Elements of a theory for
  multiparton interactions in QCD}},
  \href{http://dx.doi.org/10.1007/JHEP03(2012)089}{\emph{JHEP} {\bf 03} (2012)
  089}, [\href{http://arxiv.org/abs/1111.0910}{{\tt 1111.0910}}].

\bibitem{Diehl:2015bca}
M.~Diehl, J.~R. Gaunt, D.~Ostermeier, P.~Pl{\"o}{\ss}l and A.~Sch{\"a}fer,
  \emph{{Cancellation of Glauber gluon exchange in the double Drell-Yan
  process}}, \href{http://dx.doi.org/10.1007/JHEP01(2016)076}{\emph{JHEP} {\bf
  01} (2016) 076}, [\href{http://arxiv.org/abs/1510.08696}{{\tt 1510.08696}}].

\bibitem{Gaunt:2014ska}
J.~R. Gaunt, \emph{{Glauber Gluons and Multiple Parton Interactions}},
  \href{http://dx.doi.org/10.1007/JHEP07(2014)110}{\emph{JHEP} {\bf 07} (2014)
  110}, [\href{http://arxiv.org/abs/1405.2080}{{\tt 1405.2080}}].

\bibitem{Collins:1989gx}
J.~C. Collins, D.~E. Soper and G.~F. Sterman, \emph{{Factorization of Hard
  Processes in QCD}},
  \href{http://dx.doi.org/10.1142/9789814503266_0001}{\emph{Adv. Ser. Direct.
  High Energy Phys.} {\bf 5} (1989) 1--91},
  [\href{http://arxiv.org/abs/hep-ph/0409313}{{\tt hep-ph/0409313}}].

\bibitem{Sterman:1995fz}
G.~F. Sterman, \emph{{Partons, factorization and resummation}},  in \emph{{QCD
  and beyond. Proceedings, Theoretical Advanced Study Institute in Elementary
  Particle Physics, TASI-95, Boulder, USA, June 4-30, 1995}}, pp.~327--408,
  1995.
\newblock \href{http://arxiv.org/abs/hep-ph/9606312}{{\tt hep-ph/9606312}}.

\bibitem{Dominguez:2011saa}
F.~Dominguez, \emph{{Unintegrated Gluon Distributions at Small-$x$}}.
\newblock PhD thesis, Columbia U., 2011.

\bibitem{Rogers:2015sqa}
T.~C. Rogers, \emph{{An overview of transverse-momentum-dependent factorization
  and evolution}},
  \href{http://dx.doi.org/10.1140/epja/i2016-16153-7}{\emph{Eur. Phys. J.} {\bf
  A52} (2016) 153}, [\href{http://arxiv.org/abs/1509.04766}{{\tt 1509.04766}}].

\bibitem{Bonocore:2016wur}
D.~Bonocore, \emph{{Next-to-Soft Factorization and Unitarity in Drell-Yan
  Processes}}.
\newblock PhD thesis, Amsterdam U., 2016.

\bibitem{Qiu:1990xxa}
J.~Qiu and G.~F. Sterman, \emph{{Power corrections in hadronic scattering. 1.
  Leading $1/Q^2$ corrections to the Drell-Yan cross-section}},
  \href{http://dx.doi.org/10.1016/0550-3213(91)90503-P}{\emph{Nucl. Phys.} {\bf
  B353} (1991) 105--136}.

\bibitem{Qiu:1990xy}
J.~Qiu and G.~F. Sterman, \emph{{Power corrections to hadronic scattering. 2.
  Factorization}},
  \href{http://dx.doi.org/10.1016/0550-3213(91)90504-Q}{\emph{Nucl. Phys.} {\bf
  B353} (1991) 137--164}.

\bibitem{Collins:1981uk}
J.~C. Collins and D.~E. Soper, \emph{{Back-To-Back Jets in QCD}},
  \href{http://dx.doi.org/10.1016/0550-3213(81)90339-4}{\emph{Nucl. Phys.} {\bf
  B193} (1981) 381}.

\bibitem{Ji:2004wu}
X.~Ji, J.-P. Ma and F.~Yuan, \emph{{QCD factorization for semi-inclusive
  deep-inelastic scattering at low transverse momentum}},
  \href{http://dx.doi.org/10.1103/PhysRevD.71.034005}{\emph{Phys. Rev.} {\bf
  D71} (2005) 034005}, [\href{http://arxiv.org/abs/hep-ph/0404183}{{\tt
  hep-ph/0404183}}].

\bibitem{Buffing:2017mqm}
M.~G.~A. Buffing, M.~Diehl and T.~Kasemets, \emph{{Transverse momentum in
  double parton scattering: factorisation, evolution and matching}},
  \href{http://dx.doi.org/10.1007/JHEP01(2018)044}{\emph{JHEP} {\bf 01} (2018)
  044}, [\href{http://arxiv.org/abs/1708.03528}{{\tt 1708.03528}}].

\bibitem{Drell:1970wh}
S.~D. Drell and T.-M. Yan, \emph{{Massive Lepton Pair Production in
  Hadron-Hadron Collisions at High-Energies}},
  \href{http://dx.doi.org/10.1103/PhysRevLett.25.316}{\emph{Phys. Rev. Lett.}
  {\bf 25} (1970) 316--320}.

\bibitem{Boer:1999mm}
D.~Boer, \emph{{Investigating the origins of transverse spin asymmetries at
  RHIC}}, \href{http://dx.doi.org/10.1103/PhysRevD.60.014012}{\emph{Phys. Rev.}
  {\bf D60} (1999) 014012}, [\href{http://arxiv.org/abs/hep-ph/9902255}{{\tt
  hep-ph/9902255}}].

\bibitem{Boer:2002ju}
D.~Boer, S.~J. Brodsky and D.~S. Hwang, \emph{{Initial state interactions in
  the unpolarized Drell-Yan process}},
  \href{http://dx.doi.org/10.1103/PhysRevD.67.054003}{\emph{Phys. Rev.} {\bf
  D67} (2003) 054003}, [\href{http://arxiv.org/abs/hep-ph/0211110}{{\tt
  hep-ph/0211110}}].

\bibitem{Collins:1977iv}
J.~C. Collins and D.~E. Soper, \emph{{Angular Distribution of Dileptons in
  High-Energy Hadron Collisions}},
  \href{http://dx.doi.org/10.1103/PhysRevD.16.2219}{\emph{Phys. Rev.} {\bf D16}
  (1977) 2219}.

\bibitem{Libby:1978bx}
S.~B. Libby and G.~F. Sterman, \emph{{Mass Divergences in Two Particle
  Inelastic Scattering}},
  \href{http://dx.doi.org/10.1103/PhysRevD.18.4737}{\emph{Phys. Rev.} {\bf D18}
  (1978) 4737}.

\bibitem{Sterman:1978bi}
G.~F. Sterman, \emph{{Mass Divergences in Annihilation Processes. 1. Origin and
  Nature of Divergences in Cut Vacuum Polarization Diagrams}},
  \href{http://dx.doi.org/10.1103/PhysRevD.17.2773}{\emph{Phys. Rev.} {\bf D17}
  (1978) 2773}.

\bibitem{Landau:1959fi}
L.~D. Landau, \emph{{On analytic properties of vertex parts in quantum field
  theory}}, \href{http://dx.doi.org/10.1016/0029-5582(59)90154-3}{\emph{Nucl.
  Phys.} {\bf 13} (1959) 181--192}.

\bibitem{Coleman:1965xm}
S.~Coleman and R.~E. Norton, \emph{{Singularities in the physical region}},
  \href{http://dx.doi.org/10.1007/BF02750472}{\emph{Nuovo Cim.} {\bf 38} (1965)
  438--442}.

\bibitem{Rothstein:2016bsq}
I.~Z. Rothstein and I.~W. Stewart, \emph{{An Effective Field Theory for Forward
  Scattering and Factorization Violation}},
  \href{http://dx.doi.org/10.1007/JHEP08(2016)025}{\emph{JHEP} {\bf 08} (2016)
  025}, [\href{http://arxiv.org/abs/1601.04695}{{\tt 1601.04695}}].

\bibitem{Manohar:2006nz}
A.~V. Manohar and I.~W. Stewart, \emph{{The Zero-Bin and Mode Factorization in
  Quantum Field Theory}},
  \href{http://dx.doi.org/10.1103/PhysRevD.76.074002}{\emph{Phys. Rev.} {\bf
  D76} (2007) 074002}, [\href{http://arxiv.org/abs/hep-ph/0605001}{{\tt
  hep-ph/0605001}}].

\bibitem{Chiu:2011qc}
J.-Y. Chiu, A.~Jain, D.~Neill and I.~Z. Rothstein, \emph{{The Rapidity
  Renormalization Group}},
  \href{http://dx.doi.org/10.1103/PhysRevLett.108.151601}{\emph{Phys. Rev.
  Lett.} {\bf 108} (2012) 151601}, [\href{http://arxiv.org/abs/1104.0881}{{\tt
  1104.0881}}].

\bibitem{Chiu:2012ir}
J.-Y. Chiu, A.~Jain, D.~Neill and I.~Z. Rothstein, \emph{{A Formalism for the
  Systematic Treatment of Rapidity Logarithms in Quantum Field Theory}},
  \href{http://dx.doi.org/10.1007/JHEP05(2012)084}{\emph{JHEP} {\bf 05} (2012)
  084}, [\href{http://arxiv.org/abs/1202.0814}{{\tt 1202.0814}}].

\bibitem{Becher:2014oda}
T.~Becher, A.~Broggio and A.~Ferroglia, \emph{{Introduction to Soft-Collinear
  Effective Theory}},
  \href{http://dx.doi.org/10.1007/978-3-319-14848-9}{\emph{Lect. Notes Phys.}
  {\bf 896} (2015) pp.1--206}, [\href{http://arxiv.org/abs/1410.1892}{{\tt
  1410.1892}}].

\bibitem{Grammer:1973db}
G.~Grammer, Jr. and D.~R. Yennie, \emph{{Improved treatment for the infrared
  divergence problem in quantum electrodynamics}},
  \href{http://dx.doi.org/10.1103/PhysRevD.8.4332}{\emph{Phys. Rev.} {\bf D8}
  (1973) 4332--4344}.

\bibitem{Luo:1994np}
M.~Luo, J.~Qiu and G.~F. Sterman, \emph{{Anomalous nuclear enhancement in
  deeply inelastic scattering and photoproduction}},
  \href{http://dx.doi.org/10.1103/PhysRevD.50.1951}{\emph{Phys. Rev.} {\bf D50}
  (1994) 1951--1971}.

\bibitem{Boer:1999si}
D.~Boer and P.~J. Mulders, \emph{{Color gauge invariance in the Drell-Yan
  process}}, \href{http://dx.doi.org/10.1016/S0550-3213(99)00719-1}{\emph{Nucl.
  Phys.} {\bf B569} (2000) 505--526},
  [\href{http://arxiv.org/abs/hep-ph/9906223}{{\tt hep-ph/9906223}}].

\bibitem{Cutkosky:1960sp}
R.~E. Cutkosky, \emph{{Singularities and discontinuities of Feynman
  amplitudes}}, \href{http://dx.doi.org/10.1063/1.1703676}{\emph{J. Math.
  Phys.} {\bf 1} (1960) 429--433}.

\bibitem{Veltman:1994wz}
M.~J.~G. Veltman, \emph{{Diagrammatica: The Path to Feynman rules}},
  {\emph{Cambridge Lect. Notes Phys.} {\bf 4} (1994) 1--284}.

\bibitem{Rietkerk:2016elf}
R.~J. Rietkerk, \emph{{Unitarity Methods and On-Shell Particles in Scattering
  Amplitudes}}.
\newblock PhD thesis, Amsterdam U., 2016.

\bibitem{Angeles-Martinez:2015bla}
R.~Ángeles Martínez, \emph{{Coulomb gluons and the ordering variable}}.
\newblock PhD thesis, The University of Manchester, 2015-12-27.

\bibitem{Collins:2004nx}
J.~C. Collins and A.~Metz, \emph{{Universality of soft and collinear factors in
  hard-scattering factorization}},
  \href{http://dx.doi.org/10.1103/PhysRevLett.93.252001}{\emph{Phys. Rev.
  Lett.} {\bf 93} (2004) 252001},
  [\href{http://arxiv.org/abs/hep-ph/0408249}{{\tt hep-ph/0408249}}].

\bibitem{Vladimirov:2017ksc}
A.~Vladimirov, \emph{{Structure of rapidity divergences in multi-parton
  scattering soft factors}},
  \href{http://dx.doi.org/10.1007/JHEP04(2018)045}{\emph{JHEP} {\bf 04} (2018)
  045}, [\href{http://arxiv.org/abs/1707.07606}{{\tt 1707.07606}}].

\bibitem{Collins:2007nk}
J.~Collins and J.~Qiu, \emph{{$k_{T}$ factorization is violated in production
  of high-transverse-momentum particles in hadron-hadron collisions}},
  \href{http://dx.doi.org/10.1103/PhysRevD.75.114014}{\emph{Phys. Rev.} {\bf
  D75} (2007) 114014}, [\href{http://arxiv.org/abs/0705.2141}{{\tt
  0705.2141}}].

\bibitem{Collins:2007jp}
J.~Collins, \emph{{2-soft-gluon exchange and factorization breaking}},
  \href{http://arxiv.org/abs/0708.4410}{{\tt 0708.4410}}.

\bibitem{Rogers:2010dm}
T.~C. Rogers and P.~J. Mulders, \emph{{No Generalized TMD-Factorization in
  Hadro-Production of High Transverse Momentum Hadrons}},
  \href{http://dx.doi.org/10.1103/PhysRevD.81.094006}{\emph{Phys. Rev.} {\bf
  D81} (2010) 094006}, [\href{http://arxiv.org/abs/1001.2977}{{\tt
  1001.2977}}].

\bibitem{Collins:1992kk}
J.~C. Collins, \emph{{Fragmentation of transversely polarized quarks probed in
  transverse momentum distributions}},
  \href{http://dx.doi.org/10.1016/0550-3213(93)90262-N}{\emph{Nucl. Phys.} {\bf
  B396} (1993) 161--182}, [\href{http://arxiv.org/abs/hep-ph/9208213}{{\tt
  hep-ph/9208213}}].

\bibitem{Musch:2011er}
B.~U. Musch, P.~Hägler, M.~Engelhardt, J.~W. Negele and A.~Schäfer,
  \emph{{Sivers and Boer-Mulders observables from lattice QCD}},
  \href{http://dx.doi.org/10.1103/PhysRevD.85.094510}{\emph{Phys. Rev.} {\bf
  D85} (2012) 094510}, [\href{http://arxiv.org/abs/1111.4249}{{\tt
  1111.4249}}].

\bibitem{Engelhardt:2015xja}
M.~Engelhardt, P.~Hägler, B.~Musch, J.~Negele and A.~Schäfer, \emph{{Lattice
  QCD study of the Boer-Mulders effect in a pion}},
  \href{http://dx.doi.org/10.1103/PhysRevD.93.054501}{\emph{Phys. Rev.} {\bf
  D93} (2016) 054501}, [\href{http://arxiv.org/abs/1506.07826}{{\tt
  1506.07826}}].

\bibitem{Gamberg:2003ey}
L.~P. Gamberg, G.~R. Goldstein and K.~A. Oganessyan, \emph{{Novel transversity
  properties in semi-inclusive deep inelastic scattering}},
  \href{http://dx.doi.org/10.1103/PhysRevD.67.071504}{\emph{Phys. Rev.} {\bf
  D67} (2003) 071504}, [\href{http://arxiv.org/abs/hep-ph/0301018}{{\tt
  hep-ph/0301018}}].

\bibitem{Lu:2004hu}
Z.~Lu and B.-Q. Ma, \emph{{Non-zero transversity distribution of the pion in a
  quark-spectator-antiquark model}},
  \href{http://dx.doi.org/10.1103/PhysRevD.70.094044}{\emph{Phys. Rev.} {\bf
  D70} (2004) 094044}, [\href{http://arxiv.org/abs/hep-ph/0411043}{{\tt
  hep-ph/0411043}}].

\bibitem{Gamberg:2005ip}
L.~P. Gamberg and G.~R. Goldstein, \emph{{T-odd effects in unpolarized
  Drell-Yan scattering}},
  \href{http://dx.doi.org/10.1016/j.physletb.2007.05.043}{\emph{Phys. Lett.}
  {\bf B650} (2007) 362--368}, [\href{http://arxiv.org/abs/hep-ph/0506127}{{\tt
  hep-ph/0506127}}].

\bibitem{Lu:2005rq}
Z.~Lu and B.-Q. Ma, \emph{{Azimuthal asymmetry in unpolarized $\pi N$ Drell-Yan
  process}},
  \href{http://dx.doi.org/10.1016/j.physletb.2005.04.002}{\emph{Phys. Lett.}
  {\bf B615} (2005) 200--206}, [\href{http://arxiv.org/abs/hep-ph/0504184}{{\tt
  hep-ph/0504184}}].

\bibitem{Lu:2006ew}
Z.~Lu, B.-Q. Ma and I.~Schmidt, \emph{{Flavor separation of the Boer-Mulders
  function from unpolarized $\pi^- p$ and $\pi^- D$ Drell-Yan processes}},
  \href{http://dx.doi.org/10.1016/j.physletb.2006.06.053}{\emph{Phys. Lett.}
  {\bf B639} (2006) 494--498}, [\href{http://arxiv.org/abs/hep-ph/0702006}{{\tt
  hep-ph/0702006}}].

\bibitem{Gamberg:2007wm}
L.~P. Gamberg, G.~R. Goldstein and M.~Schlegel, \emph{{Transverse Quark Spin
  Effects and the Flavor Dependence of the Boer-Mulders Function}},
  \href{http://dx.doi.org/10.1103/PhysRevD.77.094016}{\emph{Phys. Rev.} {\bf
  D77} (2008) 094016}, [\href{http://arxiv.org/abs/0708.0324}{{\tt
  0708.0324}}].

\bibitem{Bacchetta:2008af}
A.~Bacchetta, F.~Conti and M.~Radici, \emph{{Transverse-momentum distributions
  in a diquark spectator model}},
  \href{http://dx.doi.org/10.1103/PhysRevD.78.074010}{\emph{Phys. Rev.} {\bf
  D78} (2008) 074010}, [\href{http://arxiv.org/abs/0807.0323}{{\tt
  0807.0323}}].

\bibitem{Courtoy:2009pc}
A.~Courtoy, S.~Scopetta and V.~Vento, \emph{{Analyzing the Boer-Mulders
  function within different quark models}},
  \href{http://dx.doi.org/10.1103/PhysRevD.80.074032}{\emph{Phys. Rev.} {\bf
  D80} (2009) 074032}, [\href{http://arxiv.org/abs/0909.1404}{{\tt
  0909.1404}}].

\bibitem{Pasquini:2010af}
B.~Pasquini and F.~Yuan, \emph{{Sivers and Boer-Mulders functions in Light-Cone
  Quark Models}},
  \href{http://dx.doi.org/10.1103/PhysRevD.81.114013}{\emph{Phys. Rev.} {\bf
  D81} (2010) 114013}, [\href{http://arxiv.org/abs/1001.5398}{{\tt
  1001.5398}}].

\bibitem{Pasquini:2014ppa}
B.~Pasquini and P.~Schweitzer, \emph{{Pion transverse momentum dependent parton
  distributions in a light-front constituent approach, and the Boer-Mulders
  effect in the pion-induced Drell-Yan process}},
  \href{http://dx.doi.org/10.1103/PhysRevD.90.014050}{\emph{Phys. Rev.} {\bf
  D90} (2014) 014050}, [\href{http://arxiv.org/abs/1406.2056}{{\tt
  1406.2056}}].

\bibitem{Wang:2017onm}
Z.~Wang, X.~Wang and Z.~Lu, \emph{{Boer-Mulders function of pion meson and
  $q_T$-weighted $\cos2\phi$ asymmetry in the unpolarized $\pi^- p$ Drell-Yan
  at COMPASS}}, \href{http://dx.doi.org/10.1103/PhysRevD.95.094004}{\emph{Phys.
  Rev.} {\bf D95} (2017) 094004}, [\href{http://arxiv.org/abs/1702.03637}{{\tt
  1702.03637}}].

\bibitem{Wang:2018naw}
X.~Wang, W.~Mao and Z.~Lu, \emph{{Boer-Mulders effect in the unpolarized pion
  induced Drell–Yan process at COMPASS within TMD factorization}},
  \href{http://dx.doi.org/10.1140/epjc/s10052-018-6114-4}{\emph{Eur. Phys. J.}
  {\bf C78} (2018) 643}, [\href{http://arxiv.org/abs/1805.03017}{{\tt
  1805.03017}}].

\bibitem{Falciano:1986wk}
{\scshape NA10} collaboration, S.~Falciano et~al., \emph{{Angular Distributions
  of Muon Pairs Produced by $194$-{GeV}/$c$ Negative Pions}},
  \href{http://dx.doi.org/10.1007/BF01551072}{\emph{Z. Phys.} {\bf C31} (1986)
  513}.

\bibitem{Guanziroli:1987rp}
{\scshape NA10} collaboration, M.~Guanziroli et~al., \emph{{Angular
  Distributions of Muon Pairs Produced by Negative Pions on Deuterium and
  Tungsten}}, \href{http://dx.doi.org/10.1007/BF01549713}{\emph{Z. Phys.} {\bf
  C37} (1988) 545}.

\bibitem{Conway:1989fs}
J.~S. Conway et~al., \emph{{Experimental Study of Muon Pairs Produced by
  $252$-GeV Pions on Tungsten}},
  \href{http://dx.doi.org/10.1103/PhysRevD.39.92}{\emph{Phys. Rev.} {\bf D39}
  (1989) 92--122}.

\bibitem{Zhu:2006gx}
{\scshape NuSea} collaboration, L.~Y. Zhu et~al., \emph{{Measurement of Angular
  Distributions of Drell-Yan Dimuons in $p$-$d$ Interactions at
  $800$-GeV/$c$}},
  \href{http://dx.doi.org/10.1103/PhysRevLett.99.082301}{\emph{Phys. Rev.
  Lett.} {\bf 99} (2007) 082301},
  [\href{http://arxiv.org/abs/hep-ex/0609005}{{\tt hep-ex/0609005}}].

\bibitem{Zhu:2008sj}
{\scshape NuSea} collaboration, L.~Y. Zhu et~al., \emph{{Measurement of Angular
  Distributions of Drell-Yan Dimuons in $p$-$p$ Interactions at
  $800$-GeV/$c$}},
  \href{http://dx.doi.org/10.1103/PhysRevLett.102.182001}{\emph{Phys. Rev.
  Lett.} {\bf 102} (2009) 182001}, [\href{http://arxiv.org/abs/0811.4589}{{\tt
  0811.4589}}].

\bibitem{Peng:2008sp}
J.-C. Peng, \emph{{TMDs and Drell-Yan Experiments at Fermilab and J-PARC}},  in
  \emph{{Transversity 2008: 2nd International Workshop on Transverse
  Polarization Phenomena in Hard Processes Ferrara, Italy, May 28-31, 2008}}.
\newblock \href{http://arxiv.org/abs/0811.4596}{{\tt 0811.4596}}.

\bibitem{Nakano:2016jdm}
{\scshape E906/SeaQuest} collaboration, K.~Nakano, \emph{{Measurement of
  Boer-Mulders Function via Drell-Yan Process by SeaQuest Experiment at
  Fermilab}}, \href{http://dx.doi.org/10.1142/S2010194516600417}{\emph{Int. J.
  Mod. Phys. Conf. Ser.} {\bf 40} (2016) 1660041}.

\bibitem{Zhang:2008nu}
B.~Zhang, Z.~Lu, B.-Q. Ma and I.~Schmidt, \emph{{Extracting Boer-Mulders
  functions from $p$-$D$ Drell-Yan processes}},
  \href{http://dx.doi.org/10.1103/PhysRevD.77.054011}{\emph{Phys. Rev.} {\bf
  D77} (2008) 054011}, [\href{http://arxiv.org/abs/0803.1692}{{\tt
  0803.1692}}].

\bibitem{Lu:2009ip}
Z.~Lu and I.~Schmidt, \emph{{Updating Boer-Mulders functions from unpolarized
  $pd$ and $pp$ Drell-Yan data}},
  \href{http://dx.doi.org/10.1103/PhysRevD.81.034023}{\emph{Phys. Rev.} {\bf
  D81} (2010) 034023}, [\href{http://arxiv.org/abs/0912.2031}{{\tt
  0912.2031}}].

\bibitem{Barone:2010gk}
V.~Barone, S.~Melis and A.~Prokudin, \emph{{Azimuthal asymmetries in
  unpolarized Drell-Yan processes and the Boer-Mulders distributions of
  antiquarks}}, \href{http://dx.doi.org/10.1103/PhysRevD.82.114025}{\emph{Phys.
  Rev.} {\bf D82} (2010) 114025}, [\href{http://arxiv.org/abs/1009.3423}{{\tt
  1009.3423}}].

\bibitem{Lambertsen:2016wgj}
M.~Lambertsen and W.~Vogelsang, \emph{{Drell-Yan lepton angular distributions
  in perturbative QCD}},
  \href{http://dx.doi.org/10.1103/PhysRevD.93.114013}{\emph{Phys. Rev.} {\bf
  D93} (2016) 114013}, [\href{http://arxiv.org/abs/1605.02625}{{\tt
  1605.02625}}].

\bibitem{Barone:2009hw}
V.~Barone, S.~Melis and A.~Prokudin, \emph{{The Boer-Mulders effect in
  unpolarized SIDIS: An Analysis of the COMPASS and HERMES data on the $\cos
  2\phi$ asymmetry}},
  \href{http://dx.doi.org/10.1103/PhysRevD.81.114026}{\emph{Phys. Rev.} {\bf
  D81} (2010) 114026}, [\href{http://arxiv.org/abs/0912.5194}{{\tt
  0912.5194}}].

\bibitem{Barone:2015ksa}
V.~Barone, M.~Boglione, J.~O. Gonzalez~Hernandez and S.~Melis,
  \emph{{Phenomenological analysis of azimuthal asymmetries in unpolarized
  semi-inclusive deep inelastic scattering}},
  \href{http://dx.doi.org/10.1103/PhysRevD.91.074019}{\emph{Phys. Rev.} {\bf
  D91} (2015) 074019}, [\href{http://arxiv.org/abs/1502.04214}{{\tt
  1502.04214}}].

\bibitem{Airapetian:2012ki}
{\scshape HERMES} collaboration, A.~Airapetian et~al., \emph{{Multiplicities of
  charged pions and kaons from semi-inclusive deep-inelastic scattering by the
  proton and the deuteron}},
  \href{http://dx.doi.org/10.1103/PhysRevD.87.074029}{\emph{Phys. Rev.} {\bf
  D87} (2013) 074029}, [\href{http://arxiv.org/abs/1212.5407}{{\tt
  1212.5407}}].

\bibitem{Airapetian:2012yg}
{\scshape HERMES} collaboration, A.~Airapetian et~al., \emph{{Azimuthal
  distributions of charged hadrons, pions, and kaons produced in deep-inelastic
  scattering off unpolarized protons and deuterons}},
  \href{http://dx.doi.org/10.1103/PhysRevD.87.012010}{\emph{Phys. Rev.} {\bf
  D87} (2013) 012010}, [\href{http://arxiv.org/abs/1204.4161}{{\tt
  1204.4161}}].

\bibitem{Adolph:2013stb}
{\scshape COMPASS} collaboration, C.~Adolph et~al., \emph{{Hadron Transverse
  Momentum Distributions in Muon Deep Inelastic Scattering at $160$ GeV/$c$}},
  \href{http://dx.doi.org/10.1140/epjc/s10052-013-2531-6,
  10.1140/epjc/s10052-014-3255-y}{\emph{Eur. Phys. J.} {\bf C73} (2013) 2531},
  [\href{http://arxiv.org/abs/1305.7317}{{\tt 1305.7317}}].

\bibitem{Adolph:2014pwc}
{\scshape COMPASS} collaboration, C.~Adolph et~al., \emph{{Measurement of
  azimuthal hadron asymmetries in semi-inclusive deep inelastic scattering off
  unpolarised nucleons}},
  \href{http://dx.doi.org/10.1016/j.nuclphysb.2014.07.019}{\emph{Nucl. Phys.}
  {\bf B886} (2014) 1046--1077}, [\href{http://arxiv.org/abs/1401.6284}{{\tt
  1401.6284}}].

\bibitem{Lu:2016pdp}
Z.~Lu, \emph{{Spin physics through unpolarized processes}},
  \href{http://dx.doi.org/10.1007/s11467-015-0525-6}{\emph{Front.
  Phys.(Beijing)} {\bf 11} (2016) 111204}.

\bibitem{Jakob:1997wg}
R.~Jakob, P.~J. Mulders and J.~Rodrigues, \emph{{Modeling quark distribution
  and fragmentation functions}},
  \href{http://dx.doi.org/10.1016/S0375-9474(97)00588-5}{\emph{Nucl. Phys.}
  {\bf A626} (1997) 937--965}, [\href{http://arxiv.org/abs/hep-ph/9704335}{{\tt
  hep-ph/9704335}}].

\bibitem{Bacchetta:2003rz}
A.~Bacchetta, A.~Sch{\"a}fer and J.-J. Yang, \emph{{Sivers function in a
  spectator model with axial vector diquarks}},
  \href{http://dx.doi.org/10.1016/j.physletb.2003.10.045}{\emph{Phys. Lett.}
  {\bf B578} (2004) 109--118}, [\href{http://arxiv.org/abs/hep-ph/0309246}{{\tt
  hep-ph/0309246}}].

\bibitem{Meissner:2007rx}
S.~Mei{\ss}ner, A.~Metz and K.~Goeke, \emph{{Relations between generalized and
  transverse momentum dependent parton distributions}},
  \href{http://dx.doi.org/10.1103/PhysRevD.76.034002}{\emph{Phys. Rev.} {\bf
  D76} (2007) 034002}, [\href{http://arxiv.org/abs/hep-ph/0703176}{{\tt
  hep-ph/0703176}}].

\bibitem{Kasemets:2013nma}
T.~Kasemets, \emph{{Double parton scattering -- a tale of two partons}}.
\newblock PhD thesis, Hamburg U., 2013.

\bibitem{Passarino:1978jh}
G.~Passarino and M.~J.~G. Veltman, \emph{{One Loop Corrections for $e^+ e^-$
  Annihilation Into $\mu^+ \mu^-$ in the Weinberg Model}},
  \href{http://dx.doi.org/10.1016/0550-3213(79)90234-7}{\emph{Nucl. Phys.} {\bf
  B160} (1979) 151--207}.

\bibitem{Goldstein:2002vv}
G.~R. Goldstein and L.~Gamberg, \emph{{Transversity and meson
  photoproduction}},  in \emph{{High energy physics. Proceedings, 31st
  International Conference, ICHEP 2002, Amsterdam, Netherlands, July 25-31,
  2002}}, pp.~452--454.
\newblock \href{http://arxiv.org/abs/hep-ph/0209085}{{\tt hep-ph/0209085}}.

\bibitem{Kasemets:2012pr}
T.~Kasemets and M.~Diehl, \emph{{Angular correlations in the double Drell-Yan
  process}}, \href{http://dx.doi.org/10.1007/JHEP01(2013)121}{\emph{JHEP} {\bf
  01} (2013) 121}, [\href{http://arxiv.org/abs/1210.5434}{{\tt 1210.5434}}].

\bibitem{Arnold:2008kf}
S.~Arnold, A.~Metz and M.~Schlegel, \emph{{Dilepton production from polarized
  hadron-hadron collisions}},
  \href{http://dx.doi.org/10.1103/PhysRevD.79.034005}{\emph{Phys. Rev.} {\bf
  D79} (2009) 034005}, [\href{http://arxiv.org/abs/0809.2262}{{\tt
  0809.2262}}].

\bibitem{Boer:2011vq}
D.~Boer, W.~J. den Dunnen and A.~Kotzinian, \emph{{Double Sivers effect
  asymmetries and their impact on transversity measurements at RHIC}},
  \href{http://dx.doi.org/10.1103/PhysRevD.83.114032}{\emph{Phys. Rev.} {\bf
  D83} (2011) 114032}, [\href{http://arxiv.org/abs/1103.0908}{{\tt
  1103.0908}}].

\bibitem{Mulders:2000sh}
P.~J. Mulders and J.~Rodrigues, \emph{{Transverse momentum dependence in gluon
  distribution and fragmentation functions}},
  \href{http://dx.doi.org/10.1103/PhysRevD.63.094021}{\emph{Phys. Rev.} {\bf
  D63} (2001) 094021}, [\href{http://arxiv.org/abs/hep-ph/0009343}{{\tt
  hep-ph/0009343}}].

\bibitem{Jaffe:1989xy}
R.~L. Jaffe and A.~Manohar, \emph{{Nuclear Gluonometry}},
  \href{http://dx.doi.org/10.1016/0370-2693(89)90242-6}{\emph{Phys. Lett.} {\bf
  B223} (1989) 218--224}.

\bibitem{Artru:1989zv}
X.~Artru and M.~Mekhfi, \emph{{Transversely Polarized Parton Densities, their
  Evolution and their Measurement}},
  \href{http://dx.doi.org/10.1007/BF01556280}{\emph{Z. Phys.} {\bf C45} (1990)
  669}.

\bibitem{Bomhof:2007xt}
C.~J. Bomhof and P.~J. Mulders, \emph{{Non-universality of transverse momentum
  dependent parton distribution functions}},
  \href{http://dx.doi.org/10.1016/j.nuclphysb.2007.11.024}{\emph{Nucl. Phys.}
  {\bf B795} (2008) 409--427}, [\href{http://arxiv.org/abs/0709.1390}{{\tt
  0709.1390}}].

\bibitem{Cotogno:2017mwy}
S.~Cotogno, \emph{{Parametrization of the Transverse Momentum Dependent
  Light-Front Correlator for Gluons}},
  \href{http://dx.doi.org/10.1007/s00601-017-1254-x}{\emph{Few Body Syst.} {\bf
  58} (2017) 92}.

\bibitem{Soffer:1994ww}
J.~Soffer, \emph{{Positivity constraints for spin dependent parton
  distributions}},
  \href{http://dx.doi.org/10.1103/PhysRevLett.74.1292}{\emph{Phys. Rev. Lett.}
  {\bf 74} (1995) 1292--1294}, [\href{http://arxiv.org/abs/hep-ph/9409254}{{\tt
  hep-ph/9409254}}].

\bibitem{Vogelsang:1997ak}
W.~Vogelsang, \emph{{Next-to-leading order evolution of transversity
  distributions and Soffer's inequality}},
  \href{http://dx.doi.org/10.1103/PhysRevD.57.1886}{\emph{Phys. Rev.} {\bf D57}
  (1998) 1886--1894}, [\href{http://arxiv.org/abs/hep-ph/9706511}{{\tt
  hep-ph/9706511}}].

\bibitem{Bourrely:1997bx}
C.~Bourrely, J.~Soffer and O.~V. Teryaev, \emph{{The $Q^2$ evolution of Soffer
  inequality}},
  \href{http://dx.doi.org/10.1016/S0370-2693(97)01538-4}{\emph{Phys. Lett.}
  {\bf B420} (1998) 375--381}, [\href{http://arxiv.org/abs/hep-ph/9710224}{{\tt
  hep-ph/9710224}}].

\bibitem{Martin:1997rz}
O.~Martin, A.~Schäfer, M.~Stratmann and W.~Vogelsang, \emph{{Soffer's
  inequality and the transversely polarized Drell-Yan process at
  next-to-leading order}},
  \href{http://dx.doi.org/10.1103/PhysRevD.57.3084}{\emph{Phys. Rev.} {\bf D57}
  (1998) 3084--3090}, [\href{http://arxiv.org/abs/hep-ph/9710300}{{\tt
  hep-ph/9710300}}].

\bibitem{Echevarria:2014rua}
M.~G. Echevarría, A.~Idilbi and I.~Scimemi, \emph{{Unified treatment of the
  QCD evolution of all (un-)polarized transverse momentum dependent functions:
  Collins function as a study case}},
  \href{http://dx.doi.org/10.1103/PhysRevD.90.014003}{\emph{Phys. Rev.} {\bf
  D90} (2014) 014003}, [\href{http://arxiv.org/abs/1402.0869}{{\tt
  1402.0869}}].

\bibitem{Echevarria:2015uaa}
M.~G. Echevarría, T.~Kasemets, P.~J. Mulders and C.~Pisano, \emph{{QCD
  evolution of (un)polarized gluon TMDPDFs and the Higgs $q_T$-distribution}},
  \href{http://dx.doi.org/10.1007/JHEP07(2015)158,
  10.1007/JHEP05(2017)073}{\emph{JHEP} {\bf 07} (2015) 158},
  [\href{http://arxiv.org/abs/1502.05354}{{\tt 1502.05354}}].

\bibitem{Boer:2013fca}
D.~Boer, W.~J. den Dunnen, C.~Pisano and M.~Schlegel, \emph{{Determining the
  Higgs spin and parity in the diphoton decay channel}},
  \href{http://dx.doi.org/10.1103/PhysRevLett.111.032002}{\emph{Phys. Rev.
  Lett.} {\bf 111} (2013) 032002}, [\href{http://arxiv.org/abs/1304.2654}{{\tt
  1304.2654}}].

\bibitem{Pisano:2013cya}
C.~Pisano, D.~Boer, S.~J. Brodsky, M.~G.~A. Buffing and P.~J. Mulders,
  \emph{{Linear polarization of gluons and photons in unpolarized collider
  experiments}}, \href{http://dx.doi.org/10.1007/JHEP10(2013)024}{\emph{JHEP}
  {\bf 10} (2013) 024}, [\href{http://arxiv.org/abs/1307.3417}{{\tt
  1307.3417}}].

\bibitem{Boer:2016fqd}
D.~Boer, P.~J. Mulders, C.~Pisano and J.~Zhou, \emph{{Asymmetries in Heavy
  Quark Pair and Dijet Production at an EIC}},
  \href{http://dx.doi.org/10.1007/JHEP08(2016)001}{\emph{JHEP} {\bf 08} (2016)
  001}, [\href{http://arxiv.org/abs/1605.07934}{{\tt 1605.07934}}].

\bibitem{Boer:2015vso}
D.~Boer, C.~Lorcé, C.~Pisano and J.~Zhou, \emph{{The gluon Sivers
  distribution: status and future prospects}},
  \href{http://dx.doi.org/10.1155/2015/371396}{\emph{Adv. High Energy Phys.}
  {\bf 2015} (2015) 371396}, [\href{http://arxiv.org/abs/1504.04332}{{\tt
  1504.04332}}].

\bibitem{Kharzeev:2003wz}
D.~Kharzeev, Y.~V. Kovchegov and K.~Tuchin, \emph{{Cronin effect and high-$p_T$
  suppression in $pA$ collisions}},
  \href{http://dx.doi.org/10.1103/PhysRevD.68.094013}{\emph{Phys. Rev.} {\bf
  D68} (2003) 094013}, [\href{http://arxiv.org/abs/hep-ph/0307037}{{\tt
  hep-ph/0307037}}].

\bibitem{Dominguez:2010xd}
F.~Dominguez, B.-W. Xiao and F.~Yuan, \emph{{$k_T$-factorization for Hard
  Processes in Nuclei}},
  \href{http://dx.doi.org/10.1103/PhysRevLett.106.022301}{\emph{Phys. Rev.
  Lett.} {\bf 106} (2011) 022301}, [\href{http://arxiv.org/abs/1009.2141}{{\tt
  1009.2141}}].

\bibitem{Dominguez:2011wm}
F.~Dominguez, C.~Marquet, B.-W. Xiao and F.~Yuan, \emph{{Universality of
  Unintegrated Gluon Distributions at small $x$}},
  \href{http://dx.doi.org/10.1103/PhysRevD.83.105005}{\emph{Phys. Rev.} {\bf
  D83} (2011) 105005}, [\href{http://arxiv.org/abs/1101.0715}{{\tt
  1101.0715}}].

\bibitem{Kotko:2015ura}
P.~Kotko, K.~Kutak, C.~Marquet, E.~Petreska, S.~Sapeta and A.~van Hameren,
  \emph{{Improved TMD factorization for forward dijet production in
  dilute-dense hadronic collisions}},
  \href{http://dx.doi.org/10.1007/JHEP09(2015)106}{\emph{JHEP} {\bf 09} (2015)
  106}, [\href{http://arxiv.org/abs/1503.03421}{{\tt 1503.03421}}].

\bibitem{Brodsky:2012vg}
S.~J. Brodsky, F.~Fleuret, C.~Hadjidakis and J.-P. Lansberg, \emph{{Physics
  Opportunities of a Fixed-Target Experiment using the LHC Beams}},
  \href{http://dx.doi.org/10.1016/j.physrep.2012.10.001}{\emph{Phys. Rept.}
  {\bf 522} (2013) 239--255}, [\href{http://arxiv.org/abs/1202.6585}{{\tt
  1202.6585}}].

\bibitem{Lansberg:2017dzg}
J.-P. Lansberg, C.~Pisano, F.~Scarpa and M.~Schlegel, \emph{{Pinning down the
  linearly-polarised gluons inside unpolarised protons using quarkonium-pair
  production at the LHC}},
  \href{http://dx.doi.org/10.1016/j.physletb.2018.08.004}{\emph{Phys. Lett.}
  {\bf B784} (2018) 217--222}, [\href{http://arxiv.org/abs/1710.01684}{{\tt
  1710.01684}}].

\bibitem{Schmidt:2005gv}
I.~Schmidt, J.~Soffer and J.-J. Yang, \emph{{Transverse single spin asymmetries
  in photon production}},
  \href{http://dx.doi.org/10.1016/j.physletb.2005.03.008}{\emph{Phys. Lett.}
  {\bf B612} (2005) 258--262}, [\href{http://arxiv.org/abs/hep-ph/0503127}{{\tt
  hep-ph/0503127}}].

\bibitem{Bacchetta:2007sz}
A.~Bacchetta, C.~Bomhof, U.~D'Alesio, P.~J. Mulders and F.~Murgia, \emph{{The
  Sivers single-spin asymmetry in photon-jet production}},
  \href{http://dx.doi.org/10.1103/PhysRevLett.99.212002}{\emph{Phys. Rev.
  Lett.} {\bf 99} (2007) 212002},
  [\href{http://arxiv.org/abs/hep-ph/0703153}{{\tt hep-ph/0703153}}].

\bibitem{Dunnen:2014eta}
W.~J. den Dunnen, J.-P. Lansberg, C.~Pisano and M.~Schlegel, \emph{{Accessing
  the Transverse Dynamics and Polarization of Gluons inside the Proton at the
  LHC}}, \href{http://dx.doi.org/10.1103/PhysRevLett.112.212001}{\emph{Phys.
  Rev. Lett.} {\bf 112} (2014) 212001},
  [\href{http://arxiv.org/abs/1401.7611}{{\tt 1401.7611}}].

\bibitem{Lansberg:2014myg}
J.-P. Lansberg et~al., \emph{{Spin physics and TMD studies at A Fixed-Target
  ExpeRiment at the LHC (AFTER@LHC)}},
  \href{http://dx.doi.org/10.1051/epjconf/20158502038}{\emph{EPJ Web Conf.}
  {\bf 85} (2015) 02038}, [\href{http://arxiv.org/abs/1410.1962}{{\tt
  1410.1962}}].

\bibitem{Lansberg:2015lva}
J.-P. Lansberg and H.-S. Shao, \emph{{Double-quarkonium production at a
  fixed-target experiment at the LHC (AFTER@LHC)}},
  \href{http://dx.doi.org/10.1016/j.nuclphysb.2015.09.005}{\emph{Nucl. Phys.}
  {\bf B900} (2015) 273--294}, [\href{http://arxiv.org/abs/1504.06531}{{\tt
  1504.06531}}].

\bibitem{Abeyratne:2012ah}
S.~Abeyratne et~al., \emph{{Science Requirements and Conceptual Design for a
  Polarized Medium Energy Electron-Ion Collider at Jefferson Lab}},
  \href{http://arxiv.org/abs/1209.0757}{{\tt 1209.0757}}.

\bibitem{Abeyratne:2015pma}
S.~Abeyratne et~al., \emph{{MEIC Design Summary}},
  \href{http://arxiv.org/abs/1504.07961}{{\tt 1504.07961}}.

\bibitem{Ball:2006zz}
J.~Ball et~al., \emph{{On the large COMPASS polarized deuteron target}},
  {\emph{Czech. J. Phys.} {\bf 56} (2006) F295--F305}.

\bibitem{DetmoldJLab}
W.~Detmold, R.~Jaffe, J.~Maxwell, R.~Milner, D.~Crabb, D.~Day et~al.,
  \emph{{Letter of intent}}, {\emph{Jefferson Lab} {\bf LOI12-14-001} (2014) }.

\bibitem{Boer:2010zf}
D.~Boer, S.~J. Brodsky, P.~J. Mulders and C.~Pisano, \emph{{Direct Probes of
  Linearly Polarized Gluons inside Unpolarized Hadrons}},
  \href{http://dx.doi.org/10.1103/PhysRevLett.106.132001}{\emph{Phys. Rev.
  Lett.} {\bf 106} (2011) 132001}, [\href{http://arxiv.org/abs/1011.4225}{{\tt
  1011.4225}}].

\bibitem{Qiu:2011ai}
J.~Qiu, M.~Schlegel and W.~Vogelsang, \emph{{Probing Gluonic Spin-Orbit
  Correlations in Photon Pair Production}},
  \href{http://dx.doi.org/10.1103/PhysRevLett.107.062001}{\emph{Phys. Rev.
  Lett.} {\bf 107} (2011) 062001}, [\href{http://arxiv.org/abs/1103.3861}{{\tt
  1103.3861}}].

\bibitem{Boer:2012bt}
D.~Boer and C.~Pisano, \emph{{Polarized gluon studies with charmonium and
  bottomonium at LHCb and AFTER}},
  \href{http://dx.doi.org/10.1103/PhysRevD.86.094007}{\emph{Phys. Rev.} {\bf
  D86} (2012) 094007}, [\href{http://arxiv.org/abs/1208.3642}{{\tt
  1208.3642}}].

\bibitem{Boer:2015uqa}
D.~Boer, \emph{{Linearly polarized gluon effects in unpolarized collisions}},
  {\emph{PoS} {\bf QCDEV2015} (2015) 023},
  [\href{http://arxiv.org/abs/1510.05915}{{\tt 1510.05915}}].

\bibitem{Boer:2016jnn}
D.~Boer, \emph{{Gluon TMD studies at EIC}},
  \href{http://dx.doi.org/10.1051/epjconf/201611202003}{\emph{EPJ Web Conf.}
  {\bf 112} (2016) 02003}, [\href{http://arxiv.org/abs/1601.01813}{{\tt
  1601.01813}}].

\bibitem{Lansberg:2017tlc}
J.-P. Lansberg, C.~Pisano and M.~Schlegel, \emph{{Associated production of a
  dilepton and a $\Upsilon(J/\psi)$ at the LHC as a probe of gluon transverse
  momentum dependent distributions}},
  \href{http://dx.doi.org/10.1016/j.nuclphysb.2017.04.011}{\emph{Nucl. Phys.}
  {\bf B920} (2017) 192--210}, [\href{http://arxiv.org/abs/1702.00305}{{\tt
  1702.00305}}].

\bibitem{Godbole:2018mmh}
R.~M. Godbole, A.~Kaushik, A.~Misra and S.~Padval, \emph{{Probing the Gluon
  Sivers Function through direct photon production at RHIC}},
  \href{http://arxiv.org/abs/1810.07113}{{\tt 1810.07113}}.

\bibitem{Boer:2015pni}
D.~Boer, M.~G. Echevarría, P.~Mulders and J.~Zhou, \emph{{Single spin
  asymmetries from a single Wilson loop}},
  \href{http://dx.doi.org/10.1103/PhysRevLett.116.122001}{\emph{Phys. Rev.
  Lett.} {\bf 116} (2016) 122001}, [\href{http://arxiv.org/abs/1511.03485}{{\tt
  1511.03485}}].

\bibitem{Lorce:2013pza}
C.~Lorcé and B.~Pasquini, \emph{{Structure analysis of the generalized
  correlator of quark and gluon for a spin-$\tfrac{1}{2}$ target}},
  \href{http://dx.doi.org/10.1007/JHEP09(2013)138}{\emph{JHEP} {\bf 09} (2013)
  138}, [\href{http://arxiv.org/abs/1307.4497}{{\tt 1307.4497}}].

\bibitem{Hatta:2016dxp}
Y.~Hatta, B.-W. Xiao and F.~Yuan, \emph{{Probing the Small-$x$ Gluon Tomography
  in Correlated Hard Diffractive Dijet Production in Deep Inelastic
  Scattering}},
  \href{http://dx.doi.org/10.1103/PhysRevLett.116.202301}{\emph{Phys. Rev.
  Lett.} {\bf 116} (2016) 202301}, [\href{http://arxiv.org/abs/1601.01585}{{\tt
  1601.01585}}].

\bibitem{Hatta:2005as}
Y.~Hatta, E.~Iancu, K.~Itakura and L.~McLerran, \emph{{Odderon in the color
  glass condensate}},
  \href{http://dx.doi.org/10.1016/j.nuclphysa.2005.05.163}{\emph{Nucl. Phys.}
  {\bf A760} (2005) 172--207}, [\href{http://arxiv.org/abs/hep-ph/0501171}{{\tt
  hep-ph/0501171}}].

\bibitem{Lukaszuk:1973nt}
L.~{\L}ukaszuk and B.~Nicolescu, \emph{{A Possible interpretation of $pp$
  rising total cross-sections}},
  \href{http://dx.doi.org/10.1007/BF02824484}{\emph{Lett. Nuovo Cim.} {\bf 8}
  (1973) 405--413}.

\bibitem{Levin:1990gg}
E.~M. Levin and M.~G. Ryskin, \emph{{High-energy hadron collisions in QCD}},
  \href{http://dx.doi.org/10.1016/0370-1573(90)90016-U}{\emph{Phys. Rept.} {\bf
  189} (1990) 267--382}.

\bibitem{Martynov:2017zjz}
E.~Martynov and B.~Nicolescu, \emph{{Did TOTEM experiment discover the
  Odderon?}},
  \href{http://dx.doi.org/10.1016/j.physletb.2018.01.054}{\emph{Phys. Lett.}
  {\bf B778} (2018) 414--418}, [\href{http://arxiv.org/abs/1711.03288}{{\tt
  1711.03288}}].

\bibitem{Antchev:2017yns}
{\scshape TOTEM} collaboration, G.~Antchev et~al., \emph{{First determination
  of the $\rho$ parameter at $\sqrt{s} = 13$ TeV -- probing the existence of a
  colourless three-gluon bound state}}, {\emph{Submitted to: Phys. Rev.} (2017)
  }.

\bibitem{Csorgo:2018uyp}
T.~Cs{\"o}rg\H{o}, R.~Pasechnik and A.~Ster, \emph{{Odderon and proton
  substructure from a model-independent L\'evy imaging of elastic $pp$ and
  $p\bar{p}$ collisions}},  \href{http://arxiv.org/abs/1807.02897}{{\tt
  1807.02897}}.

\bibitem{Martynov:2018sga}
E.~Martynov and B.~Nicolescu, \emph{{Odderon effects in the differential
  cross-sections at Tevatron and LHC energies}},
  \href{http://arxiv.org/abs/1808.08580}{{\tt 1808.08580}}.

\bibitem{Ball:2017otu}
R.~D. Ball, V.~Bertone, M.~Bonvini, S.~Marzani, J.~Rojo and L.~Rottoli,
  \emph{{Parton distributions with small-$x$ resummation: evidence for BFKL
  dynamics in HERA data}},
  \href{http://dx.doi.org/10.1140/epjc/s10052-018-5774-4}{\emph{Eur. Phys. J.}
  {\bf C78} (2018) 321}, [\href{http://arxiv.org/abs/1710.05935}{{\tt
  1710.05935}}].

\bibitem{Ball:2016spl}
R.~D. Ball, E.~R. Nocera and J.~Rojo, \emph{{The asymptotic behaviour of parton
  distributions at small and large $x$}},
  \href{http://dx.doi.org/10.1140/epjc/s10052-016-4240-4}{\emph{Eur. Phys. J.}
  {\bf C76} (2016) 383}, [\href{http://arxiv.org/abs/1604.00024}{{\tt
  1604.00024}}].

\bibitem{Marquet:2016cgx}
C.~Marquet, E.~Petreska and C.~Roiesnel, \emph{{Transverse-momentum-dependent
  gluon distributions from JIMWLK evolution}},
  \href{http://dx.doi.org/10.1007/JHEP10(2016)065}{\emph{JHEP} {\bf 10} (2016)
  065}, [\href{http://arxiv.org/abs/1608.02577}{{\tt 1608.02577}}].

\bibitem{Kovchegov:2017lsr}
Y.~V. Kovchegov, D.~Pitonyak and M.~D. Sievert, \emph{{Small-$x$ Asymptotics of
  the Gluon Helicity Distribution}},
  \href{http://dx.doi.org/10.1007/JHEP10(2017)198}{\emph{JHEP} {\bf 10} (2017)
  198}, [\href{http://arxiv.org/abs/1706.04236}{{\tt 1706.04236}}].

\bibitem{Kovchegov:2018znm}
Y.~V. Kovchegov and M.~D. Sievert, \emph{{Small-$x$ Helicity Evolution: an
  Operator Treatment}},  \href{http://arxiv.org/abs/1808.09010}{{\tt
  1808.09010}}.

\bibitem{Ryskin:1987ya}
M.~G. Ryskin, \emph{{Odderon and Polarization Phenomena in QCD}}, {\emph{Sov.
  J. Nucl. Phys.} {\bf 46} (1987) 337--342}.

\bibitem{Buttimore:1998rj}
N.~H. Buttimore, B.~Z. Kopeliovich, E.~Leader, J.~Soffer and T.~L. Trueman,
  \emph{{The spin dependence of high-energy proton scattering}},
  \href{http://dx.doi.org/10.1103/PhysRevD.59.114010}{\emph{Phys. Rev.} {\bf
  D59} (1999) 114010}, [\href{http://arxiv.org/abs/hep-ph/9901339}{{\tt
  hep-ph/9901339}}].

\bibitem{Leader:1999ua}
E.~Leader and T.~L. Trueman, \emph{{The Odderon and spin dependence of
  high-energy proton-proton scattering}},
  \href{http://dx.doi.org/10.1103/PhysRevD.61.077504}{\emph{Phys. Rev.} {\bf
  D61} (2000) 077504}, [\href{http://arxiv.org/abs/hep-ph/9908221}{{\tt
  hep-ph/9908221}}].

\bibitem{Zhou:2013gsa}
J.~Zhou, \emph{{Transverse single spin asymmetries at small $x$ and the
  anomalous magnetic moment}},
  \href{http://dx.doi.org/10.1103/PhysRevD.89.074050}{\emph{Phys. Rev.} {\bf
  D89} (2014) 074050}, [\href{http://arxiv.org/abs/1308.5912}{{\tt
  1308.5912}}].

\bibitem{Iancu:2002xk}
E.~Iancu, A.~Leonidov and L.~D. McLerran, \emph{{The Color glass condensate: An
  Introduction}},  in \emph{{QCD perspectives on hot and dense matter.
  Proceedings, NATO Advanced Study Institute, Summer School, Cargese, France,
  August 6-18, 2001}}, pp.~73--145.
\newblock \href{http://arxiv.org/abs/hep-ph/0202270}{{\tt hep-ph/0202270}}.

\bibitem{Iancu:2003xm}
E.~Iancu and R.~Venugopalan, \emph{{The Color glass condensate and high-energy
  scattering in QCD}},  in \emph{Quark-gluon plasma 4} (R.~C. Hwa and X.-N.
  Wang, eds.), pp.~249--3363.
\newblock 2003.
\newblock \href{http://arxiv.org/abs/hep-ph/0303204}{{\tt hep-ph/0303204}}.

\bibitem{Wessels:2009ona}
E.~Wessels, \emph{{Signatures of gluon saturation in high energy scattering}}.
\newblock PhD thesis, Vrije U., 2009.

\bibitem{Gelis:2010nm}
F.~Gelis, E.~Iancu, J.~Jalilian-Marian and R.~Venugopalan, \emph{{The Color
  Glass Condensate}},
  \href{http://dx.doi.org/10.1146/annurev.nucl.010909.083629}{\emph{Ann. Rev.
  Nucl. Part. Sci.} {\bf 60} (2010) 463--489},
  [\href{http://arxiv.org/abs/1002.0333}{{\tt 1002.0333}}].

\bibitem{Kovchegov:2012mbw}
Y.~V. Kovchegov and E.~Levin, \emph{{Quantum chromodynamics at high energy}},
  vol.~33.
\newblock Cambridge University Press, 2012.

\bibitem{Taels:2017shj}
P.~Taels, \emph{{Quantum Chromodynamics at small Bjorken-$x$}}.
\newblock PhD thesis, Antwerp U., 2017.
\newblock \href{http://arxiv.org/abs/1711.03928}{{\tt 1711.03928}}.

\bibitem{McLerran:1993ka}
L.~D. McLerran and R.~Venugopalan, \emph{{Gluon distribution functions for very
  large nuclei at small transverse momentum}},
  \href{http://dx.doi.org/10.1103/PhysRevD.49.3352}{\emph{Phys. Rev.} {\bf D49}
  (1994) 3352--3355}, [\href{http://arxiv.org/abs/hep-ph/9311205}{{\tt
  hep-ph/9311205}}].

\bibitem{McLerran:1993ni}
L.~D. McLerran and R.~Venugopalan, \emph{{Computing quark and gluon
  distribution functions for very large nuclei}},
  \href{http://dx.doi.org/10.1103/PhysRevD.49.2233}{\emph{Phys. Rev.} {\bf D49}
  (1994) 2233--2241}, [\href{http://arxiv.org/abs/hep-ph/9309289}{{\tt
  hep-ph/9309289}}].

\bibitem{McLerran:1994vd}
L.~D. McLerran and R.~Venugopalan, \emph{{Green's functions in the color field
  of a large nucleus}},
  \href{http://dx.doi.org/10.1103/PhysRevD.50.2225}{\emph{Phys. Rev.} {\bf D50}
  (1994) 2225--2233}, [\href{http://arxiv.org/abs/hep-ph/9402335}{{\tt
  hep-ph/9402335}}].

\bibitem{Kovchegov:2014kua}
Y.~V. Kovchegov, \emph{{Brief Review of Saturation Physics}},
  \href{http://dx.doi.org/10.5506/APhysPolB.45.2241}{\emph{Acta Phys. Polon.}
  {\bf B45} (2014) 2241--2256}, [\href{http://arxiv.org/abs/1410.7722}{{\tt
  1410.7722}}].

\bibitem{Yang:1954ek}
C.-N. Yang and R.~L. Mills, \emph{{Conservation of Isotopic Spin and Isotopic
  Gauge Invariance}},
  \href{http://dx.doi.org/10.1103/PhysRev.96.191}{\emph{Phys. Rev.} {\bf 96}
  (1954) 191--195}.

\bibitem{Iancu:2000hn}
E.~Iancu, A.~Leonidov and L.~D. McLerran, \emph{{Nonlinear gluon evolution in
  the color glass condensate. 1.}},
  \href{http://dx.doi.org/10.1016/S0375-9474(01)00642-X}{\emph{Nucl. Phys.}
  {\bf A692} (2001) 583--645}, [\href{http://arxiv.org/abs/hep-ph/0011241}{{\tt
  hep-ph/0011241}}].

\bibitem{Gelis:2001da}
F.~Gelis and A.~Peshier, \emph{{Probing colored glass via $q\bar{q}$
  photoproduction}},
  \href{http://dx.doi.org/10.1016/S0375-9474(01)01264-7}{\emph{Nucl. Phys.}
  {\bf A697} (2002) 879--901}, [\href{http://arxiv.org/abs/hep-ph/0107142}{{\tt
  hep-ph/0107142}}].

\bibitem{JalilianMarian:1997jx}
J.~Jalilian-Marian, A.~Kovner, A.~Leonidov and H.~Weigert, \emph{{The BFKL
  equation from the Wilson renormalization group}},
  \href{http://dx.doi.org/10.1016/S0550-3213(97)00440-9}{\emph{Nucl. Phys.}
  {\bf B504} (1997) 415--431}, [\href{http://arxiv.org/abs/hep-ph/9701284}{{\tt
  hep-ph/9701284}}].

\bibitem{JalilianMarian:1997gr}
J.~Jalilian-Marian, A.~Kovner, A.~Leonidov and H.~Weigert, \emph{{The Wilson
  renormalization group for low $x$ physics: Towards the high density regime}},
  \href{http://dx.doi.org/10.1103/PhysRevD.59.014014}{\emph{Phys. Rev.} {\bf
  D59} (1998) 014014}, [\href{http://arxiv.org/abs/hep-ph/9706377}{{\tt
  hep-ph/9706377}}].

\bibitem{JalilianMarian:1997dw}
J.~Jalilian-Marian, A.~Kovner and H.~Weigert, \emph{{The Wilson renormalization
  group for low $x$ physics: Gluon evolution at finite parton density}},
  \href{http://dx.doi.org/10.1103/PhysRevD.59.014015}{\emph{Phys. Rev.} {\bf
  D59} (1998) 014015}, [\href{http://arxiv.org/abs/hep-ph/9709432}{{\tt
  hep-ph/9709432}}].

\bibitem{Weigert:2000gi}
H.~Weigert, \emph{{Unitarity at small Bjorken $x$}},
  \href{http://dx.doi.org/10.1016/S0375-9474(01)01668-2}{\emph{Nucl. Phys.}
  {\bf A703} (2002) 823--860}, [\href{http://arxiv.org/abs/hep-ph/0004044}{{\tt
  hep-ph/0004044}}].

\bibitem{Iancu:2001ad}
E.~Iancu, A.~Leonidov and L.~D. McLerran, \emph{{The Renormalization group
  equation for the color glass condensate}},
  \href{http://dx.doi.org/10.1016/S0370-2693(01)00524-X}{\emph{Phys. Lett.}
  {\bf B510} (2001) 133--144}, [\href{http://arxiv.org/abs/hep-ph/0102009}{{\tt
  hep-ph/0102009}}].

\bibitem{Ferreiro:2001qy}
E.~Ferreiro, E.~Iancu, A.~Leonidov and L.~D. McLerran, \emph{{Nonlinear gluon
  evolution in the color glass condensate. 2.}},
  \href{http://dx.doi.org/10.1016/S0375-9474(01)01329-X}{\emph{Nucl. Phys.}
  {\bf A703} (2002) 489--538}, [\href{http://arxiv.org/abs/hep-ph/0109115}{{\tt
  hep-ph/0109115}}].

\bibitem{Jeon:2005cf}
S.~Jeon and R.~Venugopalan, \emph{{A Classical Odderon in QCD at high
  energies}}, \href{http://dx.doi.org/10.1103/PhysRevD.71.125003}{\emph{Phys.
  Rev.} {\bf D71} (2005) 125003},
  [\href{http://arxiv.org/abs/hep-ph/0503219}{{\tt hep-ph/0503219}}].

\bibitem{Jeon:2004rk}
S.~Jeon and R.~Venugopalan, \emph{{Random walks of partons in $\text{SU}(N_c)$
  and classical representations of color charges in QCD at small $x$}},
  \href{http://dx.doi.org/10.1103/PhysRevD.70.105012}{\emph{Phys. Rev.} {\bf
  D70} (2004) 105012}, [\href{http://arxiv.org/abs/hep-ph/0406169}{{\tt
  hep-ph/0406169}}].

\bibitem{Mueller:1989st}
A.~H. Mueller, \emph{{Small $x$ Behavior and Parton Saturation: A QCD Model}},
  \href{http://dx.doi.org/10.1016/0550-3213(90)90173-B}{\emph{Nucl. Phys.} {\bf
  B335} (1990) 115--137}.

\bibitem{Kovchegov:2012ga}
Y.~V. Kovchegov and M.~D. Sievert, \emph{{A New Mechanism for Generating a
  Single Transverse Spin Asymmetry}},
  \href{http://dx.doi.org/10.1103/PhysRevD.86.034028,
  10.1103/PhysRevD.86.079906}{\emph{Phys. Rev.} {\bf D86} (2012) 034028},
  [\href{http://arxiv.org/abs/1201.5890}{{\tt 1201.5890}}].

\bibitem{Kovchegov:2003dm}
Y.~V. Kovchegov, L.~Szymanowski and S.~Wallon, \emph{{Perturbative odderon in
  the dipole model}},
  \href{http://dx.doi.org/10.1016/j.physletb.2004.02.036}{\emph{Phys. Lett.}
  {\bf B586} (2004) 267--281}, [\href{http://arxiv.org/abs/hep-ph/0309281}{{\tt
  hep-ph/0309281}}].

\bibitem{Abelev:2012ola}
{\scshape ALICE} collaboration, B.~Abelev et~al., \emph{{Long-range angular
  correlations on the near and away side in $p$-$Pb$ collisions at
  $\sqrt{s_{NN}}=5.02$ TeV}},
  \href{http://dx.doi.org/10.1016/j.physletb.2013.01.012}{\emph{Phys. Lett.}
  {\bf B719} (2013) 29--41}, [\href{http://arxiv.org/abs/1212.2001}{{\tt
  1212.2001}}].

\bibitem{ABELEV:2013wsa}
{\scshape ALICE} collaboration, B.~B. Abelev et~al., \emph{{Long-range angular
  correlations of $\pi$, $K$ and $p$ in $p$-$Pb$ collisions at $\sqrt{s_{NN}} =
  5.02$ TeV}},
  \href{http://dx.doi.org/10.1016/j.physletb.2013.08.024}{\emph{Phys. Lett.}
  {\bf B726} (2013) 164--177}, [\href{http://arxiv.org/abs/1307.3237}{{\tt
  1307.3237}}].

\bibitem{Abelev:2014mda}
{\scshape ALICE} collaboration, B.~B. Abelev et~al., \emph{{Multiparticle
  azimuthal correlations in $p$-$Pb$ and $Pb$-$Pb$ collisions at the CERN Large
  Hadron Collider}},
  \href{http://dx.doi.org/10.1103/PhysRevC.90.054901}{\emph{Phys. Rev.} {\bf
  C90} (2014) 054901}, [\href{http://arxiv.org/abs/1406.2474}{{\tt
  1406.2474}}].

\bibitem{Adam:2015bka}
{\scshape ALICE} collaboration, J.~Adam et~al., \emph{{Forward-central
  two-particle correlations in $p$-$Pb$ collisions at $\sqrt{s_{NN}} = 5.02$
  TeV}}, \href{http://dx.doi.org/10.1016/j.physletb.2015.12.010}{\emph{Phys.
  Lett.} {\bf B753} (2016) 126--139},
  [\href{http://arxiv.org/abs/1506.08032}{{\tt 1506.08032}}].

\bibitem{Acharya:2018dxy}
{\scshape ALICE} collaboration, S.~Acharya et~al., \emph{{Azimuthal anisotropy
  of heavy-flavour decay electrons in $p$-$Pb$ collisions at $\sqrt{s_{NN}} =
  5.02$ TeV}},  \href{http://arxiv.org/abs/1805.04367}{{\tt 1805.04367}}.

\bibitem{Aad:2012gla}
{\scshape ATLAS} collaboration, G.~Aad et~al., \emph{{Observation of Associated
  Near-Side and Away-Side Long-Range Correlations in $\sqrt{s_{NN}}=5.02$ TeV
  Proton-Lead Collisions with the ATLAS Detector}},
  \href{http://dx.doi.org/10.1103/PhysRevLett.110.182302}{\emph{Phys. Rev.
  Lett.} {\bf 110} (2013) 182302}, [\href{http://arxiv.org/abs/1212.5198}{{\tt
  1212.5198}}].

\bibitem{Aad:2013fja}
{\scshape ATLAS} collaboration, G.~Aad et~al., \emph{{Measurement with the
  ATLAS detector of multi-particle azimuthal correlations in $p$-$Pb$
  collisions at $\sqrt{s_{NN}} = 5.02$ TeV}},
  \href{http://dx.doi.org/10.1016/j.physletb.2013.06.057}{\emph{Phys. Lett.}
  {\bf B725} (2013) 60--78}, [\href{http://arxiv.org/abs/1303.2084}{{\tt
  1303.2084}}].

\bibitem{Aad:2014lta}
{\scshape ATLAS} collaboration, G.~Aad et~al., \emph{{Measurement of long-range
  pseudorapidity correlations and azimuthal harmonics in $\sqrt{s_{NN}}=5.02$
  TeV proton-lead collisions with the ATLAS detector}},
  \href{http://dx.doi.org/10.1103/PhysRevC.90.044906}{\emph{Phys. Rev.} {\bf
  C90} (2014) 044906}, [\href{http://arxiv.org/abs/1409.1792}{{\tt
  1409.1792}}].

\bibitem{Aad:2015gqa}
{\scshape ATLAS} collaboration, G.~Aad et~al., \emph{{Observation of Long-Range
  Elliptic Azimuthal Anisotropies in $\sqrt{s}= 13$ and $2.76$ TeV $pp$
  Collisions with the ATLAS Detector}},
  \href{http://dx.doi.org/10.1103/PhysRevLett.116.172301}{\emph{Phys. Rev.
  Lett.} {\bf 116} (2016) 172301}, [\href{http://arxiv.org/abs/1509.04776}{{\tt
  1509.04776}}].

\bibitem{Aaboud:2016yar}
{\scshape ATLAS} collaboration, M.~Aaboud et~al., \emph{{Measurements of
  long-range azimuthal anisotropies and associated Fourier coefficients for
  $p$-$p$ collisions at $\sqrt{s}=5.02$ and $13$ TeV and $p$-$Pb$ collisions at
  $\sqrt{s_{\mathrm{NN}}}=5.02$ TeV with the ATLAS detector}},
  \href{http://dx.doi.org/10.1103/PhysRevC.96.024908}{\emph{Phys. Rev.} {\bf
  C96} (2017) 024908}, [\href{http://arxiv.org/abs/1609.06213}{{\tt
  1609.06213}}].

\bibitem{Aaboud:2017acw}
{\scshape ATLAS} collaboration, M.~Aaboud et~al., \emph{{Measurement of
  multi-particle azimuthal correlations in $p$-$p$, $p$-$Pb$ and
  low-multiplicity $Pb$-$Pb$ collisions with the ATLAS detector}},
  \href{http://dx.doi.org/10.1140/epjc/s10052-017-4988-1}{\emph{Eur. Phys. J.}
  {\bf C77} (2017) 428}, [\href{http://arxiv.org/abs/1705.04176}{{\tt
  1705.04176}}].

\bibitem{Aaboud:2017blb}
{\scshape ATLAS} collaboration, M.~Aaboud et~al., \emph{{Measurement of
  long-range multiparticle azimuthal correlations with the subevent cumulant
  method in $p$-$p$ and $p$-$Pb$ collisions with the ATLAS detector at the CERN
  Large Hadron Collider}},
  \href{http://dx.doi.org/10.1103/PhysRevC.97.024904}{\emph{Phys. Rev.} {\bf
  C97} (2018) 024904}, [\href{http://arxiv.org/abs/1708.03559}{{\tt
  1708.03559}}].

\bibitem{Khachatryan:2010gv}
{\scshape CMS} collaboration, V.~Khachatryan et~al., \emph{{Observation of
  Long-Range Near-Side Angular Correlations in Proton-Proton Collisions at the
  LHC}}, \href{http://dx.doi.org/10.1007/JHEP09(2010)091}{\emph{JHEP} {\bf 09}
  (2010) 091}, [\href{http://arxiv.org/abs/1009.4122}{{\tt 1009.4122}}].

\bibitem{CMS:2012qk}
{\scshape CMS} collaboration, S.~Chatrchyan et~al., \emph{{Observation of
  long-range near-side angular correlations in proton-lead collisions at the
  LHC}}, \href{http://dx.doi.org/10.1016/j.physletb.2012.11.025}{\emph{Phys.
  Lett.} {\bf B718} (2013) 795--814},
  [\href{http://arxiv.org/abs/1210.5482}{{\tt 1210.5482}}].

\bibitem{Chatrchyan:2013nka}
{\scshape CMS} collaboration, S.~Chatrchyan et~al., \emph{{Multiplicity and
  transverse momentum dependence of two- and four-particle correlations in
  $p$-$Pb$ and $Pb$-$Pb$ collisions}},
  \href{http://dx.doi.org/10.1016/j.physletb.2013.06.028}{\emph{Phys. Lett.}
  {\bf B724} (2013) 213--240}, [\href{http://arxiv.org/abs/1305.0609}{{\tt
  1305.0609}}].

\bibitem{Khachatryan:2015lva}
{\scshape CMS} collaboration, V.~Khachatryan et~al., \emph{{Measurement of
  long-range near-side two-particle angular correlations in pp collisions at
  $\sqrt s =13$ TeV}},
  \href{http://dx.doi.org/10.1103/PhysRevLett.116.172302}{\emph{Phys. Rev.
  Lett.} {\bf 116} (2016) 172302}, [\href{http://arxiv.org/abs/1510.03068}{{\tt
  1510.03068}}].

\bibitem{Khachatryan:2015oea}
{\scshape CMS} collaboration, V.~Khachatryan et~al., \emph{{Evidence for
  transverse momentum and pseudorapidity dependent event plane fluctuations in
  $Pb$-$Pb$ and $p$-$Pb$ collisions}},
  \href{http://dx.doi.org/10.1103/PhysRevC.92.034911}{\emph{Phys. Rev.} {\bf
  C92} (2015) 034911}, [\href{http://arxiv.org/abs/1503.01692}{{\tt
  1503.01692}}].

\bibitem{Khachatryan:2015waa}
{\scshape CMS} collaboration, V.~Khachatryan et~al., \emph{{Evidence for
  Collective Multiparticle Correlations in $p$-$Pb$ Collisions}},
  \href{http://dx.doi.org/10.1103/PhysRevLett.115.012301}{\emph{Phys. Rev.
  Lett.} {\bf 115} (2015) 012301}, [\href{http://arxiv.org/abs/1502.05382}{{\tt
  1502.05382}}].

\bibitem{Khachatryan:2016txc}
{\scshape CMS} collaboration, V.~Khachatryan et~al., \emph{{Evidence for
  collectivity in $pp$ collisions at the LHC}},
  \href{http://dx.doi.org/10.1016/j.physletb.2016.12.009}{\emph{Phys. Lett.}
  {\bf B765} (2017) 193--220}, [\href{http://arxiv.org/abs/1606.06198}{{\tt
  1606.06198}}].

\bibitem{Sirunyan:2017igb}
{\scshape CMS} collaboration, A.~M. Sirunyan et~al., \emph{{Pseudorapidity and
  transverse momentum dependence of flow harmonics in $p$-$Pb$ and $Pb$-$Pb$
  collisions}}, \href{http://dx.doi.org/10.1103/PhysRevC.98.044902}{\emph{Phys.
  Rev.} {\bf C98} (2018) 044902}, [\href{http://arxiv.org/abs/1710.07864}{{\tt
  1710.07864}}].

\bibitem{Sirunyan:2017uyl}
{\scshape CMS} collaboration, A.~M. Sirunyan et~al., \emph{{Observation of
  Correlated Azimuthal Anisotropy Fourier Harmonics in $p$-$p$ and $p$-$Pb$
  Collisions at the LHC}},
  \href{http://dx.doi.org/10.1103/PhysRevLett.120.092301}{\emph{Phys. Rev.
  Lett.} {\bf 120} (2018) 092301}, [\href{http://arxiv.org/abs/1709.09189}{{\tt
  1709.09189}}].

\bibitem{Aaij:2015qcq}
{\scshape LHCb} collaboration, R.~Aaij et~al., \emph{{Measurements of
  long-range near-side angular correlations in $\sqrt{s_{\text{NN}}}=5$ TeV
  proton-lead collisions in the forward region}},
  \href{http://dx.doi.org/10.1016/j.physletb.2016.09.064}{\emph{Phys. Lett.}
  {\bf B762} (2016) 473--483}, [\href{http://arxiv.org/abs/1512.00439}{{\tt
  1512.00439}}].

\bibitem{Adare:2013piz}
{\scshape PHENIX} collaboration, A.~Adare et~al., \emph{{Quadrupole Anisotropy
  in Dihadron Azimuthal Correlations in Central $d$-$Au$ Collisions at
  $\sqrt{s_{NN}}=200$ GeV}},
  \href{http://dx.doi.org/10.1103/PhysRevLett.111.212301}{\emph{Phys. Rev.
  Lett.} {\bf 111} (2013) 212301}, [\href{http://arxiv.org/abs/1303.1794}{{\tt
  1303.1794}}].

\bibitem{Adare:2014keg}
{\scshape PHENIX} collaboration, A.~Adare et~al., \emph{{Measurement of
  long-range angular correlation and quadrupole anisotropy of pions and
  (anti)protons in central $d$-$Au$ collisions at $\sqrt{s_{NN}}=200$ GeV}},
  \href{http://dx.doi.org/10.1103/PhysRevLett.114.192301}{\emph{Phys. Rev.
  Lett.} {\bf 114} (2015) 192301}, [\href{http://arxiv.org/abs/1404.7461}{{\tt
  1404.7461}}].

\bibitem{Belmont:2017lse}
{\scshape PHENIX} collaboration, R.~Belmont, \emph{{PHENIX results on
  multiparticle correlations in small systems}},
  \href{http://dx.doi.org/10.1016/j.nuclphysa.2017.04.044}{\emph{Nucl. Phys.}
  {\bf A967} (2017) 341--344}, [\href{http://arxiv.org/abs/1704.04570}{{\tt
  1704.04570}}].

\bibitem{Adamczyk:2014fcx}
{\scshape STAR} collaboration, L.~Adamczyk et~al., \emph{{Effect of event
  selection on jetlike correlation measurement in $d$-$Au$ collisions at
  $\sqrt{s_{NN}}=200$ GeV}},
  \href{http://dx.doi.org/10.1016/j.physletb.2015.02.068}{\emph{Phys. Lett.}
  {\bf B743} (2015) 333--339}, [\href{http://arxiv.org/abs/1412.8437}{{\tt
  1412.8437}}].

\bibitem{Adamczyk:2015xjc}
{\scshape STAR} collaboration, L.~Adamczyk et~al., \emph{{Long-range
  pseudorapidity dihadron correlations in $d$-$Au$ collisions at
  $\sqrt{s_{NN}}=200$ GeV}},
  \href{http://dx.doi.org/10.1016/j.physletb.2015.05.075}{\emph{Phys. Lett.}
  {\bf B747} (2015) 265--271}, [\href{http://arxiv.org/abs/1502.07652}{{\tt
  1502.07652}}].

\bibitem{dEnterria:2010xip}
D.~d'Enterria, G.~K. Eyyubova, V.~L. Korotkikh, I.~P. Lokhtin, S.~V.
  Petrushanko, L.~I. Sarycheva et~al., \emph{{Estimates of hadron azimuthal
  anisotropy from multiparton interactions in proton-proton collisions at
  $\sqrt{s}=14$ TeV}},
  \href{http://dx.doi.org/10.1140/epjc/s10052-009-1232-7}{\emph{Eur. Phys. J.}
  {\bf C66} (2010) 173--185}, [\href{http://arxiv.org/abs/0910.3029}{{\tt
  0910.3029}}].

\bibitem{Bozek:2010pb}
P.~Bo{\.z}ek, \emph{{Elliptic flow in proton-proton collisions at $\sqrt{s} =
  7$ TeV}}, \href{http://dx.doi.org/10.1140/epjc/s10052-010-1530-0}{\emph{Eur.
  Phys. J.} {\bf C71} (2011) 1530}, [\href{http://arxiv.org/abs/1010.0405}{{\tt
  1010.0405}}].

\bibitem{Bozek:2012gr}
P.~Bo{\.z}ek and W.~Broniowski, \emph{{Correlations from hydrodynamic flow in
  $p$-$Pb$ collisions}},
  \href{http://dx.doi.org/10.1016/j.physletb.2012.12.051}{\emph{Phys. Lett.}
  {\bf B718} (2013) 1557--1561}, [\href{http://arxiv.org/abs/1211.0845}{{\tt
  1211.0845}}].

\bibitem{Qin:2013bha}
G.-Y. Qin and B.~Müller, \emph{{Elliptic and triangular flow anisotropy in
  deuteron-gold collisions at $\sqrt{s_{NN}}=200$ GeV at RHIC and in
  proton-lead collisions at $\sqrt{s_{NN}}=5.02$ TeV at the LHC}},
  \href{http://dx.doi.org/10.1103/PhysRevC.89.044902}{\emph{Phys. Rev.} {\bf
  C89} (2014) 044902}, [\href{http://arxiv.org/abs/1306.3439}{{\tt
  1306.3439}}].

\bibitem{Werner:2013ipa}
K.~Werner, M.~Bleicher, B.~Guiot, I.~Karpenko and T.~Pierog, \emph{{Evidence
  for Flow from Hydrodynamic Simulations of $p$-$Pb$ Collisions at $5.02$ TeV
  from $\nu_2$ Mass Splitting}},
  \href{http://dx.doi.org/10.1103/PhysRevLett.112.232301}{\emph{Phys. Rev.
  Lett.} {\bf 112} (2014) 232301}, [\href{http://arxiv.org/abs/1307.4379}{{\tt
  1307.4379}}].

\bibitem{Bozek:2013ska}
P.~Bo{\.z}ek, W.~Broniowski and G.~Torrieri, \emph{{Mass hierarchy in
  identified particle distributions in proton-lead collisions}},
  \href{http://dx.doi.org/10.1103/PhysRevLett.111.172303}{\emph{Phys. Rev.
  Lett.} {\bf 111} (2013) 172303}, [\href{http://arxiv.org/abs/1307.5060}{{\tt
  1307.5060}}].

\bibitem{Bozek:2013uha}
P.~Bo{\.z}ek and W.~Broniowski, \emph{{Collective dynamics in high-energy
  proton-nucleus collisions}},
  \href{http://dx.doi.org/10.1103/PhysRevC.88.014903}{\emph{Phys. Rev.} {\bf
  C88} (2013) 014903}, [\href{http://arxiv.org/abs/1304.3044}{{\tt
  1304.3044}}].

\bibitem{Bzdak:2014dia}
A.~Bzdak and G.-L. Ma, \emph{{Elliptic and triangular flow in $p$-$Pb$ and
  peripheral $Pb$-$Pb$ collisions from parton scatterings}},
  \href{http://dx.doi.org/10.1103/PhysRevLett.113.252301}{\emph{Phys. Rev.
  Lett.} {\bf 113} (2014) 252301}, [\href{http://arxiv.org/abs/1406.2804}{{\tt
  1406.2804}}].

\bibitem{Weller:2017tsr}
R.~D. Weller and P.~Romatschke, \emph{{One fluid to rule them all: viscous
  hydrodynamic description of event-by-event central $p$-$p$, $p$-$Pb$ and
  $Pb$-$Pb$ collisions at $\sqrt{s}=5.02$ TeV}},
  \href{http://dx.doi.org/10.1016/j.physletb.2017.09.077}{\emph{Phys. Lett.}
  {\bf B774} (2017) 351--356}, [\href{http://arxiv.org/abs/1701.07145}{{\tt
  1701.07145}}].

\bibitem{Dusling:2015gta}
K.~Dusling, W.~Li and B.~Schenke, \emph{{Novel collective phenomena in
  high-energy proton-proton and proton-nucleus collisions}},
  \href{http://dx.doi.org/10.1142/S0218301316300022}{\emph{Int. J. Mod. Phys.}
  {\bf E25} (2016) 1630002}, [\href{http://arxiv.org/abs/1509.07939}{{\tt
  1509.07939}}].

\bibitem{Avsar:2010rf}
E.~Avsar, C.~Flensburg, Y.~Hatta, J.-Y. Ollitrault and T.~Ueda,
  \emph{{Eccentricity and elliptic flow in proton-proton collisions from parton
  evolution}},
  \href{http://dx.doi.org/10.1016/j.physletb.2011.07.031}{\emph{Phys. Lett.}
  {\bf B702} (2011) 394--397}, [\href{http://arxiv.org/abs/1009.5643}{{\tt
  1009.5643}}].

\bibitem{Dumitru:2010iy}
A.~Dumitru, K.~Dusling, F.~Gelis, J.~Jalilian-Marian, T.~Lappi and
  R.~Venugopalan, \emph{{The Ridge in proton-proton collisions at the LHC}},
  \href{http://dx.doi.org/10.1016/j.physletb.2011.01.024}{\emph{Phys. Lett.}
  {\bf B697} (2011) 21--25}, [\href{http://arxiv.org/abs/1009.5295}{{\tt
  1009.5295}}].

\bibitem{Kovner:2010xk}
A.~Kovner and M.~Lublinsky, \emph{{Angular Correlations in Gluon Production at
  High Energy}},
  \href{http://dx.doi.org/10.1103/PhysRevD.83.034017}{\emph{Phys. Rev.} {\bf
  D83} (2011) 034017}, [\href{http://arxiv.org/abs/1012.3398}{{\tt
  1012.3398}}].

\bibitem{Levin:2011fb}
E.~Levin and A.~H. Rezaeian, \emph{{The Ridge from the BFKL evolution and
  beyond}}, \href{http://dx.doi.org/10.1103/PhysRevD.84.034031}{\emph{Phys.
  Rev.} {\bf D84} (2011) 034031}, [\href{http://arxiv.org/abs/1105.3275}{{\tt
  1105.3275}}].

\bibitem{Iancu:2011ns}
E.~Iancu and D.~N. Triantafyllopoulos, \emph{{Higher-point correlations from
  the JIMWLK evolution}},
  \href{http://dx.doi.org/10.1007/JHEP11(2011)105}{\emph{JHEP} {\bf 11} (2011)
  105}, [\href{http://arxiv.org/abs/1109.0302}{{\tt 1109.0302}}].

\bibitem{Kovner:2011pe}
A.~Kovner and M.~Lublinsky, \emph{{On Angular Correlations and High Energy
  Evolution}}, \href{http://dx.doi.org/10.1103/PhysRevD.84.094011}{\emph{Phys.
  Rev.} {\bf D84} (2011) 094011}, [\href{http://arxiv.org/abs/1109.0347}{{\tt
  1109.0347}}].

\bibitem{Dusling:2012iga}
K.~Dusling and R.~Venugopalan, \emph{{Azimuthal collimation of long range
  rapidity correlations by strong color fields in high multiplicity
  hadron-hadron collisions}},
  \href{http://dx.doi.org/10.1103/PhysRevLett.108.262001}{\emph{Phys. Rev.
  Lett.} {\bf 108} (2012) 262001}, [\href{http://arxiv.org/abs/1201.2658}{{\tt
  1201.2658}}].

\bibitem{Kovner:2012jm}
A.~Kovner and M.~Lublinsky, \emph{{Angular and long range rapidity correlations
  in particle production at high energy}},
  \href{http://dx.doi.org/10.1142/S0218301313300014}{\emph{Int. J. Mod. Phys.}
  {\bf E22} (2013) 1330001}, [\href{http://arxiv.org/abs/1211.1928}{{\tt
  1211.1928}}].

\bibitem{Dusling:2013qoz}
K.~Dusling and R.~Venugopalan, \emph{{Comparison of the color glass condensate
  to dihadron correlations in proton-proton and proton-nucleus collisions}},
  \href{http://dx.doi.org/10.1103/PhysRevD.87.094034}{\emph{Phys. Rev.} {\bf
  D87} (2013) 094034}, [\href{http://arxiv.org/abs/1302.7018}{{\tt
  1302.7018}}].

\bibitem{Kovner:2014qea}
A.~Kovner and A.~H. Rezaeian, \emph{{Diphoton production in high-energy $pA$
  collisions}}, \href{http://dx.doi.org/10.1103/PhysRevD.90.014031}{\emph{Phys.
  Rev.} {\bf D90} (2014) 014031}, [\href{http://arxiv.org/abs/1404.5632}{{\tt
  1404.5632}}].

\bibitem{Dumitru:2014yza}
A.~Dumitru, L.~D. McLerran and V.~Skokov, \emph{{Azimuthal asymmetries and the
  emergence of `collectivity' from multi-particle correlations in high-energy
  $pA$ collisions}},
  \href{http://dx.doi.org/10.1016/j.physletb.2015.02.046}{\emph{Phys. Lett.}
  {\bf B743} (2015) 134--137}, [\href{http://arxiv.org/abs/1410.4844}{{\tt
  1410.4844}}].

\bibitem{Schenke:2015aqa}
B.~Schenke, S.~Schlichting and R.~Venugopalan, \emph{{Azimuthal anisotropies in
  $p$-$Pb$ collisions from classical Yang-Mills dynamics}},
  \href{http://dx.doi.org/10.1016/j.physletb.2015.05.051}{\emph{Phys. Lett.}
  {\bf B747} (2015) 76--82}, [\href{http://arxiv.org/abs/1502.01331}{{\tt
  1502.01331}}].

\bibitem{Altinoluk:2015dpi}
T.~Altinoluk, N.~Armesto, G.~Beuf and A.~H. Rezaeian, \emph{{Diffractive Dijet
  Production in Deep Inelastic Scattering and Photon-Hadron Collisions in the
  Color Glass Condensate}},
  \href{http://dx.doi.org/10.1016/j.physletb.2016.05.032}{\emph{Phys. Lett.}
  {\bf B758} (2016) 373--383}, [\href{http://arxiv.org/abs/1511.07452}{{\tt
  1511.07452}}].

\bibitem{Dusling:2015rja}
K.~Dusling, P.~Tribedy and R.~Venugopalan, \emph{{Energy dependence of the
  ridge in high multiplicity proton-proton collisions}},
  \href{http://dx.doi.org/10.1103/PhysRevD.93.014034}{\emph{Phys. Rev.} {\bf
  D93} (2016) 014034}, [\href{http://arxiv.org/abs/1509.04410}{{\tt
  1509.04410}}].

\bibitem{Kovner:2015rna}
A.~Kovner and A.~H. Rezaeian, \emph{{Diphoton ridge in $pp$ and $pA$ collisions
  at RHIC and the LHC}},
  \href{http://dx.doi.org/10.1103/PhysRevD.92.074045}{\emph{Phys. Rev.} {\bf
  D92} (2015) 074045}, [\href{http://arxiv.org/abs/1508.02412}{{\tt
  1508.02412}}].

\bibitem{Schenke:2016lrs}
B.~Schenke, S.~Schlichting, P.~Tribedy and R.~Venugopalan, \emph{{Mass ordering
  of spectra from fragmentation of saturated gluon states in high multiplicity
  proton-proton collisions}},
  \href{http://dx.doi.org/10.1103/PhysRevLett.117.162301}{\emph{Phys. Rev.
  Lett.} {\bf 117} (2016) 162301}, [\href{http://arxiv.org/abs/1607.02496}{{\tt
  1607.02496}}].

\bibitem{Rezaeian:2016szi}
A.~H. Rezaeian, \emph{{Photon-jet ridge at RHIC and the LHC}},
  \href{http://dx.doi.org/10.1103/PhysRevD.93.094030}{\emph{Phys. Rev.} {\bf
  D93} (2016) 094030}, [\href{http://arxiv.org/abs/1603.07354}{{\tt
  1603.07354}}].

\bibitem{Dusling:2017aot}
K.~Dusling, M.~Mace and R.~Venugopalan, \emph{{Parton model description of
  multiparticle azimuthal correlations in $pA$ collisions}},
  \href{http://dx.doi.org/10.1103/PhysRevD.97.016014}{\emph{Phys. Rev.} {\bf
  D97} (2018) 016014}, [\href{http://arxiv.org/abs/1706.06260}{{\tt
  1706.06260}}].

\bibitem{Dusling:2017dqg}
K.~Dusling, M.~Mace and R.~Venugopalan, \emph{{Multiparticle collectivity from
  initial state correlations in high energy proton-nucleus collisions}},
  \href{http://dx.doi.org/10.1103/PhysRevLett.120.042002}{\emph{Phys. Rev.
  Lett.} {\bf 120} (2018) 042002}, [\href{http://arxiv.org/abs/1705.00745}{{\tt
  1705.00745}}].

\bibitem{Fukushima:2017mko}
K.~Fukushima and Y.~Hidaka, \emph{{General formulae for dipole Wilson line
  correlators with the Color Glass Condensate}},
  \href{http://dx.doi.org/10.1007/JHEP11(2017)114}{\emph{JHEP} {\bf 11} (2017)
  114}, [\href{http://arxiv.org/abs/1708.03051}{{\tt 1708.03051}}].

\bibitem{Altinoluk:2018hcu}
T.~Altinoluk, N.~Armesto and D.~E. Wertepny, \emph{{Correlations and the ridge
  in the Color Glass Condensate beyond the glasma graph approximation}},
  \href{http://dx.doi.org/10.1007/JHEP05(2018)207}{\emph{JHEP} {\bf 05} (2018)
  207}, [\href{http://arxiv.org/abs/1804.02910}{{\tt 1804.02910}}].

\bibitem{Davy:2018hsl}
M.~K. Davy, C.~Marquet, Y.~Shi, B.-W. Xiao and C.~Zhang, \emph{{Two particle
  azimuthal harmonics in $pA$ collisions}},
  \href{http://arxiv.org/abs/1808.09851}{{\tt 1808.09851}}.

\bibitem{Bzdak:2013zma}
A.~Bzdak, B.~Schenke, P.~Tribedy and R.~Venugopalan, \emph{{Initial state
  geometry and the role of hydrodynamics in proton-proton, proton-nucleus and
  deuteron-nucleus collisions}},
  \href{http://dx.doi.org/10.1103/PhysRevC.87.064906}{\emph{Phys. Rev.} {\bf
  C87} (2013) 064906}, [\href{http://arxiv.org/abs/1304.3403}{{\tt
  1304.3403}}].

\bibitem{Dumitru:2014dra}
A.~Dumitru and A.~V. Giannini, \emph{{Initial state angular asymmetries in high
  energy $pA$ collisions: spontaneous breaking of rotational symmetry by a
  color electric field and C-odd fluctuations}},
  \href{http://dx.doi.org/10.1016/j.nuclphysa.2014.10.037}{\emph{Nucl. Phys.}
  {\bf A933} (2015) 212--228}, [\href{http://arxiv.org/abs/1406.5781}{{\tt
  1406.5781}}].

\bibitem{Dumitru:2014vka}
A.~Dumitru and V.~Skokov, \emph{{Anisotropy of the semiclassical gluon field of
  a large nucleus at high energy}},
  \href{http://dx.doi.org/10.1103/PhysRevD.91.074006}{\emph{Phys. Rev.} {\bf
  D91} (2015) 074006}, [\href{http://arxiv.org/abs/1411.6630}{{\tt
  1411.6630}}].

\bibitem{Lappi:2015vha}
T.~Lappi, \emph{{Azimuthal harmonics of color fields in a high energy
  nucleus}},
  \href{http://dx.doi.org/10.1016/j.physletb.2015.04.015}{\emph{Phys. Lett.}
  {\bf B744} (2015) 315--319}, [\href{http://arxiv.org/abs/1501.05505}{{\tt
  1501.05505}}].

\bibitem{Lappi:2015vta}
T.~Lappi, B.~Schenke, S.~Schlichting and R.~Venugopalan, \emph{{Tracing the
  origin of azimuthal gluon correlations in the color glass condensate}},
  \href{http://dx.doi.org/10.1007/JHEP01(2016)061}{\emph{JHEP} {\bf 01} (2016)
  061}, [\href{http://arxiv.org/abs/1509.03499}{{\tt 1509.03499}}].

\bibitem{Kovner:2016jfp}
A.~Kovner, M.~Lublinsky and V.~Skokov, \emph{{Exploring correlations in the CGC
  wave function: odd azimuthal anisotropy}},
  \href{http://dx.doi.org/10.1103/PhysRevD.96.016010}{\emph{Phys. Rev.} {\bf
  D96} (2017) 016010}, [\href{http://arxiv.org/abs/1612.07790}{{\tt
  1612.07790}}].

\bibitem{Kovchegov:2018jun}
Y.~V. Kovchegov and V.~V. Skokov, \emph{{How classical gluon fields generate
  odd azimuthal harmonics for the two-gluon correlation function in high-energy
  collisions}}, \href{http://dx.doi.org/10.1103/PhysRevD.97.094021}{\emph{Phys.
  Rev.} {\bf D97} (2018) 094021}, [\href{http://arxiv.org/abs/1802.08166}{{\tt
  1802.08166}}].

\bibitem{Hagiwara:2017ofm}
Y.~Hagiwara, Y.~Hatta, B.-W. Xiao and F.~Yuan, \emph{{Elliptic Flow in Small
  Systems due to Elliptic Gluon Distributions?}},
  \href{http://dx.doi.org/10.1016/j.physletb.2017.05.083}{\emph{Phys. Lett.}
  {\bf B771} (2017) 374--378}, [\href{http://arxiv.org/abs/1701.04254}{{\tt
  1701.04254}}].

\bibitem{Iancu:2017fzn}
E.~Iancu and A.~H. Rezaeian, \emph{{Elliptic flow from color-dipole orientation
  in $pp$ and $pA$ collisions}},
  \href{http://dx.doi.org/10.1103/PhysRevD.95.094003}{\emph{Phys. Rev.} {\bf
  D95} (2017) 094003}, [\href{http://arxiv.org/abs/1702.03943}{{\tt
  1702.03943}}].

\bibitem{Petreska:2018cbf}
E.~Petreska, \emph{{TMD gluon distributions at small $x$ in the CGC theory}},
  \href{http://dx.doi.org/10.1142/S0218301318300035}{\emph{Int. J. Mod. Phys.}
  {\bf E27} (2018) 1830003}, [\href{http://arxiv.org/abs/1804.04981}{{\tt
  1804.04981}}].

\bibitem{Strikman:2010bg}
M.~Strikman and W.~Vogelsang, \emph{{Multiple parton interactions and forward
  double pion production in $pp$ and $dA$ scattering}},
  \href{http://dx.doi.org/10.1103/PhysRevD.83.034029}{\emph{Phys. Rev.} {\bf
  D83} (2011) 034029}, [\href{http://arxiv.org/abs/1009.6123}{{\tt
  1009.6123}}].

\bibitem{Stasto:2011ru}
A.~Stasto, B.-W. Xiao and F.~Yuan, \emph{{Back-to-Back Correlations of
  Di-hadrons in $d$-$Au$ Collisions at RHIC}},
  \href{http://dx.doi.org/10.1016/j.physletb.2012.08.044}{\emph{Phys. Lett.}
  {\bf B716} (2012) 430--434}, [\href{http://arxiv.org/abs/1109.1817}{{\tt
  1109.1817}}].

\bibitem{Lappi:2012nh}
T.~Lappi and H.~Mäntysaari, \emph{{Forward dihadron correlations in
  deuteron-gold collisions with the Gaussian approximation of JIMWLK}},
  \href{http://dx.doi.org/10.1016/j.nuclphysa.2013.03.017}{\emph{Nucl. Phys.}
  {\bf A908} (2013) 51--72}, [\href{http://arxiv.org/abs/1209.2853}{{\tt
  1209.2853}}].

\bibitem{Diehl:2017wew}
M.~Diehl and J.~R. Gaunt, \emph{{Double parton scattering theory overview}},
  \href{http://arxiv.org/abs/1710.04408}{{\tt 1710.04408}}.

\bibitem{Dumitru:2005gt}
A.~Dumitru, A.~Hayashigaki and J.~Jalilian-Marian, \emph{{The Color glass
  condensate and hadron production in the forward region}},
  \href{http://dx.doi.org/10.1016/j.nuclphysa.2005.11.014}{\emph{Nucl. Phys.}
  {\bf A765} (2006) 464--482}, [\href{http://arxiv.org/abs/hep-ph/0506308}{{\tt
  hep-ph/0506308}}].

\bibitem{Hagiwara:2016kam}
Y.~Hagiwara, Y.~Hatta and T.~Ueda, \emph{{Wigner, Husimi, and generalized
  transverse momentum dependent distributions in the color glass condensate}},
  \href{http://dx.doi.org/10.1103/PhysRevD.94.094036}{\emph{Phys. Rev.} {\bf
  D94} (2016) 094036}, [\href{http://arxiv.org/abs/1609.05773}{{\tt
  1609.05773}}].

\bibitem{Zhou:2016rnt}
J.~Zhou, \emph{{Elliptic gluon generalized transverse-momentum-dependent
  distribution inside a large nucleus}},
  \href{http://dx.doi.org/10.1103/PhysRevD.94.114017}{\emph{Phys. Rev.} {\bf
  D94} (2016) 114017}, [\href{http://arxiv.org/abs/1611.02397}{{\tt
  1611.02397}}].

\bibitem{Hatta:2017cte}
Y.~Hatta, B.-W. Xiao and F.~Yuan, \emph{{Gluon Tomography from Deeply Virtual
  Compton Scattering at Small-$x$}},
  \href{http://dx.doi.org/10.1103/PhysRevD.95.114026}{\emph{Phys. Rev.} {\bf
  D95} (2017) 114026}, [\href{http://arxiv.org/abs/1703.02085}{{\tt
  1703.02085}}].

\bibitem{Hagiwara:2017fye}
Y.~Hagiwara, Y.~Hatta, R.~Pasechnik, M.~Tasevsky and O.~Teryaev,
  \emph{{Accessing the gluon Wigner distribution in ultraperipheral $pA$
  collisions}}, \href{http://dx.doi.org/10.1103/PhysRevD.96.034009}{\emph{Phys.
  Rev.} {\bf D96} (2017) 034009}, [\href{http://arxiv.org/abs/1706.01765}{{\tt
  1706.01765}}].

\bibitem{Blok:2010ge}
B.~Blok, {\relax Yu}.~Dokshitzer, L.~Frankfurt and M.~Strikman, \emph{{The Four
  jet production at LHC and Tevatron in QCD}},
  \href{http://dx.doi.org/10.1103/PhysRevD.83.071501}{\emph{Phys. Rev.} {\bf
  D83} (2011) 071501}, [\href{http://arxiv.org/abs/1009.2714}{{\tt
  1009.2714}}].

\bibitem{Voloshin:1994mz}
S.~Voloshin and Y.~Zhang, \emph{{Flow study in relativistic nuclear collisions
  by Fourier expansion of Azimuthal particle distributions}},
  \href{http://dx.doi.org/10.1007/s002880050141}{\emph{Z. Phys.} {\bf C70}
  (1996) 665--672}, [\href{http://arxiv.org/abs/hep-ph/9407282}{{\tt
  hep-ph/9407282}}].

\bibitem{Woods:1954zz}
R.~D. Woods and D.~S. Saxon, \emph{{Diffuse Surface Optical Model for
  Nucleon-Nuclei Scattering}},
  \href{http://dx.doi.org/10.1103/PhysRev.95.577}{\emph{Phys. Rev.} {\bf 95}
  (1954) 577--578}.

\bibitem{Diehl:2004cx}
M.~Diehl, T.~Feldmann, R.~Jakob and P.~Kroll, \emph{{Generalized parton
  distributions from nucleon form-factor data}},
  \href{http://dx.doi.org/10.1140/epjc/s2004-02063-4}{\emph{Eur. Phys. J.} {\bf
  C39} (2005) 1--39}, [\href{http://arxiv.org/abs/hep-ph/0408173}{{\tt
  hep-ph/0408173}}].

\bibitem{Diehl:2014vaa}
M.~Diehl, T.~Kasemets and S.~Keane, \emph{{Correlations in double parton
  distributions: effects of evolution}},
  \href{http://dx.doi.org/10.1007/JHEP05(2014)118}{\emph{JHEP} {\bf 05} (2014)
  118}, [\href{http://arxiv.org/abs/1401.1233}{{\tt 1401.1233}}].

\bibitem{Belitsky:1997ay}
A.~V. Belitsky, \emph{{Leading order analysis of the twist-three space- and
  time-like cut vertices in QCD}},  in \emph{{Annual 31st PNPI Winter School on
  Nuclear and Particle Physics St. Petersburg, Russia, February 24-March 2,
  1997}}.
\newblock \href{http://arxiv.org/abs/hep-ph/9703432}{{\tt hep-ph/9703432}}.

\bibitem{Peskin:1995ev}
M.~E. Peskin and D.~V. Schroeder, \emph{{An Introduction to quantum field
  theory}}.
\newblock Addison-Wesley, Reading, USA, 1995.

\bibitem{Romao:2012pq}
J.~C. Rom{\~ a}o and J.~P. Silva, \emph{{A resource for signs and Feynman
  diagrams of the Standard Model}},
  \href{http://dx.doi.org/10.1142/S0217751X12300256}{\emph{Int. J. Mod. Phys.}
  {\bf A27} (2012) 1230025}, [\href{http://arxiv.org/abs/1209.6213}{{\tt
  1209.6213}}].

\bibitem{Boer:2011xd}
D.~Boer, L.~Gamberg, B.~Musch and A.~Prokudin, \emph{{Bessel-Weighted
  Asymmetries in Semi Inclusive Deep Inelastic Scattering}},
  \href{http://dx.doi.org/10.1007/JHEP10(2011)021}{\emph{JHEP} {\bf 10} (2011)
  021}, [\href{http://arxiv.org/abs/1107.5294}{{\tt 1107.5294}}].

\end{thebibliography}\endgroup

\end{fmffile} 
\end{document}